\documentclass[twocolumn,tighten]{aastex6}
\usepackage{apjfonts}
\DeclareGraphicsExtensions{.png,.pdf,.jpg,.ps}

\newcommand{\rstar}{\ensuremath{\,{R}_{*}}}
\newcommand{\teff}{\ensuremath{T_{\rm eff}}}
\newcommand{\fbol}{\ensuremath{F_{\rm bol}}}

\newcommand{\feh}{{\rm [Fe/H]}}

\newcommand{\loggstar}{\ensuremath{\log{g_\star}}}
\newcommand{\msun}{\ensuremath{\,{M}_\Sun}}
\newcommand{\rsun}{\ensuremath{\,{R}_\Sun}}

\newcommand{\avmax}{\ensuremath{\,A_{V,\rm max}}}
\usepackage{color}
\usepackage[normalem]{ulem}

\begin{document}

\title{Empirical Bolometric Fluxes and Angular Diameters of 1.6 Million Tycho-2 Stars and Radii of 350,000 Stars with \emph{Gaia} DR1 Parallaxes}
\author{Daniel J.\ Stevens\altaffilmark{1}, Keivan G.\ Stassun\altaffilmark{2,3}, B.\ Scott Gaudi\altaffilmark{1,4}}

\altaffiltext{1}{The Ohio State University, Department of Astronomy, 140 W. 18th Ave., Columbus, OH 43210, USA}

\altaffiltext{2}{Vanderbilt University, Department of Physics \& Astronomy, 6301 Stevenson Center Ln., Nashville, TN 37235, USA}

\altaffiltext{3}{Fisk University, Department of Physics, 1000 17th Ave.\ N., Nashville, TN 37208, USA}

\altaffiltext{4}{Jet Propulsion Laboratory, California Institute of Technology, 4800 Oak Grove Dr., Pasadena, CA 91109, USA}

\begin{abstract}
We present bolometric fluxes and angular diameters for over 1.6 million stars in the {\it Tycho-2} catalog, determined using previously-determined empirical color-temperature and color-flux relations. We vet these relations via full fits to the full broadband spectral energy distributions for a subset of benchmark stars, and perform quality checks against the large set of stars for which spectroscopically-determined parameters are available from LAMOST, RAVE, and/or APOGEE. We then estimate radii for the 355,502 {\it Tycho-2} stars in our sample whose \emph{Gaia} DR1 parallaxes are precise to $\lesssim$10\%. 
For the 64,960 of these stars with external spectroscopic information, we achieve median uncertainties on the effective temperatures, bolometric fluxes, angular diameters, and radii of $1.0\%$, $1.4\%$, $2.4\%$, and $7.5\%$, respectively. For the 290,542 remaining stars, we achieve median uncertainties of $0.9\%$, $1.3\%$, $2.2\%$, and $7.5\%$, respectively. These stellar parameters are shown to be reliable for stars with \teff$\lesssim$7000~K. The over half a million bolometric fluxes and angular diameters presented here will serve as an immediate trove of empirical stellar radii with the {\it Gaia\/} second data release, at which point effective temperature uncertainties will dominate the radius uncertainties. Already, dwarf, subgiant, and giant populations are readily identifiable in our purely empirical luminosity-effective temperature (theoretical) Hertzsprung-Russell diagrams.
\end{abstract}

\section{Introduction}\label{sec:intro}
Measurements of stellar radii are paramount to our understanding of stellar evolution. Different physical prescriptions in stellar evolution models for, e.g., winds, mass-loss, and convective overshoot, predict different masses and radii for stars of the same mass, age, and metallicity.  Similarly, stars with different elemental abundance ratios will have significantly different evolutionary paths in the luminosity-effective temperature plane, even if they have the same mass and overall metal abundance.   Thus, placing precise constraints on these parameters are our only way of constraining the wide range of plausible stellar evolution models. 

For example, in sparsely populated areas of the Hertzsprung-Russel (HR) diagram -- e.g., the Hertzsprung gap, wherein massive ($M_{\rm ZAMS} \gtrsim 1.5 \msun$) stars have ceased core hydrogen fusion but have not yet ignited hydrogen fusion in their shells -- stellar evolution models are poorly constrained.  Thus improving the precision with which we measure the fundamental parameters of the few stars in these regimes provides us with the most promising way of constraining short-lived phases of stellar evolution. 

To date, double-lined eclipsing binaries and stars with angular radii measured interferometrically and distances measured by parallax provide the most robustly determined model-independent stellar radii. The canonical \citet{Torres2010} sample contains double-lined eclipsing binaries (and $\alpha$ Centauri A and B) with masses and radii determined to $<3\%$, but their sample contains only four M dwarfs. \citet{Birkby2012} lists a few dozen M dwarfs in eclipsing binaries or with radii known from interferometry, but the radius uncertainty for this sample is as large as $6.4\%$. Interferometry provides radii (via angular diameters) to $\sim 1.5\%$ for AFG stars \citep{Boyajian2012a} and $\sim 5\%$ for K and M dwarfs \citep{Boyajian2012b}, but this technique is limited to very bright (and thus nearby) stars.  Among young, low-mass pre-main-sequence stars, there is a severe paucity of benchmark-quality eclipsing binaries, limiting empirical tests of star formation and evolution models \citep[e.g.,][]{Stassun:2014}. 

This paucity of precise radius measurements for isolated low-mass stars hinders our ability to make progress on several long-standing puzzles.  For example, there is strong evidence that magnetic activity affects the structure of low-mass stars. Measured K and M dwarf radii have been shown to exceed model-predicted radii at fixed \teff\ by 10-15\% (cf. \citealt{Mann2015,Birkby2012}). This  ``radius inflation" problem of K- and M-dwarfs has yet to be fully captured in stellar models \citep[see, e.g.,][and references therein]{Stassun:2012,Somers:2016}. 

The solution to the aforementioned problems is twofold: we must increase the sample of stars for which radii are emprically measured, and we must improve the precision and accuracy of these radius measurements. This requires precise parallaxes; \emph{Gaia} DR1 \citep{Gaia_main,GaiaDR1} provides astrometry at roughly the precision of \emph{Hipparcos}, including parallaxes, but for about two million bright stars \citep{gaia_astrometry}, as compared to the roughly 100,000 stars in the Hipparos catalog \citep{Perryman:1997}. To obtain similarly precise radii, we must know the effective temperatures and bolometric fluxes to high precision as well; moreover, we now also need to know the \emph{extinction} to high precision, since assuming zero extinction for stars outside the immediate solar neighborhood introduces an uncertainty that is now non-negligible in comparison to the uncertainties in the other quantities.

A methodology for determining empirical radii of stars has been demonstrated by \citet{StassunGaiaPlanets:2016} for some 500 planet-host stars, in which empirical measurements of stellar bolometric fluxes and temperatures permitted determination of accurate, empirical angular diameters, which with the {\it Gaia\/} DR1 parallaxes permitted accurate and empirical measurement of the stellar radii. Extending this approach to a much larger sample of stars across the entire sky would be of great value in particular for improving the selection of promising targets for the upcoming {\it TESS\/} \citep{Ricker:2015} and {\it PLATO\/} \citep{Rauer:2014} missions, which rely on accurate estimates of stellar radii and other parameters to optimize the target samples for finding small transiting planets \citep[see][]{StassunTESS:2014,Campante:2016,StassunTIC:2017}.

So motivated, we present estimated extinctions, effective temperatures, bolometric fluxes (and thus angular diameters), for over 1.6 million \emph{Tycho-2} stars. We also present radii for 355,502 of these stars that have {\it Gaia\/} DR1 parallaxes with reported precisions of less than 10\%. In Section \ref{sec:meas}, we describe the broad-band flux measurements we use to derive the temperatures and bolometric fluxes as well as the spectroscopic parameters that we adopt as priors. In Section \ref{sec:method}, we describe the iterative method for de-reddening the literature magnitude measurements and obtaining effective temperatures and bolometric fluxes, and we furthermore validate the method against a large subset of stars with spectroscopic parameter determinations. As the fundamental products of this work, we present our bolometric fluxes, as well as angular and physical radii, in Section \ref{sec:results}. In Section \ref{sec:disc}, we compare our radii and effective temperatures to the \teff-radius relation of \citet{Boyajian2012b} and note the limitations of our technique. Finally, we conclude with a summary of our results and the expected improvements that will be possible with future \emph{Gaia} data releases in Section \ref{sec:conclusion}.

\section{Data}\label{sec:data}

We begin with the {\it Tycho-2} catalog as our base sample, providing $\sim$2 million stars with $B_T$ and  $V_T$ magnitudes as well as astrometry that the {\it Gaia} team has used to provide parallaxes in its first data release (DR1). In this section, we describe the additional catalogs and other literature sources that we combine with the {\it Tycho-2} base catalog to form our study sample. A summary of the parent sample and the various subsamples employed in this work is provided in Table \ref{tab:summary}.

\subsection{Literature Photometry}\label{sec:meas}
    We queried VizieR \citep{Ochsenbein2000} for photometry in the $U$, $B$ and $V$ bands from \citet{Mermilliod:2006}; in Str\"{o}mgren $b$ and $y$ from \citet{Paunzen2015}; in $B_T$ and $V_T$ from the Tycho-2 catalog \citep{Hog2000}; and in $J$, $H$, and $K_S$ from the Two-Micron All Sky Survey (2MASS; \citealt{Skrutskie2006}, \citealt{Cutri2003}). We queried by Tycho-2 ID, taking the closest match that lied within 0.1\arcmin\ of the VizieR-calculated sky position. We excluded stars Tycho stars that appear in the original Tycho catalog but not in the Tycho-2 catalog, and we exclude photometric measurements for which no uncertainties are listed. If a quoted uncertainty is less than 0.01 mag, then we inflate the uncertainty to 0.01 mag to be conservative. We thus have an initial catalog of 2,539,914 stars with at least one flux measurement.
    
    \subsection{Literature Spectroscopic Parameters}
    We also queried VizieR for spectroscopically determined effective temperatures \teff\ and metallicities to use as priors in our analysis. We began by searching for matches from the first LAMOST data release catalog \citep{Luo2015}, choosing the source with the closest match within 0.1\arcmin. If no LAMOST match was found, we then searched the fourth RAVE data release catalog \citep{Kordopatis2013}. If a given RAVE match had multiple sets of spectroscopic parameters, then we adopted the parameters with the largest uncertainties 
    (to avoid over-constraining our fits with possibly unrealistically small catalog uncertainties). 
    If no RAVE match was found, we then searched the Apache Point Observatory Galactic Evolution Experiment (APOGEE) 13th data release \citep{Blanton2017} from the Sloan Digital Sky Survey \citep{SDSS2016}, including the post-hoc correction\footnote{\url{http://www.sdss.org/dr13/irspec/parameters/}} to effective temperatures when available, and use the initial pipeline effective temperature otherwise (Holtzman et al., in preparation). Finally, if no SDSS spectroscopic [Fe/H] was found, then we assume the star has the median \feh\ of the measured distribution of late-type, solar neighborhood stars from the Geneva-Copenhagen Survey \citep{Casagrande2011}, and an `uncertainty' equal to the 1-$\sigma$ dispersion in this distribution; specifically, we adopt a metallicity \feh = -0.05 and uncertainty $\sigma_{\feh} = 0.22$.
    
    Of our initial 2,539,913 stars, 74,515 have LAMOST DR1 temperatures and metallicities; an additional 239,017 have RAVE DR4 temperatures and metallicities; an additional 24,029 have APOGEE temperatures and metallicities; and the remaining 2,202,352 do not have temperatures and metallicities from the aforementioned datasets. We do not exclude giants \emph{a priori}, nor do we identify blends, binaries, or multiple stellar systems. 
    Our methods of estimating the temperature and bolometric flux (presented below) were successful for 1,600,080 stars; we list the photometry and spectroscopic parameters we used in Table \ref{tab:allphot}.  
    For the remainder of our analyses, we will separately consider these four subgroups---those with spectroscopic \teff\ from (1) LAMOST, (2) RAVE, (3) APOGEE, and (4) those without spectroscopic \teff---where appropriate.

\subsection{{\it Gaia\/} DR1 Parallaxes}

We adopt the parallax measurements from {\it Gaia\/} \citep{Gaia_main}. We adopt the {\it Gaia\/} parallax, $\pi$, and its uncertainty as provided by the first {\it Gaia\/} data release (DR1)\footnote{Accessed on 14 September 2016.} (see Table \ref{tab:allresults2}). A total of 424,489 stars in our sample have parallaxes in the Tycho-Gaia Astrometric Solution (TGAS) sample.

We note that, at the time of this writing, the {\it Gaia\/} team reports that the DR1 $\pi$ values have systematic errors that are $\sim$0.2~mas on small scales and zero-point variations as large as $\sim$0.3~mas on large spatial scales.\footnote{See \url{http://www.cosmos.esa.int/web/gaia/dr1}}
Preliminary assessments suggest a global offset of $\sim$-0.25~mas \footnote{Where the negative sign indicates that the {\it Gaia\/} parallaxes are underestimated.} for $\pi \gtrsim 1$~mas \citep{StassunGaiaError:2016}, corroborating the {\it Gaia\/} claim, based on comparison to directly-measured distances to well-studied eclipsing binaries by \citet{StassunGaiaEB:2016}, which itself is based on the sample of \citet{Torres2010}.  
\citet{Gould2016} similarly claim a systematic uncertainty of $0.12$mas.
\citet{Casertano:2016} used a large sample of Cepheids to show that there is likely little to no systematic error in the {\it Gaia\/} parallaxes for $\pi \lesssim 1$~mas, but find evidence for an offset at larger $\pi$ consistent with \citet{StassunGaiaError:2016}.

Thus the available evidence suggests that any systematic errors in the {\it Gaia\/} parallaxes are likely to be small. Therefore, for the purposes of this work, we simply use and propagate the reported {\it random} uncertainties on $\pi$ only, emphasizing that (a) the fundamental \fbol\ and $\Theta$ measurements that we report are independent of $\pi$, and (b) additional (or different) choices of statistical and systematic uncertainties in $\pi$ may be applied to our \fbol\ and $\Theta$ measurements following the methodology, equations, and error propagation coefficients supplied below.

\begin{deluxetable}{lrrr}
\tabletypesize{\scriptsize}
\tablecaption{\label{tab:summary} Stellar Sample Overview}
\tablewidth{0pt}
\setlength\tabcolsep{5pt}
\tablehead{
\colhead{Description} & \colhead{Spectroscopic} & \colhead{No Spectroscopic} & \colhead{Total} \\ & \colhead{Priors} & \colhead{Priors} & }
\startdata
Stars in initial sample & 337,561 & 2,202,353 & 2,539,914 \\
Stars with iterative IRFM parameters & 293,412 & 1,306,668 & 1,600,080 \\
\indent Stars with TGAS solutions & 212,025 & 1,049,641 & 1,261,666 \\
\indent Stars with $\leq$ 10\% parallaxes & 64,960 & 290,542 & 355,502 \\ %these are correct
\enddata
\end{deluxetable}
\section{Method}\label{sec:method}

The basic precepts for the methodology used here are from \citet{StassunGaiaEB:2016} and \citet{StassunGaiaPlanets:2016}. Briefly,
to calculate the radius of each star, we calculate its distance $d$ from its \emph{Gaia} parallax. The radius is then given by the equation $\rstar = (\theta d)/2$, where
\begin{equation}
\theta \equiv 2\left(\frac{\fbol}{\sigma\teff^4}\right)^{1/2}.
\label{eq:sigr}
\end{equation}
We then calculate the uncertainty on the radius as
\begin{equation}
\frac{\sigma_{\rstar}}{\rstar} = \sqrt{\left(\frac{\sigma_d}{d}\right)^{2} + \left(\frac{\sigma_{\theta}}{\theta}\right)^{2}},
\label{eq:sigtheta}
\end{equation}
where
\begin{equation}
\left(\frac{\sigma_{\theta}}{\theta}\right)^2 =  \frac{1}{4}\left(\frac{\sigma_{\fbol}}{\fbol}\right)^{2} + 4\left(\frac{\sigma_{\teff}}{\teff}\right)^{2} - 2\rho\left(\frac{\sigma_{\teff}}{\teff}\frac{\sigma_{\fbol}}{\fbol}\right)
\end{equation}
and $\rho$ is Pearson's correlation coefficient between \teff\ and \fbol. 

Thus, in the following subsections, we detail our procedures for measuring \teff\ and \fbol\ for the stars in our study sample.

We do require some modifications in procedure compared to \citet{StassunGaiaEB:2016,StassunGaiaPlanets:2016} due to the very large size of our sample.
In particular, rather than perform full broadband SED fits for all 2,539,913 stars for which we have photometry, we use empirical color--\teff\ and color--\fbol\ relations, and perform full broadband SED fits for a subset of the stars in order to assess the reliability of the empirical relations. 

We adopt this approach in this paper primarily out of practicality and convenience.  The very large number of stars in our sample with available photometry makes full SED fitting very time consuming.  Furthermore, we believe that the DR1 parallaxes are sufficiently imprecise that they do not warrant the more direct approach of fitting the SEDs, despite the deficiencies we encounter using the empirical color--\teff\ and color--\fbol\ relations, as described below. Furthermore, the \emph{Gaia} spectrophotometry, which will be released at a later date, will allow for much stronger constraints on the SEDs of the target stars. Therefore, in a future paper, we plan to perform full SED fits to all available photometry, including \emph{Gaia} spectrophotometry, for all stars with sufficiently precise \emph{Gaia} parallaxes. 

In addition, we determine $\rho$ in the equations above for each of our four subsamples separately.   We note that these equations for the uncertainties on $\rstar$ and $\theta$ fundamentally assume that the mathematical relations between the inferred quantities and the observables are linear.  In fact, this is not true in most cases, which implies that these relations are only accurate when the uncertainty in the measured quantity is small compared to the absolute value of the quantity itself.  This can be particularly problematic in the case of the measured parallax $\pi \propto d^{-1}$, which is often measured to precision that is comparable to the value of $\pi$ itself. Using parallaxes with uncertainties comparable to their magnitudes is also complicated by Lutz-Kelker bias \citep{Lutz1973}. In order to avoid these complications, we only include stars with $\sigma_\pi/\pi \la 10\%$.

\subsection{Stellar Parameters}

\subsubsection{Effective Temperature}\label{subsec:teff}

In principle, if were were only interested in inferring the angular diameters of our sample of stars (and from these diameters inferring radii using the {\it Gaia} parallaxes), we could simply adopt empirically-calibrated color-angular diameter relations (e.g., \citealt{Boyajian:2014}). However, we chose to instead derive \teff\ and \fbol\ individually from separate empirical color--\teff\ and color--\fbol\ relations, for two reasons.  First, we can compare our inferred estimates of \teff\ with spectroscopic measurements, thus validating our inferred values and allowing us to use the spectroscopic \teff\ as constraints.  Second, we do not know the extinction to the stars in our sample {\it a priori}.  A significant extinction would bias the broadband photometry we use to infer the angular diameters, thereby leading to a bias in the angular diameters and radii. We must therefore estimate the extinction as well\footnote{We note, as is well known, that color-angular diameter relations are fairly insensitive to extinction.  However, they are not {\it completely} insensitive to extinction, and some color indices are more sensitive to extinction than others.  Furthermore, a significant fraction of our sample are giant stars, which may be quite far from the Sun, and for which the extinction may be significant, particularly near the Galactic plane.}.

In order to infer \teff, we apply the \citet{Casagrande2010} infrared flux method (IRFM) relations, 
to obtain effective temperatures for the Tycho-2 stars.
As noted above, to do so, however, it is first necessary to de-redden the photometric measurements. Since we do not know the extinction $A_V$ \emph{a priori}, we estimate the extinction and the effective temperature as follows:
\begin{enumerate}

\item We step through $N_i=\avmax/\Delta A_V$ extinction values $A_{V,i}$ in increments of $\Delta A_V = 0.01$ over the range $A_V \in \{0,\avmax\}$, where $\avmax\ \equiv E_{B-V}R_V$ is the maximum line-of-sight total extinction, estimated from is the maximum line-of-sight color excess (selective extinction) $E_{B-V}$ determined from the \citet{Schlegel1998} dust maps, and adopting $R_V = 3.1$ for the ratio of total to selective extinction.  We note that, despite the caution indicated by \citet{Schlegel1998}, we adopt their maximum color excess even for stars within 10$^\circ$ of the Galactic plane.  We will provide qualitative tests of the validity of our inferred extinctions for stars in this region of the sky in a later section.  

\item At each value of $A_{V,i}$, we de-redden the photometric magnitudes using the \citep{Cardelli:1989} extinction law. We then calculate $T_{{\rm eff},j}$ from each of $j=0,1,...N_j$ applicable \citet{Casagrande2010} empirical relations for which the de-reddened color and the metallicity are within the applicable ranges. The uncertainty $\sigma_{T_{{\rm eff},j}}$ on the $T_{{\rm eff},j}$ derived from each relation is calculated as the square-root of the quadrature sum of the standard deviation about the empirical relation listed in Table 4 of \citet{Casagrande2010} and the (linearly) propagated uncertainty due to the (assumed to be independent) errors on \feh\ and the photometric measurements. If no empirical relations apply -- e.g. because all de-reddened colors lie outside the suitable ranges for all IRFM relations -- then we skip to the next $A_{V,i}$.

\item We calculate the weighted mean $T_{{\rm eff,mean},i}$ for each $A_{V,i}$ from all $N_j$ applied relations, where we weight each $T_{{\rm eff},j}$ by the square of the uncertainty in each relation $\sigma_{T_{{\rm eff},j}}$.  We then reject $T_{{\rm eff},j}$ if $T_{{\rm eff},j} - T_{{\rm eff,mean},i} > 3\sigma_{T_{{\rm eff},j}}$. We then re-calculate the weighted mean $T_{{\rm eff,mean},i}$ and iterate until no outliers remain or until only one relation remains.

\item We calculate a "merit function" $\chi^2_{\rm IRFM} \equiv \sum_{j=0}^{N_j} (T_{{\rm eff},j} - T_{{\rm eff,mean},i})^2/\sigma_{T_{{\rm eff},j}}^2$, which essentially quantifies how well values all the $N_j$ inferred values of $T_{{\rm eff},j}$ are consistent with a constant value of \teff, given their respective uncertainties $\sigma_{T_{{\rm eff},j}}$ and the assumed value of $A_{V,i}$. 

\item We add a penalty term $\chi_{\rm spec}^2\equiv (T_{{\rm eff,mean},i}-T_{\rm eff,spec})^2/\sigma_{T_{\rm eff,spec}}^2$ if a spectroscopic \teff\ exists. If $\chi^2 =0$ because there is no spectroscopic effective temperature for this star and only one calibration relation applies, then we skip to the next $A_{V,i}$.

\item We select the value $A_{V,i}$ and the error-weighted mean $T_{{\rm eff,mean},i}$ corresponding to the minimum $\chi^2 \equiv \chi^2_{\rm IRFM} + \chi^2_{\rm spec}$. If no merit function was calculated for any value of $A_{V,i}$, then we have insufficient information to inform our choice of effective temperature; therefore, we drop this star from the sample and move onto the next star.

\item We re-scale the uncertainty on $T_{{\rm eff,mean},i}$ such that $\chi^2_{\nu}=1$ for the minimum $\chi^2$, where the $\chi^2_{\nu}$ is reduced $\chi^2$ with $\nu$ degrees of freedom. Explicitly, the scale factor is 
\begin{equation}
\sqrt{\frac{1}{N_j+1}\left[
(T_{{\rm eff,mean},i}-T_{\rm eff,spec})^2+
\sum_{j=0}^{N_j}  (T_{{\rm eff},j} - T_{{\rm eff,mean},i})^2\right]}
\end{equation}
if there is a spectroscopic \teff\ and 
\begin{equation}
\sqrt{\sigma_{T_{{\rm eff,mean}}} \equiv \frac{1}{N_j}\sum_{j=0}^{N_j} \left[(T_{{\rm eff},j} - T_{{\rm eff,mean},i})^2\right]}
\end{equation} 
if not. We re-calculate $T_{{\rm eff,mean},i}$ using these rescaled uncertainties.

\item We then estimate the uncertainty on the extinction by taking the range of extinctions corresponding to $\Delta \chi^2 = 1$, using this scaled uncertainty. If no extinctions lie within this range, then we adopt our stepsize, $\Delta A_V = 0.01$, as the uncertainty.

If no IRFM relations were applied for any extinction, we drop the star from the sample. Additionally, for the stars without spectroscopic temperatures, if there exists an extinction for which only one IRFM relation applies, then the merit function $\chi^2 = 0$ by definition for that extinction. In these cases, we infer that we do not have enough information to determine both extinction and \teff, and so we also drop such stars from our sample.

\end{enumerate}

\subsubsection{Bolometric Flux}\label{subsec:fbol}
We estimate the unextincted bolometric flux \fbol\ as follows:

\begin{enumerate}

\item We de-redden the magnitudes with the extinctions obtained in Section \ref{subsec:teff} and apply all $N_{\fbol}$ applicable \citet{Casagrande2010} bolometric flux relations.

\item For each relation, we calculate the mean error-weighted bolometric flux, \fbol, and the weighted uncertainty on the flux, $\sigma_{\fbol}$, where the weights are the quadrature sums of the scatter about each relation as cited by \citet{Casagrande2010} and the (linearly) propagated uncertainties due to the uncertainties on the magnitudes, extinction, and \feh, assuming all uncertainties are independent. As with the effective temperature procedure, if no IRFM relations were applicable -- e.g. because the de-reddened color lies outside the ranges of all relations -- we drop the star from our sample. 

\item We calculate $\chi^2_{\fbol} \equiv \sum_{j=0}^{N_{\fbol}} (F_{\rm bol,j} - F_{\rm bol,mean})^2/\sigma_{F_{\rm bol,j}}^2$ and scale the uncertainties by a constant factor such that $\chi^2_{\nu, \fbol} = 1$, where $\nu = N_{\fbol}$. We then re-calculate $F_{\rm bol,mean}$ and $\sigma_{F_{\rm bol,mean}}$ using these rescaled uncertainties.

\end{enumerate}

\subsection{Validating \teff\ and \fbol\ Determined via the Iterative IRFM Technique}\label{sec:comp}

Before we apply the {\it Gaia} parallaxes to infer radii for the $\sim356,0000$ stars in our final sample, we first validate our technique for inferring the \teff\ and \fbol\ for the over 1.6 million stars that remain in our sample after applying the IRFM to infer \teff\ and \fbol.

\subsubsection{Extinctions}
\begin{figure*}
\begin{center}
\includegraphics[width=0.5\linewidth]{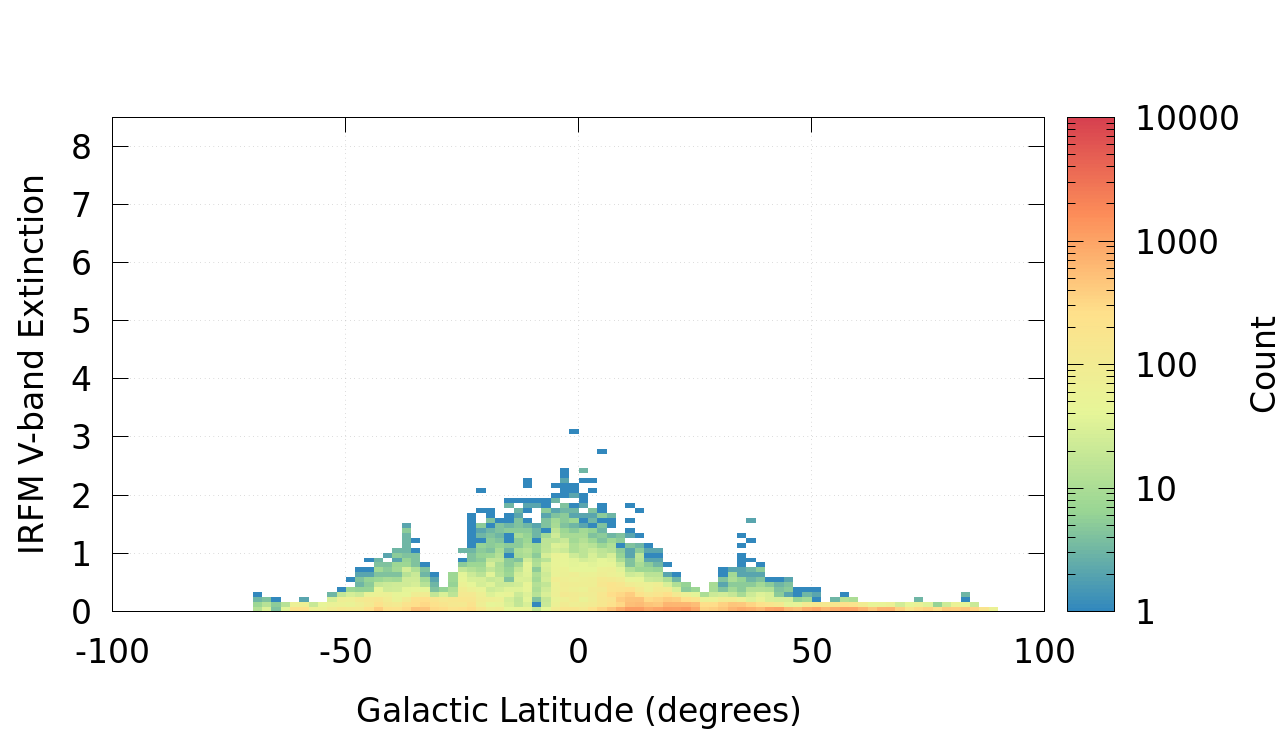}\includegraphics[width=0.5\linewidth]{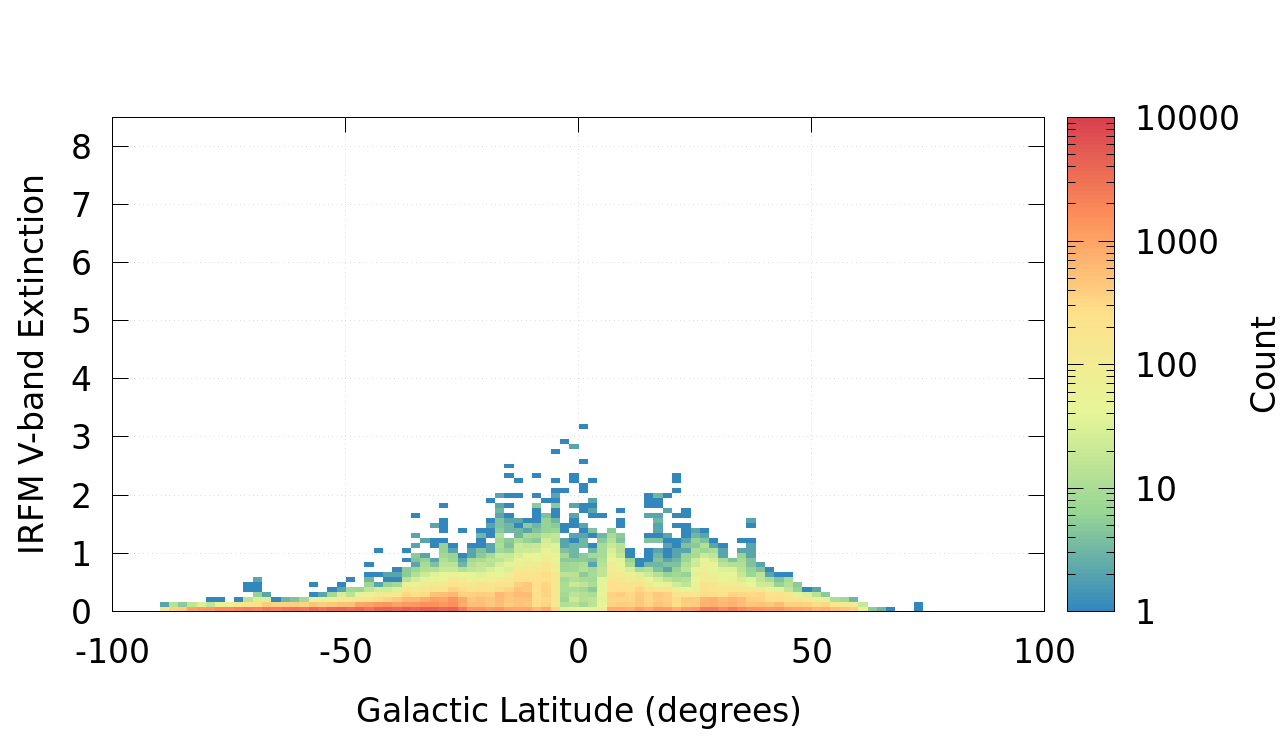}
\includegraphics[width=0.5\linewidth]{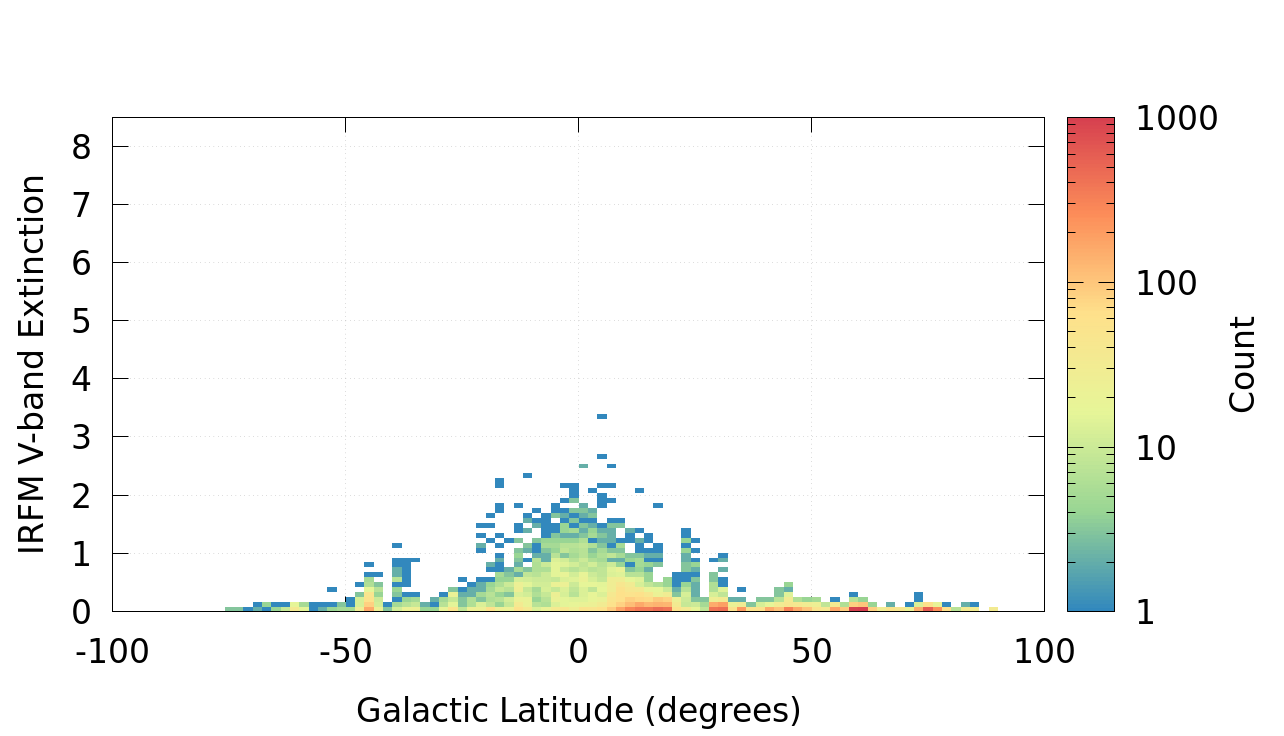}\includegraphics[width=0.5\linewidth]{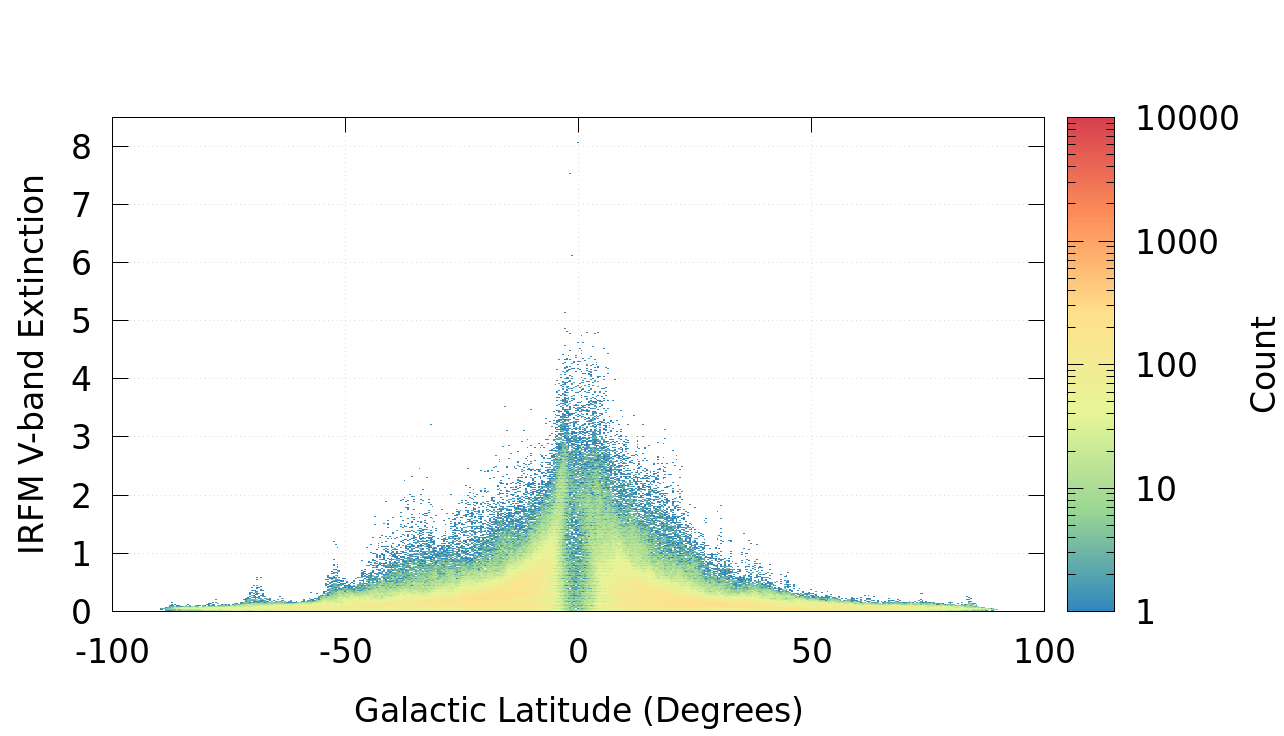}
\caption{\label{fig:avb} Best-fit extinction values $A_v$ as a function of Galactic latitude $b$ for the stars in the LAMOST (\emph{top-left}), RAVE (\emph{top-right}), APOGEE (\emph{bottom-left}), and non-spectroscopic (\emph{bottom-right}) samples.}
\end{center}
\end{figure*}

As one way of validating the results of our iterative procedure outlined in Section \ref{sec:method}, we check that our method prefers reasonable extinction values. First, we examine the distribution of extinctions across Galactic latitude $b$; most of the highly extincted stars should lie in the Galactic disk -- roughly $|b| < 10\arcdeg$ -- where there is more dust along a typical line of sight. Figure \ref{fig:avb} shows these distributions for the four subsamples; indeed, the most highly extincted stars are those in the disk.

\subsubsection{Comparison to Spectroscopic Effective Temperatures}\label{subsec:specteff}

\begin{figure*}
\includegraphics[width=0.5\linewidth]{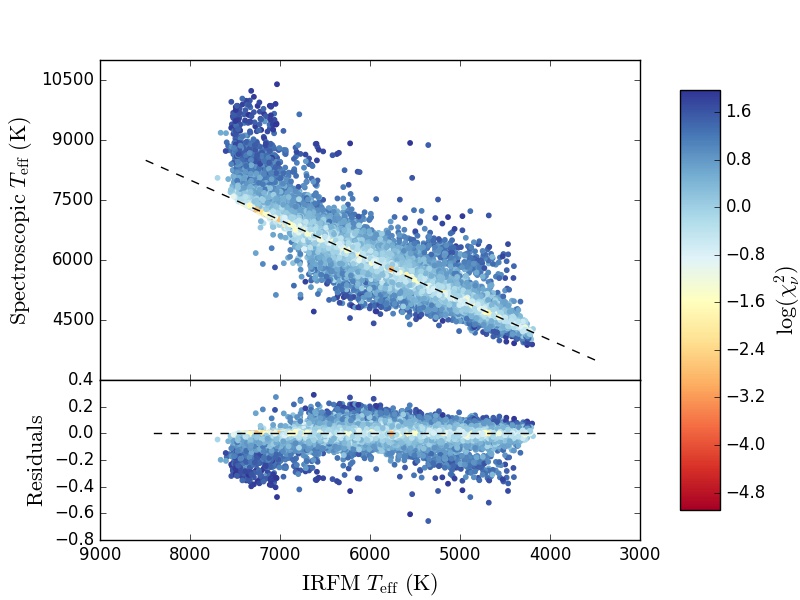}\includegraphics[width=0.5\linewidth]{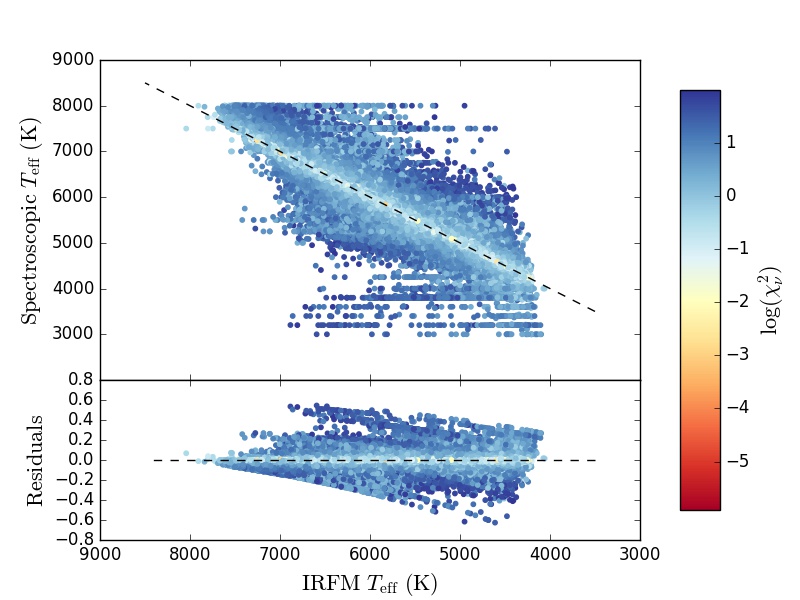}
\begin{center}
\includegraphics[width=0.5\linewidth]{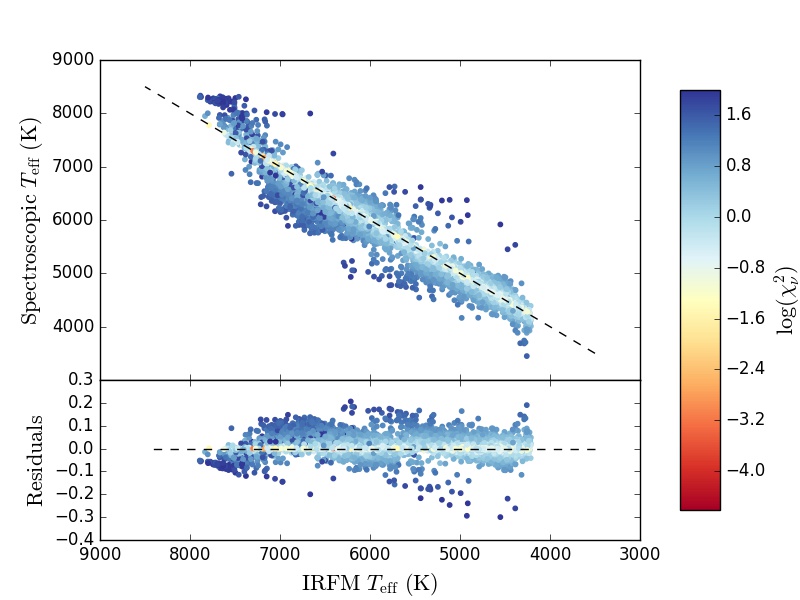}
\end{center}
\caption{\label{fig:teffcomp1} Spectroscopic versus best-fit effective temperatures \teff\ (top sub-panels) and the fractional deviations (bottom sub-panels) for the stars in our LAMOST (\emph{top-left}), RAVE (\emph{top-right}), and APOGEE (\emph{bottom}) samples. The colors denote the unscaled $\log \chi_{\nu}^2$. }
\end{figure*}

Figure \ref{fig:teffcomp1} show the IRFM-derived effective temperatures versus the spectroscopic values for the resulting sample of 64,345 LAMOST stars, 214,707 RAVE stars, and 14,360 APOGEE stars. 

This excludes stars with reduced $\chi_{\nu}^2 > 100$, which effectively removes stars for which the IRFM temperatures and spectroscopic temperatures differ by ${\cal O}(10^3~{\rm K})$.  Notably, these include many giants, for which the IRFM relations were not calibrated. The IRFM effective temperatures are positively correlated with the spectroscopic temperatures; stars with discrepant temperatures tend to have higher $\chi^2$ values, while the IRFM temperatures for stars with lower $\chi^2$ values tend to agree with the spectroscopic values. Moreover, our iterative IRFM technique appears to systematically underestimate the effective temperatures for stars with spectroscopic temperatures above 7,000 K.

\begin{figure*}
\begin{center}
\includegraphics[width=0.5\linewidth]{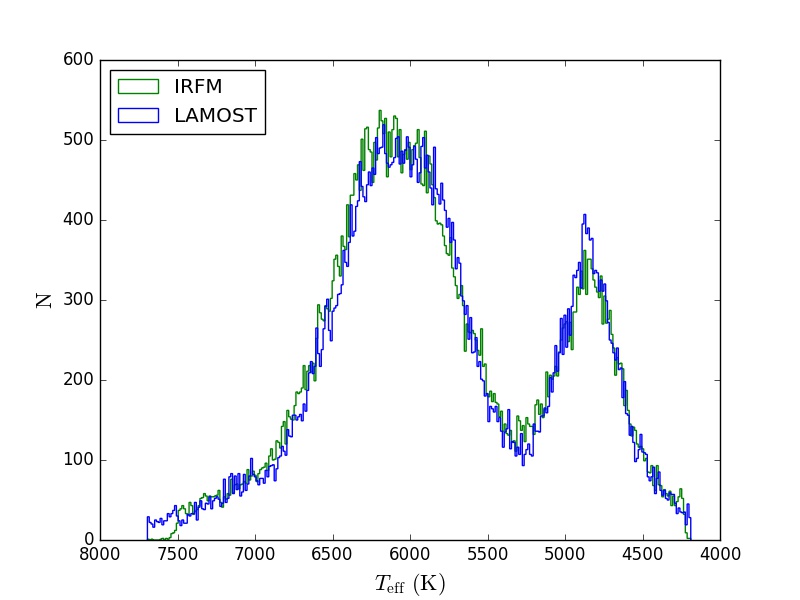}\includegraphics[width=0.5\linewidth]{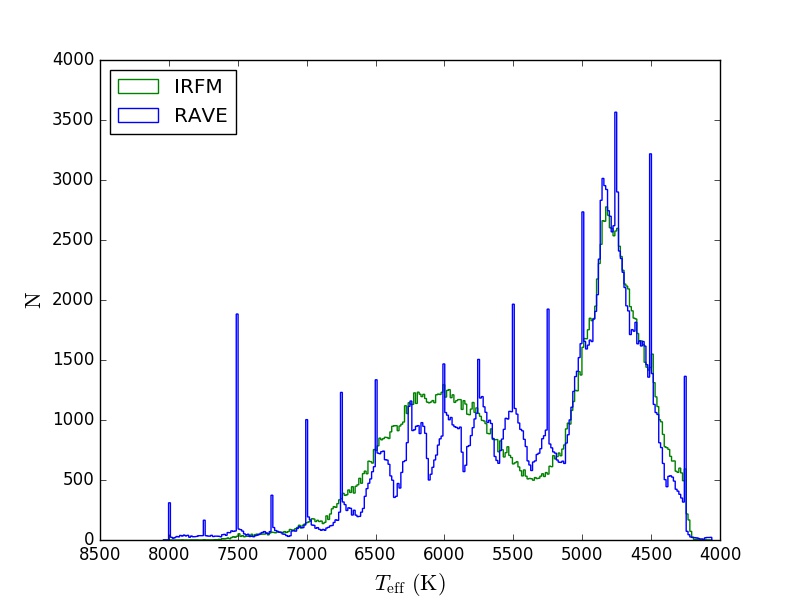}
\includegraphics[width=0.5\linewidth]{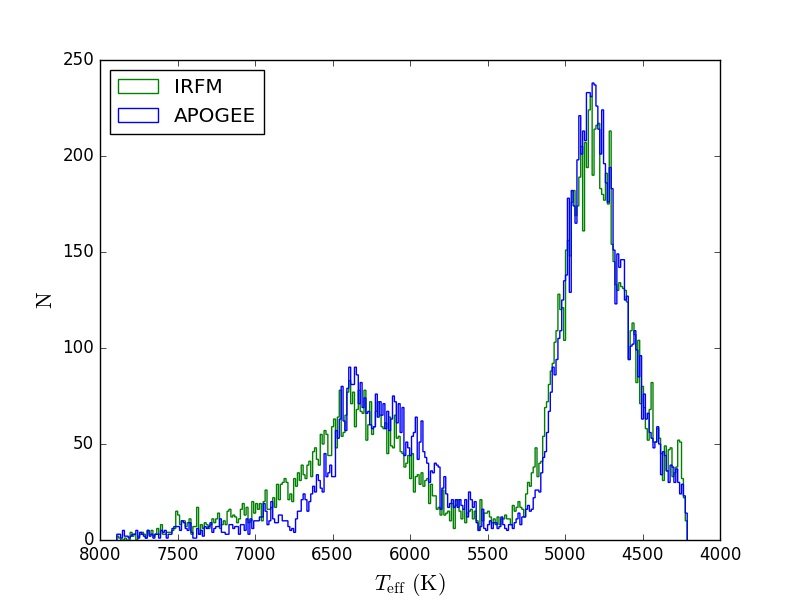}
\end{center}

\caption{\label{fig:teffcomp2} Spectroscopic versus best-fit effective temperatures \teff\ for the stars in our LAMOST (\emph{top-left}), RAVE (\emph{top-right}), and APOGEE (\emph{bottom}) samples. In all three cases, our iterative IRFM procedure appears to be slightly biased towards hotter effective temperatures. The peaks in the RAVE histogram correspond to the grid resolution of synthetic spectra used by \citet{Kordopatis2013} in the stellar parameter pipeline pipeline: the grid steps in $\Delta \teff=250$K increments.}
\end{figure*}

Figure \ref{fig:teffcomp2} shows our effective temperature distributions inferred using the IRFM for the stars in our sample that also have spectroscopically-measured effective temperatures. 
In the RAVE sample, \citet{Kordopatis2013} use a grid of synthetic stellar spectra to derive the stellar parameters; hence, the peaks in the histogram correspond to the \teff\ grid points and are separated by the grid resolution of 250 K.
As shown in Figure \ref{fig:teffcomp1}, the iterative IRFM technique infers an excess of stars with $\teff > 7,000$ K relative to the spectroscopically-determined temperature distributions for the stars with LAMOST, RAVE, and APOGEE spectra. These stars typically have very few photometric measurements with which to infer \teff, \fbol, and $A_V$, and our inferences about their properties are thus highly suspect. Therefore, we urge the reader to use extreme caution when applying our results for stars hotter than about 7,000 K. 

As discussed below, the uncertainty on the \emph{Gaia} parallax distances begin to dominate the radius error budget beyond $\sim$100~pc \citep[see also][]{StassunGaiaPlanets:2016}. The uncertainty in radius due to the error on the projected end-of-mission \emph{Gaia} parallax should be greatly improved; for example, at 100~pc, the parallax signal is 10 milliarcseconds, so a projected 10 microarcsecond uncertainty translates to a 0.1\% uncertainty on the radius.

\subsubsection{Effect of Metallicity on IRFM Temperatures and Fluxes}

\begin{figure*}
\includegraphics[width=0.5\linewidth]{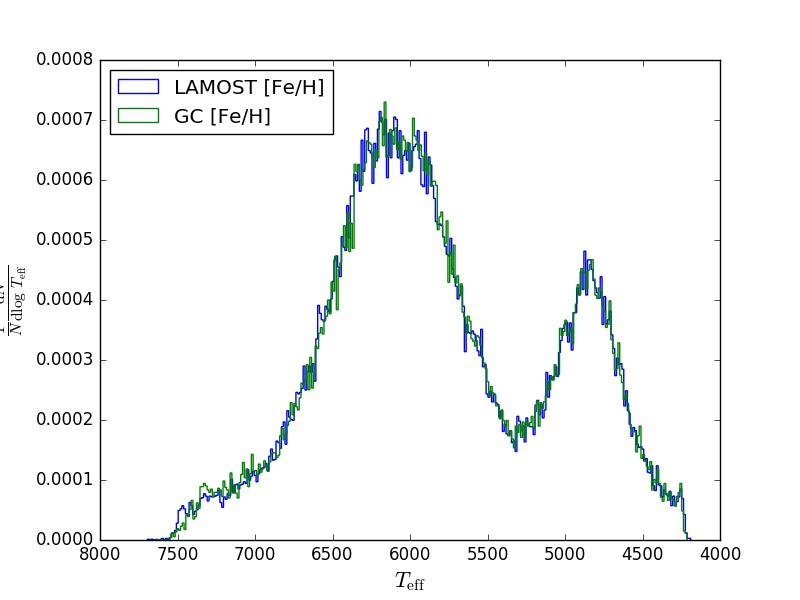}\includegraphics[width=0.5\linewidth]{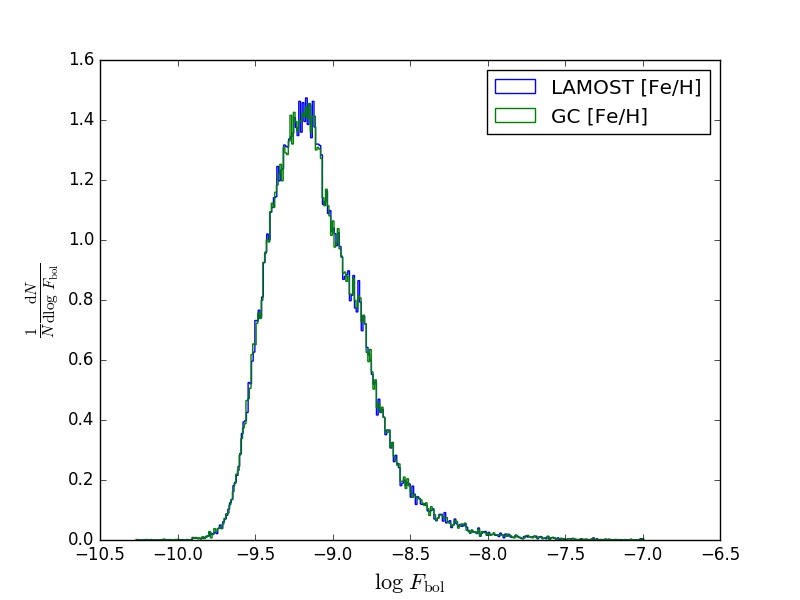}
\includegraphics[width=0.5\linewidth]{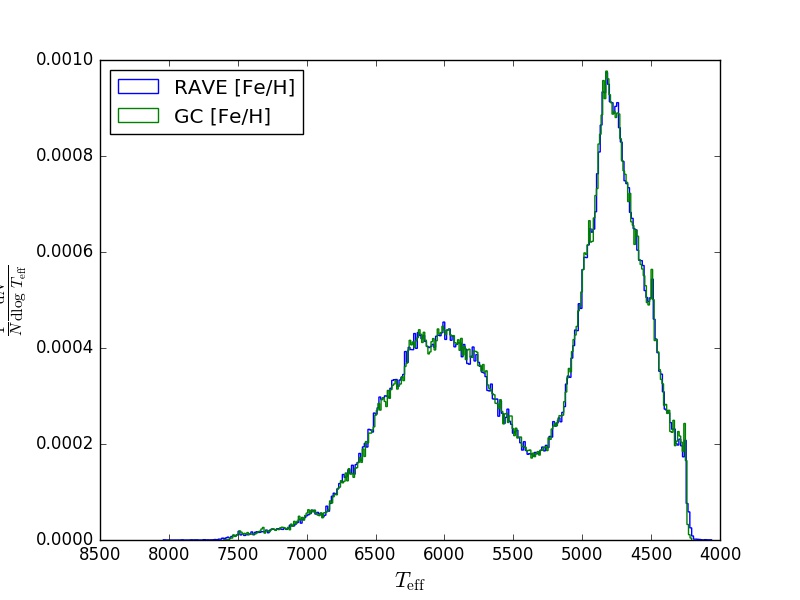}\includegraphics[width=0.5\linewidth]{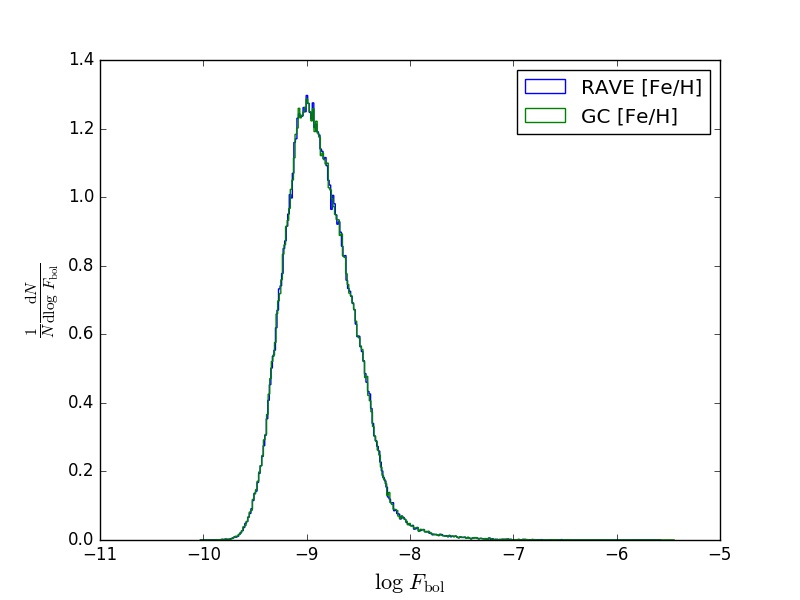}
\includegraphics[width=0.5\linewidth]{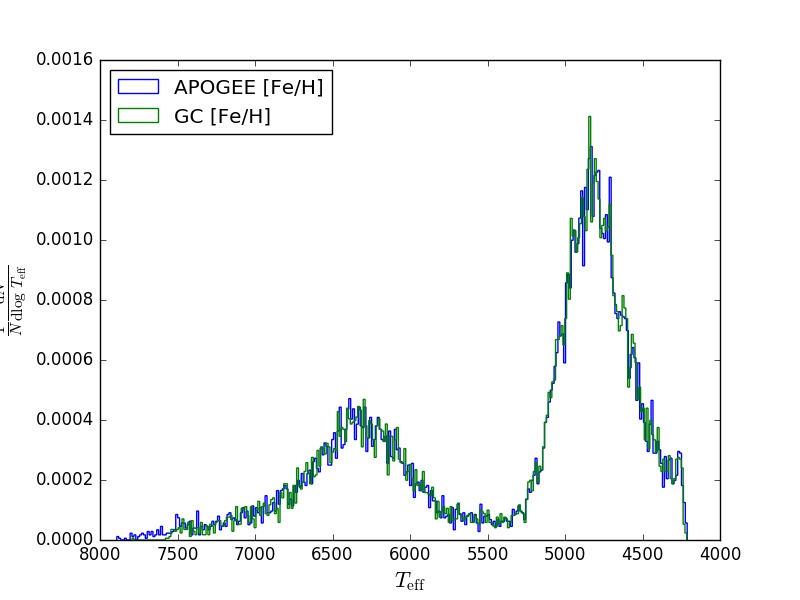}\includegraphics[width=0.5\linewidth]{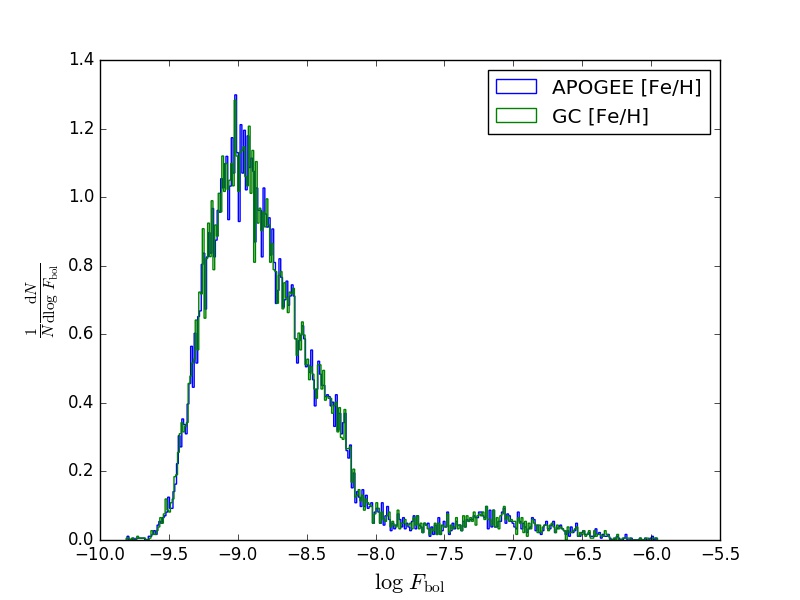}
\caption{\label{fig:fehcomp} IRFM effective temperature (left) and bolometric flux (right) histograms using spectroscopically-determined metallicities of individual stars (blue) and assuming a metallicity and uncertainty for all stars equal to median and dispersion of the distribution of \feh\ of stars from the Geneva-Copenhagen Survey \citep{Casagrande2011} (green) for the LAMOST (top row), RAVE (middle row), and APOGEE (bottom row) stars. Overall, metallicity has a negligible effect on the recovered temperatures and fluxes.}
\end{figure*}

As mentioned in Section \ref{sec:meas}, for the stars without measured metallicities, we adopt a metallicity and uncertainty equal to the median \feh\ and dispersion from the Geneva-Copenhagen Survey of late-type, solar neighborhood stars \citep{Casagrande2011}, as described in Section \ref{sec:method}. To determine what effect our choice of metallicity has on the recovered effective temperatures and bolometric fluxes, we repeat our method for the subset of stars with spectroscopic parameters, this time using the Geneva-Copenhagen median and $1\sigma$ metallicity instead of the LAMOST metallicities. As Figure \ref{fig:fehcomp} illustrates, our decision to use the median and dispersion of the Geneva-Copenhagen survey as a proxy for the metallicity of stars without directly-measured metallicities has a negligible effect on the temperature and bolometric flux we infer for these stars from our iterative IRFM procedure.

\subsubsection{Comparison to Full Broad-band Spectral Energy Distributions}\label{sec:SED}
Nominally, the IRFM relations of \citet{Casagrande2010} are parameterizations of stellar spectral energy distributions (SEDs). As another check on our effective temperatures, bolometric fluxes, and angular diameters, we generate SEDs for 132 stars in common between our sample and the \citet{Casagrande2010} sample. We use Kurucz model atmospheres \citep{Kurucz2013} to fit SEDs to Tycho-2, \citet{Mermilliod:2006}, \citet{Paunzen2015}, 2MASS, GALEX NUV \citep{Bianchi2011} and WISE \citep{Cutri2014} photometry when available. As before, we adopt the listed measurement uncertainties unless these uncertainties are smaller than 0.01~mag (or 0.1~mag for the GALEX NUV and WISE4 bands); in these cases, we adopt 0.01~mag (0.1~mag) as the measurement uncertainties. In addition, to account for an artifact in the Kurucz atmospheres at $10\mu$m, we artificially inflate the WISE3 uncertainty to 0.3 mag unless the reported uncertainty was already larger than 0.3 mag. Additionally, we iteratively clip $5\sigma$ outlier measurements.

\begin{figure*}
\begin{center}
\includegraphics[width=0.75\linewidth]{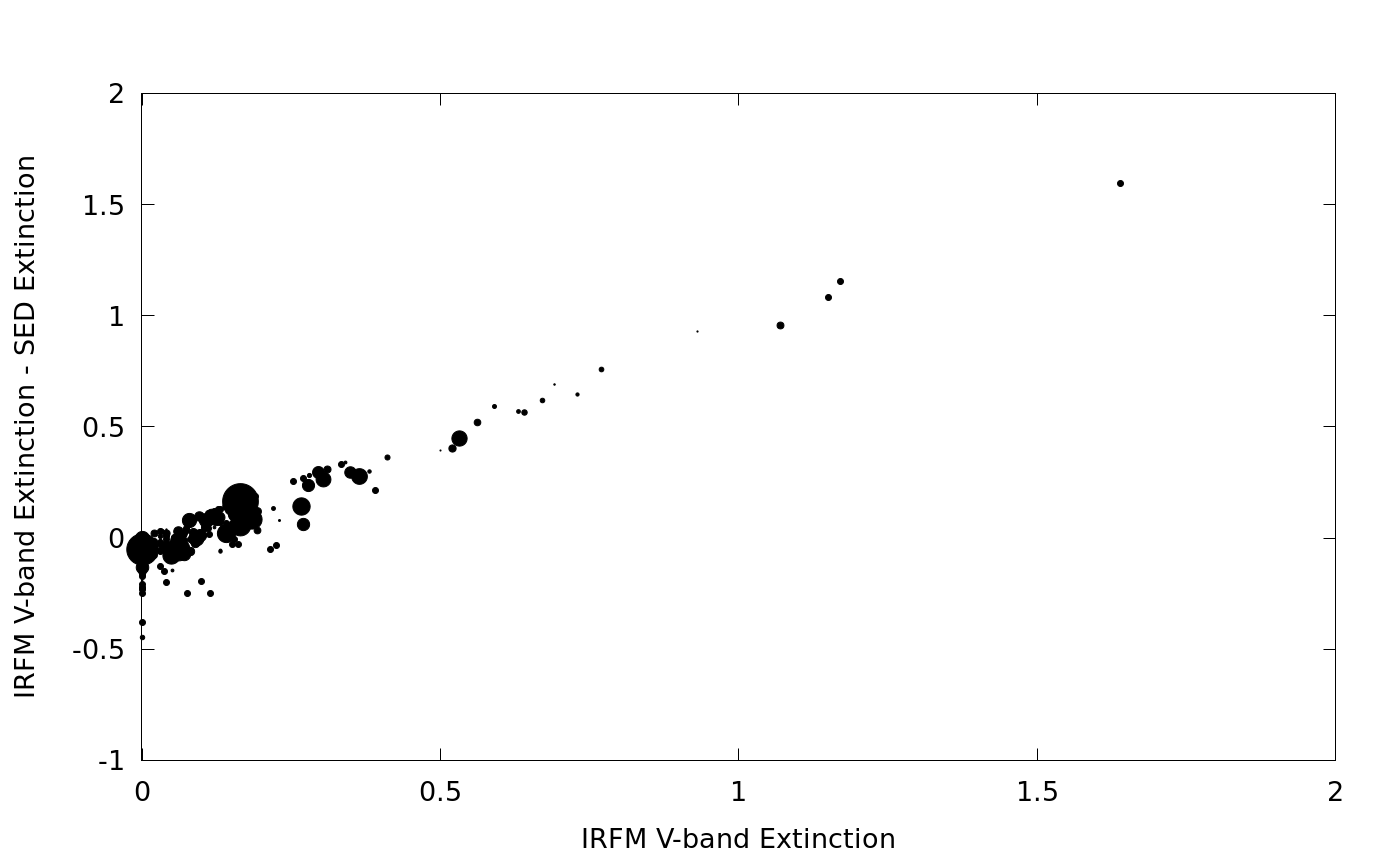}
\includegraphics[width=0.75\linewidth]{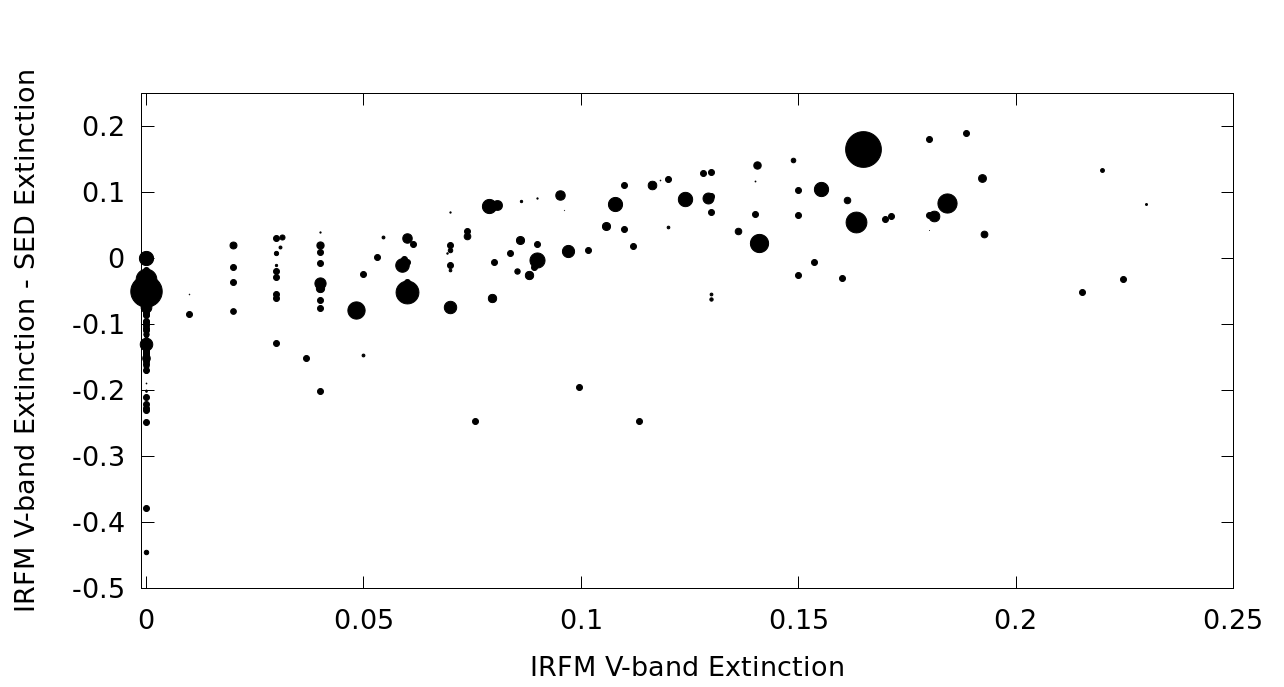}
\includegraphics[width=0.75\linewidth]{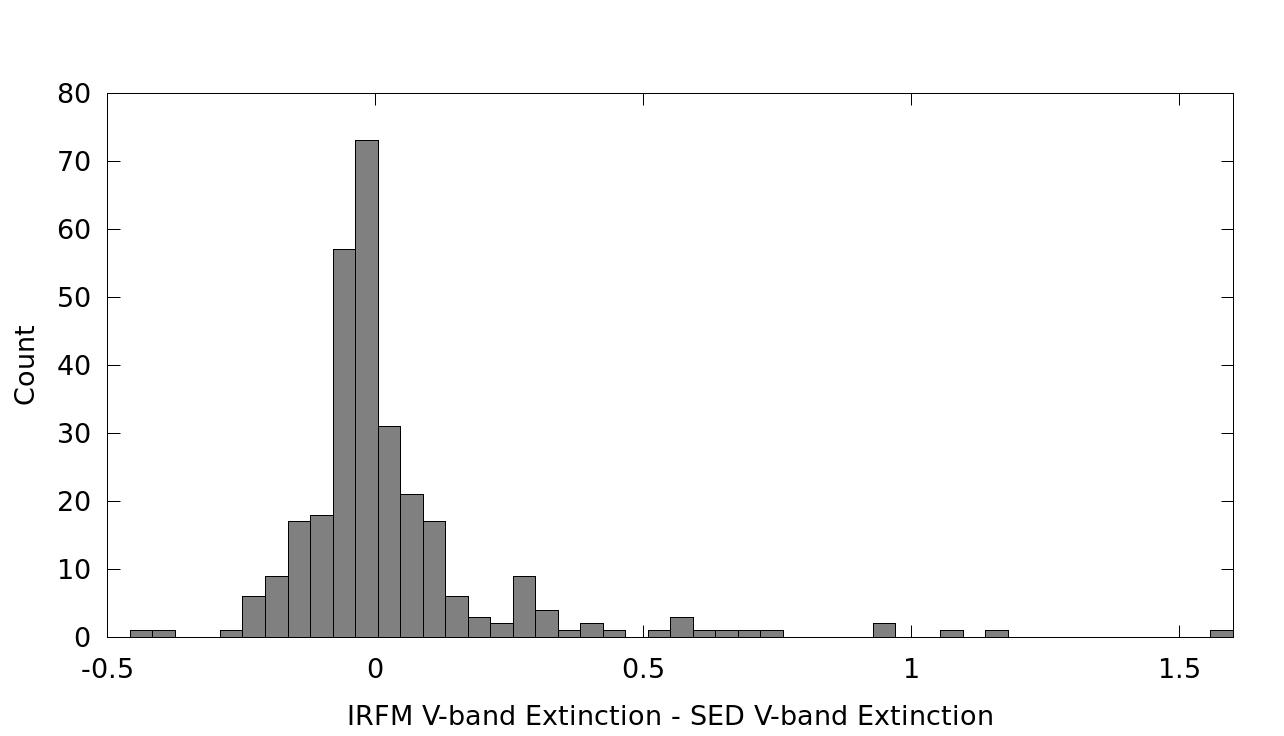}

\caption{\label{fig:sedavcomp2} \emph{Top:} Difference between IRFM and SED extinctions. In both panels, larger points denote higher $\chi^2_{\nu}$ values. The largest fractional deviations are for the lowest extinctions.\emph{Middle:} Zoom-in of the top panel showing that the IRFM and SED extinctions agree reasonably well for the majority of stars in this comparison. \emph{Bottom:} Histogram showing the absolute difference between the IRFM and SED extinctions. The median and standard deviation of the absolute differences between the IRFM and SED extinctions are, respectively, $-4.5\times10^{-3}$ and $2.2\times10^{-1}$.}
\end{center}
\end{figure*}

\begin{figure*}
\begin{center}
\includegraphics[width=0.75\linewidth]{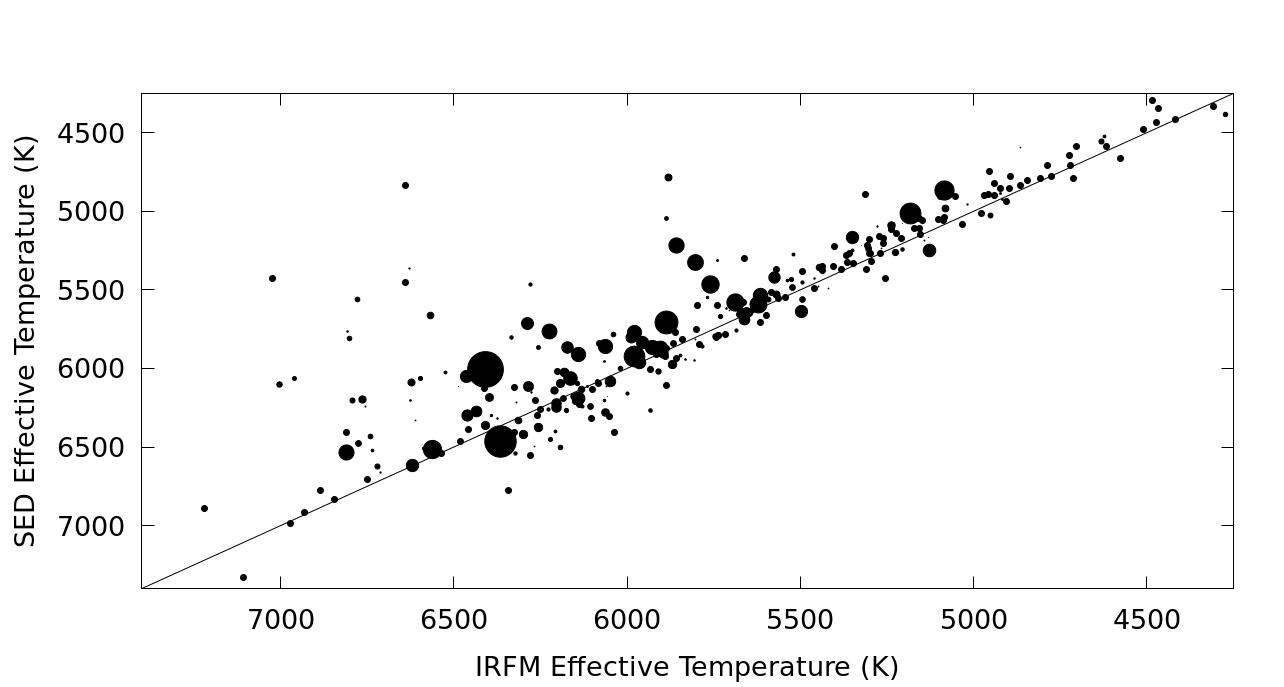}
\includegraphics[width=0.75\linewidth]{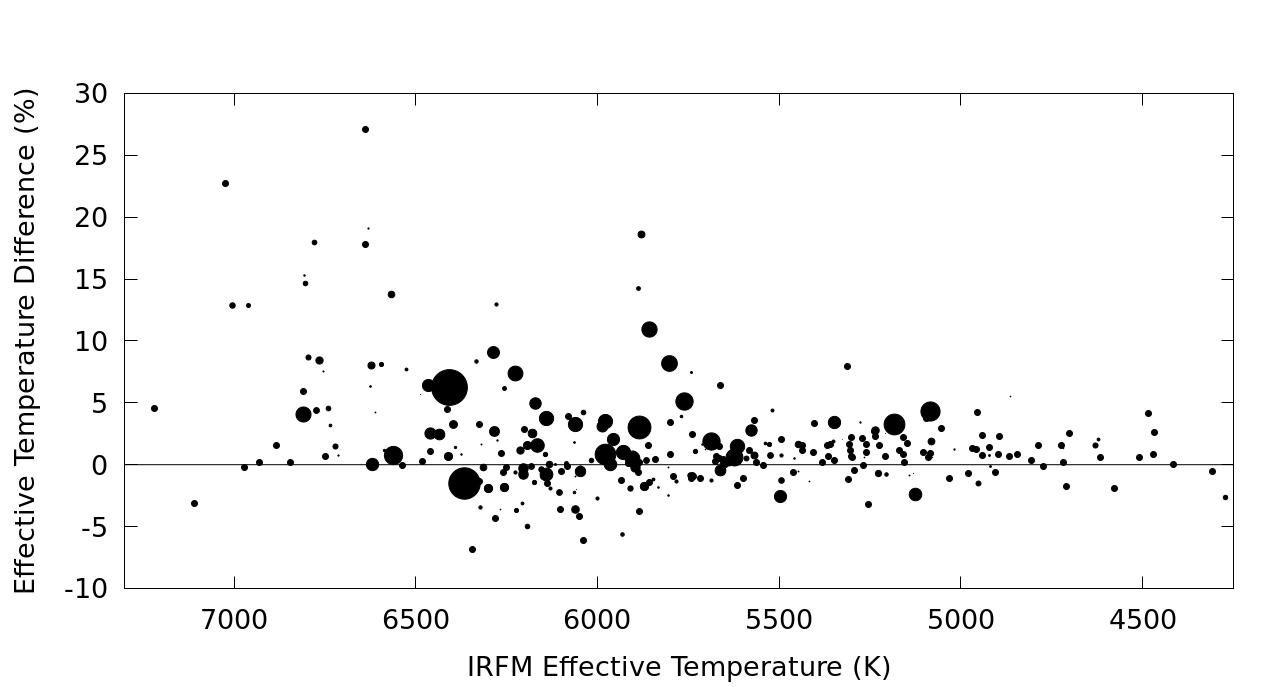}
\includegraphics[width=0.75\linewidth]{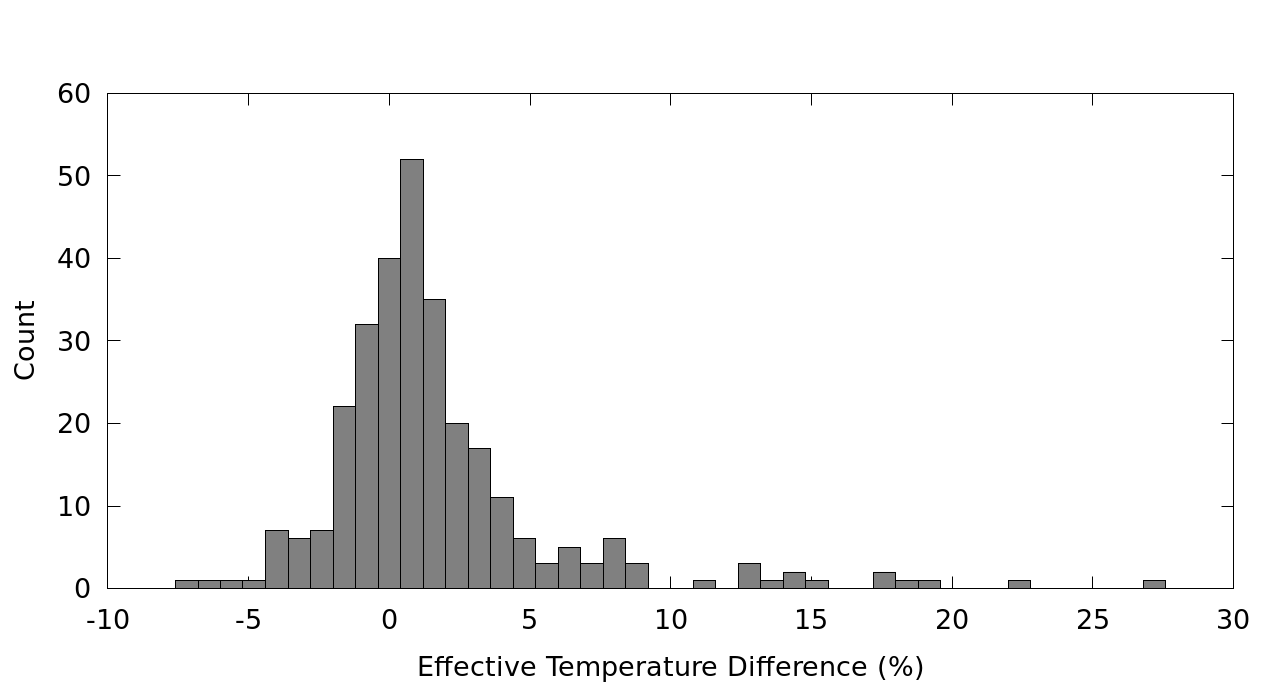}
\caption{\label{fig:sedteffcomp} \emph{Top:} SED versus iterative IRFM effective temperatures (right). The black line shows what would be perfect agreement between the two methods. \emph{Middle:} Fractional difference between the IRFM and SED values relative to the IRFM values. In both panels, larger points denote higher $\chi^2_{\nu}$ values. The median and standard deviation of the fractional differences are, respectively, 0.81\% and 19\%.\emph{Bottom:} Histogram showing the absolute difference between the IRFM and SED effective temperatures.}
\end{center}
\end{figure*}

\begin{figure*}
\begin{center}
\includegraphics[width=0.75\linewidth]{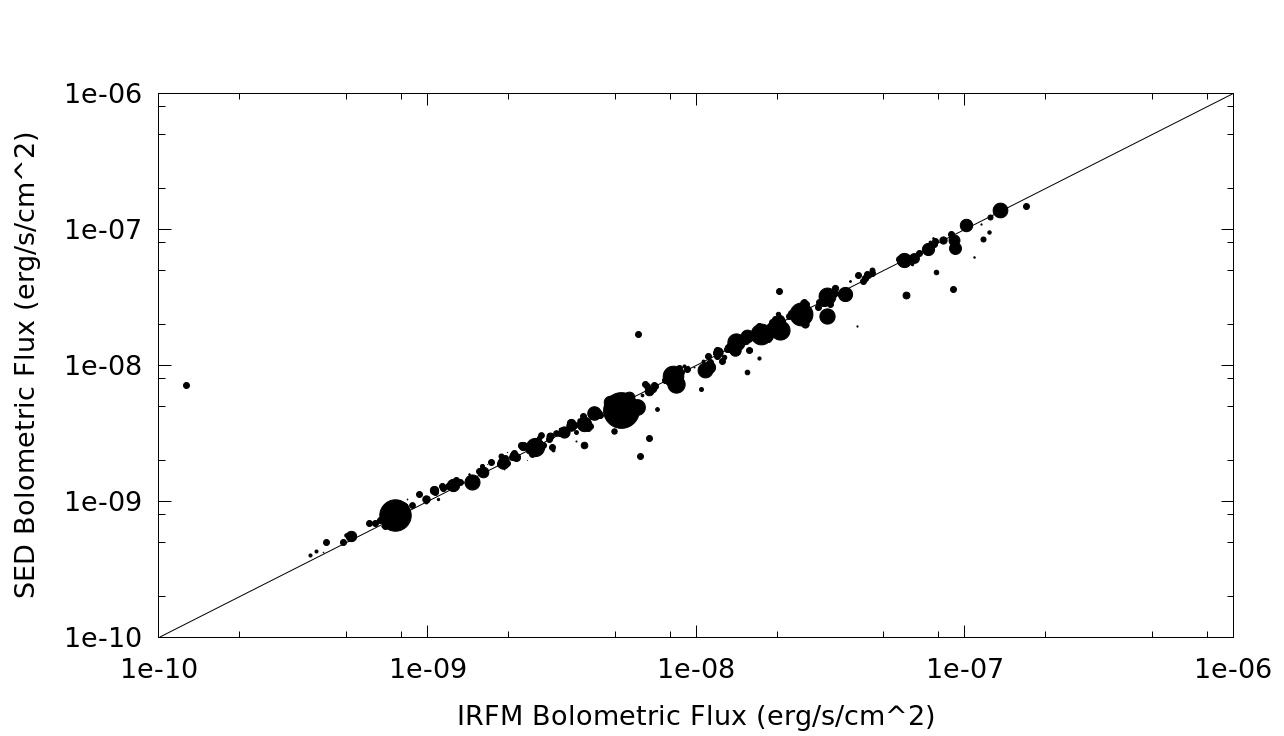}
\includegraphics[width=0.75\linewidth]{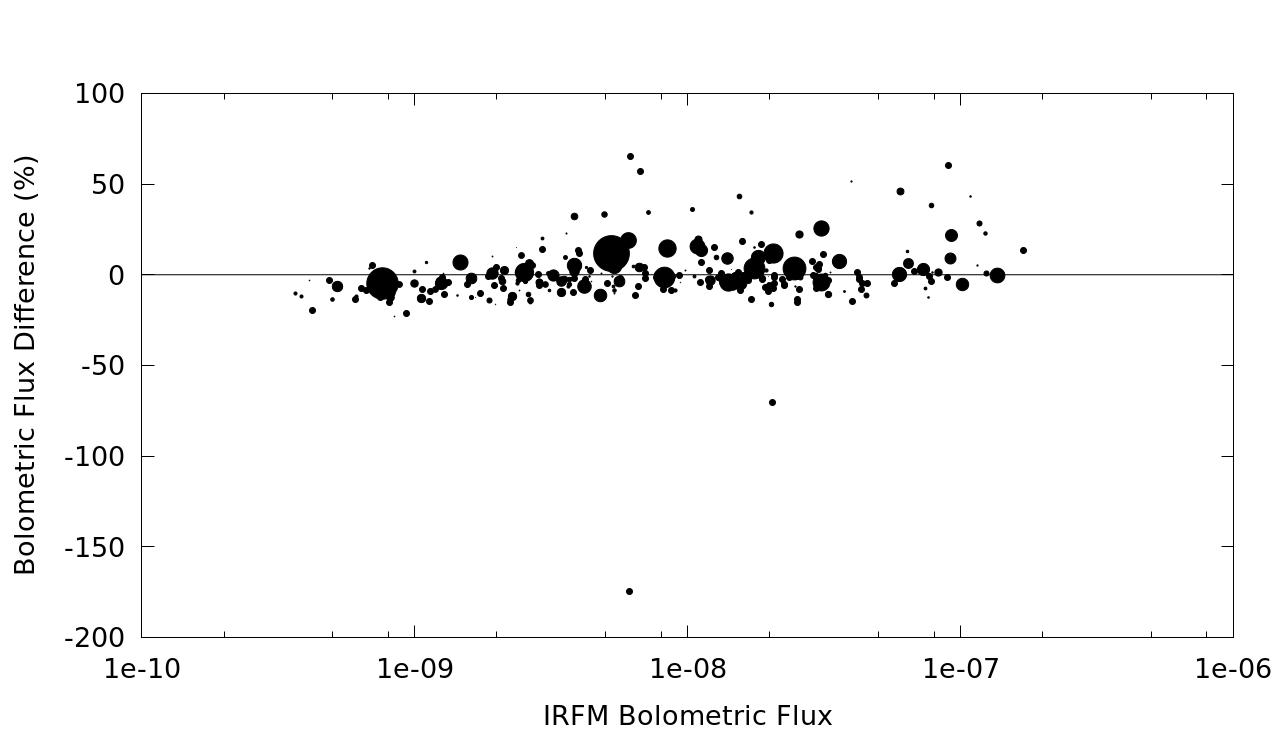}
\includegraphics[width=0.75\linewidth]{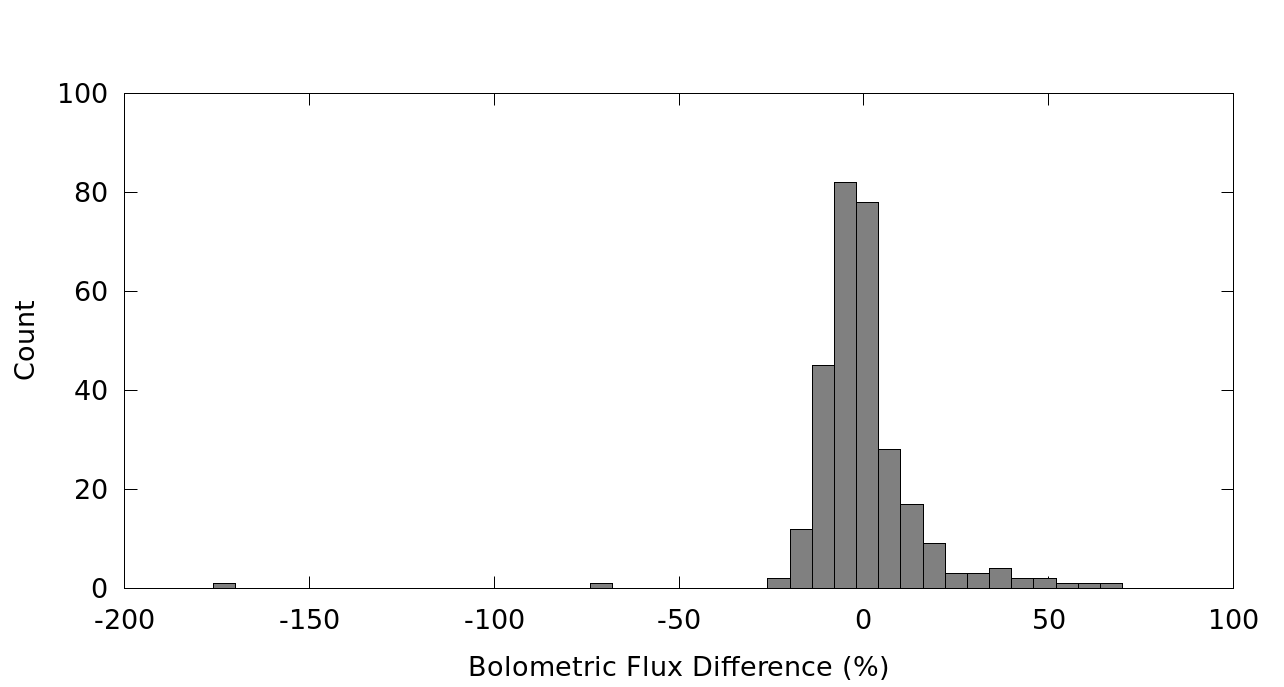}
\caption{\label{fig:sedfbolcomp} \emph{Top:} SED versus iterative IRFM bolometric fluxes. The black line shows what would be perfect agreement between the two methods. \emph{Middle:} Fractional difference between the IRFM and SED values relative to the IRFM values; negative values indicate smaller SED values. The median and standard deviation of the fractional differences are, respectively, -1.8\% and 17\%. \emph{Bottom:} Histogram showing the absolute difference between the IRFM and SED bolometric fluxes.}
\end{center}
\end{figure*}

\begin{figure*}
\begin{center}
\includegraphics[width=0.75\linewidth]{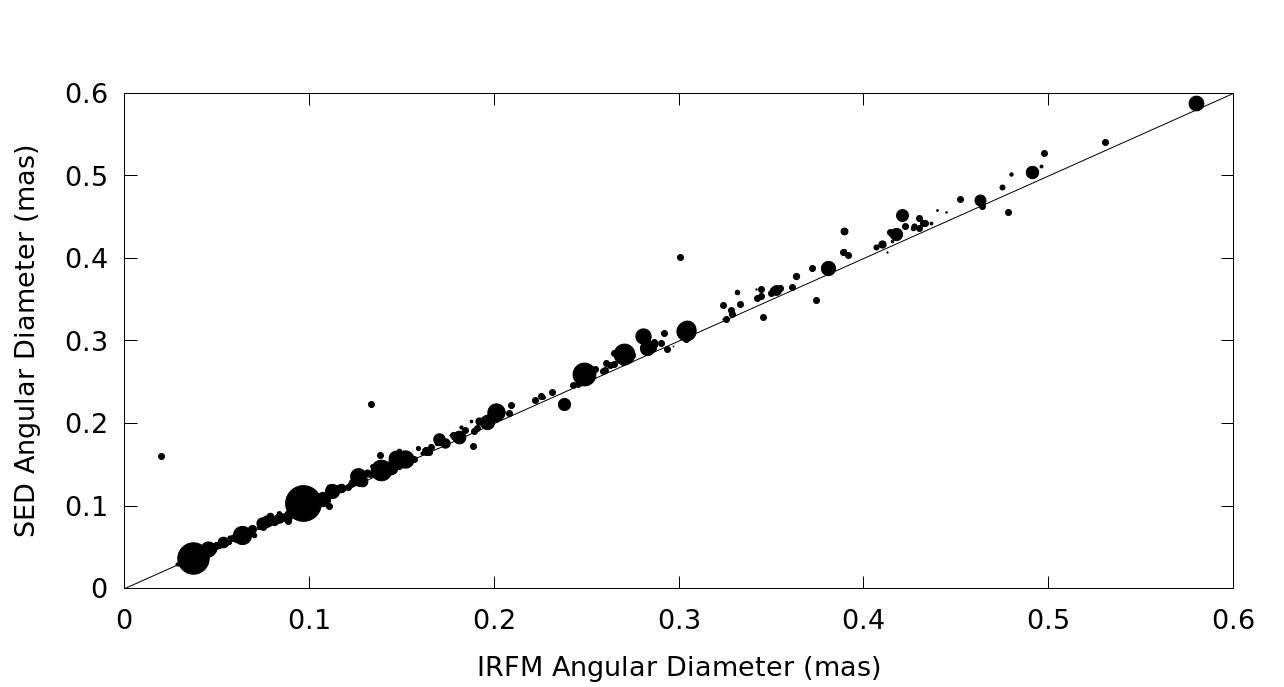}
\includegraphics[width=0.75\linewidth]{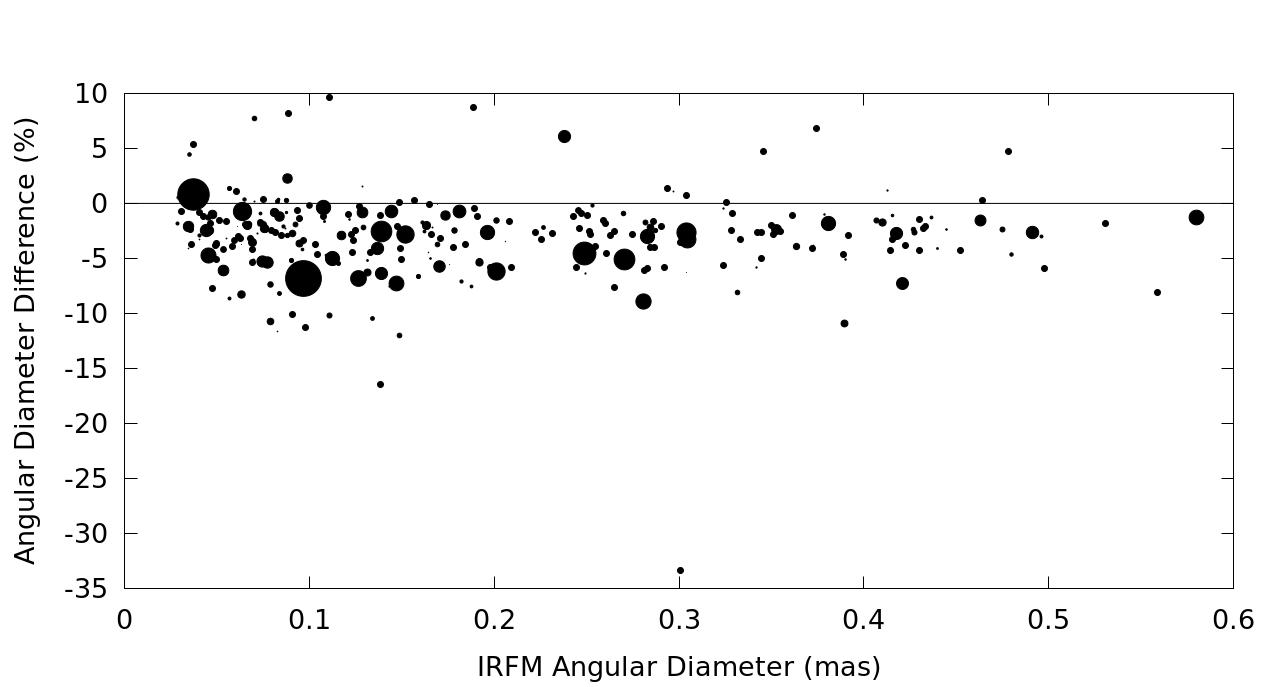}
\includegraphics[width=0.75\linewidth]{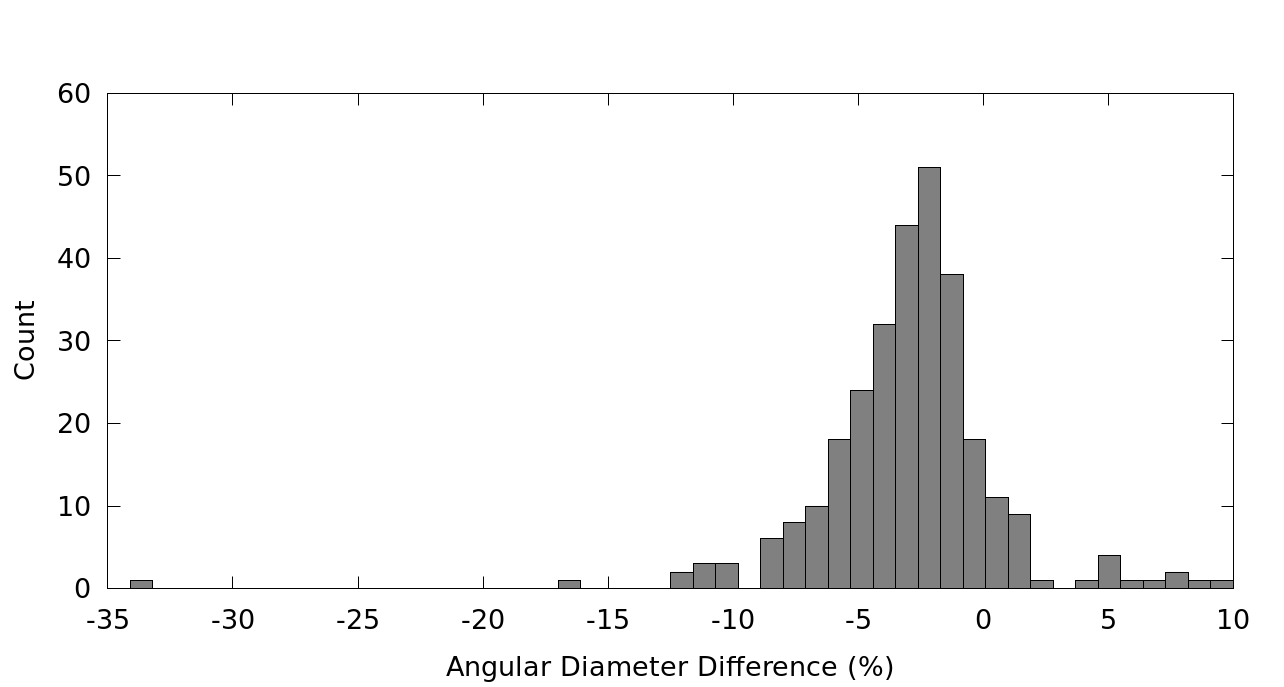}
\caption{\label{fig:sedangcomp} \emph{Top:} SED versus iterative IRFM angular diameters. The black line shows what would be perfect agreement between the two methods. \emph{Middle:} Fractional difference between the IRFM and SED values relative to the IRFM values. In both panels, larger points denote higher $\chi^2_{\nu}$ values. The median and standard deviation of the fractional differences are, respectively, -2.7\% and 3.6\%.\emph{Bottom:} Histogram showing the absolute difference between the IRFM and SED angular diameters.}
\end{center}
\end{figure*}

We sample the effective temperature at the \citet{Casagrande2010} listed value as well as $\pm 1\sigma$. We fix $\loggstar$ and \feh\ to the \citet{Casagrande2010} values, rounded to the nearest values for which a Kurucz model atmosphere exists. We sample ten extinctions from $A_V =0$ to the maximum line-of-sight extinction from the \citet{Schlegel1998} dust maps. We show our SEDs in Appendix \ref{sec:sed_appendix}. 
Figures \ref{fig:sedavcomp1}, \ref{fig:sedavcomp2},\ref{fig:sedteffcomp}, \ref{fig:sedfbolcomp}, \ref{fig:sedangcomp} compare the iterative IRFM extinctions, effective temperatures, bolometric fluxes, and angular diameters, respectively, to those from the SEDs. We find generally good agreement between the two methods. We note that, for the majority of stars, the SED and iterative IRFM extinctions agree quite well -- as seen in the bottom two panels of Figure \ref{fig:sedavcomp2} -- though we overestimate the extinctions relative to the SED values for a few systems in the tail of the aforementioned histogram. Correspondingly, we overestimate the effective temperatures and bolometric fluxes for these stars and underestimate the angular diameters relative to the model SEDs at the level of a few percent. We see no trend with unscaled reduced $\chi^2$. The model SED fits to the photometry are shown in Figure \ref{fig:seds} and can be seen to fit the data well overall. 

\begin{figure*}
\begin{center}
\includegraphics[width=0.75\linewidth]{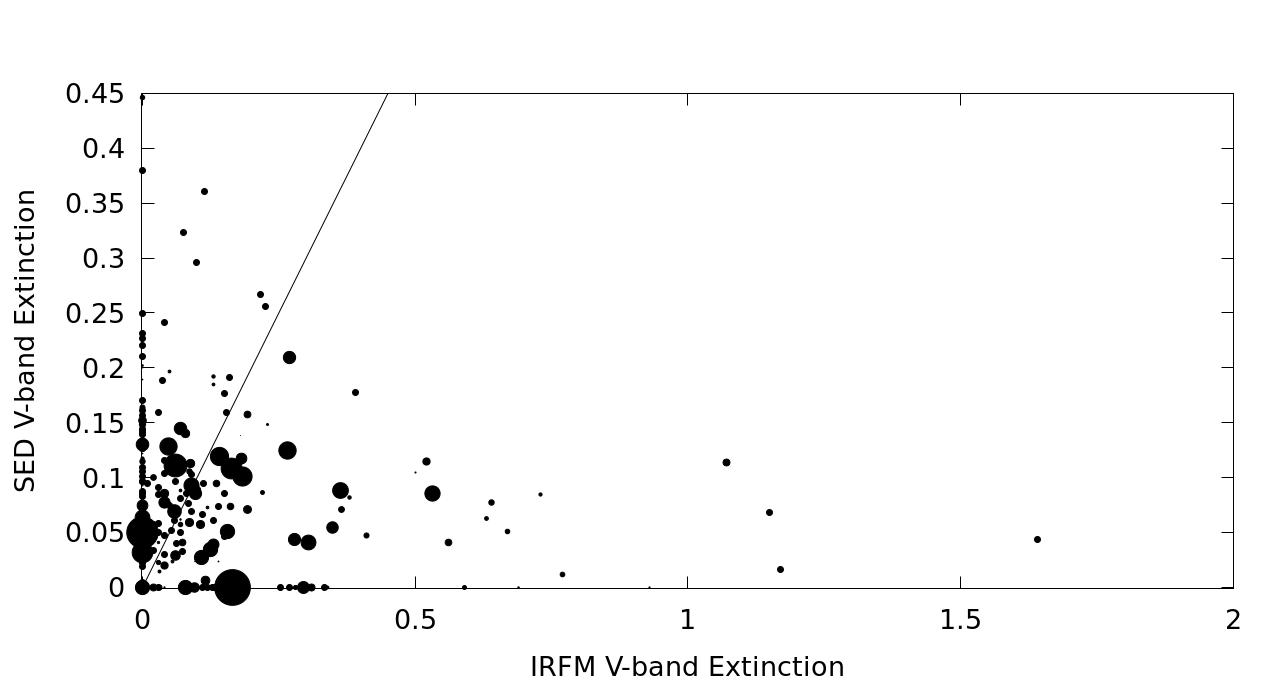}
\caption{\label{fig:sedavcomp1} SED versus iterative IRFM extinction. The black line shows what would be perfect agreement between the two methods.}
\end{center}
\end{figure*}

\section{Results}\label{sec:results}

\subsection{Effective Temperatures, Bolometric Fluxes, and Angular Diameters}\label{sec:params}
For  each subsample, Tables \ref{tab:bulk} and \ref{tab:bulk2} list the number of stars and median values plus 68\% confidence interval values for the unscaled $\chi^2_{\nu}$ and the fractional precision on the effective temperature, bolometric flux, and angular diameter for the full sample of stars and the subset of stars with radii, respectively. We note that the larger median \teff\ uncertainties for the spectroscopic sample, which lead to larger median angular diameter and radius uncertainties, are the result of including the nonzero $\chi^2$ penalty when re-scaling the uncertainties. This suggests that the uncertainties are likely underestimated.

We attempt to inflate these lower uncertainties by repeating the iterative IRFM method on the stars with spectroscopic parameters without applying the \teff\ prior. We then bin the spectroscopic stars according to the IRFM \teff\ without the prior, in bins of 100K. In each bin, we calculate the median $A_{\emph{V}}$, \teff\ and \fbol uncertainties with the prior and calculate the ratio of these median uncertainties to the median uncertainties without the prior. Finally, we bin up the non-spectroscopic stars in the same \teff\ bins and, for each star in each bin, we multiply the $A_{\emph{V}}$, \teff, and \fbol uncertainties by these ratios. The scaled uncertainties on \teff\ and \fbol as well as the propagated uncertainties on the angular diameter and linear radius are given as the last columns in Tables \ref{tab:bulk} and \ref{tab:bulk2}. We emphasize again that we \emph{post-hoc} inflate the parameter uncertainties for the stars without spectroscopic priors to try to compensate for our method's underestimation of these uncertainties. The full SED fits that we will perform in future work will elucidate the accuracy of these uncertainties.

\begin{figure*}
\includegraphics[width=0.5\linewidth]{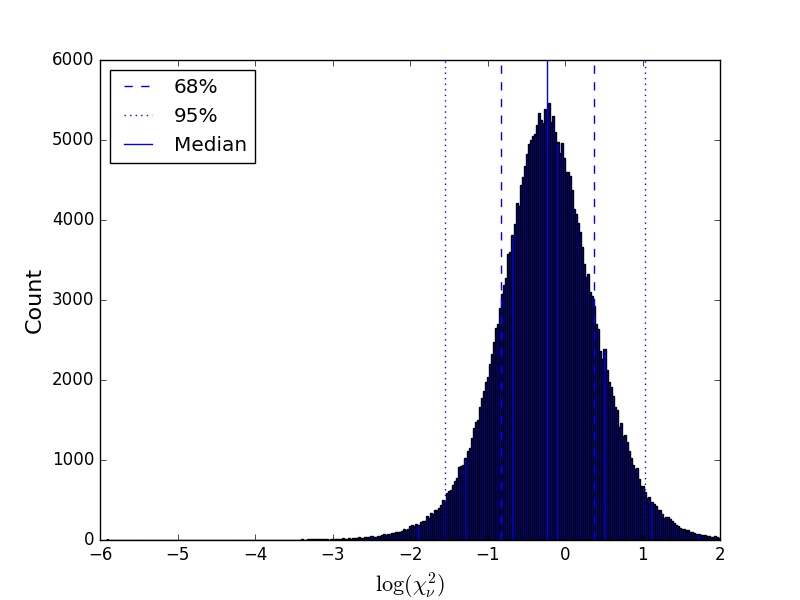}\includegraphics[width=0.5\linewidth]{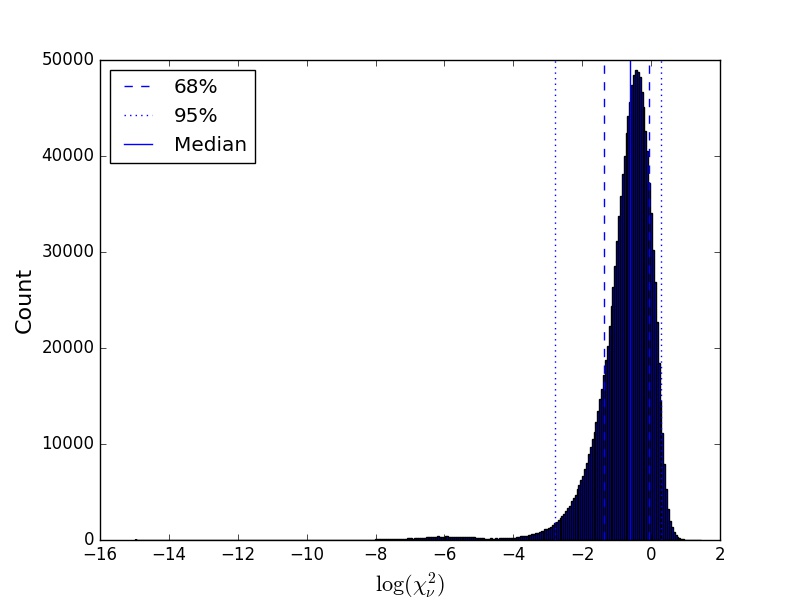}
\caption{\label{fig:chi2} $\log \chi^2_{\nu}$ distributions for the stars with spectroscopic parameters (left) and without (right). In each panel, the solid line denoted the median value, the dashed lines denote the 68\% interval, and the dotted lines denote the 95\% interval. }
\end{figure*}

\begin{figure*}
\includegraphics[width=0.5\linewidth]{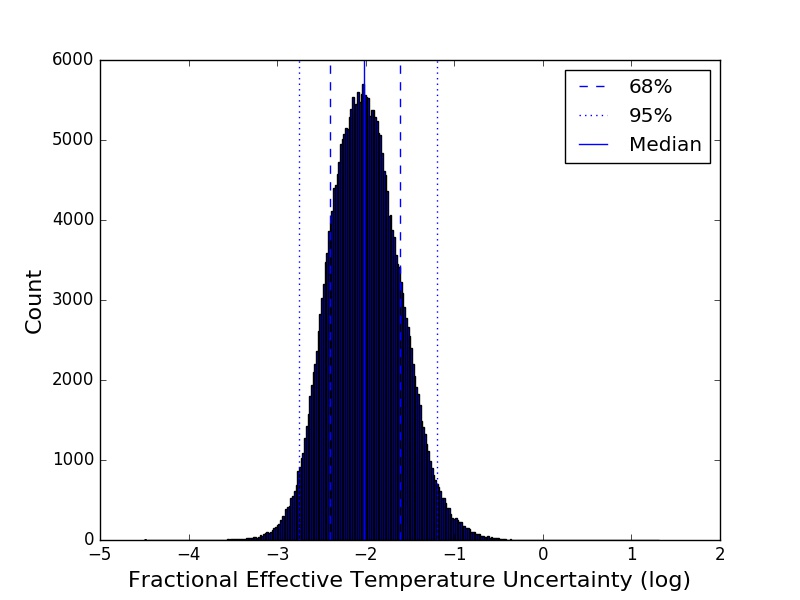}\includegraphics[width=0.5\linewidth]{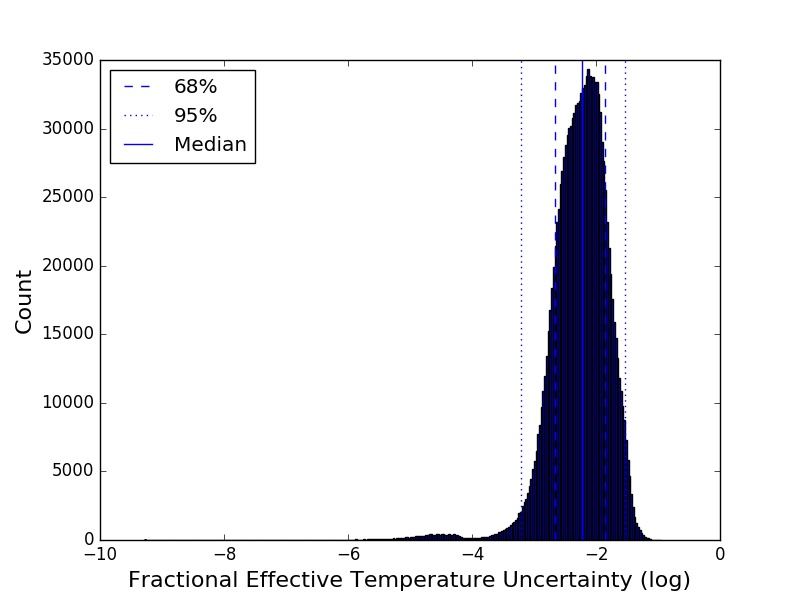}
\caption{\label{fig:sigteff} Histograms showing the fractional precision on the effective temperature for the stars with spectroscopic parameters (left) and without (right). In each panel, the solid line denoted the median value, the dashed lines denote the 68\% interval, and the dotted lines denote the 95\% interval.}
\end{figure*}

\begin{figure*}
\includegraphics[width=0.5\linewidth]{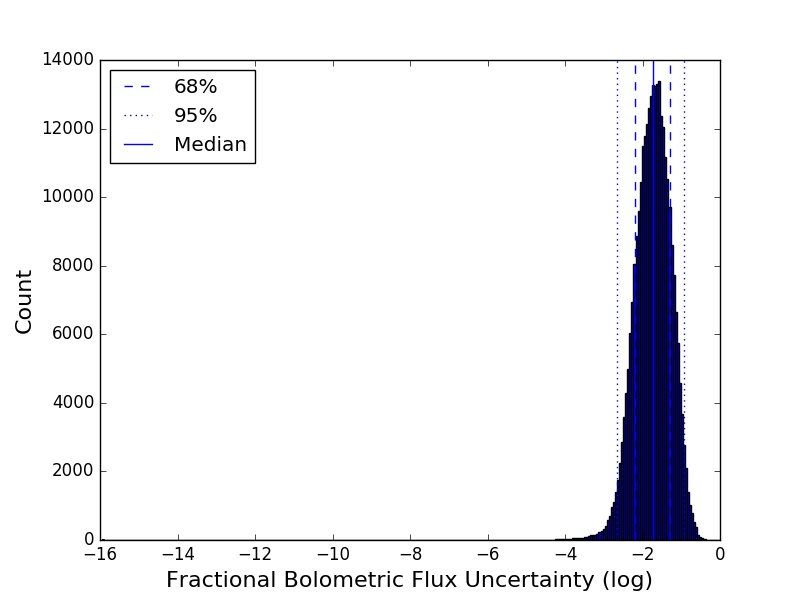}\includegraphics[width=0.5\linewidth]{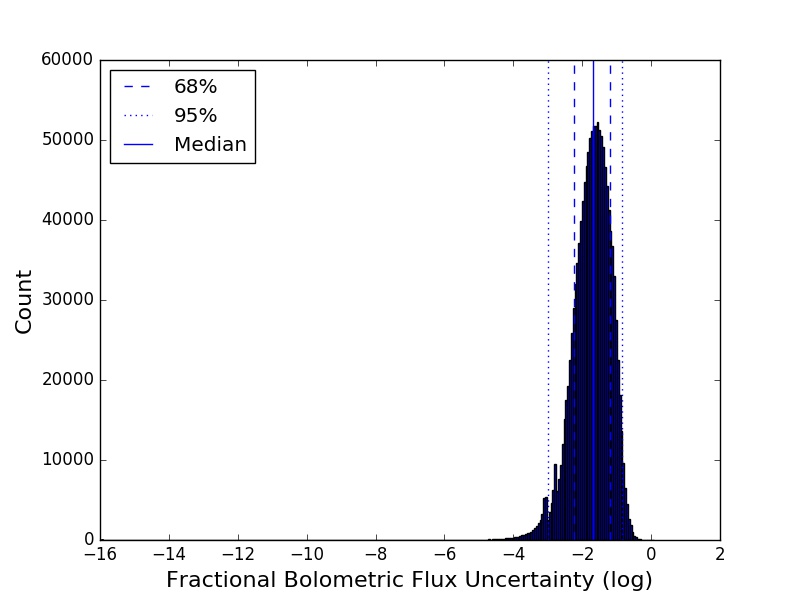}
\caption{\label{fig:sigfbol} Histograms showing the fractional precision on the bolometric flux for the stars with spectroscopic parameters (left) and without (right). In each panel, the solid line denoted the median value, the dashed lines denote the 68\% interval, and the dotted lines denote the 95\% interval.}
\end{figure*}

\begin{figure*}
\includegraphics[width=0.5\linewidth]{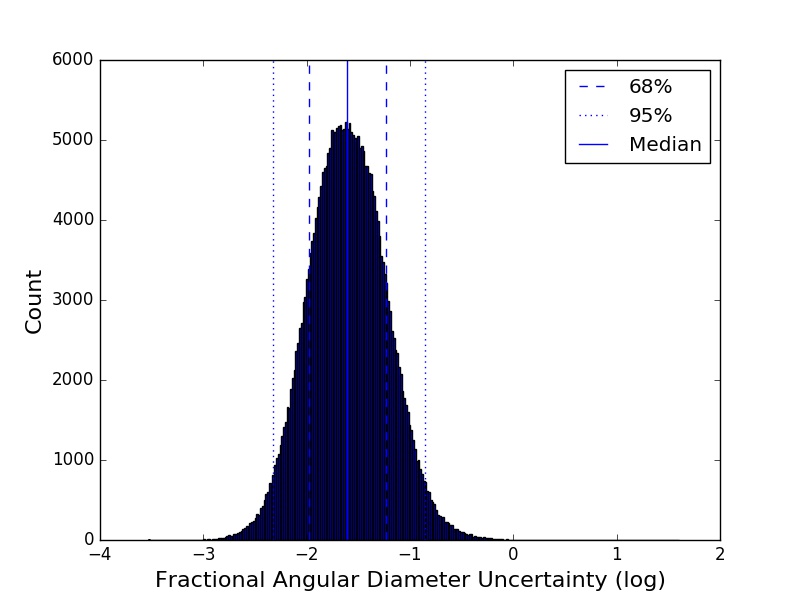}\includegraphics[width=0.5\linewidth]{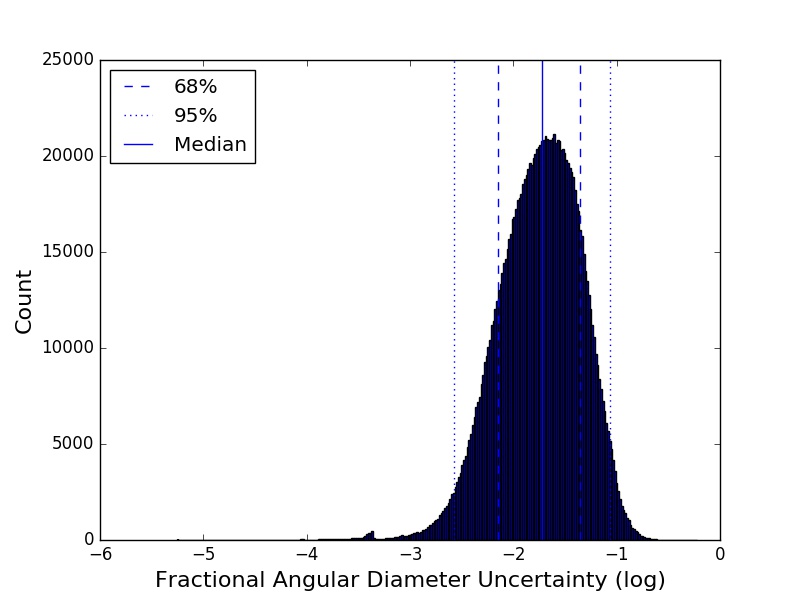}
\caption{\label{fig:sigtheta} Histograms showing the fractional precision on the angular diameter for the stars with spectroscopic parameters (left) and without (right). In each panel, the solid line denoted the median value, the dashed lines denote the 68\% interval, and the dotted lines denote the 95\% interval.}
\end{figure*}

Figures \ref{fig:chi2} shows the $\chi^2$ distributions, while Figures \ref{fig:sigteff}, \ref{fig:sigfbol}, and \ref{fig:sigtheta} show the fractional effective temperature precision, fractional bolometric flux precision, and fractional angular diameter distributions for the four subsamples, respectively.

We emphasize that these bolometric fluxes and angular diameters are fundamental, empirical products of this study that can be utilized to determine linear radii and other quantities as the {\it Gaia\/} parallaxes improve with upcoming data releases.
We present the determined extinctions, temperatures, bolometric fluxes, and angular diameters for these stars in Table~\ref{tab:allresults}.

\subsection{Radii}\label{sec:radii}
We queried the \emph{Gaia} DR1 archive for parallaxes and proper motions from the Tycho-Gaia Astrometric Solution for our $\sim$1,600,080 stars. Of this sample, 1,289,988 stars have \emph{Gaia} parallaxes and proper motions. 1,280,289 of these stars have non-negative parallaxes; we suspect that the negative parallaxes quoted for some stars are due to orbital motion in binary and multiple stellar systems, as \emph{Gaia} DR1 treats all sources as single stars. Of these, 1,153,804 of these stars have fractional parallax uncertainties better than 50\%, while 355,502 have fractional parallax uncertainties better than 10\%.

\begin{figure*}
\includegraphics[width=0.5\linewidth]{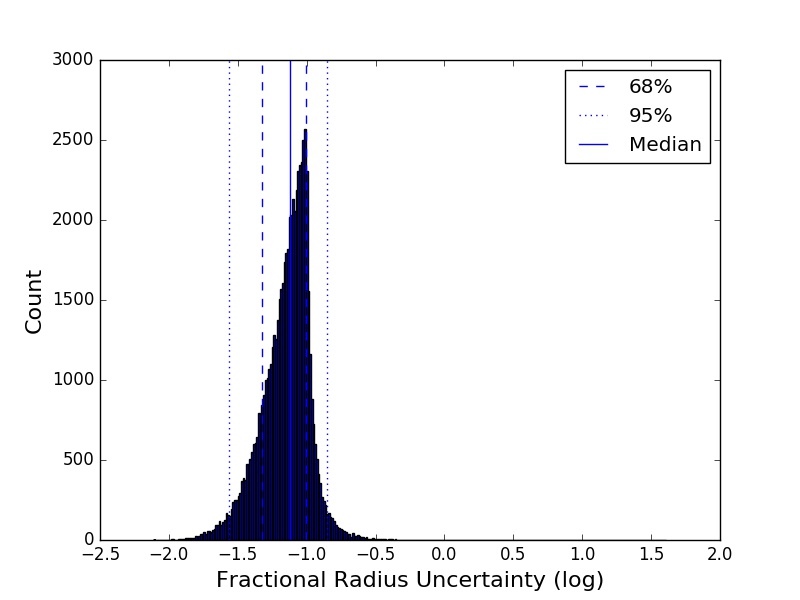}\includegraphics[width=0.5\linewidth]{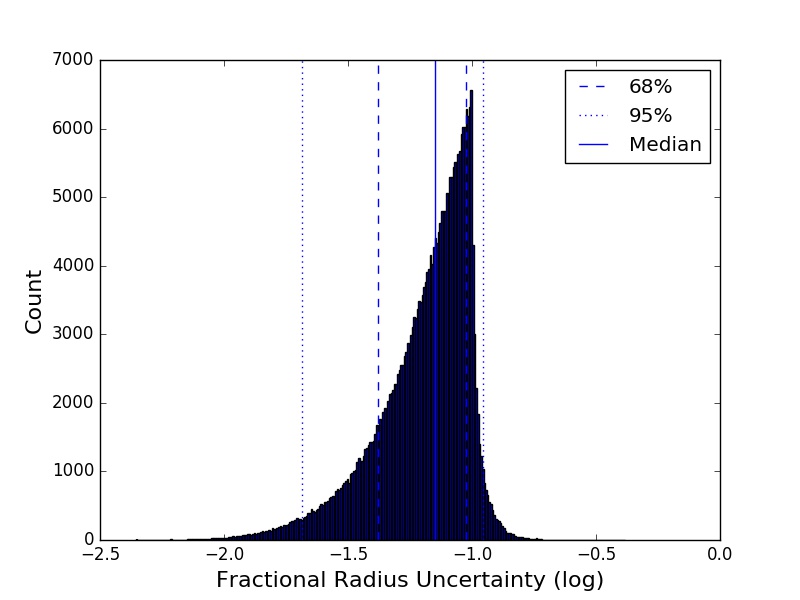}
\caption{\label{fig:sigr} Histograms showing the fractional precision on the radius for the subset of stars with spectroscopic parameters (left) and without (right) that have precise TGAS parallaxes. In each panel, the solid line denoted the median value, the dashed lines denote the 68\% interval, and the dotted lines denote the 95\% interval. The effect of our parallax precision cut manifests as a large drop-off around 10\%.}
\end{figure*}

We restrict ourselves to stars with parallax uncertainties $\la 10\%$ not least of all because stars with worse parallax precision are more subject to Lutz-Kelker bias \citep{Lutz1973}: specifically, the observed parallax systematically exceeds the true parallax by an amount that increases with increasing fractional parallax uncertainty. The effect of our precision cut can be seen in Figure \ref{fig:sigr} as a steep decline in the number of stars with fractional radius uncertainties larger than 10\%.

\begin{figure*}
\includegraphics[width=1.0\linewidth]{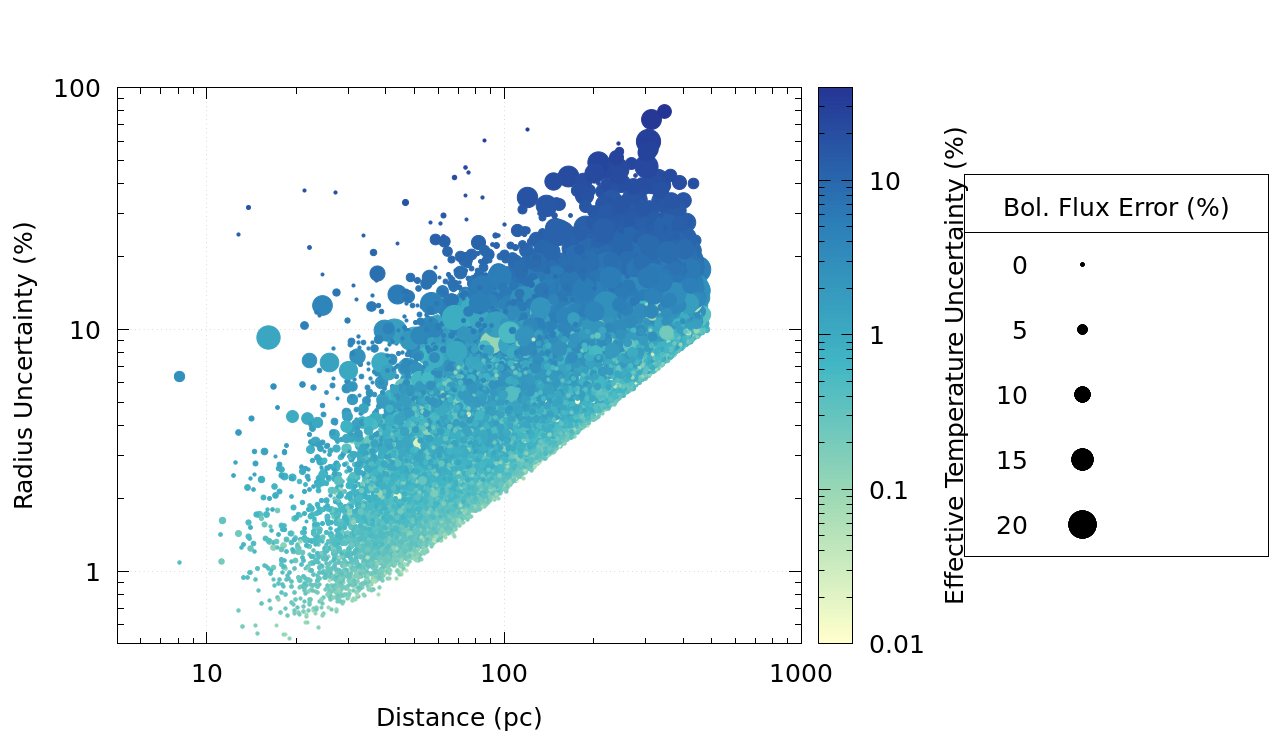}
\includegraphics[width=1.0\linewidth]{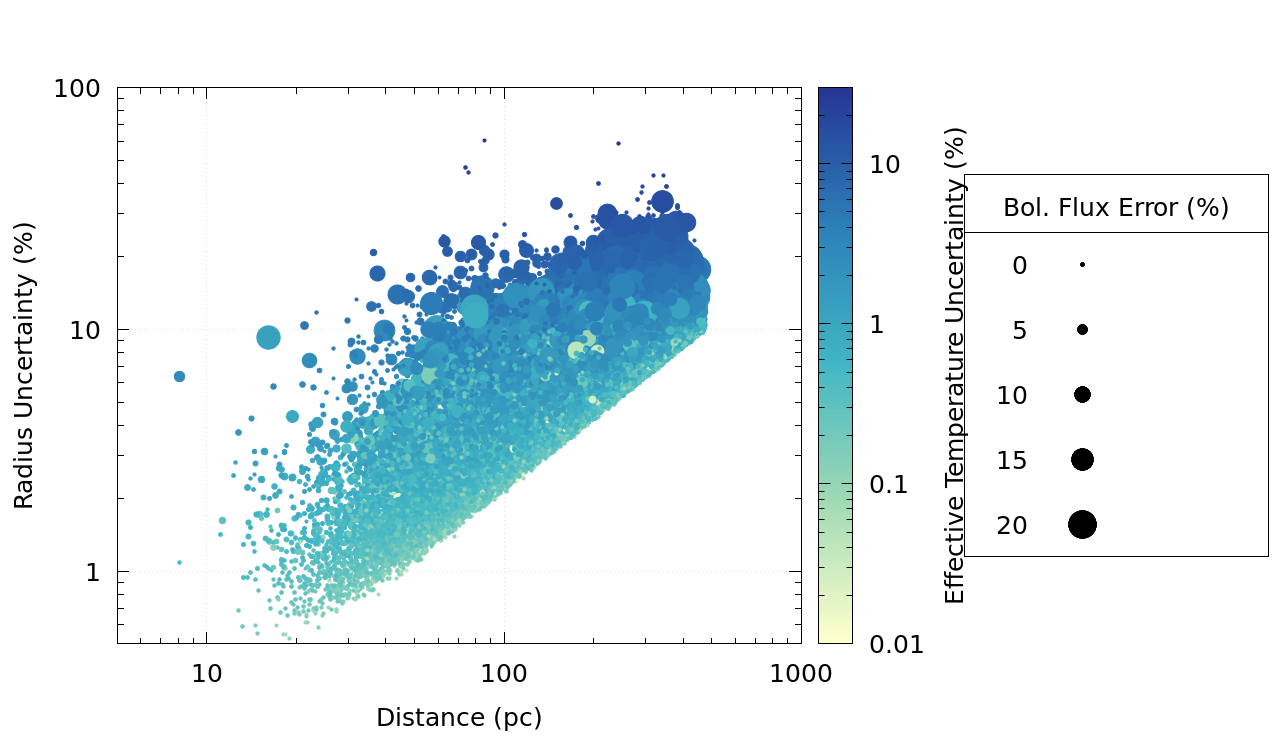}
\caption{\label{fig:rd} Fractional radius uncertainty as a function of \emph{Gaia} parallax distance for the stars with spectroscopic parameters (top) and without (bottom). The points are color-coded by the fractional effective temperature uncertainty and sized by the fractional bolometric flux uncertainty; for presentation purposes, we only plot stars with $<20\%$ bolometric flux uncertainties. The parallax uncertainties set the floor of the radius uncertainties, and a strong vertical color gradient highlights the effective temperature uncertainty's dominance on the radius error budget.}
\end{figure*}

We calculate the distances and radii for these stars. We list the median and 68\% confidence interval values for the reduced $\chi^2$ and fractional effective temperature, bolometric flux, angular diameter, and radius uncertainties in Table \ref{tab:bulk2}, and we list the values themselves in Table \ref{tab:allresults2}. We achieve $<10\%$ median uncertainties on the stellar radius in all four subsamples. Following Equations \ref{eq:sigr} and \ref{eq:sigtheta} and as shown in Figure \ref{fig:rd}, we see that our quoted effective temperature uncertainties of a couple percent -- along with the several-percent uncertainties on the parallaxes -- dominate the radius error budget.

In both panels of Figure \ref{fig:rd}, we see that the radius uncertainty is fundamentally bounded by the increasing parallax uncertainty at greater distances. If \emph{Gaia} achieves the predicted end-of-mission uncertainty of $\sim 20~\mu$as for bright stars, then obtaining sufficiently precise (and, given the required precision, accurate) effective temperatures becomes the paramount challenge to obtaining precise radii, particularly at the $3-5\%$ level.

\subsection{Empirical Hertzsprung-Russell Diagram}

\begin{figure*}
\includegraphics[width=1.0\linewidth]{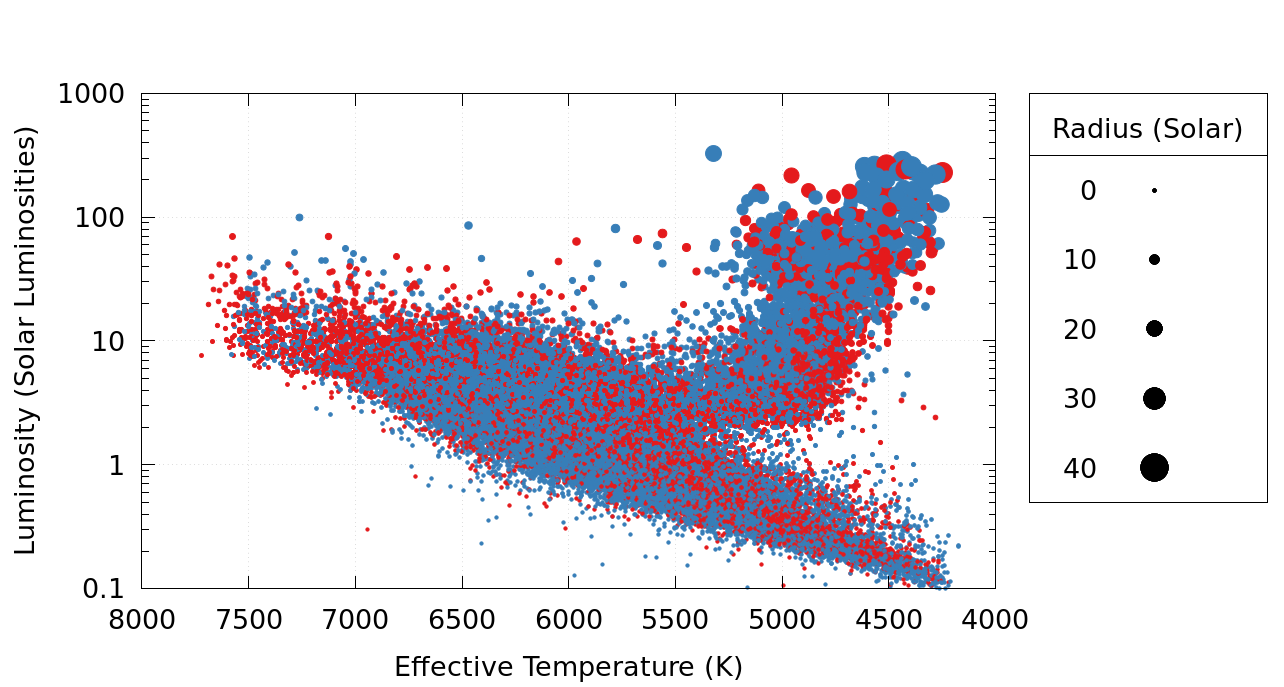}
\includegraphics[width=1.0\linewidth]{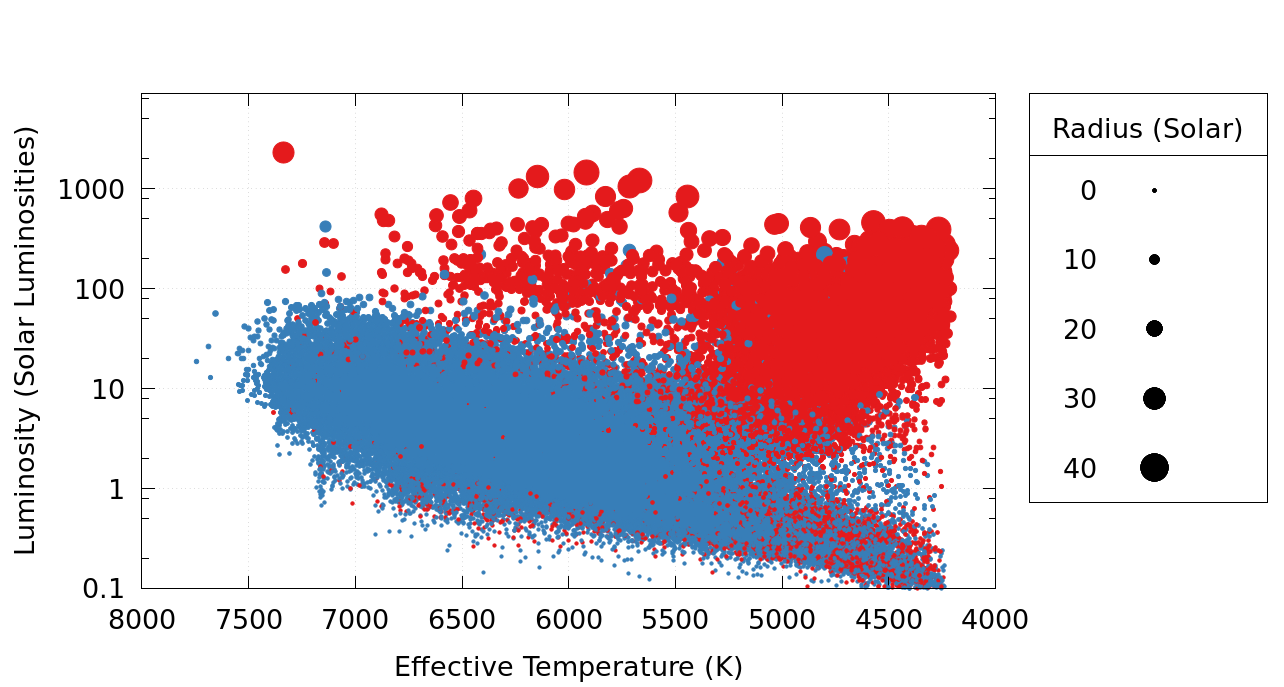}
\caption{\label{fig:hrd} Theoretical HR diagrams for the subsample of stars with $<10\%$ \emph{Gaia} parallaxes, with (top) and without (bottom) spectroscopic parameters. The top panel is color-coded by metallicity such that stars with \feh > 0 are red and stars with \feh < 0 are blue. In the bottom panel, the stars that the reduced proper motion cut identifies as dwarfs are in blue, while the rest are in red. Point size increases with radius. For presentation purposes, we show only stars with $L_{\rm bol}>0.1~L_{{\rm bol},\odot}$. In both panels, a cluster of cool, evolved stars distinguishes itself from the main sequence. We note that the cluster of hot, luminous stars to the upper-left of the non-spectroscopic HR diagram, which reflects the limitations of our technique for stars > 7,000K.}
\end{figure*}

Figure \ref{fig:hrd} shows luminosity-temperature diagrams (`theorists' HR diagrams) for our stars. Dwarf, subgiant, and giant populations are coarsely identifiable by eye.  The top panel, corresponding to the spectroscopic sample, also shows distinct groupings according to metallicity. Specifically, the metal-poor stars sit below the main sequence and to the left of the giant branch, as expected: metal-poor main-sequence stars tend to be slightly hotter and smaller than their metal-rich counterparts.  The relative abundance of early-type and evolved stars versus late-type non-evolved stars is likely the result of Malmquist bias. 

\begin{deluxetable}{lrrrrr}
\tablecaption{\label{tab:bulk} Median Values of and 68\% Confidence Intervals for 1,600,080 Tycho-2 Stars with Iterative-IRFM Temperatures, Fluxes, and Angular Diameters}
\tablehead{
\colhead{Parameter} & \multicolumn{4}{c}{Value}\\
 & \colhead{LAMOST} & \colhead{RAVE} & \colhead{APOGEE} & \multicolumn{2}{c}{No Prior}\\ & & & & \colhead{(Unscaled)} & \colhead{(Scaled)}
}
\startdata
Number of stars & 64,345 & 214,707 & 14,360 & 1,306,668 & 1,306,668\\ %230,809\\
Median Unscaled $\chi^2_{\nu}$ & $0.42_{-0.32}^{+1.21}$ &$0.64_{-0.47}^{+1.94}$ &$0.49_{-0.37}^{+2.04}$ &  $0.25_{-0.21}^{+0.60}$ &  $0.25_{-0.21}^{+0.60}$\\ 
Median $\sigma_{\teff}/\teff\ (\%) $ & $ 1.14_{-0.67}^{+1.63}$ &$ 0.90_{-0.52}^{+1.44}$ &  $ 0.75_{-0.44}^{+1.62}$ & $ 0.59_{-0.37}^{+0.83}$ & $1.05_{-0.68}^{1.61}$\\ 
Median $\sigma_{F_{\rm bol}}/F_{\rm bol}\ (\%)$ & $2.69_{-1.95}^{+4.83}$ & $1.77_{-1.13}^{+2.87}$ & $1.64_{-1.08}^{+2.99}$ & $2.07_{-1.50}^{+4.31}$ & $1.86_{-1.35}^{+3.94}$ \\
Median $\sigma_{\theta}/\theta\ (\%)$ &$3.19_{-1.89}^{+3.91}$ &$2.28_{-1.27}^{+3.14}$ & $2.04_{-1.19}^{+3.69}$ &$1.88_{-1.18}^{+2.57}$ & $2.69_{-1.69}^{+3.63}$\\ 
\enddata
\end{deluxetable}

\begin{deluxetable}{lrrrrr}
\tablecaption{\label{tab:bulk2} Median Values of and 68\% Confidence Intervals for 355,502 Tycho-2 Stars with \emph{Gaia} Astrometry}
\tablehead{
\colhead{Parameter} & \multicolumn{5}{c}{Value}\\
 & \colhead{LAMOST} & \colhead{RAVE} & \colhead{APOGEE} & \multicolumn{2}{c}{No Prior}\\ & & & & \colhead{(Unscaled)} & \colhead{(Scaled)}
}
\startdata
Number of stars & 12,874 & 49,405 & 2,681 & 290,542 & 290,542 \\
Median Unscaled $\chi^2_{\nu}$ & $0.38_{-0.28}^{+0.97}$ & $1.05_{-0.78}^{+2.72}$ & $0.72_{-0.53}^{+2.63}$ & $0.28_{-0.23}^{+0.59}$ & $0.28_{-0.23}^{+0.59}$ \\
Median $\sigma_{\teff}/\teff\ (\%) $ & $0.93_{-0.54}^{+1.16}$ & $1.07_{-0.60}^{+1.32}$ & $ 0.69_{-0.41}^{+1.45}$ &$ 0.50_{-0.29}^{+0.67}$ &$ 0.90_{-0.55}^{+1.27}$ \\
Median $\sigma_{F_{\rm bol}}/F_{\rm bol}\ (\%)$ & $2.20_{-1.57}^{+3.85}$ & $1.33_{-0.87}^{+2.43}$ & $0.90_{-0.58}^{+1.93}$ &$1.40_{-0.97}^{+3.09}$ & $1.30_{-0.91}^{+2.92}$ \\
Median $\sigma_{\theta}/\theta\ (\%)$ & $2.38_{-1.42}^{+2.85}$ & $2.42_{-1.35}^{+2.87}$ & $1.77_{-1.02}^{+3.45}$ & $1.42_{-0.86}^{+2.10}$ & $2.17_{-1.32}^{+3.03}$ \\
Median $\sigma_{\rstar}/\rstar\ (\%) $& $7.95_{-2.68}^{+2.08}$ & $7.40_{-2.76}^{+2.46}$ & $6.86_{-3.11}^{+2.97}$ & $7.08_{-2.88}^{+2.36}$& $7.47_{-3.01}^{+2.41}$ \\
\enddata
\end{deluxetable}

\section{Discussion}\label{sec:disc}

\subsection{Radii of Low-mass Stars}\label{subsec:inter}
\begin{figure*}
\includegraphics[width=1.0\linewidth]{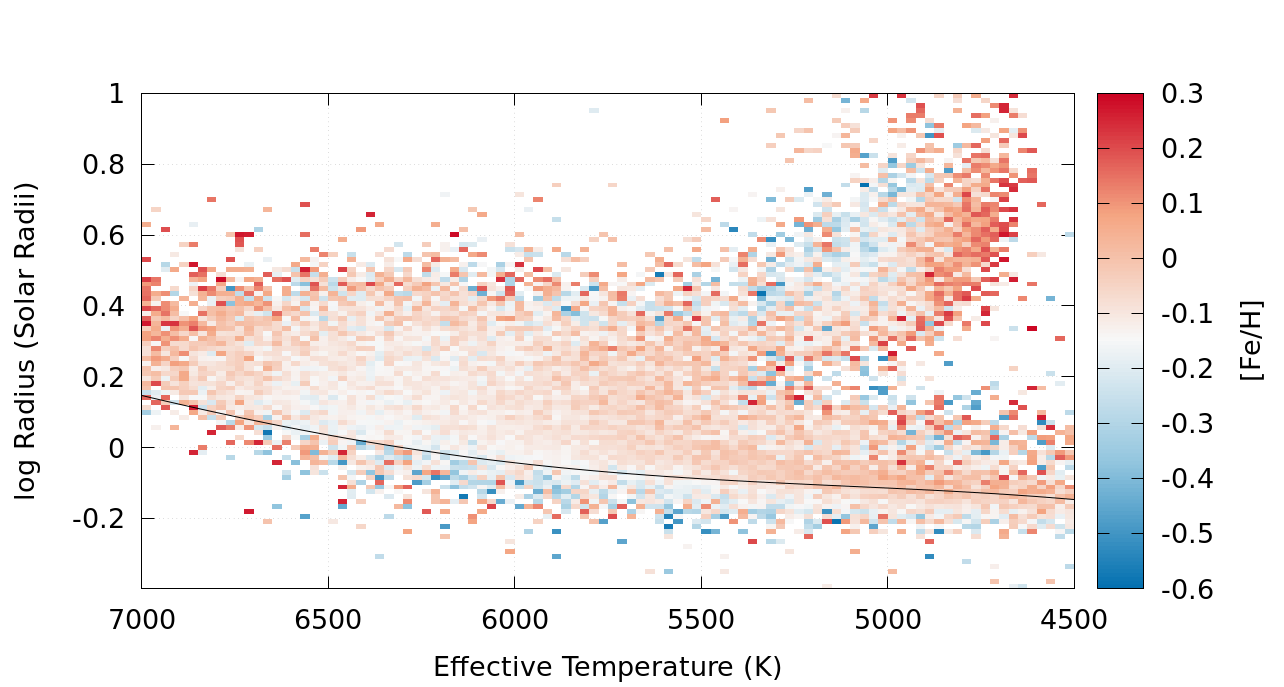}
\includegraphics[width=1.0\linewidth]{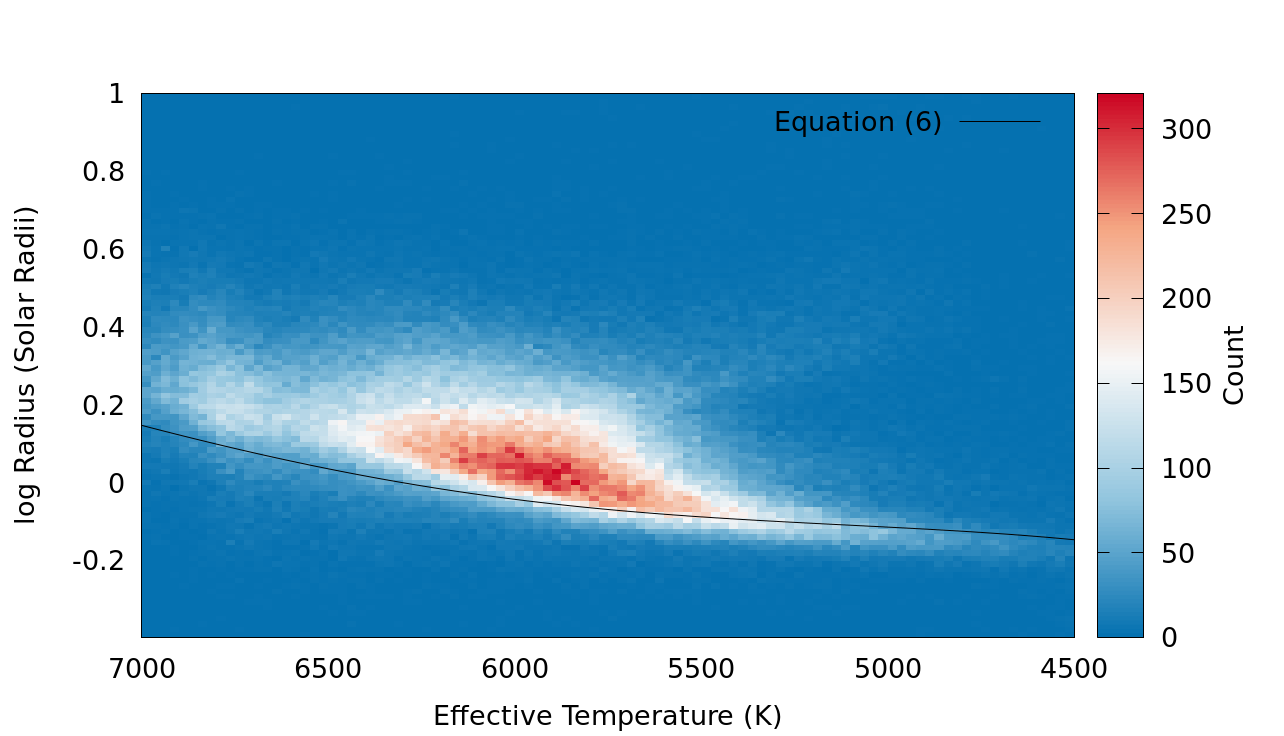}
    \caption{\label{fig:rteff} Radius as a function of effective temperature for the spectroscopic (top) and non-spectroscopic (bottom) star samples. The top panel is restricted to stars with a $\loggstar > 3$, while the bottom panel is restricted to stars identified as dwarfs by the reduced proper motion cut. The black line is the relation given by Equation \ref{eq:rteff} (Equation 9 of \citealt{Boyajian2012b}).}
    \end{figure*}

We compare the radii derived for our cool stars to the radii predicted by the \teff-radius relation given in Equation 9 of \citep{Boyajian2012b}. This relation is calibrated on a sample of 33 K- and M-dwarfs ($3200{\rm K} \leq \teff \leq 5500{\rm K}$) plus the Sun, which is used as a bridge to extrapolate the relation to hotter stars. The relation, for stars with \teff\ > 3,200K, is
\begin{eqnarray}
\label{eq:rteff}
\rstar/R_{\odot} &	= 	&	-8.133 (\pm 0.226) + 5.09342 (\pm 0.16745) \times 10^{-3} \teff \nonumber \\
					&		&	- 9.86602 (\pm 0.40672) \times 10^{-7} \teff^2 \nonumber \\
					&       &   + 6.47963 (\pm 0.32429) \times 10^{-11} \teff^3
\end{eqnarray}
Figure \ref{fig:rteff} shows radius as a function of effective temperature for our stars, with the aforementioned relation plotted as the black line. We excluded giants from our comparison that were identified by either the $\loggstar$ cut or the reduced proper motion cut described in Section \ref{subsec:limits}. We find good agreement over the calibrated temperature range between our dwarfs and this relation, even out to dwarfs hotter than the Sun. Additionally, the top panel shows the metal-poor stars sitting below the metal-rich stars on the main sequence and to the left of the metal-rich stars on the giant branch.

\subsection{Limitations of the Iterative IRFM Technique}\label{subsec:limits}
We note that our iterative IRFM technique has several limitations, and we caution the reader against unscrupulous application of the results presented in this paper. First and foremost, the IRFM relations from \citet{Casagrande2010} were calibrated using dwarfs and subgiants, so the resulting effective temperatures and bolometric fluxes for giants have not been verified. To this end, we identify stars with spectroscopic $\loggstar \leq 3$ as giants. For the stars without spectroscopic gravities, we use the \emph{Gaia} DR1 proper motions and 2MASS $J$- and $H$-band photometry to calculate reduced proper motions. We then apply the reduced proper motion cut \citep{Gould2003} as described by \citep{Collier2007} to flag giants.

We also note that one could derive more accurate results for a given star if the extinction was known \emph{a priori}, and that having extinctions for the nearly one million Tycho-2 stars that were cut from our sample would enable us to estimate their temperatures, bolometric fluxes, angular diameters, and radii -- given a precise enough TGAS astrometric solution. 

Finally, as discussed in Section \ref{sec:params}, the unscaled parameter uncertainties for the stars without spectroscopic priors are likely understated, given that the median uncertainties on \teff (and thus the angular diameter) were a factor of two lower than for the stars with spectroscopic priors. This is because our $\chi^2$ merit function includes a nonnegative penalty for the stars with spectroscopic \teff\ priors, so when we re-scale the \teff\ uncertainties to force $\chi^2_{\nu} = 1$, this scale factor is larger for the stars with \teff\ priors than for those without. Thus, for each star with a spectroscopic \teff\, we run our analysis with and without applying a \teff\ prior (and penalty); bin the stars according to the IRFM \teff\ determined without applying the prior; find the ratio of the median uncertainties on the parameters with and without the prior; partition the stars without spectroscopic \teff\ values into the same \teff\ bins; and scale the uncertainties by these ratios. We can compare this approach to the uncertainties determined from full SED fits, which we will perform in future work.

\section{Conclusion}\label{sec:conclusion}
We have determined effective temperatures, bolometric fluxes and angular diameters for over 1.6 million \emph{Tycho-2} stars plus linear radii for 355,502 of these stars that have $<10\%$ \emph{Gaia} parallaxes. We demonstrate the ability to create the theorists' HR diagram directly from measured quantities using a substantially larger set of stars than were available in the \emph{Hipparcos} era. This is the largest collection of empirical stellar angular diameters and radii and should thus serve as a canonical sample for stellar and exoplanetary investigations as outlined in the introduction. 

Now, in the era of precision astronomy enabled by \emph{Gaia}, we have the opportunity to determine the fundamental parameters of a large sample of stars {\it empirically} and to considerably tighter precision and accuracy than has previously been possible. Because $\sigma_{\rstar}/\rstar \sim 2\sigma_{T_{\rm eff}}/T_{\rm eff}$ (when \teff\ uncertainties dominate all other sources of error) and because our knowledge of the extinction to each star is \emph{a priori} poor, the uncertainties on effective temperature and extinction will fundamentally limit our ability to measure the radii precisely and accurately if \emph{Gaia} reaches its expected end-of-mission astrometric precision of $\sim 10\mu$~as for bright stars. We need precise (several percent or better) effective temperatures -- of order the precision quoted by \citet{Brewer2016} for $\sim$1,600 F, G, and K stars -- and we need to resolve the discrepancies and systematic offsets between different methods of inferring stellar effective temperature (cf. \citealt{Boyajian2012b}).

In the next few years, there are great prospects both for dramatically increasing the sample of stars with precise radii and for improving the precision on these radii. \emph{Gaia}'s improved precision and expanded astrometric catalog will yield a considerably larger and more precise set of stellar radii, with potentially more than 600,000 stars with parameters derived from our iterative IRFM technique presented here. The release of either \emph{Gaia} effective temperatures or the blue-pass and red-pass spectrophotometry slated for 2018 will also enable us to expand our sample to include stars for which this information is made available. 

The BP/RP spectrophotometry in particular will make precise SED modeling possible, particularly for AFGK stars whose SEDs peak in the 330-1050nm range of the BP/RP filters. Combining literature broad-band photometry, \emph{Gaia} spectrophotometry, and spectrophotometry from $\sim 0.75-5\mu$m from the proposed SPHEREx mission \citep{Dore2016} will capture nearly all the flux for these stars, cover the SED peaks for later-type stars, and enable direct measurements of the line-of-sight extinction as a function of wavelength, without reliance on previously calibrated extinction laws.

\acknowledgments
Work by B.S.G. and D.J.S was partially supported by NSF CAREER Grant AST-1056524. D.J.S. thanks R. Oelkers for assistance querying the TGAS catalog; J. Birkby, C. Dressing, J. Johnson, M. Pinsonneault, J. Tayar, A. Vanderburg, and J. Yee for fruitful scientific discussion and S. Kim, M. Penny, and D. Will for computational advice.

This work has made use of NASA's Astrophysics Data System, the SIMBAD database operated at CDS, Strasbourg, France, the VizieR catalogue access tool, CDS, Strasbourg, France \citep{Ochsenbein2000}, and the Two Micron All Sky Survey, which is a joint project of the University of Massachusetts and the Infrared Processing and Analysis Center/California Institute of Technology, funded by the National Aeronautics and Space Administration and the National Science Foundation.

The research described in this paper makes use of Filtergraph, an online data visualization tool developed at Vanderbilt University through the Vanderbilt Initiative in Data-intensive Astrophysics (VIDA; \citealt{Burger2013}).

Funding for the Sloan Digital Sky Survey IV has been provided by the
Alfred P. Sloan Foundation, the U.S. Department of Energy Office of
Science, and the Participating Institutions. SDSS acknowledges
support and resources from the Center for High-Performance Computing at
the University of Utah. The SDSS web site is \url{www.sdss.org}.

Funding for RAVE has been provided by: the Australian Astronomical Observatory; the Leibniz-Institut fuer Astrophysik Potsdam (AIP); the Australian National University; the Australian Research Council; the French National Research Agency; the German Research Foundation (SPP 1177 and SFB 881); the European Research Council (ERC-StG 240271 Galactica); the Istituto Nazionale di Astrofisica at Padova; The Johns Hopkins University; the National Science Foundation of the USA (AST-0908326); the W. M. Keck foundation; the Macquarie University; the Netherlands Research School for Astronomy; the Natural Sciences and Engineering Research Council of Canada; the Slovenian Research Agency; the Swiss National Science Foundation; the Science \& Technology Facilities Council of the UK; Opticon; Strasbourg Observatory; and the Universities of Groningen, Heidelberg and Sydney.
The RAVE web site is located at

\url{https://www.rave-survey.org}.

\bibliographystyle{aasjournal.bst}
%\bibstyle{aasjournal.bst}
\bibliography{main.bib}

\begin{thebibliography}{}
\expandafter\ifx\csname natexlab\endcsname\relax\def\natexlab#1{#1}\fi

\bibitem[{{Bianchi} {et~al.}(2011){Bianchi}, {Herald}, {Efremova}, {Girardi},
  {Zabot}, {Marigo}, {Conti}, \& {Shiao}}]{Bianchi2011}
{Bianchi}, L., {Herald}, J., {Efremova}, B., {et~al.} 2011, \apss, 335, 161

\bibitem[{{Birkby} {et~al.}(2012){Birkby}, {Nefs}, {Hodgkin}, {Kov{\'a}cs},
  {Sip{\H o}cz}, {Pinfield}, {Snellen}, {Mislis}, {Murgas}, {Lodieu}, {de
  Mooij}, {Goulding}, {Cruz}, {Stoev}, {Cappetta}, {Palle}, {Barrado},
  {Saglia}, {Martin}, \& {Pavlenko}}]{Birkby2012}
{Birkby}, J., {Nefs}, B., {Hodgkin}, S., {et~al.} 2012, \mnras, 426, 1507

\bibitem[{{Blanton} {et~al.}(2017){Blanton}, {Bershady}, {Abolfathi},
  {Albareti}, {Allende Prieto}, {Almeida}, {Alonso-Garc{\'{\i}}a}, {Anders},
  {Anderson}, {Andrews}, \& et~al.}]{Blanton2017}
{Blanton}, M.~R., {Bershady}, M.~A., {Abolfathi}, B., {et~al.} 2017, ArXiv
  e-prints, arXiv:1703.00052

\bibitem[{{Boyajian} {et~al.}(2014){Boyajian}, {van Belle}, \& {von
  Braun}}]{Boyajian:2014}
{Boyajian}, T.~S., {van Belle}, G., \& {von Braun}, K. 2014, \aj, 147, 47

\bibitem[{{Boyajian} {et~al.}(2012{\natexlab{a}}){Boyajian}, {McAlister}, {van
  Belle}, {Gies}, {ten Brummelaar}, {von Braun}, {Farrington}, {Goldfinger},
  {O'Brien}, {Parks}, {Richardson}, {Ridgway}, {Schaefer}, {Sturmann},
  {Sturmann}, {Touhami}, {Turner}, \& {White}}]{Boyajian2012a}
{Boyajian}, T.~S., {McAlister}, H.~A., {van Belle}, G., {et~al.}
  2012{\natexlab{a}}, \apj, 746, 101

\bibitem[{{Boyajian} {et~al.}(2012{\natexlab{b}}){Boyajian}, {von Braun}, {van
  Belle}, {McAlister}, {ten Brummelaar}, {Kane}, {Muirhead}, {Jones}, {White},
  {Schaefer}, {Ciardi}, {Henry}, {L{\'o}pez-Morales}, {Ridgway}, {Gies}, {Jao},
  {Rojas-Ayala}, {Parks}, {Sturmann}, {Sturmann}, {Turner}, {Farrington},
  {Goldfinger}, \& {Berger}}]{Boyajian2012b}
{Boyajian}, T.~S., {von Braun}, K., {van Belle}, G., {et~al.}
  2012{\natexlab{b}}, \apj, 757, 112

\bibitem[{{Brewer} {et~al.}(2016){Brewer}, {Fischer}, {Valenti}, \&
  {Piskunov}}]{Brewer2016}
{Brewer}, J.~M., {Fischer}, D.~A., {Valenti}, J.~A., \& {Piskunov}, N. 2016,
  \apjs, 225, 32

\bibitem[{{Burger} {et~al.}(2013){Burger}, {Stassun}, {Pepper}, {Siverd},
  {Paegert}, {De Lee}, \& {Robinson}}]{Burger2013}
{Burger}, D., {Stassun}, K.~G., {Pepper}, J., {et~al.} 2013, Astronomy and
  Computing, 2, 40

\bibitem[{{Campante} {et~al.}(2016){Campante}, {Schofield}, {Kuszlewicz},
  {Bouma}, {Chaplin}, {Huber}, {Christensen-Dalsgaard}, {Kjeldsen}, {Bossini},
  {North}, {Appourchaux}, {Latham}, {Pepper}, {Ricker}, {Stassun},
  {Vanderspek}, \& {Winn}}]{Campante:2016}
{Campante}, T.~L., {Schofield}, M., {Kuszlewicz}, J.~S., {et~al.} 2016, ArXiv
  e-prints, arXiv:1608.01138

\bibitem[{{Cardelli} {et~al.}(1989){Cardelli}, {Clayton}, \&
  {Mathis}}]{Cardelli:1989}
{Cardelli}, J.~A., {Clayton}, G.~C., \& {Mathis}, J.~S. 1989, \apj, 345, 245

\bibitem[{{Casagrande} {et~al.}(2010){Casagrande}, {Ram{\'{\i}}rez},
  {Mel{\'e}ndez}, {Bessell}, \& {Asplund}}]{Casagrande2010}
{Casagrande}, L., {Ram{\'{\i}}rez}, I., {Mel{\'e}ndez}, J., {Bessell}, M., \&
  {Asplund}, M. 2010, \aap, 512, A54

\bibitem[{{Casagrande} {et~al.}(2011){Casagrande}, {Sch{\"o}nrich}, {Asplund},
  {Cassisi}, {Ram{\'{\i}}rez}, {Mel{\'e}ndez}, {Bensby}, \&
  {Feltzing}}]{Casagrande2011}
{Casagrande}, L., {Sch{\"o}nrich}, R., {Asplund}, M., {et~al.} 2011, \aap, 530,
  A138

\bibitem[{{Casertano} {et~al.}(2017){Casertano}, {Riess}, {Bucciarelli}, \&
  {Lattanzi}}]{Casertano:2016}
{Casertano}, S., {Riess}, A.~G., {Bucciarelli}, B., \& {Lattanzi}, M.~G. 2017,
  \aap, 599, A67

\bibitem[{{Collier Cameron} {et~al.}(2007){Collier Cameron}, {Wilson}, {West},
  {Hebb}, {Wang}, {Aigrain}, {Bouchy}, {Christian}, {Clarkson}, {Enoch},
  {Esposito}, {Guenther}, {Haswell}, {H{\'e}brard}, {Hellier}, {Horne},
  {Irwin}, {Kane}, {Loeillet}, {Lister}, {Maxted}, {Mayor}, {Moutou}, {Parley},
  {Pollacco}, {Pont}, {Queloz}, {Ryans}, {Skillen}, {Street}, {Udry}, \&
  {Wheatley}}]{Collier2007}
{Collier Cameron}, A., {Wilson}, D.~M., {West}, R.~G., {et~al.} 2007, \mnras,
  380, 1230

\bibitem[{{Cutri} \& {et al.}(2014)}]{Cutri2014}
{Cutri}, R.~M., \& {et al.} 2014, VizieR Online Data Catalog, 2328

\bibitem[{{Cutri} {et~al.}(2003){Cutri}, {Skrutskie}, {van Dyk}, {Beichman},
  {Carpenter}, {Chester}, {Cambresy}, {Evans}, {Fowler}, {Gizis}, {Howard},
  {Huchra}, {Jarrett}, {Kopan}, {Kirkpatrick}, {Light}, {Marsh}, {McCallon},
  {Schneider}, {Stiening}, {Sykes}, {Weinberg}, {Wheaton}, {Wheelock}, \&
  {Zacarias}}]{Cutri2003}
{Cutri}, R.~M., {Skrutskie}, M.~F., {van Dyk}, S., {et~al.} 2003, VizieR Online
  Data Catalog, 2246

\bibitem[{{Dor{\'e}} {et~al.}(2016){Dor{\'e}}, {Werner}, {Ashby}, {Banerjee},
  {Battaglia}, {Bauer}, {Benjamin}, {Bleem}, {Bock}, {Boogert}, {Bull},
  {Capak}, {Chang}, {Chiar}, {Cohen}, {Cooray}, {Crill}, {Cushing}, {de
  Putter}, {Driver}, {Eifler}, {Feng}, {Ferraro}, {Finkbeiner}, {Gaudi},
  {Greene}, {Hillenbrand}, {H{\"o}flich}, {Hsiao}, {Huffenberger}, {Jansen},
  {Jeong}, {Joshi}, {Kim}, {Kim}, {Kirkpatrick}, {Korngut}, {Krause}, {Kriek},
  {Leistedt}, {Li}, {Lisse}, {Mauskopf}, {Mechtley}, {Melnick}, {Mohr},
  {Murphy}, {Neben}, {Neufeld}, {Nguyen}, {Pierpaoli}, {Pyo}, {Rhodes},
  {Sandstrom}, {Schaan}, {Schlaufman}, {Silverman}, {Su}, {Stassun}, {Stevens},
  {Strauss}, {Tielens}, {Tsai}, {Tolls}, {Unwin}, {Viero}, {Windhorst}, \&
  {Zemcov}}]{Dore2016}
{Dor{\'e}}, O., {Werner}, M.~W., {Ashby}, M., {et~al.} 2016, ArXiv e-prints,
  arXiv:1606.07039

\bibitem[{{Gaia Collaboration}(2016)}]{Gaia_main}
{Gaia Collaboration}. 2016, ArXiv e-prints, arXiv:1609.04153

\bibitem[{{Gaia Collaboration} {et~al.}(2016){Gaia Collaboration}, {Brown},
  {Vallenari}, {Prusti}, {de Bruijne}, {Mignard}, {Drimmel}, \&
  {co-authors}}]{GaiaDR1}
{Gaia Collaboration}, {Brown}, A.~G.~A., {Vallenari}, A., {et~al.} 2016, ArXiv
  e-prints, arXiv:1609.04172

\bibitem[{{Gould} {et~al.}(2016){Gould}, {Kollmeier}, \& {Sesar}}]{Gould2016}
{Gould}, A., {Kollmeier}, J.~A., \& {Sesar}, B. 2016, ArXiv e-prints,
  arXiv:1609.06315

\bibitem[{{Gould} \& {Morgan}(2003)}]{Gould2003}
{Gould}, A., \& {Morgan}, C.~W. 2003, \apj, 585, 1056

\bibitem[{{H{\o}g} {et~al.}(2000){H{\o}g}, {Fabricius}, {Makarov}, {Urban},
  {Corbin}, {Wycoff}, {Bastian}, {Schwekendiek}, \& {Wicenec}}]{Hog2000}
{H{\o}g}, E., {Fabricius}, C., {Makarov}, V.~V., {et~al.} 2000, \aap, 355, L27

\bibitem[{{Kordopatis} {et~al.}(2013){Kordopatis}, {Gilmore}, {Steinmetz},
  {Boeche}, {Seabroke}, {Siebert}, {Zwitter}, {Binney}, {de Laverny},
  {Recio-Blanco}, {Williams}, {Piffl}, {Enke}, {Roeser}, {Bijaoui}, {Wyse},
  {Freeman}, {Munari}, {Carrillo}, {Anguiano}, {Burton}, {Campbell}, {Cass},
  {Fiegert}, {Hartley}, {Parker}, {Reid}, {Ritter}, {Russell}, {Stupar},
  {Watson}, {Bienaym{\'e}}, {Bland-Hawthorn}, {Gerhard}, {Gibson}, {Grebel},
  {Helmi}, {Navarro}, {Conrad}, {Famaey}, {Faure}, {Just}, {Kos}, {Matijevi{\v
  c}}, {McMillan}, {Minchev}, {Scholz}, {Sharma}, {Siviero}, {de Boer}, \& {{\v
  Z}erjal}}]{Kordopatis2013}
{Kordopatis}, G., {Gilmore}, G., {Steinmetz}, M., {et~al.} 2013, \aj, 146, 134

\bibitem[{{Kurucz}(2013)}]{Kurucz2013}
{Kurucz}, R.~L. 2013, {ATLAS12: Opacity sampling model atmosphere program},
  Astrophysics Source Code Library, , , ascl:1303.024

\bibitem[{{Lindegren} {et~al.}(2016){Lindegren}, {Lammers}, {Bastian},
  {Hern{\'a}ndez}, {Klioner}, {Hobbs}, {Bombrun}, {Michalik}, {Ramos-Lerate},
  {Butkevich}, {Comoretto}, {Joliet}, {Holl}, {Hutton}, {Parsons},
  {Steidelm{\"u}ller}, {Abbas}, {Altmann}, {Andrei}, {Anton}, {Bach},
  {Barache}, {Becciani}, {Berthier}, {Bianchi}, {Biermann}, {Bouquillon},
  {Bourda}, {Br{\"u}semeister}, {Bucciarelli}, {Busonero}, {Carlucci},
  {Casta{\~n}eda}, {Charlot}, {Clotet}, {Crosta}, {Davidson}, {de Felice},
  {Drimmel}, {Fabricius}, {Fienga}, {Figueras}, {Fraile}, {Gai}, {Garralda},
  {Geyer}, {Gonz{\'a}lez-Vidal}, {Guerra}, {Hambly}, {Hauser}, {Jordan},
  {Lattanzi}, {Lenhardt}, {Liao}, {L{\"o}ffler}, {McMillan}, {Mignard}, {Mora},
  {Morbidelli}, {Portell}, {Riva}, {Sarasso}, {Serraller}, {Siddiqui}, {Smart},
  {Spagna}, {Stampa}, {Steele}, {Taris}, {Torra}, {van Reeven}, {Vecchiato},
  {Zschocke}, {de Bruijne}, {Gracia}, {Raison}, {Lister}, {Marchant},
  {Messineo}, {Soffel}, {Osorio}, {de Torres}, \&
  {O'Mullane}}]{gaia_astrometry}
{Lindegren}, L., {Lammers}, U., {Bastian}, U., {et~al.} 2016, ArXiv e-prints,
  arXiv:1609.04303

\bibitem[{{Luo} {et~al.}(2015){Luo}, {Zhao}, {Zhao}, {Deng}, {Liu}, {Jing},
  {Wang}, {Zhang}, {Shi}, {Cui}, {Chu}, {Li}, {Bai}, {Wu}, {Cai}, {Cao}, {Cao},
  {Carlin}, {Chen}, {Chen}, {Chen}, {Chen}, {Chen}, {Chen}, {Chen},
  {Christlieb}, {Chu}, {Cui}, {Dong}, {Du}, {Fan}, {Feng}, {Fu}, {Gao}, {Gong},
  {Gu}, {Guo}, {Han}, {He}, {Hou}, {Hou}, {Hou}, {Hu}, {Hu}, {Hu}, {Huo},
  {Jia}, {Jiang}, {Jiang}, {Jiang}, {Jin}, {Kong}, {Kong}, {Lei}, {Li}, {Li},
  {Li}, {Li}, {Li}, {Li}, {Li}, {Li}, {Li}, {Li}, {Li}, {Li}, {Liang}, {Lin},
  {Liu}, {Liu}, {Liu}, {Liu}, {Lu}, {Luo}, {Mao}, {Newberg}, {Ni}, {Qi}, {Qi},
  {Shen}, {Shi}, {Song}, {Song}, {Su}, {Su}, {Tang}, {Tao}, {Tian}, {Wang},
  {Wang}, {Wang}, {Wang}, {Wang}, {Wang}, {Wang}, {Wang}, {Wang}, {Wang},
  {Wang}, {Wang}, {Wang}, {Wang}, {Wang}, {Wang}, {Wang}, {Wang}, {Wang},
  {Wang}, {Wei}, {Wei}, {Wu}, {Wu}, {Wu}, {Wu}, {Xing}, {Xu}, {Xu}, {Xu},
  {Yan}, {Yang}, {Yang}, {Yang}, {Yang}, {Yao}, {Yu}, {Yuan}, {Yuan}, {Yuan},
  {Yuan}, {Zhai}, {Zhang}, {Zhang}, {Zhang}, {Zhang}, {Zhang}, {Zhang},
  {Zhang}, {Zhang}, {Zhao}, {Zhou}, {Zhou}, {Zhu}, {Zhu}, {Zou}, \&
  {Zuo}}]{Luo2015}
{Luo}, A.-L., {Zhao}, Y.-H., {Zhao}, G., {et~al.} 2015, Research in Astronomy
  and Astrophysics, 15, 1095

\bibitem[{{Lutz} \& {Kelker}(1973)}]{Lutz1973}
{Lutz}, T.~E., \& {Kelker}, D.~H. 1973, \pasp, 85, 573

\bibitem[{{Mann} {et~al.}(2015){Mann}, {Feiden}, {Gaidos}, {Boyajian}, \& {von
  Braun}}]{Mann2015}
{Mann}, A.~W., {Feiden}, G.~A., {Gaidos}, E., {Boyajian}, T., \& {von Braun},
  K. 2015, \apj, 804, 64

\bibitem[{{Mermilliod}(2006)}]{Mermilliod:2006}
{Mermilliod}, J.~C. 2006, VizieR Online Data Catalog, 2168

\bibitem[{{Ochsenbein} {et~al.}(2000){Ochsenbein}, {Bauer}, \&
  {Marcout}}]{Ochsenbein2000}
{Ochsenbein}, F., {Bauer}, P., \& {Marcout}, J. 2000, \aaps, 143, 23

\bibitem[{{Paunzen}(2015)}]{Paunzen2015}
{Paunzen}, E. 2015, \aap, 580, A23

\bibitem[{{Perryman} {et~al.}(1997){Perryman}, {Lindegren}, {Kovalevsky},
  {Hoeg}, {Bastian}, {Bernacca}, {Cr{\'e}z{\'e}}, {Donati}, {Grenon},
  {Grewing}, {van Leeuwen}, {van der Marel}, {Mignard}, {Murray}, {Le Poole},
  {Schrijver}, {Turon}, {Arenou}, {Froeschl{\'e}}, \&
  {Petersen}}]{Perryman:1997}
{Perryman}, M.~A.~C., {Lindegren}, L., {Kovalevsky}, J., {et~al.} 1997, \aap,
  323, L49

\bibitem[{{Rauer} {et~al.}(2014){Rauer}, {Catala}, {Aerts}, {Appourchaux},
  {Benz}, {Brandeker}, {Christensen-Dalsgaard}, {Deleuil}, {Gizon}, {Goupil},
  {G{\"u}del}, {Janot-Pacheco}, {Mas-Hesse}, {Pagano}, {Piotto}, {Pollacco},
  {Santos}, {Smith}, {Su{\'a}rez}, {Szab{\'o}}, {Udry}, {Adibekyan}, {Alibert},
  {Almenara}, {Amaro-Seoane}, {Eiff}, {Asplund}, {Antonello}, {Barnes},
  {Baudin}, {Belkacem}, {Bergemann}, {Bihain}, {Birch}, {Bonfils}, {Boisse},
  {Bonomo}, {Borsa}, {Brand{\~a}o}, {Brocato}, {Brun}, {Burleigh}, {Burston},
  {Cabrera}, {Cassisi}, {Chaplin}, {Charpinet}, {Chiappini}, {Church},
  {Csizmadia}, {Cunha}, {Damasso}, {Davies}, {Deeg}, {D{\'{\i}}az}, {Dreizler},
  {Dreyer}, {Eggenberger}, {Ehrenreich}, {Eigm{\"u}ller}, {Erikson}, {Farmer},
  {Feltzing}, {de Oliveira Fialho}, {Figueira}, {Forveille}, {Fridlund},
  {Garc{\'{\i}}a}, {Giommi}, {Giuffrida}, {Godolt}, {Gomes da Silva},
  {Granzer}, {Grenfell}, {Grotsch-Noels}, {G{\"u}nther}, {Haswell}, {Hatzes},
  {H{\'e}brard}, {Hekker}, {Helled}, {Heng}, {Jenkins}, {Johansen},
  {Khodachenko}, {Kislyakova}, {Kley}, {Kolb}, {Krivova}, {Kupka}, {Lammer},
  {Lanza}, {Lebreton}, {Magrin}, {Marcos-Arenal}, {Marrese}, {Marques},
  {Martins}, {Mathis}, {Mathur}, {Messina}, {Miglio}, {Montalban}, {Montalto},
  {Monteiro}, {Moradi}, {Moravveji}, {Mordasini}, {Morel}, {Mortier},
  {Nascimbeni}, {Nelson}, {Nielsen}, {Noack}, {Norton}, {Ofir}, {Oshagh},
  {Ouazzani}, {P{\'a}pics}, {Parro}, {Petit}, {Plez}, {Poretti}, {Quirrenbach},
  {Ragazzoni}, {Raimondo}, {Rainer}, {Reese}, {Redmer}, {Reffert},
  {Rojas-Ayala}, {Roxburgh}, {Salmon}, {Santerne}, {Schneider}, {Schou},
  {Schuh}, {Schunker}, {Silva-Valio}, {Silvotti}, {Skillen}, {Snellen}, {Sohl},
  {Sousa}, {Sozzetti}, {Stello}, {Strassmeier}, {{\v S}vanda}, {Szab{\'o}},
  {Tkachenko}, {Valencia}, {Van Grootel}, {Vauclair}, {Ventura}, {Wagner},
  {Walton}, {Weingrill}, {Werner}, {Wheatley}, \& {Zwintz}}]{Rauer:2014}
{Rauer}, H., {Catala}, C., {Aerts}, C., {et~al.} 2014, Experimental Astronomy,
  38, 249

\bibitem[{{Ricker} {et~al.}(2015){Ricker}, {Winn}, {Vanderspek}, {Latham},
  {Bakos}, {Bean}, {Berta-Thompson}, {Brown}, {Buchhave}, {Butler}, {Butler},
  {Chaplin}, {Charbonneau}, {Christensen-Dalsgaard}, {Clampin}, {Deming},
  {Doty}, {De Lee}, {Dressing}, {Dunham}, {Endl}, {Fressin}, {Ge}, {Henning},
  {Holman}, {Howard}, {Ida}, {Jenkins}, {Jernigan}, {Johnson}, {Kaltenegger},
  {Kawai}, {Kjeldsen}, {Laughlin}, {Levine}, {Lin}, {Lissauer}, {MacQueen},
  {Marcy}, {McCullough}, {Morton}, {Narita}, {Paegert}, {Palle}, {Pepe},
  {Pepper}, {Quirrenbach}, {Rinehart}, {Sasselov}, {Sato}, {Seager},
  {Sozzetti}, {Stassun}, {Sullivan}, {Szentgyorgyi}, {Torres}, {Udry}, \&
  {Villasenor}}]{Ricker:2015}
{Ricker}, G.~R., {Winn}, J.~N., {Vanderspek}, R., {et~al.} 2015, Journal of
  Astronomical Telescopes, Instruments, and Systems, 1, 014003

\bibitem[{{Schlegel} {et~al.}(1998){Schlegel}, {Finkbeiner}, \&
  {Davis}}]{Schlegel1998}
{Schlegel}, D.~J., {Finkbeiner}, D.~P., \& {Davis}, M. 1998, \apj, 500, 525

\bibitem[{{SDSS Collaboration} {et~al.}(2016){SDSS Collaboration}, {Albareti},
  {Allende Prieto}, {Almeida}, {Anders}, {Anderson}, {Andrews},
  {Aragon-Salamanca}, {Argudo-Fernandez}, {Armengaud}, \& et~al.}]{SDSS2016}
{SDSS Collaboration}, {Albareti}, F.~D., {Allende Prieto}, C., {et~al.} 2016,
  ArXiv e-prints, arXiv:1608.02013

\bibitem[{{Skrutskie} {et~al.}(2006){Skrutskie}, {Cutri}, {Stiening},
  {Weinberg}, {Schneider}, {Carpenter}, {Beichman}, {Capps}, {Chester},
  {Elias}, {Huchra}, {Liebert}, {Lonsdale}, {Monet}, {Price}, {Seitzer},
  {Jarrett}, {Kirkpatrick}, {Gizis}, {Howard}, {Evans}, {Fowler}, {Fullmer},
  {Hurt}, {Light}, {Kopan}, {Marsh}, {McCallon}, {Tam}, {Van Dyk}, \&
  {Wheelock}}]{Skrutskie2006}
{Skrutskie}, M.~F., {Cutri}, R.~M., {Stiening}, R., {et~al.} 2006, \aj, 131,
  1163

\bibitem[{{Somers} \& {Stassun}(2016)}]{Somers:2016}
{Somers}, G., \& {Stassun}, K.~G. 2016, ArXiv e-prints, arXiv:1609.04841

\bibitem[{{Stassun} {et~al.}(2016){Stassun}, {Collins}, \&
  {Gaudi}}]{StassunGaiaPlanets:2016}
{Stassun}, K.~G., {Collins}, K.~A., \& {Gaudi}, B.~S. 2016, ArXiv e-prints,
  arXiv:1609.04389

\bibitem[{{Stassun} {et~al.}(2014{\natexlab{a}}){Stassun}, {Feiden}, \&
  {Torres}}]{Stassun:2014}
{Stassun}, K.~G., {Feiden}, G.~A., \& {Torres}, G. 2014{\natexlab{a}}, \nar,
  60, 1

\bibitem[{{Stassun} {et~al.}(2012){Stassun}, {Kratter}, {Scholz}, \&
  {Dupuy}}]{Stassun:2012}
{Stassun}, K.~G., {Kratter}, K.~M., {Scholz}, A., \& {Dupuy}, T.~J. 2012, \apj,
  756, 47

\bibitem[{{Stassun} {et~al.}(2014{\natexlab{b}}){Stassun}, {Pepper}, {Oelkers},
  {Paegert}, {De Lee}, \& {Sanchis-Ojeda}}]{StassunTESS:2014}
{Stassun}, K.~G., {Pepper}, J.~A., {Oelkers}, R., {et~al.} 2014{\natexlab{b}},
  ArXiv e-prints, arXiv:1410.6379

\bibitem[{{Stassun} \& {Torres}(2016{\natexlab{a}})}]{StassunGaiaEB:2016}
{Stassun}, K.~G., \& {Torres}, G. 2016{\natexlab{a}}, ArXiv e-prints,
  arXiv:1609.02579

\bibitem[{{Stassun} \& {Torres}(2016{\natexlab{b}})}]{StassunGaiaError:2016}
---. 2016{\natexlab{b}}, ArXiv e-prints, arXiv:1609.05390

\bibitem[{{Stassun} {et~al.}(2017){Stassun}, {Oelkers}, {Pepper}, {Paegert},
  {De Lee}, {Torres}, {Latham}, {Muirhead}, {Dressing}, {Rojas-Ayala}, {Mann},
  {Fleming}, {Levine}, {Silvotti}, {Plavchan}, \& {the TESS Target Selection
  Working Group}}]{StassunTIC:2017}
{Stassun}, K.~G., {Oelkers}, R.~J., {Pepper}, J., {et~al.} 2017, ArXiv
  e-prints, arXiv:1706.00495

\bibitem[{{Torres} {et~al.}(2010){Torres}, {Andersen}, \&
  {Gim{\'e}nez}}]{Torres2010}
{Torres}, G., {Andersen}, J., \& {Gim{\'e}nez}, A. 2010, \aapr, 18, 67

\end{thebibliography}

\floattable
\begin{deluxetable*}{lrrrrrrrrrrrrrrrrrrrr}

\rotate
\tabletypesize{\scriptsize}
\tablecaption{\label{tab:allphot} Catalog Photometry for 1,600,080 Tycho-2 Stars Used in Iterative-IRFM Routine}
\setlength\tabcolsep{5pt}
\tablehead{
\colhead{Star} &  \colhead{RA (J2000)} &  \colhead{Dec (J2000)} & \colhead{$B_T$} & \colhead{$\sigma_{B_T}$} & \colhead{$V_T$} & \colhead{$\sigma_{V_T}$} & \colhead{$J$} & \colhead{$\sigma_J$} & \colhead{$H$} & \colhead{$\sigma_H$} & \colhead{$K_S$} & \colhead{$\sigma_{K_S}$} & \colhead{$B$} & \colhead{$\sigma_B$} & \colhead{$V$} & \colhead{$\sigma_V$} & \colhead{$b$} & \colhead{$\sigma_b$} & \colhead{$y$} & \colhead{$\sigma_y$}}
\startdata
TYC 100-1166-1 & 80.01164972 & 1.12232972 & 12.512 & 0.236 & 11.850 & 0.188 & 10.262 & 0.026 & 9.748 & 0.022 & 9.639 & 0.021 & \nodata & \nodata & \nodata & \nodata & \nodata & \nodata & \nodata & \nodata \\
TYC 1001-1885-1 & 264.45634000 & 11.68801222 & 8.644 & 0.017 & 8.296 & 0.013 & 7.578 & 0.018 & 7.464 & 0.027 & 7.413 & 0.024 & \nodata & \nodata & \nodata & \nodata & 8.477 & 0.0084852814 & 8.274 & 0.006 \\
TYC 100-1394-1 & 79.92395889 & 1.58643111 & 12.071 & 0.149 & 11.166 & 0.098 & 10.106 & 0.022 & 9.677 & 0.022 & 9.597 & 0.019  & \nodata & \nodata & \nodata & \nodata & \nodata & \nodata & \nodata & \nodata \\
TYC 1003-2217-1 & 261.87815972 & 13.35782833 & 12.521 & 0.239 & 12.296 & 0.239 & 10.952 & 0.021 & 10.717 & 0.018 & 10.649 & 0.018  & \nodata & \nodata & \nodata & \nodata & \nodata & \nodata & \nodata & \nodata \\
TYC 1004-1638-1 & 264.19612222 & 14.86987694 & 12.489 & 0.220 & 11.794 & 0.173 & 9.610 & 0.022 & 8.983 & 0.020 & 8.837 & 0.017 & \nodata & \nodata & \nodata & \nodata & \nodata & \nodata & \nodata & \nodata \\
TYC 1004-1657-1 & 263.96095722 & 14.04696917 & 12.761 & 0.268 & 11.604 & 0.135 & 9.311 & 0.027 & 8.690 & 0.053 & 8.528 & 0.022  & \nodata & \nodata & \nodata & \nodata & \nodata & \nodata & \nodata & \nodata \\
TYC 1004-1680-1 & 263.91493167 & 14.72536222 & 12.986 & 0.322 & 12.013 & 0.202 & 9.762 & 0.027 & 9.048 & 0.030 & 8.919 & 0.019 & \nodata & \nodata & \nodata & \nodata & \nodata & \nodata & \nodata & \nodata \\
TYC 1004-1777-1 & 262.51756000 & 14.46246167 & 9.190 & 0.018 & 8.586 & 0.013 & 7.443 & 0.021 & 7.222 & 0.042 & 7.153 & 0.026 & \nodata & \nodata & \nodata & \nodata & 8.896 & 0.004 & 8.52 & 0 \\
TYC 1005-1574-1 & 265.36997972 & 14.72544194 & 12.604 & 0.233 & 11.451 & 0.109 & 9.563 & 0.020 & 9.033 & 0.019 & 8.877 & 0.017 & \nodata & \nodata & \nodata & \nodata & \nodata & \nodata & \nodata & \nodata \\
TYC 1005-1908-1 & 264.85883750 & 14.96200306 & 12.156 & 0.144 & 11.637 & 0.132 & 9.834 & 0.021 & 9.286 & 0.020 & 9.185 & 0.017 & \nodata & \nodata & \nodata & \nodata & \nodata & \nodata & \nodata & \nodata \\
\enddata
\tablecomments{Table \ref{tab:allphot} is published in its entirety in the machine-readable format. A portion is shown here for guidance regarding its form and content.}
\end{deluxetable*}

\floattable

\begin{deluxetable*}{lrrrrrrrrrrrrrrrrr}

\rotate
\tabletypesize{\scriptsize}
\tablecaption{\label{tab:allresults} Fundamental Parameters of 1,600,080 \emph{Tycho-2} Stars}
\tablewidth{1pt}
\setlength\tabcolsep{4pt}
\tablehead{
\colhead{Star} &  \colhead{$\log \chi^2_{\nu}$} &  \colhead{$A_v$} &  \colhead{$\sigma_{A_v}$} & \colhead{\teff} & \colhead{$\sigma_{\teff}$} &  \colhead{$\log \fbol$} &  \colhead{$\log \sigma_{\fbol}$} &  \colhead{$\theta$} & \colhead{$\sigma_{\theta}$} & \colhead{$T_{\rm eff,spec}$} &  \colhead{$\sigma_{T_{\rm eff,spec}}$} & \colhead{\feh} &  \colhead{$\sigma_{\feh}$} & \colhead{$\log (g)$} & \colhead{$\sigma_{\log (g)}$} & \colhead{Spec. Source} & \colhead{Hot Flag}\tablenotemark{1}\\  & \colhead{(unscaled)} &   &   & \colhead{(K)} & \colhead{(K)} &  \colhead{(${\rm erg\ cm^{-2} s^{-1}}$)} &  \colhead{(${\rm erg\ cm^{-2} s^{-1}}$)} &  \colhead{($\mu$as)} & \colhead{($\mu$as)} & \colhead{(K)} &  \colhead{(K)} & & & & & }
\startdata  
TYC1023-508-1 & -0.432 & 0.180 & 0.040 & 5178.050 & 11.581 & -7.391 & -9.517 & 412.046 & 2.490 & 5157 & 69 & -0.375488 & 0.0313319 & 3.33086 & 0.08 & APOGEE & 0 \\
TYC1035-1205-1 & -1.334 & 0.710 & 0.010 & 4277.800 & 15.351 & -8.689 & -10.573 & 135.396 & 1.363 & 4307 & 69 & -0.311677 & 0.030143 & 1.70132 & 0.08 & APOGEE & 0 \\
TYC1038-1289-1 & -1.045 & 1.000 & 0.030 & 4347.820 & 14.452 & -8.366 & -9.621 & 190.233 & 5.522 & 4347 & 69 & 0.0579264 & 0.0229669 & 2.24876 & 0.08 & APOGEE & 0 \\
TYC1-1024-1 & -0.231 & 0.104 & 0.054 & 6261.670 & 54.544 & -8.793 & -10.380 & 56.057 & 1.251 & 6431 & 189 & 0.044 & 0.333 & 4.116 & 0.438 & LAMOST & 0 \\
TYC1-111-1 & -0.274 & 0.000 & 0.040 & 5721.840 & 78.121 & -9.191 & -11.362 & 42.456 & 1.177 & 5665 & 218 & -0.502 & 0.582 & 4.535 & 0.535 & LAMOST & 0 \\
TYC1-1122-1 & 0.144 & 0.087 & 0.067 & 5810.800 & 78.816 & -8.815 & -10.141 & 63.478 & 2.348 & 6020 & 207 & 0.106 & 0.396 & 4.516 & 0.493 & LAMOST & 0 \\
TYC1000-1006-1 & -1.122 & 0.000 & 0.006 & 6618.690 & 35.803 & -8.775 & -11.256 & 51.211 & 0.562 & \nodata & \nodata & \nodata & \nodata & \nodata & \nodata & \nodata & 0 \\
TYC1000-1009-1 & -0.387 & 0.413 & 0.022 & 5934.640 & 134.770 & -9.218 & -10.728 & 38.247 & 1.846 & \nodata & \nodata & \nodata & \nodata & \nodata & \nodata & \nodata & 0 \\
TYC1000-1016-1 & -5.422 & 0.260 & 0.007 & 6952.350 & 1.070 & -8.742 & -11.972 & 48.240 & 0.021 & \nodata & \nodata & \nodata & \nodata & \nodata & \nodata & \nodata & 0 \\
TYC100-1166-1 & -0.125 & 0.070 & 0.110 & 5064.370 & 88.065 & -9.185 & -10.574 & 54.568 & 2.220 & 5110 & 105 & -0.08 & 0.09 & 3.34 & 0.21 & RAVE & 0 \\
TYC100-1394-1 & 1.390 & 0.000 & 0.050 & 5869.290 & 454.894 & -9.028 & -10.485 & 48.703 & 7.613 & 5148 & 66 & 0.05 & 0.09 & 3.63 & 0.1 & RAVE & 0 \\
TYC10-1118-1 & 0.299 & 0.000 & 0.020 & 6016.600 & 136.251 & -9.048 & -10.384 & 45.262 & 2.319 & 5784 & 95 & -0.01 & 0.1 & 4.16 & 0.15 & RAVE & 0 \\
\enddata
\tablenotetext{1}{"1" indicates $\teff > 7,000$~K for which the iterative IRFM parameters may not be reliable -- see Section \ref{subsec:specteff}.}
\tablecomments{Table \ref{tab:allresults} is published in its entirety in the machine-readable format. A portion is shown here for guidance regarding its form and content.}
\end{deluxetable*}

\floattable
\begin{deluxetable*}{lrrrrrrrrrrrrrrrrrrrrr}

\rotate
\tabletypesize{\scriptsize}
\tablecaption{\label{tab:allresults2} Fundamental Parameters of 355,502 TGAS Stars with Radii}
\tablewidth{1pt}
\setlength\tabcolsep{4pt}
\tablehead{
\colhead{Star} &  \colhead{$\log \chi^2_{\nu}$} &  \colhead{$A_v$} &  \colhead{$\sigma_{A_v}$} & \colhead{\teff} & \colhead{$\sigma_{\teff}$} &  \colhead{$\log \fbol$} &  \colhead{$\log \sigma_{\fbol}$} &  \colhead{$\theta$} & \colhead{$\sigma_{\theta}$} &  \colhead{Parallax} &  \colhead{Error} &  \colhead{d} &  \colhead{$\sigma_d$} &  \colhead{\rstar} & \colhead{$\sigma_{\rstar}$} & \colhead{$T_{\rm eff,spec}$} &  \colhead{$\sigma_{T_{\rm eff,spec}}$} & \colhead{\feh} &  \colhead{$\sigma_{\feh}$} & \colhead{Giant Flag}\tablenotemark{1} & \colhead{Spec. Source} \\  & \colhead{(unscaled)} &   &   & \colhead{(K)} & \colhead{(K)} &  \colhead{(${\rm erg\ cm^{-2} s^{-1}}$)} &  \colhead{(${\rm erg\ cm^{-2} s^{-1}}$)} &  \colhead{($\mu$as)} & \colhead{($\mu$as)} &  \colhead{(mas)} &  \colhead{(mas)} &  \colhead{(pc)} &  \colhead{(pc)} &  \colhead{(\rsun} & \colhead{(\rsun)} & \colhead{(K)} &  \colhead{(K)} & & & & }
\startdata
TYC1023-508-1 & -0.432 & 0.180 & 0.040 & 5178.049 & 11.581 & -7.391 & -9.517 & 412.047 & 2.819 & 12.154 & 0.369 & 82.279 & 2.497 & 3.645 & 0.113 & 5157 & 69 & -0.375488 & 0.0313319 & 0 & APOGEE \\
TYC1045-202-1 & -3.246 & 0.470 & 0.010 & 7205.317 & 3.387 & -8.283 & -11.062 & 76.162 & 0.112 & 3.902 & 0.346 & 256.282 & 22.695 & 2.099 & 0.186 & 7207 & 69 & -0.094537 & 0.00807948 & 0 & APOGEE \\
TYC1046-562-1 & -0.224 & 0.000 & 0.010 & 6714.853 & 33.323 & -8.419 & -10.473 & 74.988 & 0.924 & 4.153 & 0.289 & 240.780 & 16.776 & 1.941 & 0.137 & 6652 & 69 & -0.35377 & 0.0106944 & 0 & APOGEE \\
TYC1-1122-1 & 0.144 & 0.087 & 0.067 & 5810.803 & 78.816 & -8.815 & -10.141 & 63.478 & 2.252 & 6.687 & 0.446 & 149.537 & 9.967 & 1.021 & 0.077 & 6020 & 207 & 0.106 & 0.396 & 0 & LAMOST \\
TYC1-1167-1 & -1.050 & 0.099 & 0.019 & 6404.499 & 15.014 & -8.247 & -10.275 & 100.556 & 0.657 & 5.868 & 0.290 & 170.409 & 8.412 & 1.842 & 0.092 & 6412 & 222 & -0.539 & 0.537 & 0 & LAMOST \\
TYC1120-184-1 & 0.100 & 0.187 & 0.017 & 6382.752 & 44.454 & -8.160 & -10.484 & 111.881 & 1.574 & 10.413 & 0.260 & 96.035 & 2.401 & 1.155 & 0.033 & 6461 & 64 & 0.383 & 0.122 & 0 & LAMOST \\
TYC1000-1018-1 & -0.745 & 0.130 & 0.056 & 6204.678 & 92.669 & -9.223 & -10.690 & 34.828 & 1.253 & 2.875 & 0.269 & 347.823 & 32.578 & 1.302 & 0.131 & \nodata & \nodata & \nodata & \nodata & 0 & \nodata \\
TYC1000-1123-1 & -0.786 & 0.000 & 0.015 & 6072.540 & 56.676 & -9.033 & -10.768 & 45.210 & 0.981 & 3.933 & 0.230 & 254.245 & 14.890 & 1.236 & 0.077 & \nodata & \nodata & \nodata & \nodata & 0 & \nodata \\
TYC1000-1138-1 & -0.070 & 0.130 & 0.030 & 6222.277 & 71.453 & -8.615 & -10.121 & 69.664 & 2.032 & 7.922 & 0.727 & 126.236 & 11.587 & 0.946 & 0.091 & \nodata & \nodata & \nodata & \nodata & 0 & \nodata \\
TYC1072-2400-1 & 0.520 & 0.000 & 0.010 & 6214.807 & 47.166 & -7.217 & -9.661 & 349.431 & 5.352 & 16.276 & 0.434 & 61.441 & 1.638 & 2.308 & 0.071 & 5960 & 67 & -0.13 & 0.09 & 0 & RAVE \\
TYC1-1058-1 & 0.389 & 0.000 & 0.030 & 5578.937 & 95.636 & -8.923 & -11.155 & 60.795 & 2.095 & 7.597 & 0.504 & 131.637 & 8.727 & 0.860 & 0.064 & 5052 & 145 & -0.35 & 0.12 & 0 & RAVE \\
TYC1138-130-1 & 0.306 & 0.010 & 0.010 & 6539.489 & 44.563 & -7.046 & -9.322 & 384.006 & 5.350 & 31.803 & 0.519 & 31.444 & 0.513 & 1.298 & 0.028 & 6250 & 82 & -0.21 & 0.11 & 0 & RAVE \\
\enddata
\tablecomments{Table \ref{tab:allresults2} is published in its entirety in the machine-readable format. A portion is shown here for guidance regarding its form and content.}
\tablenotetext{1}{"0" if not identified as a giant (either via spectroscopic $\loggstar$ or reduced proper motion), "1" if so, and "2" if data are unavailable.}
\end{deluxetable*}

\appendix 

\section{Spectral Energy Distributions}\label{sec:sed_appendix}
\begin{figure*}
\includegraphics[width=0.333\linewidth]{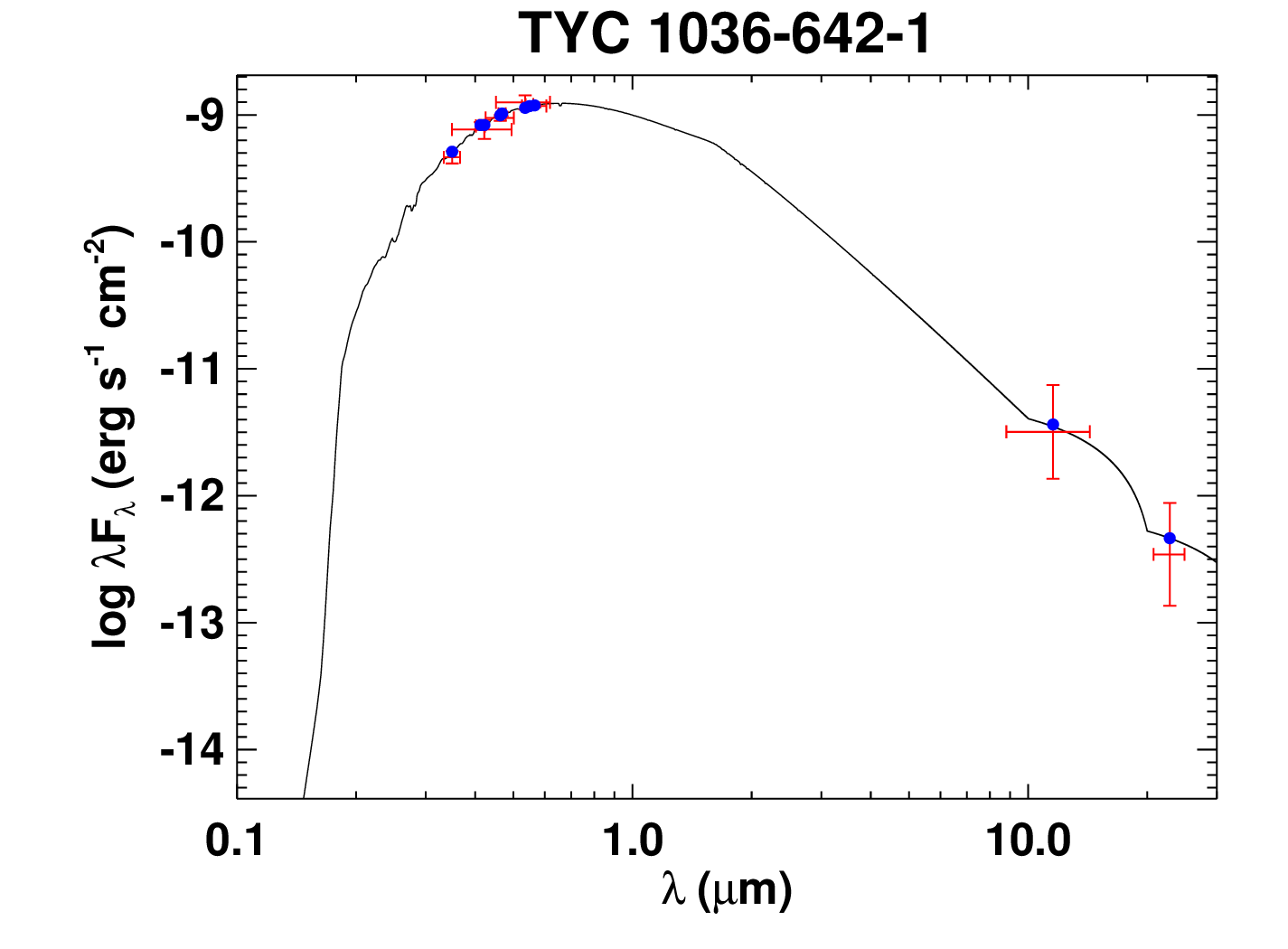}\includegraphics[width=0.333\linewidth]{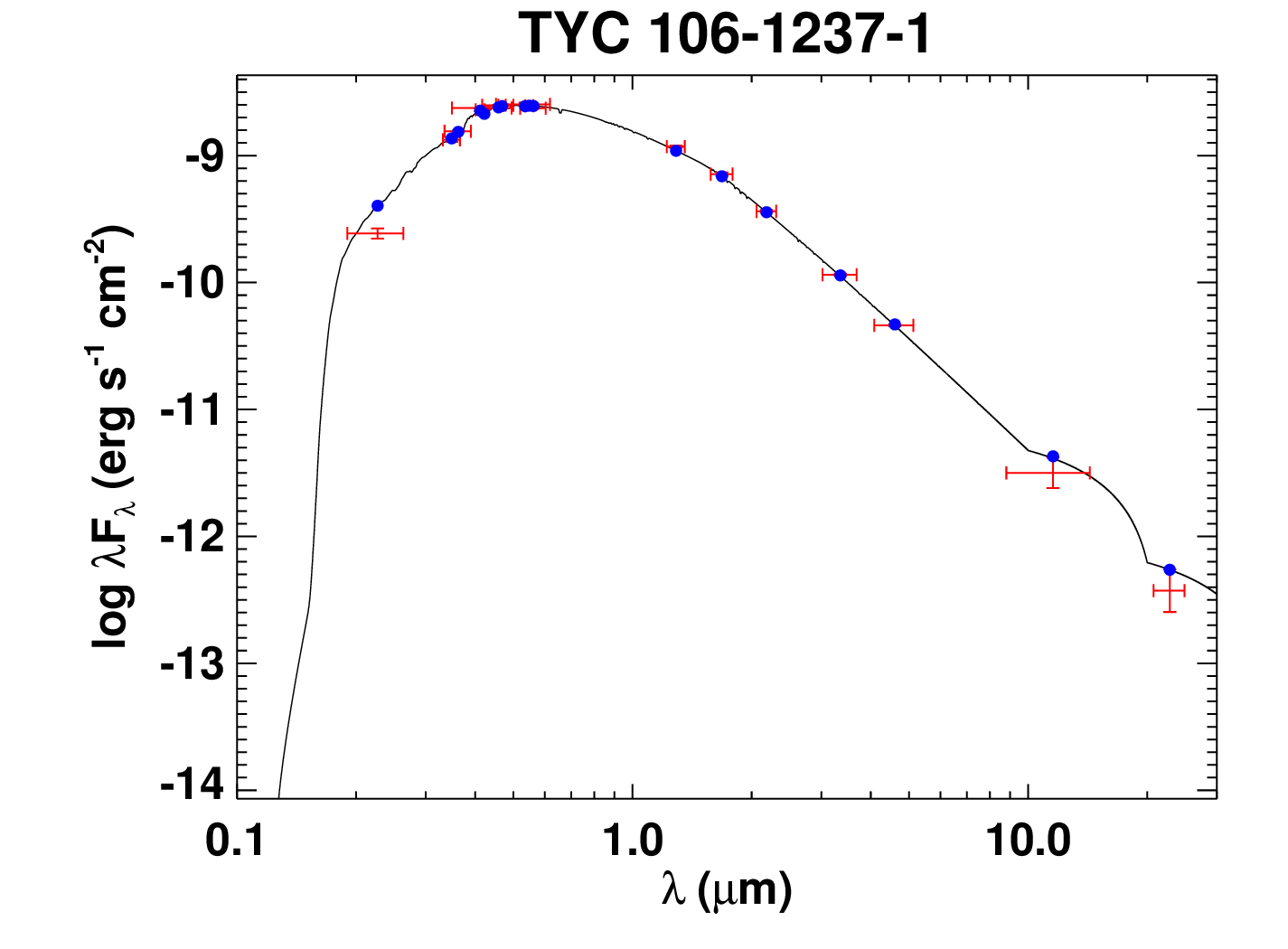}\includegraphics[width=0.333\linewidth]{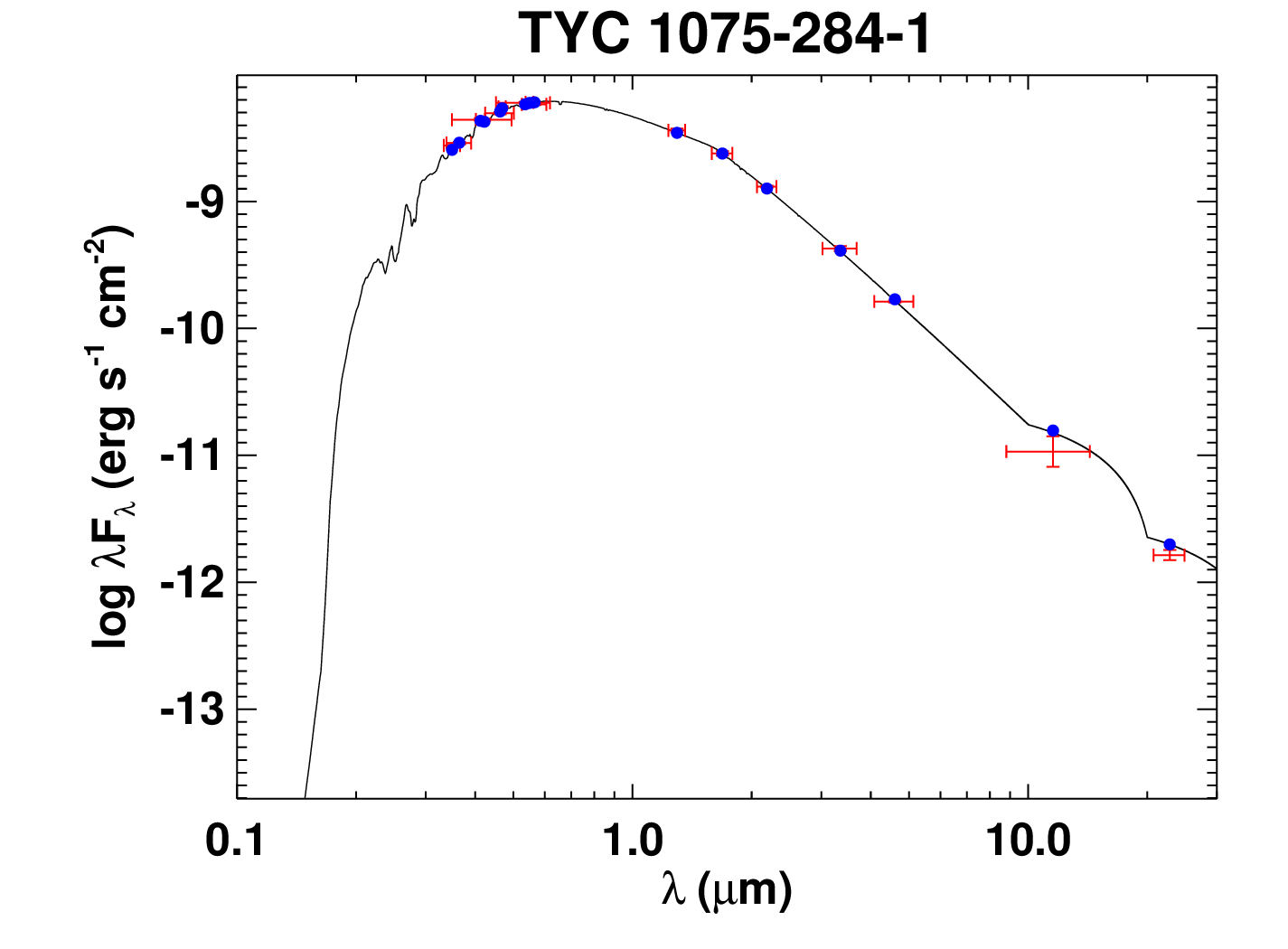}
\includegraphics[width=0.333\linewidth]{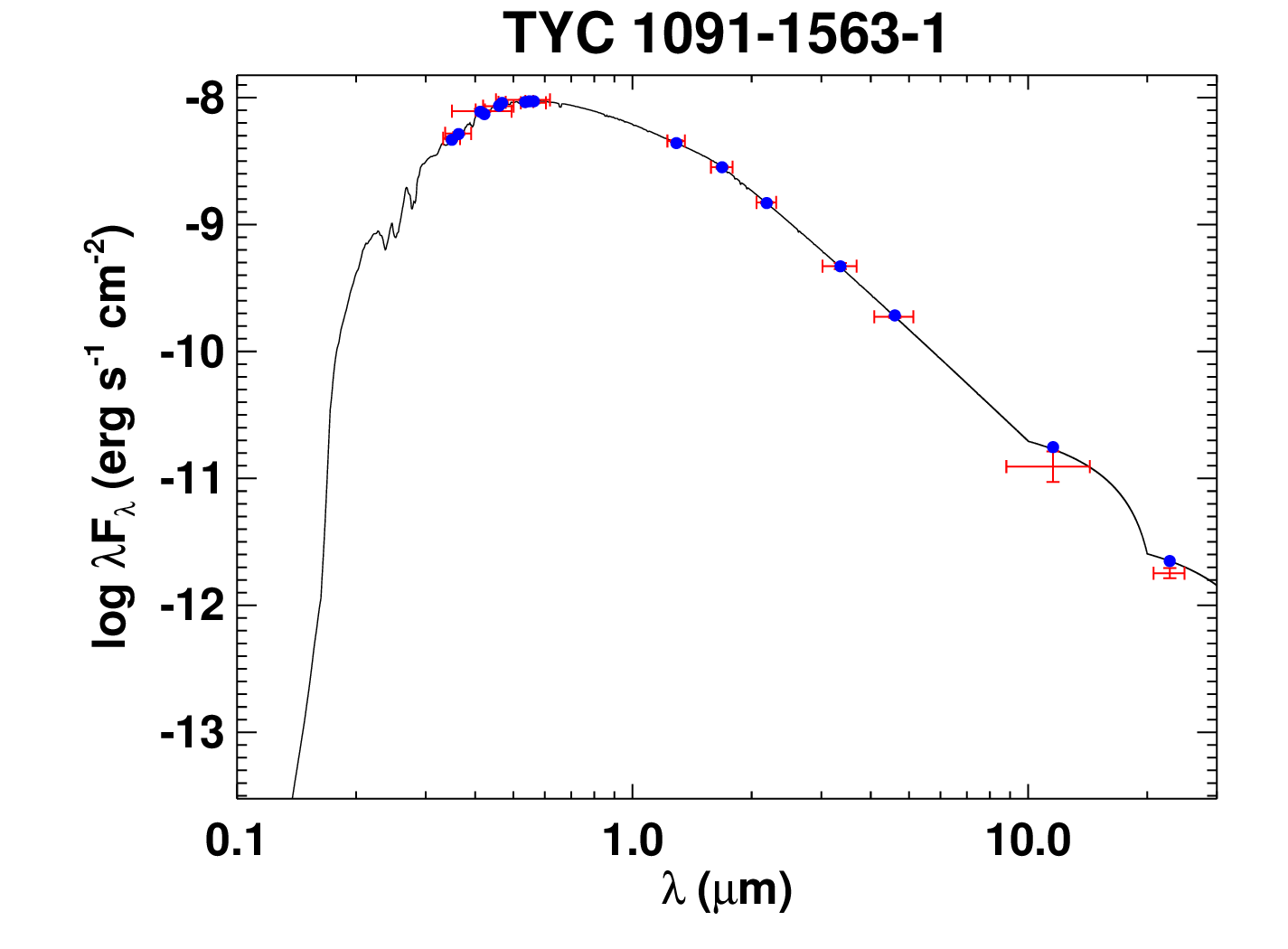}\includegraphics[width=0.333\linewidth]{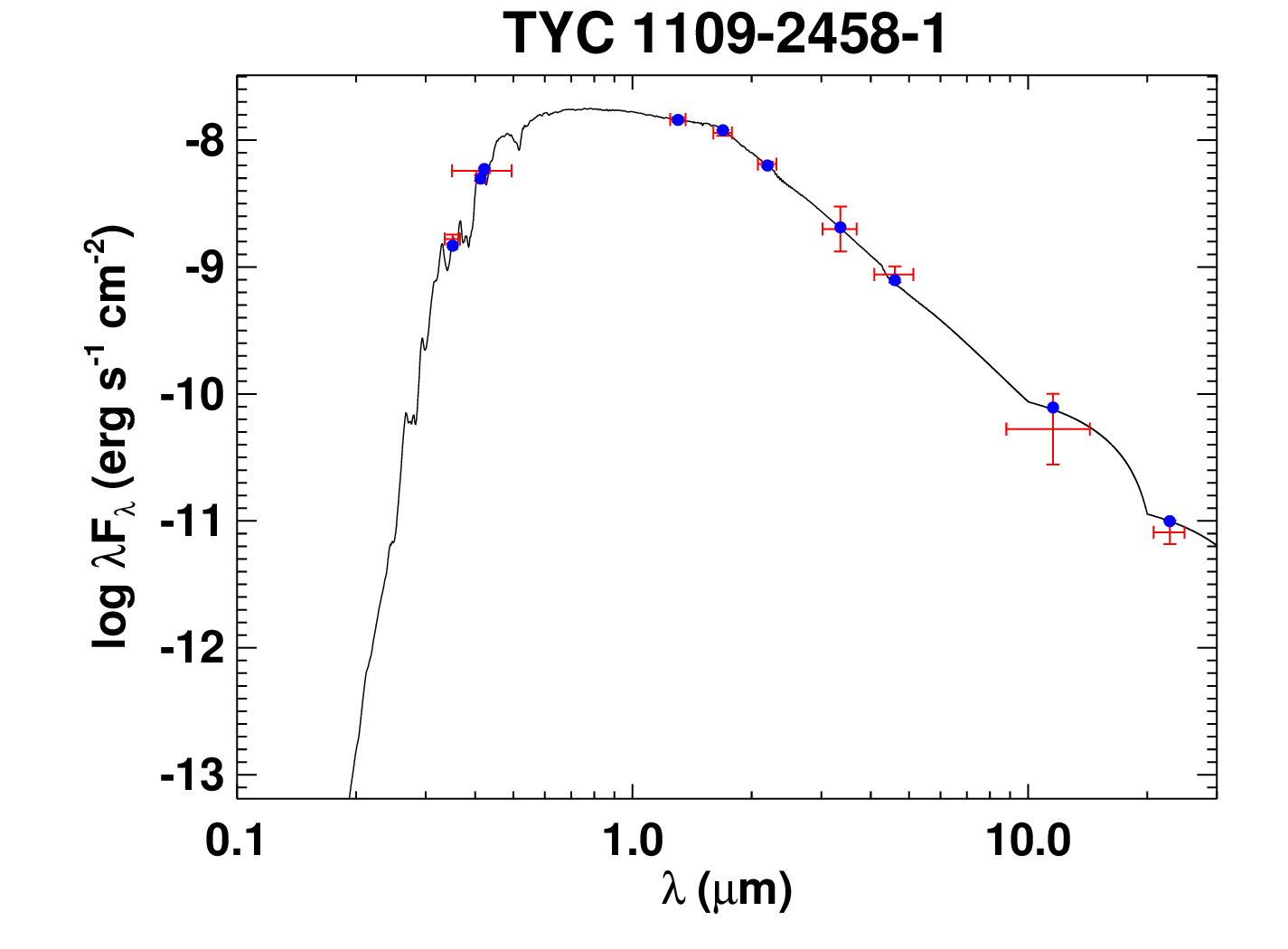}\includegraphics[width=0.333\linewidth]{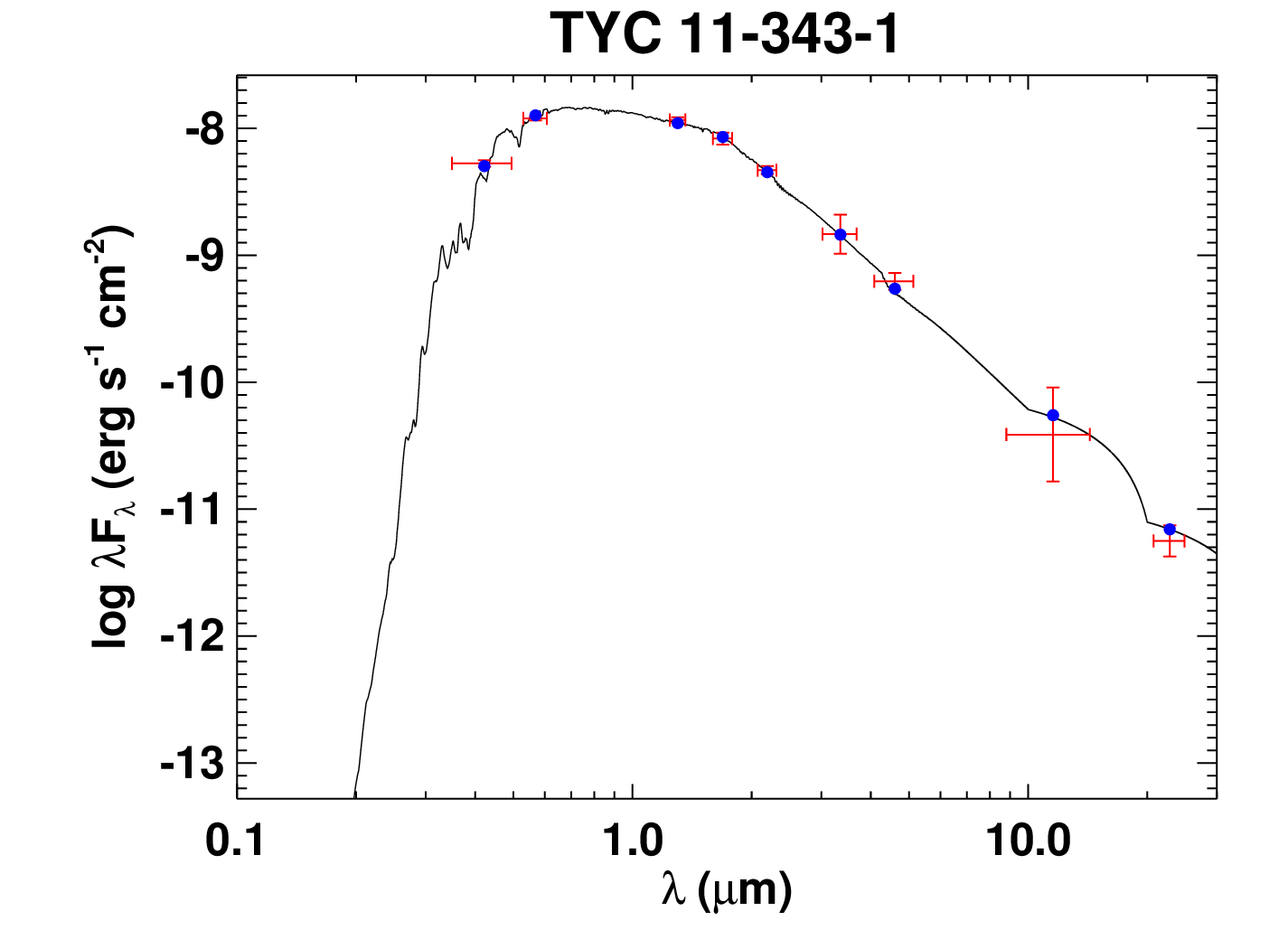}
\includegraphics[width=0.333\linewidth]{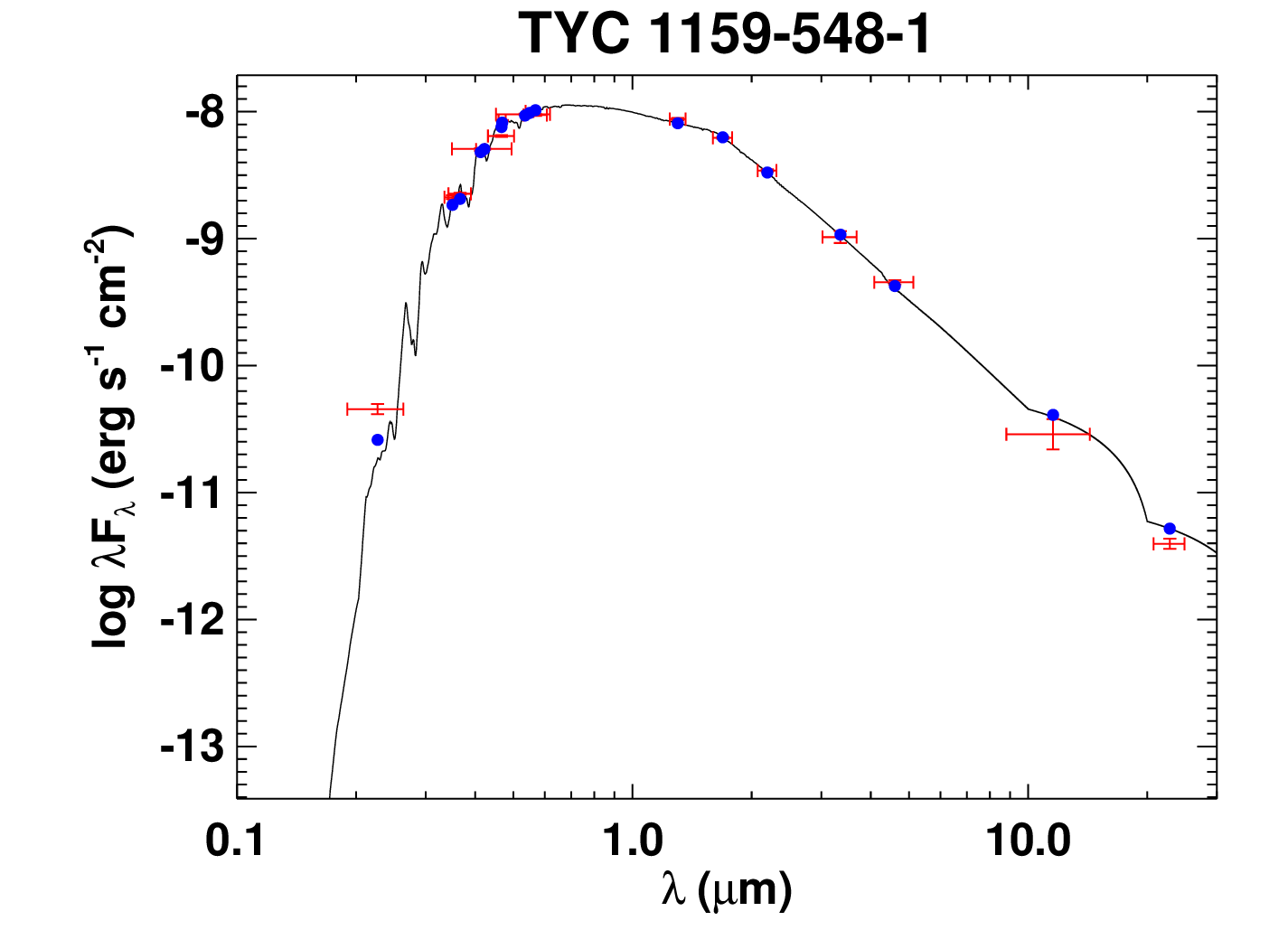}\includegraphics[width=0.333\linewidth]{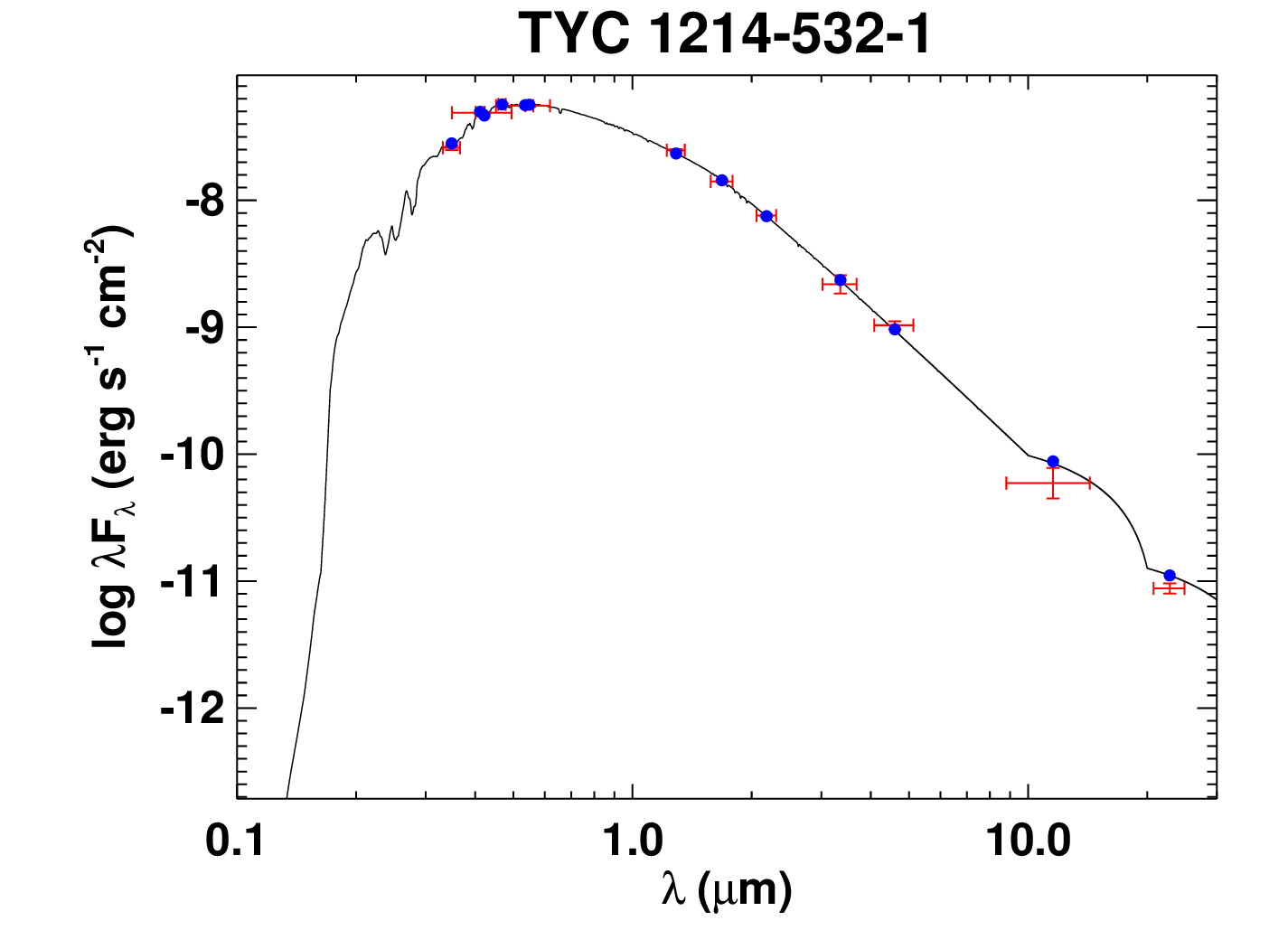}\includegraphics[width=0.333\linewidth]{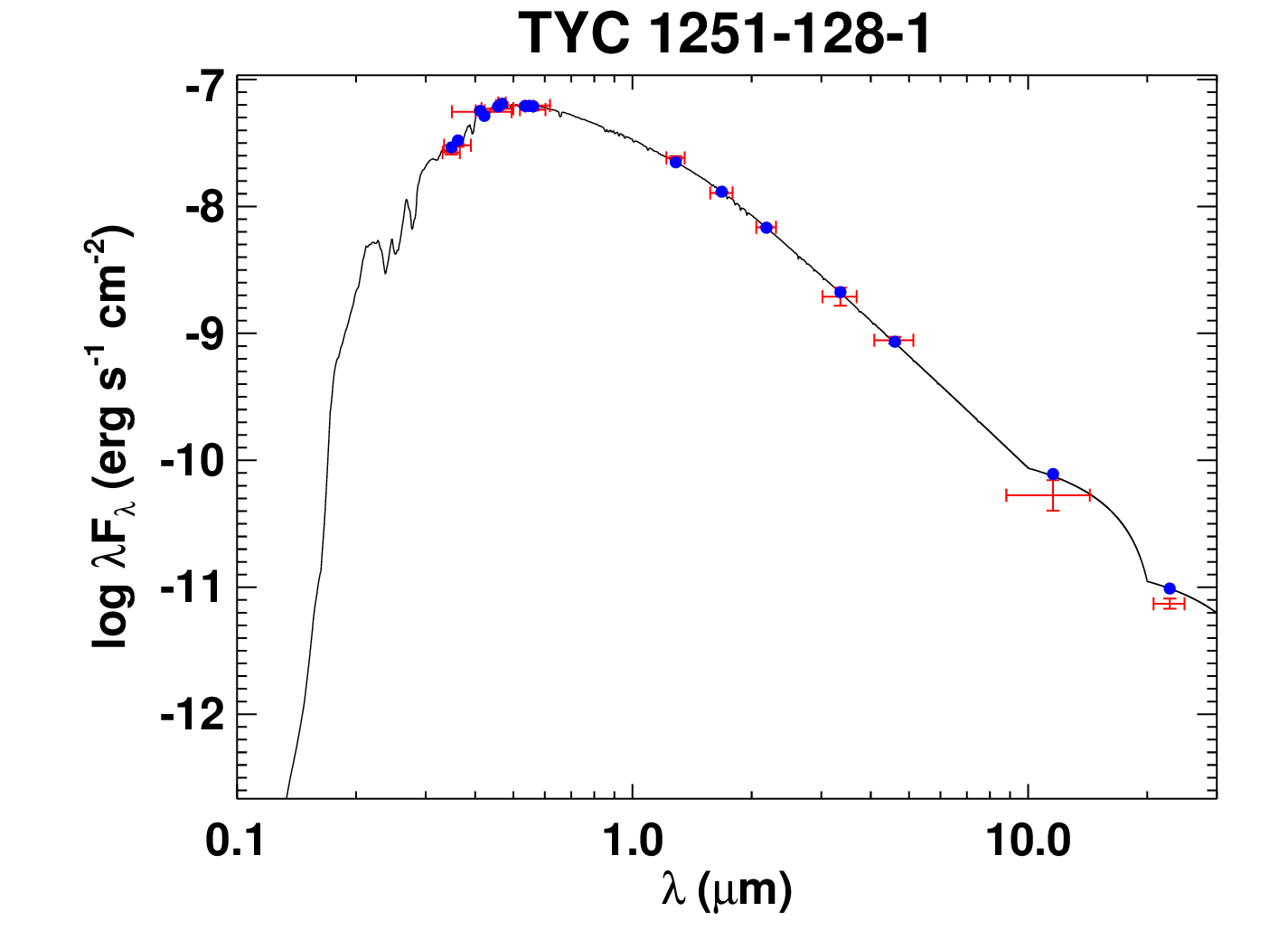}
\includegraphics[width=0.333\linewidth]{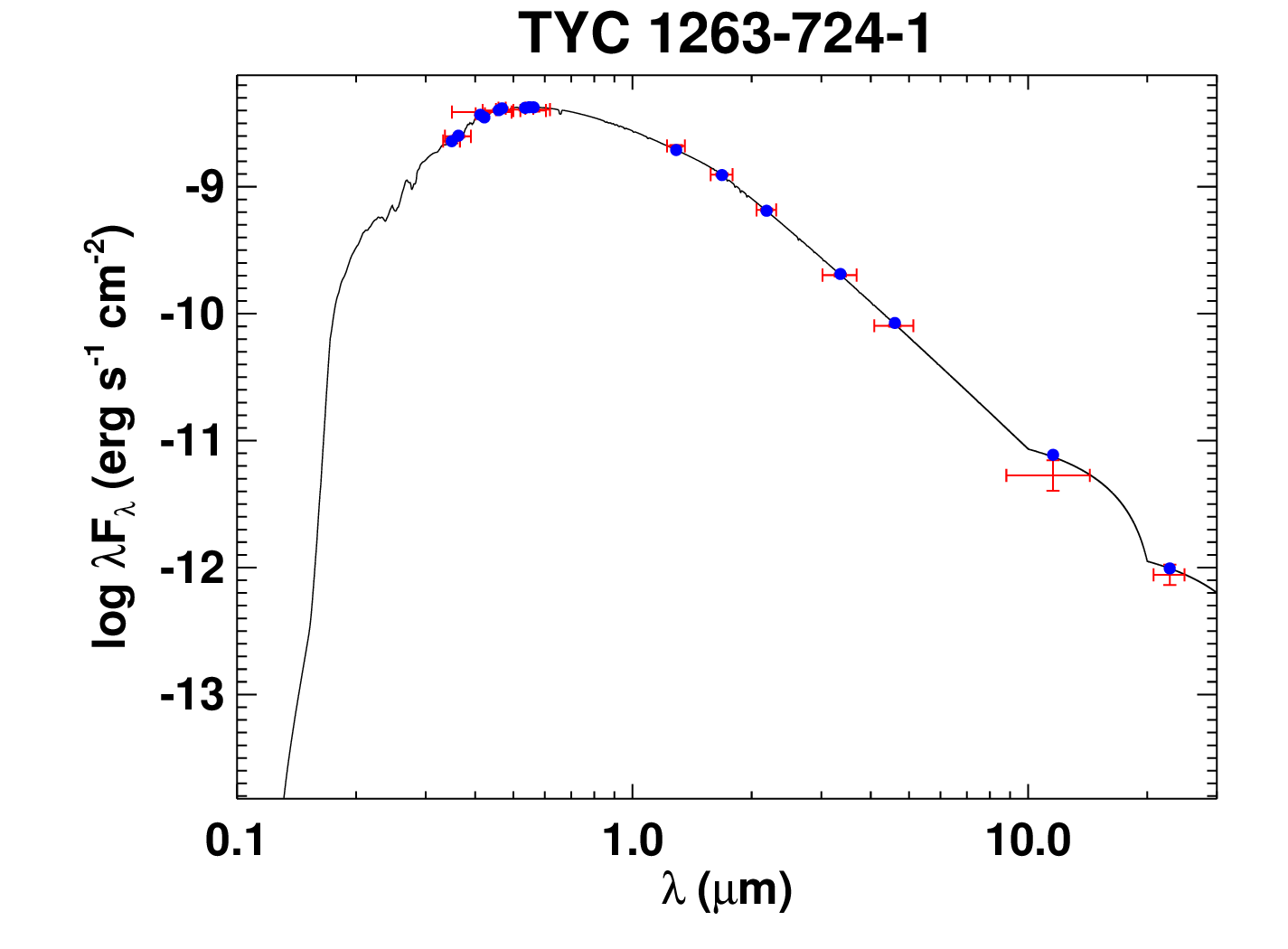}\includegraphics[width=0.333\linewidth]{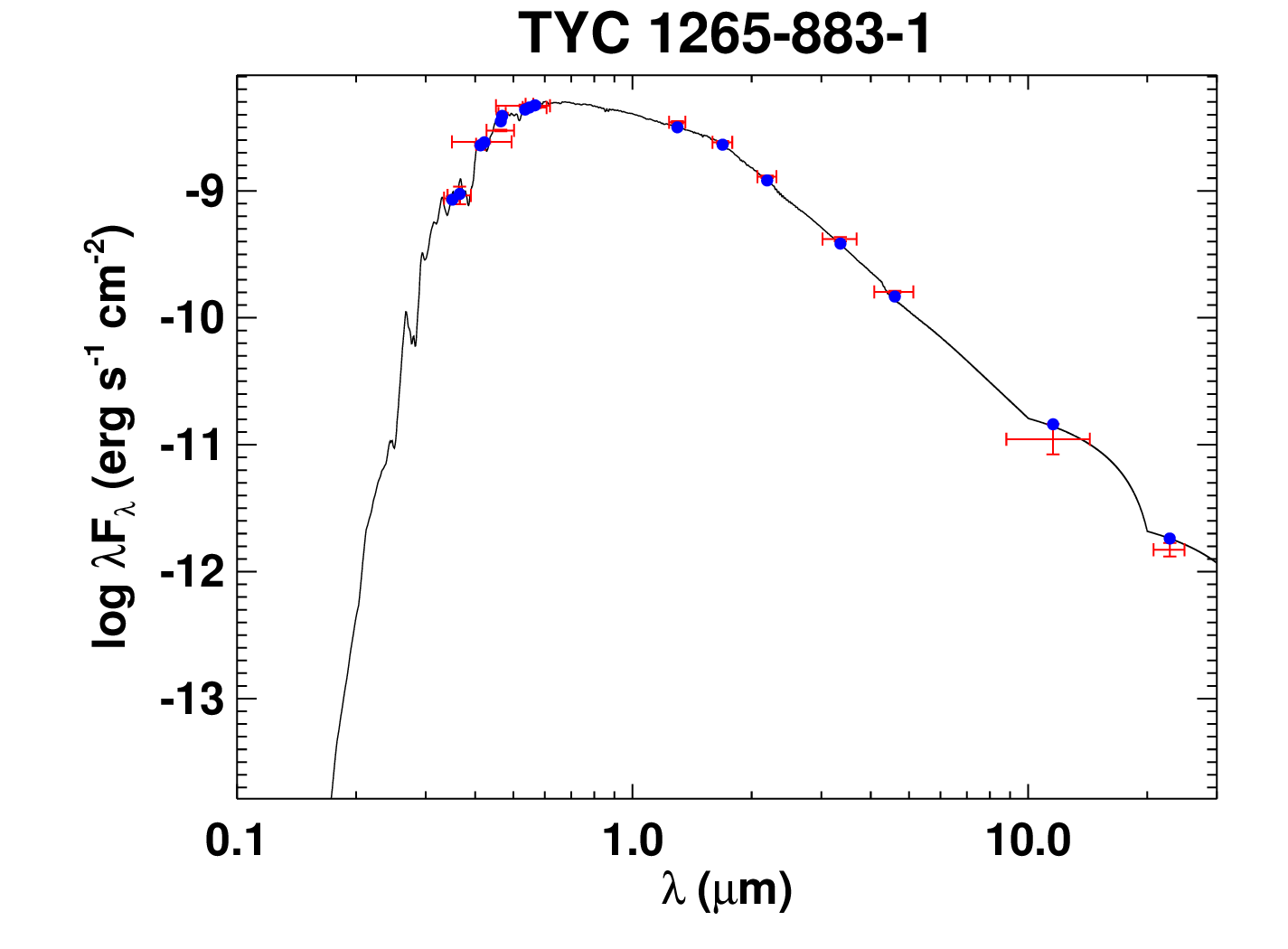}\includegraphics[width=0.333\linewidth]{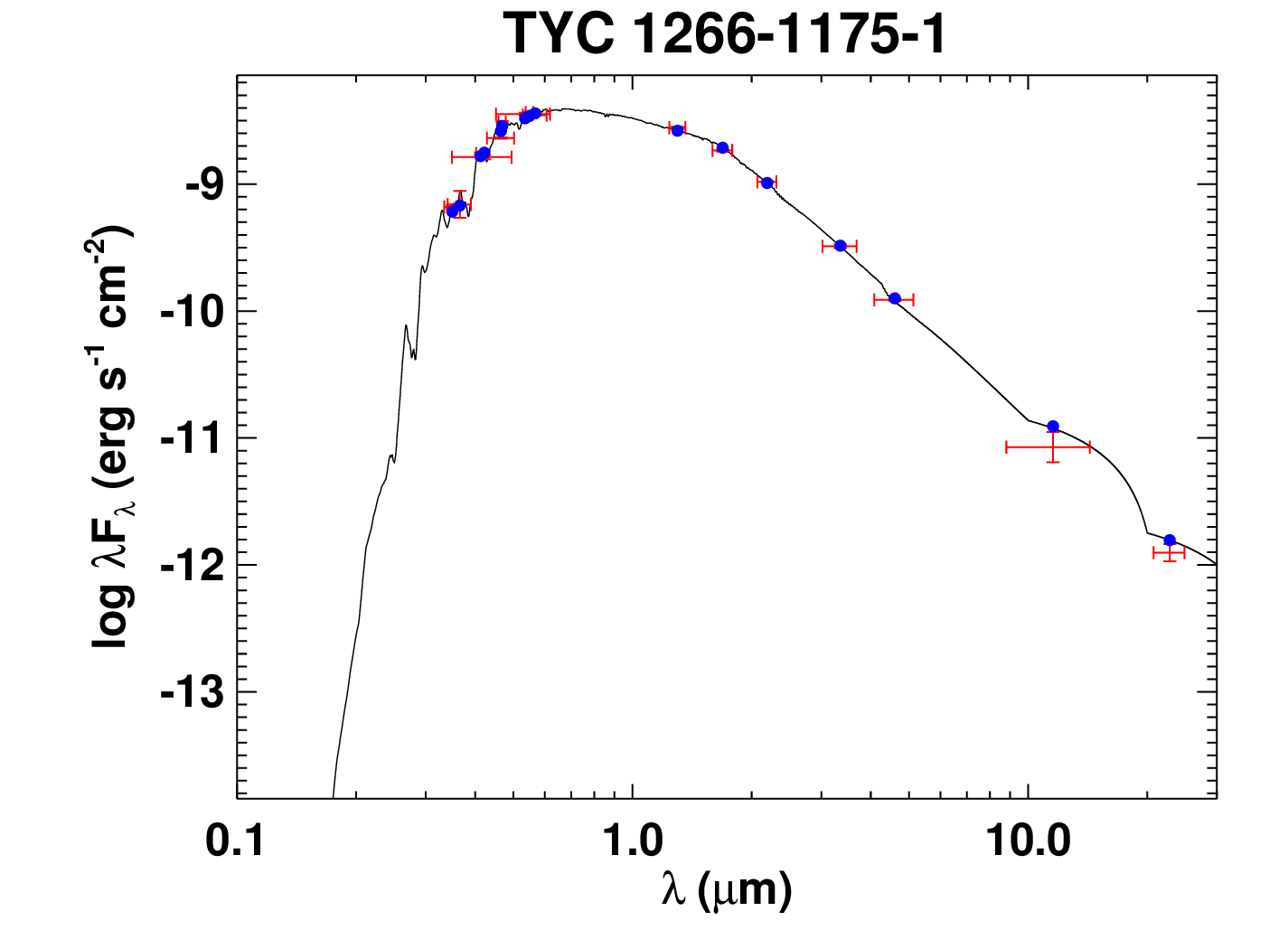}
\caption{\label{fig:seds} Spectral energy distributions (SEDs) for the 244 stars in the \citet{Casagrande2010} sample that also have iterative IRFM solutions. The black curve shows the model fit, while the blue crosses show the measured fluxes: the vertical bars denote the uncertainties on the measurements and the horizontal bars denote the width of the corresponding filter. The blue dots denote the model flux in that passband.}
\end{figure*}

In Figure Set \ref{fig:seds} we present the observed and fitted spectral energy distributions of the 244 \citet{Casagrande2010} stars with which we test the iterative IRFM method employed to determine effective temperatures, bolometric fluxes, and angular diameters for the full study sample.

\begin{figure*}
\includegraphics[width=0.333\linewidth]{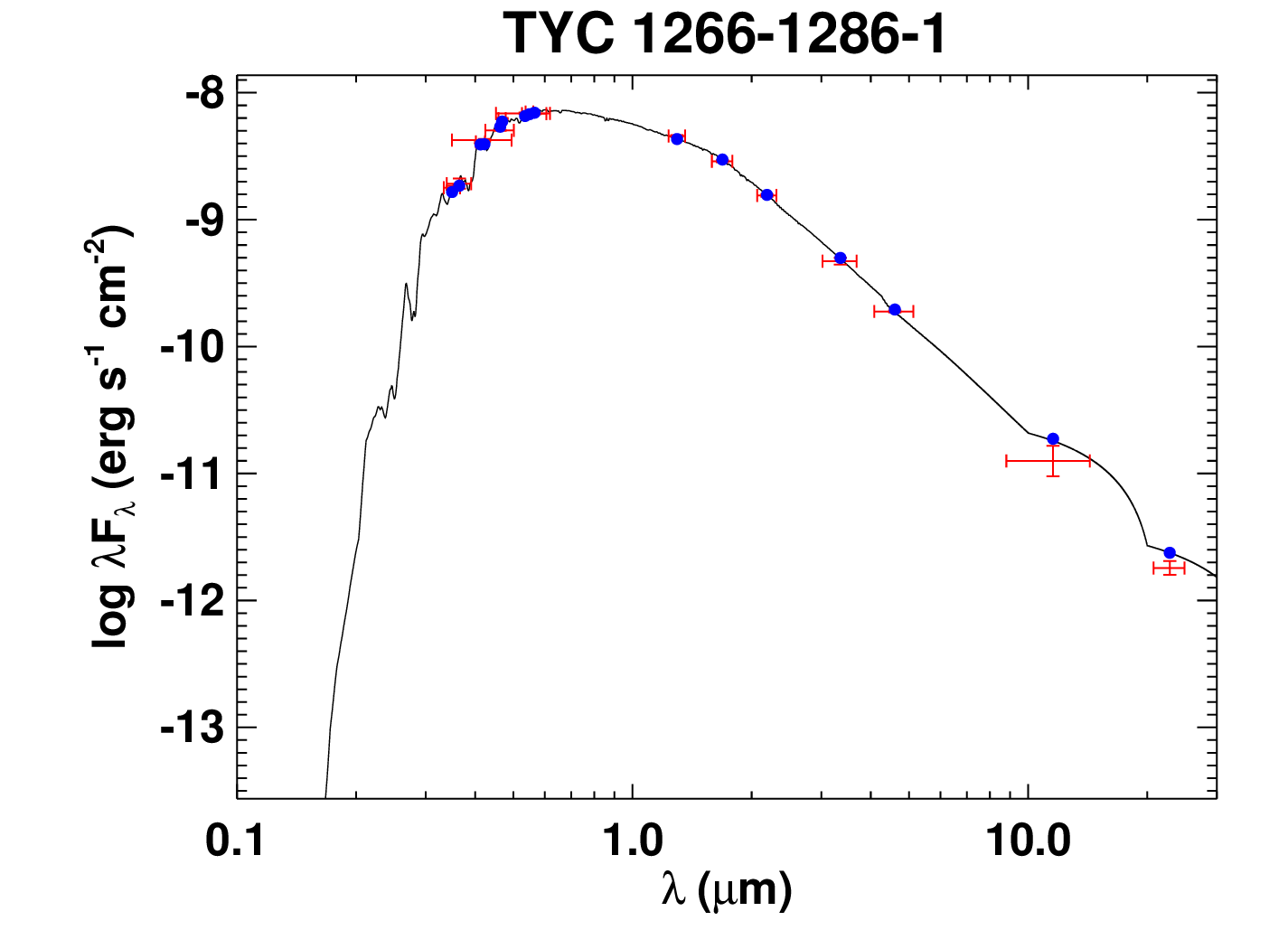}\includegraphics[width=0.333\linewidth]{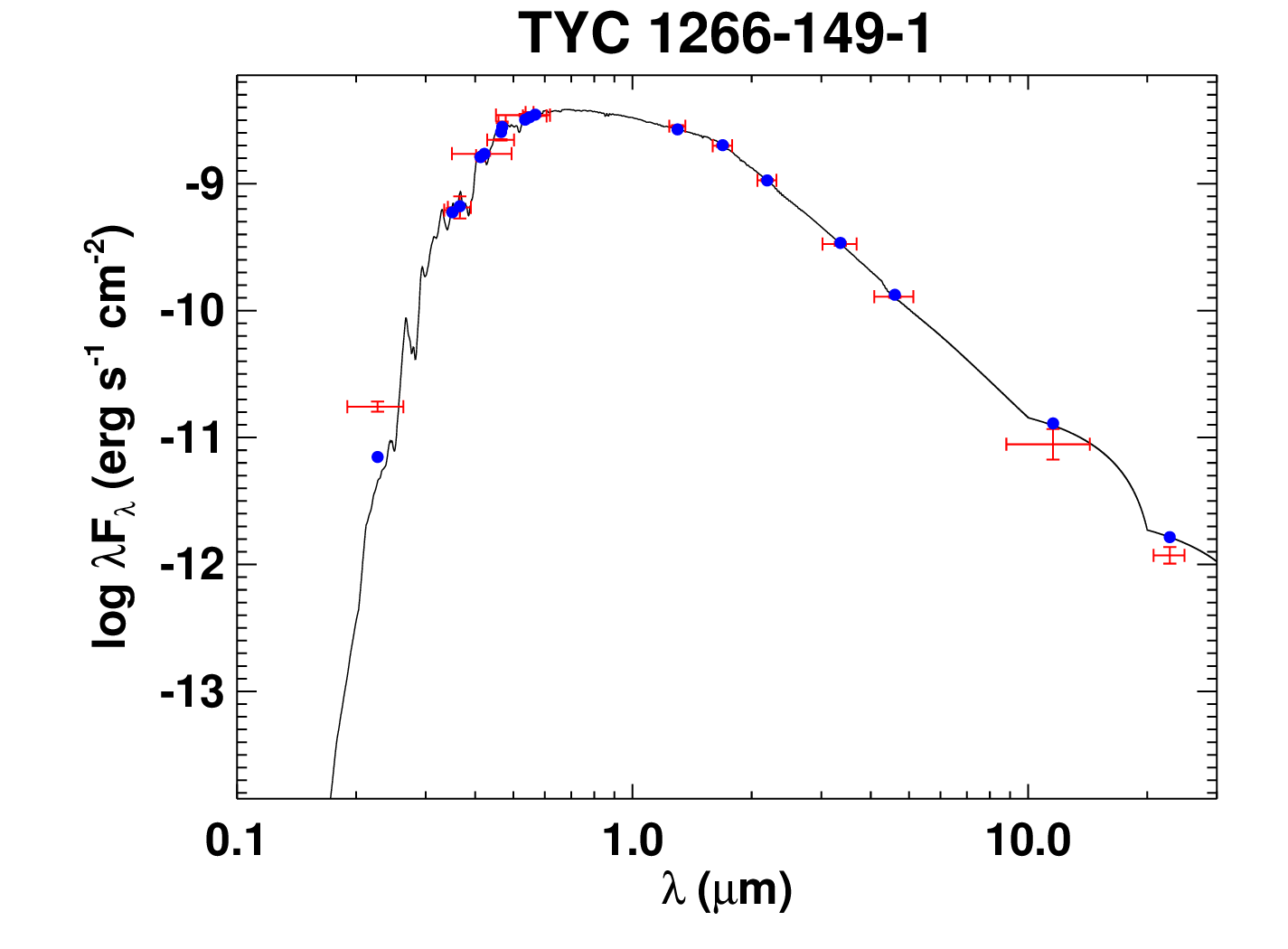}\includegraphics[width=0.333\linewidth]{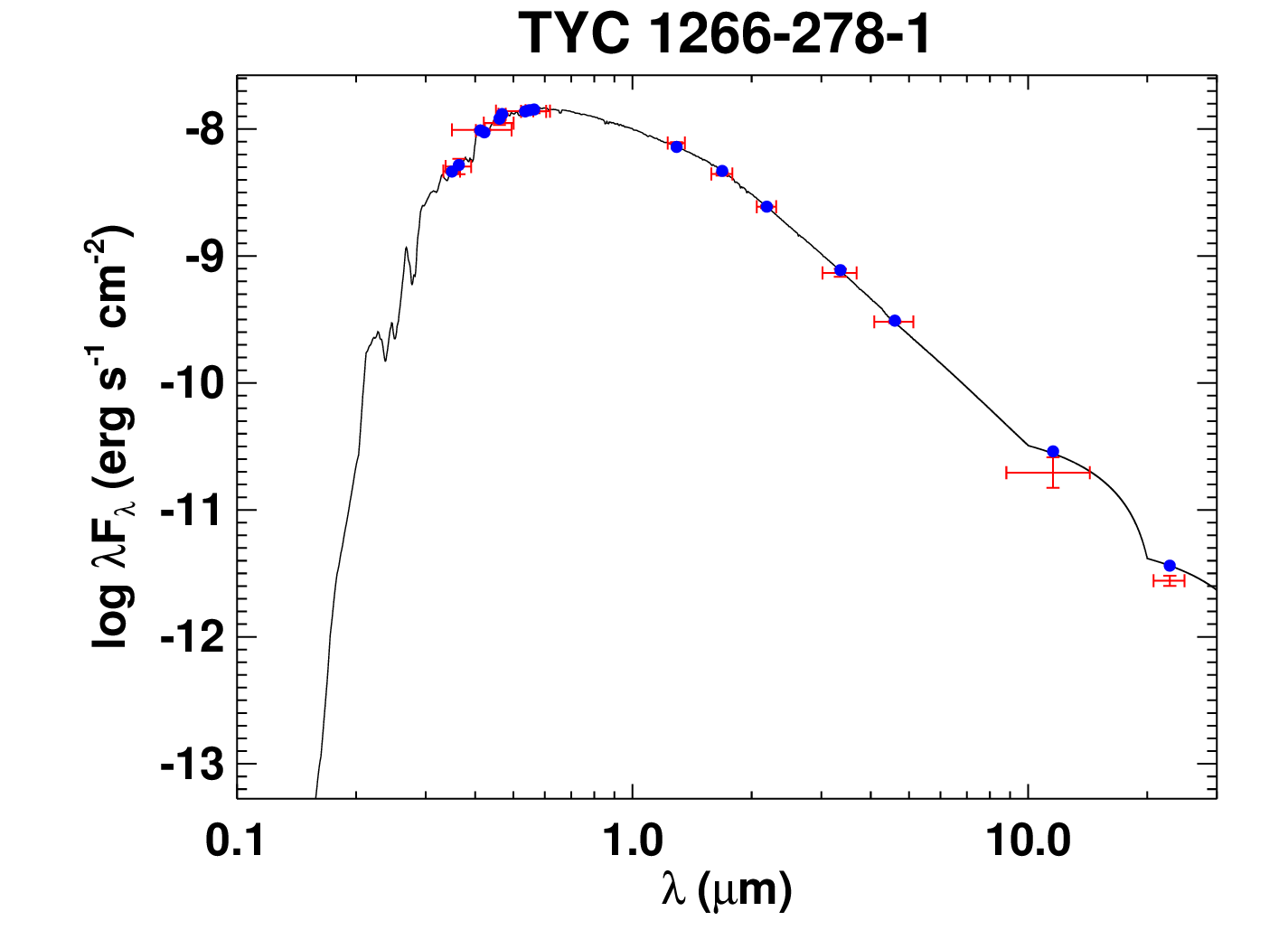}
\includegraphics[width=0.333\linewidth]{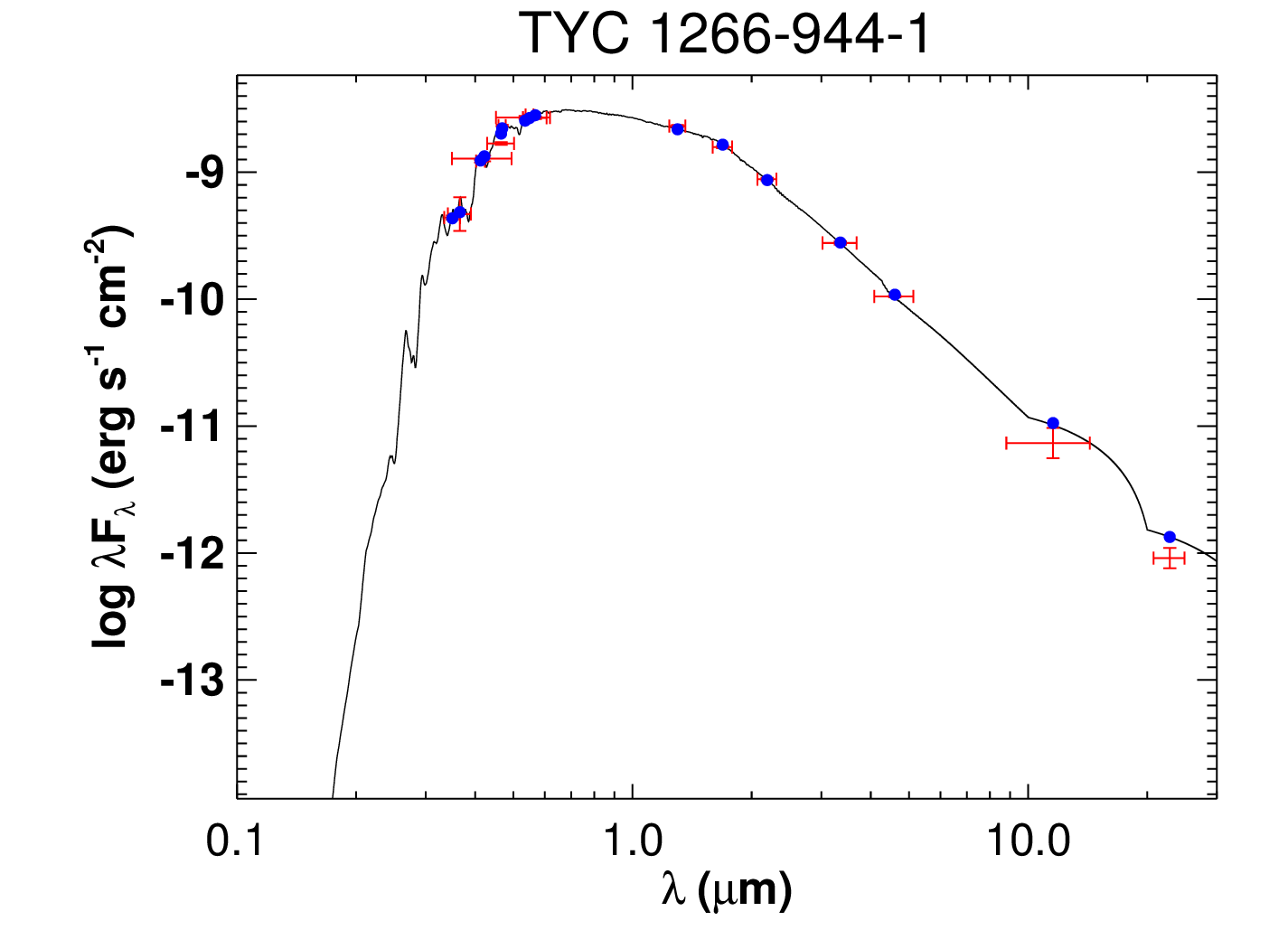}\includegraphics[width=0.333\linewidth]{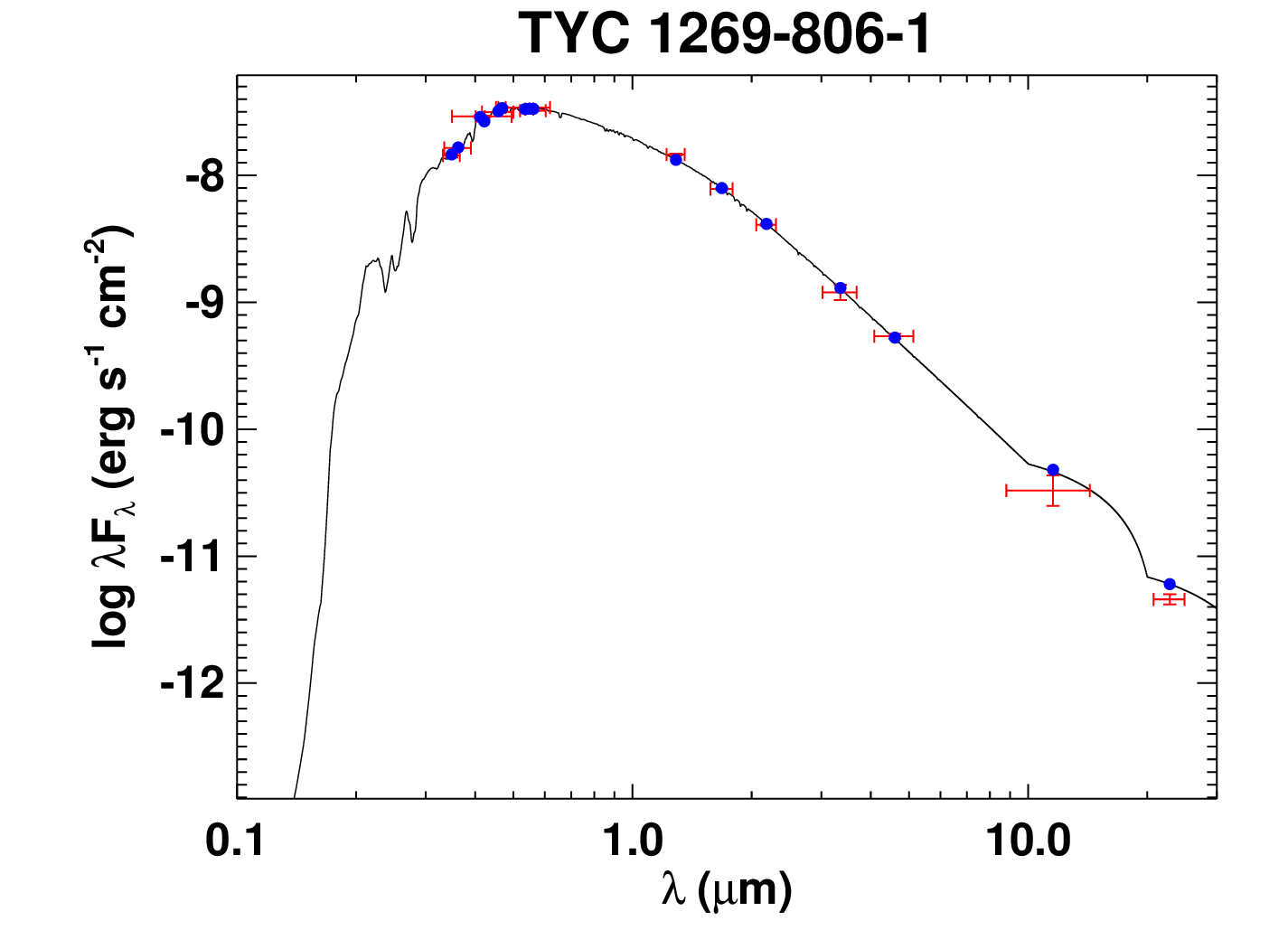}\includegraphics[width=0.333\linewidth]{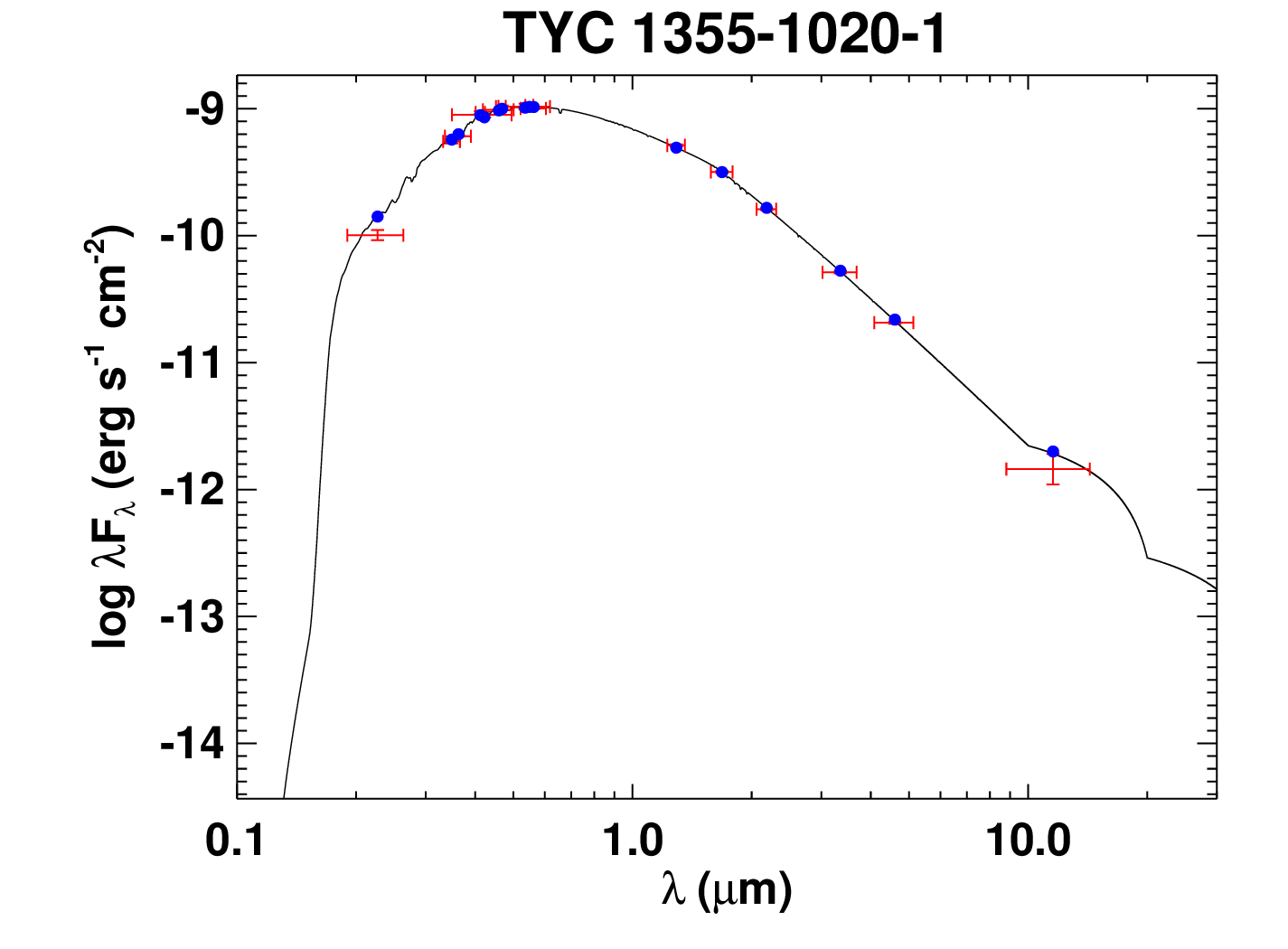}
\includegraphics[width=0.333\linewidth]{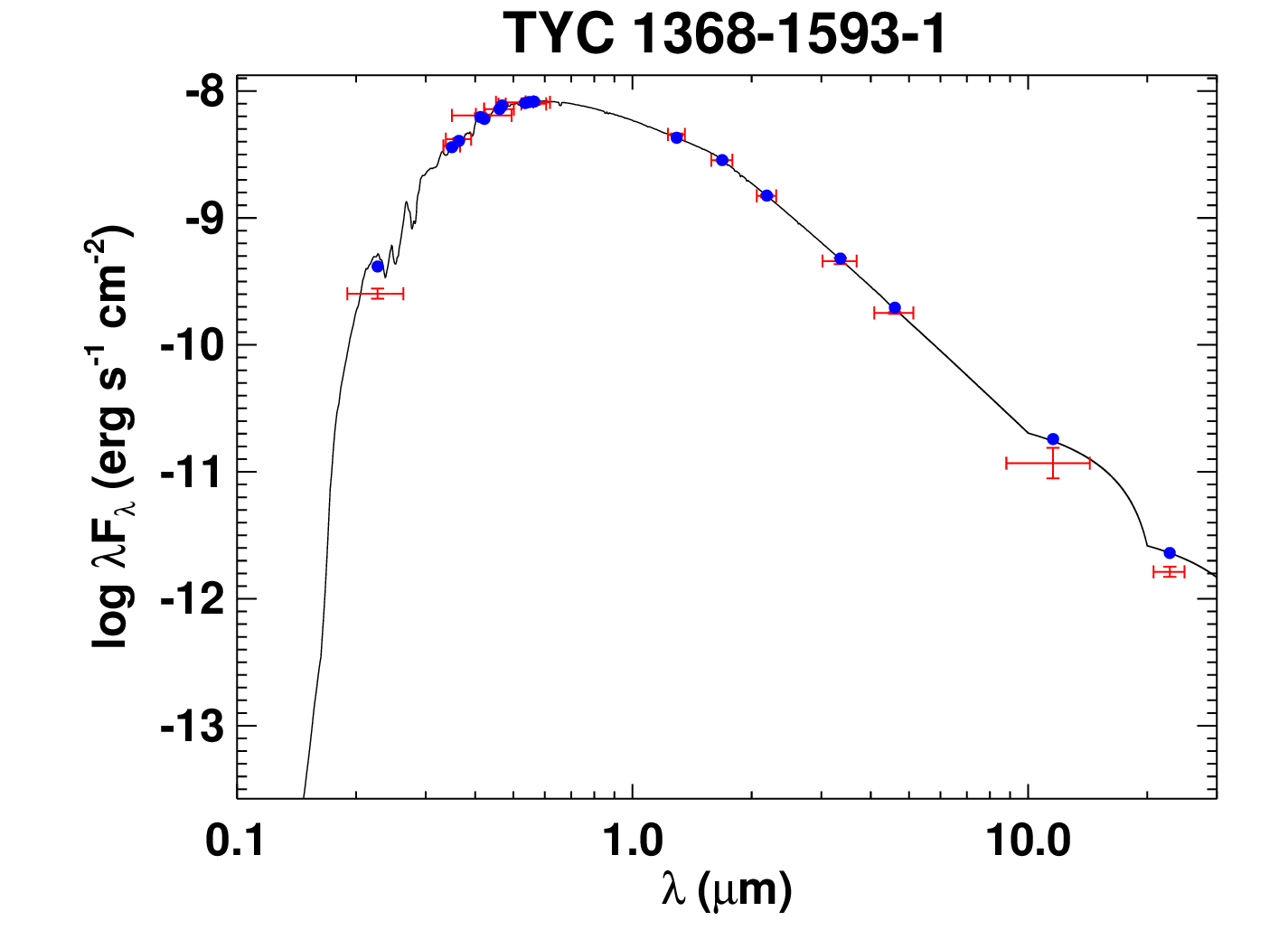}\includegraphics[width=0.333\linewidth]{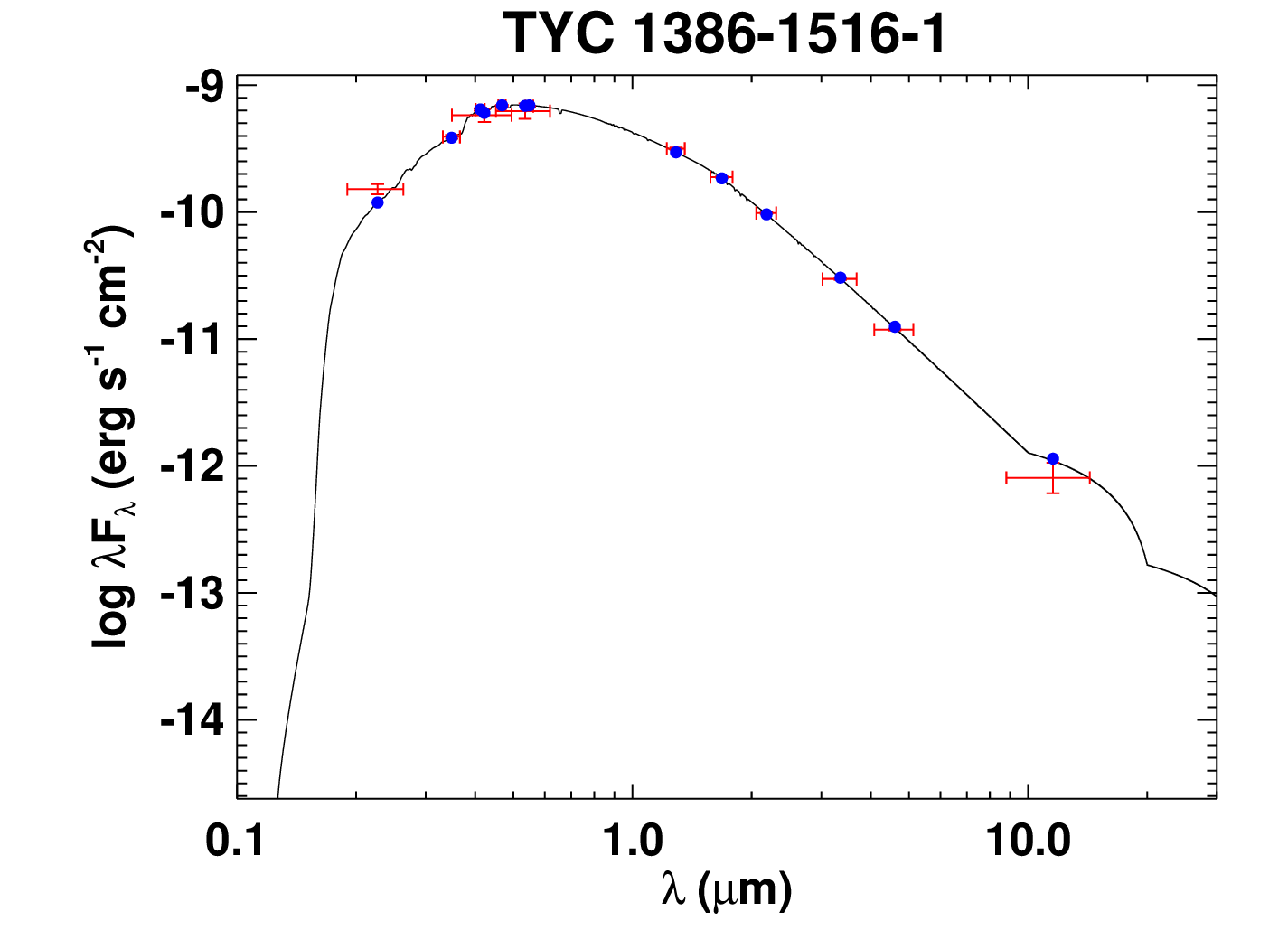}\includegraphics[width=0.333\linewidth]{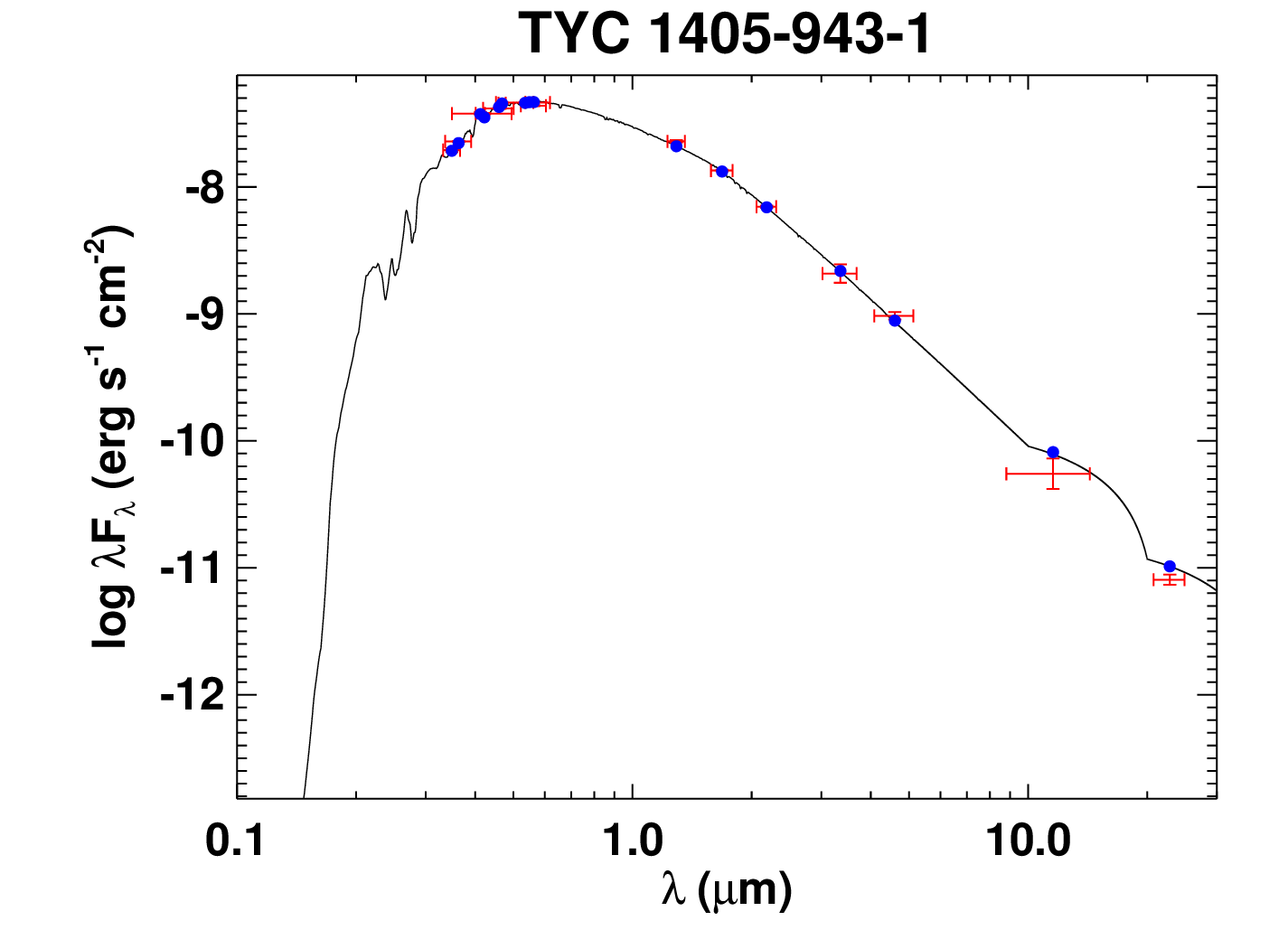}
\includegraphics[width=0.333\linewidth]{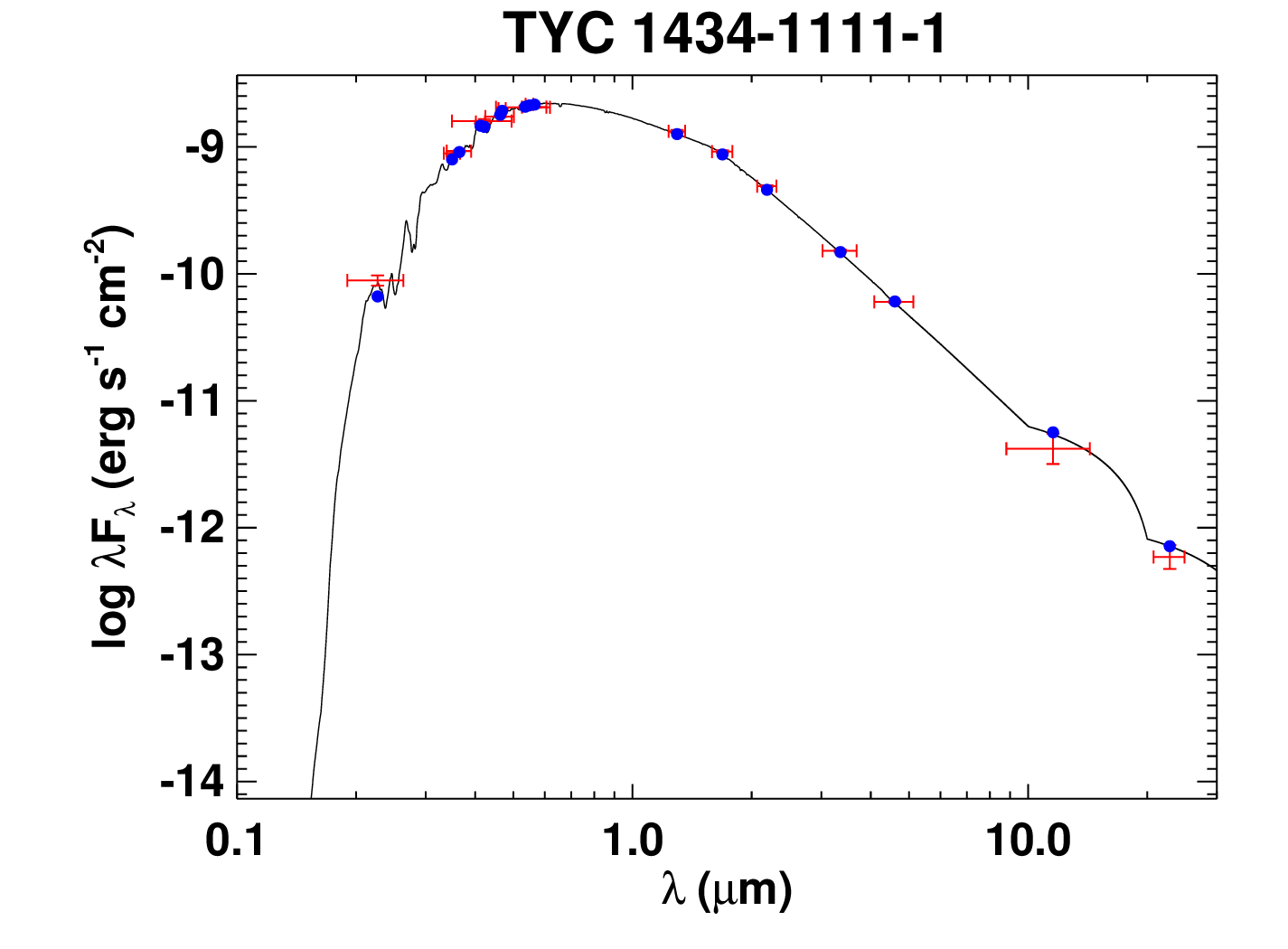}\includegraphics[width=0.333\linewidth]{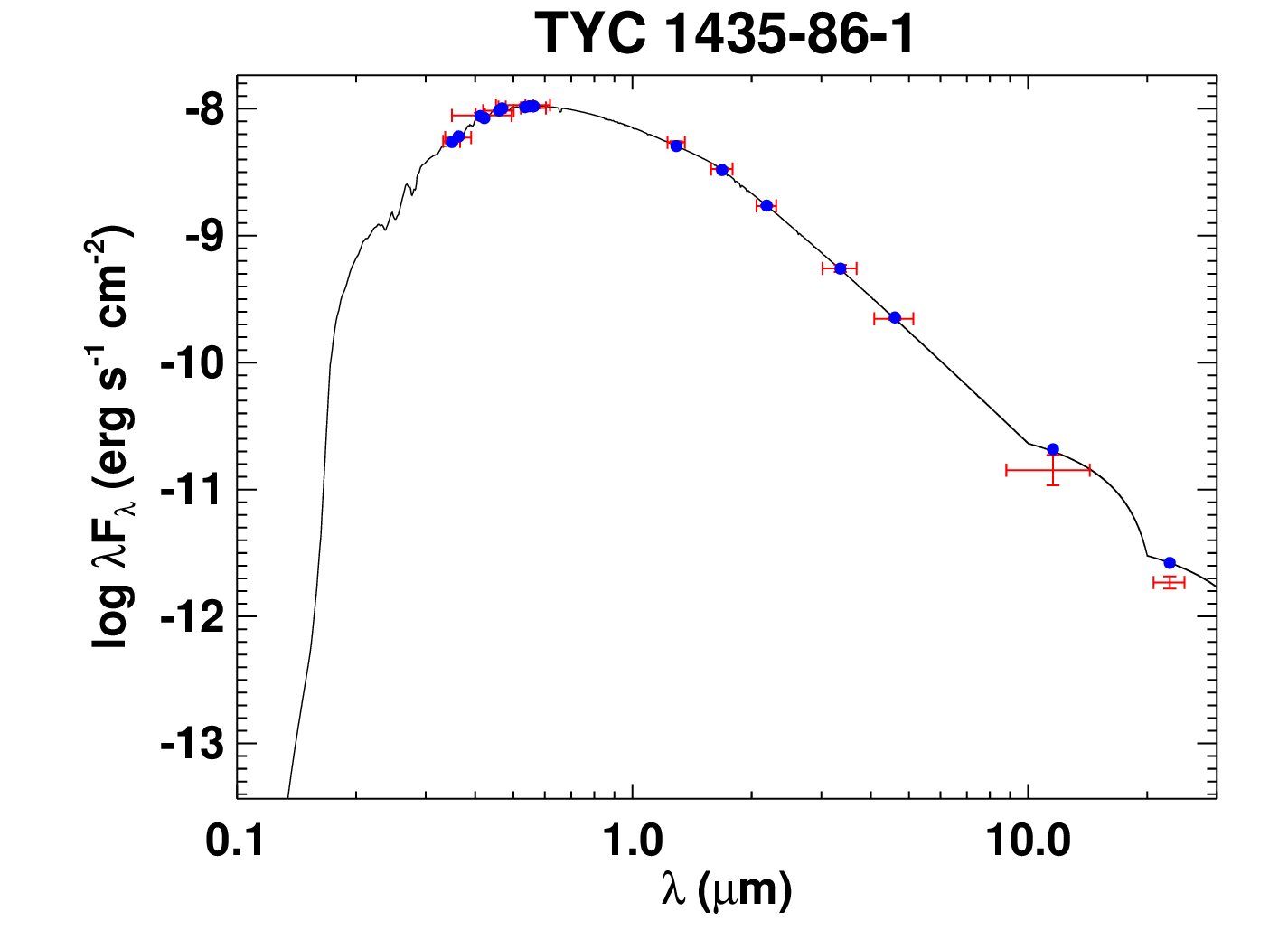}\includegraphics[width=0.333\linewidth]{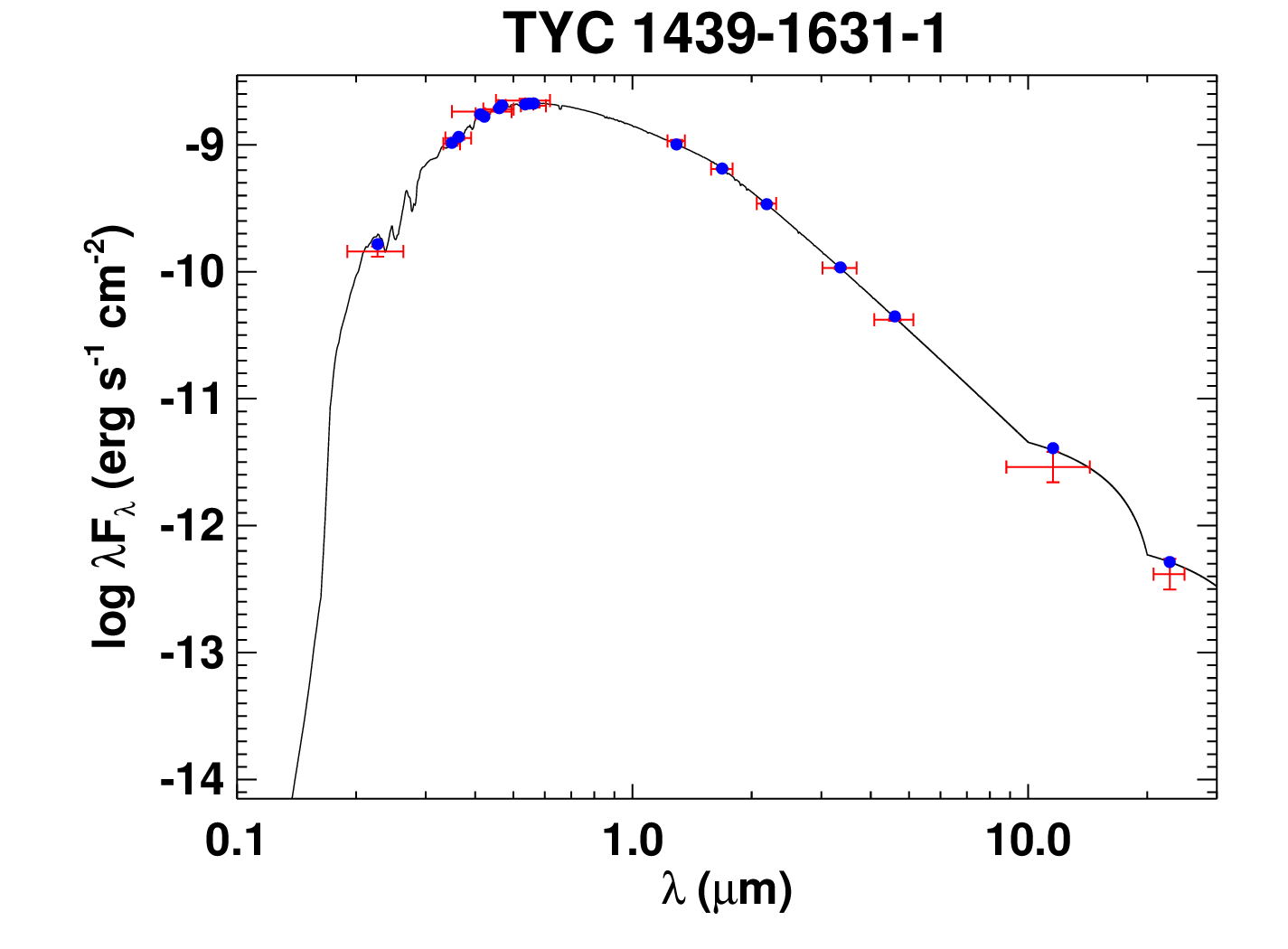}
\caption{\label{fig:seds1} All labels, lines, symbols, and colors as in Figure \ref{fig:seds}.}
\end{figure*}

\begin{figure*}
\includegraphics[width=0.333\linewidth]{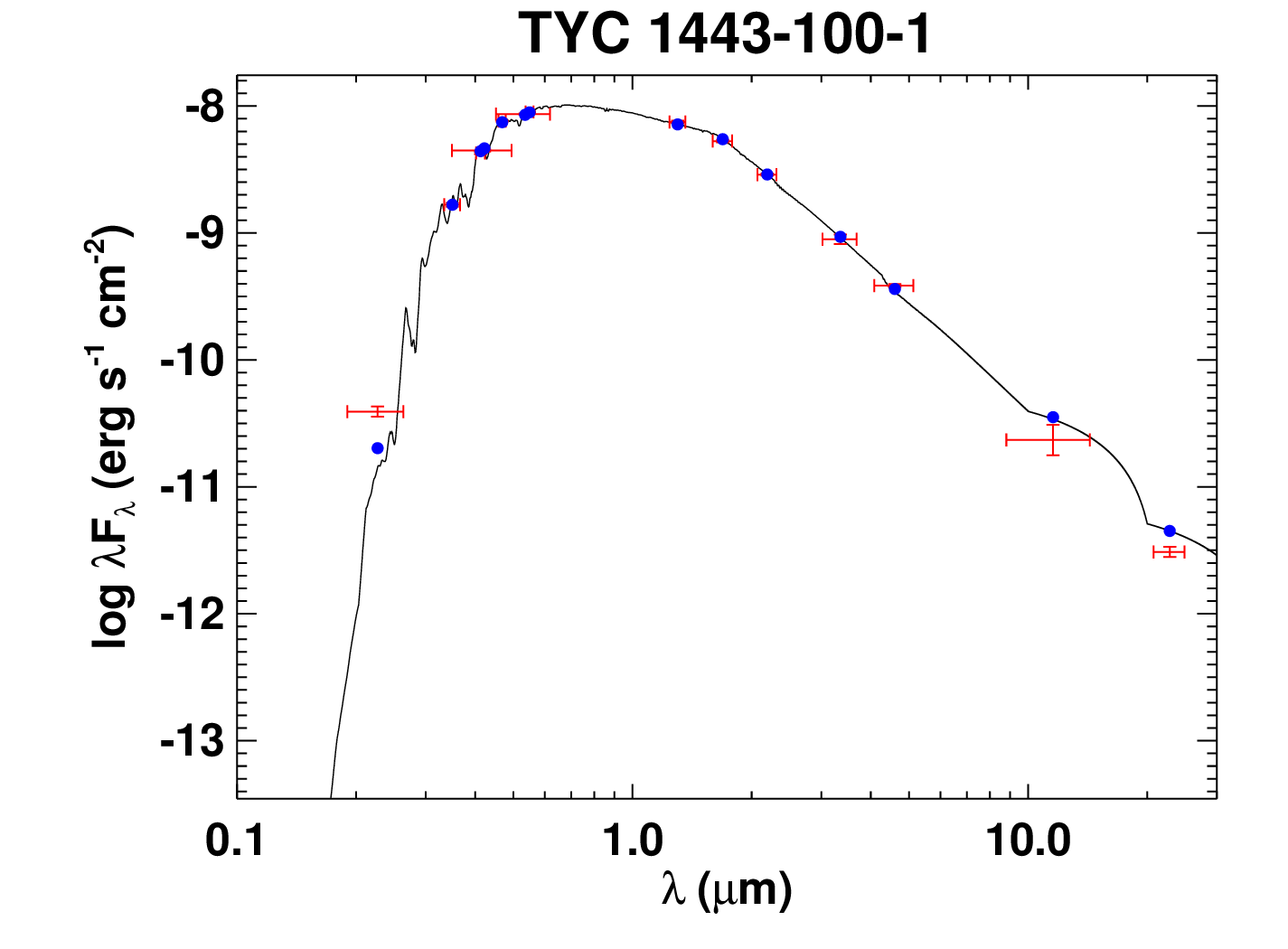}\includegraphics[width=0.333\linewidth]{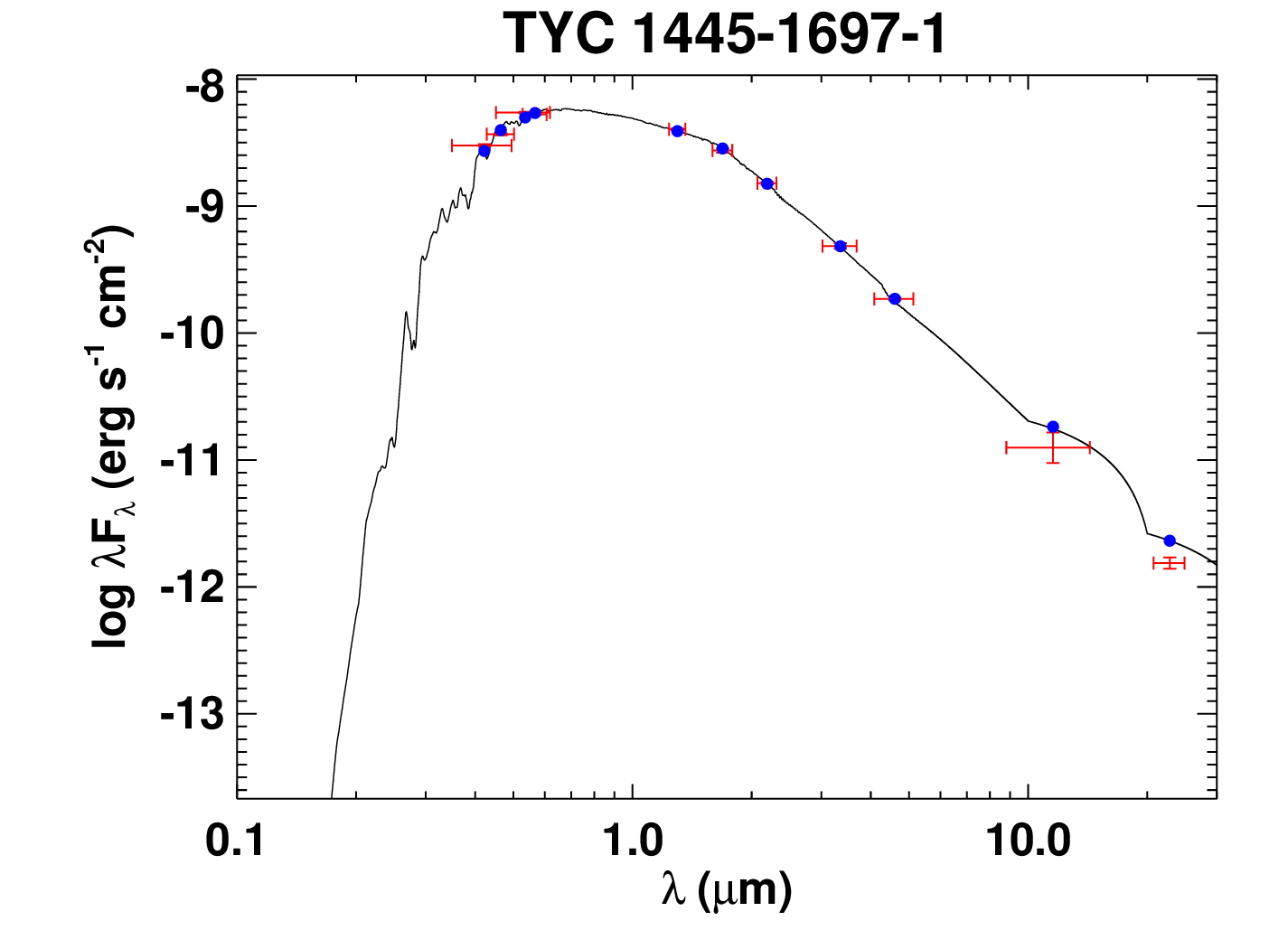}\includegraphics[width=0.333\linewidth]{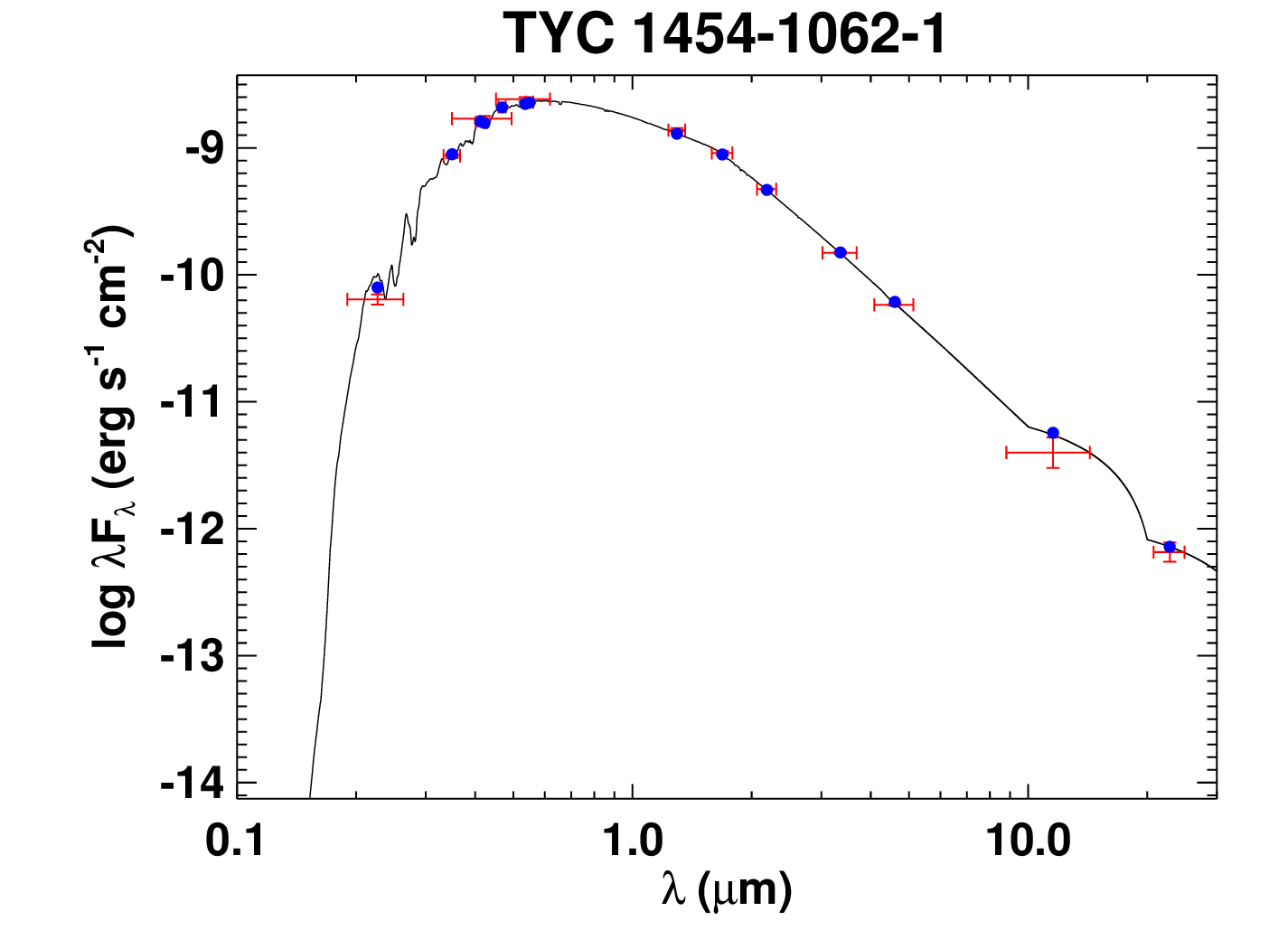}
\includegraphics[width=0.333\linewidth]{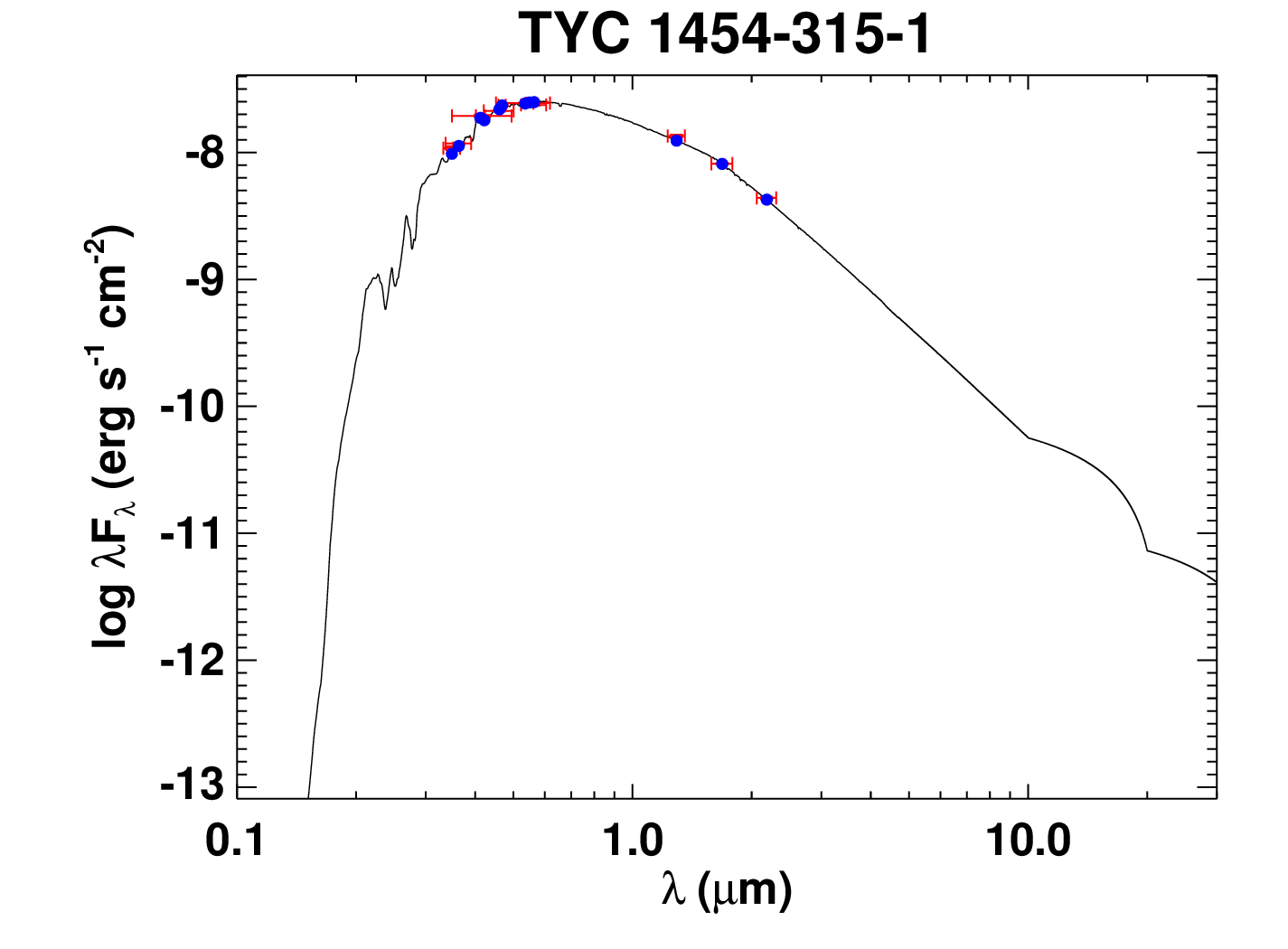}\includegraphics[width=0.333\linewidth]{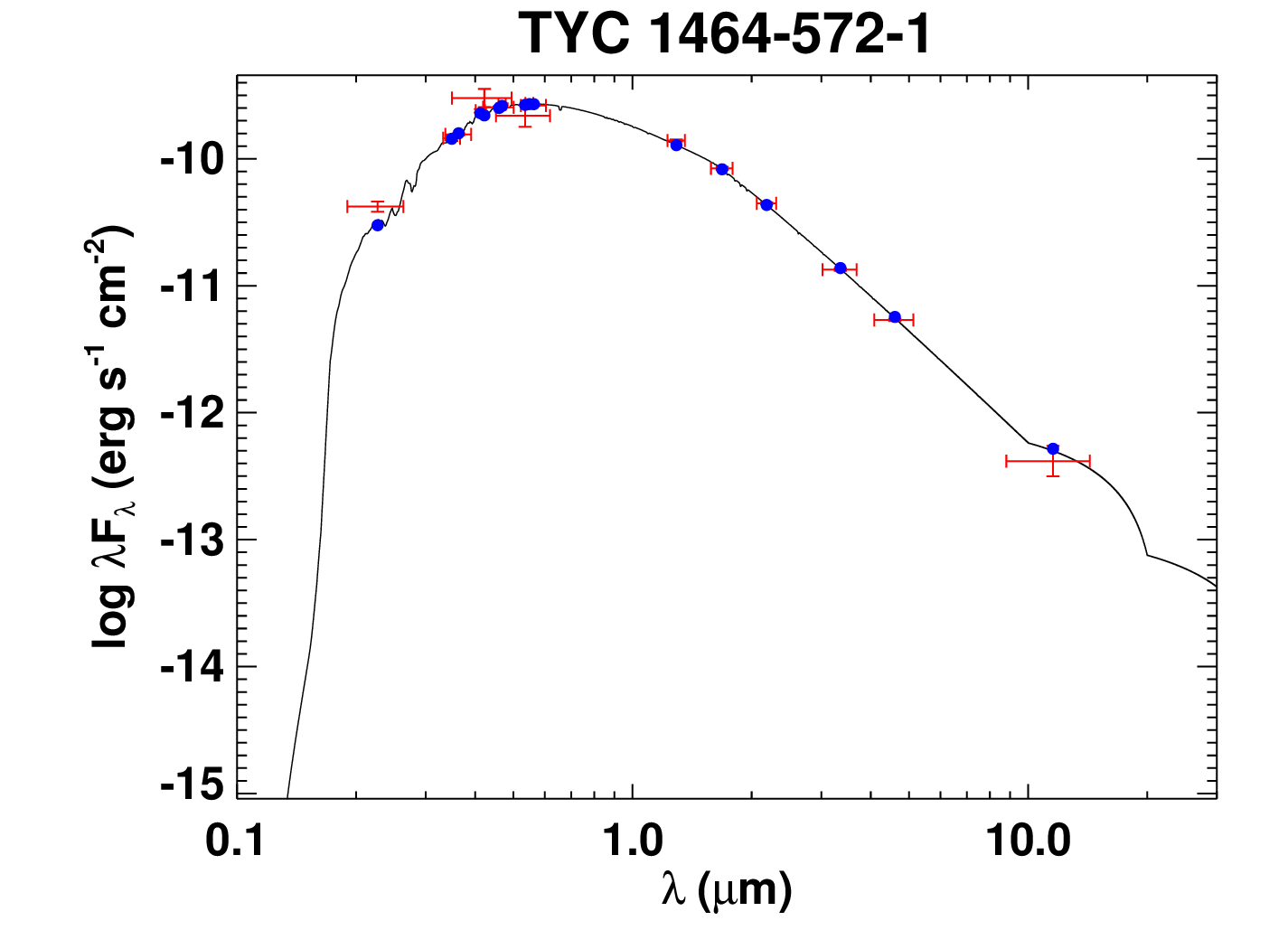}\includegraphics[width=0.333\linewidth]{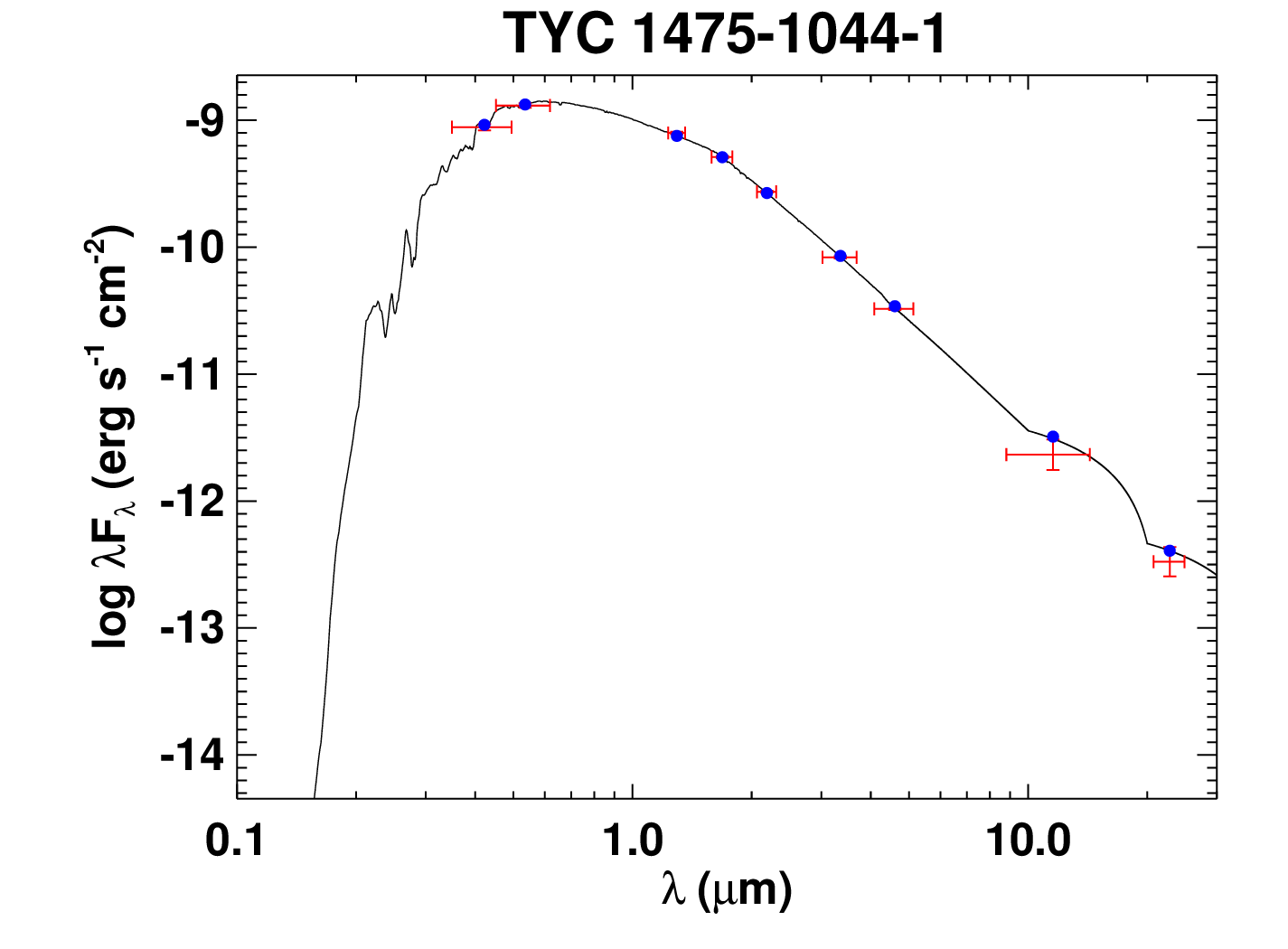}
\includegraphics[width=0.333\linewidth]{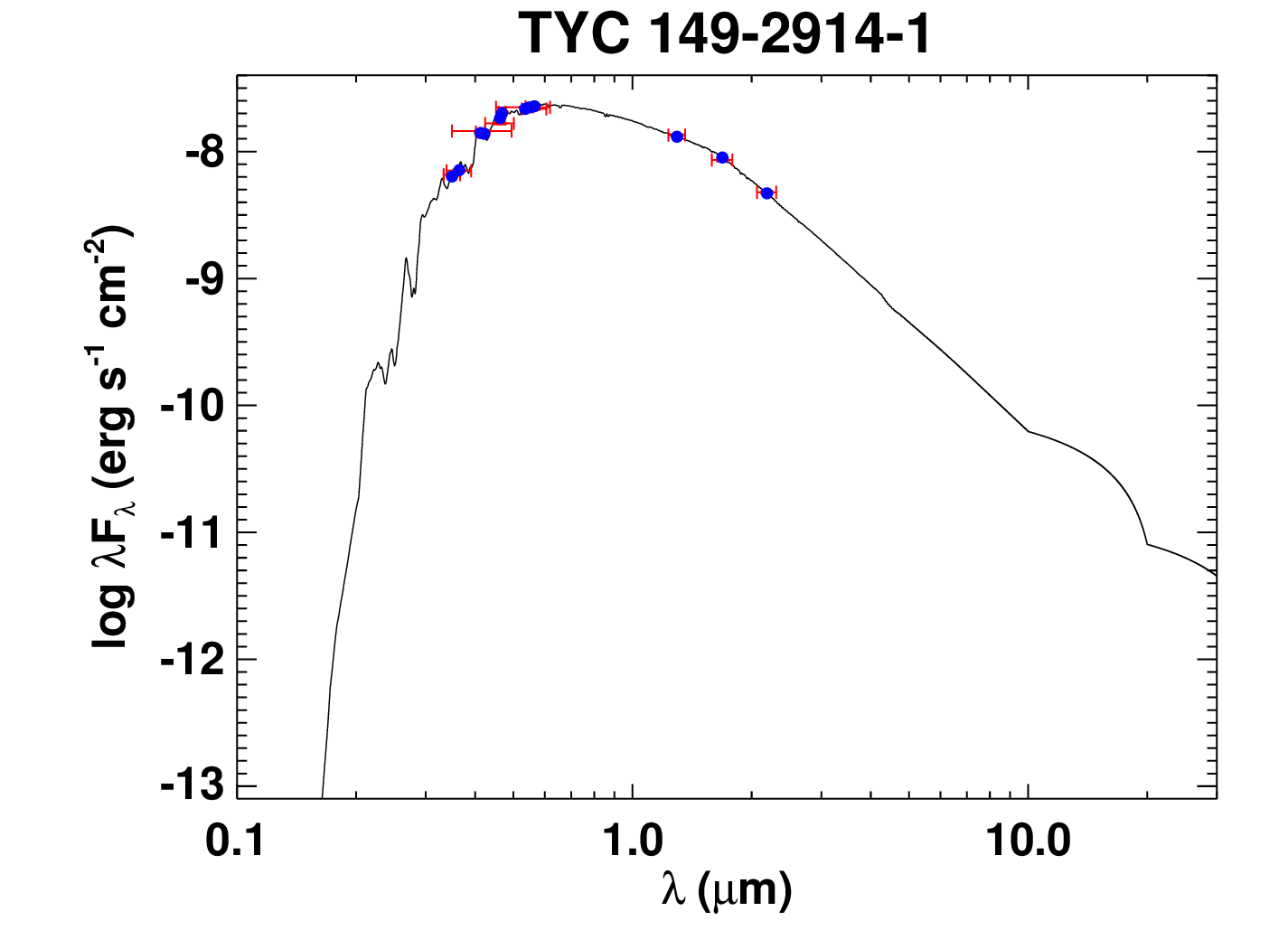}\includegraphics[width=0.333\linewidth]{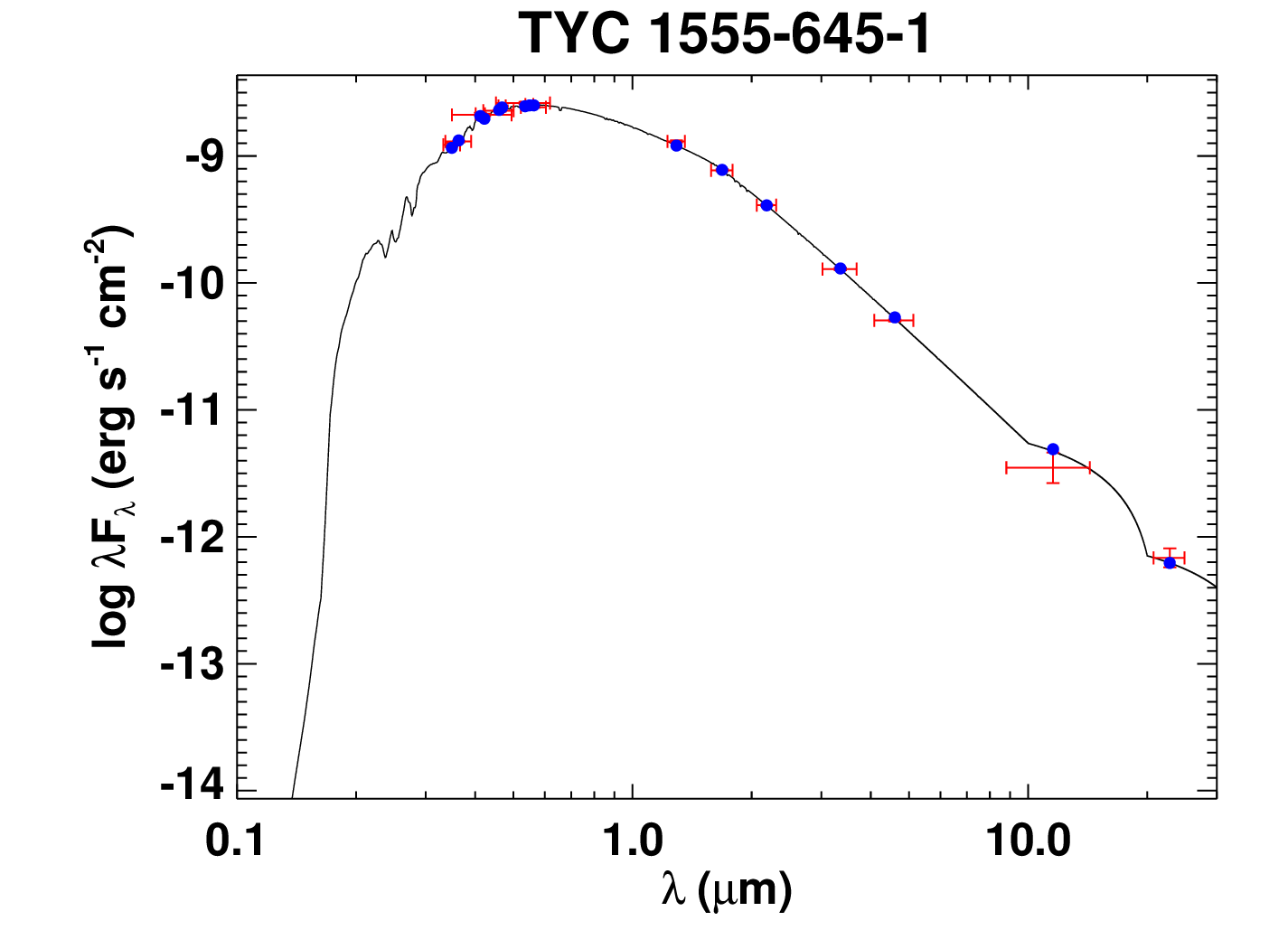}\includegraphics[width=0.333\linewidth]{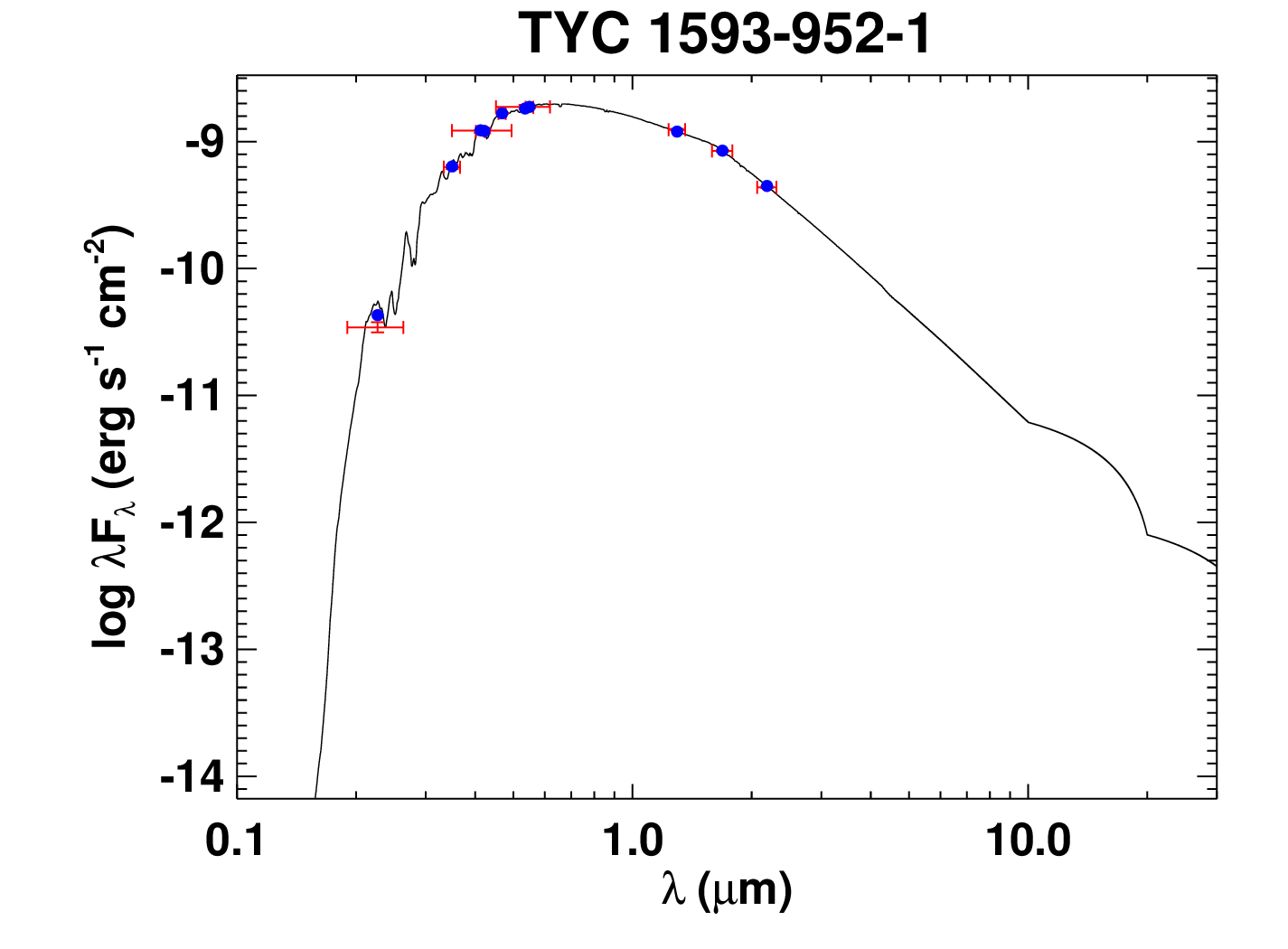}
\includegraphics[width=0.333\linewidth]{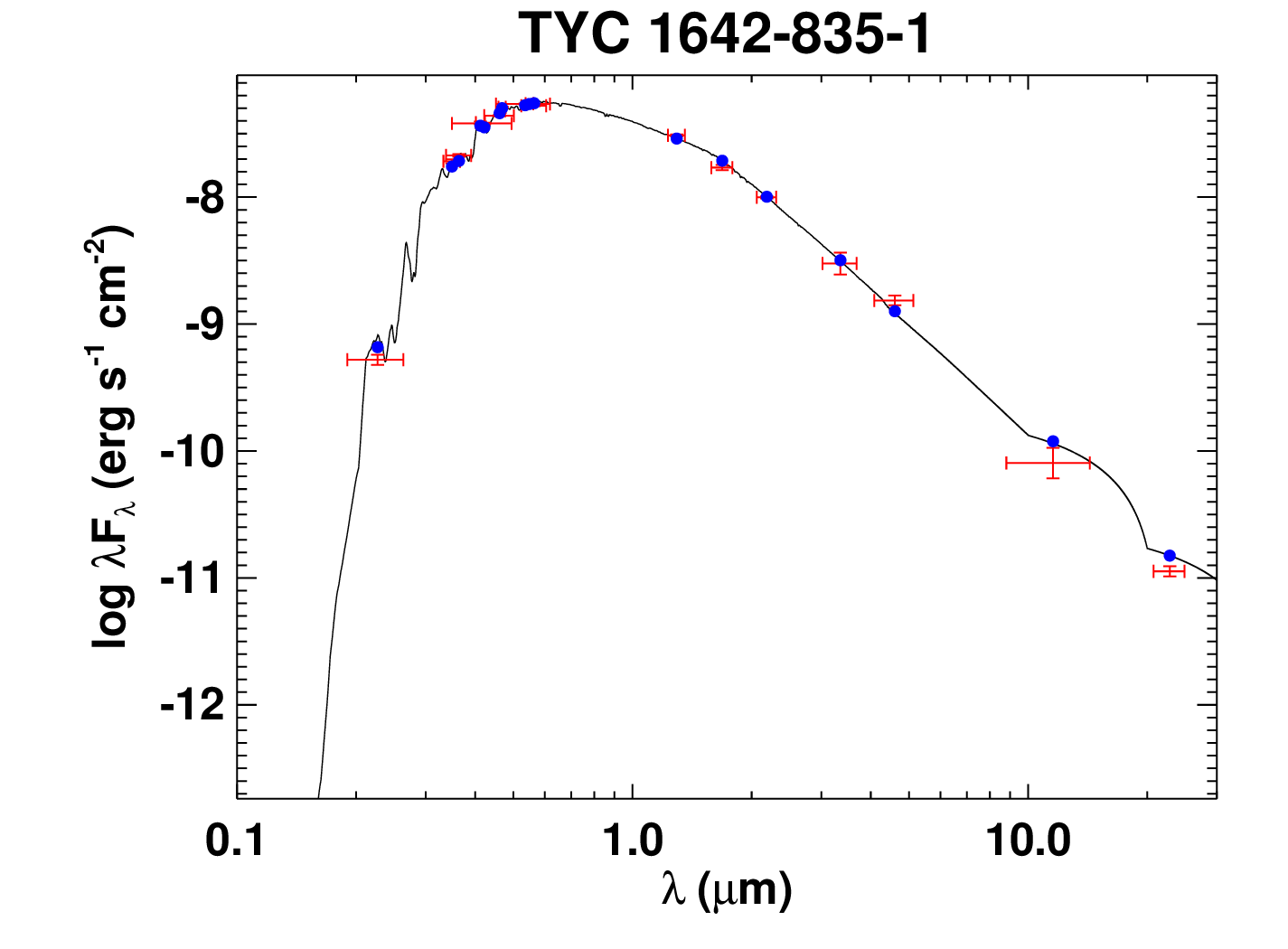}\includegraphics[width=0.333\linewidth]{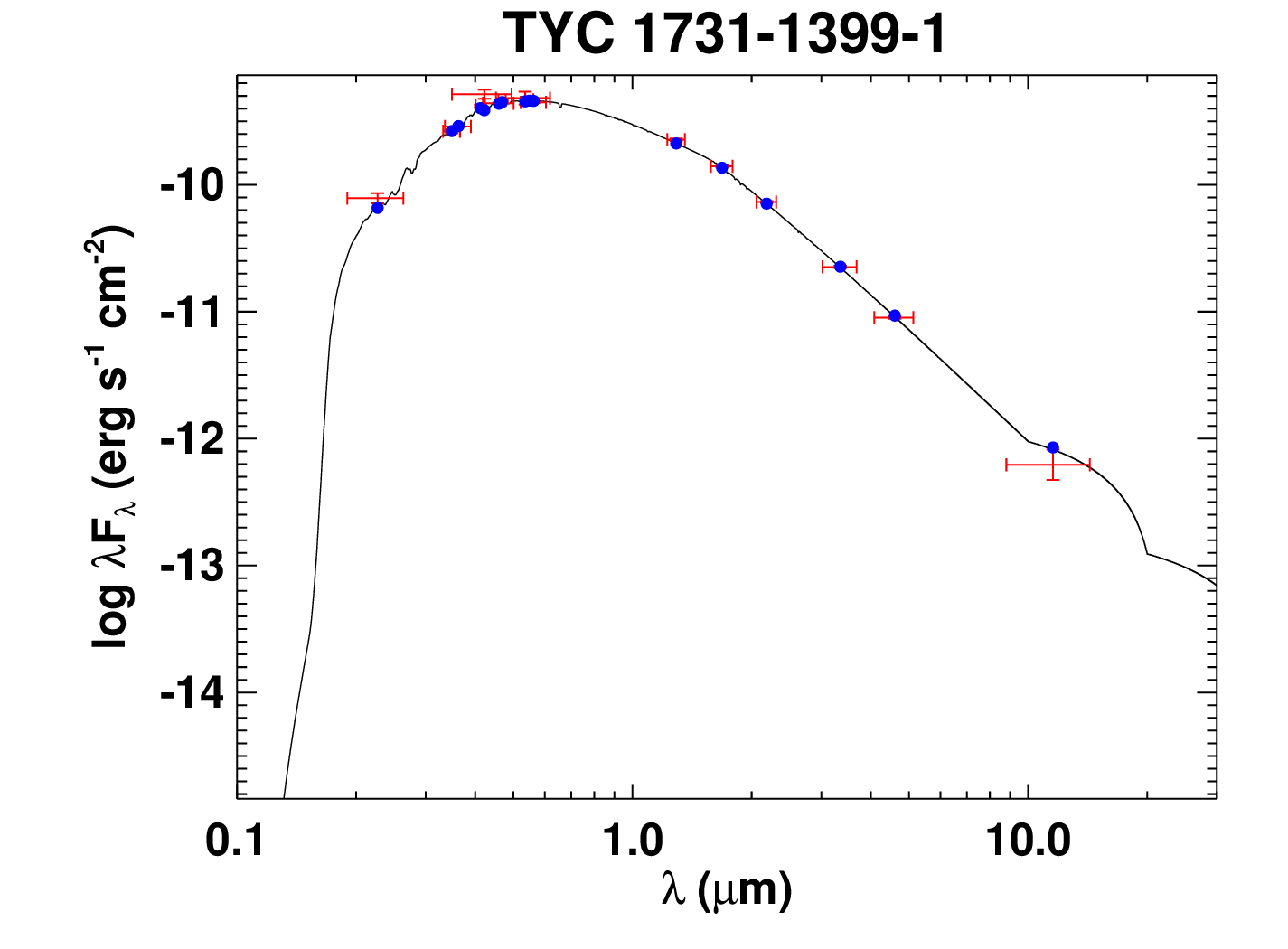}\includegraphics[width=0.333\linewidth]{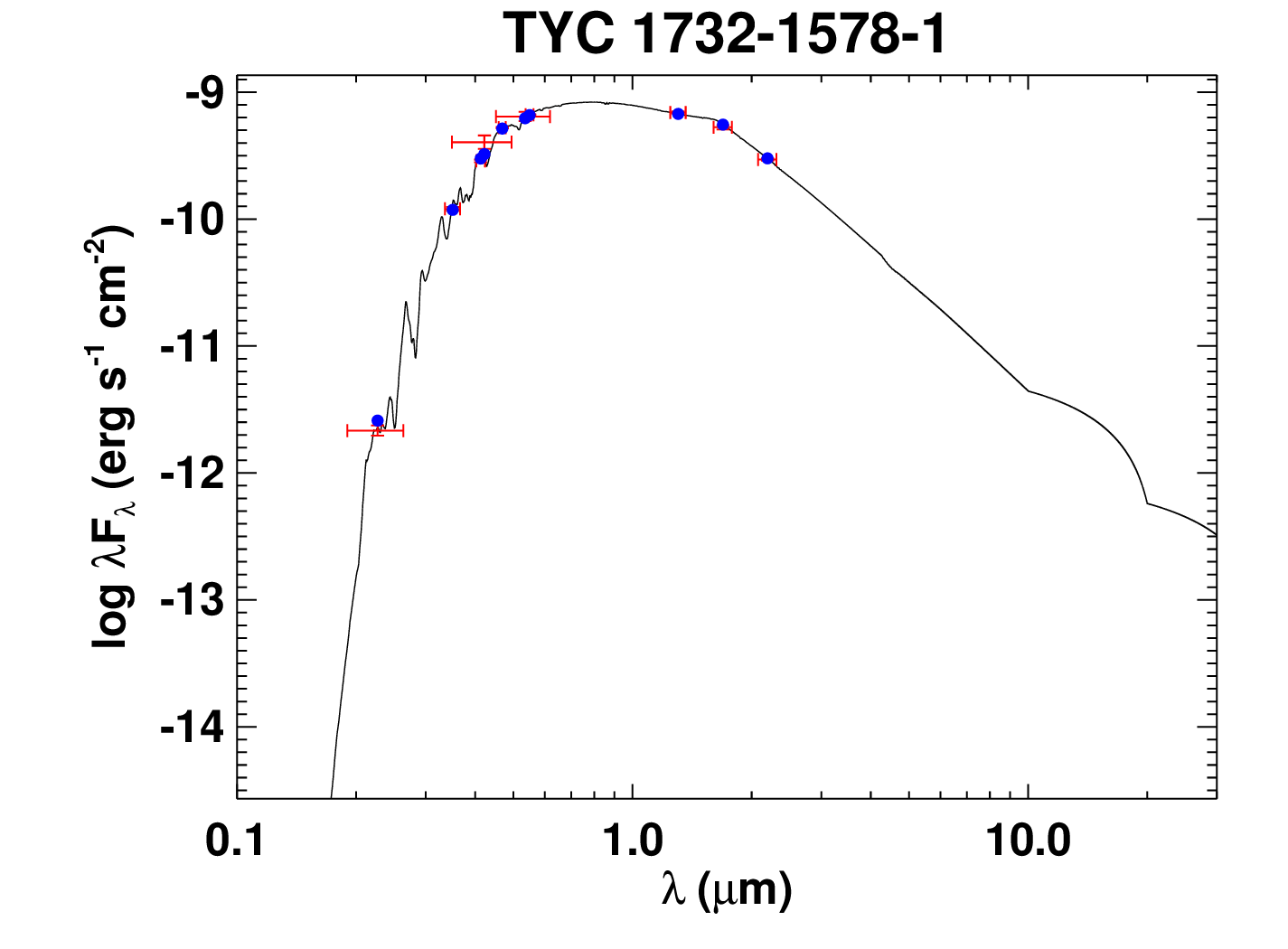}
\caption{\label{fig:seds2} All labels, lines, symbols, and colors as in Figure \ref{fig:seds}.}
\end{figure*}

\begin{figure*}
\includegraphics[width=0.333\linewidth]{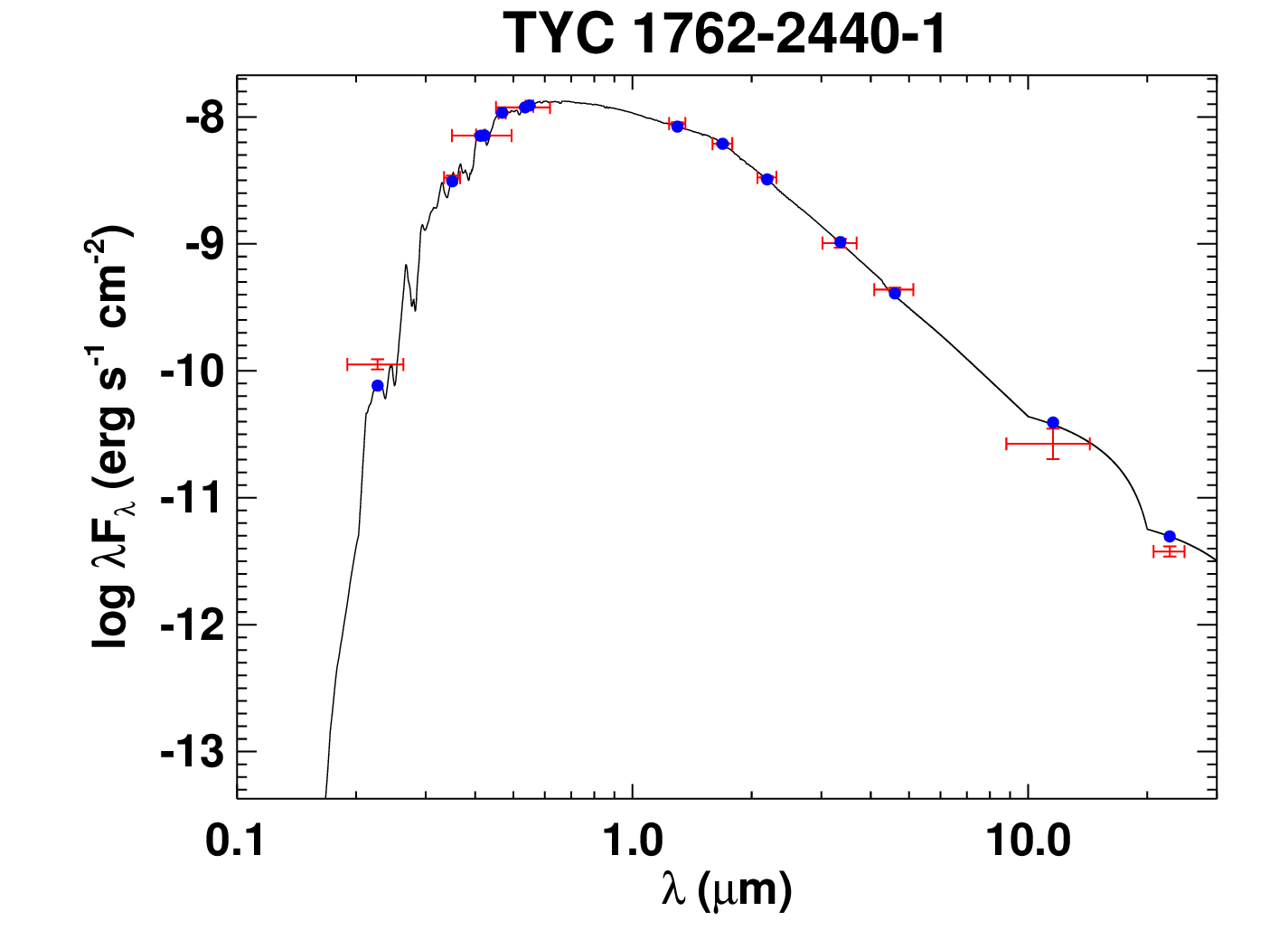}\includegraphics[width=0.333\linewidth]{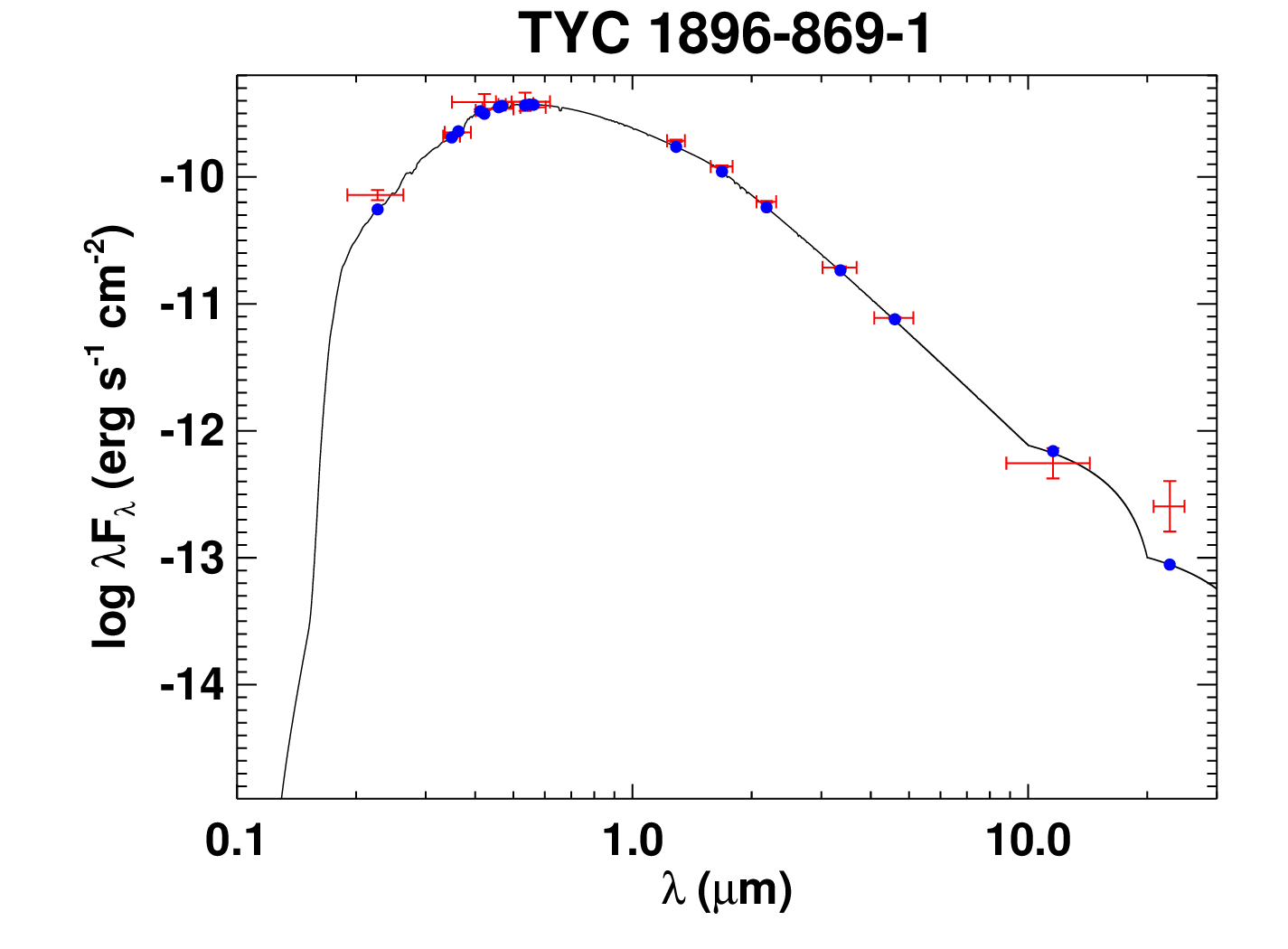}\includegraphics[width=0.333\linewidth]{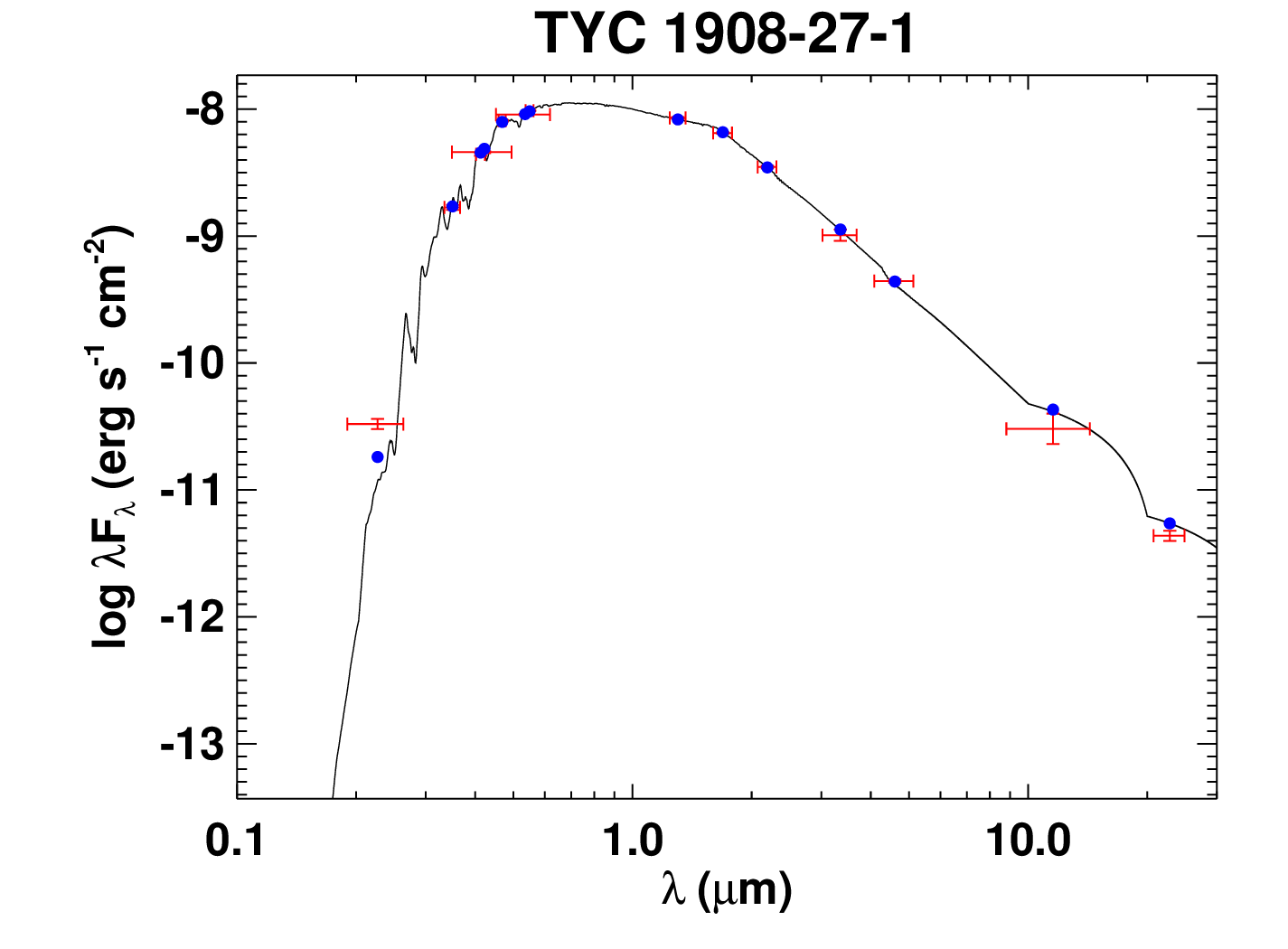}
\includegraphics[width=0.333\linewidth]{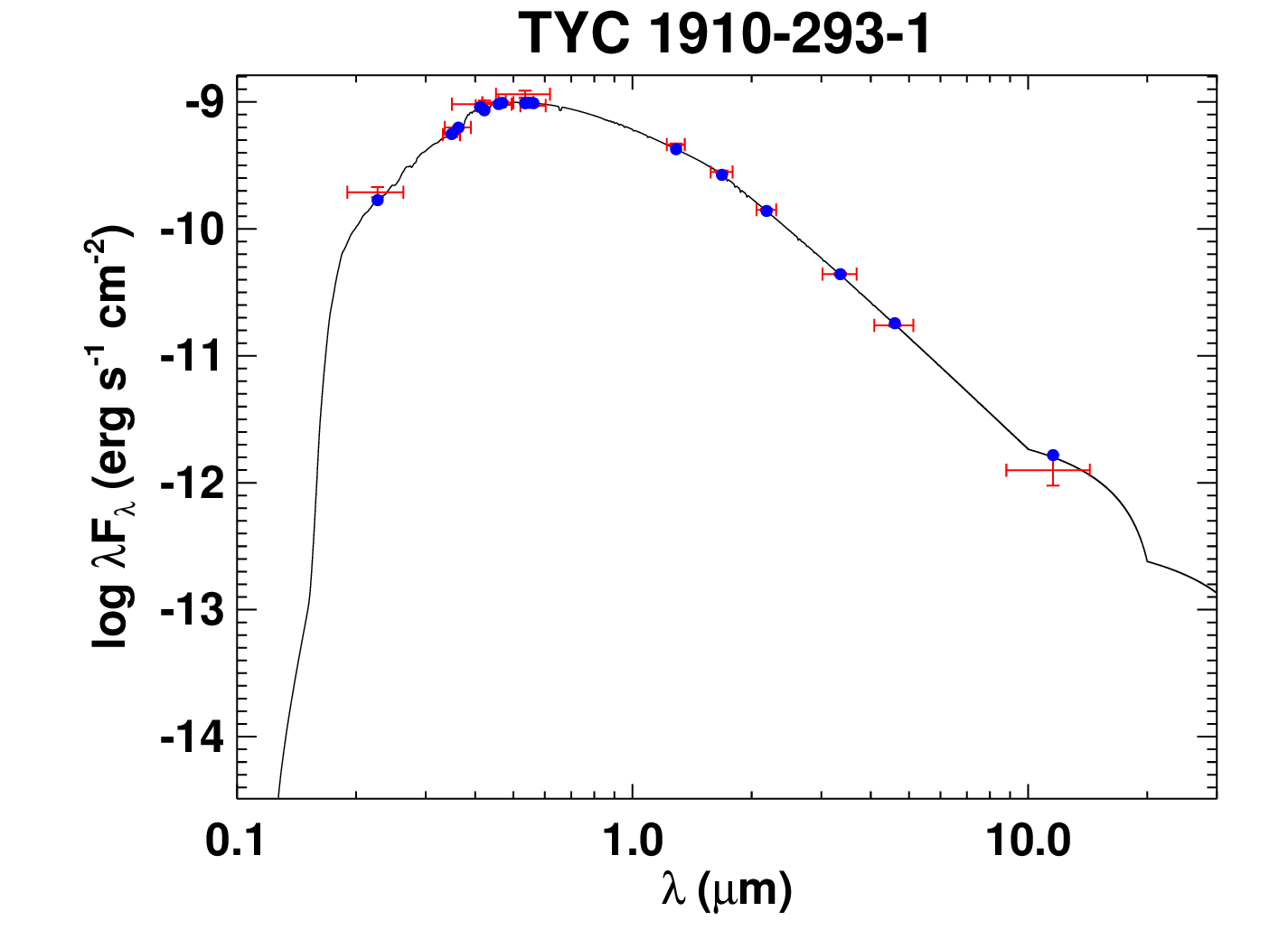}\includegraphics[width=0.333\linewidth]{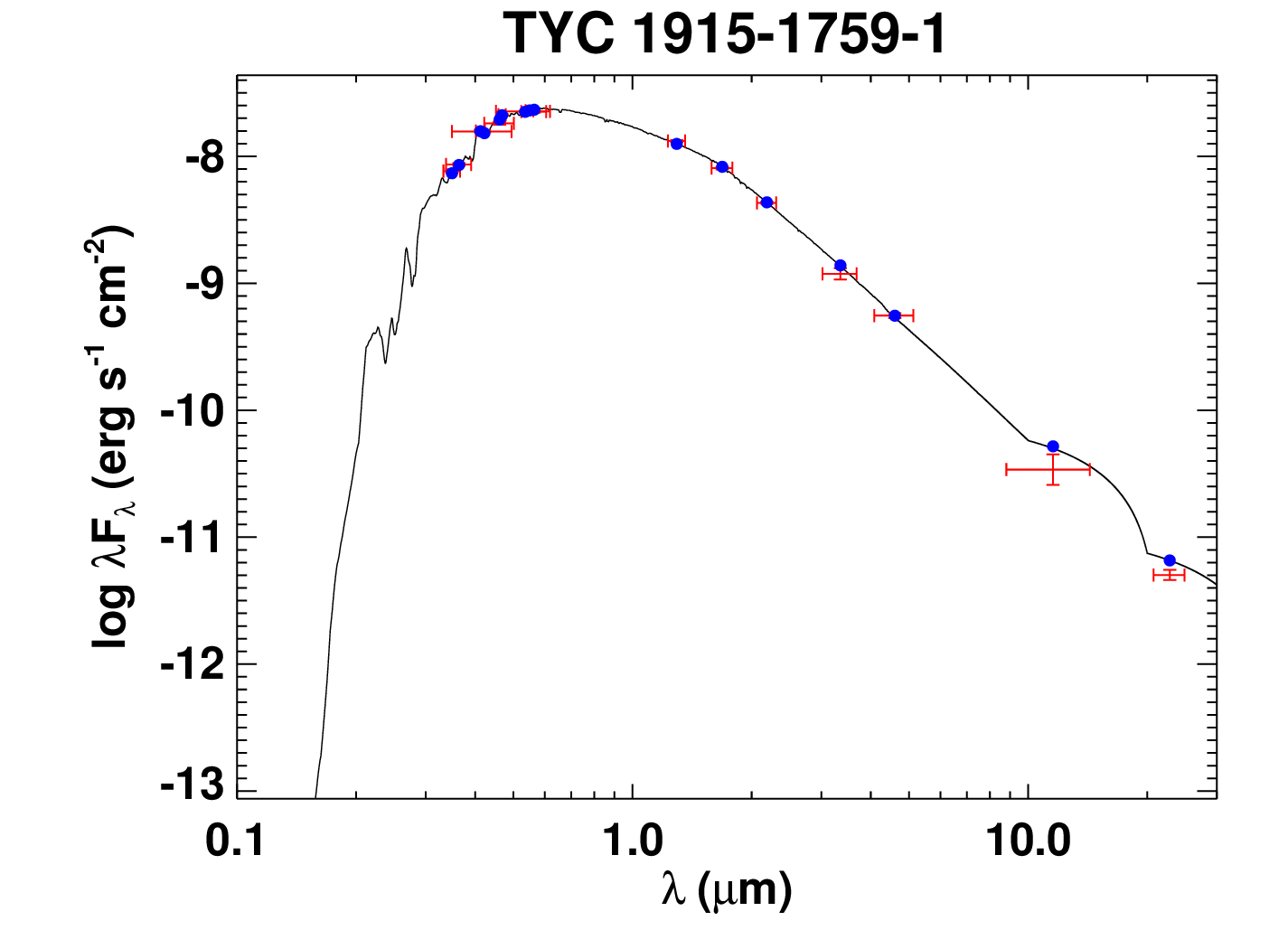}\includegraphics[width=0.333\linewidth]{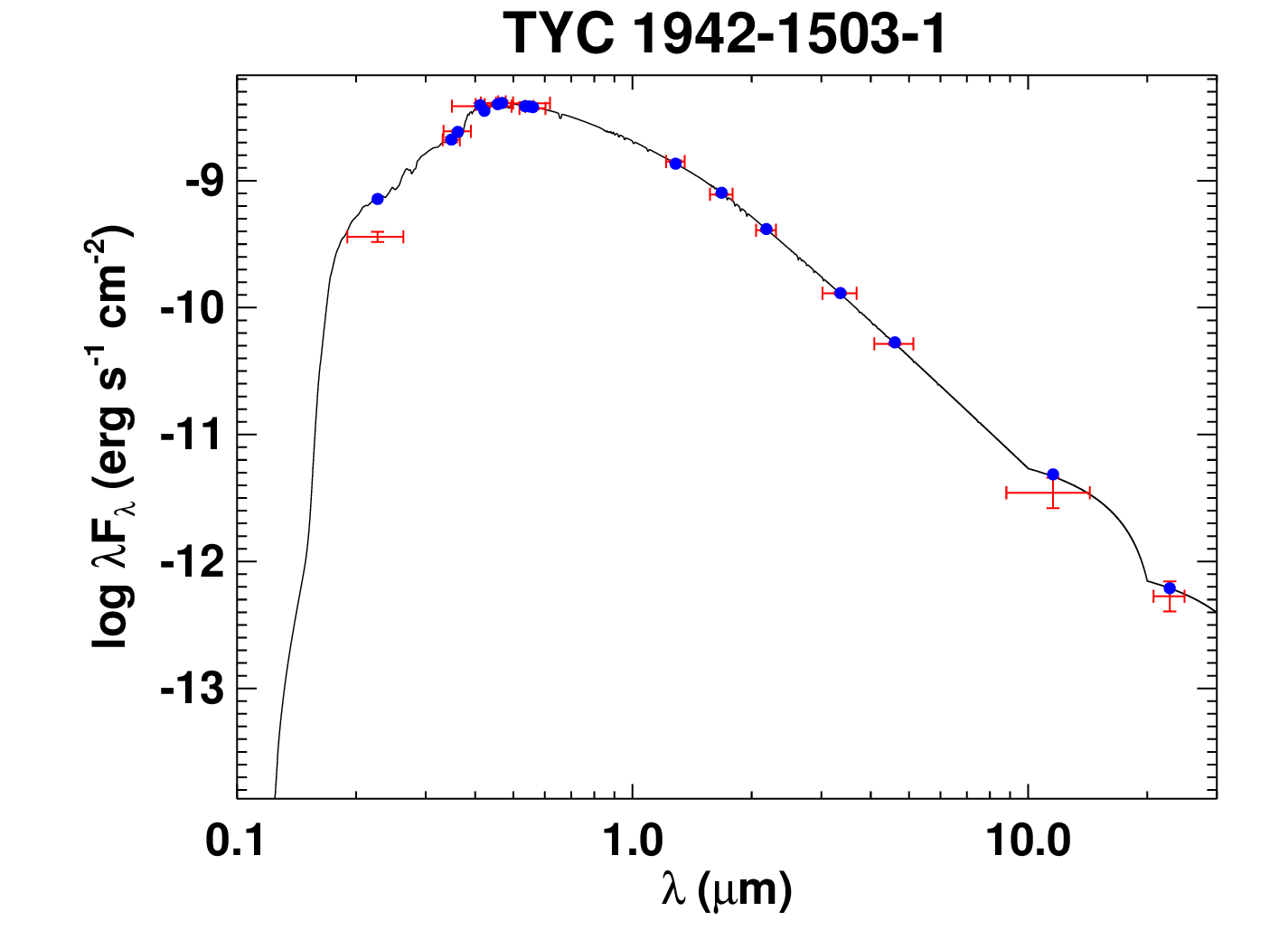}
\includegraphics[width=0.333\linewidth]{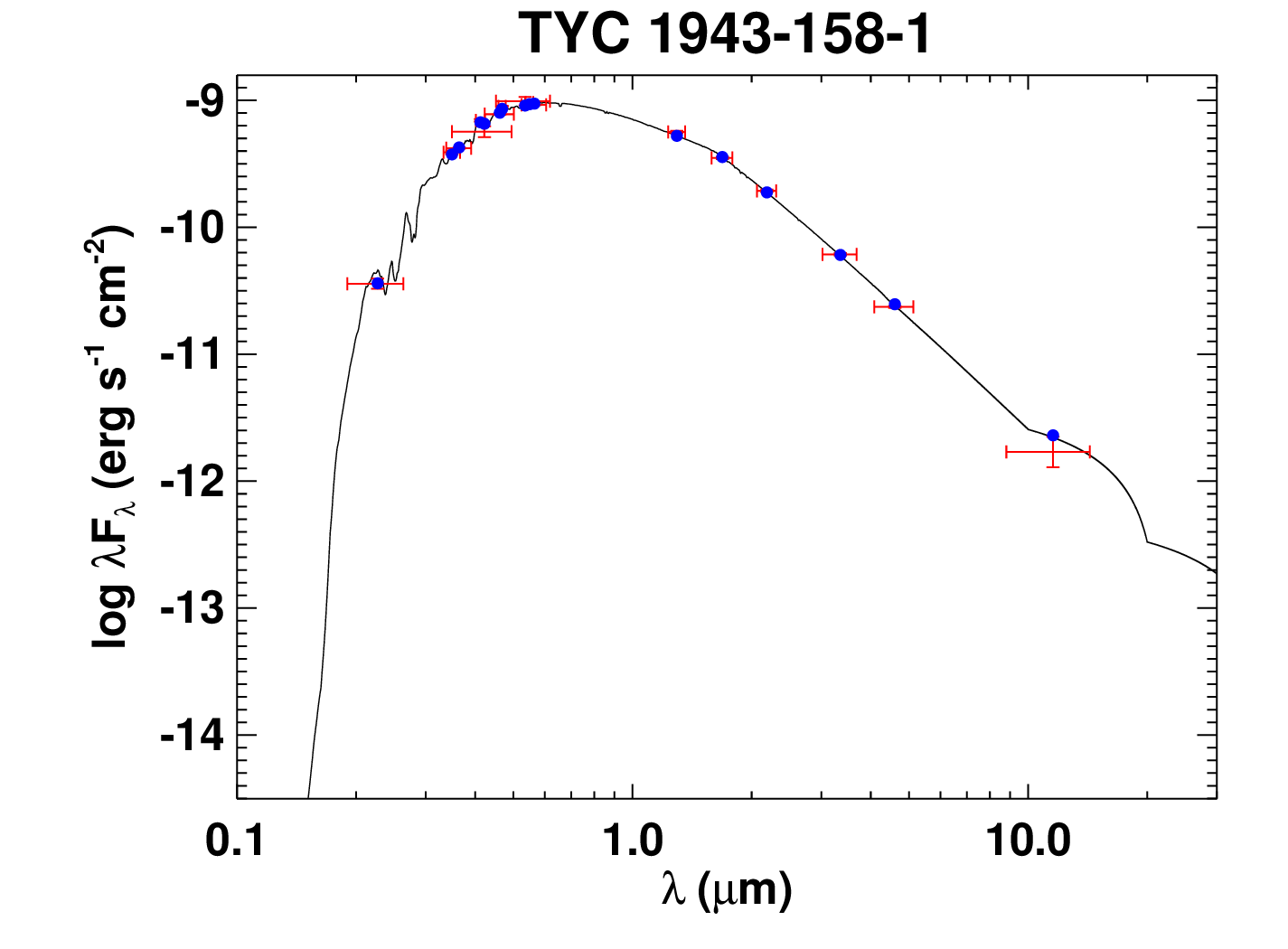}\includegraphics[width=0.333\linewidth]{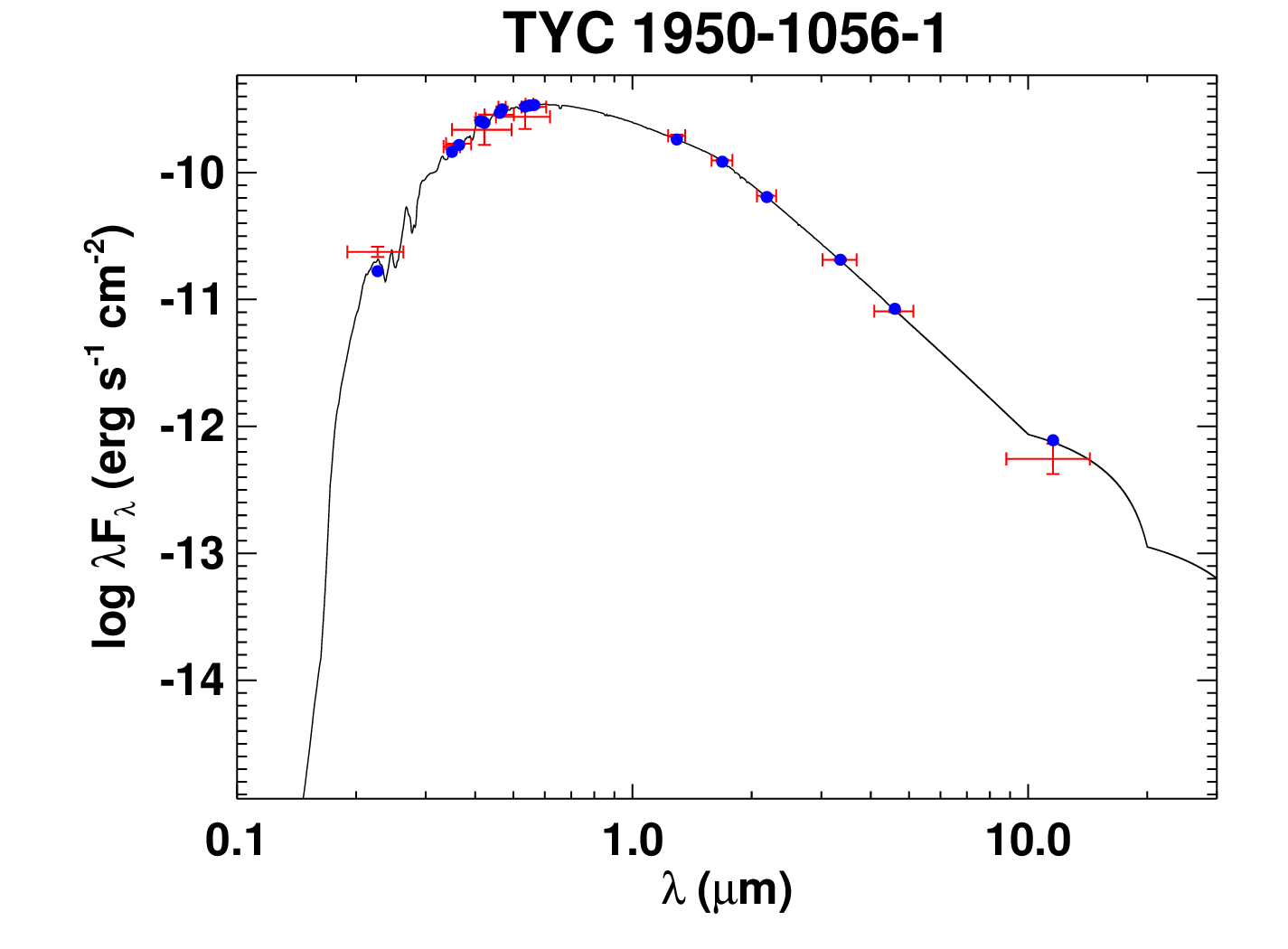}\includegraphics[width=0.333\linewidth]{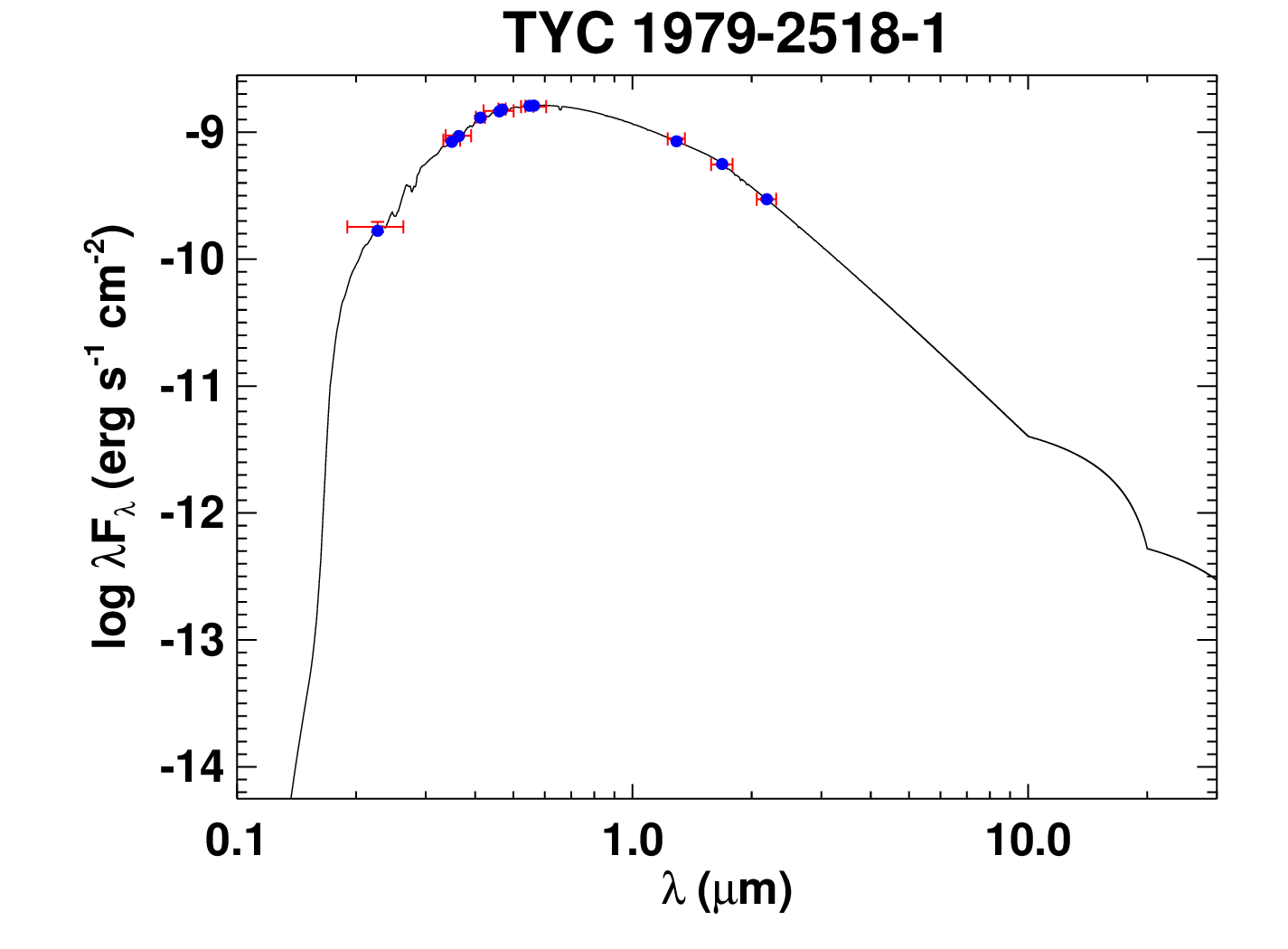}
\includegraphics[width=0.333\linewidth]{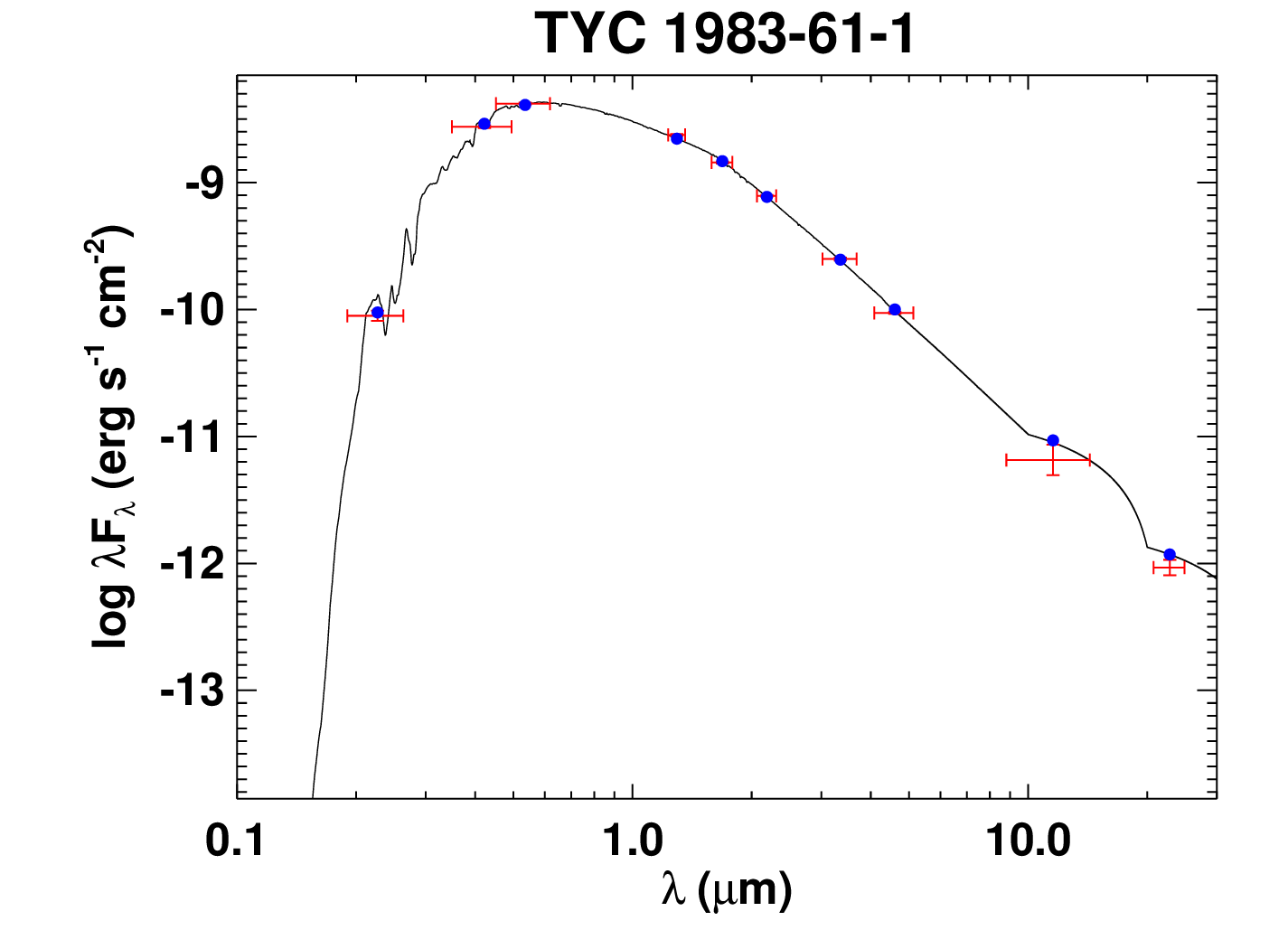}\includegraphics[width=0.333\linewidth]{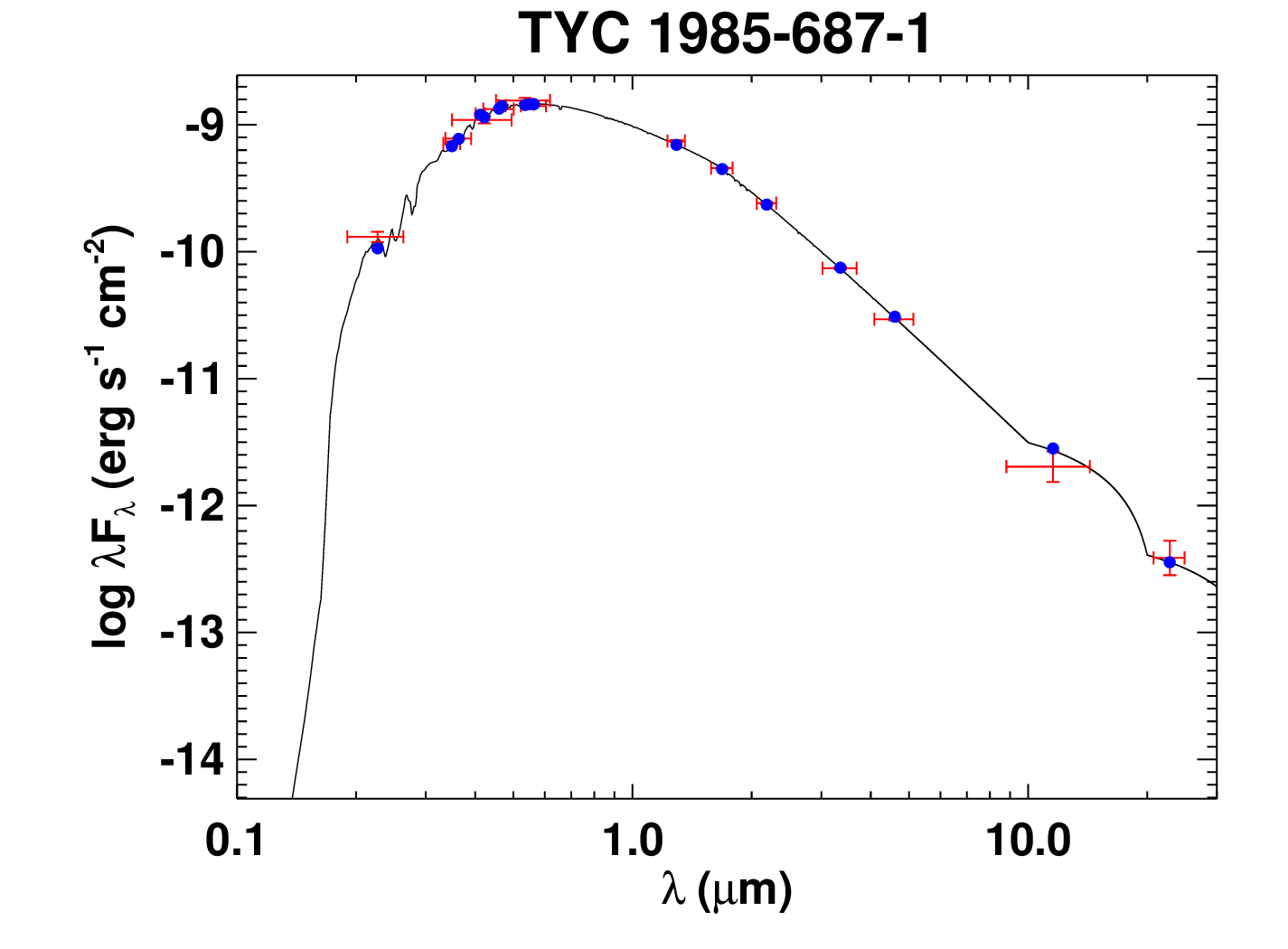}\includegraphics[width=0.333\linewidth]{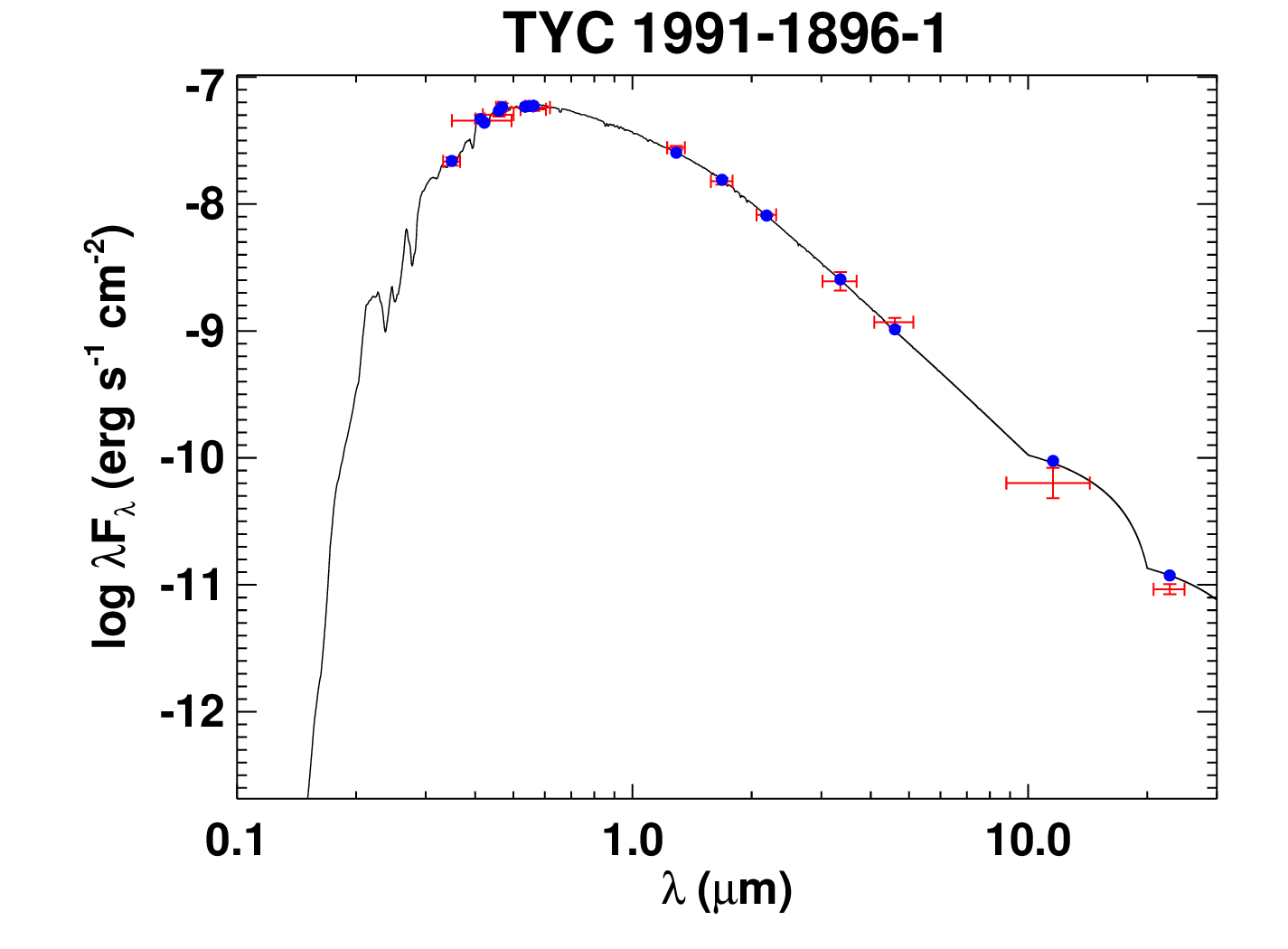}
\caption{\label{fig:seds3} All labels, lines, symbols, and colors as in Figure \ref{fig:seds}.}
\end{figure*}

\begin{figure*}
\includegraphics[width=0.333\linewidth]{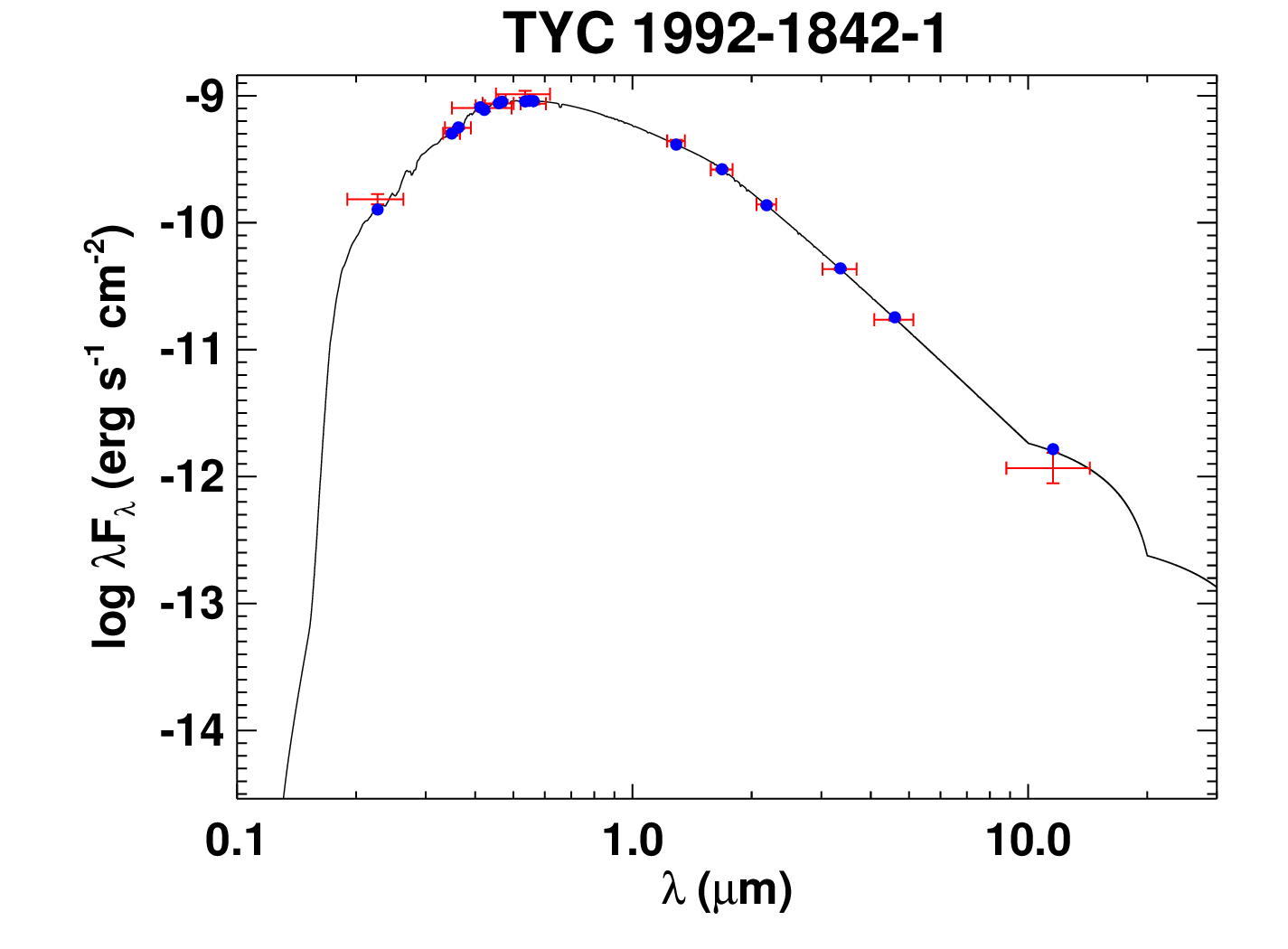}\includegraphics[width=0.333\linewidth]{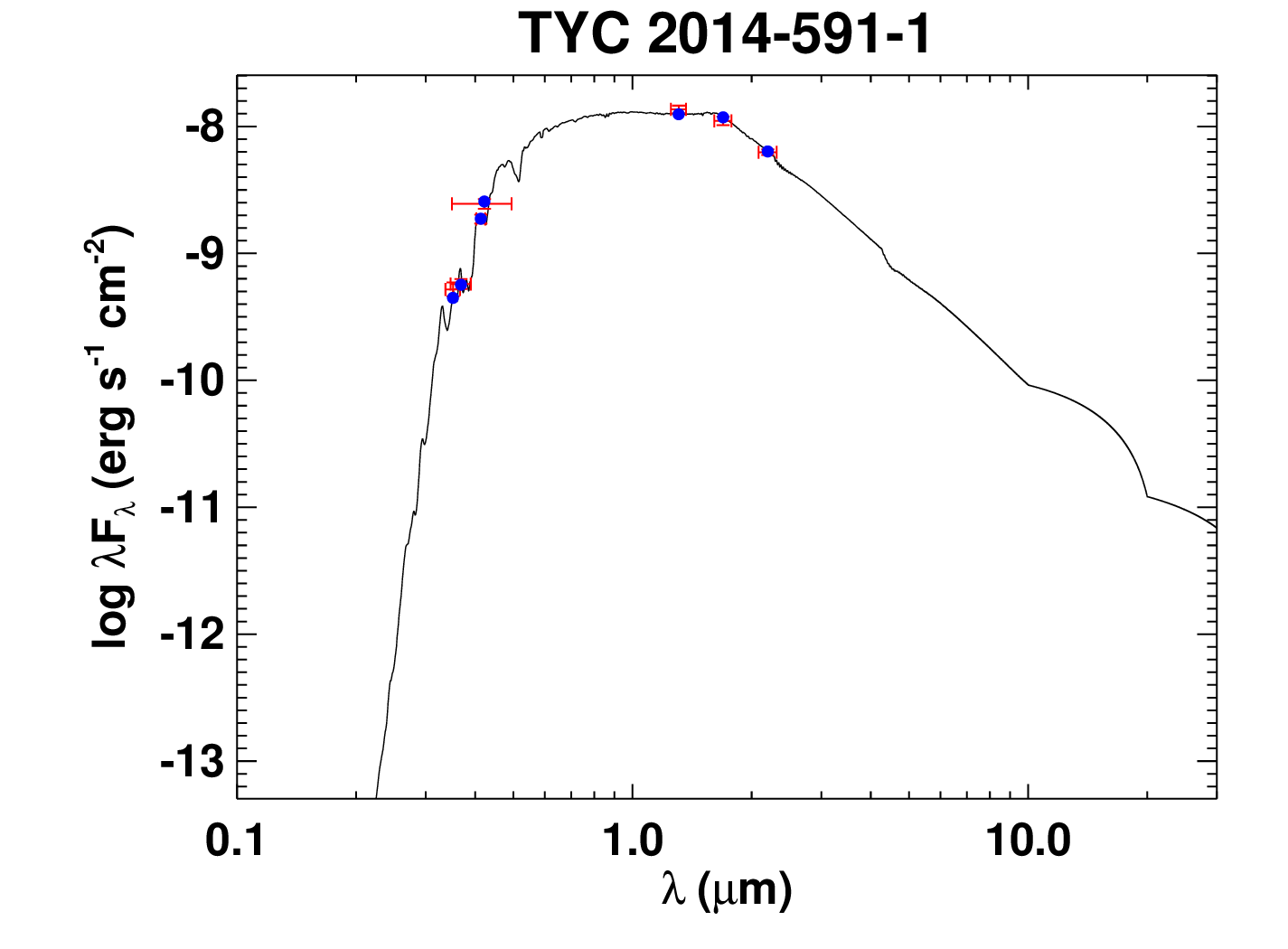}\includegraphics[width=0.333\linewidth]{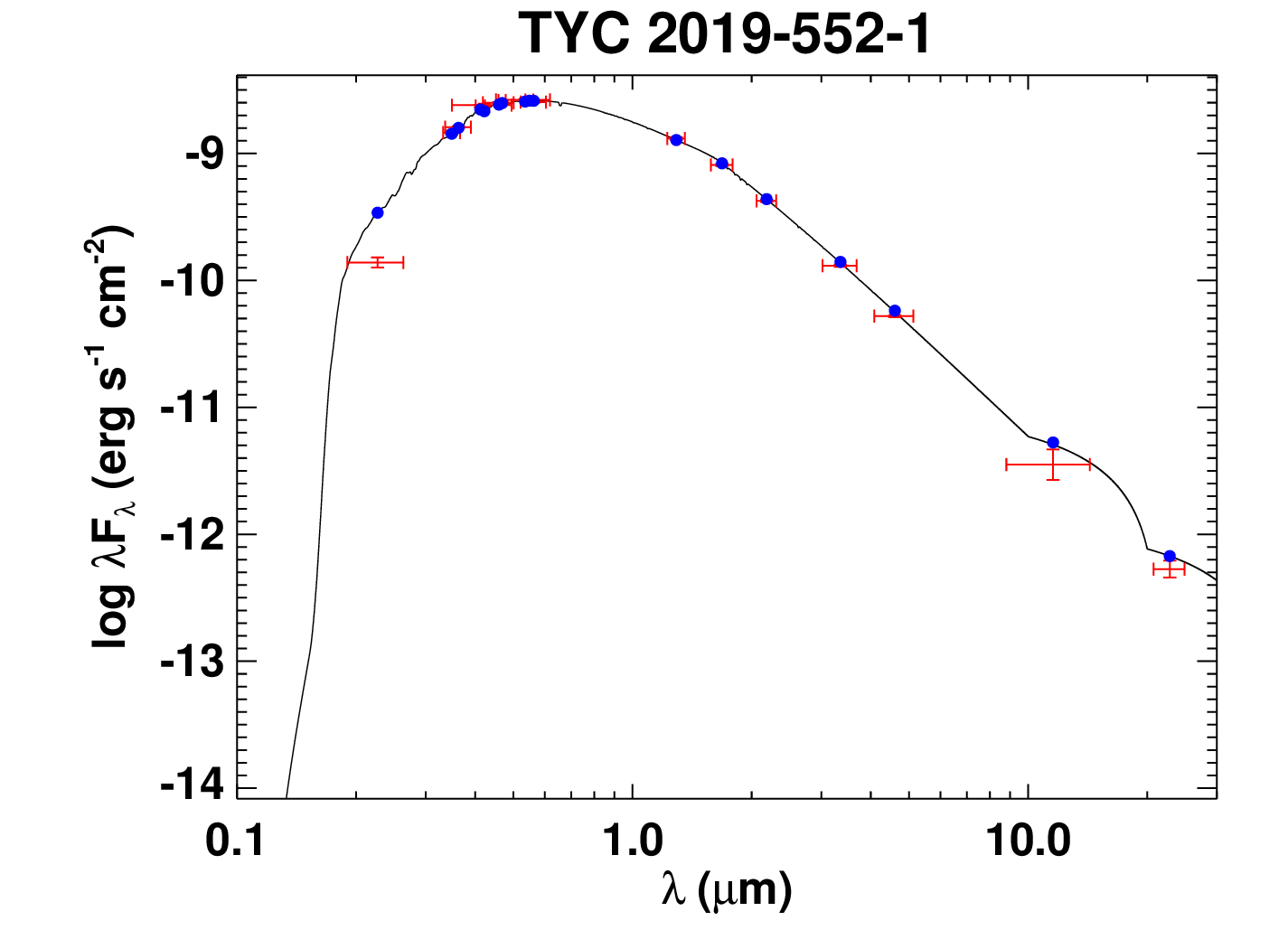}
\includegraphics[width=0.333\linewidth]{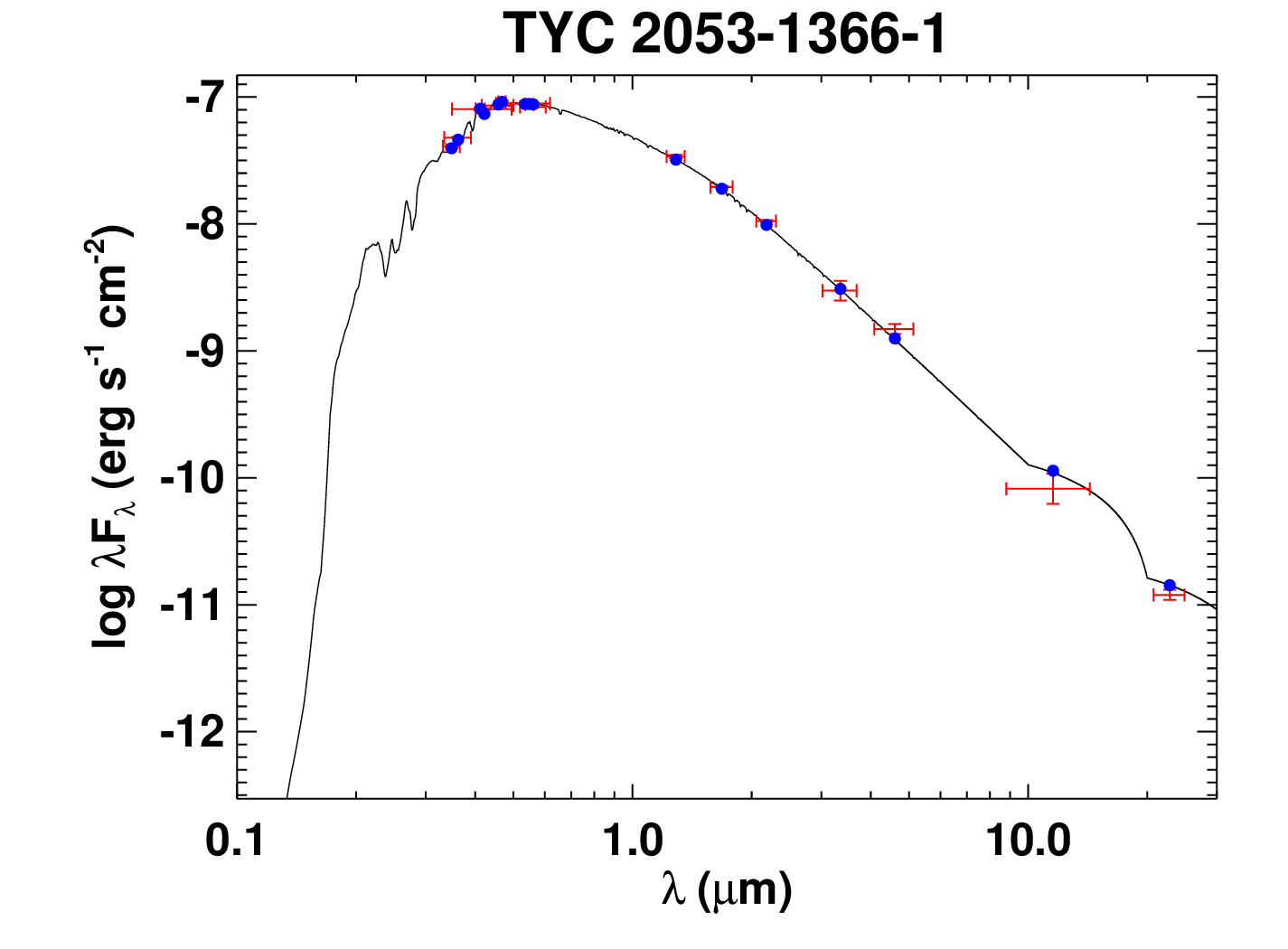}\includegraphics[width=0.333\linewidth]{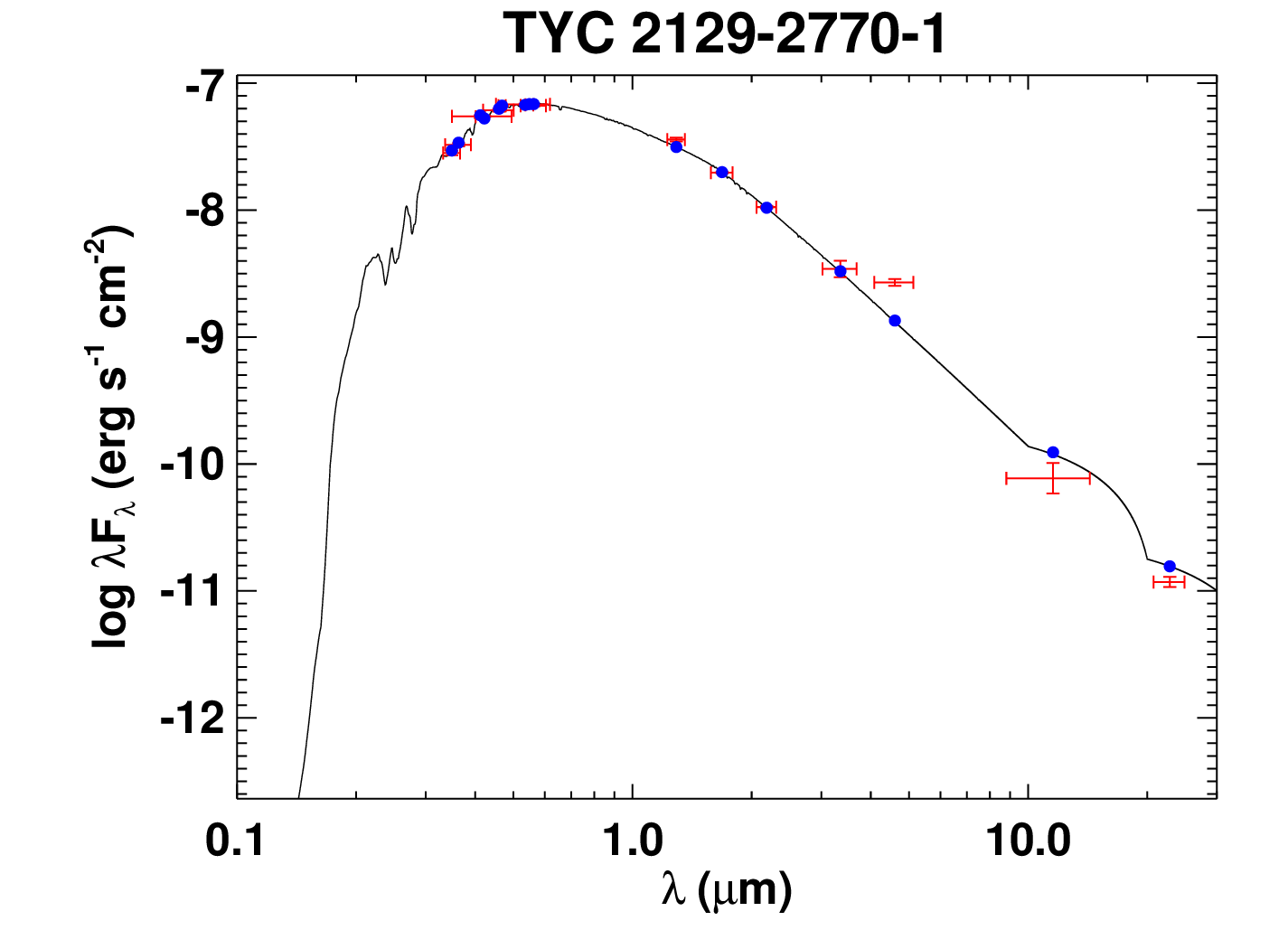}\includegraphics[width=0.333\linewidth]{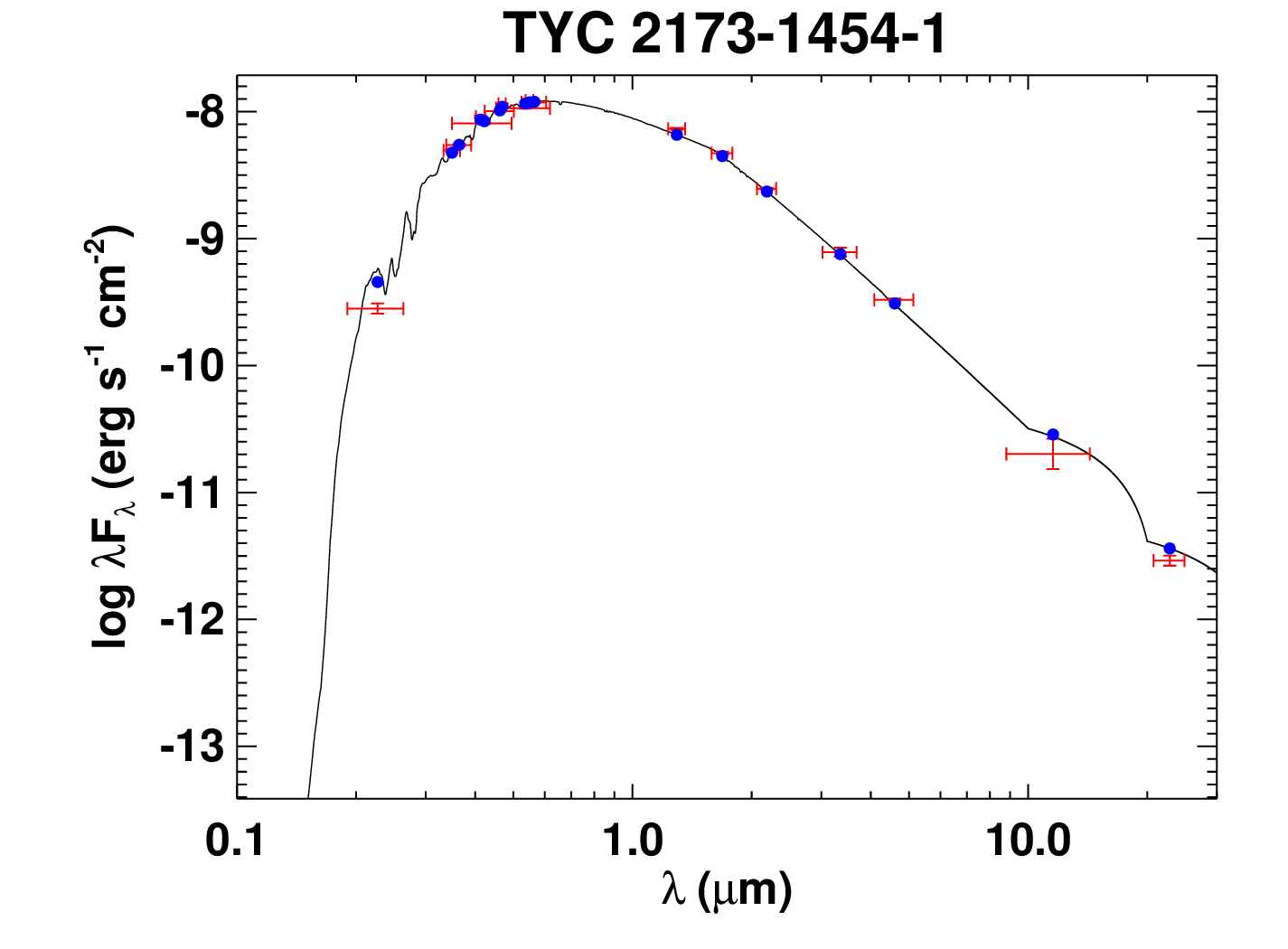}
\includegraphics[width=0.333\linewidth]{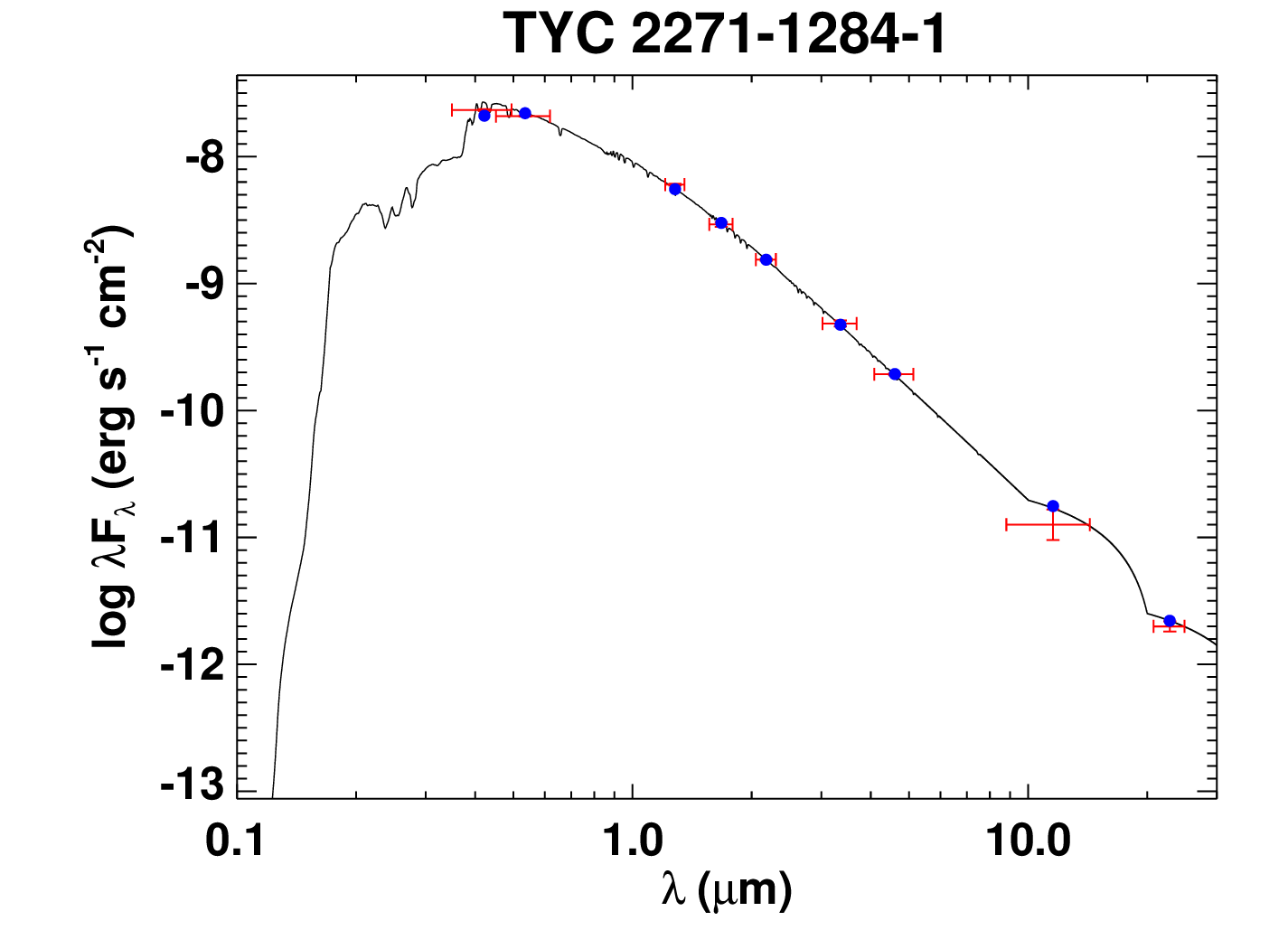}\includegraphics[width=0.333\linewidth]{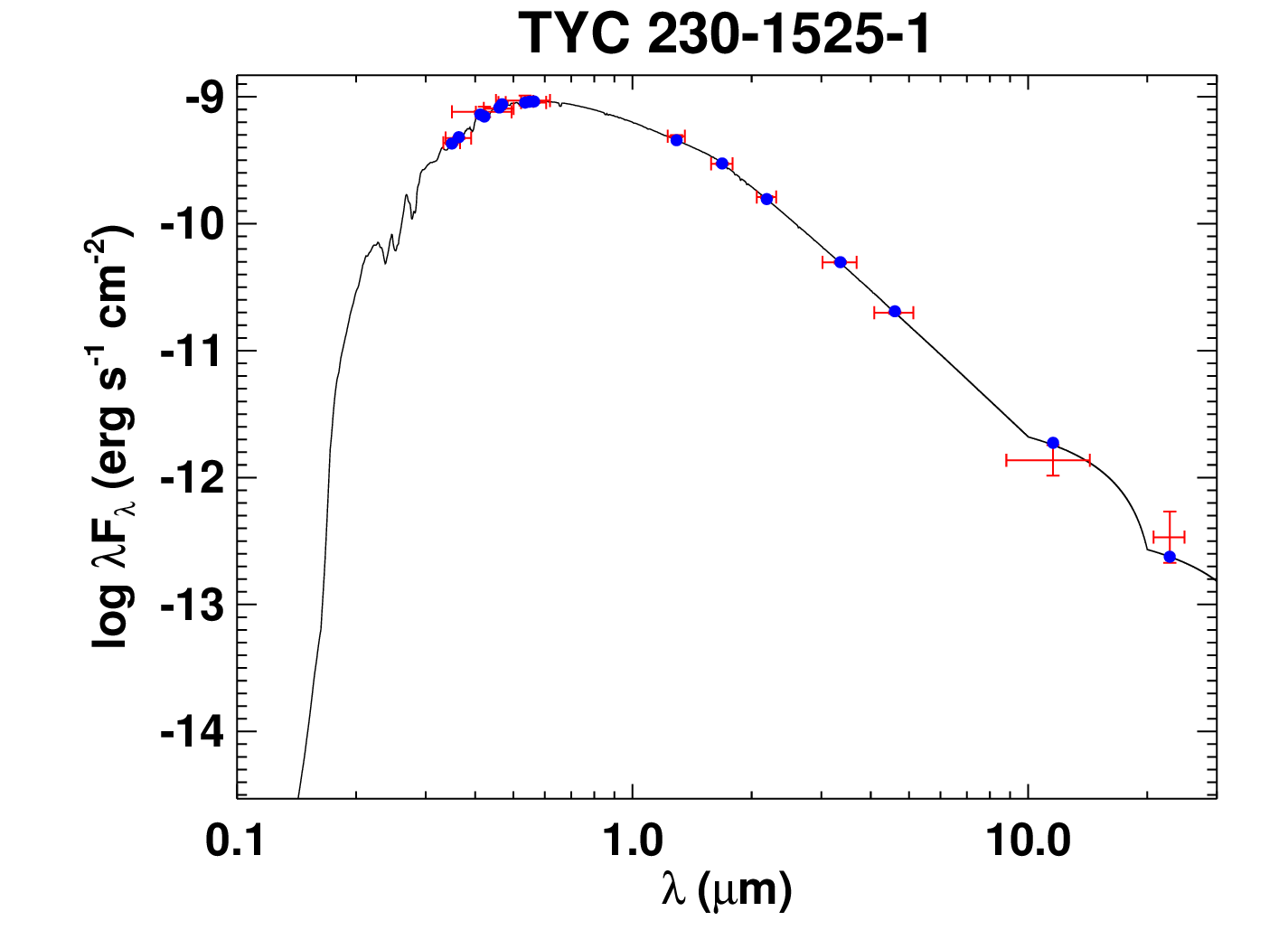}\includegraphics[width=0.333\linewidth]{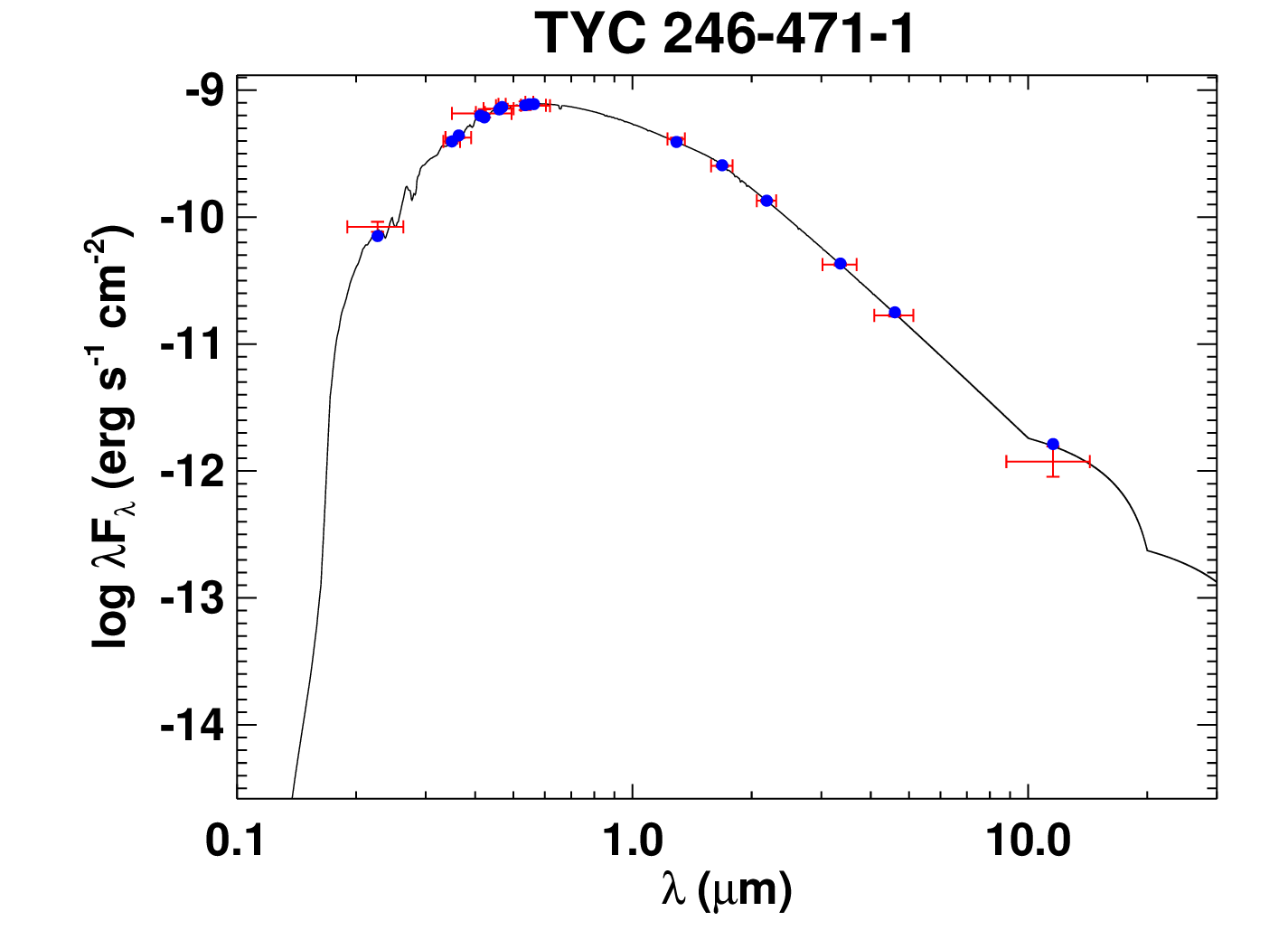}
\includegraphics[width=0.333\linewidth]{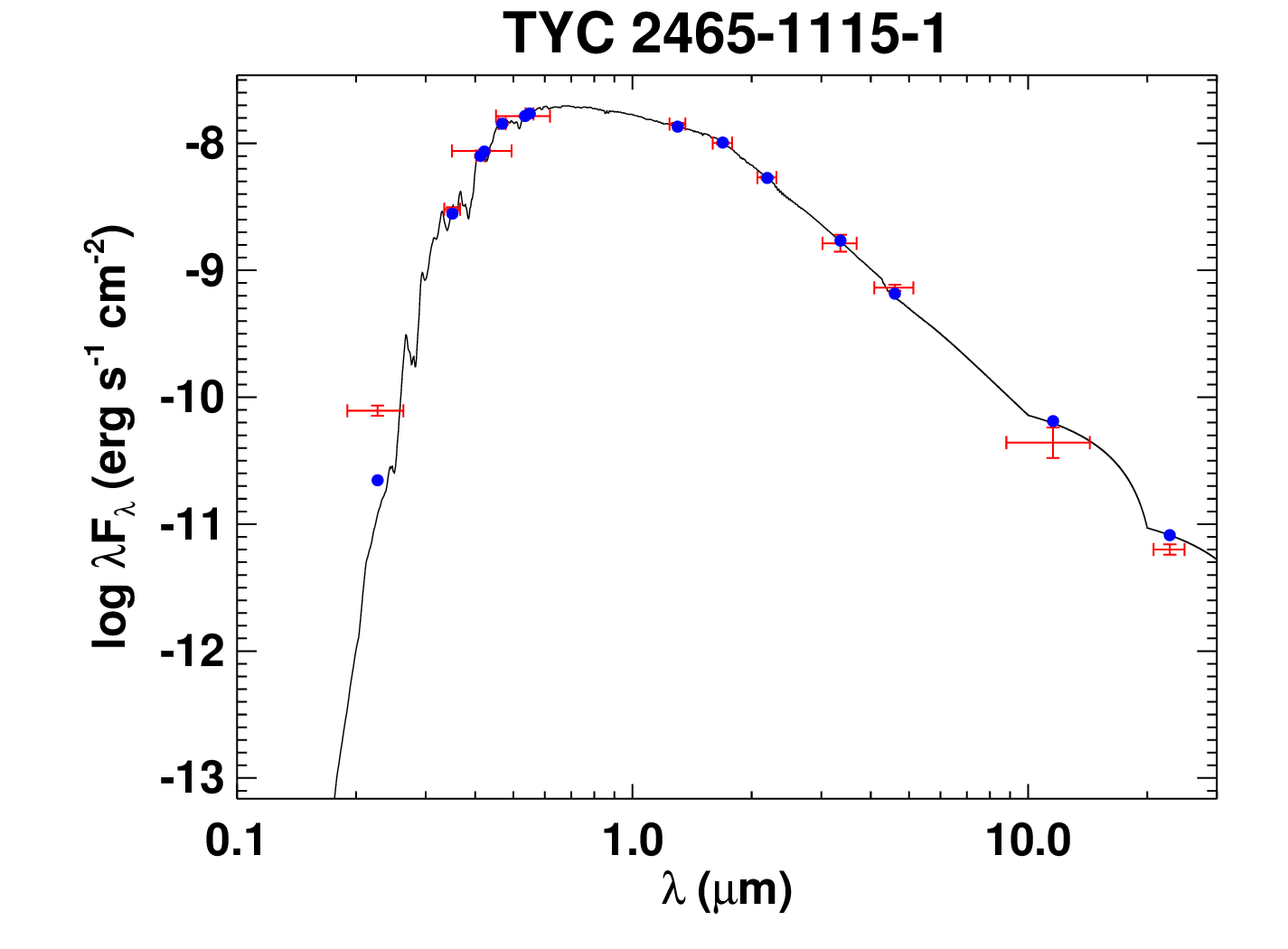}\includegraphics[width=0.333\linewidth]{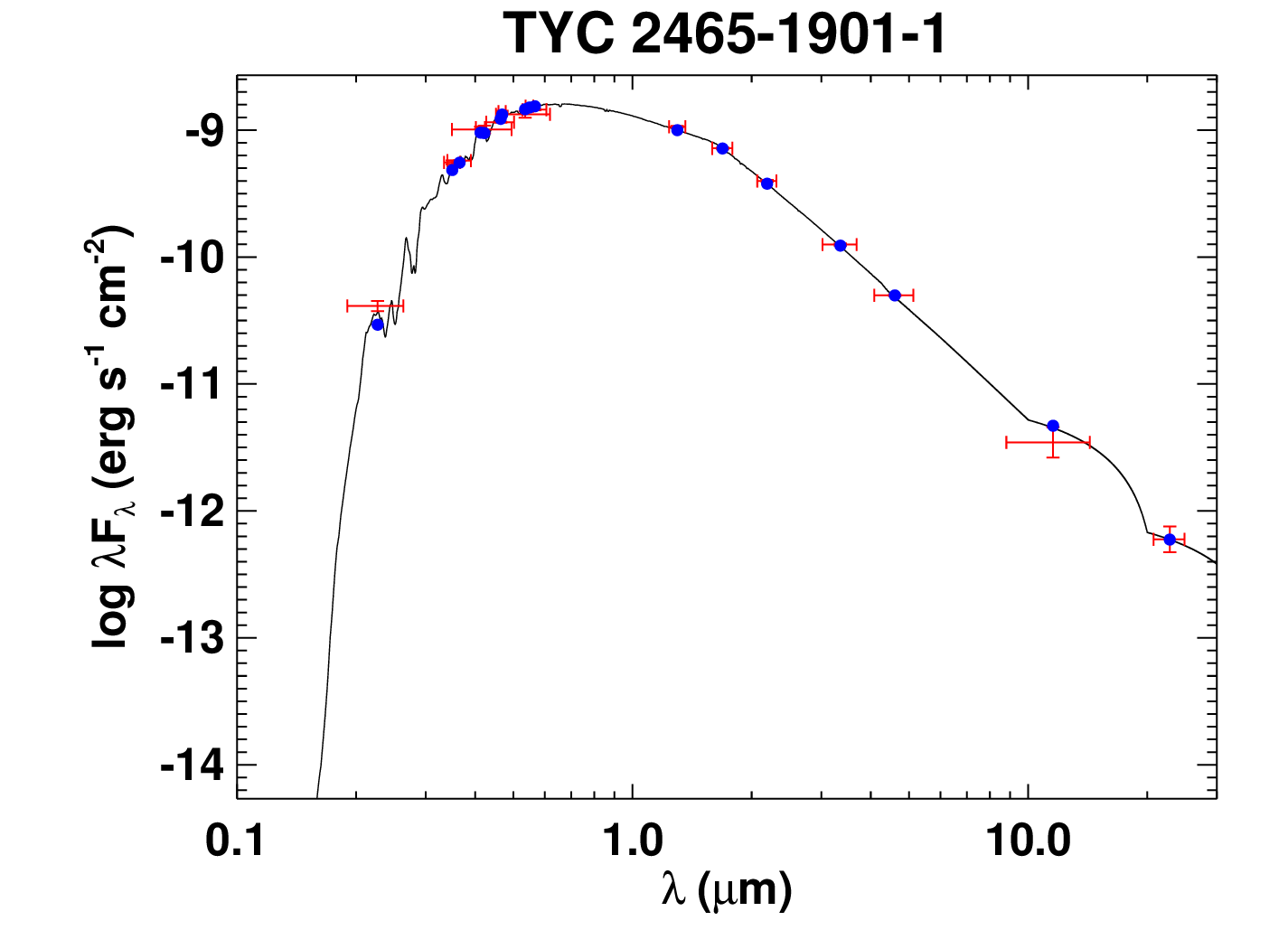}\includegraphics[width=0.333\linewidth]{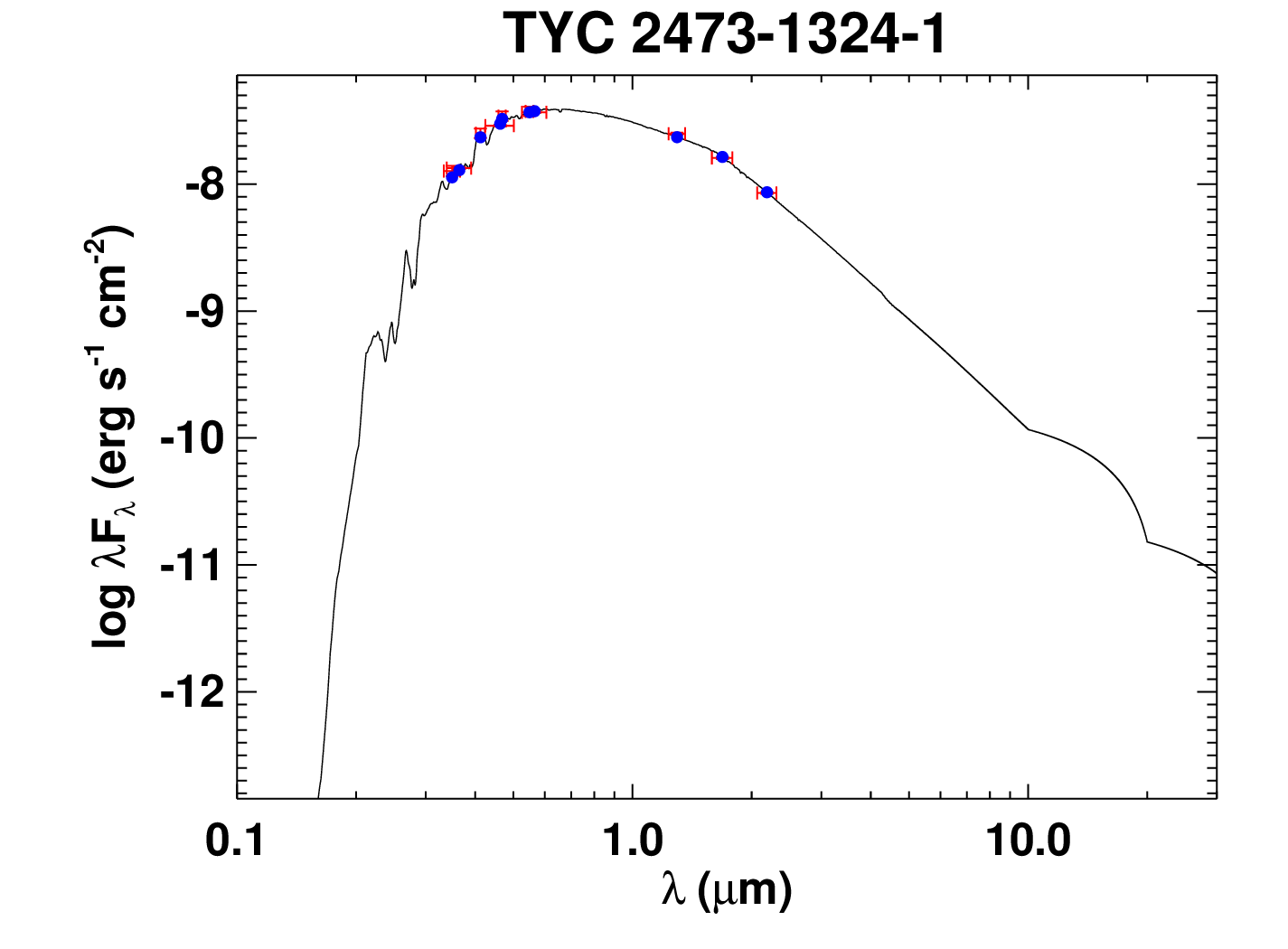}
\caption{\label{fig:seds4} All labels, lines, symbols, and colors as in Figure \ref{fig:seds}.}
\end{figure*}

\begin{figure*}
\includegraphics[width=0.333\linewidth]{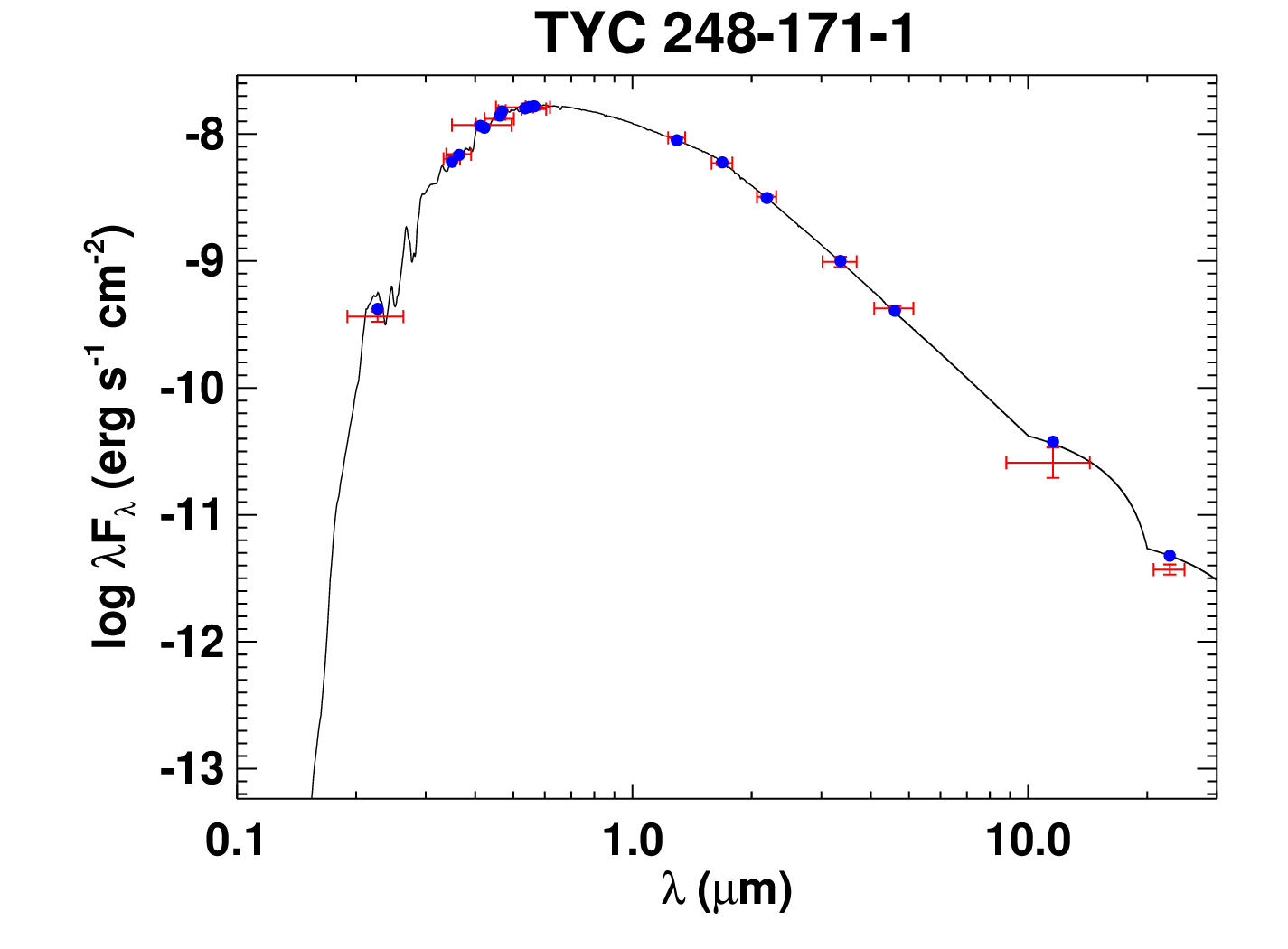}\includegraphics[width=0.333\linewidth]{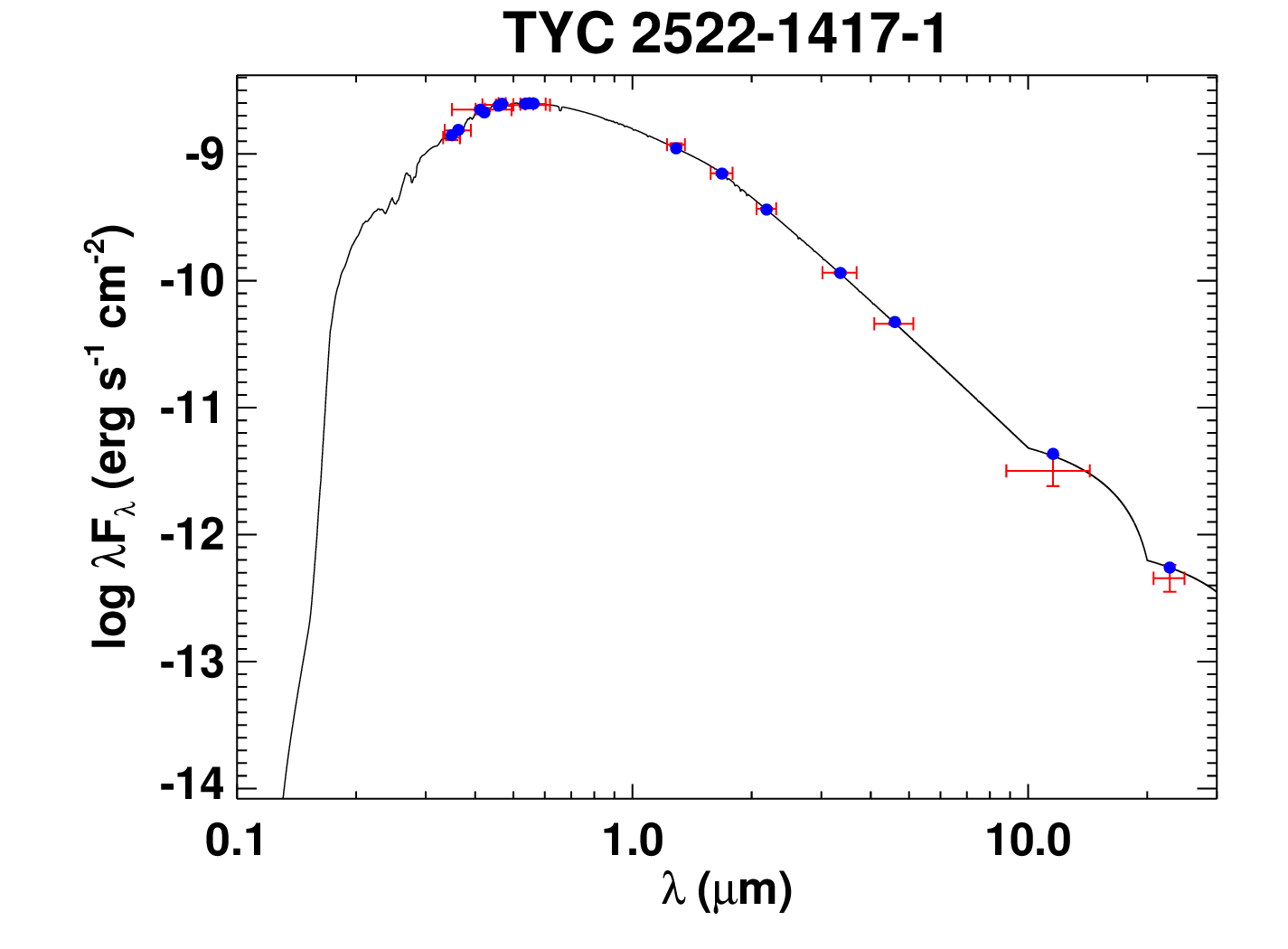}\includegraphics[width=0.333\linewidth]{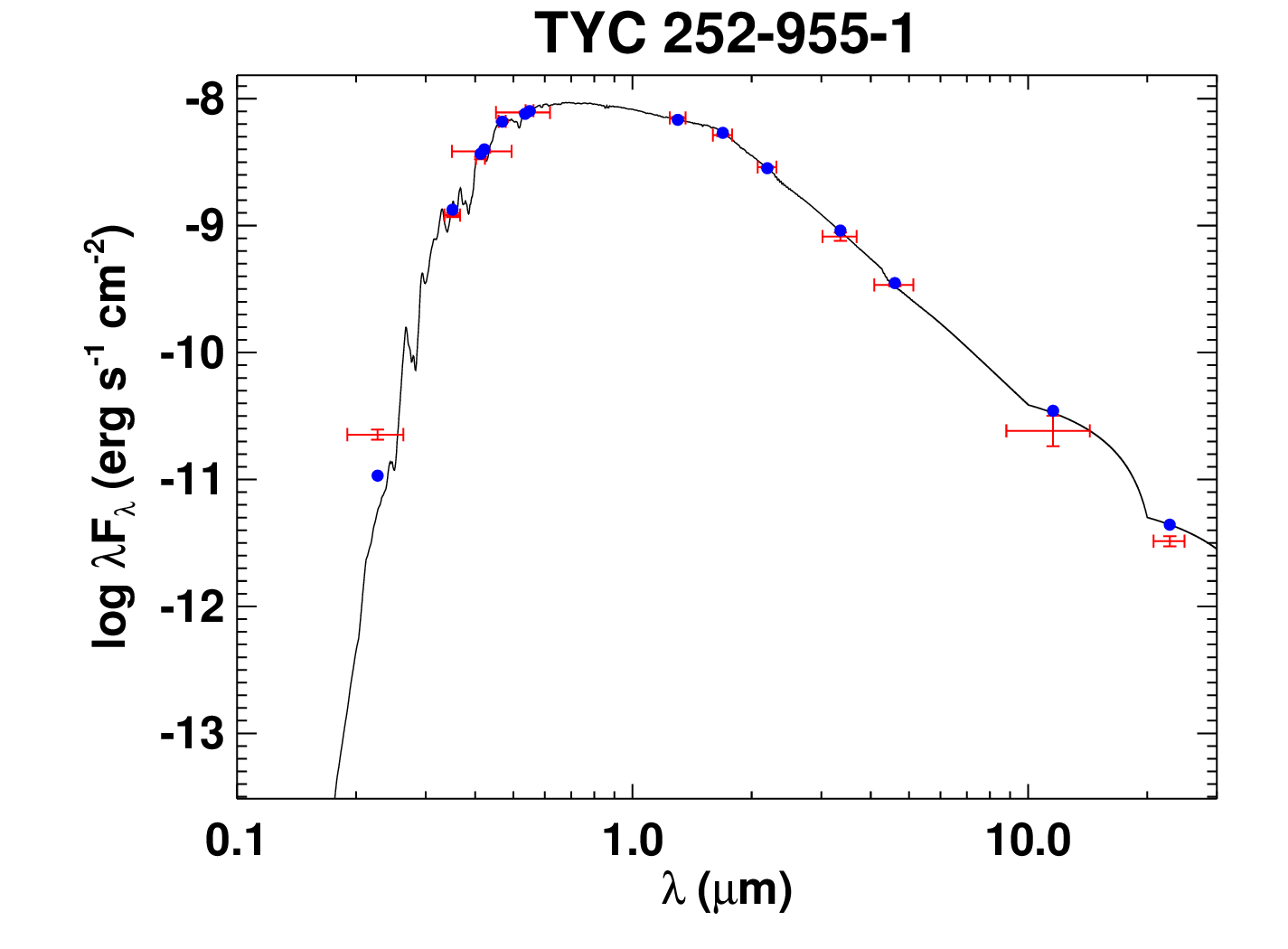}
\includegraphics[width=0.333\linewidth]{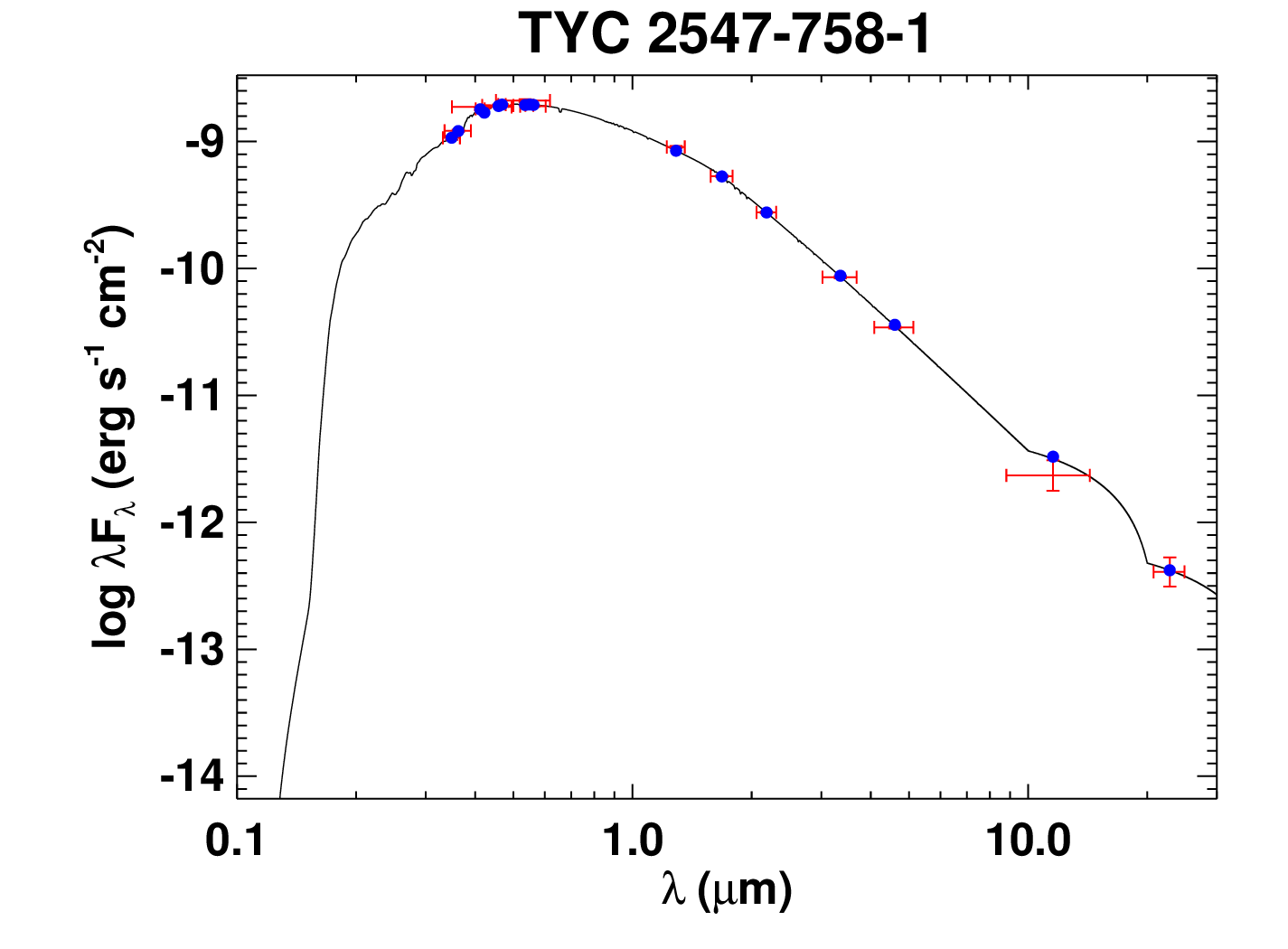}\includegraphics[width=0.333\linewidth]{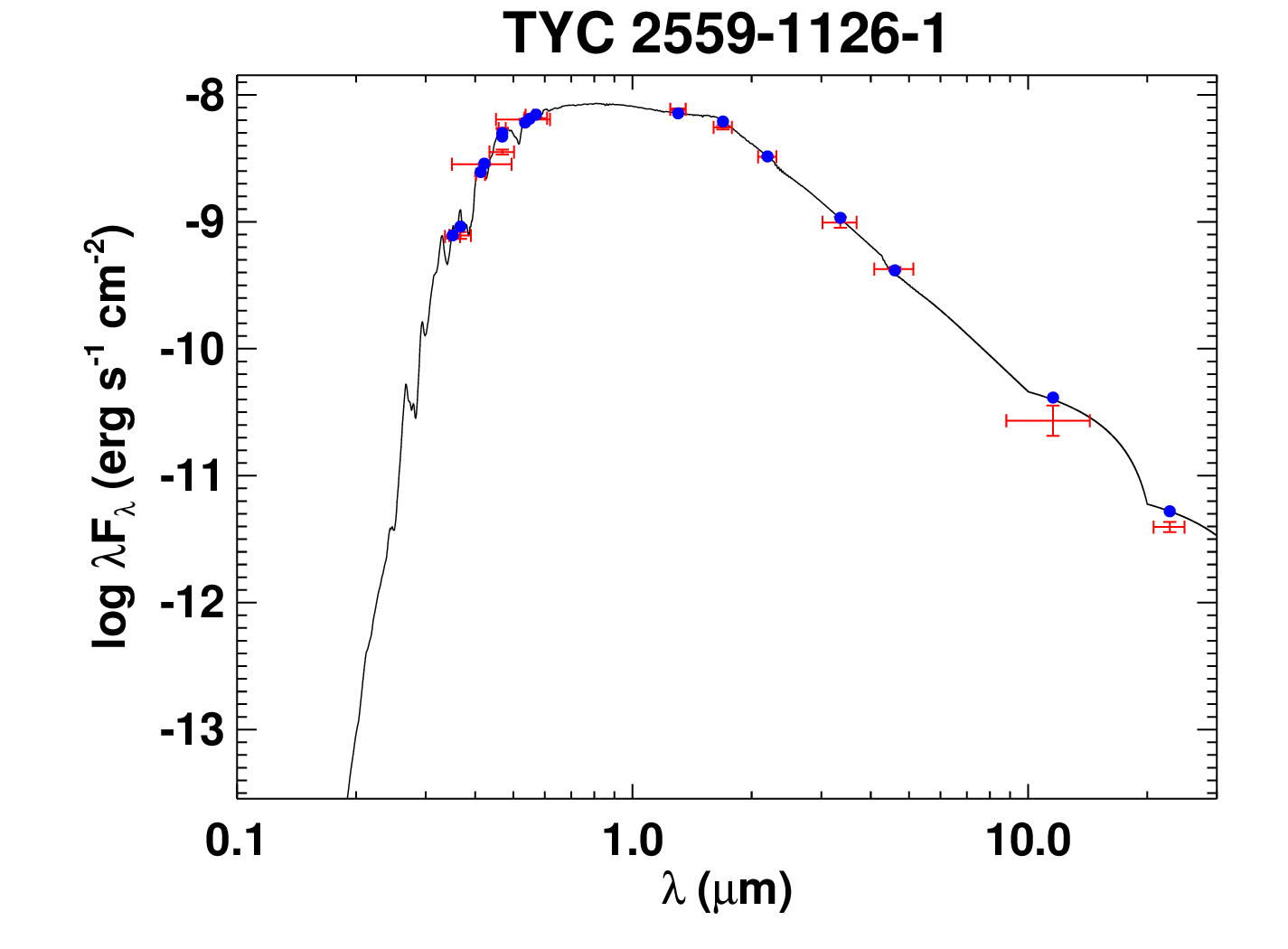}\includegraphics[width=0.333\linewidth]{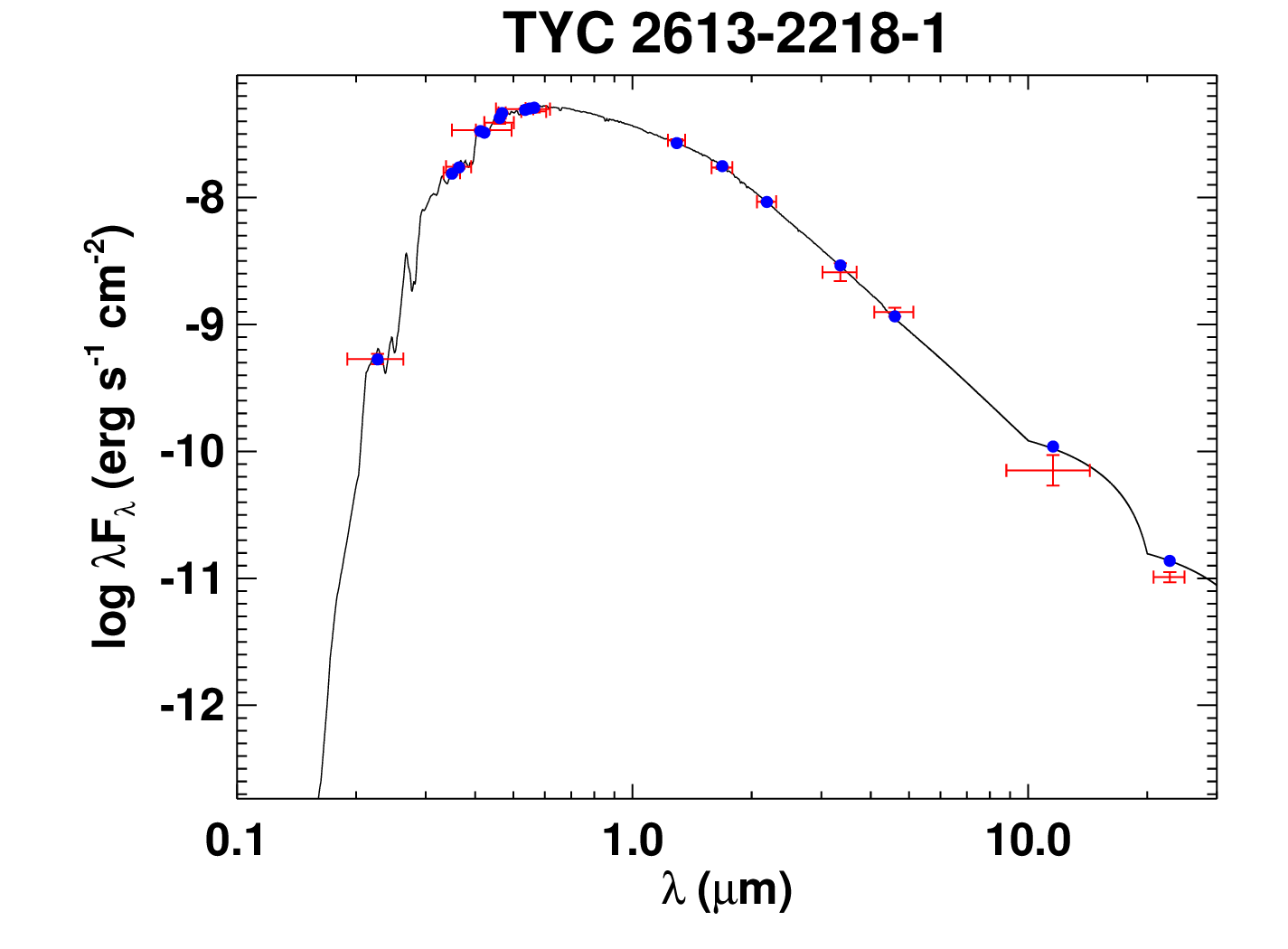}
\includegraphics[width=0.333\linewidth]{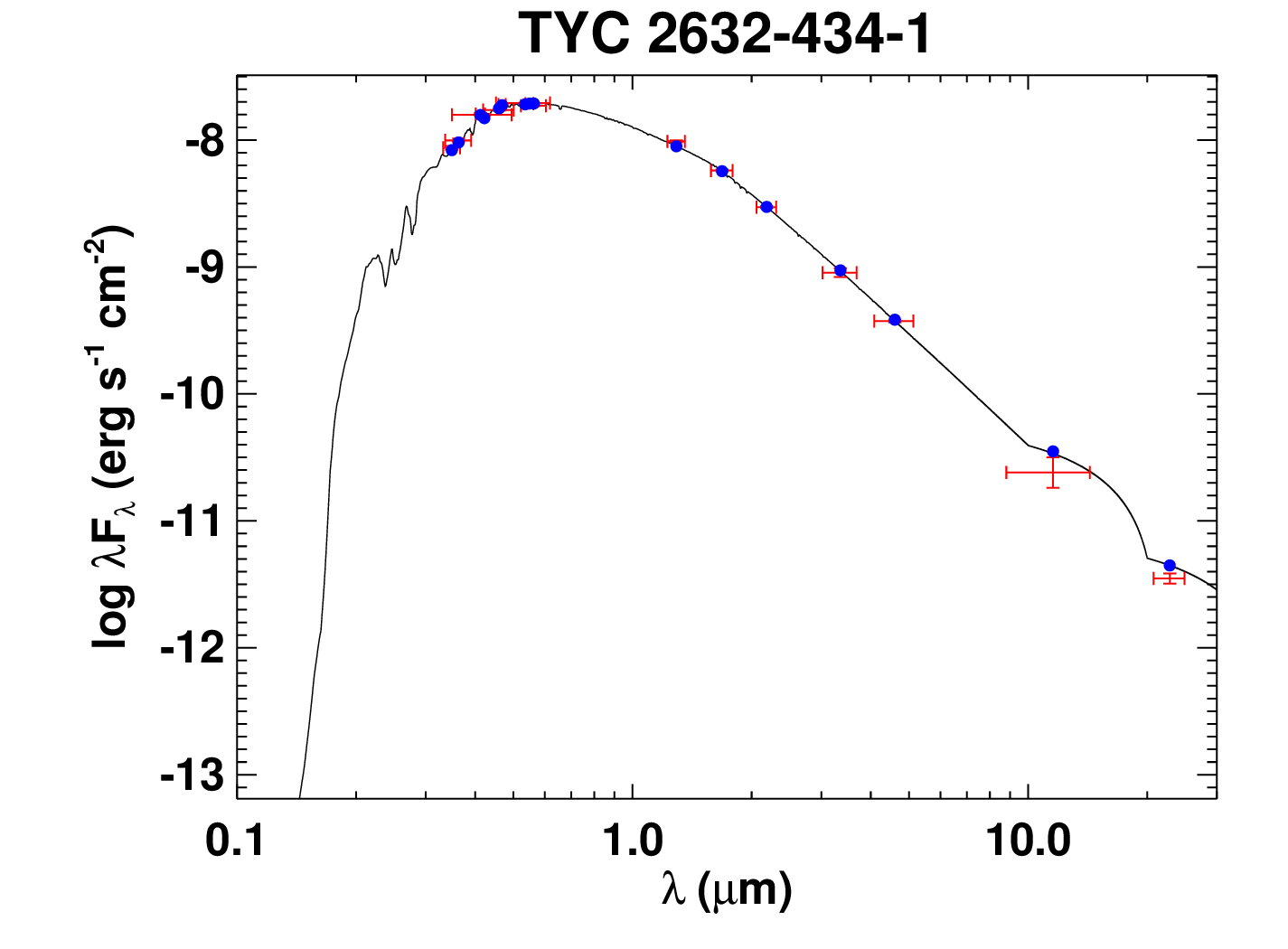}\includegraphics[width=0.333\linewidth]{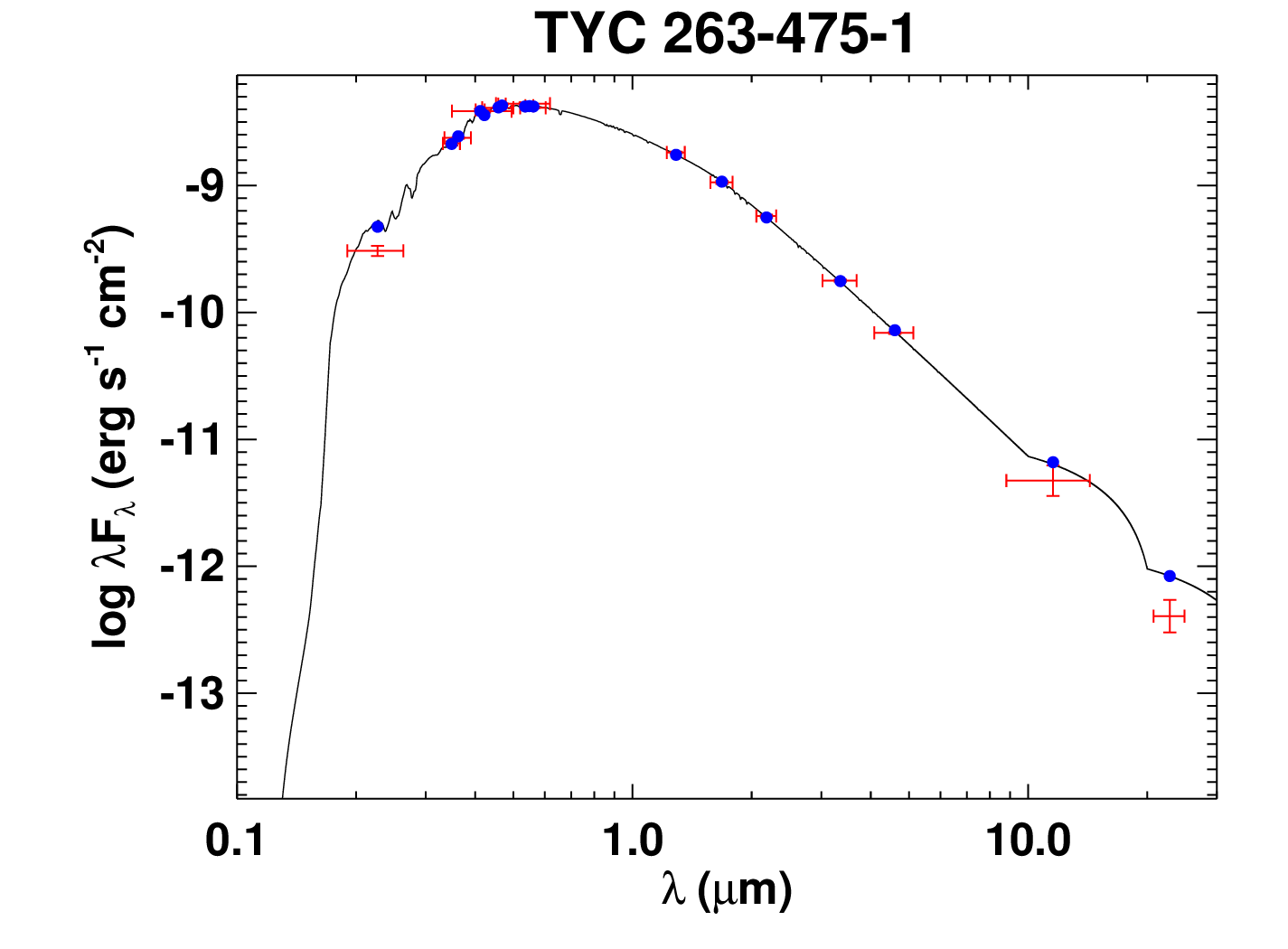}\includegraphics[width=0.333\linewidth]{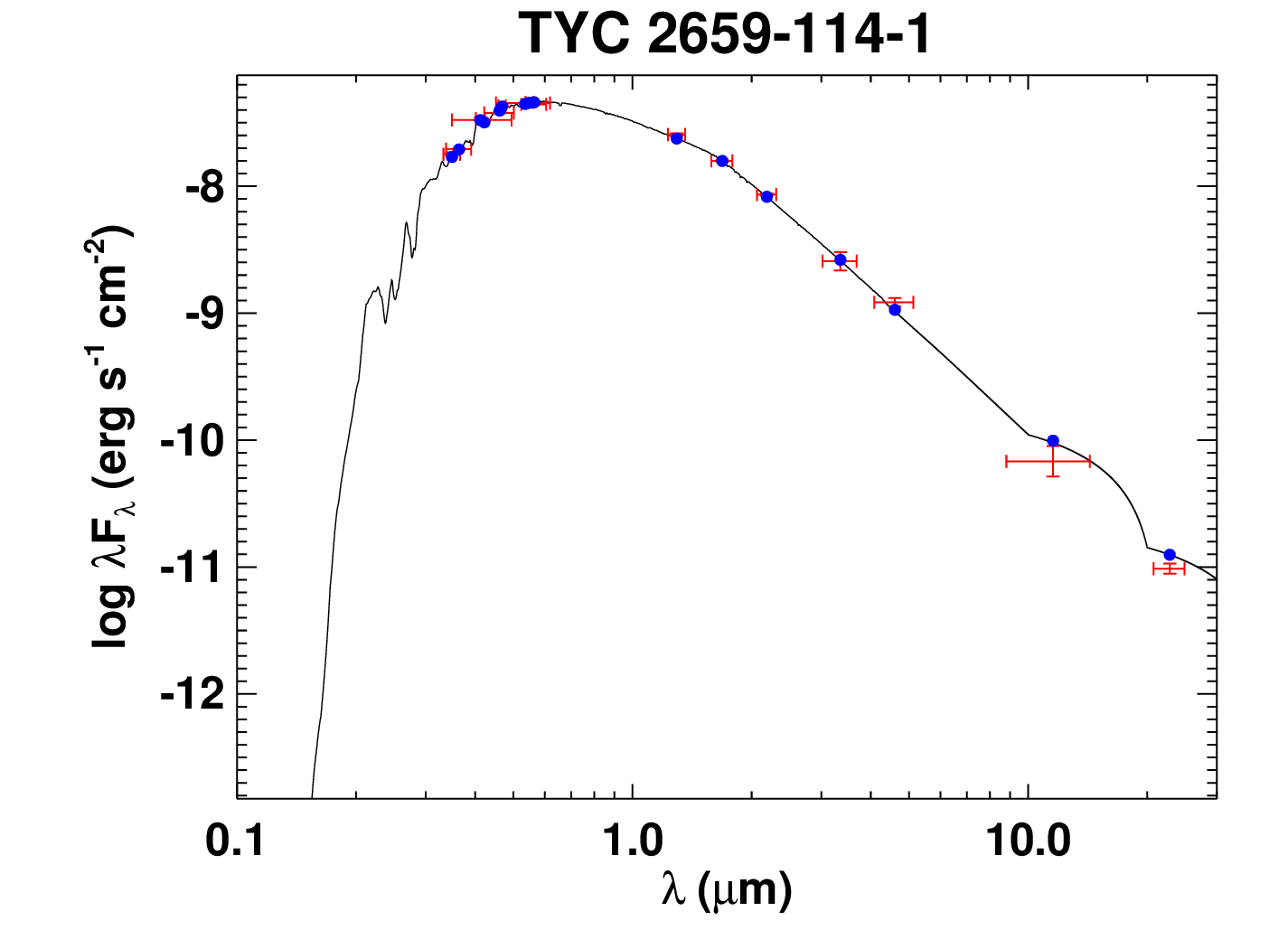}
\includegraphics[width=0.333\linewidth]{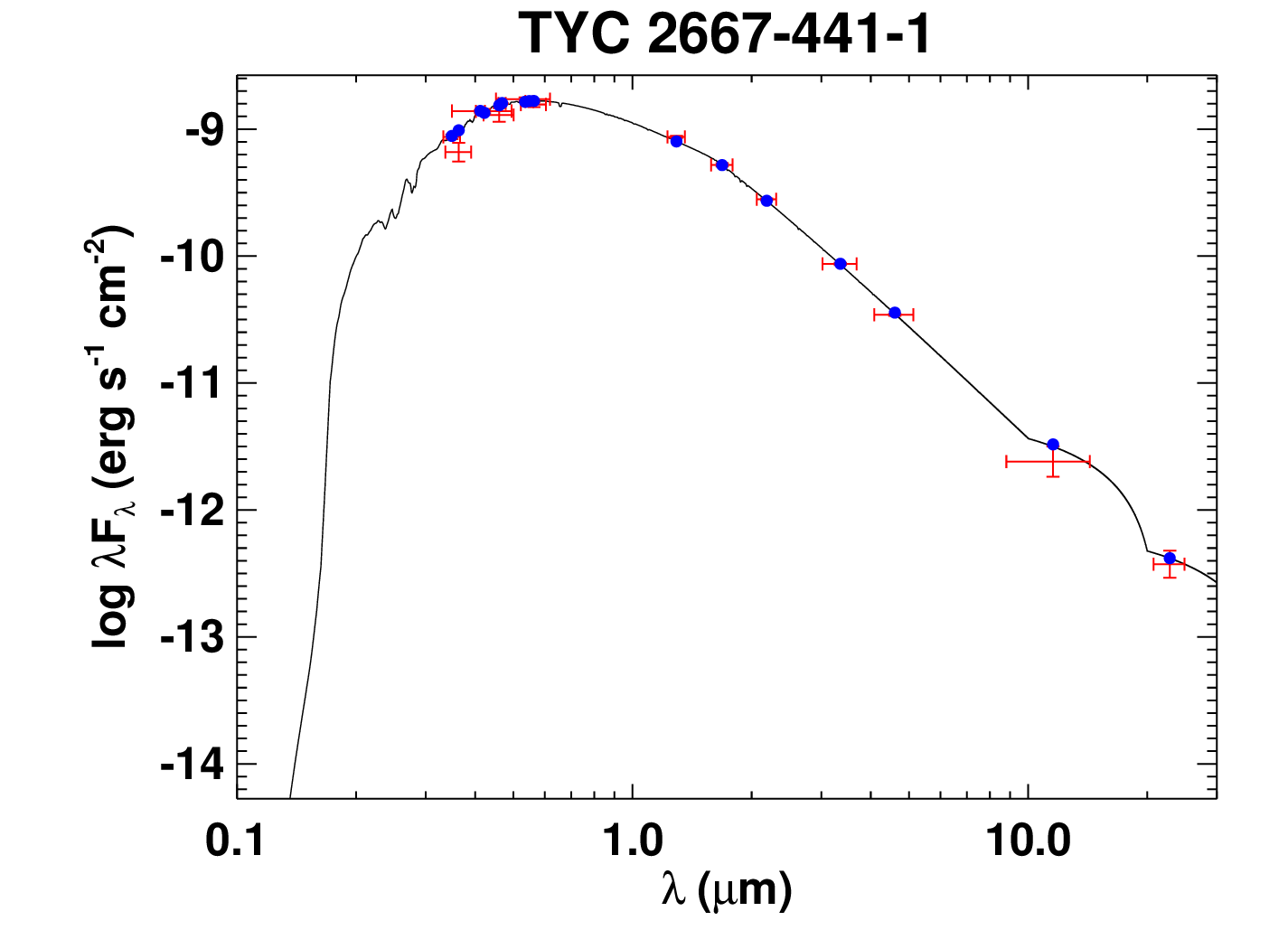}\includegraphics[width=0.333\linewidth]{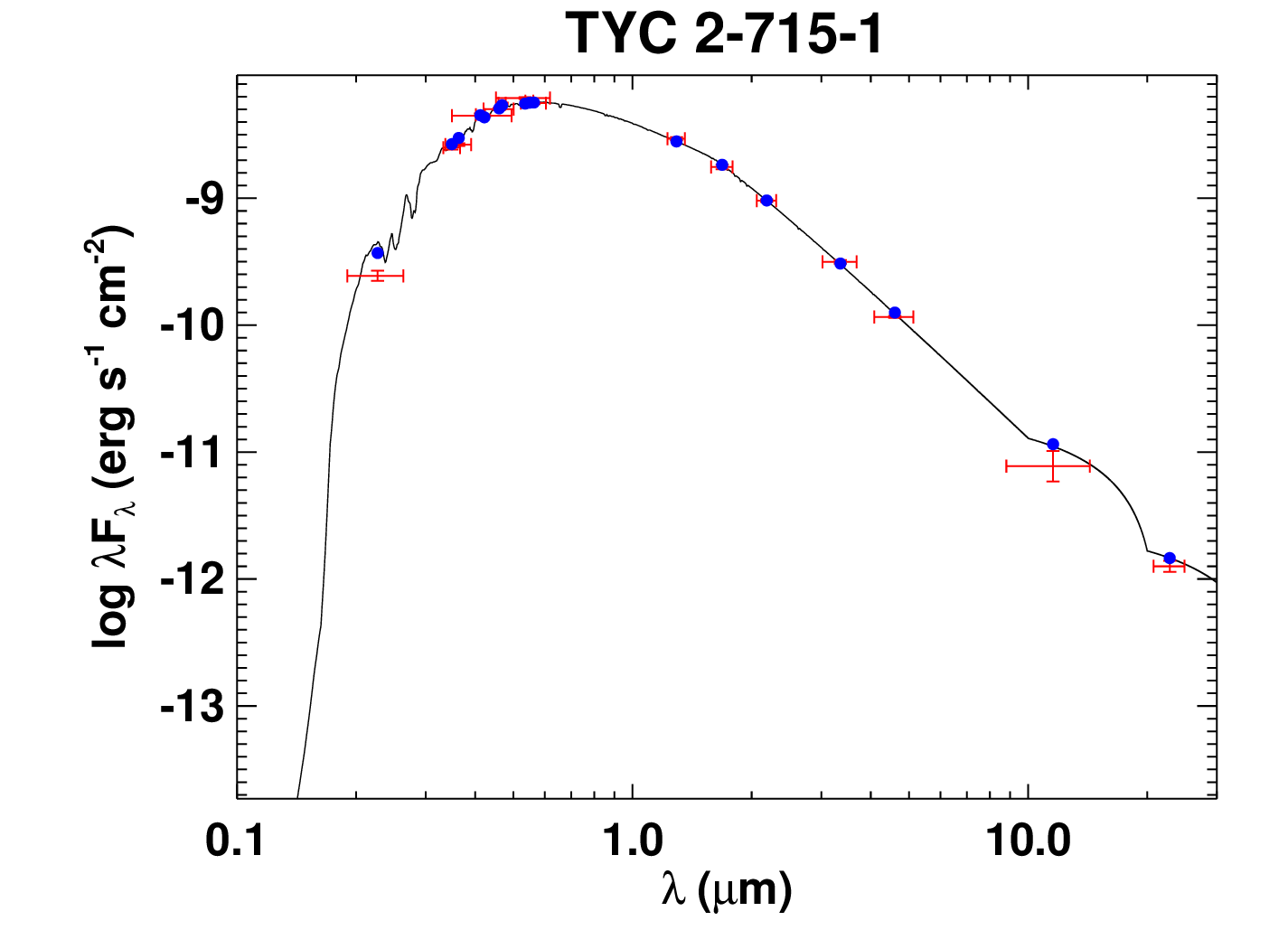}\includegraphics[width=0.333\linewidth]{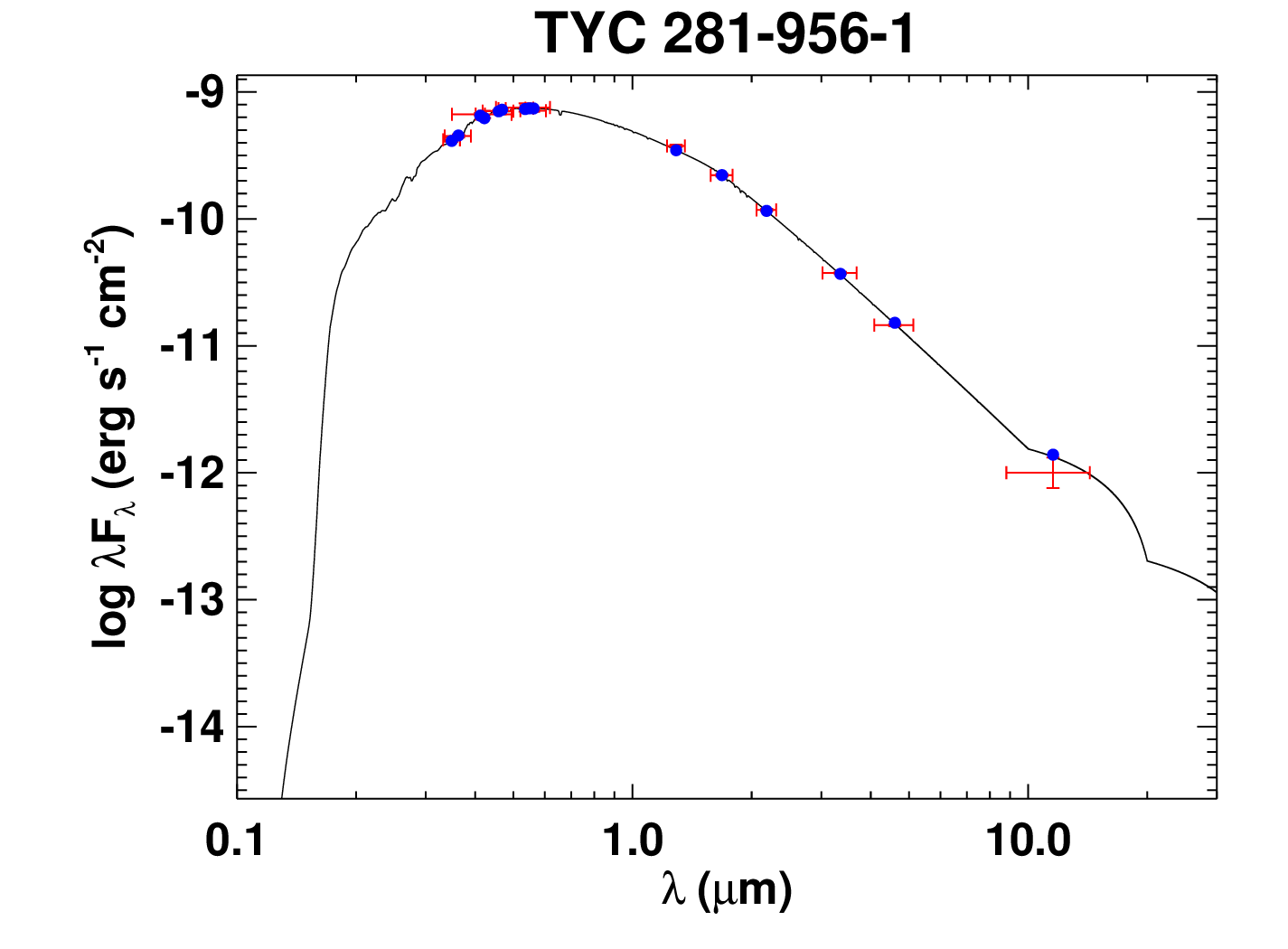}
\caption{\label{fig:seds5} All labels, lines, symbols, and colors as in Figure \ref{fig:seds}.}
\end{figure*}

\begin{figure*}
\includegraphics[width=0.333\linewidth]{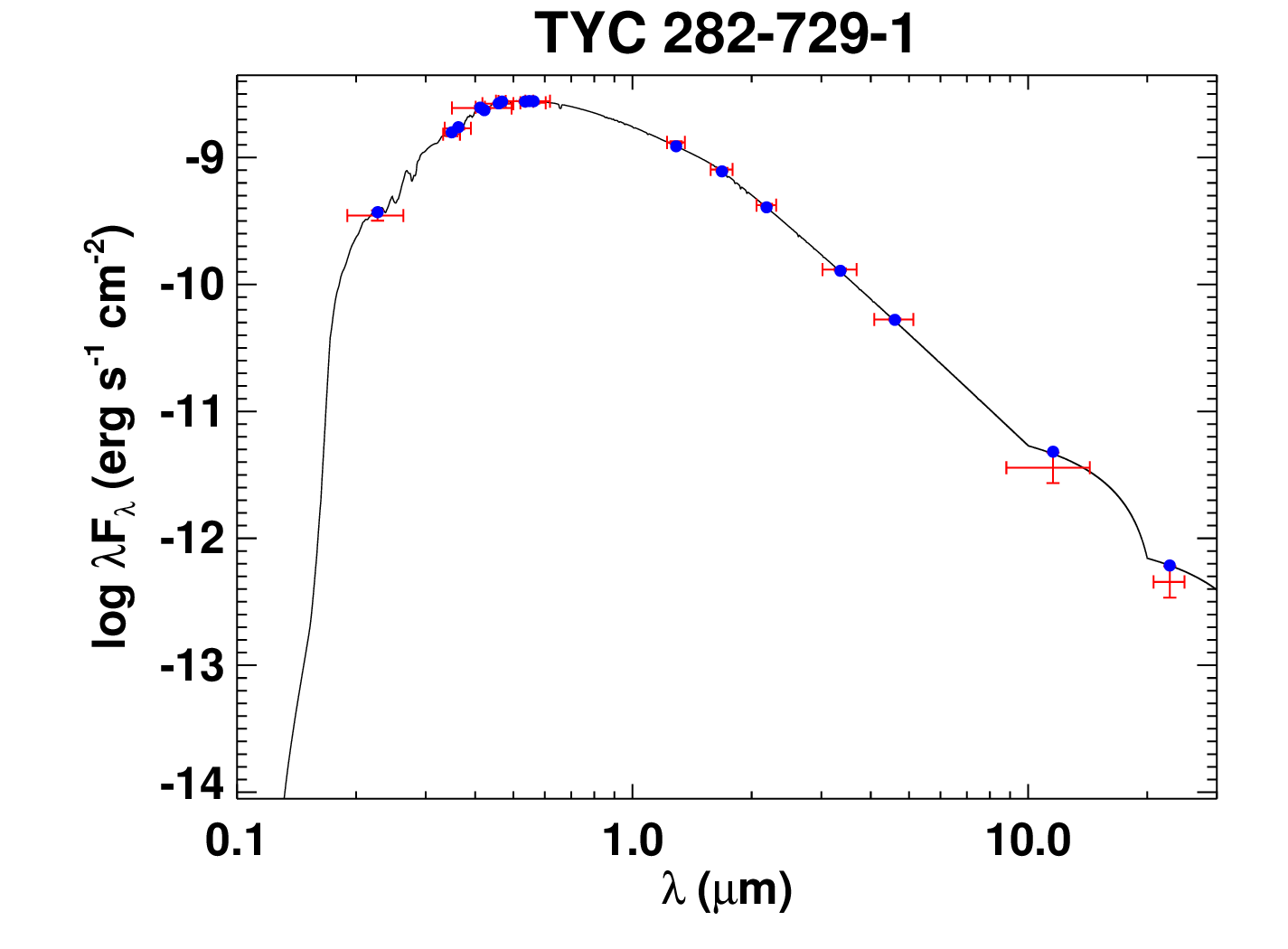}\includegraphics[width=0.333\linewidth]{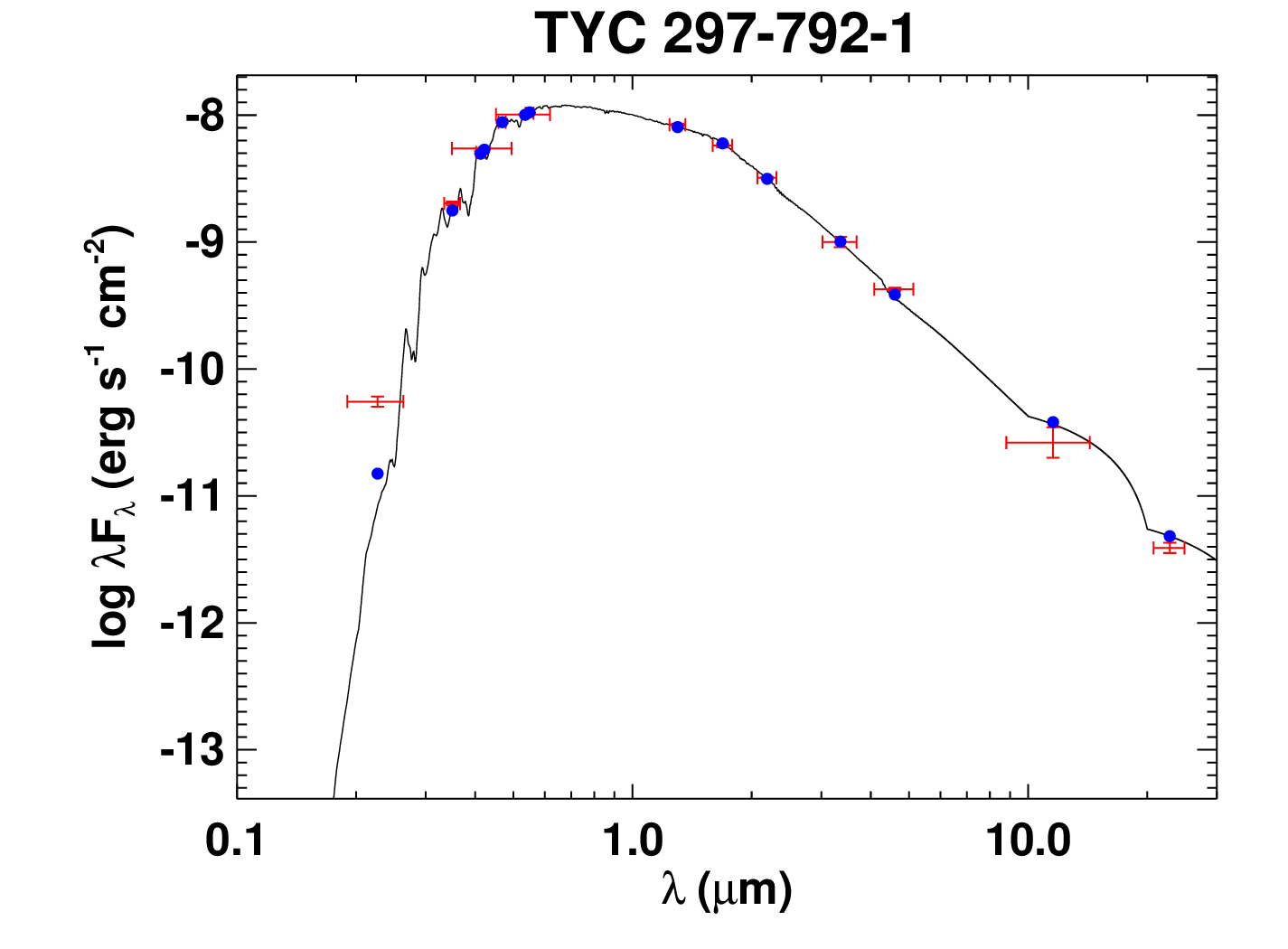}\includegraphics[width=0.333\linewidth]{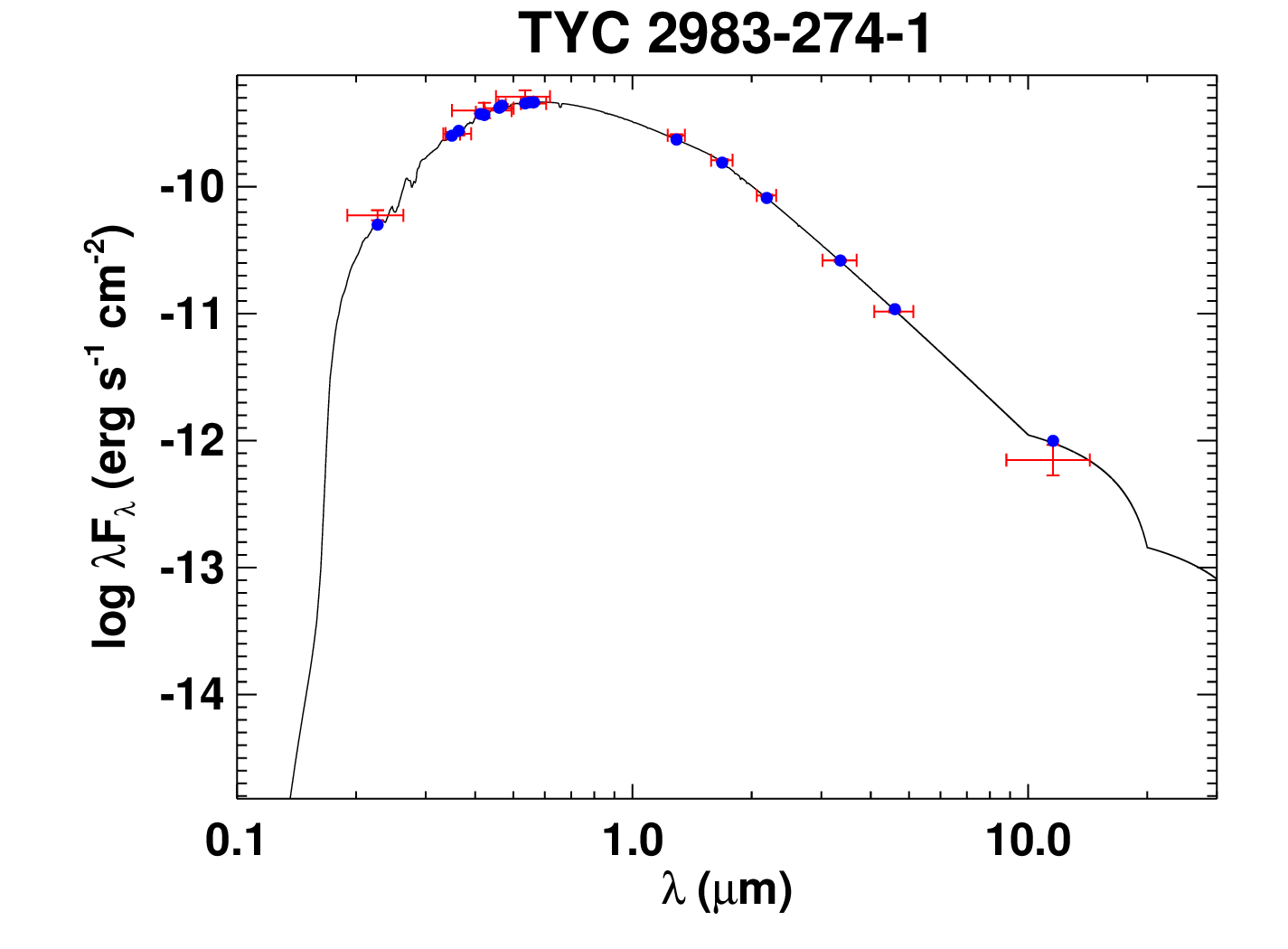}
\includegraphics[width=0.333\linewidth]{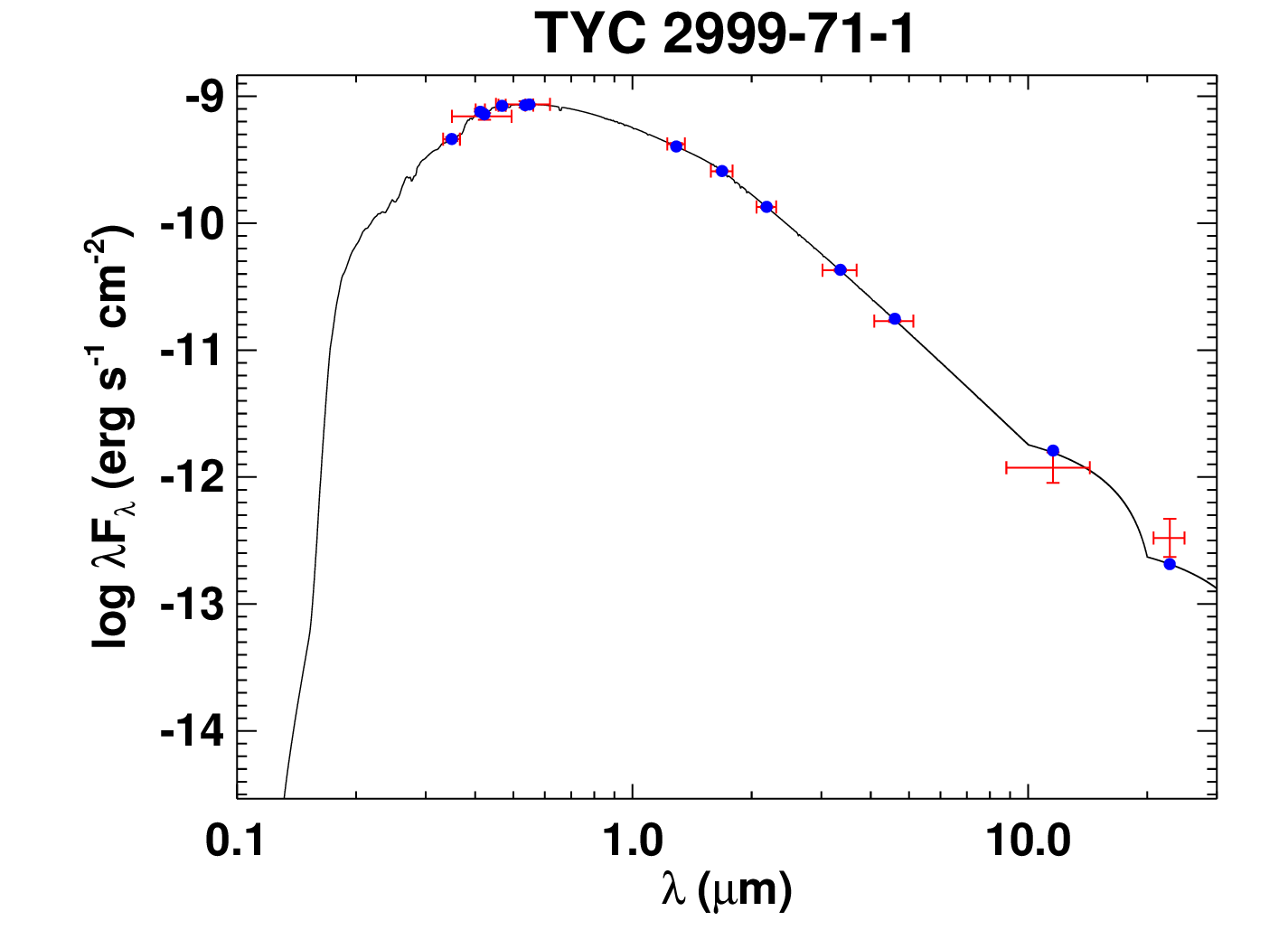}\includegraphics[width=0.333\linewidth]{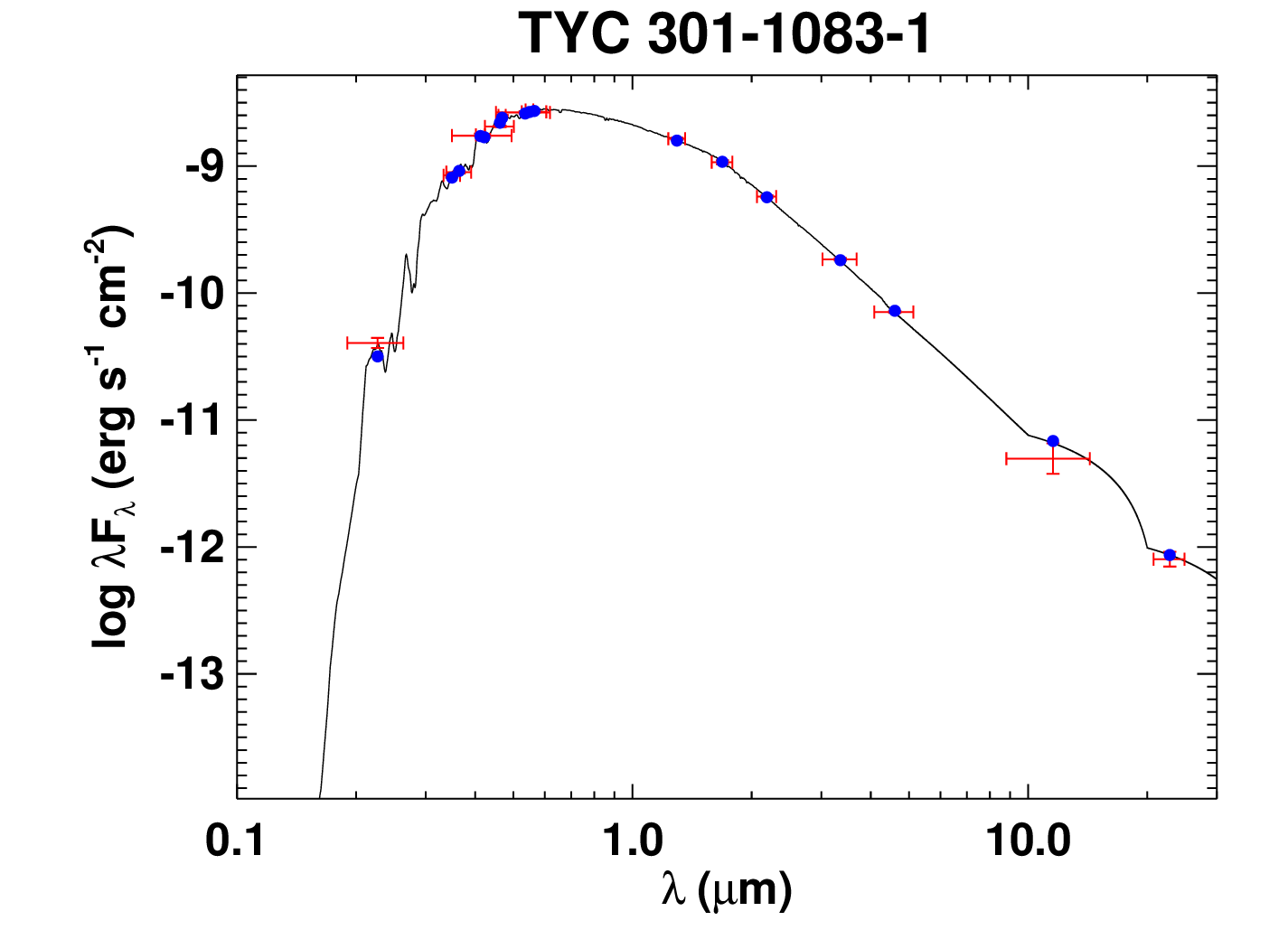}\includegraphics[width=0.333\linewidth]{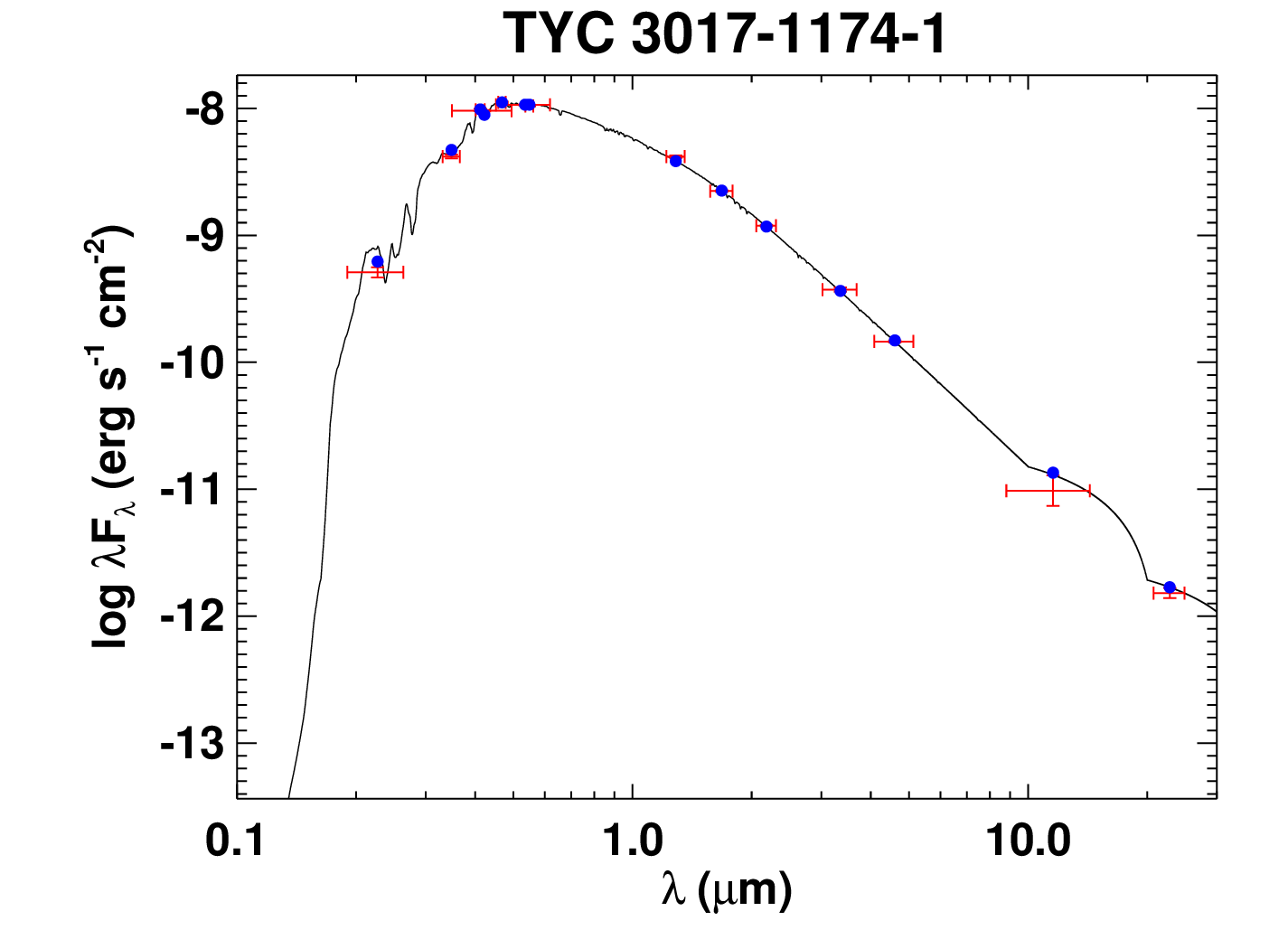}
\includegraphics[width=0.333\linewidth]{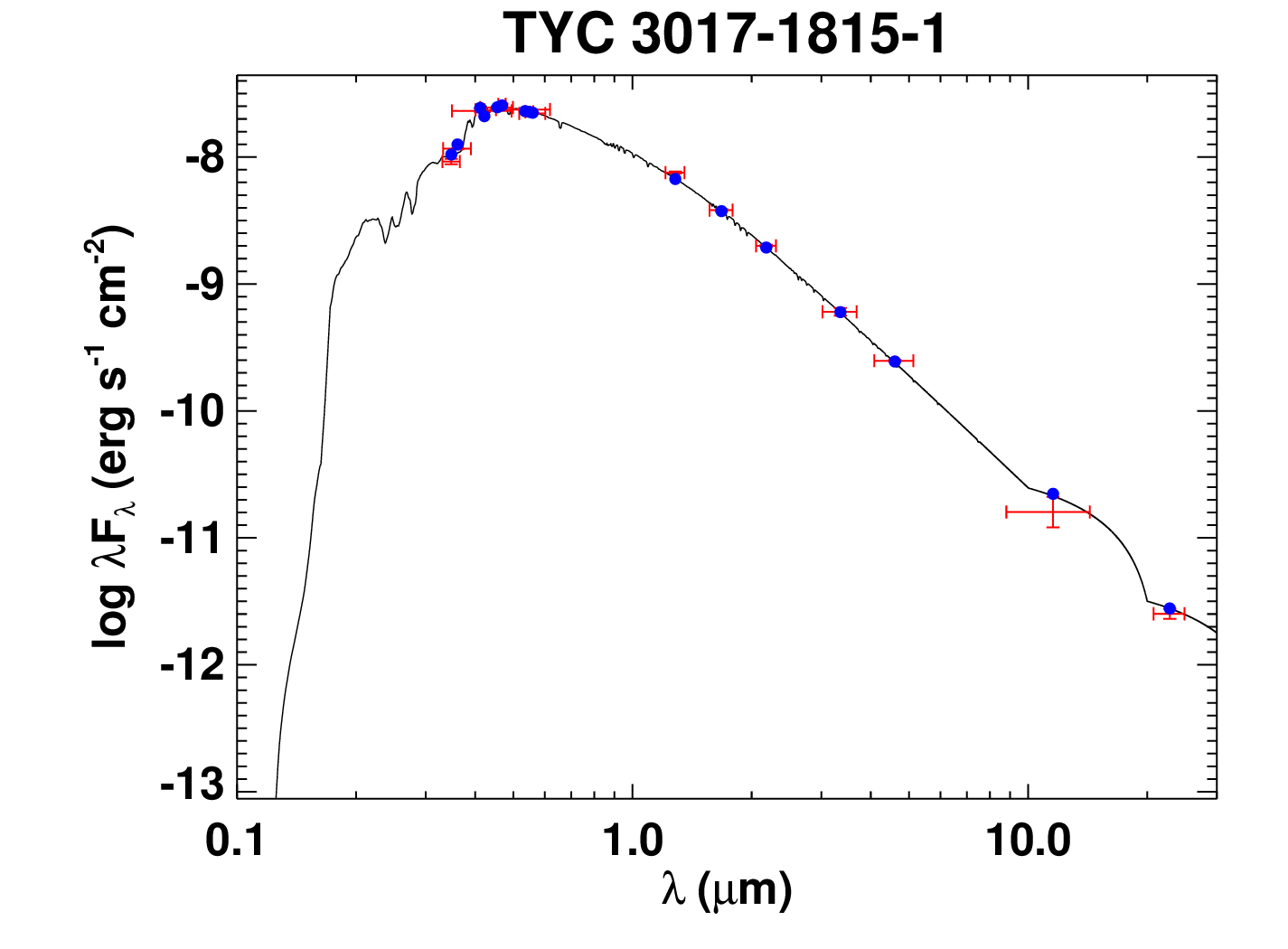}\includegraphics[width=0.333\linewidth]{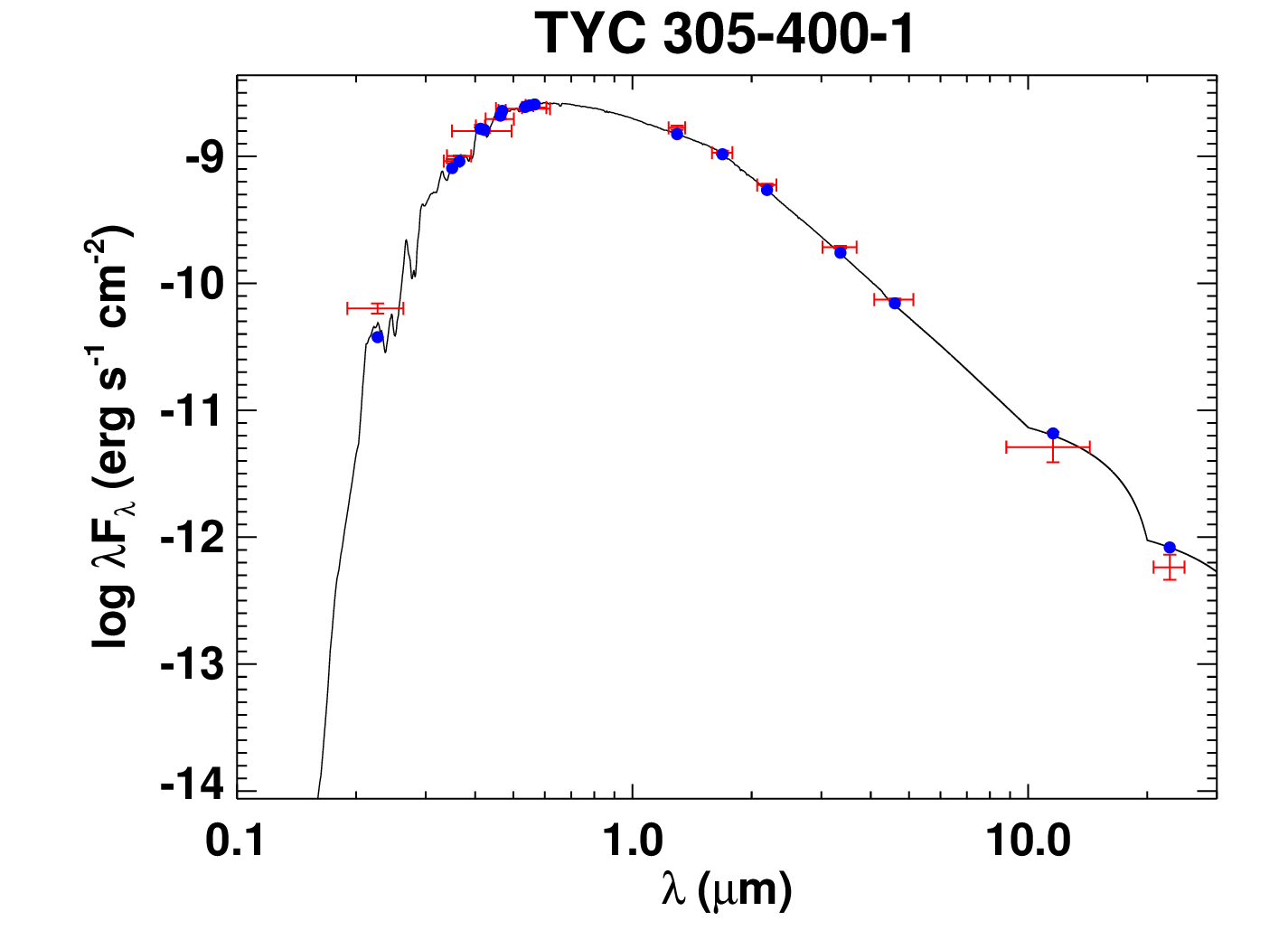}\includegraphics[width=0.333\linewidth]{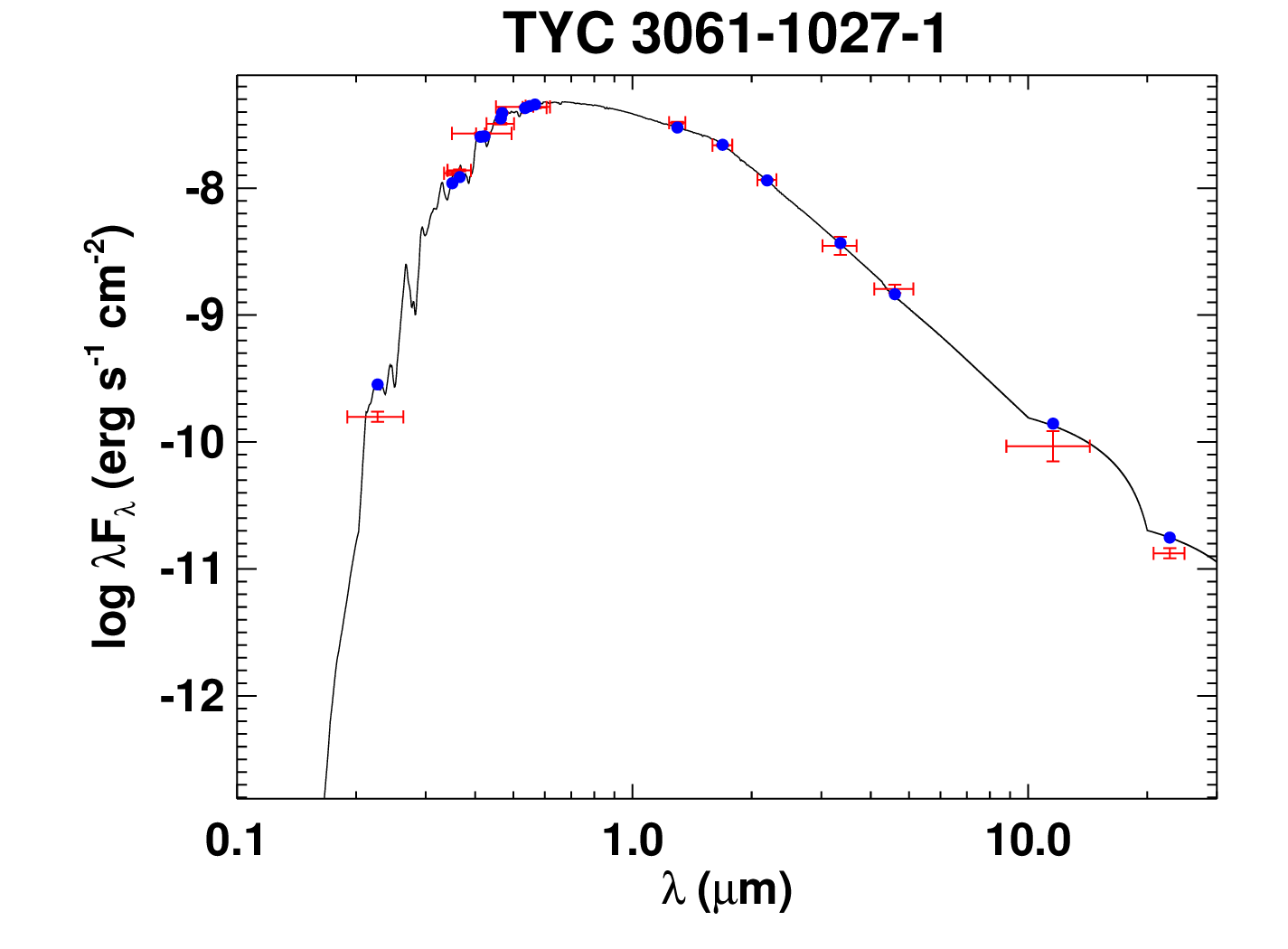}
\includegraphics[width=0.333\linewidth]{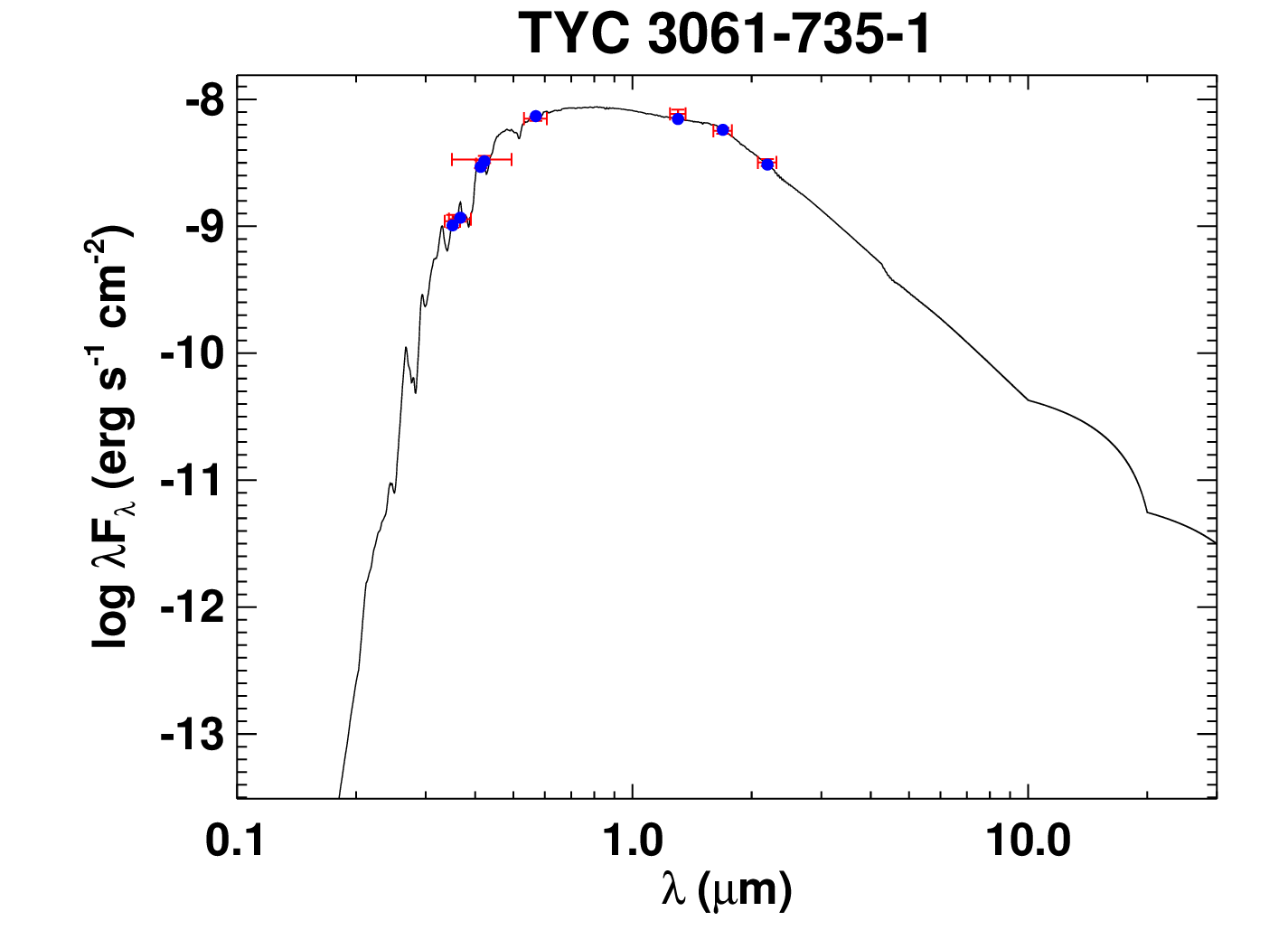}\includegraphics[width=0.333\linewidth]{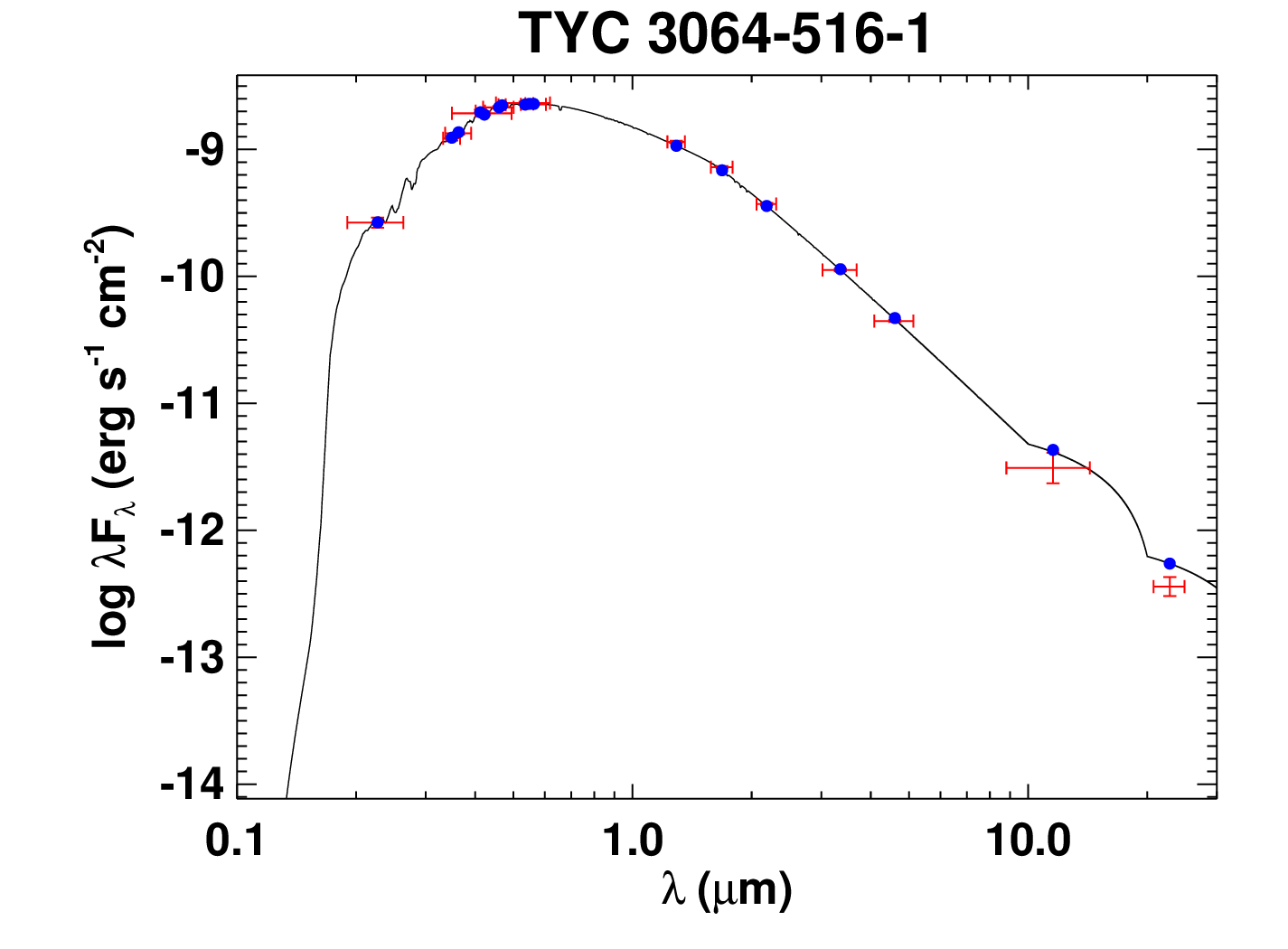}\includegraphics[width=0.333\linewidth]{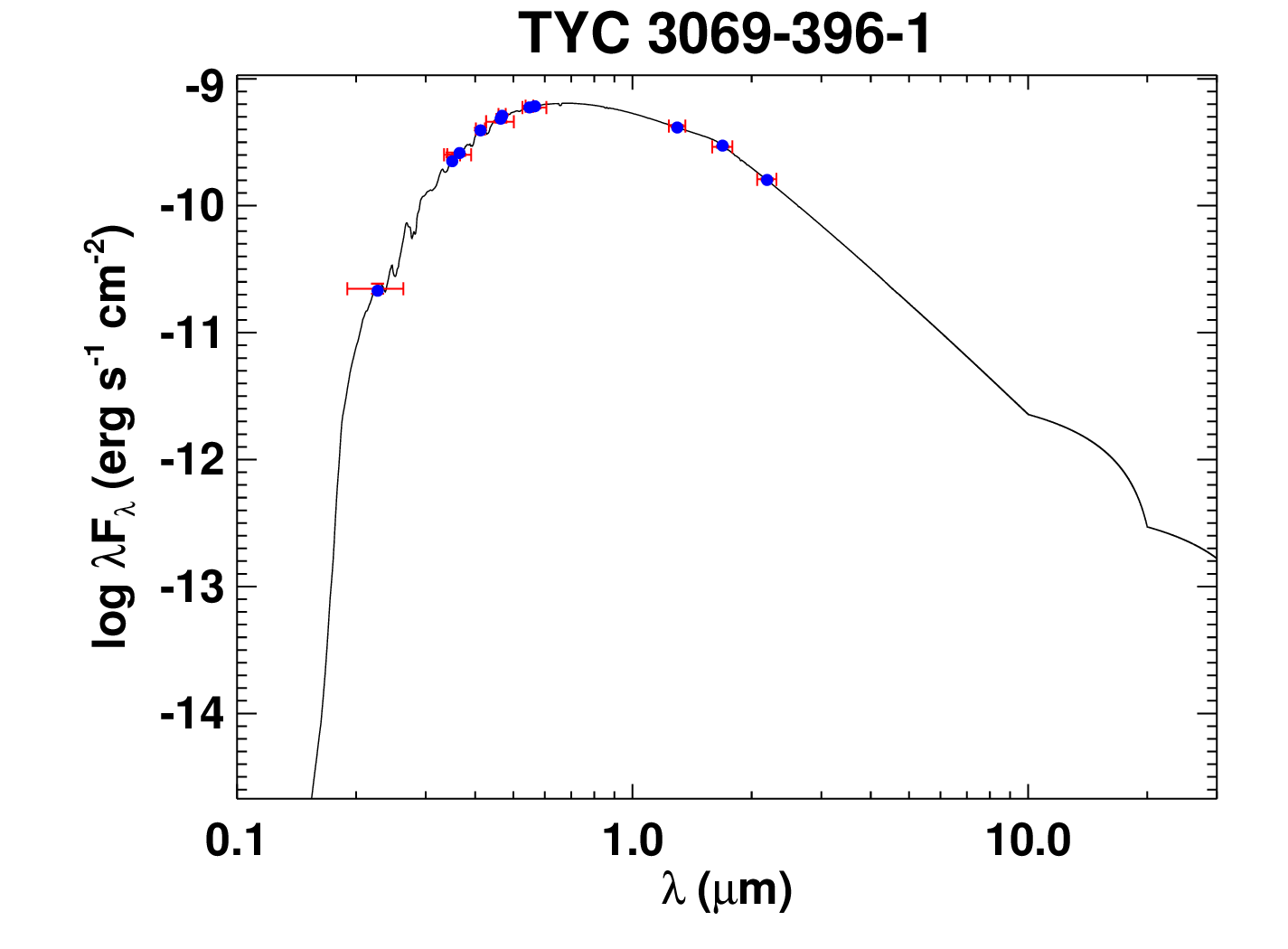}
\caption{\label{fig:seds6} All labels, lines, symbols, and colors as in Figure \ref{fig:seds}.}
\end{figure*}

\begin{figure*}
\includegraphics[width=0.333\linewidth]{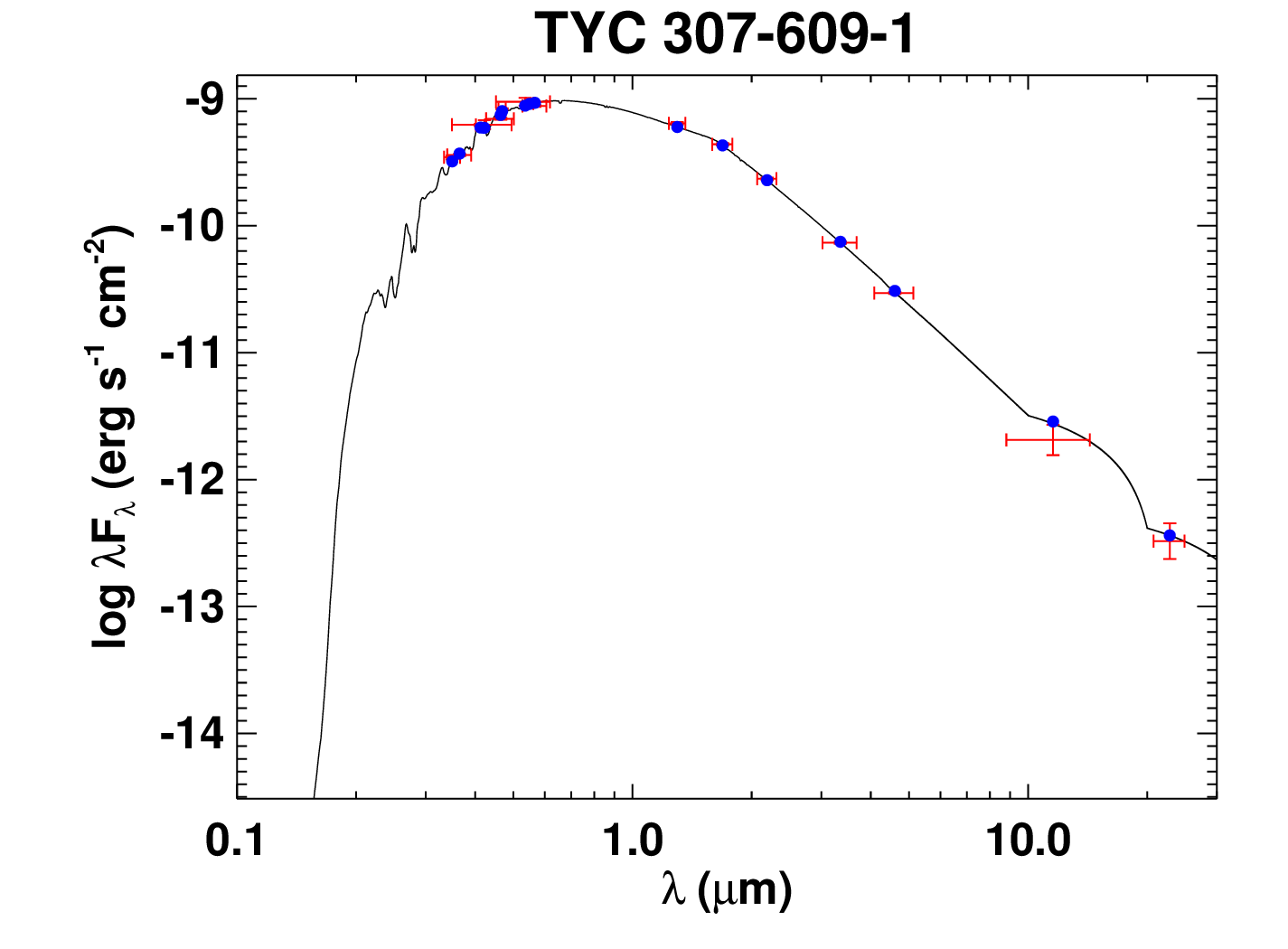}\includegraphics[width=0.333\linewidth]{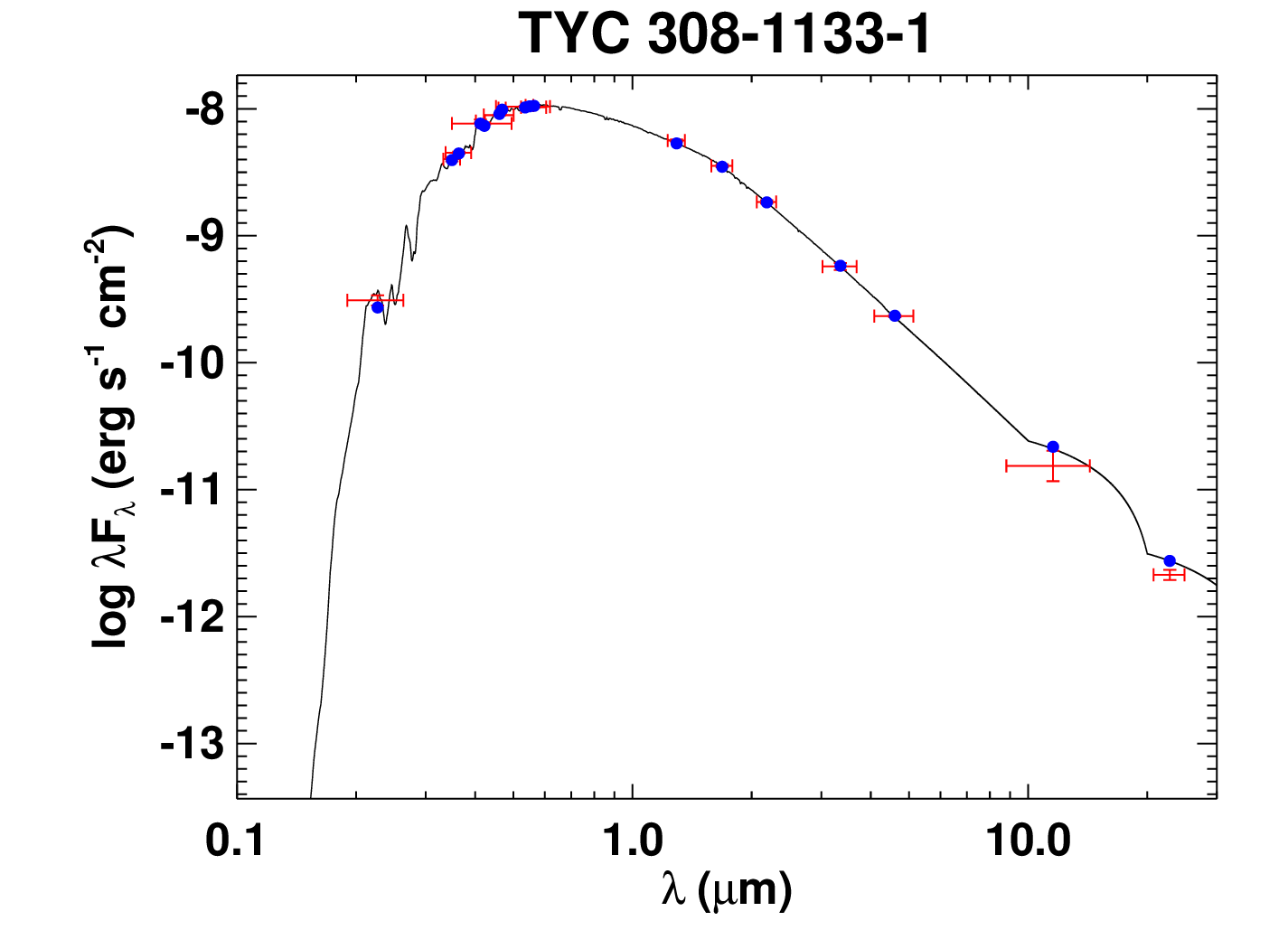}\includegraphics[width=0.333\linewidth]{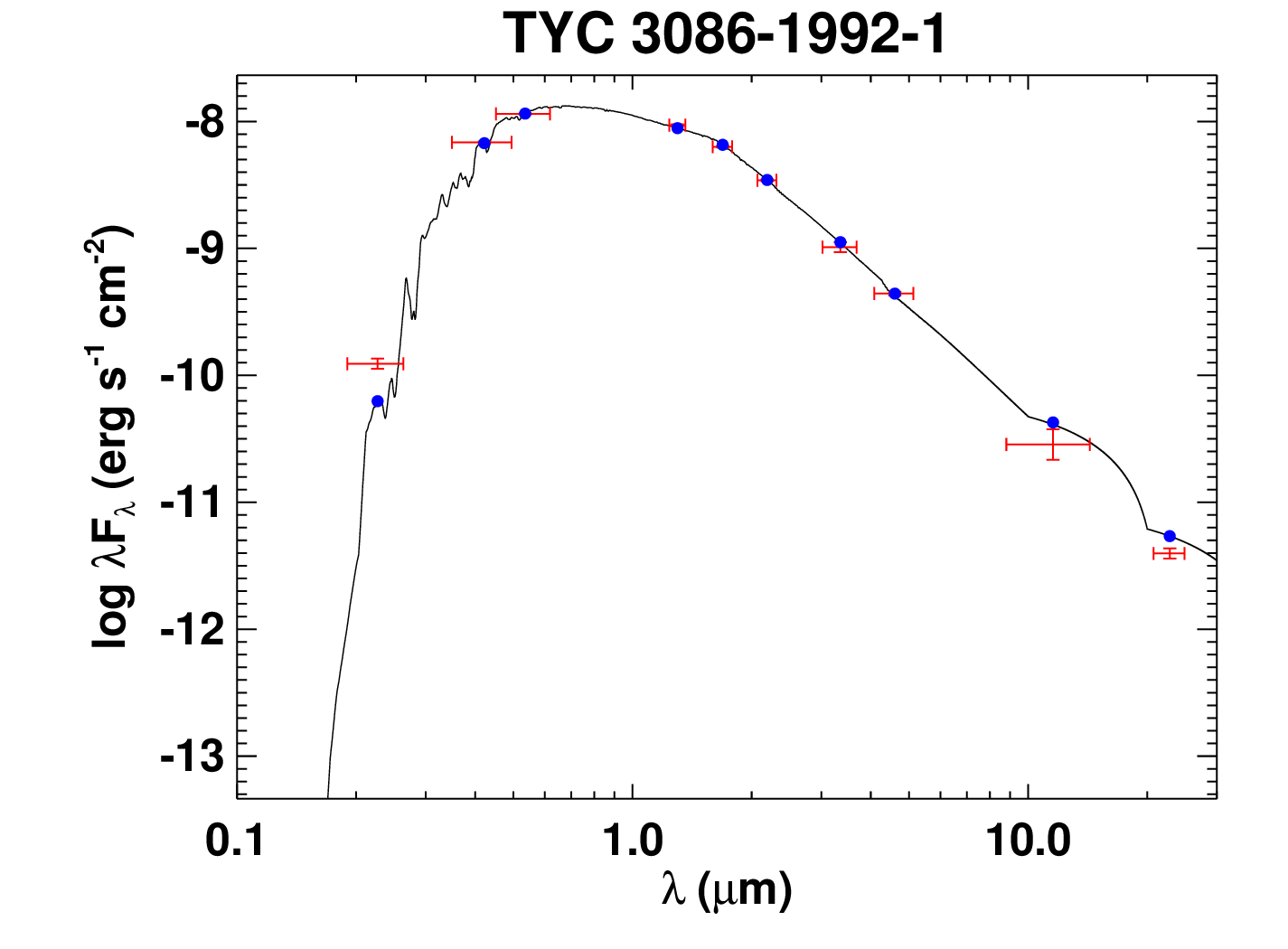}
\includegraphics[width=0.333\linewidth]{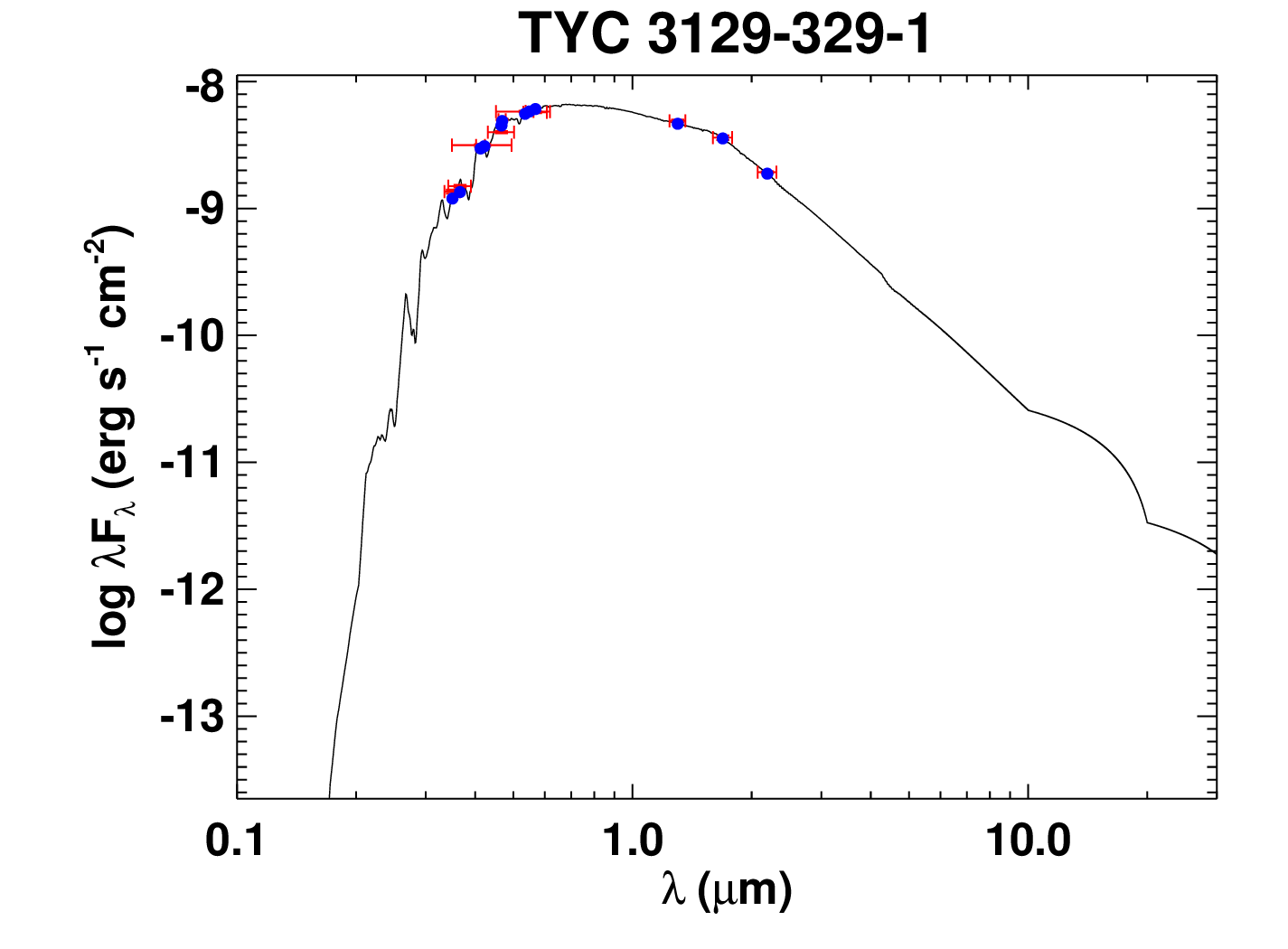}\includegraphics[width=0.333\linewidth]{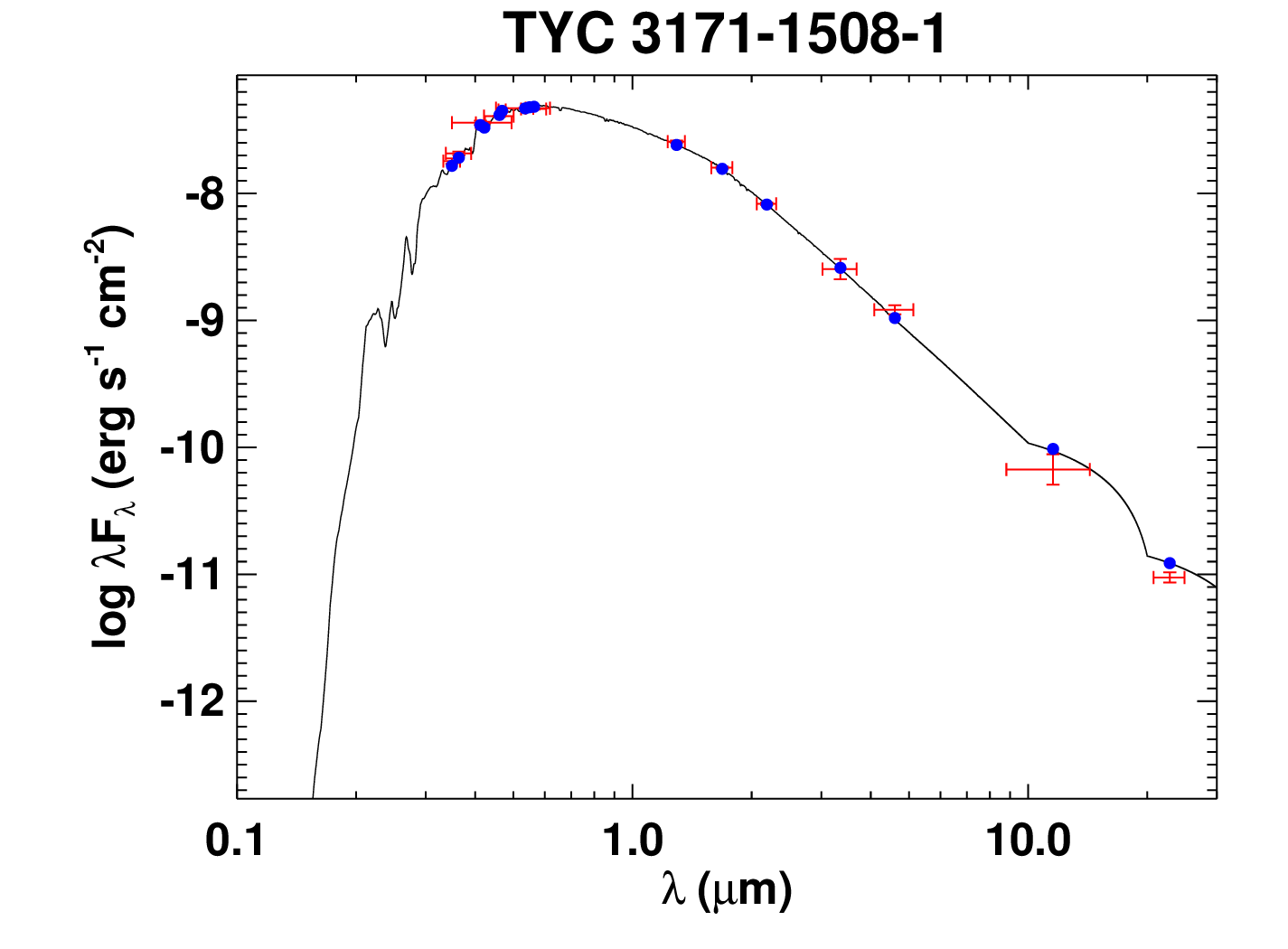}\includegraphics[width=0.333\linewidth]{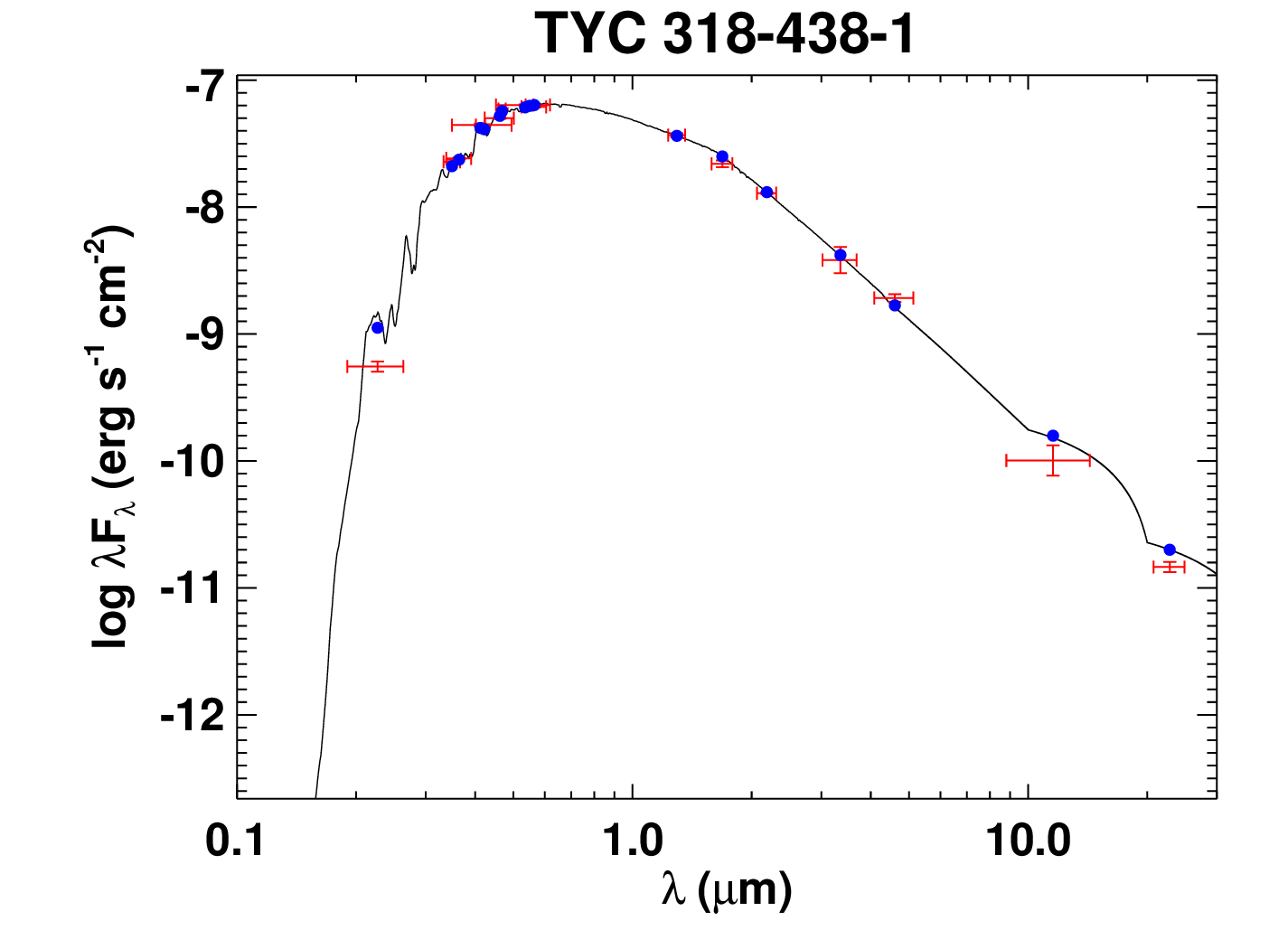}
\includegraphics[width=0.333\linewidth]{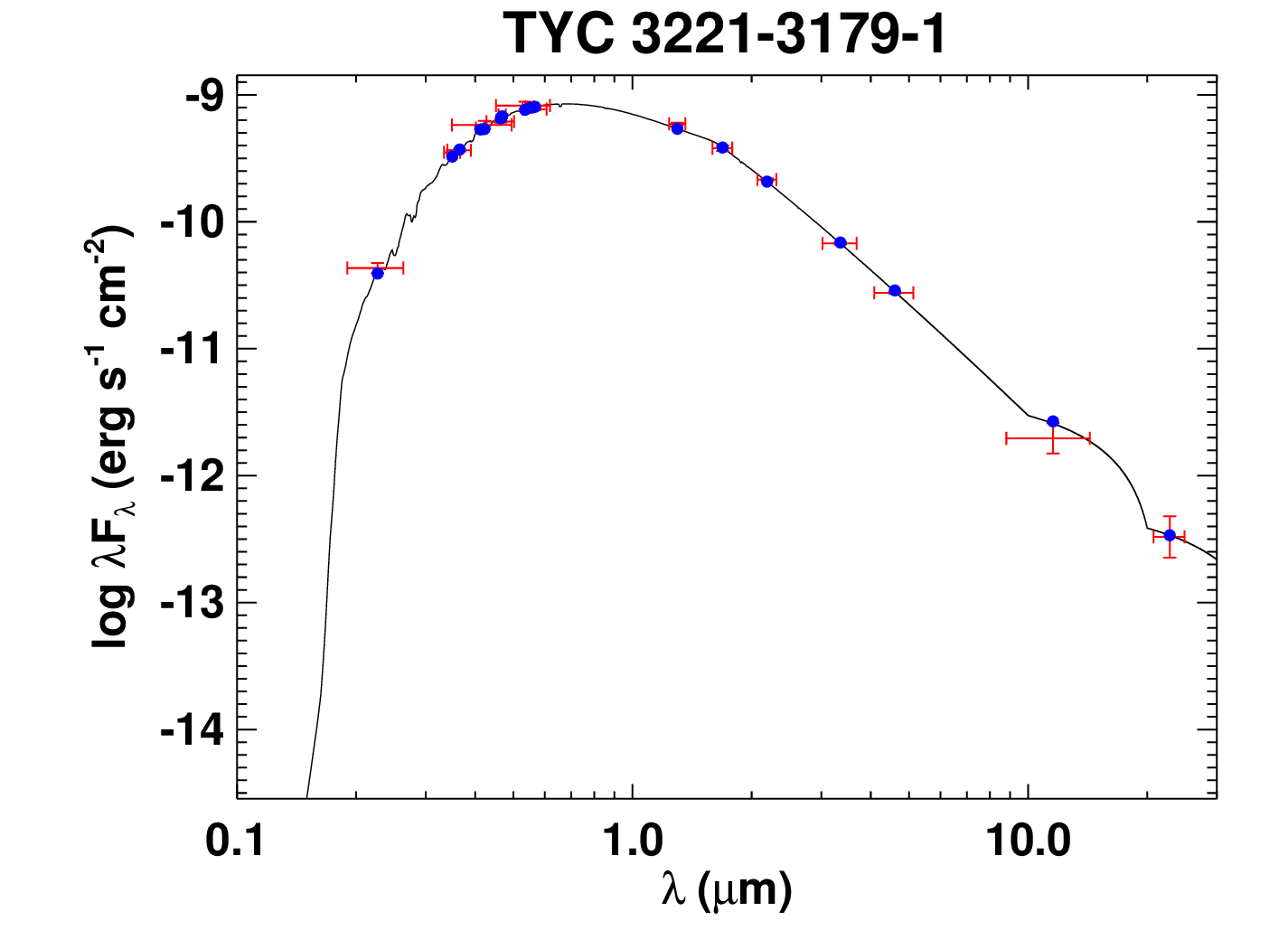}\includegraphics[width=0.333\linewidth]{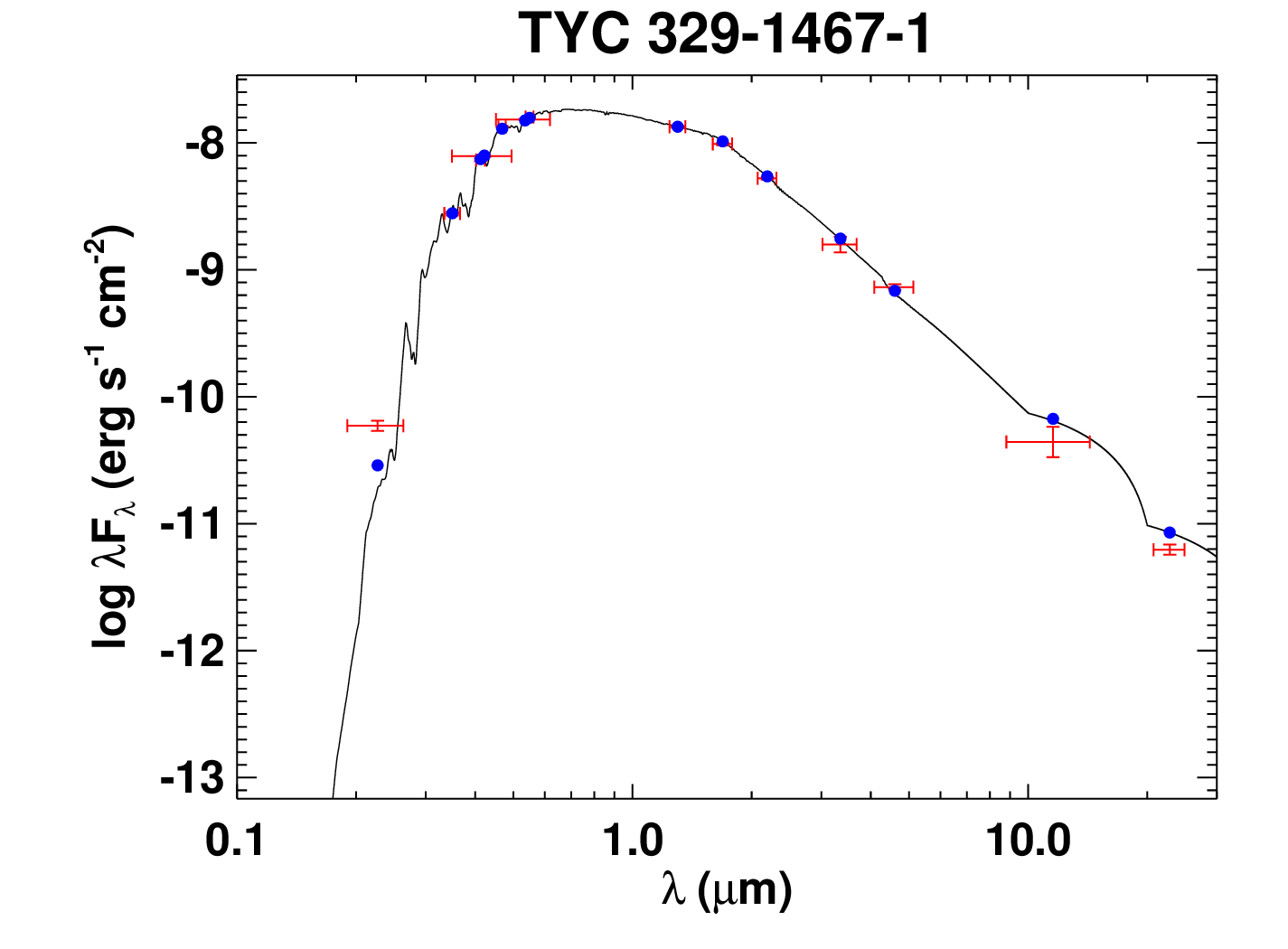}\includegraphics[width=0.333\linewidth]{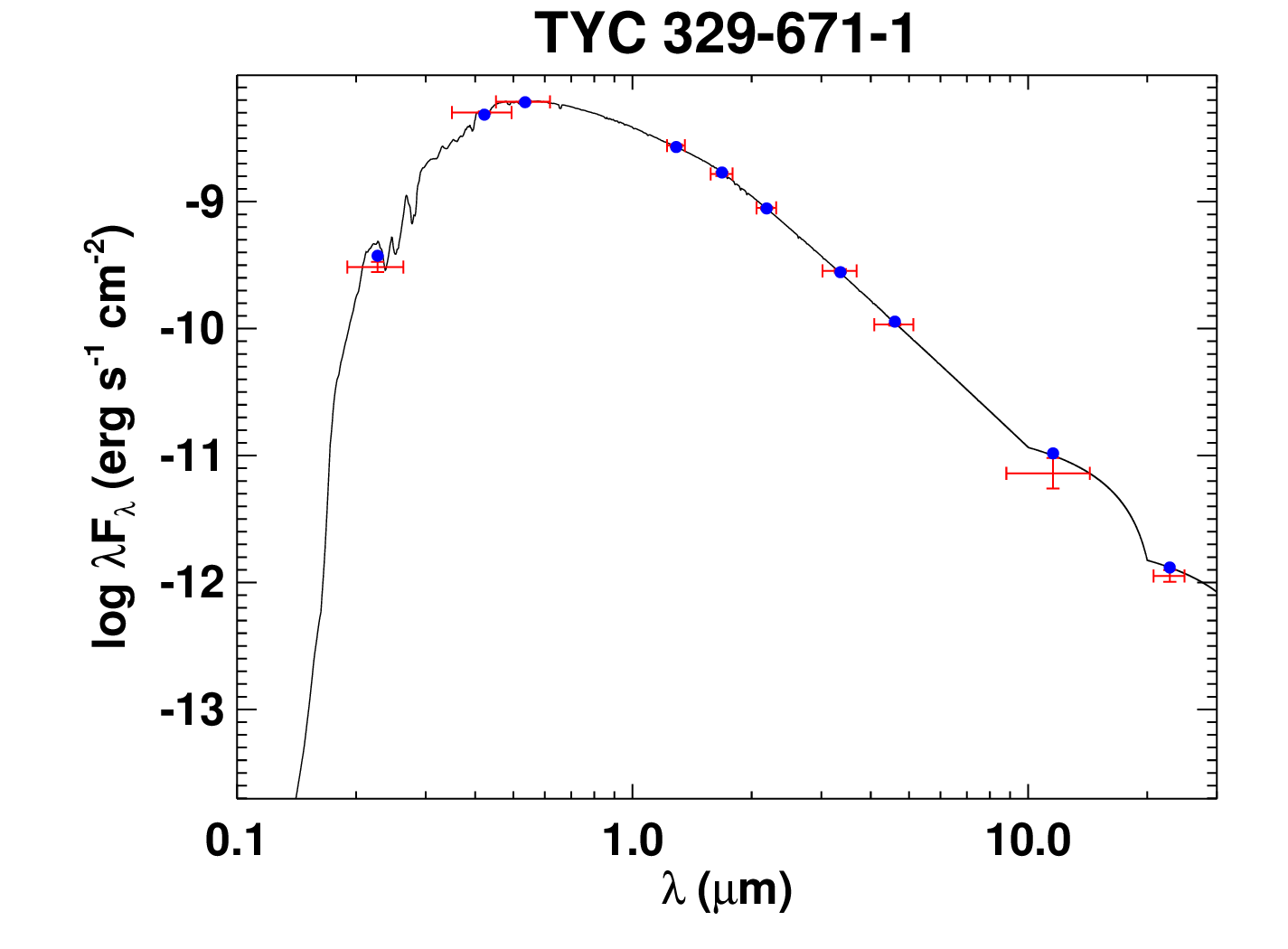}
\includegraphics[width=0.333\linewidth]{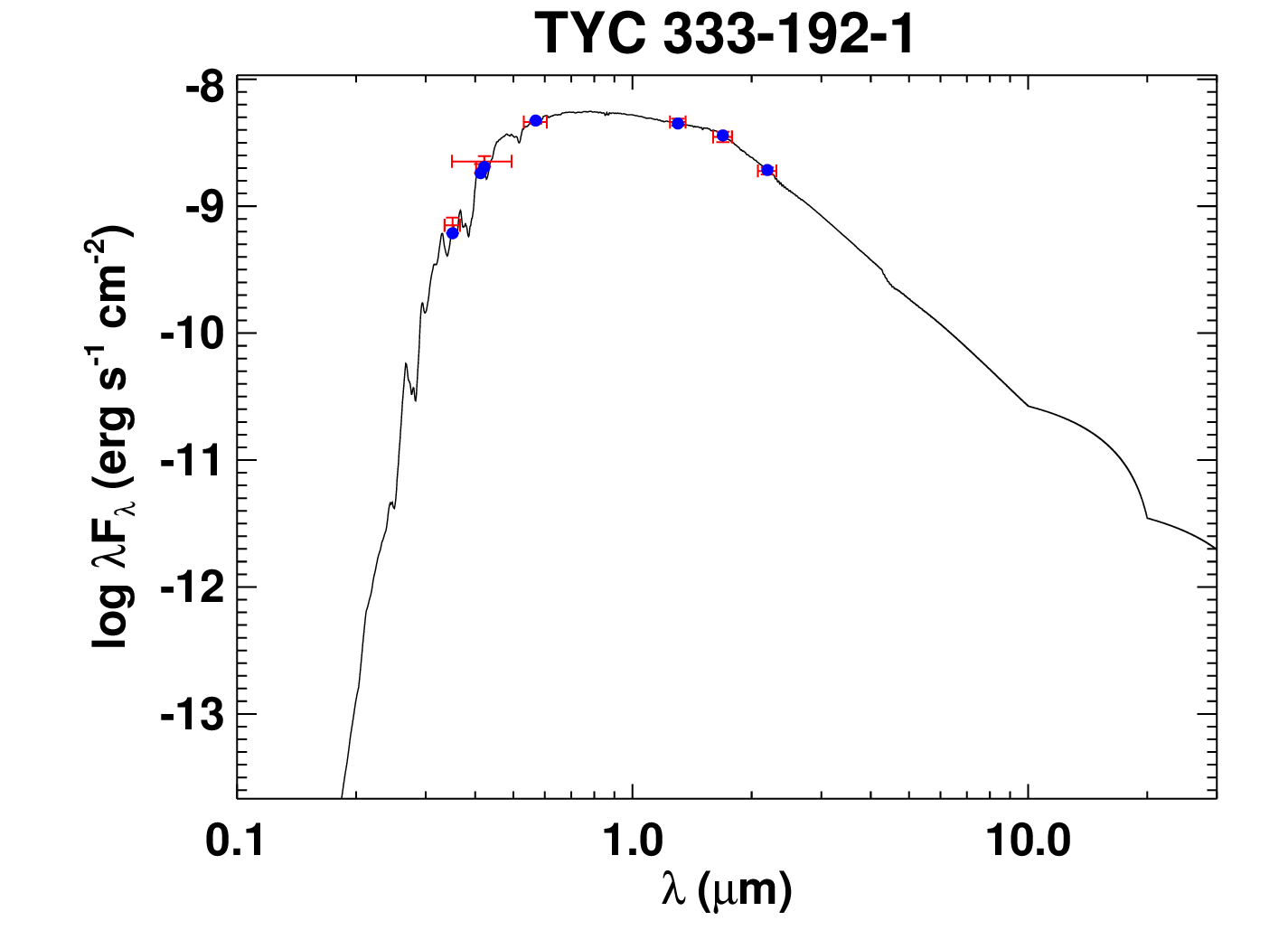}\includegraphics[width=0.333\linewidth]{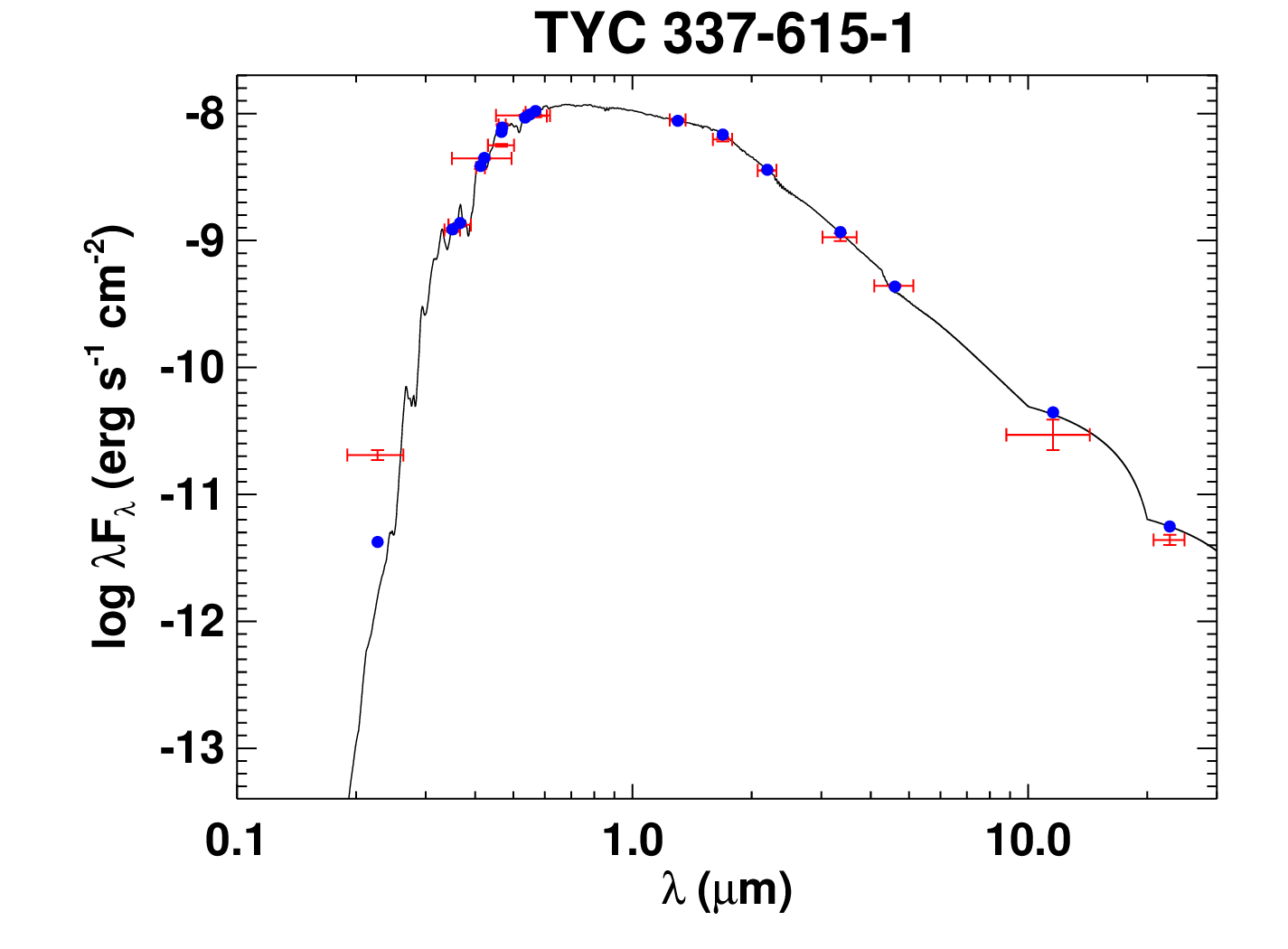}\includegraphics[width=0.333\linewidth]{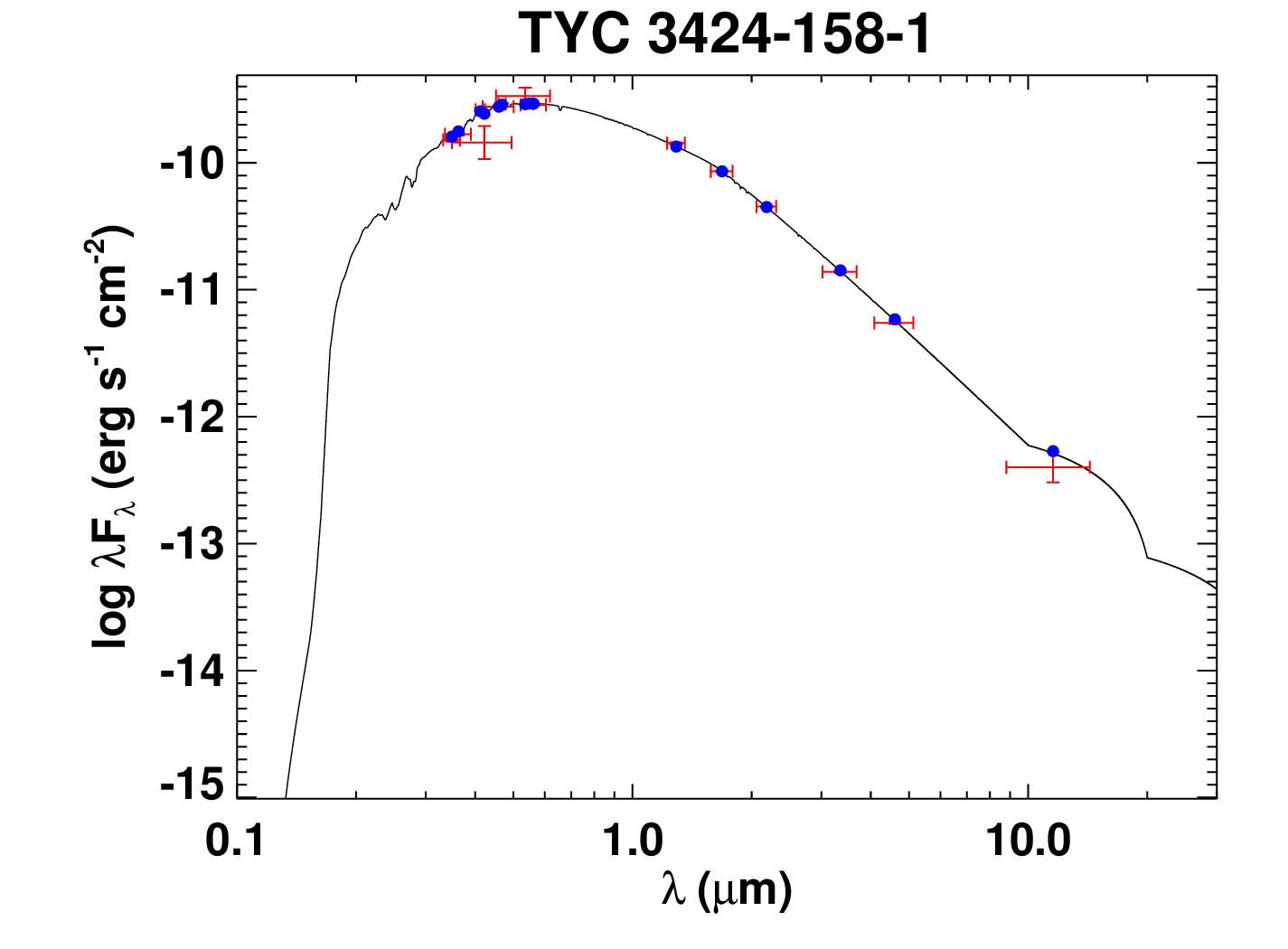}
\caption{\label{fig:seds7} All labels, lines, symbols, and colors as in Figure \ref{fig:seds}.}
\end{figure*}

\begin{figure*}
\includegraphics[width=0.333\linewidth]{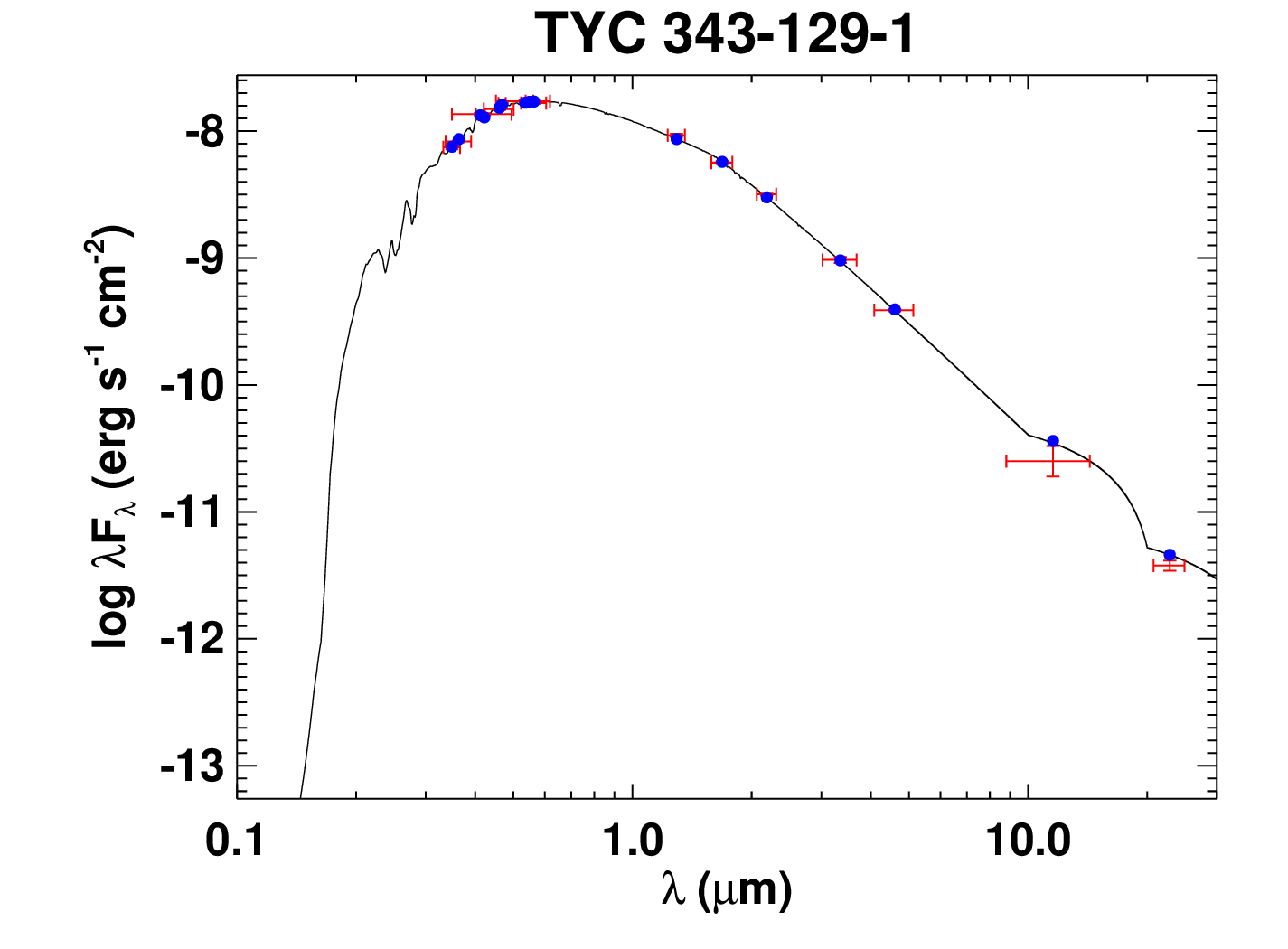}\includegraphics[width=0.333\linewidth]{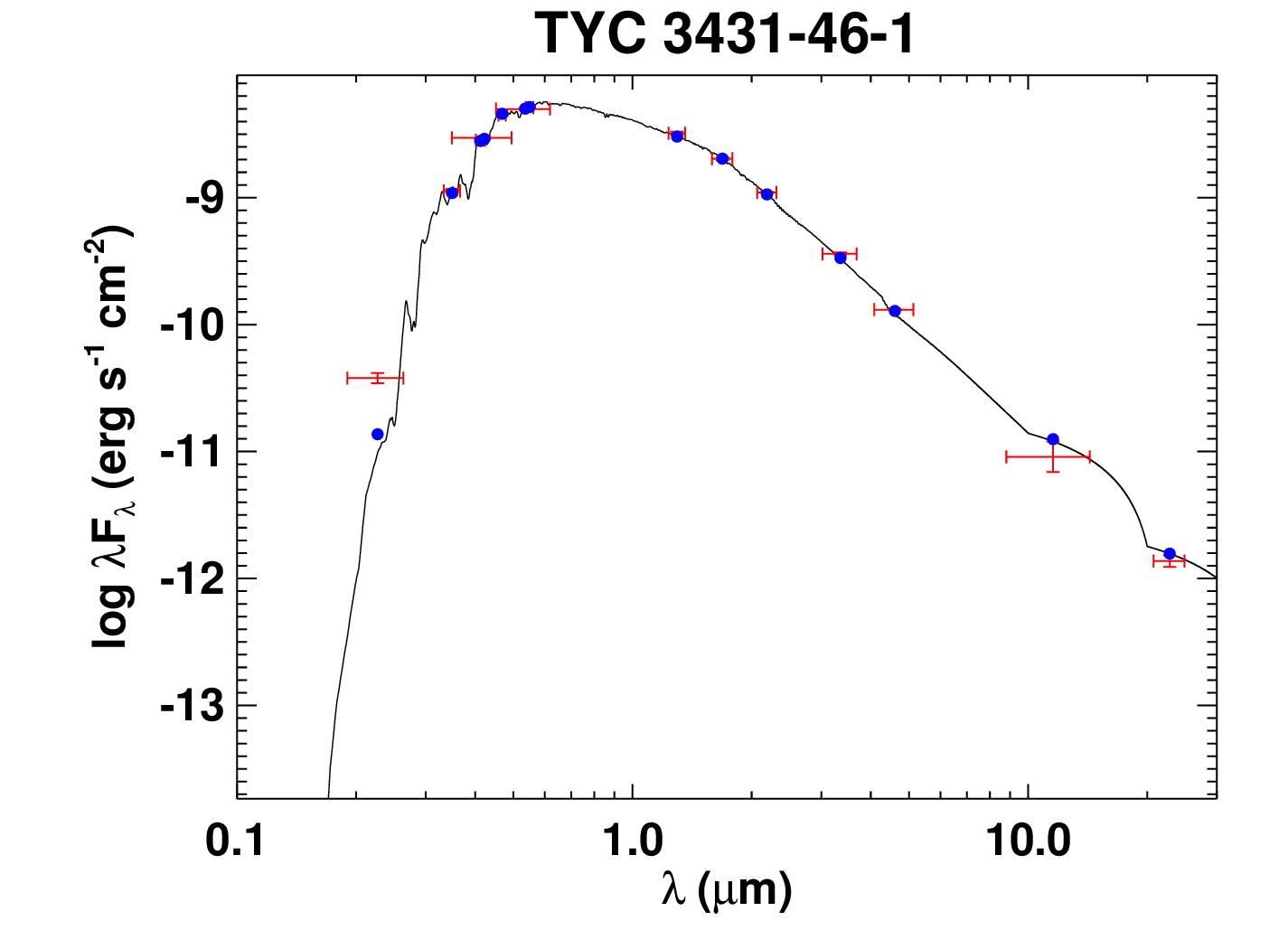}\includegraphics[width=0.333\linewidth]{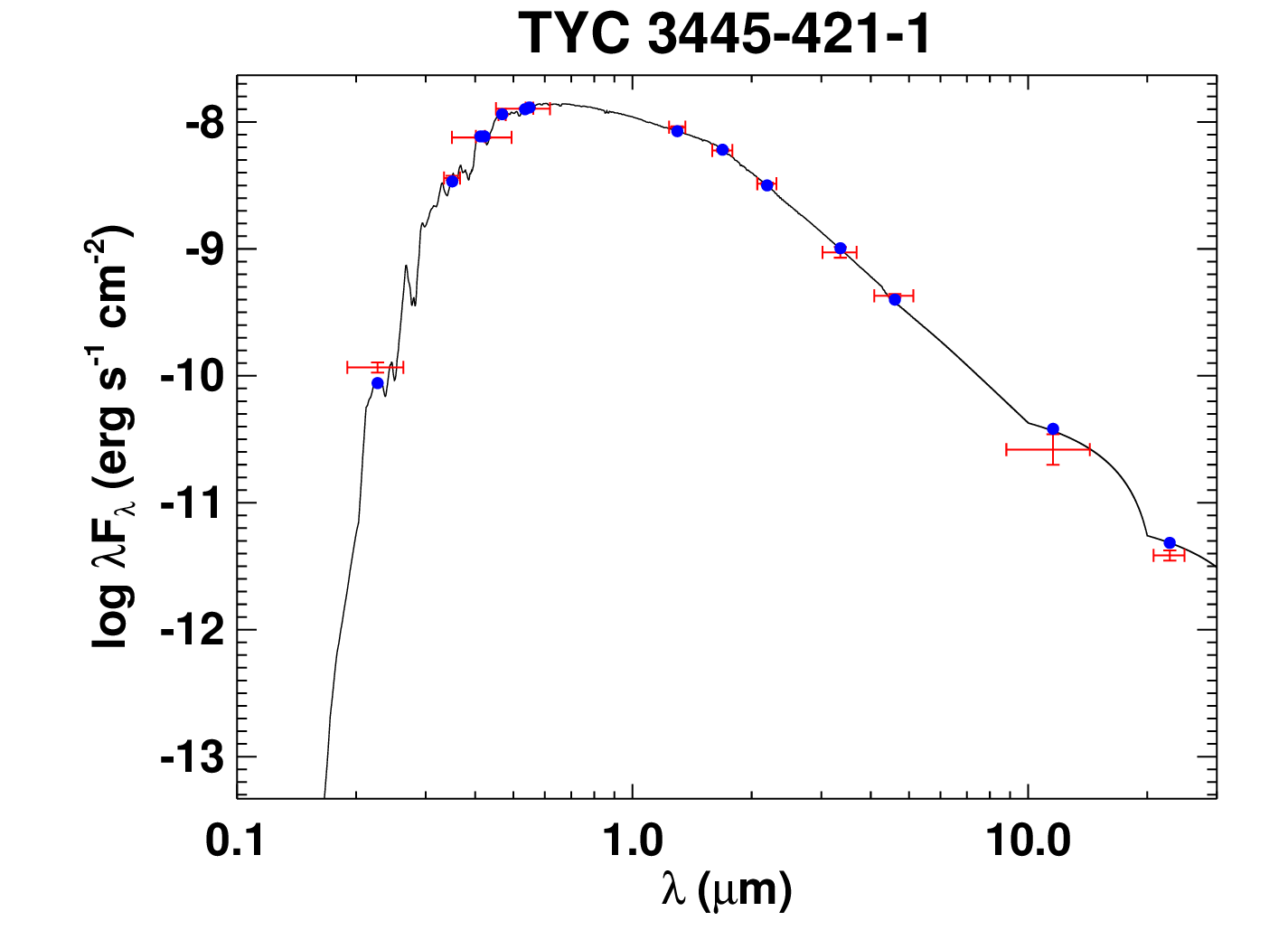}
\includegraphics[width=0.333\linewidth]{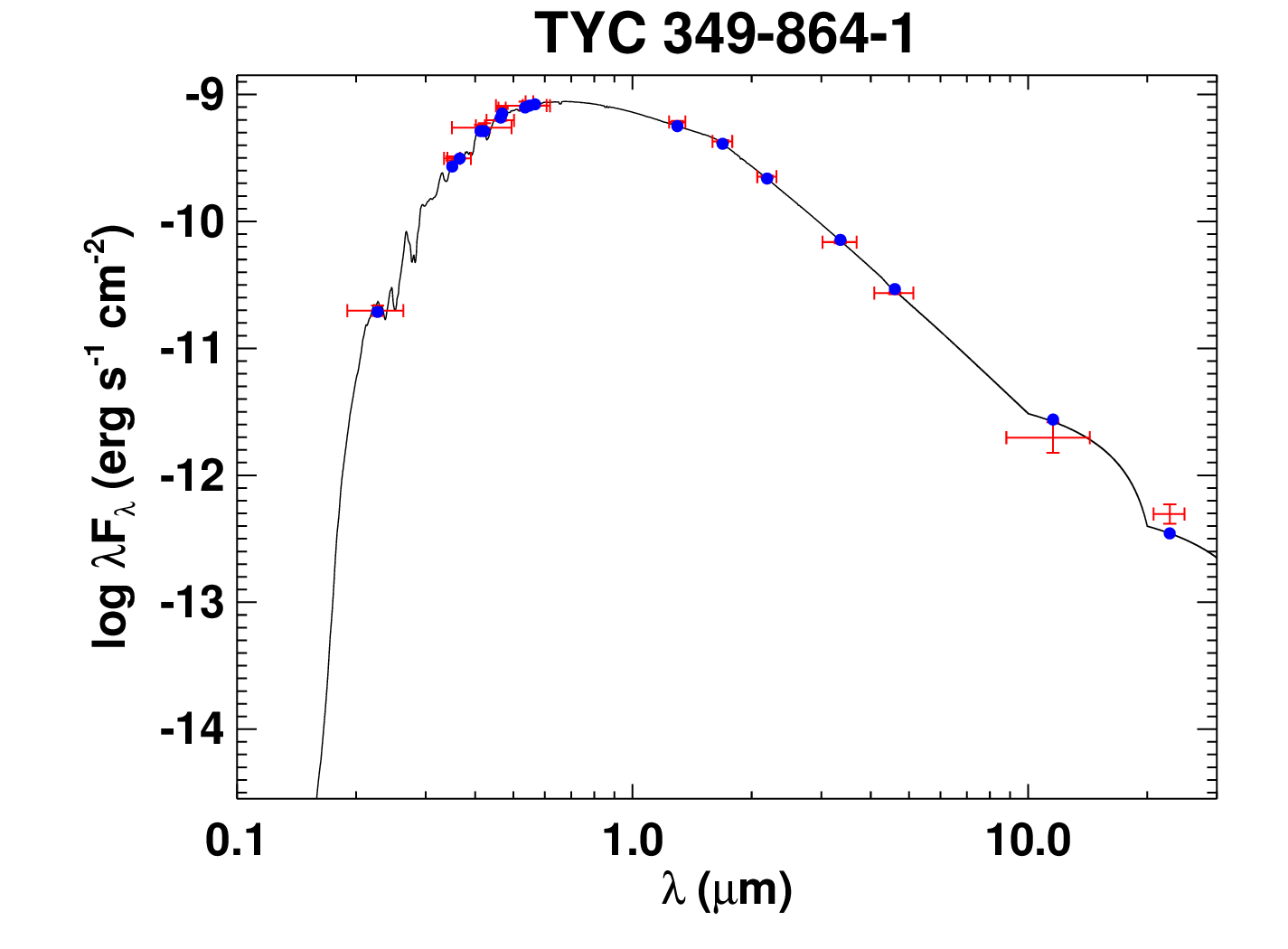}\includegraphics[width=0.333\linewidth]{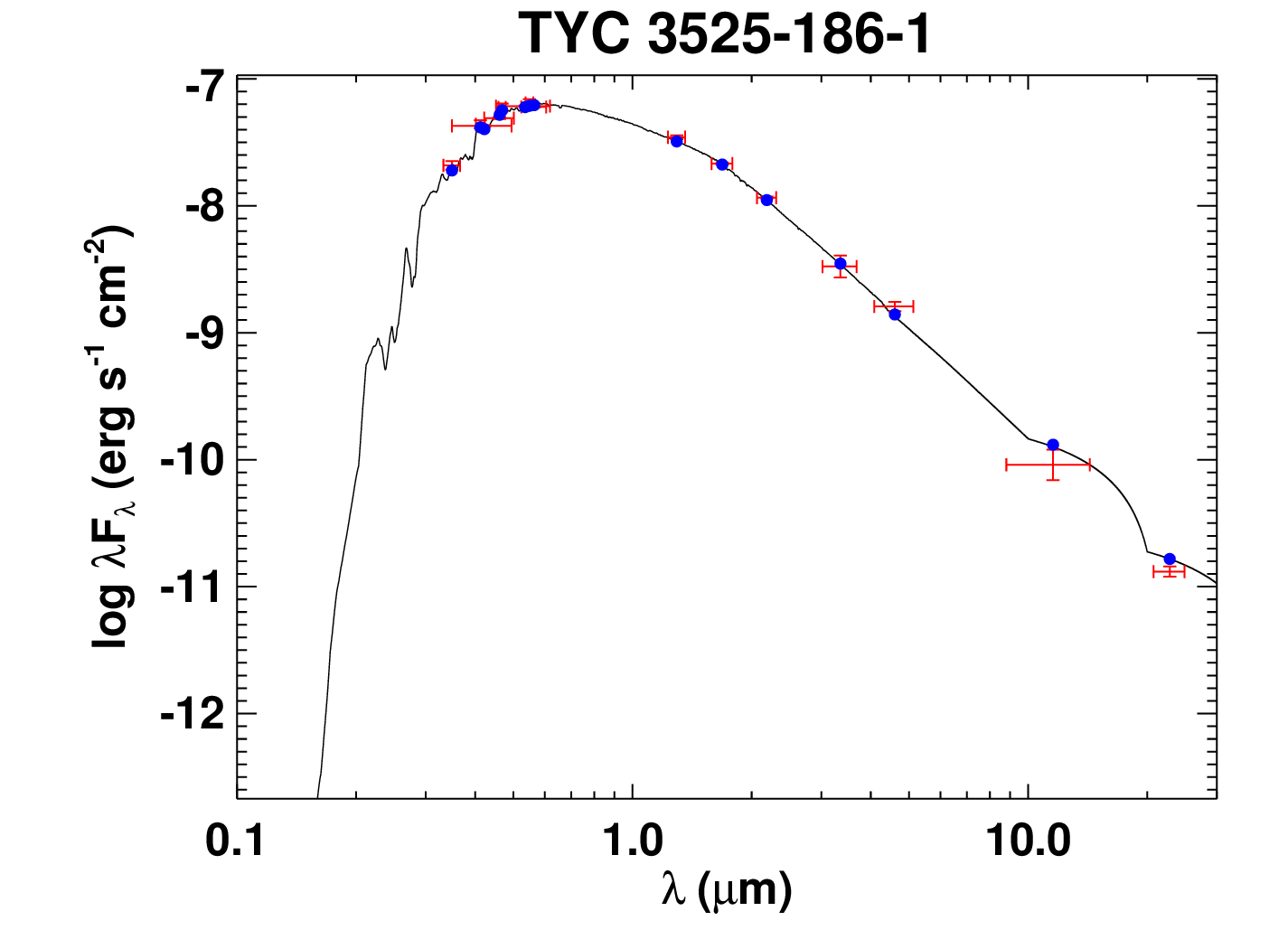}\includegraphics[width=0.333\linewidth]{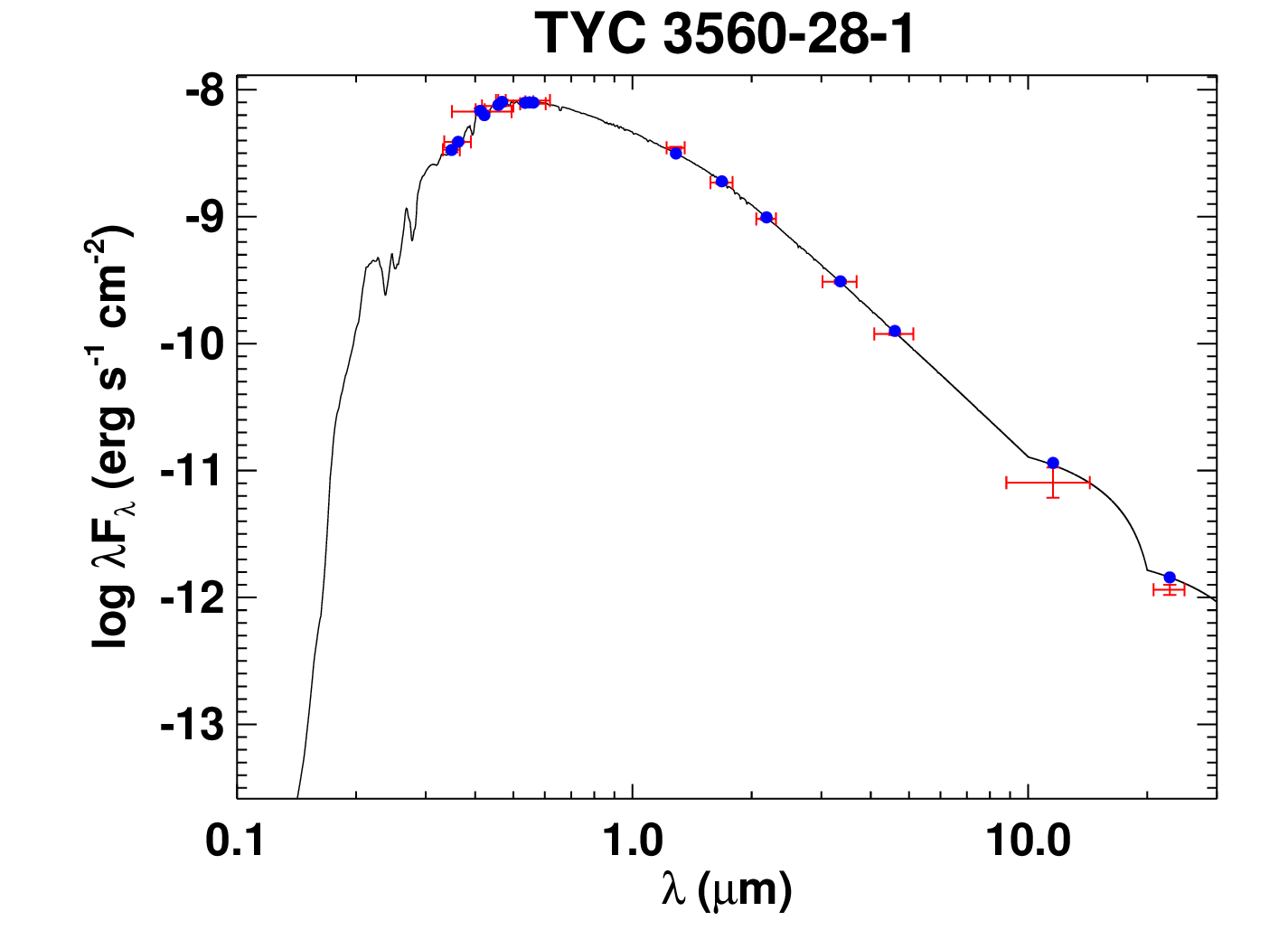}
\includegraphics[width=0.333\linewidth]{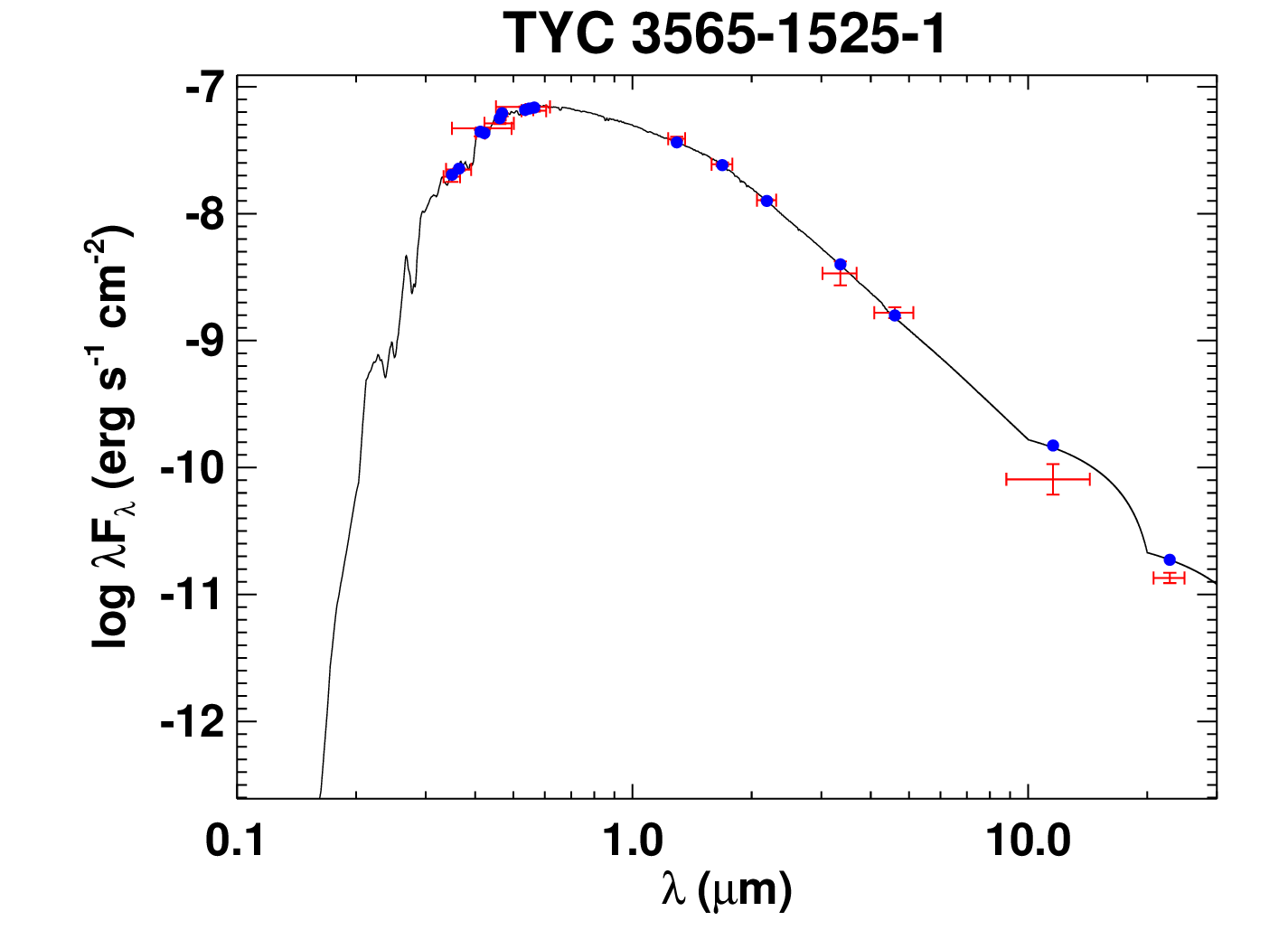}\includegraphics[width=0.333\linewidth]{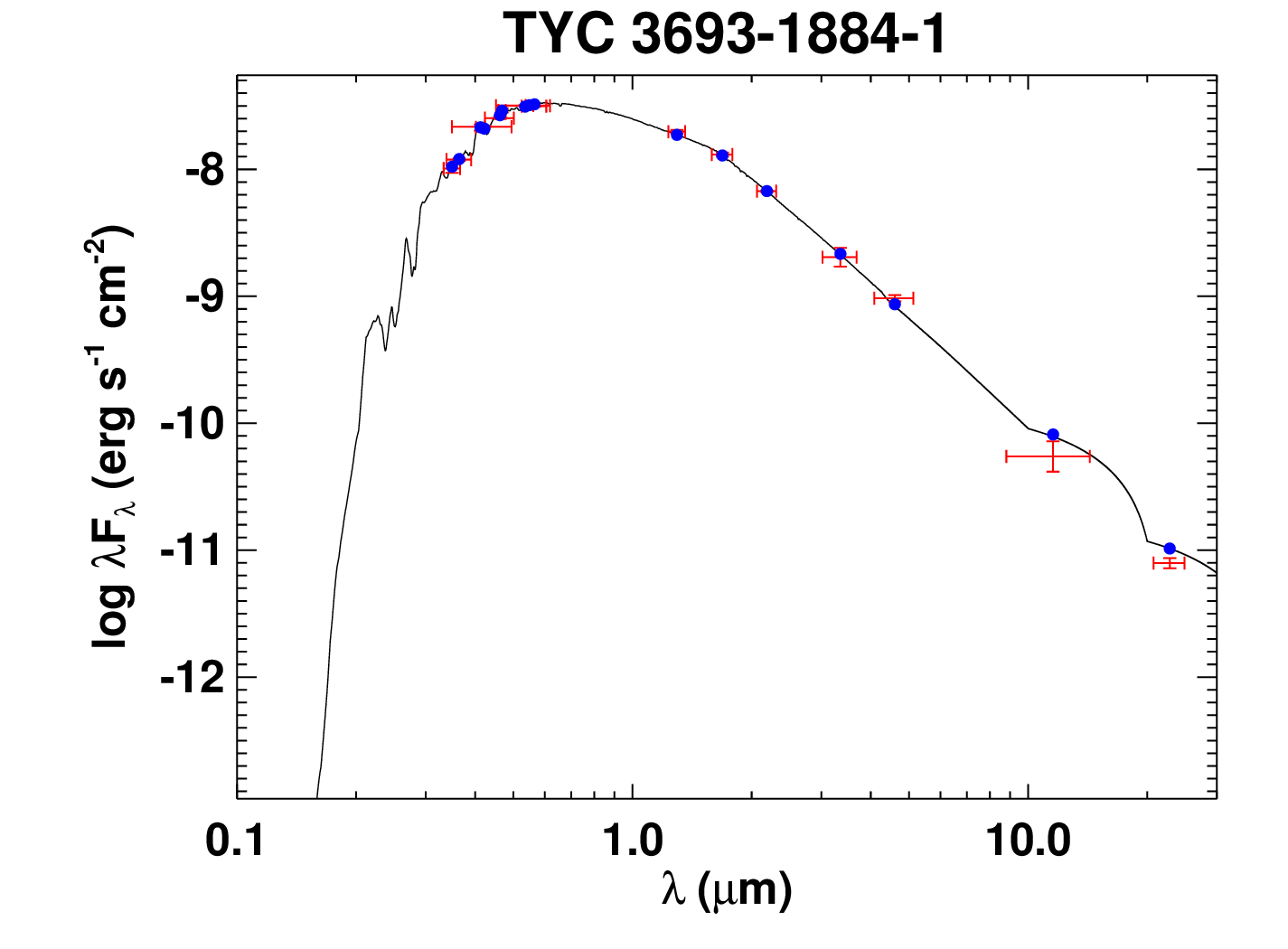}\includegraphics[width=0.333\linewidth]{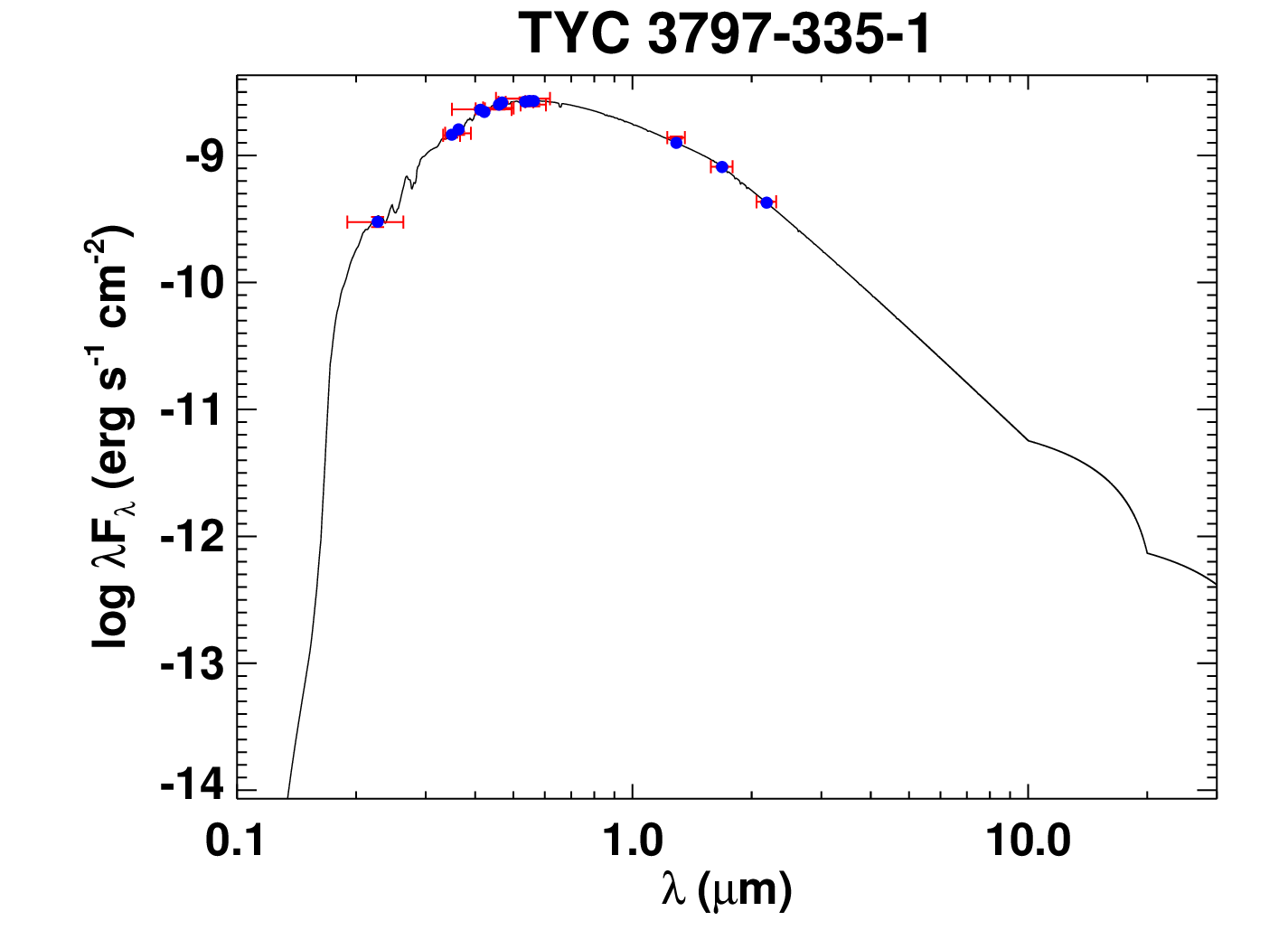}
\includegraphics[width=0.333\linewidth]{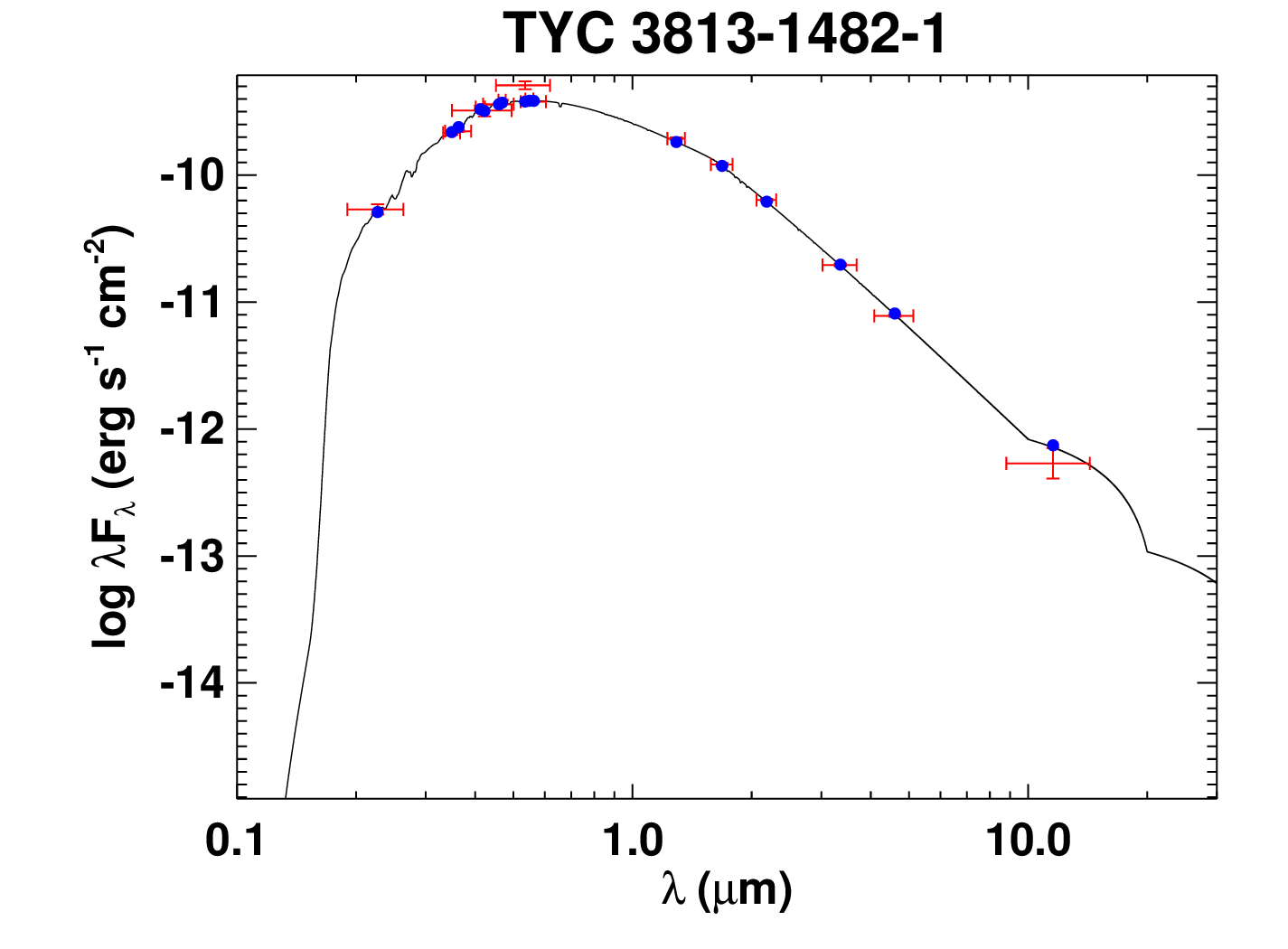}\includegraphics[width=0.333\linewidth]{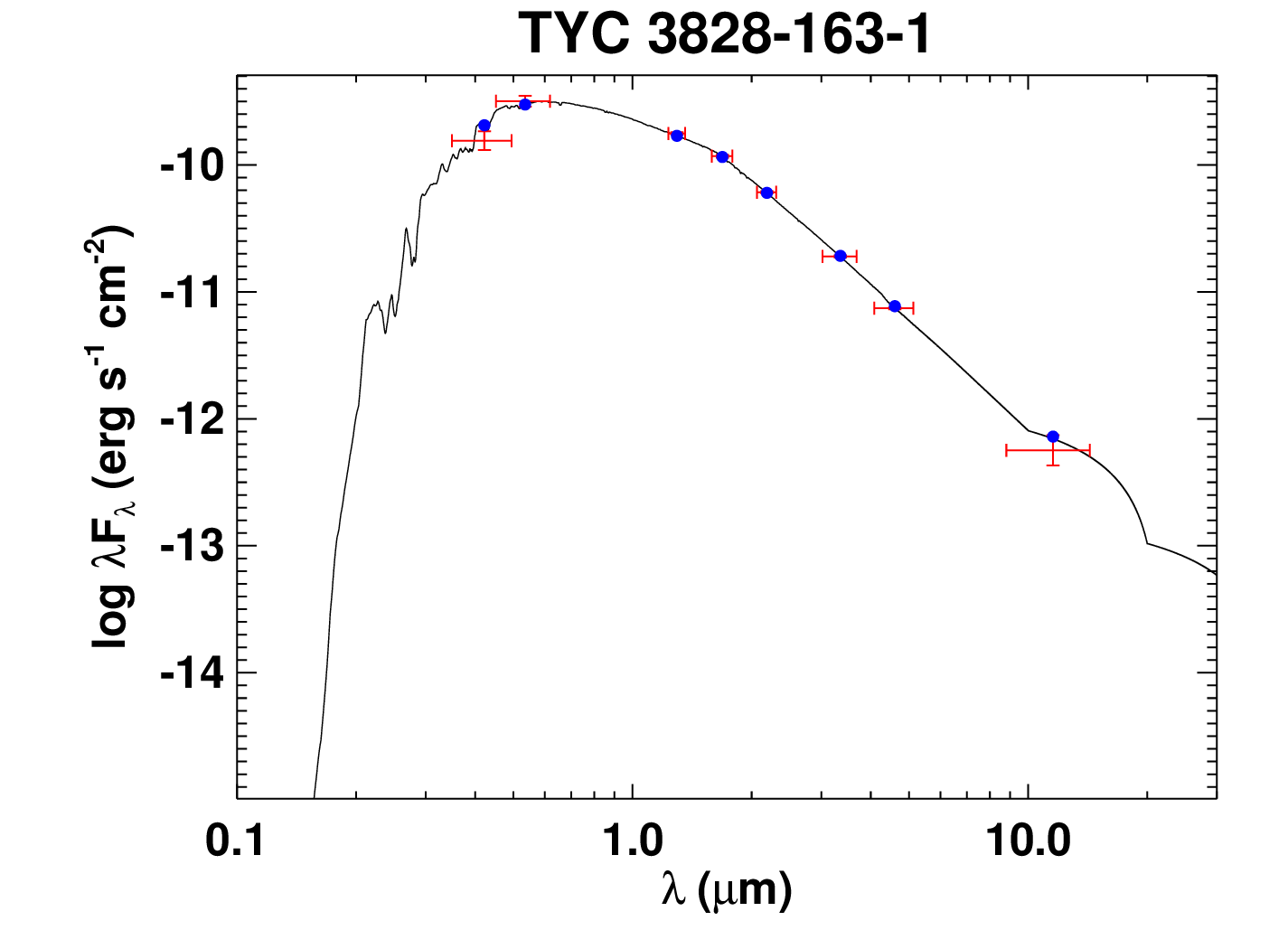}\includegraphics[width=0.333\linewidth]{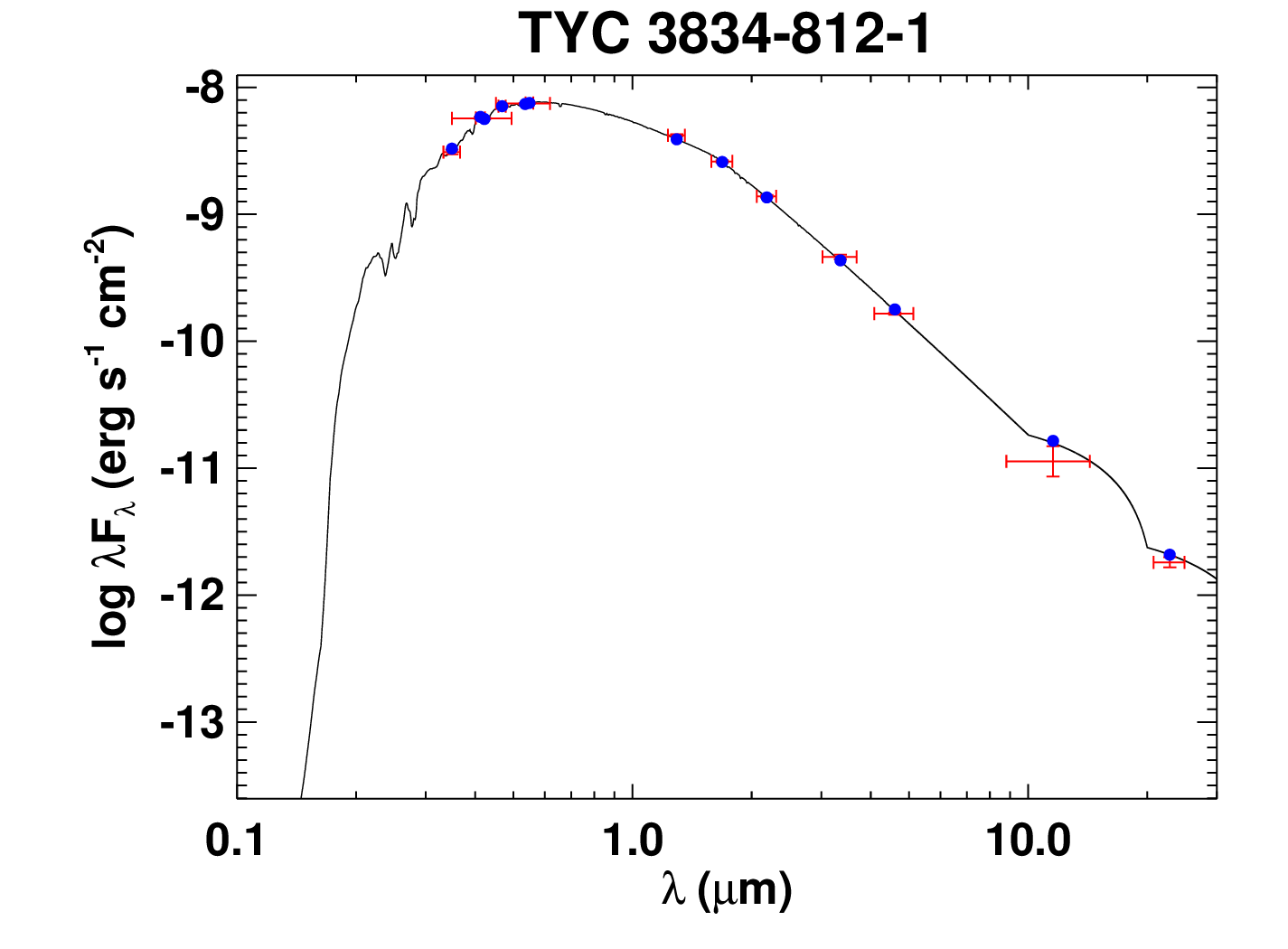}
\caption{\label{fig:seds8} All labels, lines, symbols, and colors as in Figure \ref{fig:seds}.}
\end{figure*}

\begin{figure*}
\includegraphics[width=0.333\linewidth]{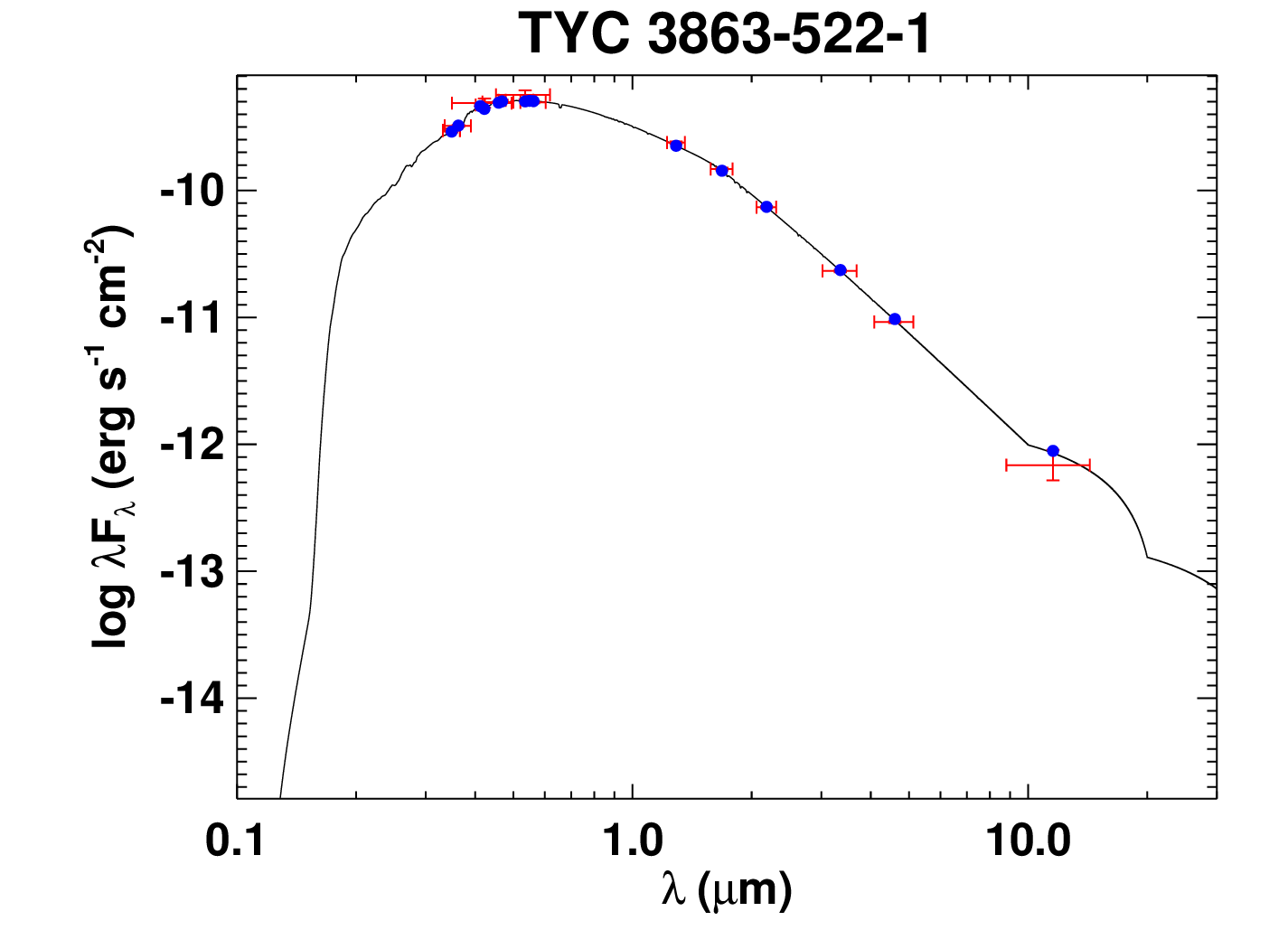}\includegraphics[width=0.333\linewidth]{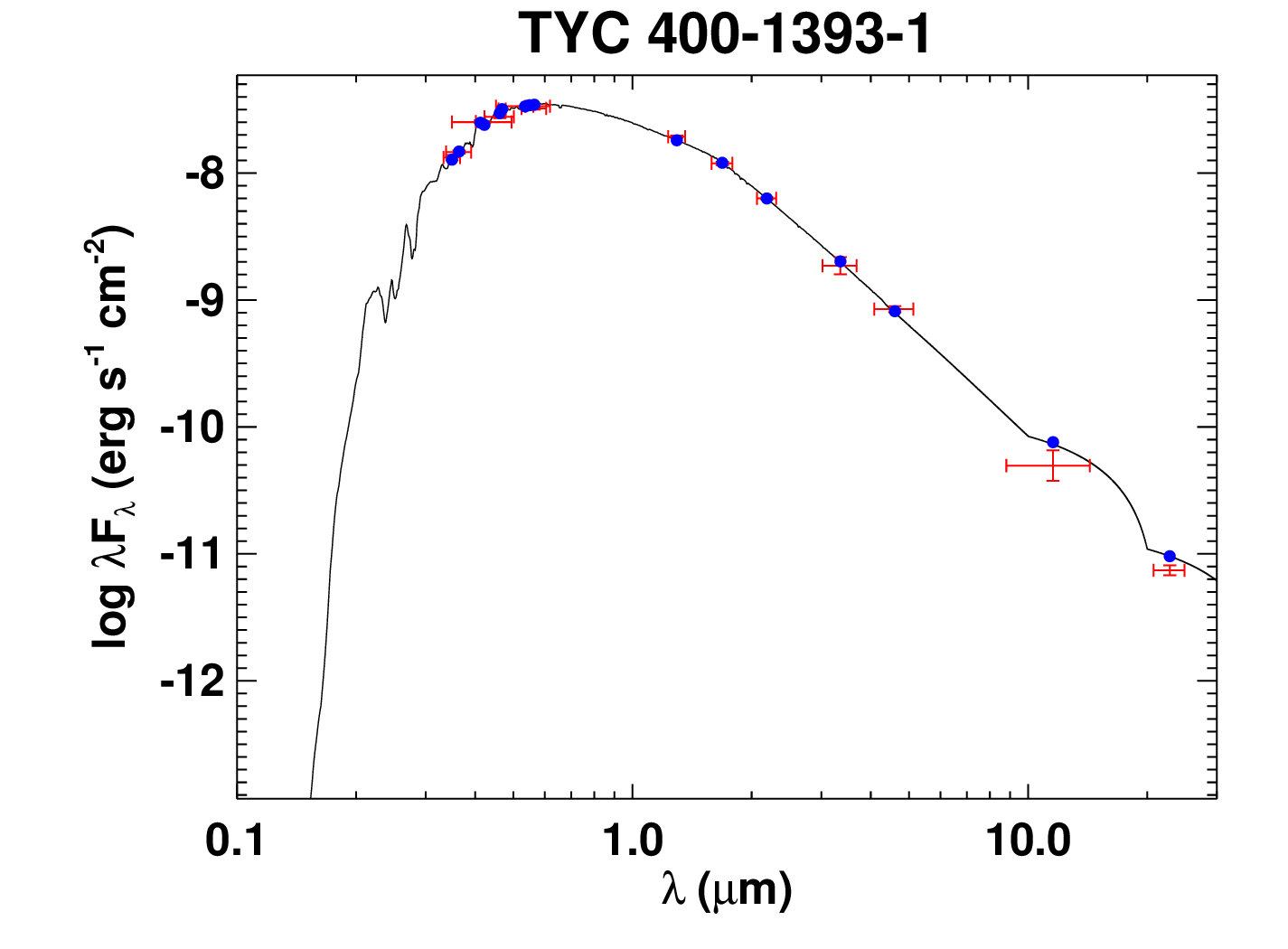}\includegraphics[width=0.333\linewidth]{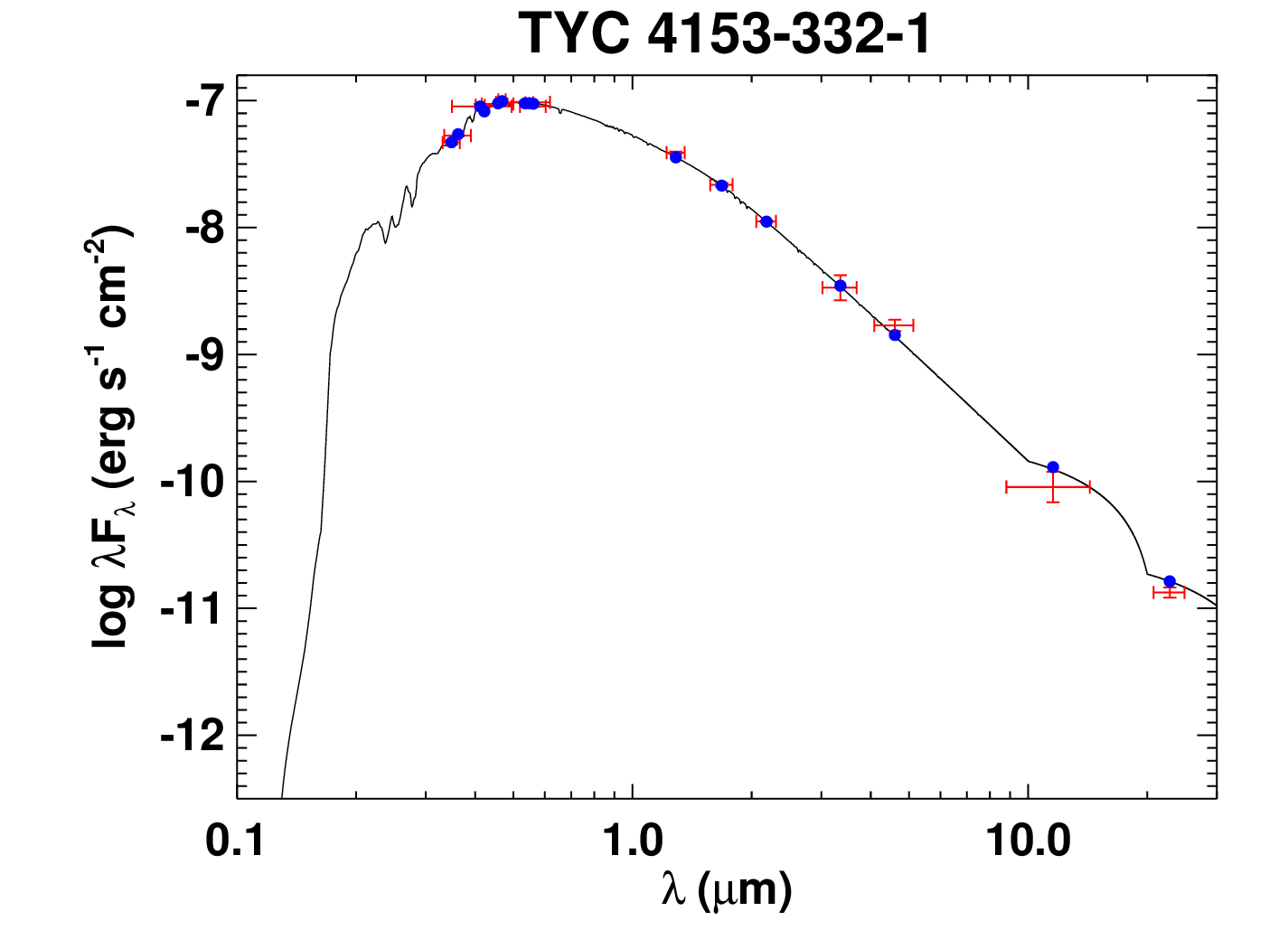}
\includegraphics[width=0.333\linewidth]{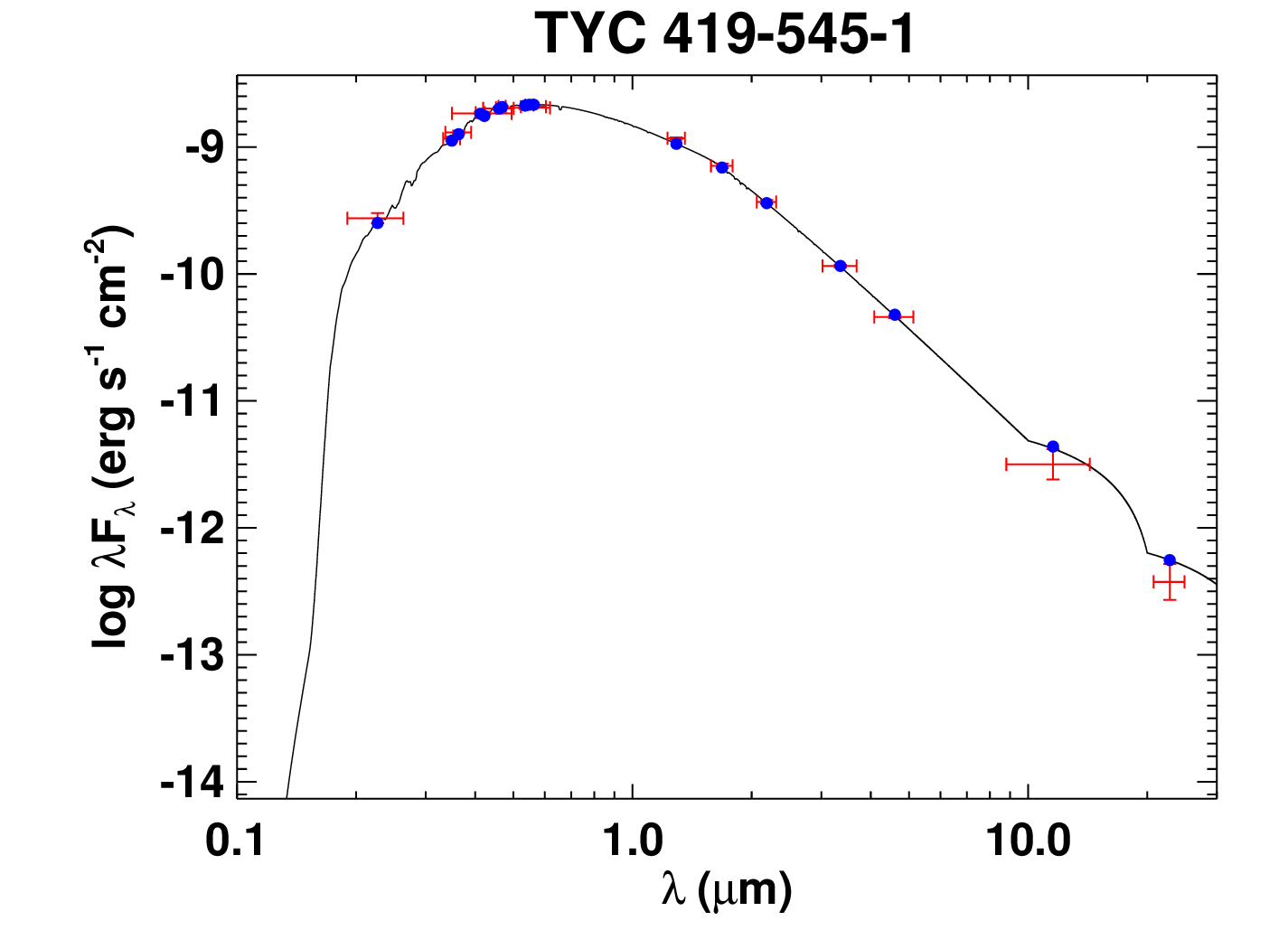}\includegraphics[width=0.333\linewidth]{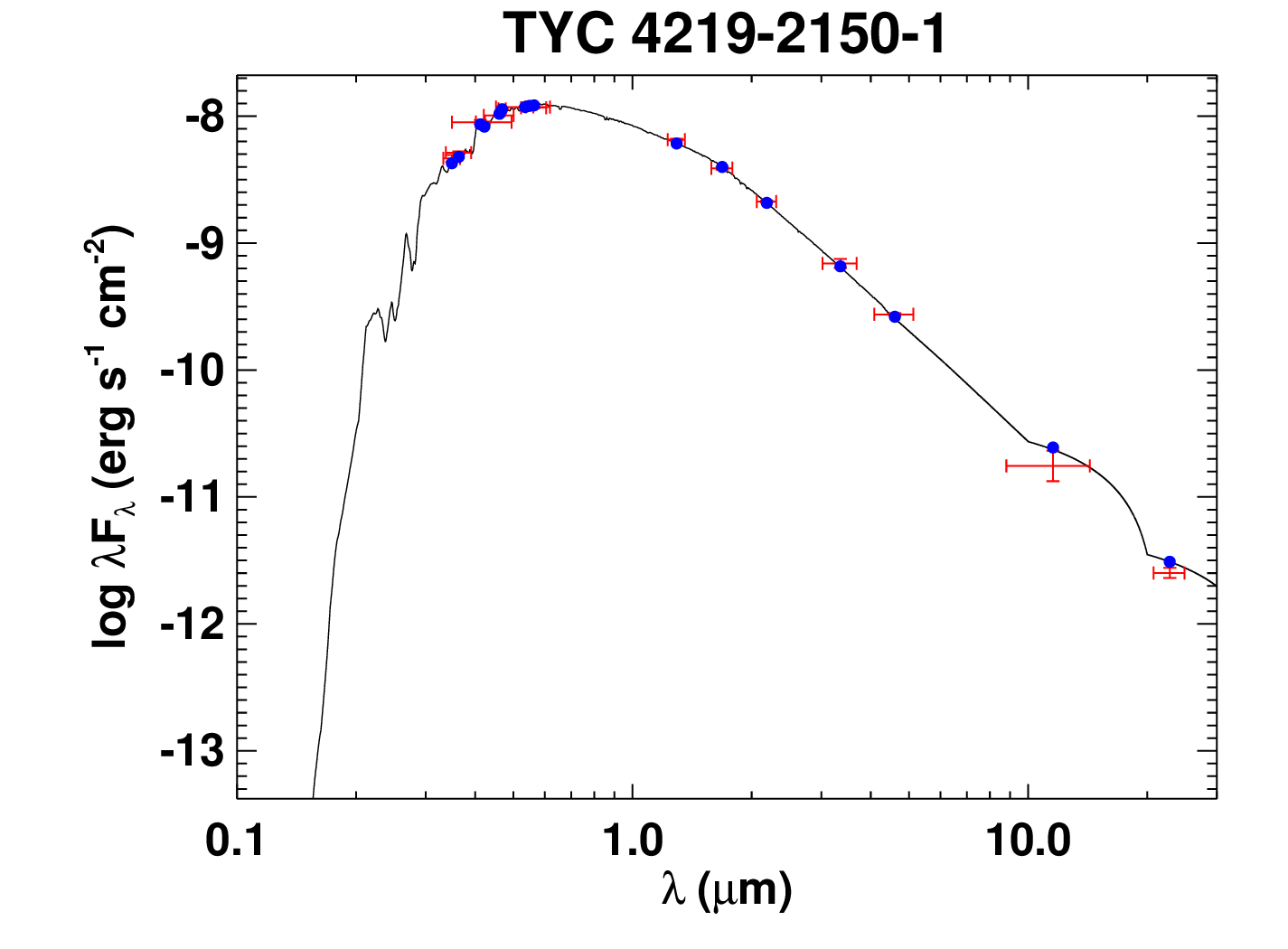}\includegraphics[width=0.333\linewidth]{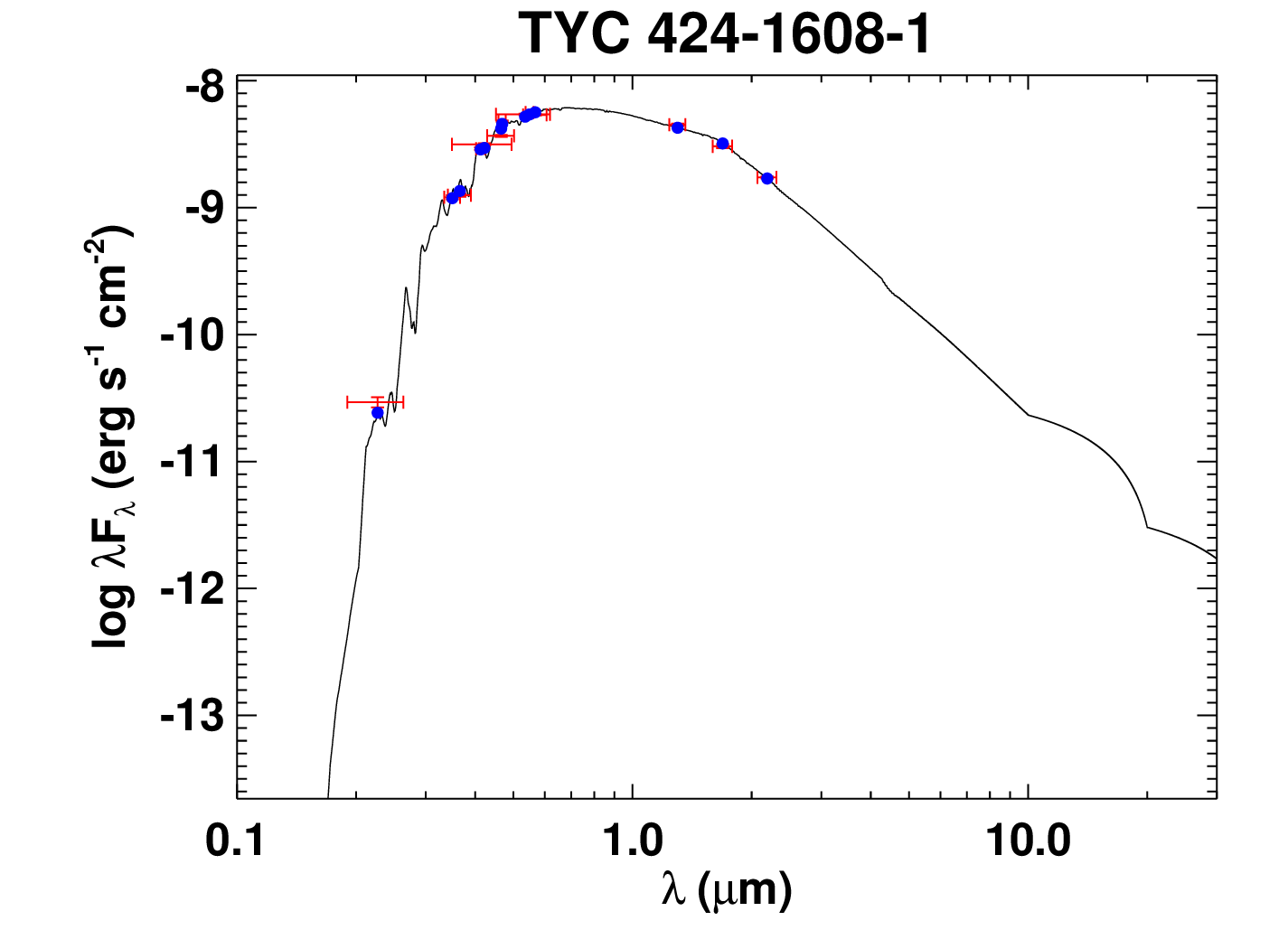}
\includegraphics[width=0.333\linewidth]{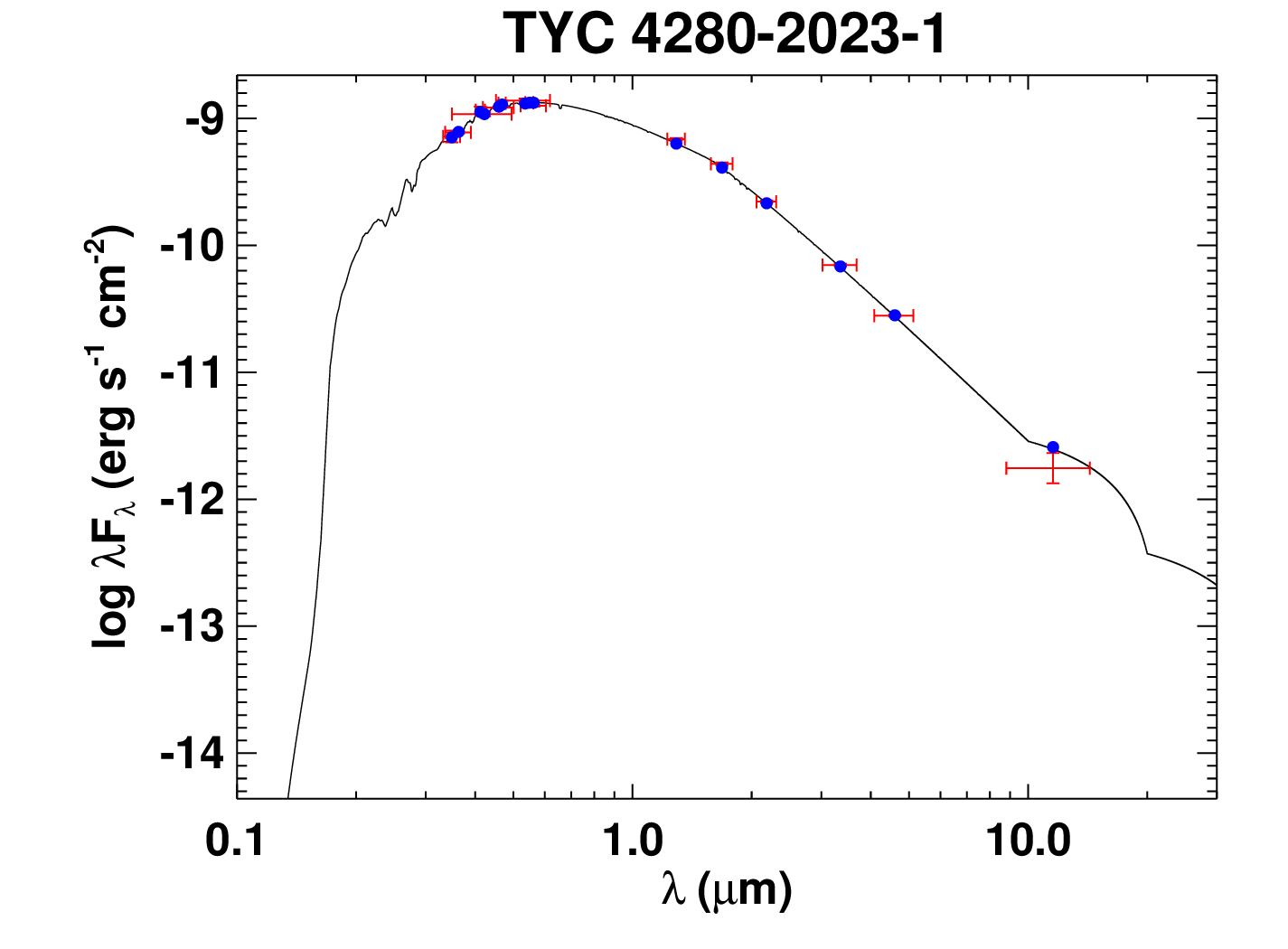}\includegraphics[width=0.333\linewidth]{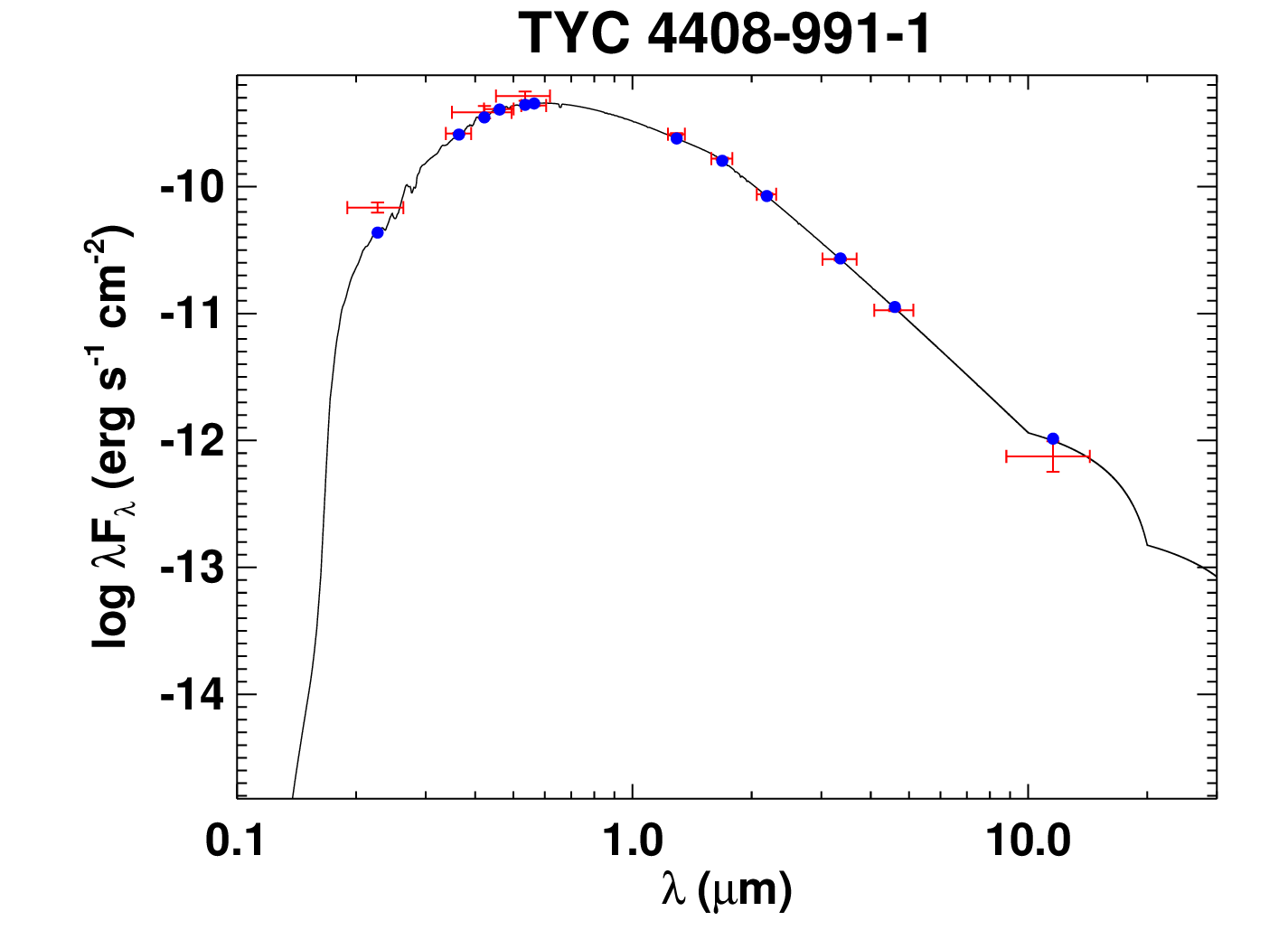}\includegraphics[width=0.333\linewidth]{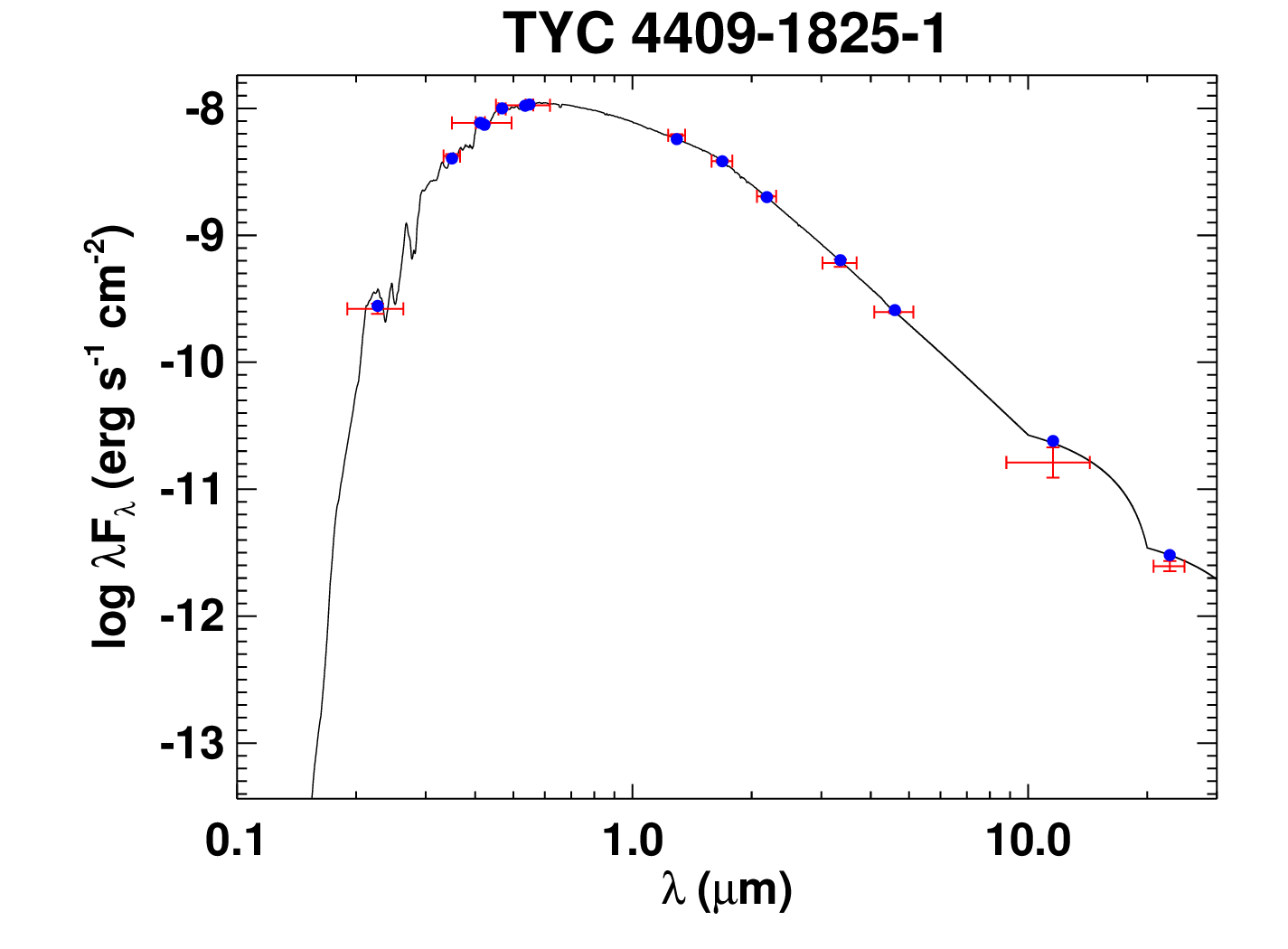}
\includegraphics[width=0.333\linewidth]{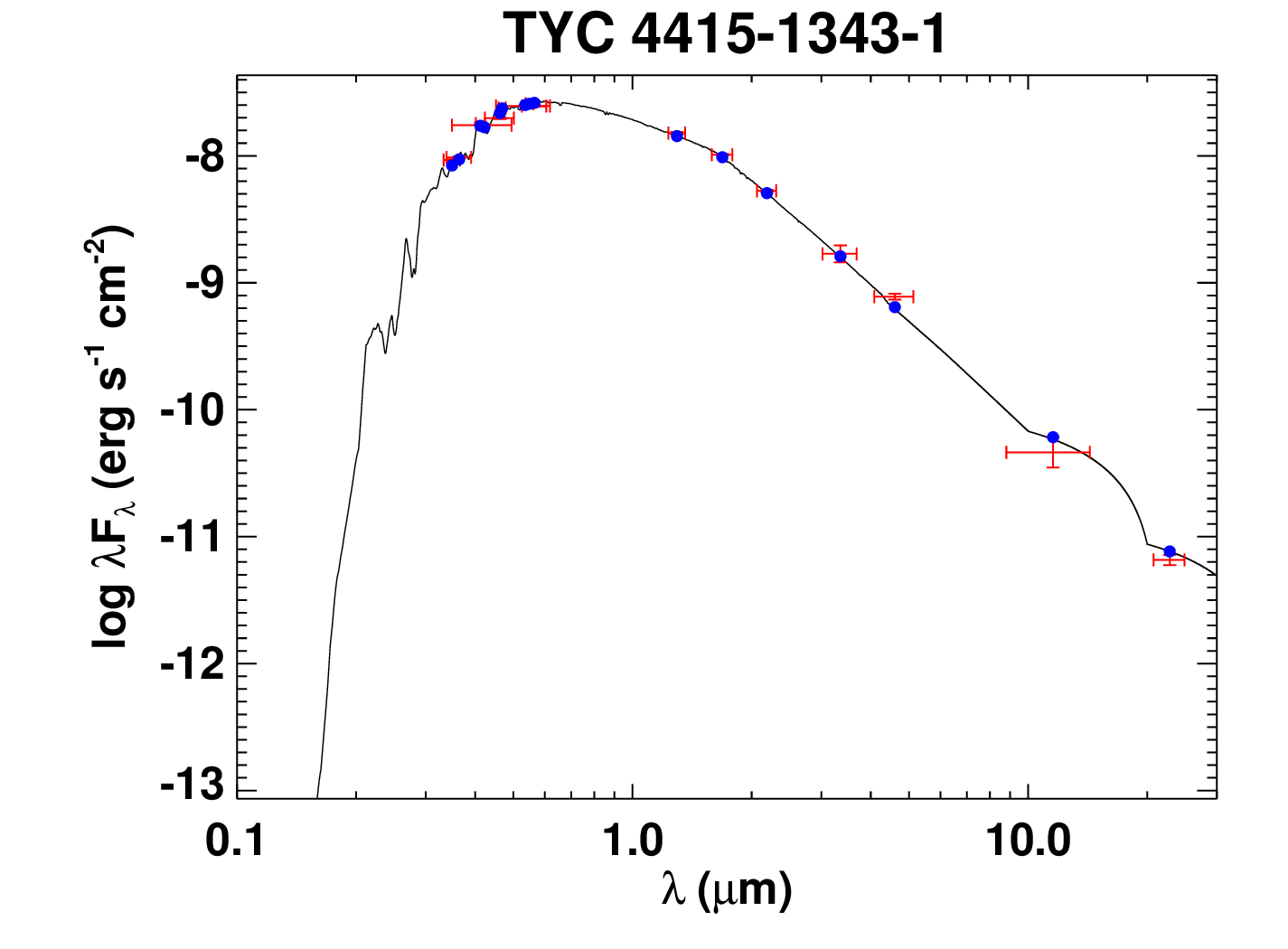}\includegraphics[width=0.333\linewidth]{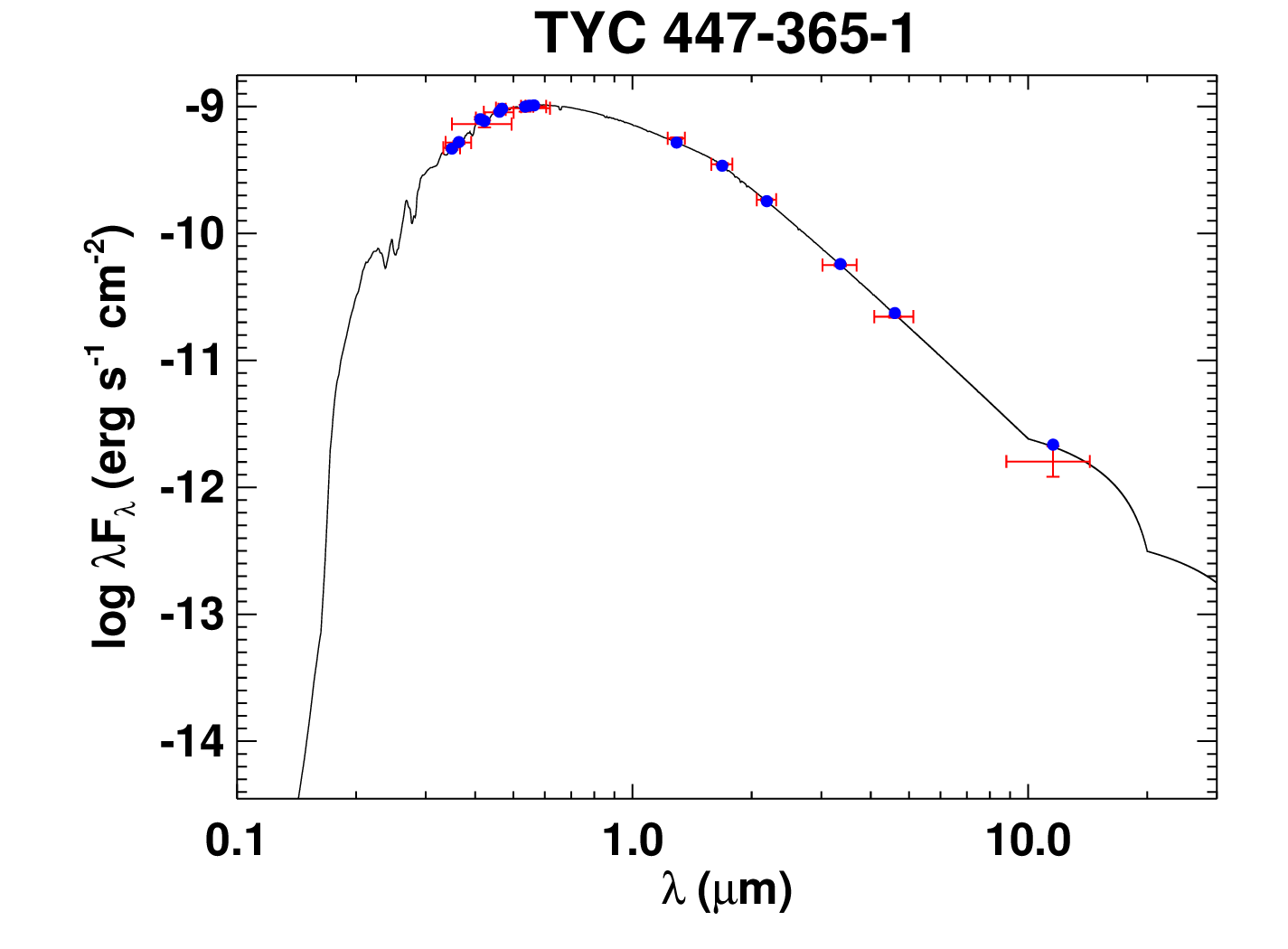}\includegraphics[width=0.333\linewidth]{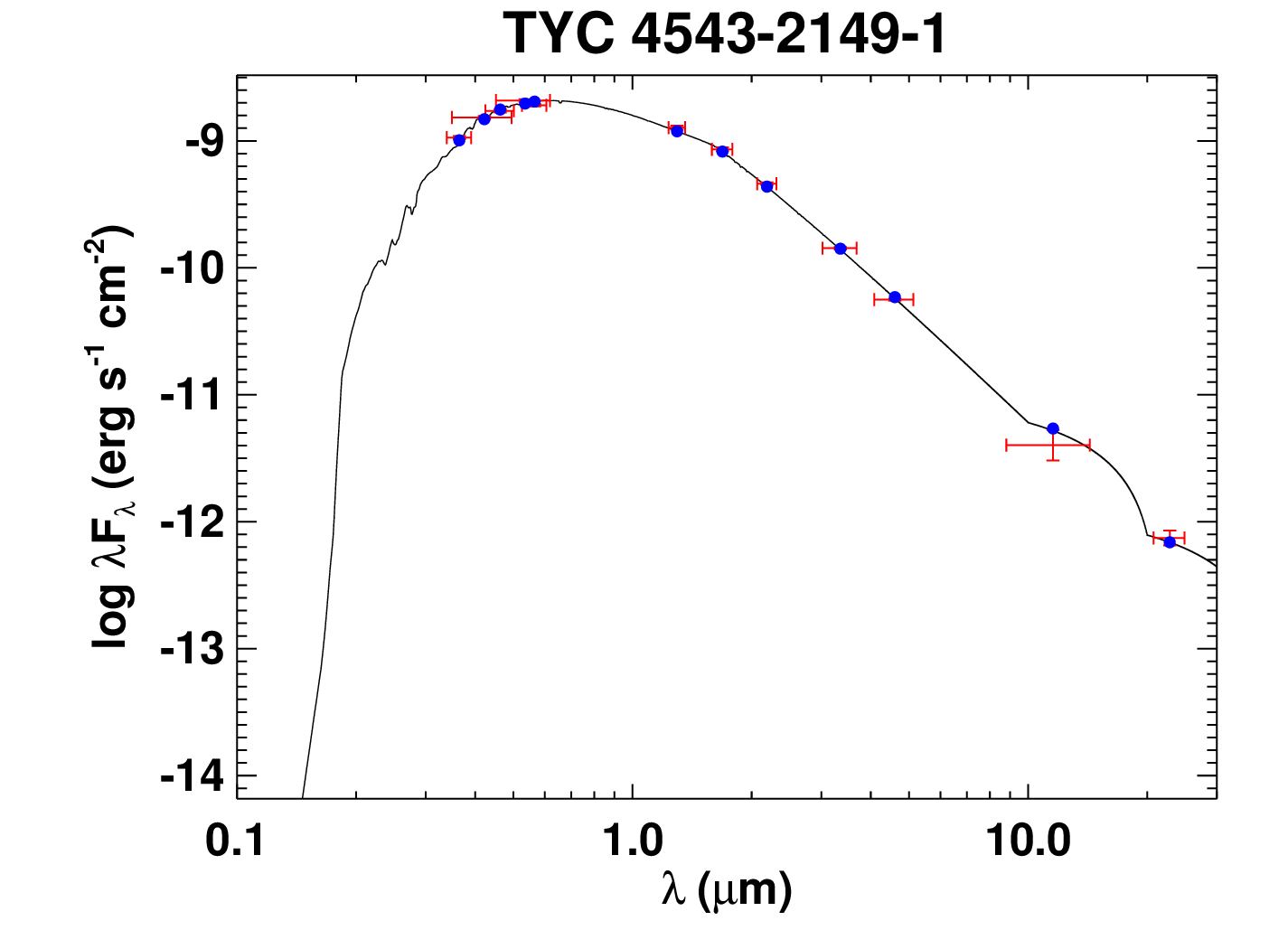}
\caption{\label{fig:seds9} All labels, lines, symbols, and colors as in Figure \ref{fig:seds}.}
\end{figure*}

\begin{figure*}
\includegraphics[width=0.333\linewidth]{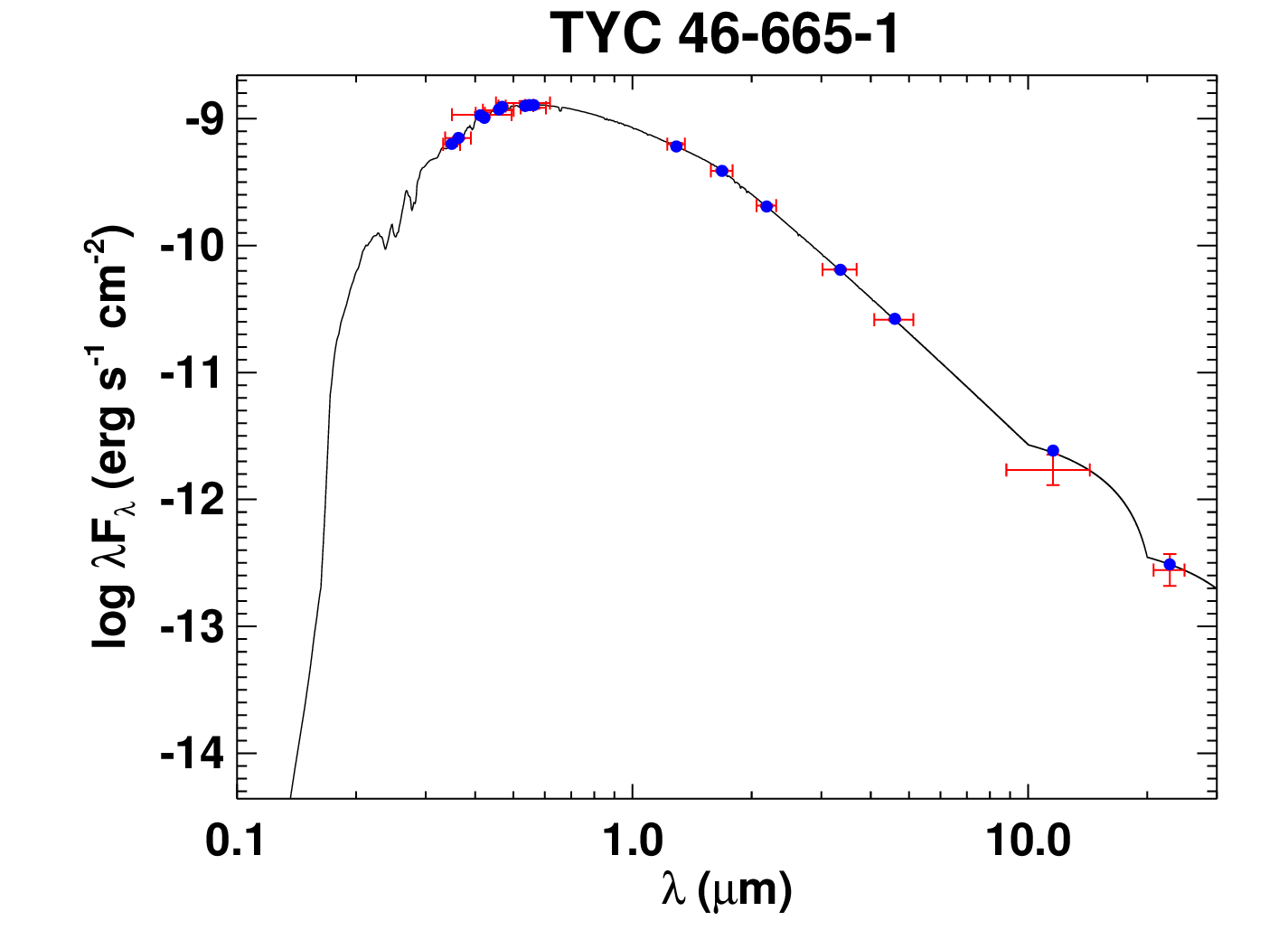}\includegraphics[width=0.333\linewidth]{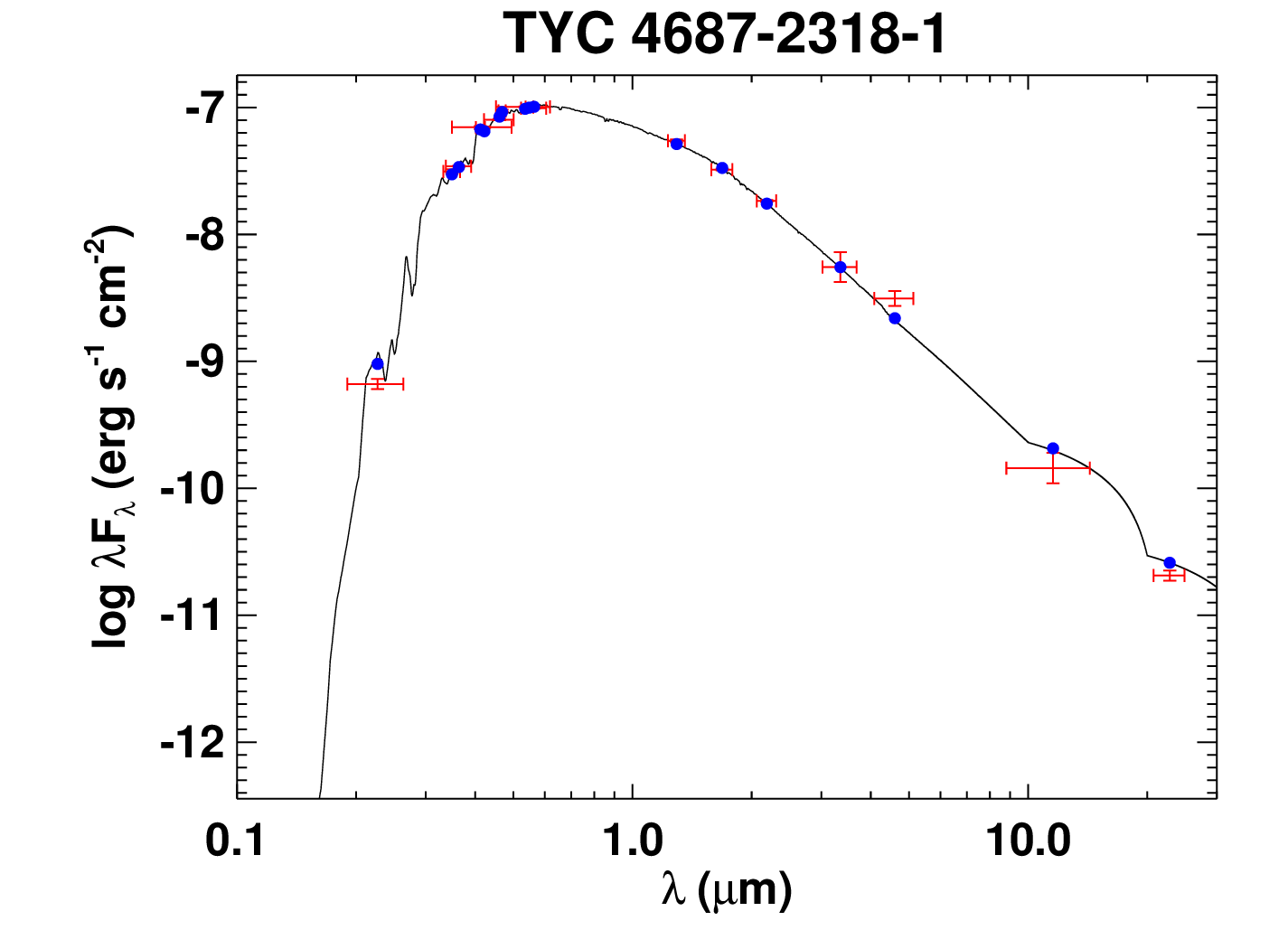}\includegraphics[width=0.333\linewidth]{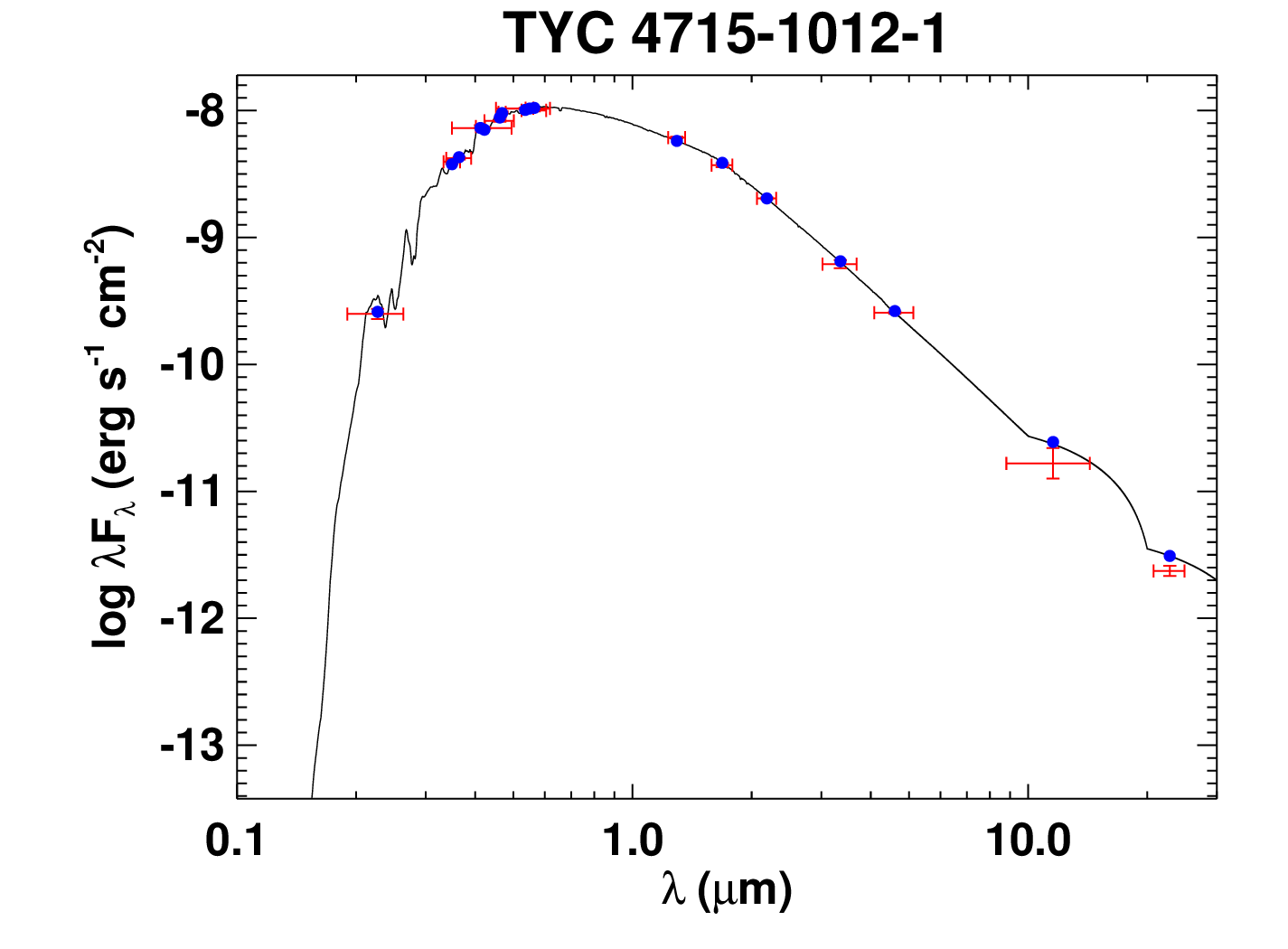}
\includegraphics[width=0.333\linewidth]{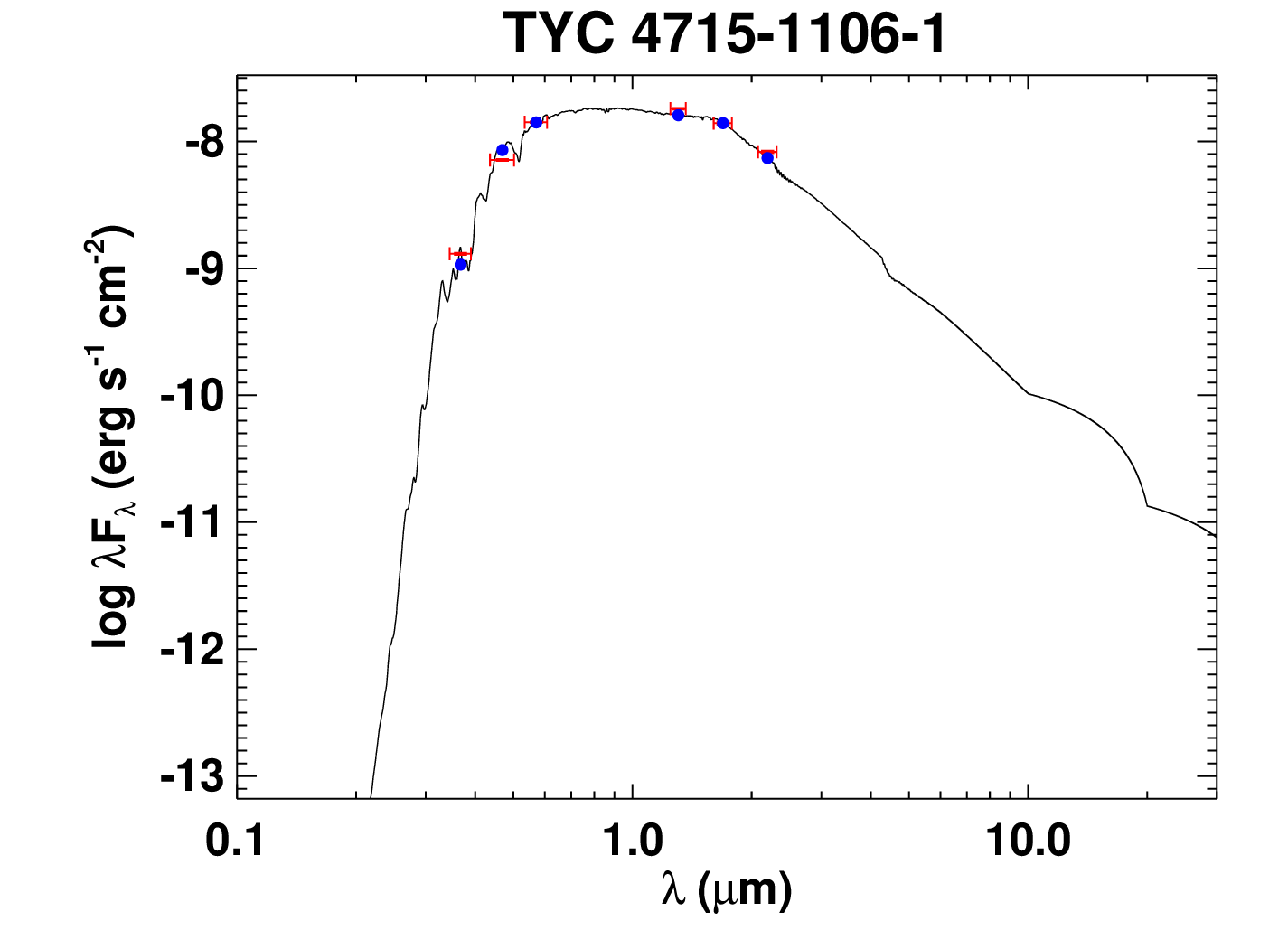}\includegraphics[width=0.333\linewidth]{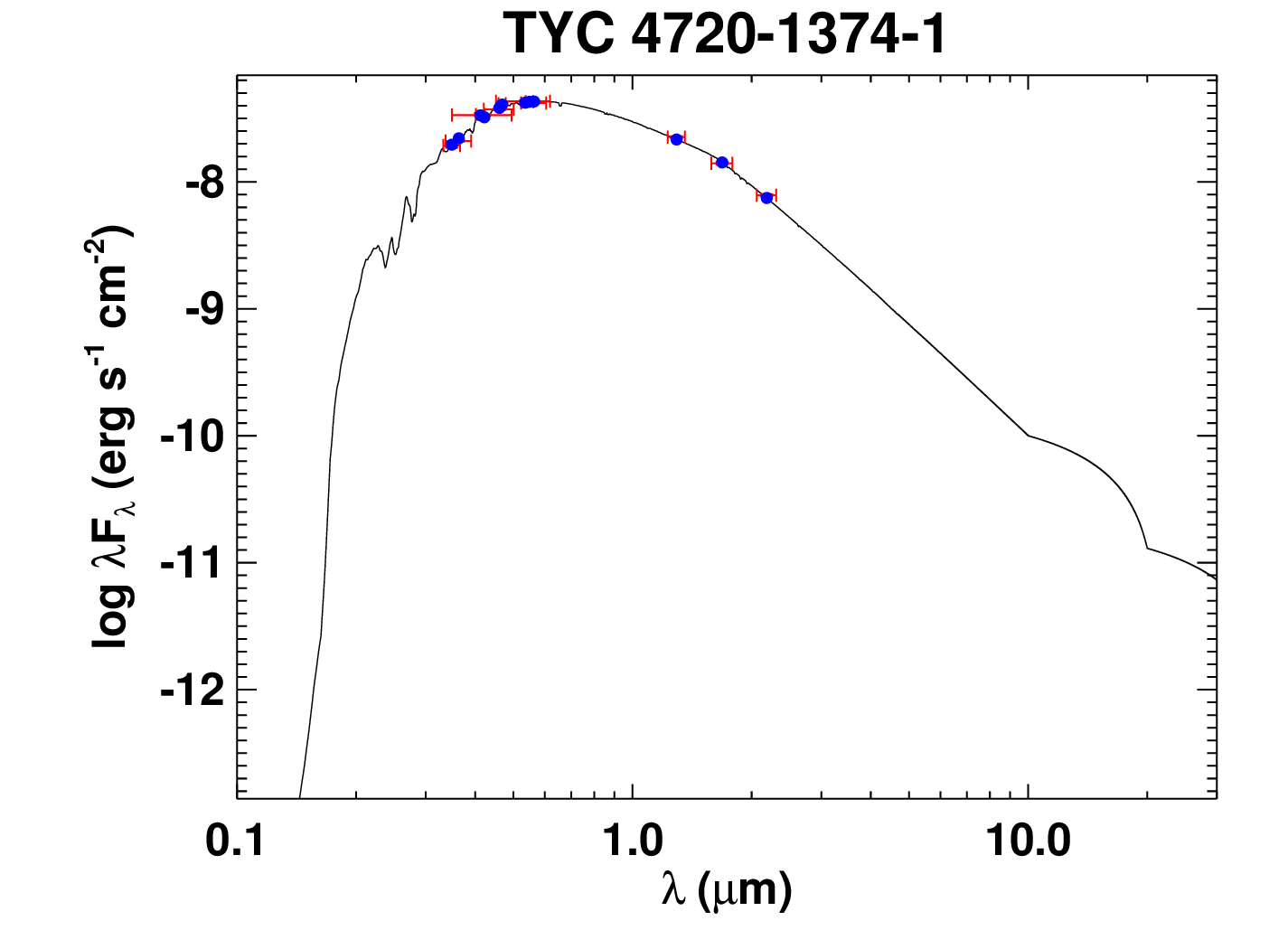}\includegraphics[width=0.333\linewidth]{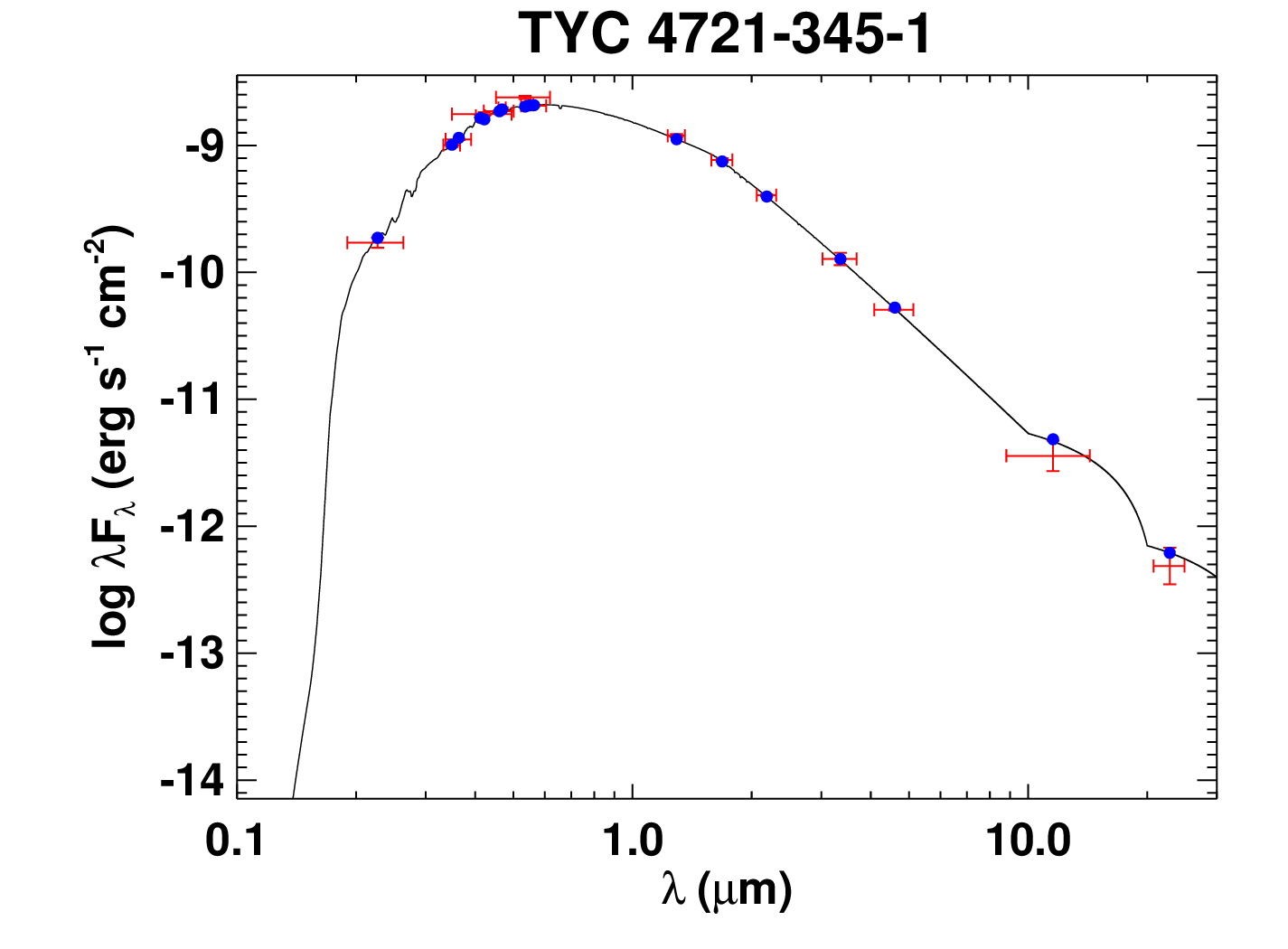}
\includegraphics[width=0.333\linewidth]{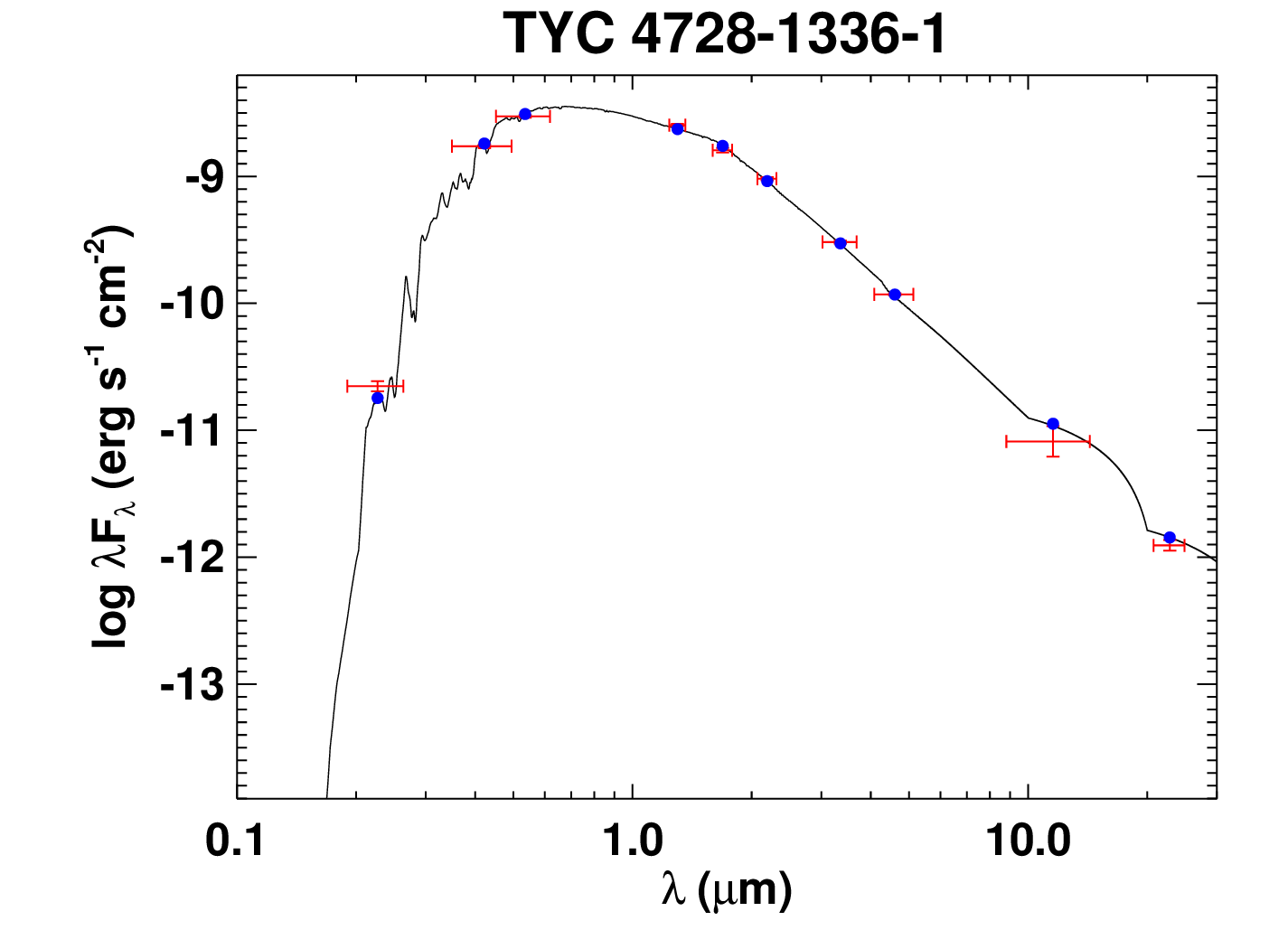}\includegraphics[width=0.333\linewidth]{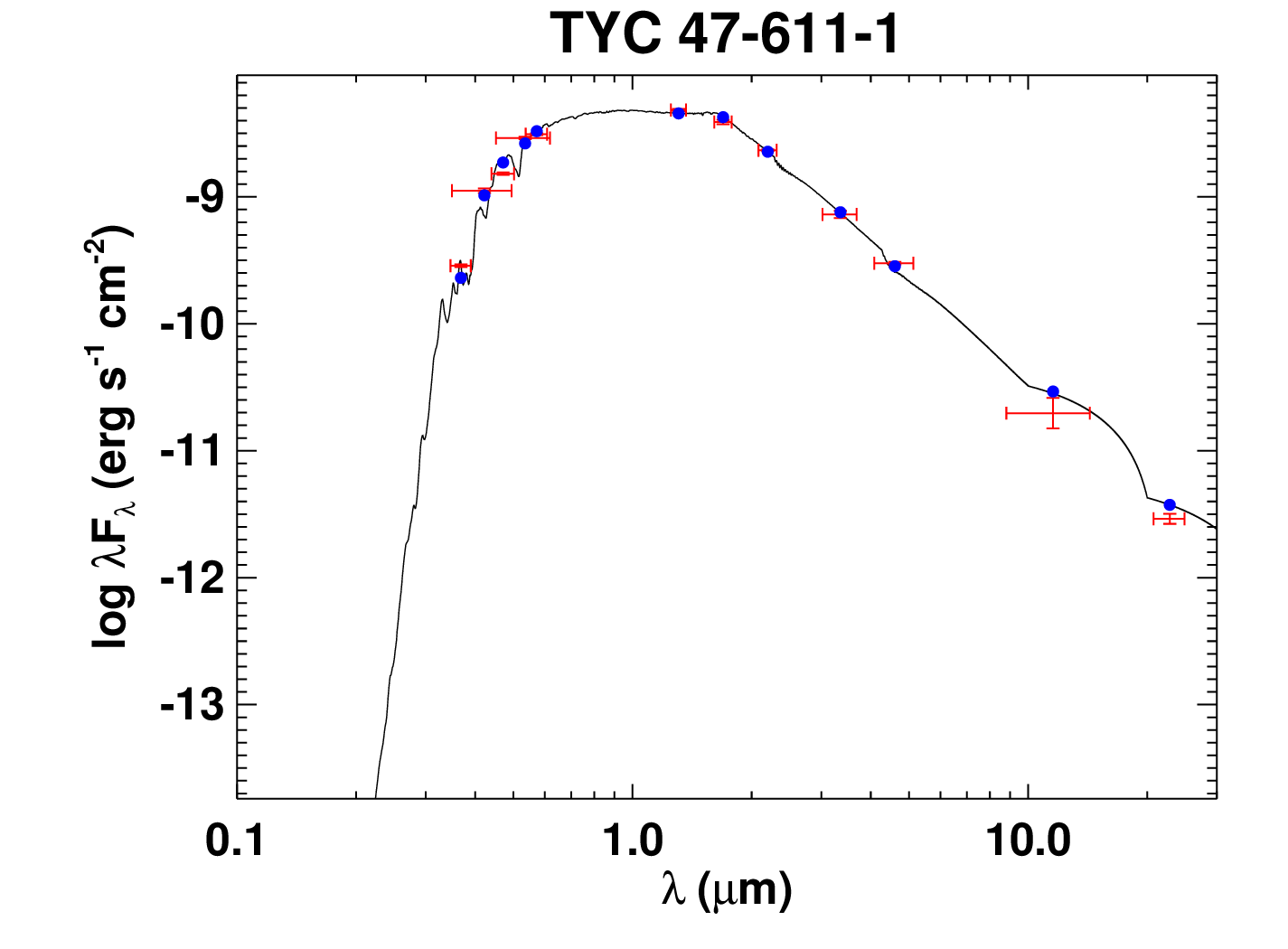}\includegraphics[width=0.333\linewidth]{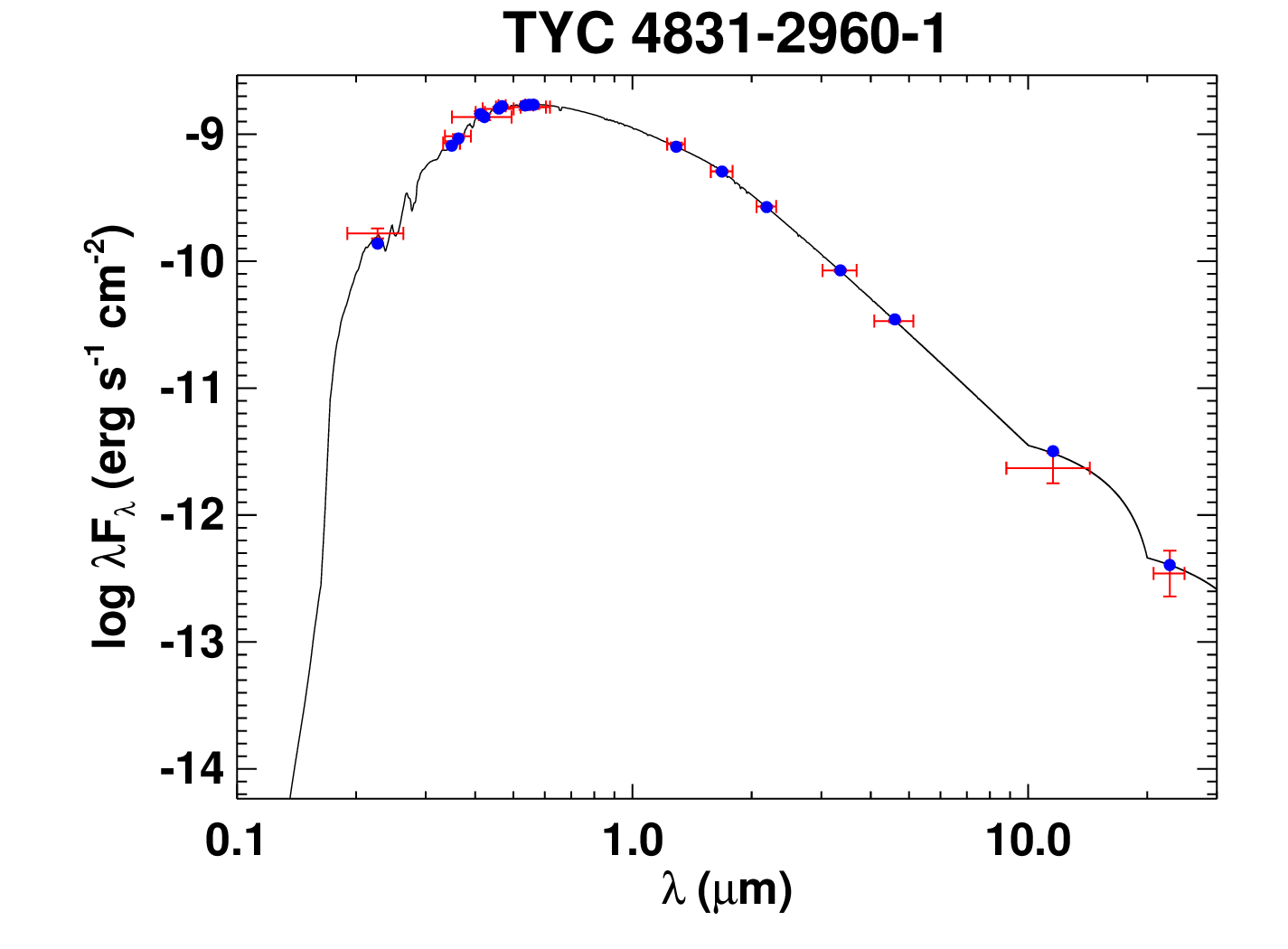}
\includegraphics[width=0.333\linewidth]{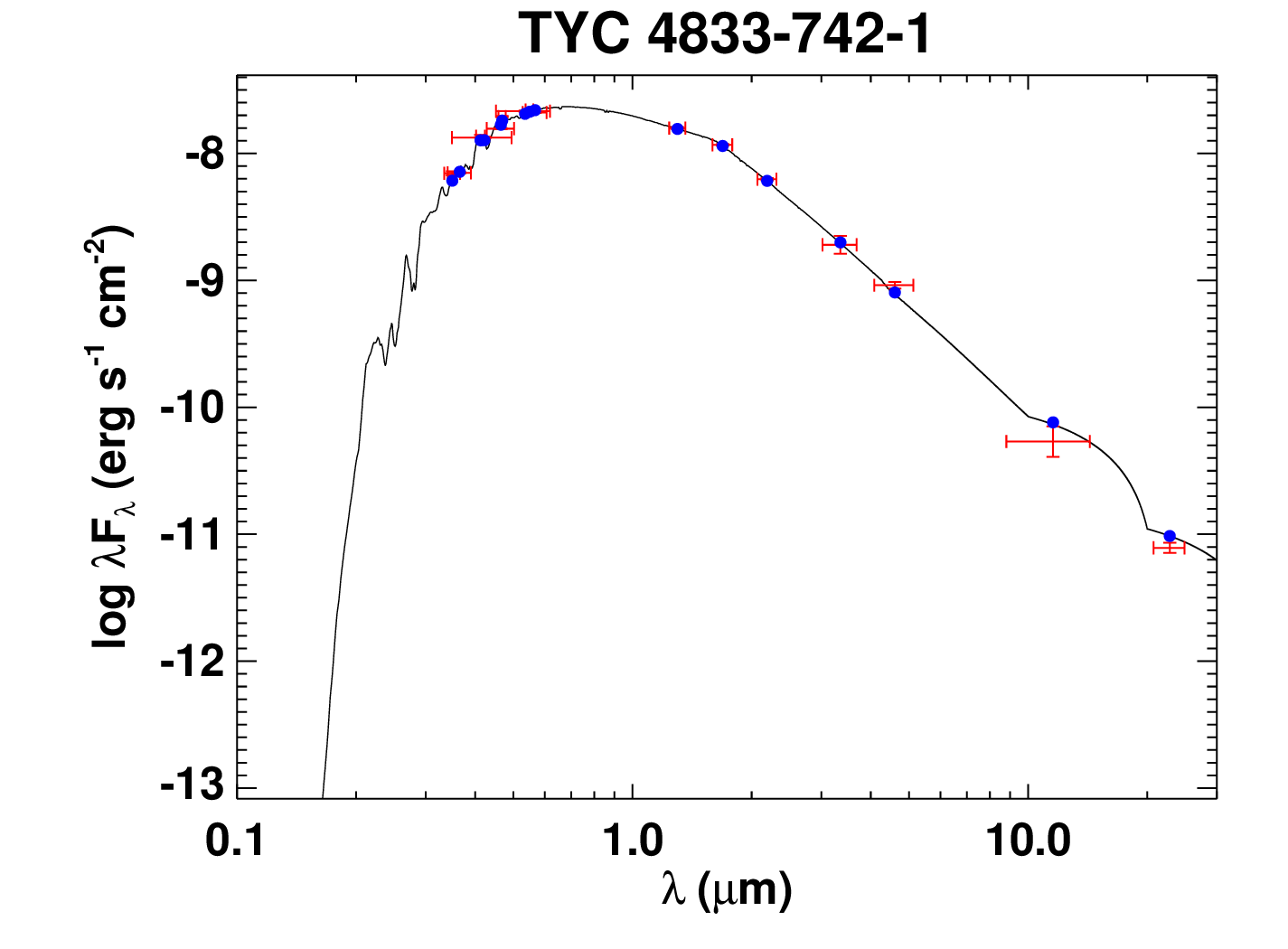}\includegraphics[width=0.333\linewidth]{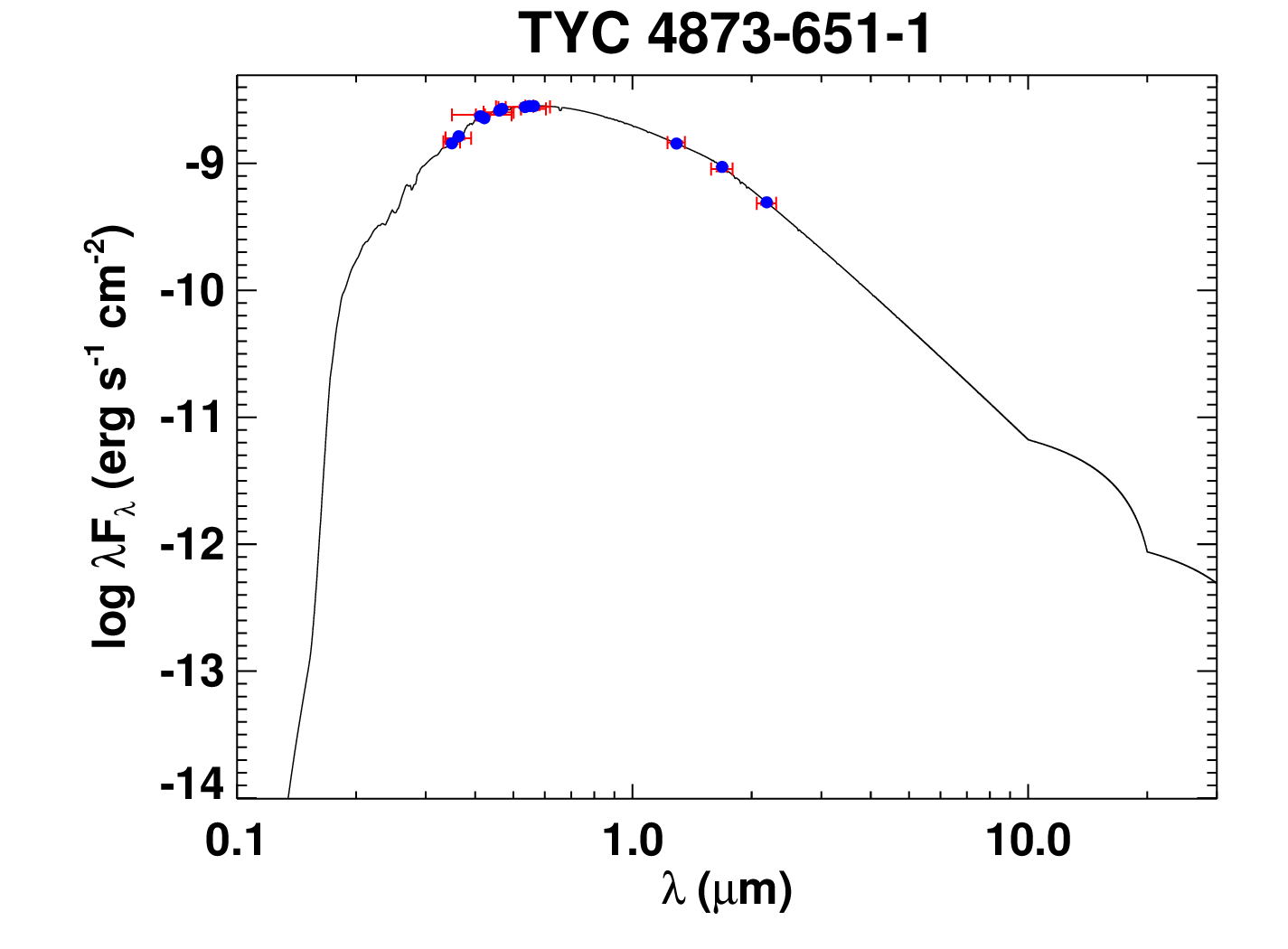}\includegraphics[width=0.333\linewidth]{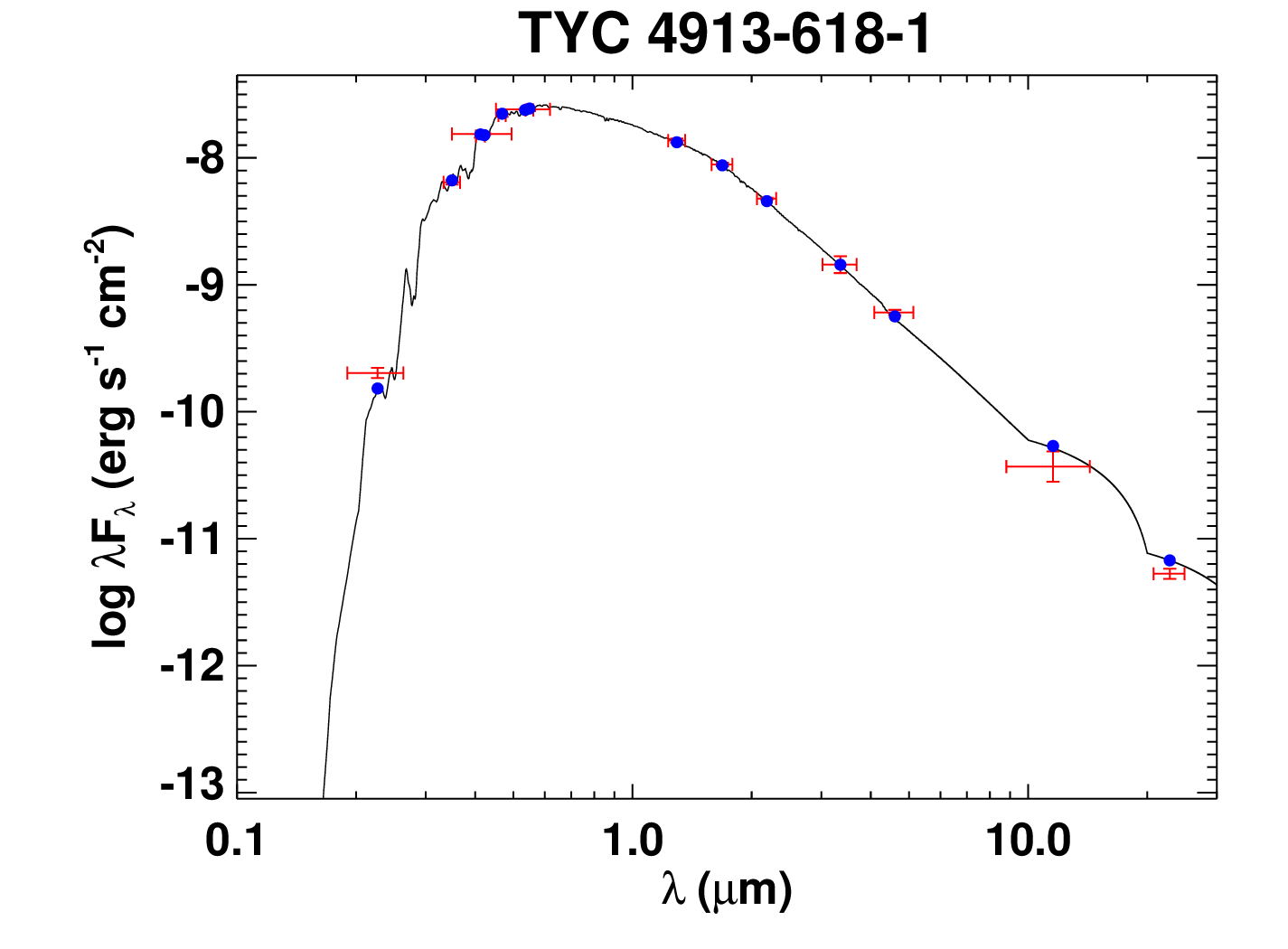}
\caption{\label{fig:seds10} All labels, lines, symbols, and colors as in Figure \ref{fig:seds}.}
\end{figure*}

\begin{figure*}
\includegraphics[width=0.333\linewidth]{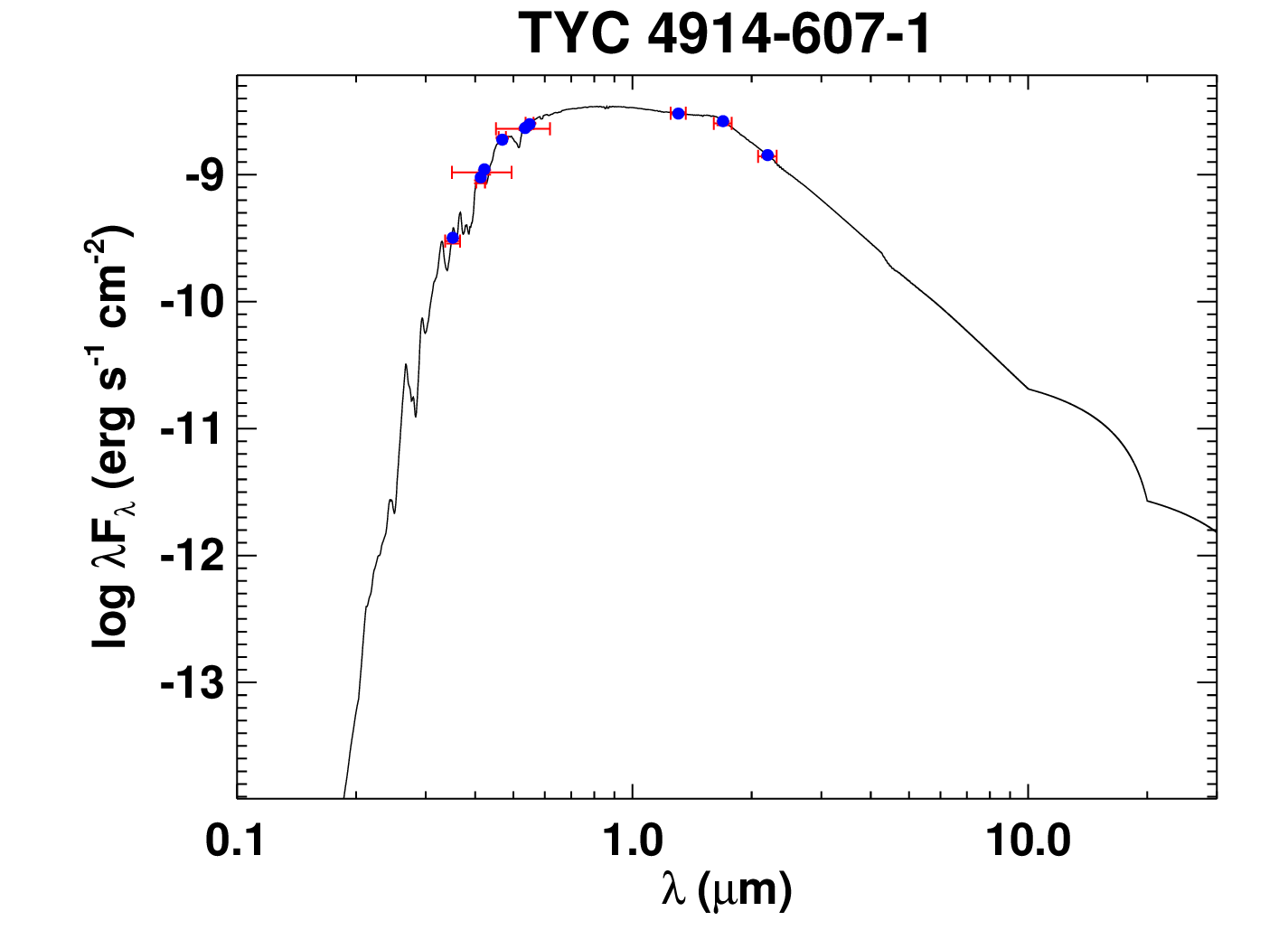}\includegraphics[width=0.333\linewidth]{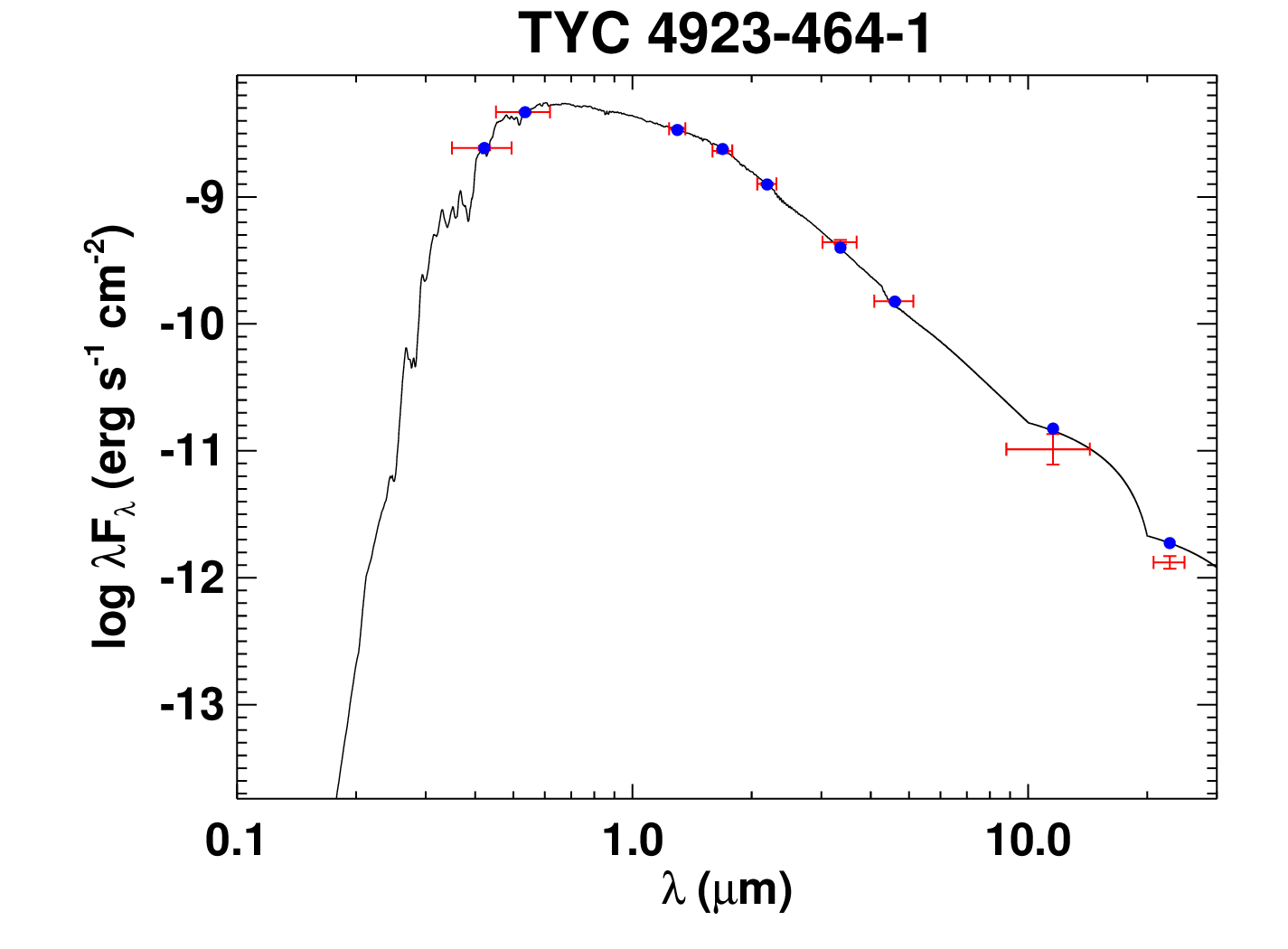}\includegraphics[width=0.333\linewidth]{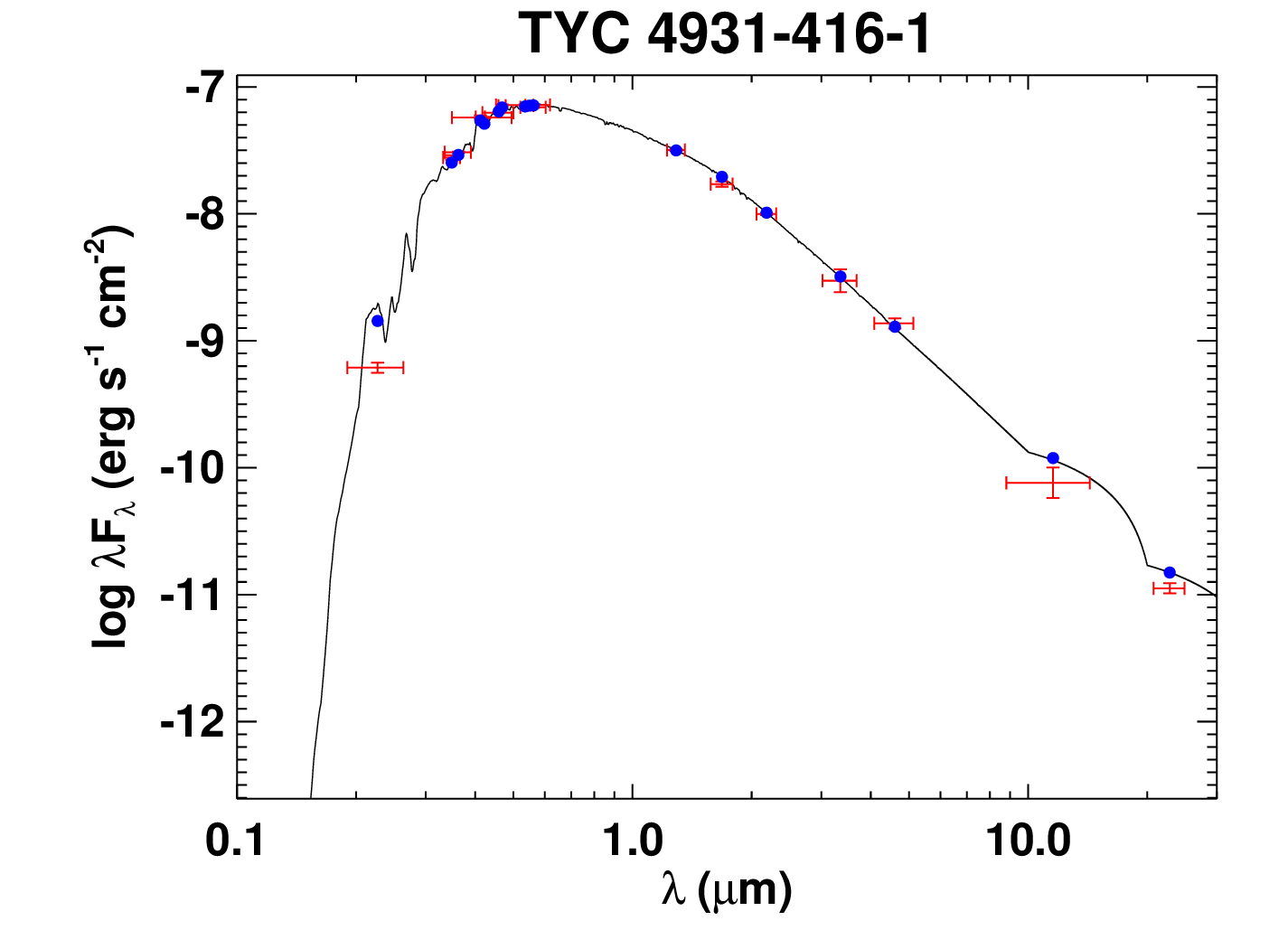}
\includegraphics[width=0.333\linewidth]{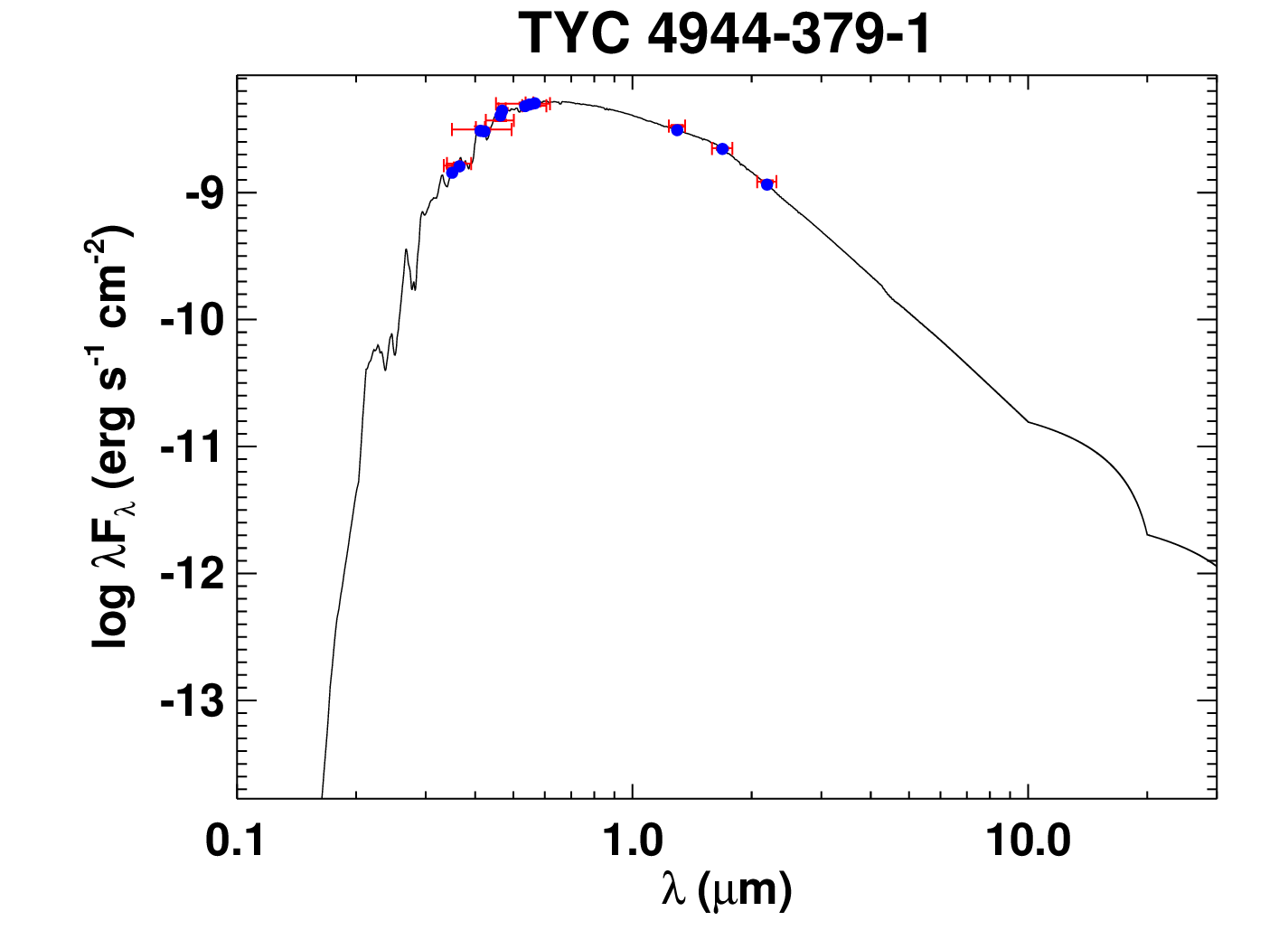}\includegraphics[width=0.333\linewidth]{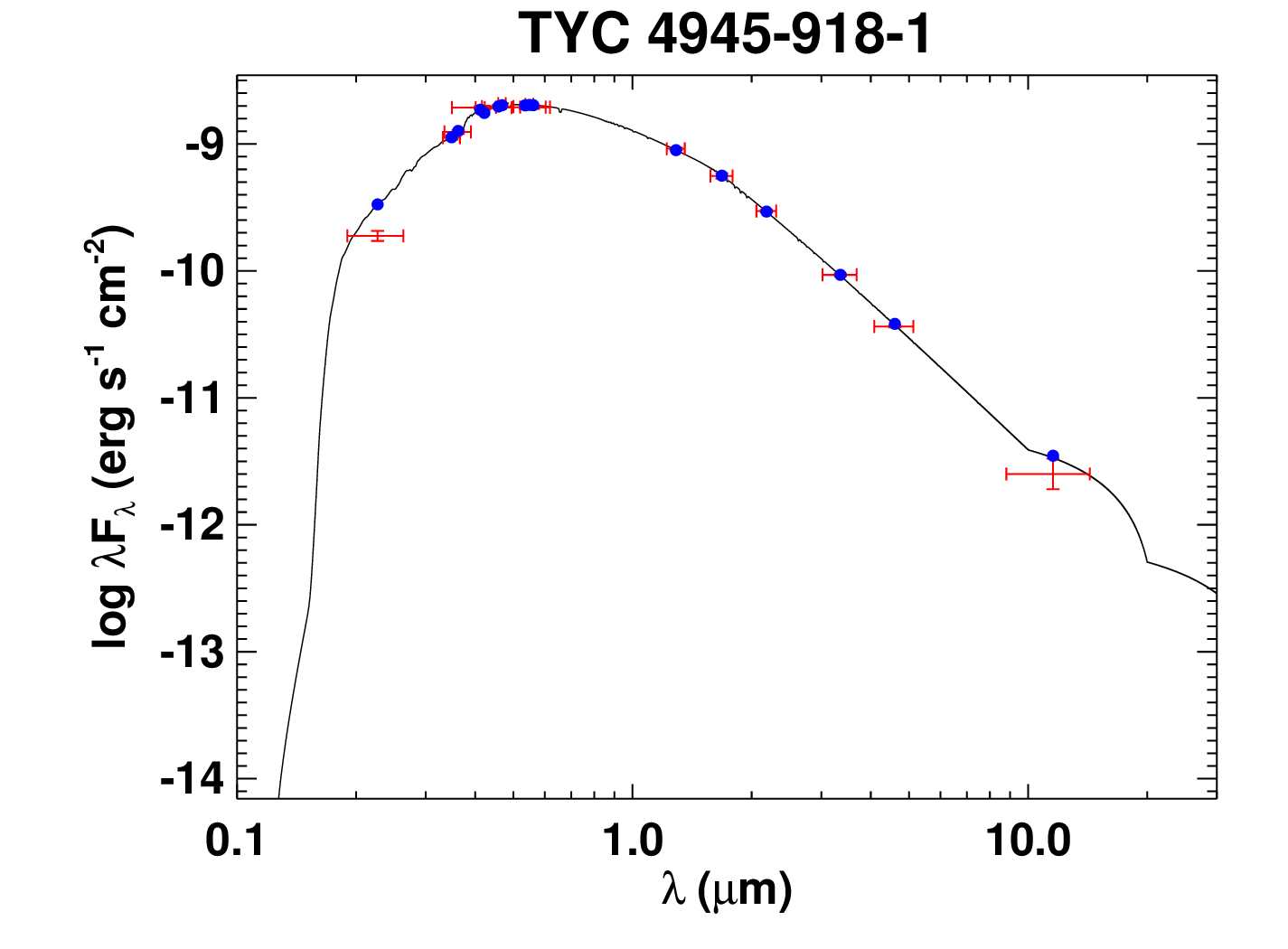}\includegraphics[width=0.333\linewidth]{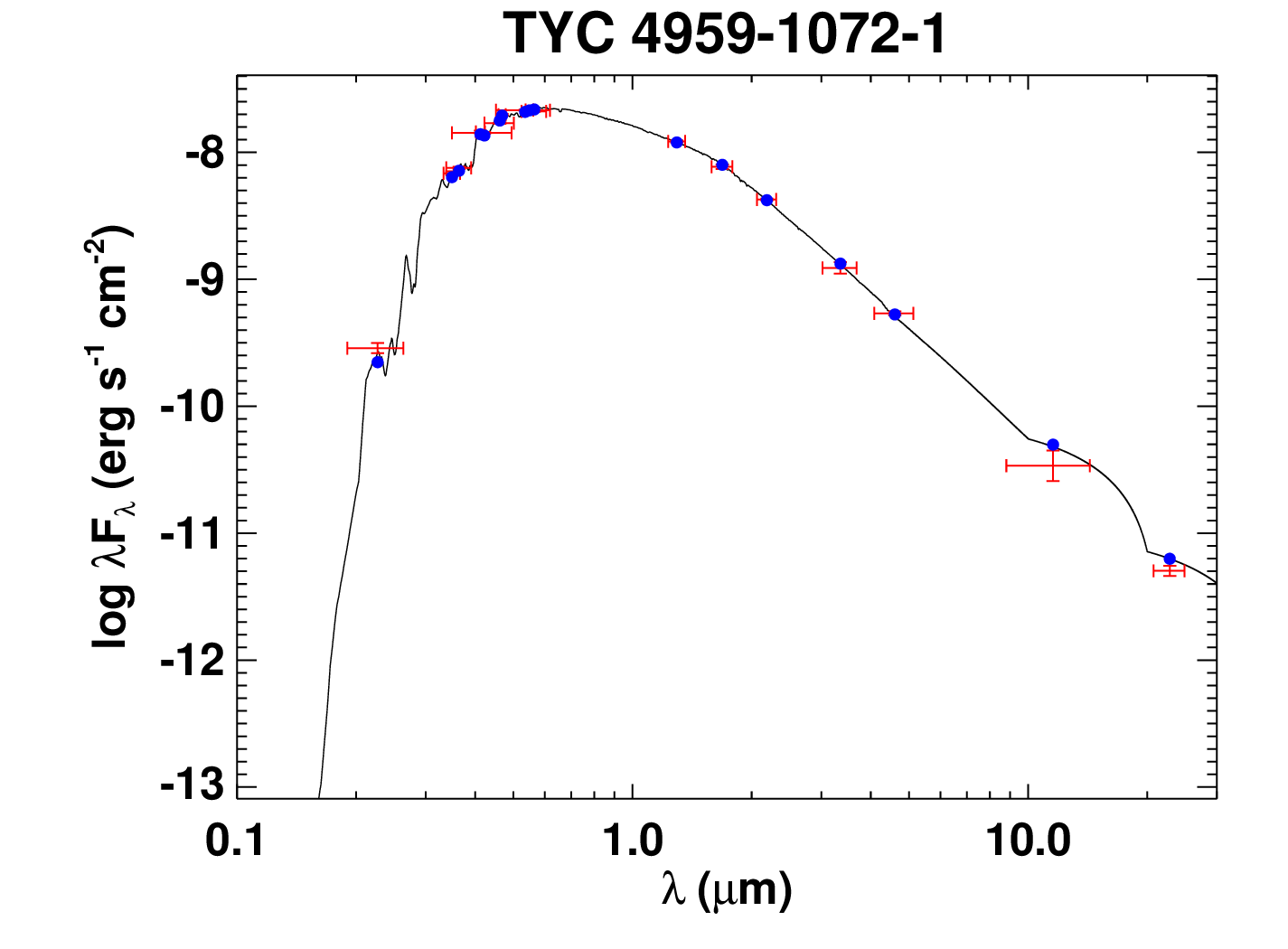}
\includegraphics[width=0.333\linewidth]{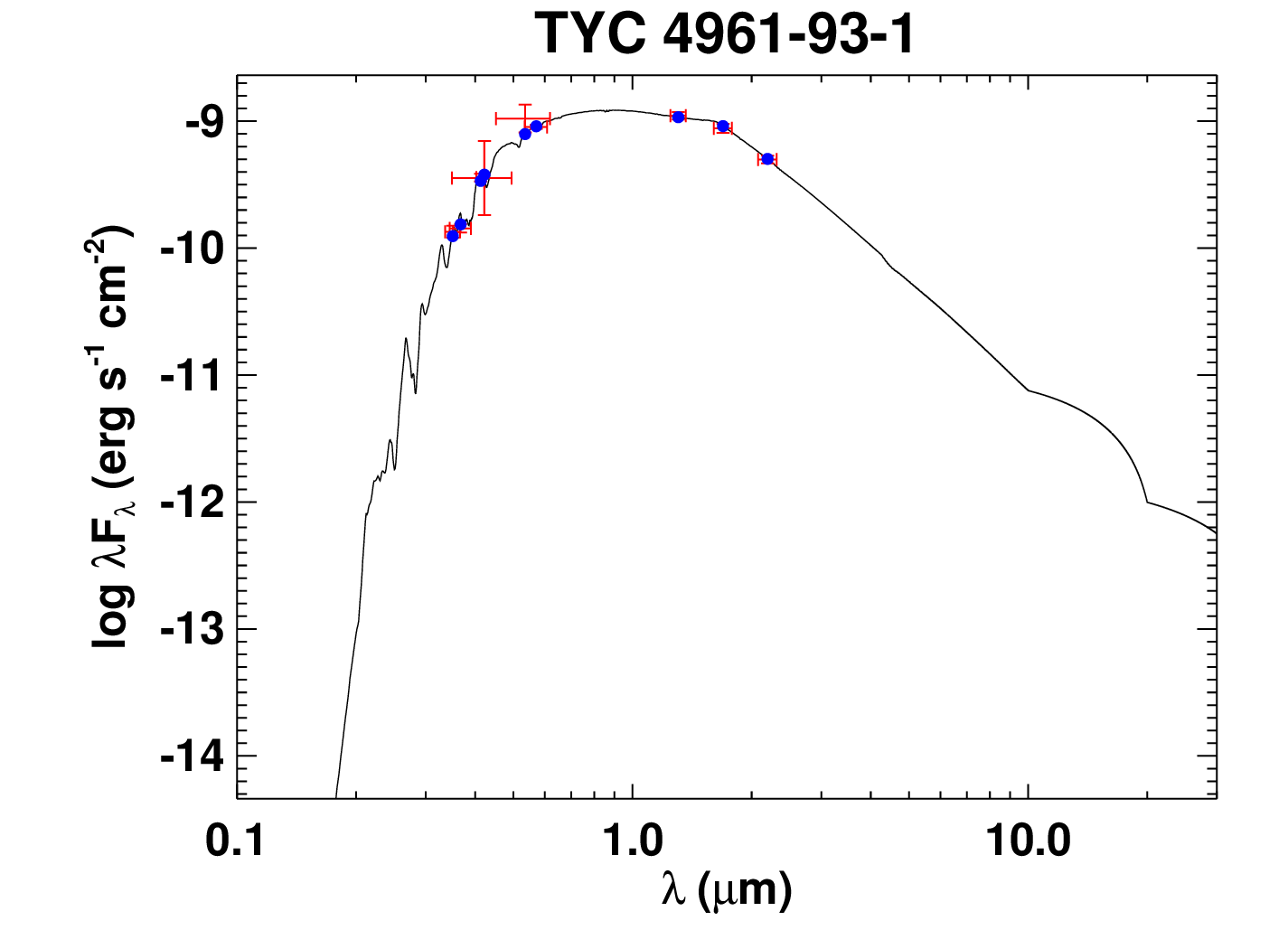}\includegraphics[width=0.333\linewidth]{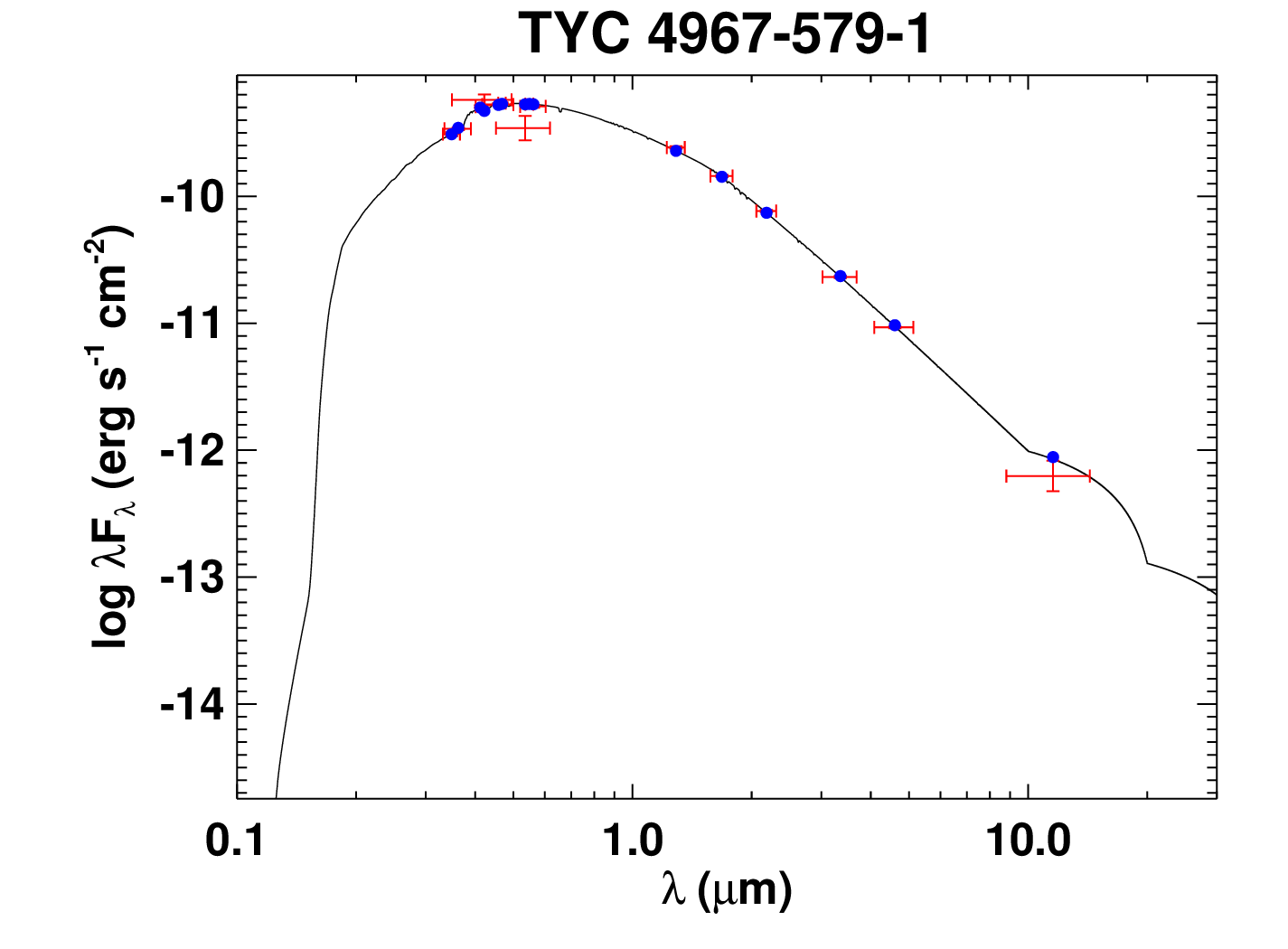}\includegraphics[width=0.333\linewidth]{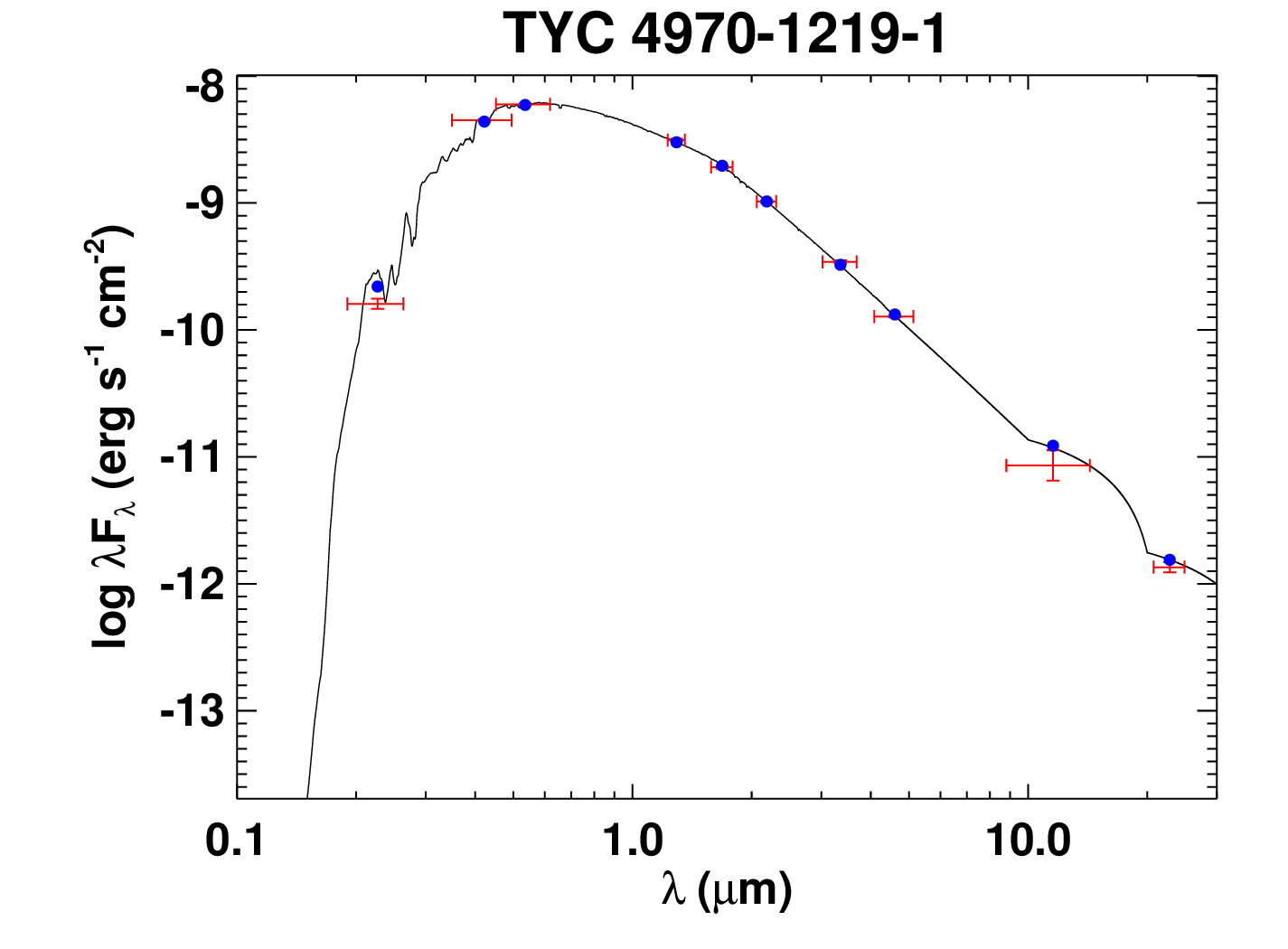}
\includegraphics[width=0.333\linewidth]{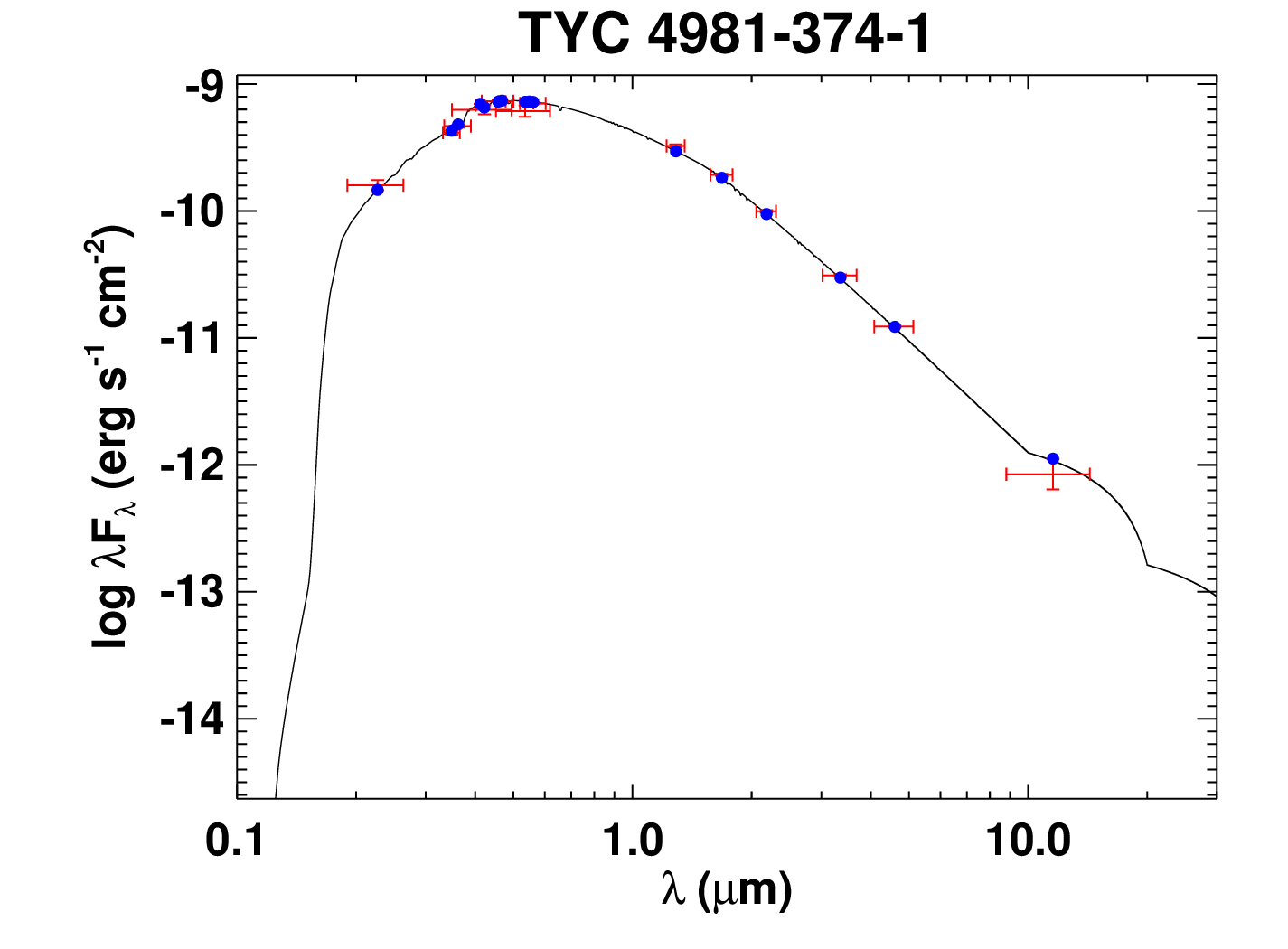}\includegraphics[width=0.333\linewidth]{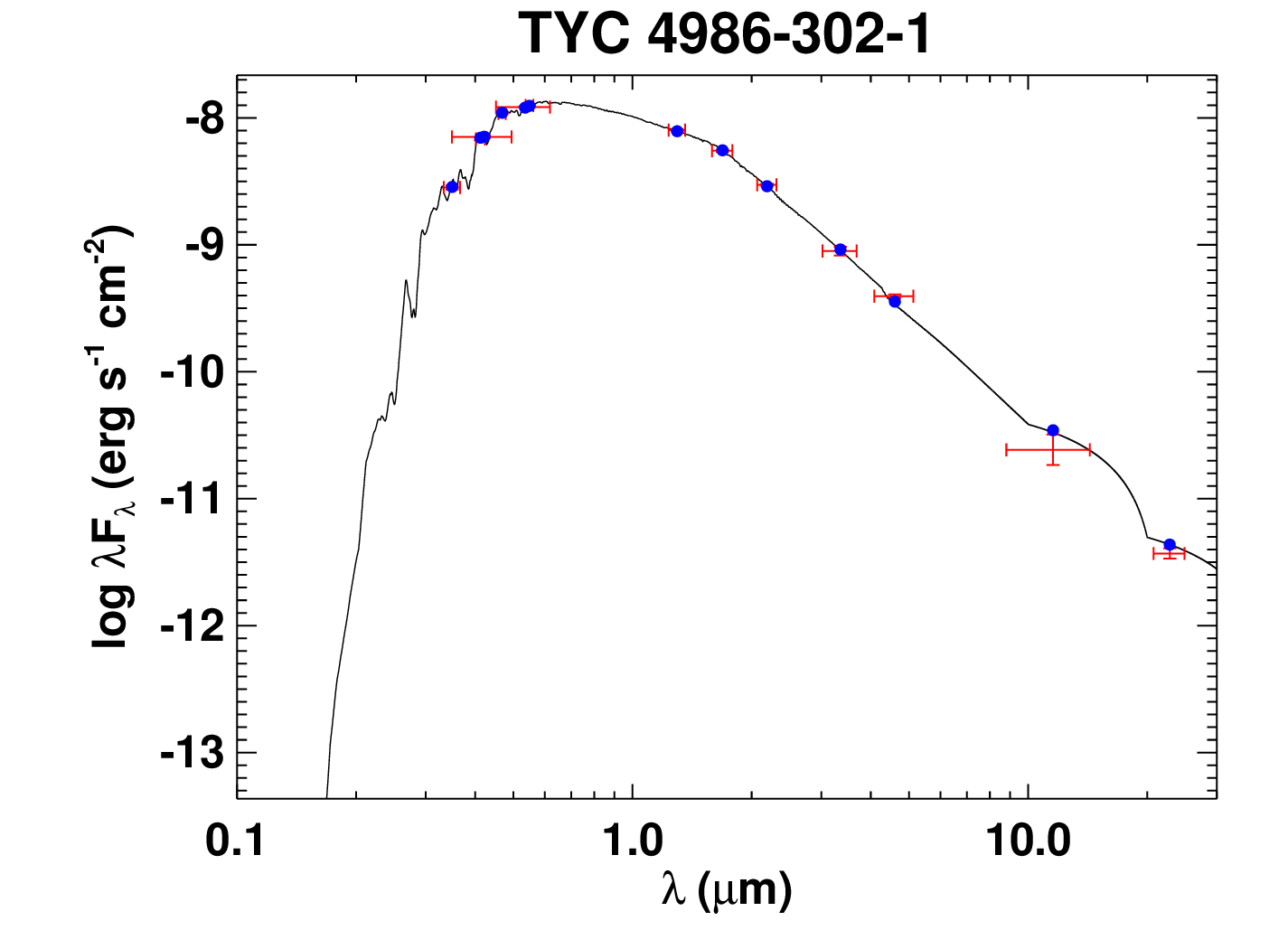}\includegraphics[width=0.333\linewidth]{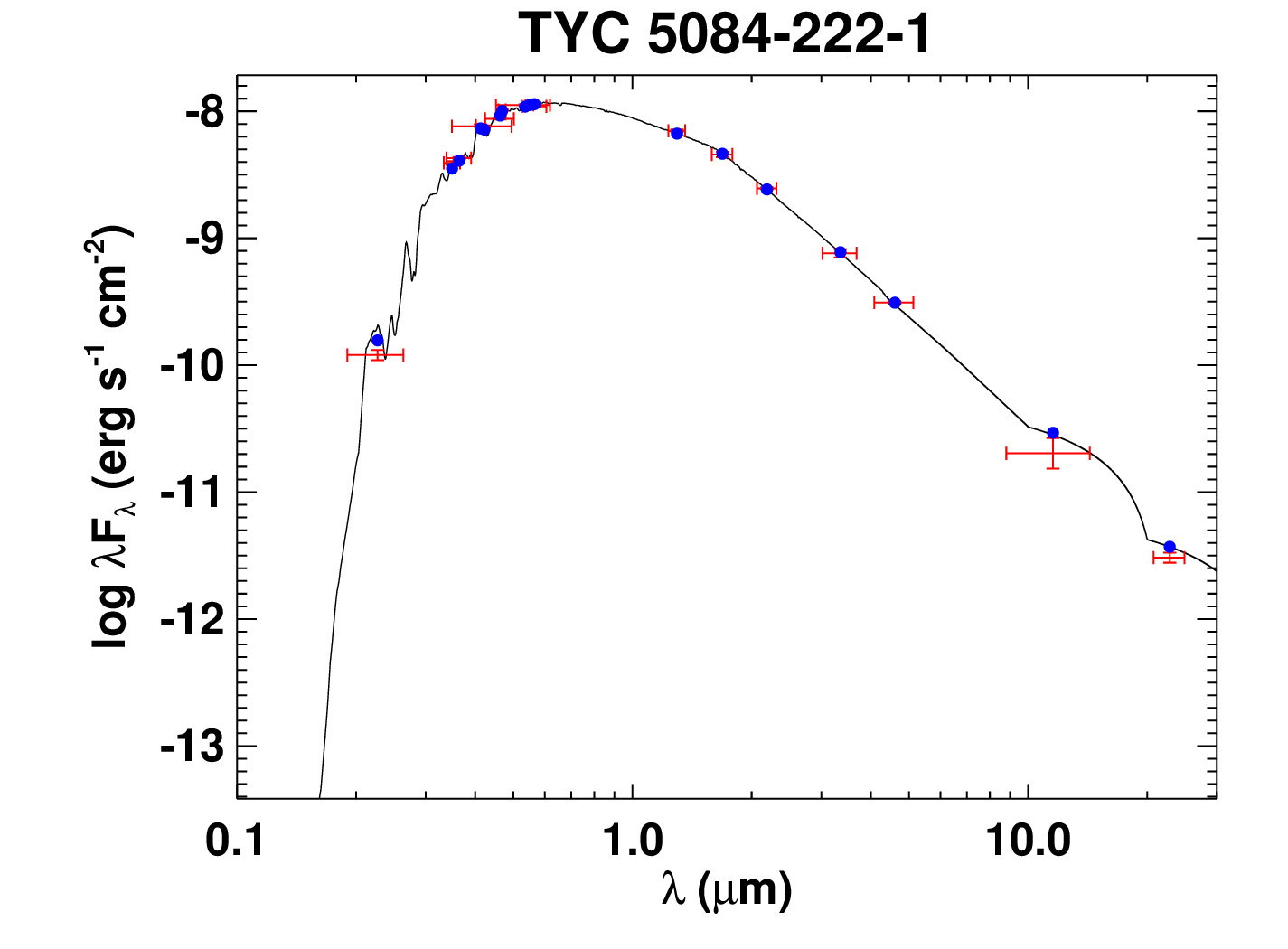}
\caption{\label{fig:seds11} All labels, lines, symbols, and colors as in Figure \ref{fig:seds}.}
\end{figure*}

\begin{figure*}
\includegraphics[width=0.333\linewidth]{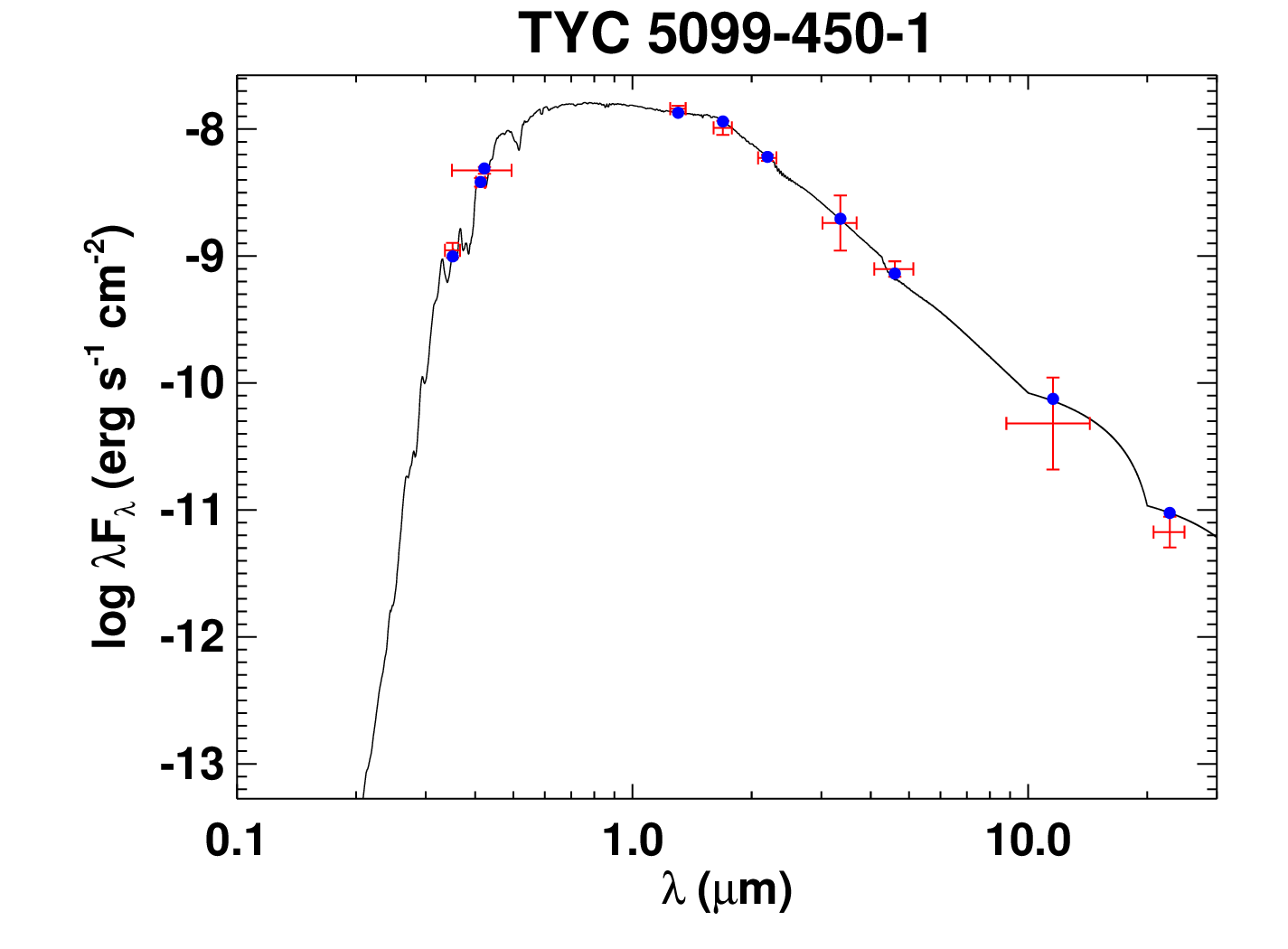}\includegraphics[width=0.333\linewidth]{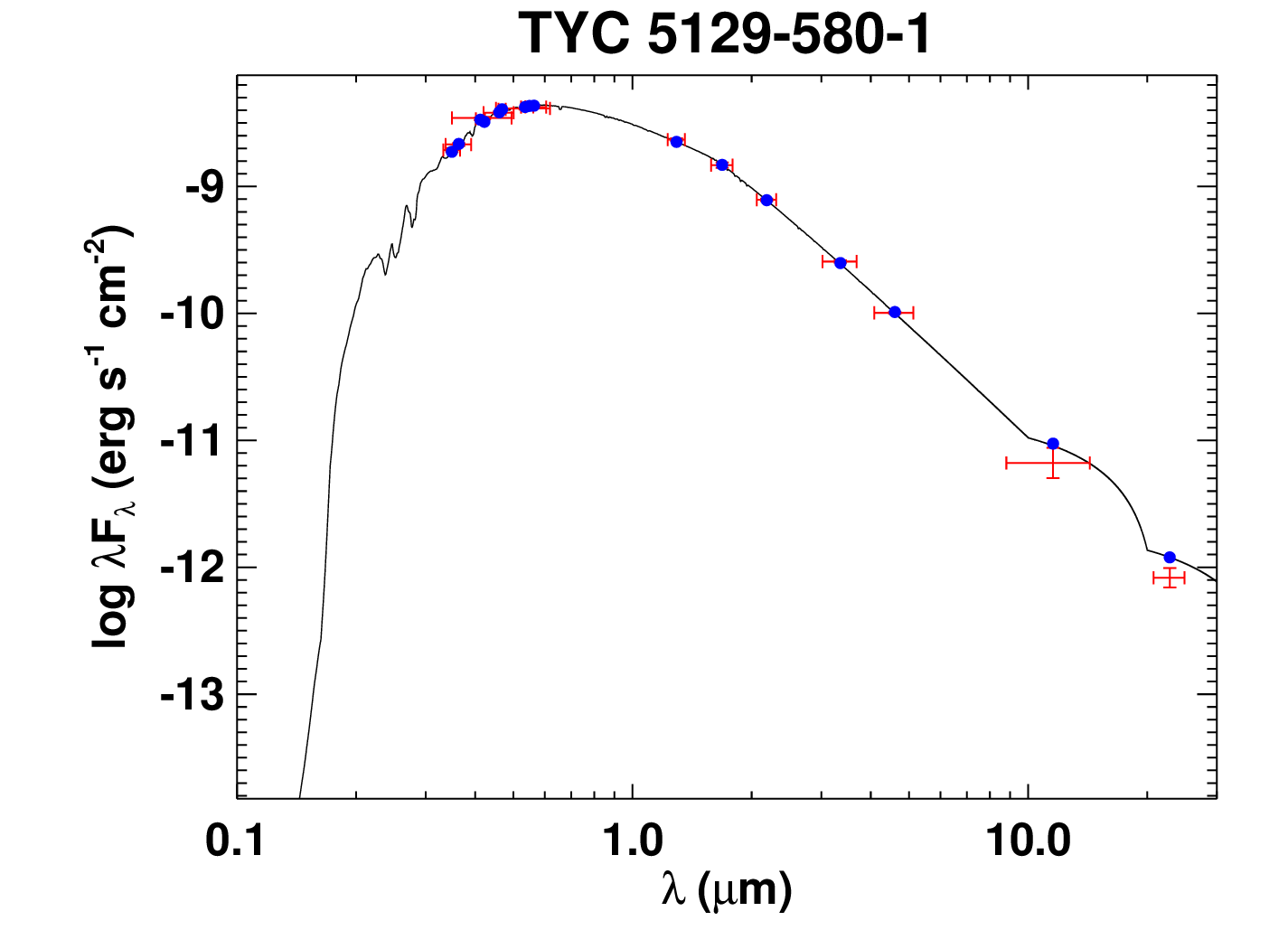}\includegraphics[width=0.333\linewidth]{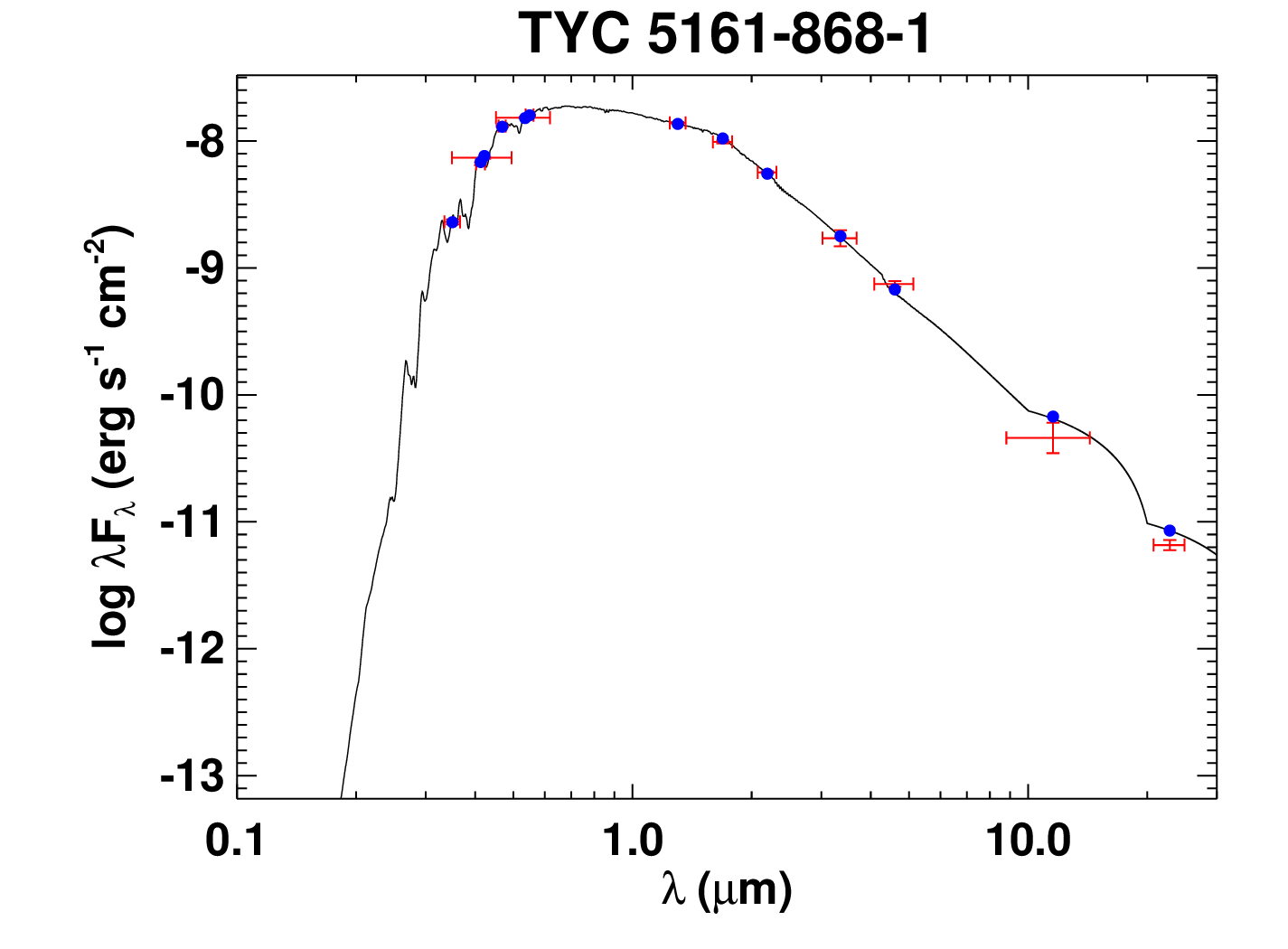}
\includegraphics[width=0.333\linewidth]{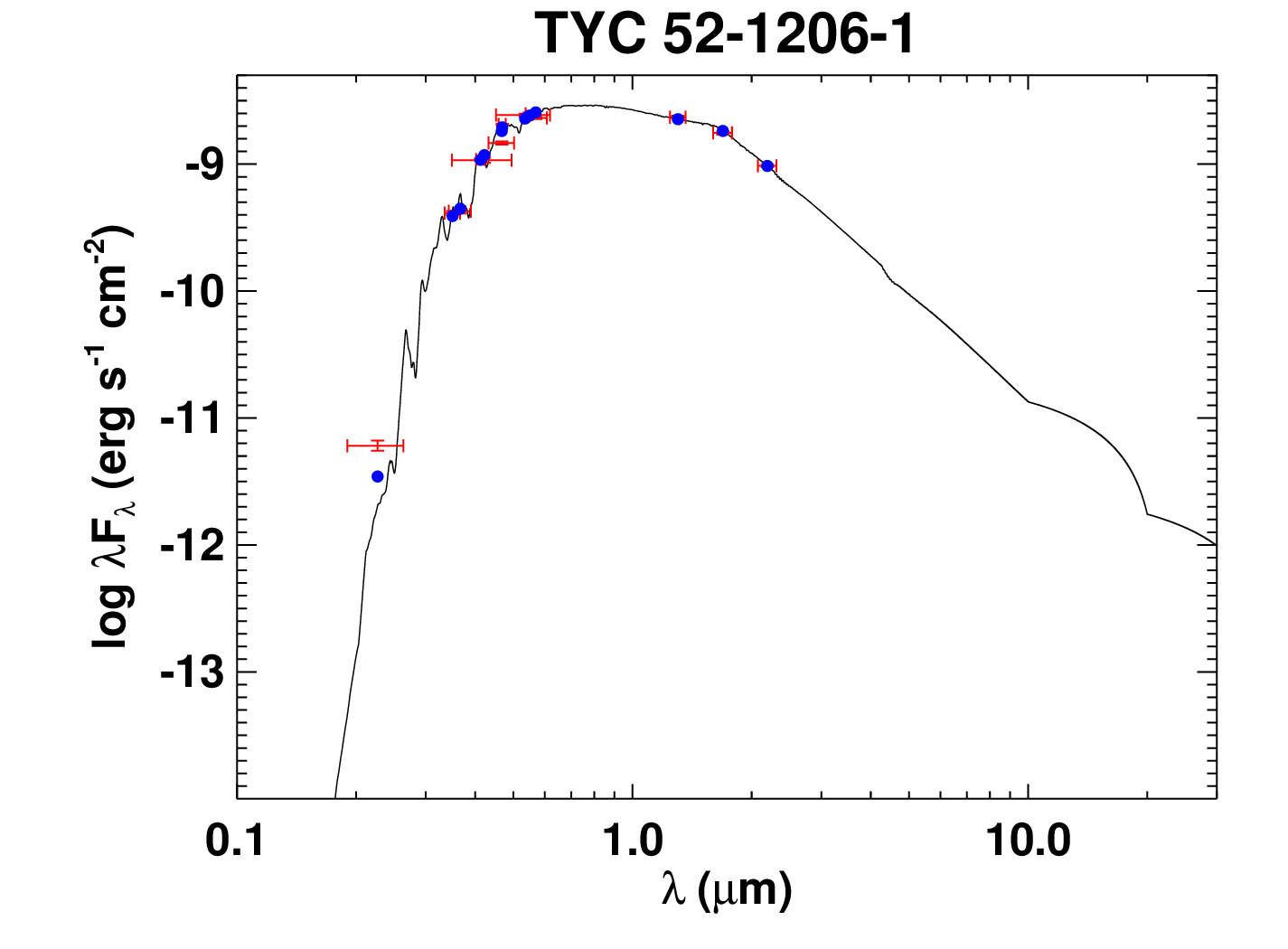}\includegraphics[width=0.333\linewidth]{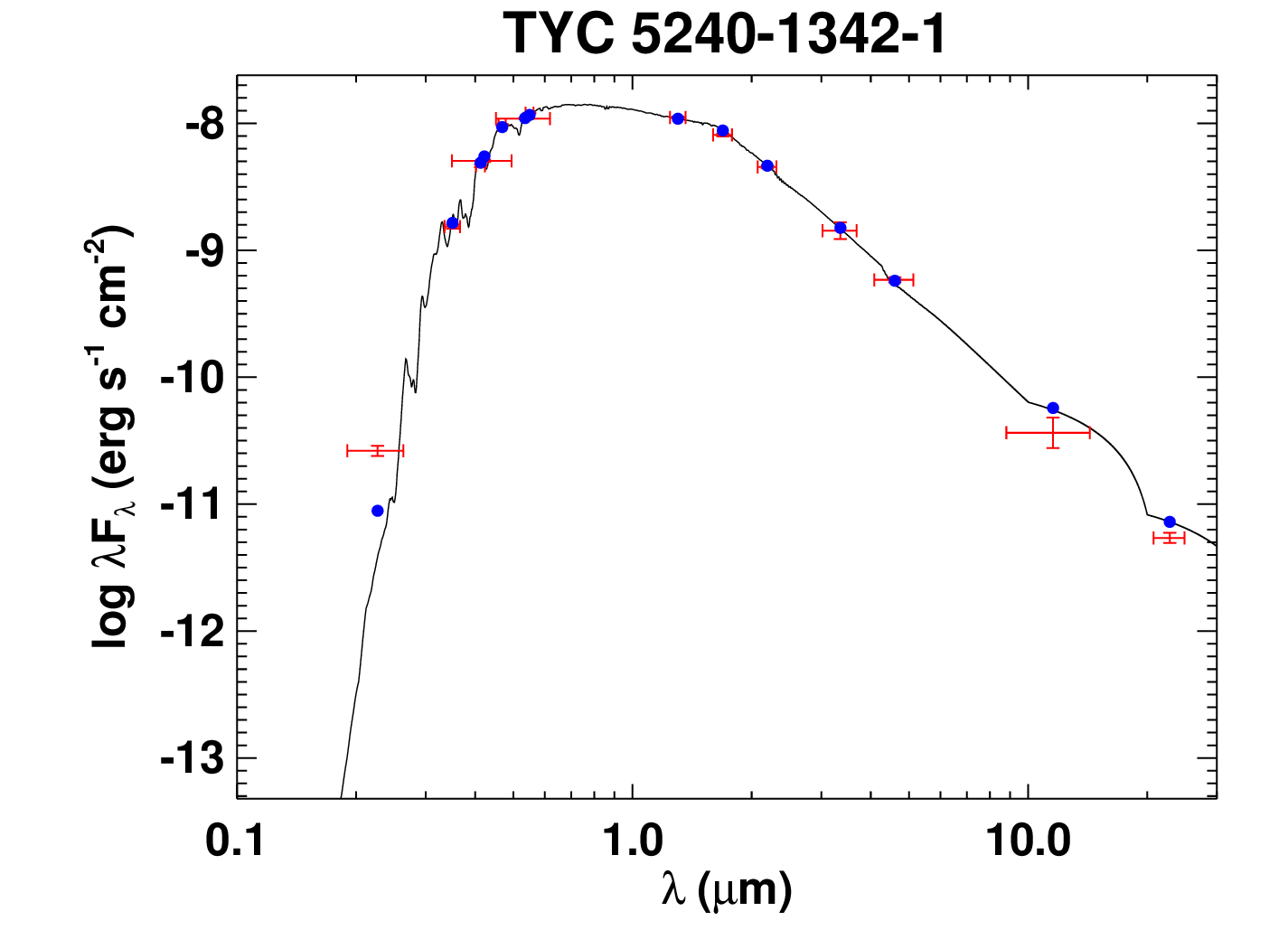}\includegraphics[width=0.333\linewidth]{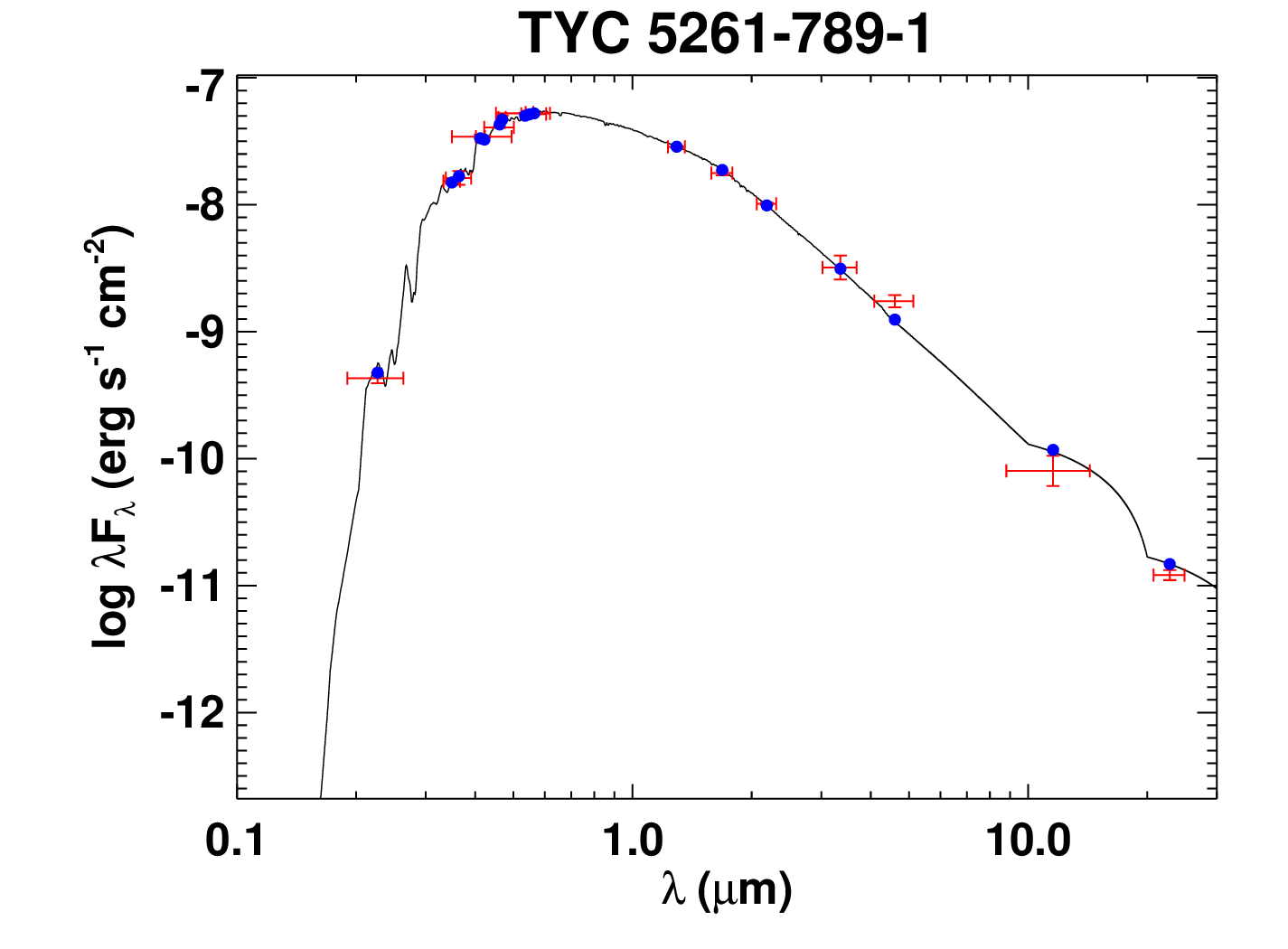}
\includegraphics[width=0.333\linewidth]{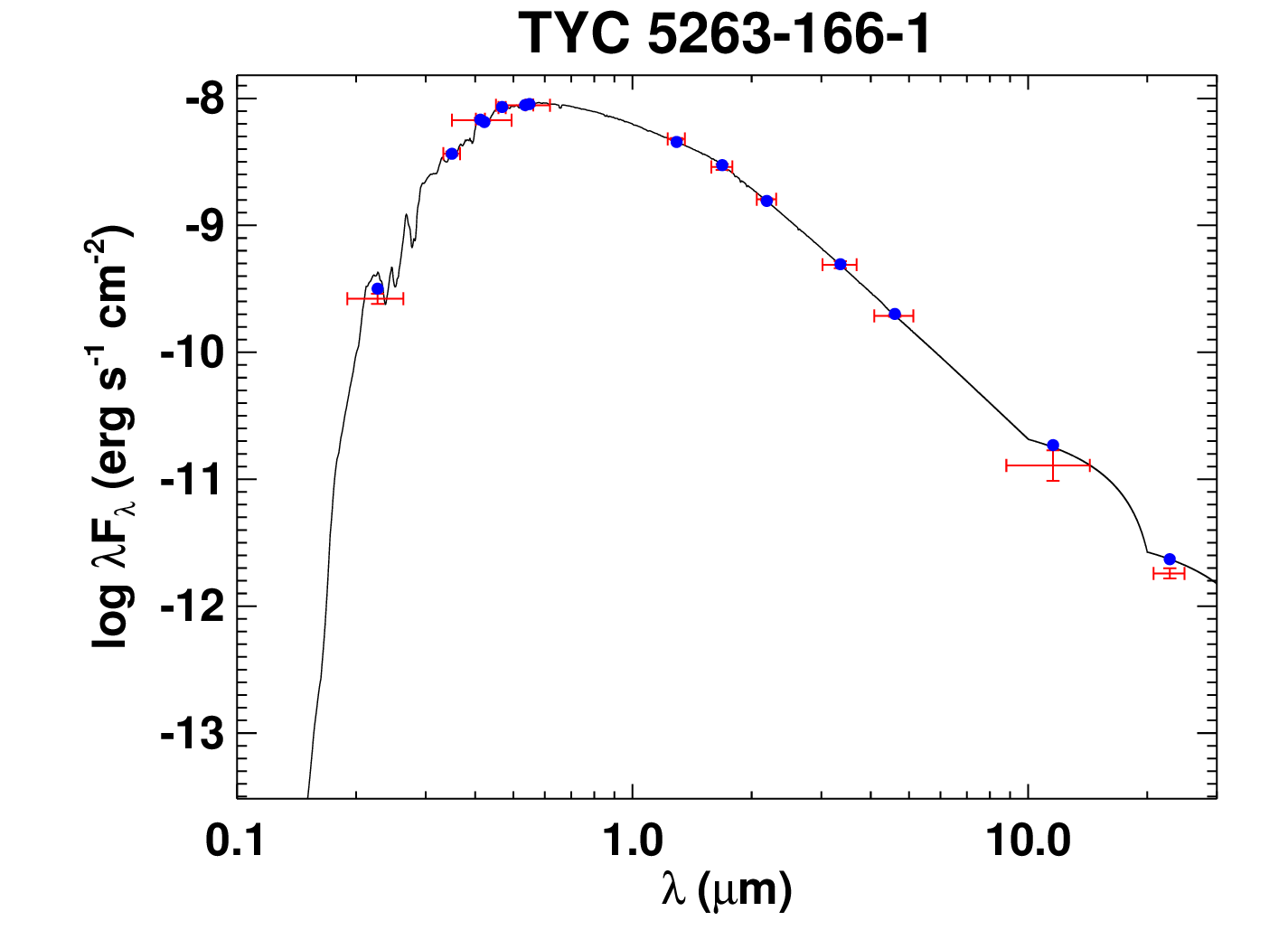}\includegraphics[width=0.333\linewidth]{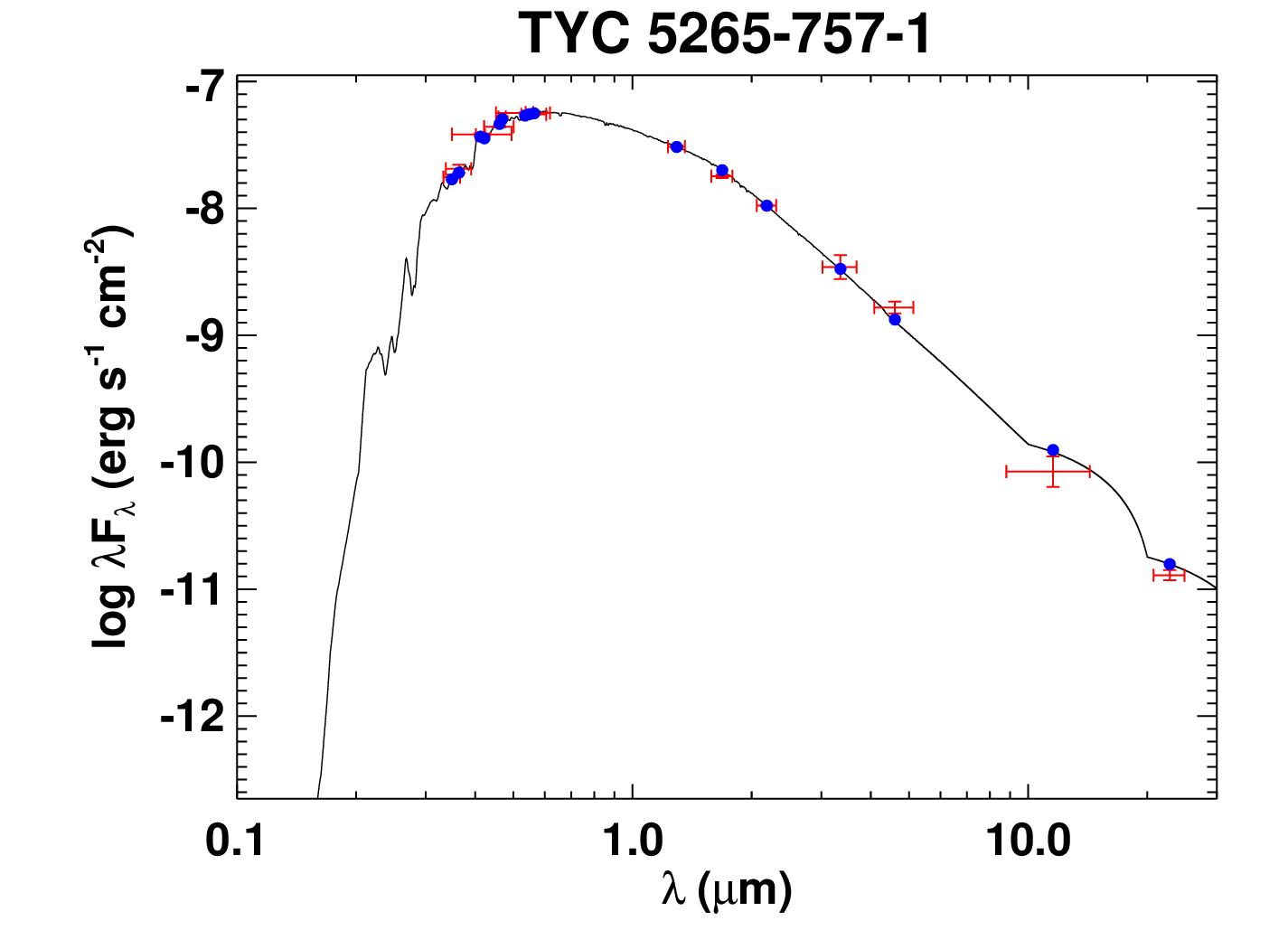}\includegraphics[width=0.333\linewidth]{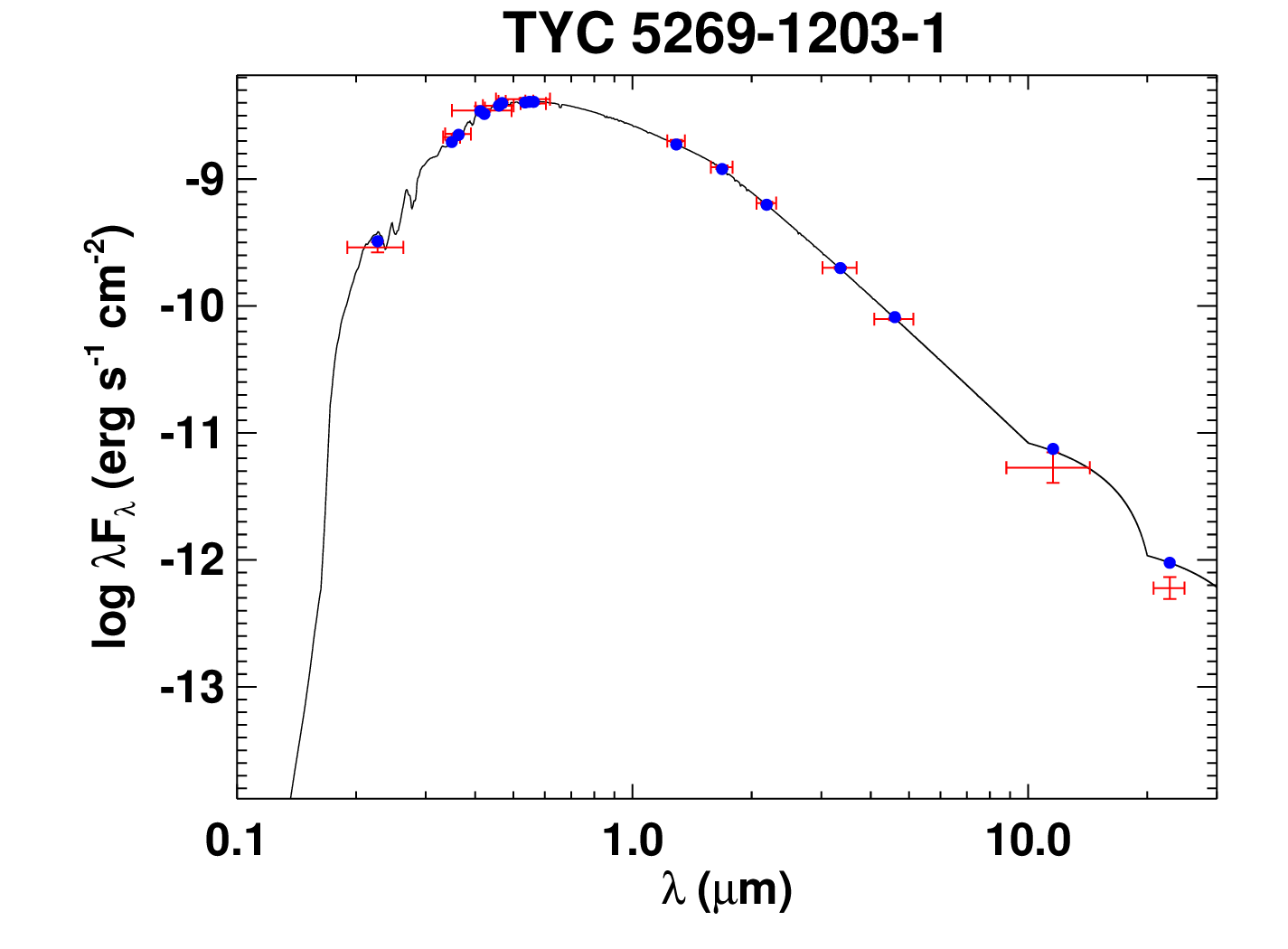}
\includegraphics[width=0.333\linewidth]{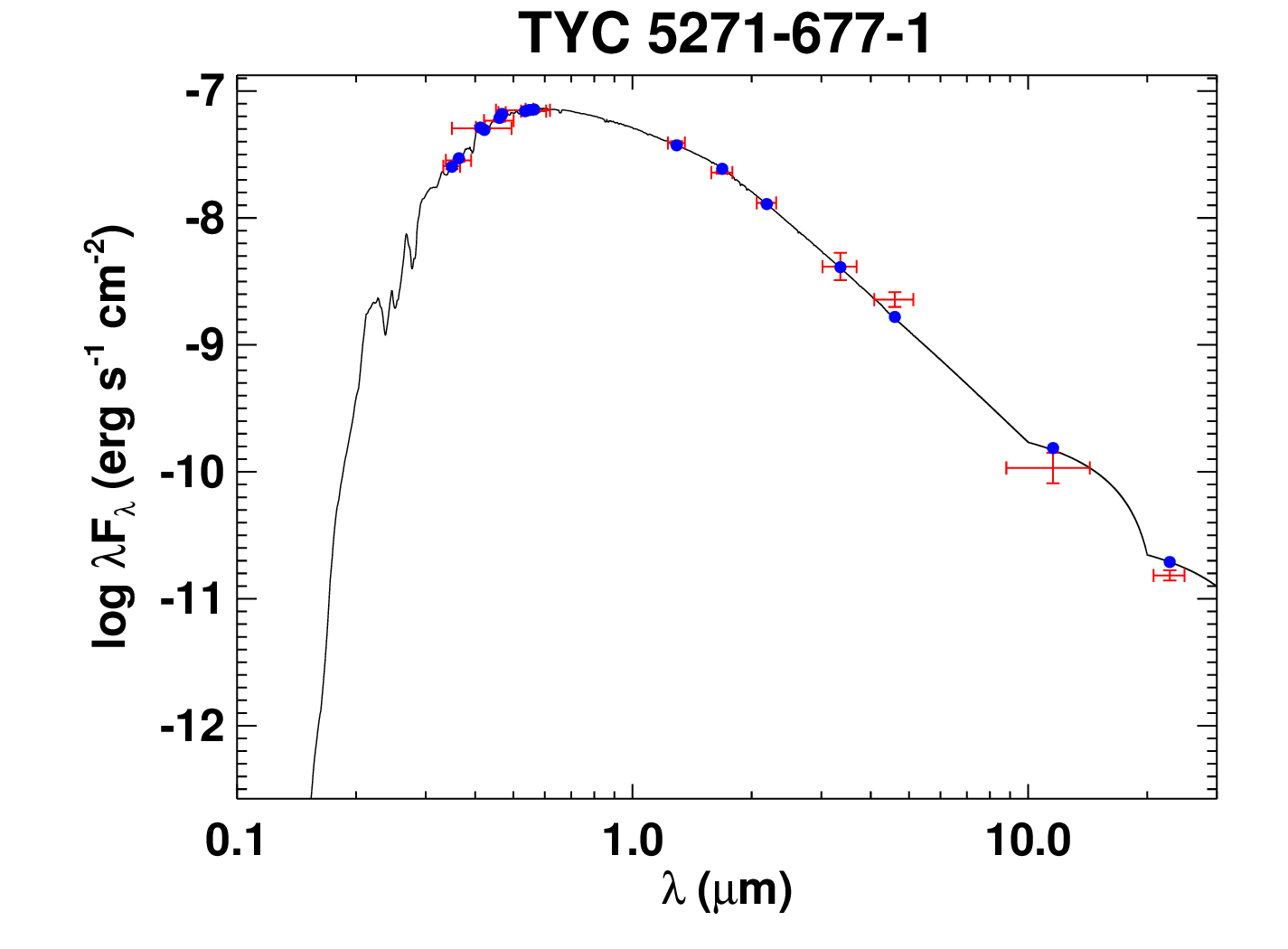}\includegraphics[width=0.333\linewidth]{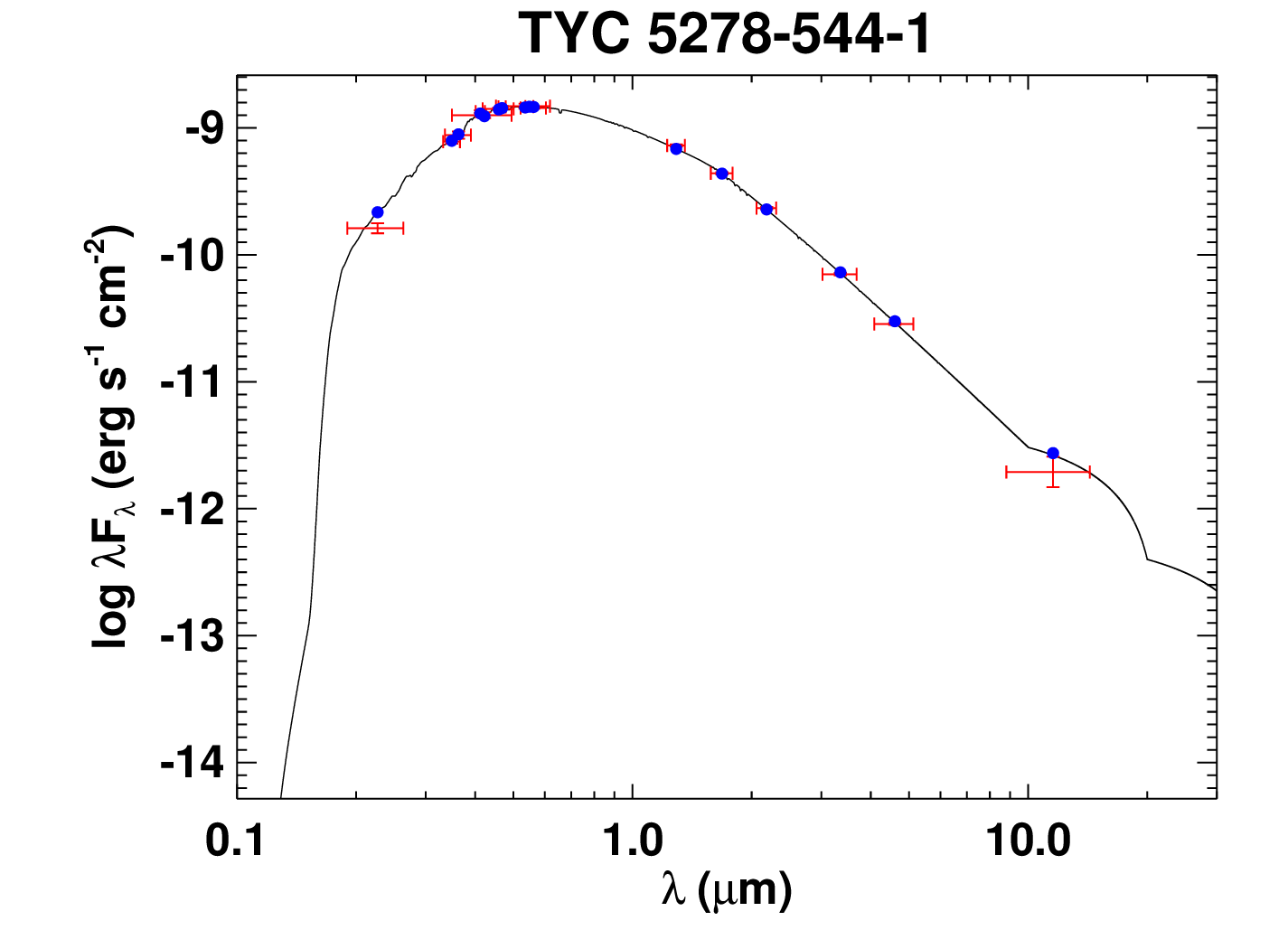}\includegraphics[width=0.333\linewidth]{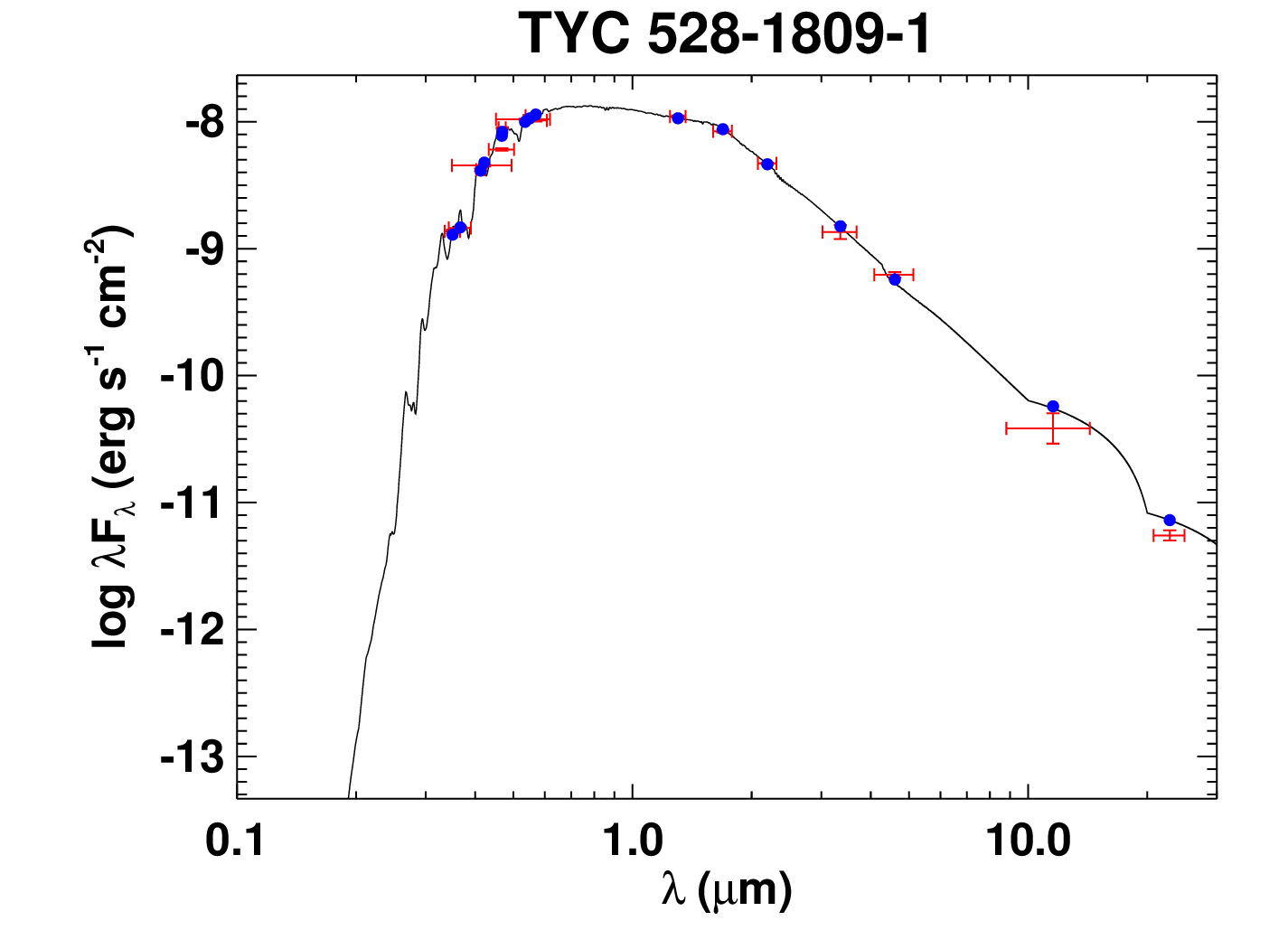}
\caption{\label{fig:seds12} All labels, lines, symbols, and colors as in Figure \ref{fig:seds}.}
\end{figure*}

\begin{figure*}
\includegraphics[width=0.333\linewidth]{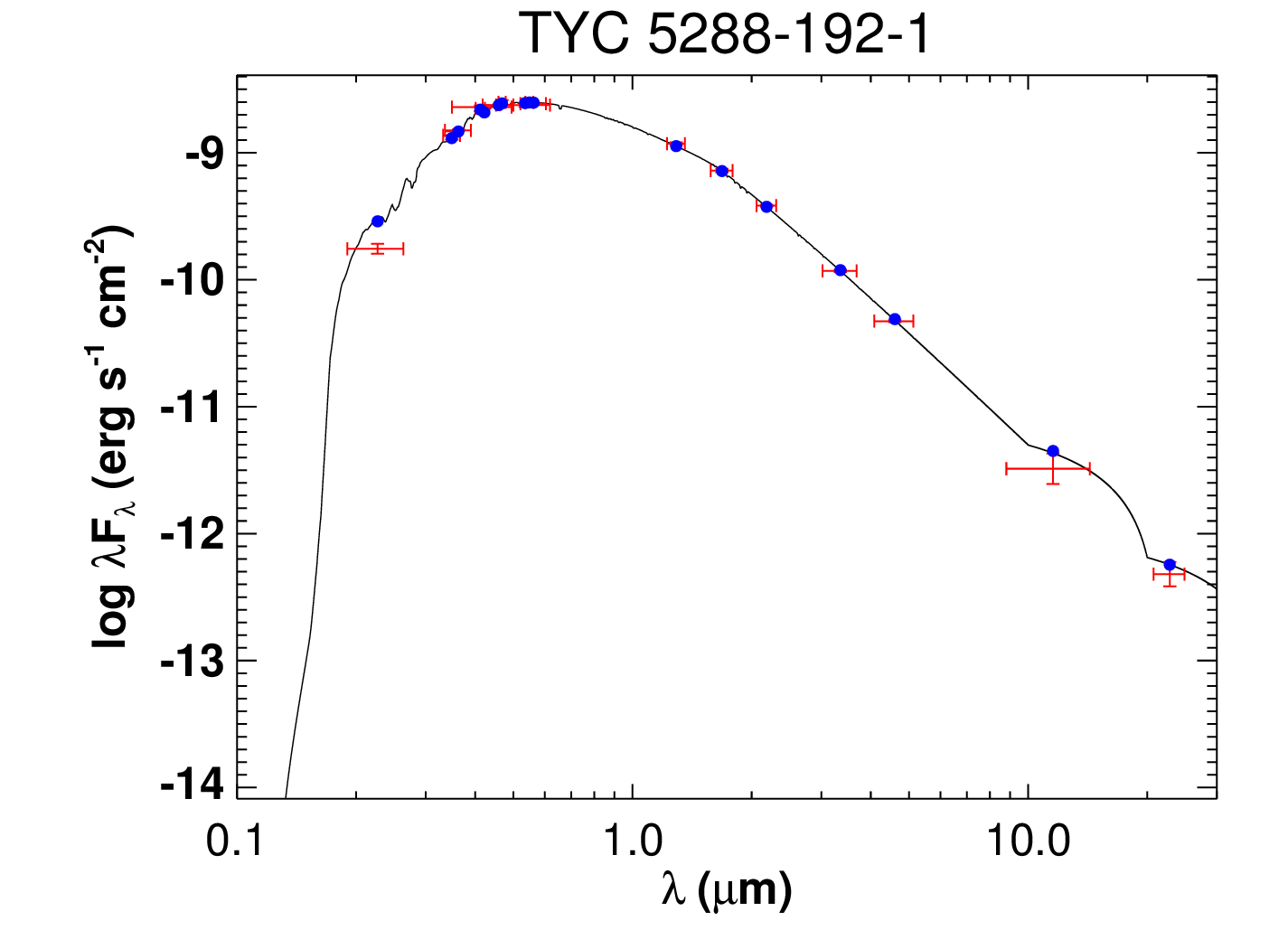}\includegraphics[width=0.333\linewidth]{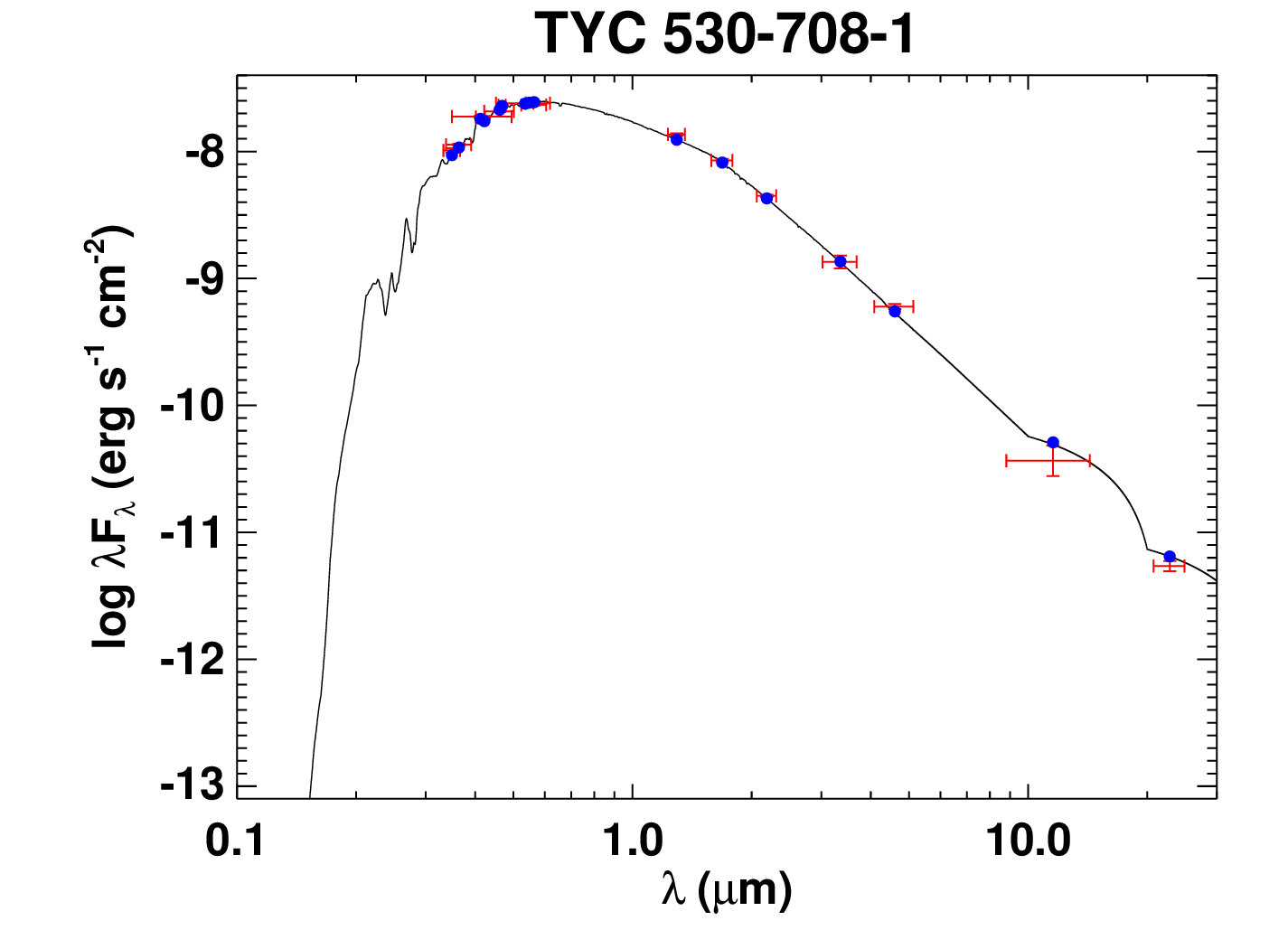}\includegraphics[width=0.333\linewidth]{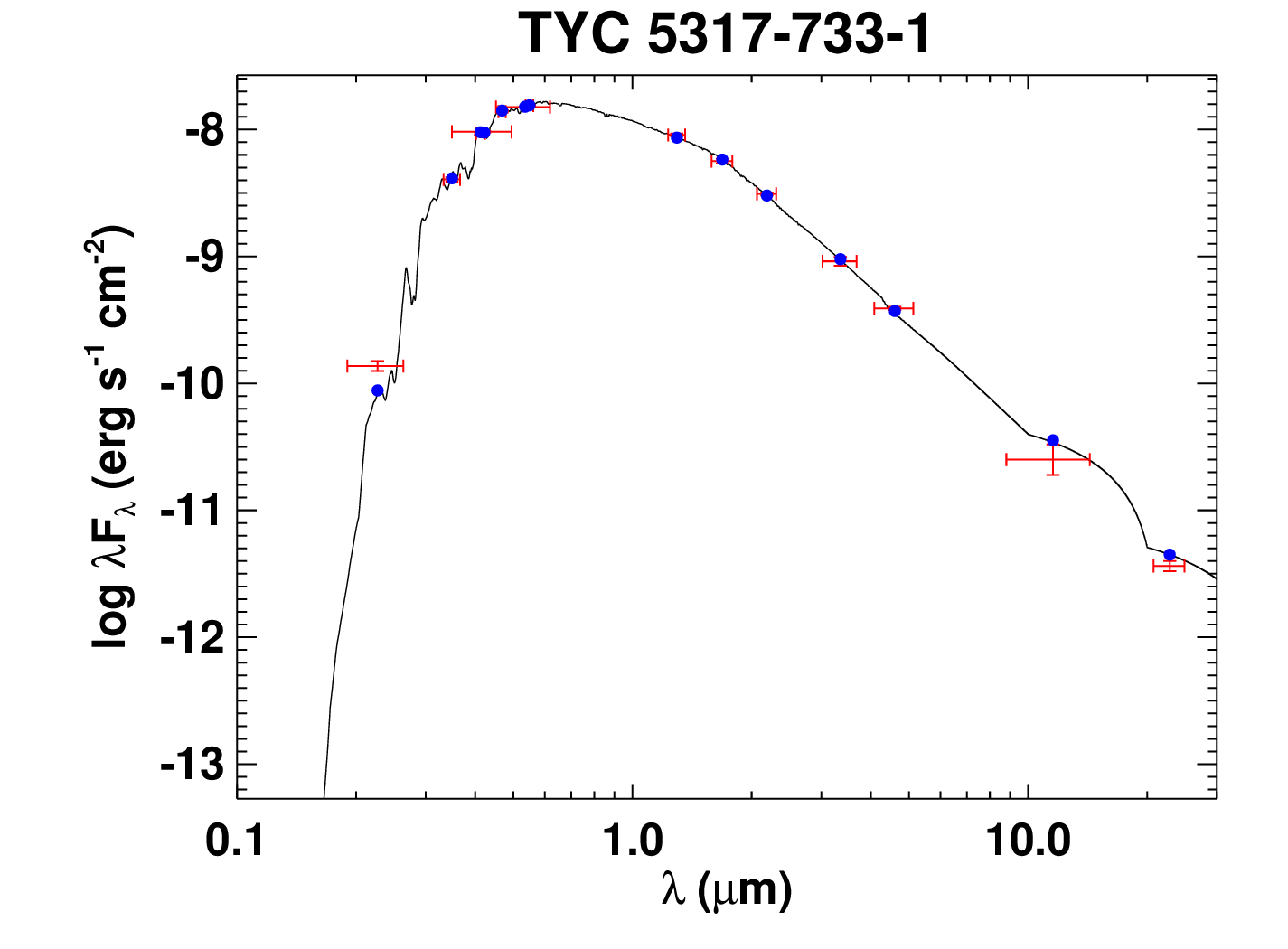}
\includegraphics[width=0.333\linewidth]{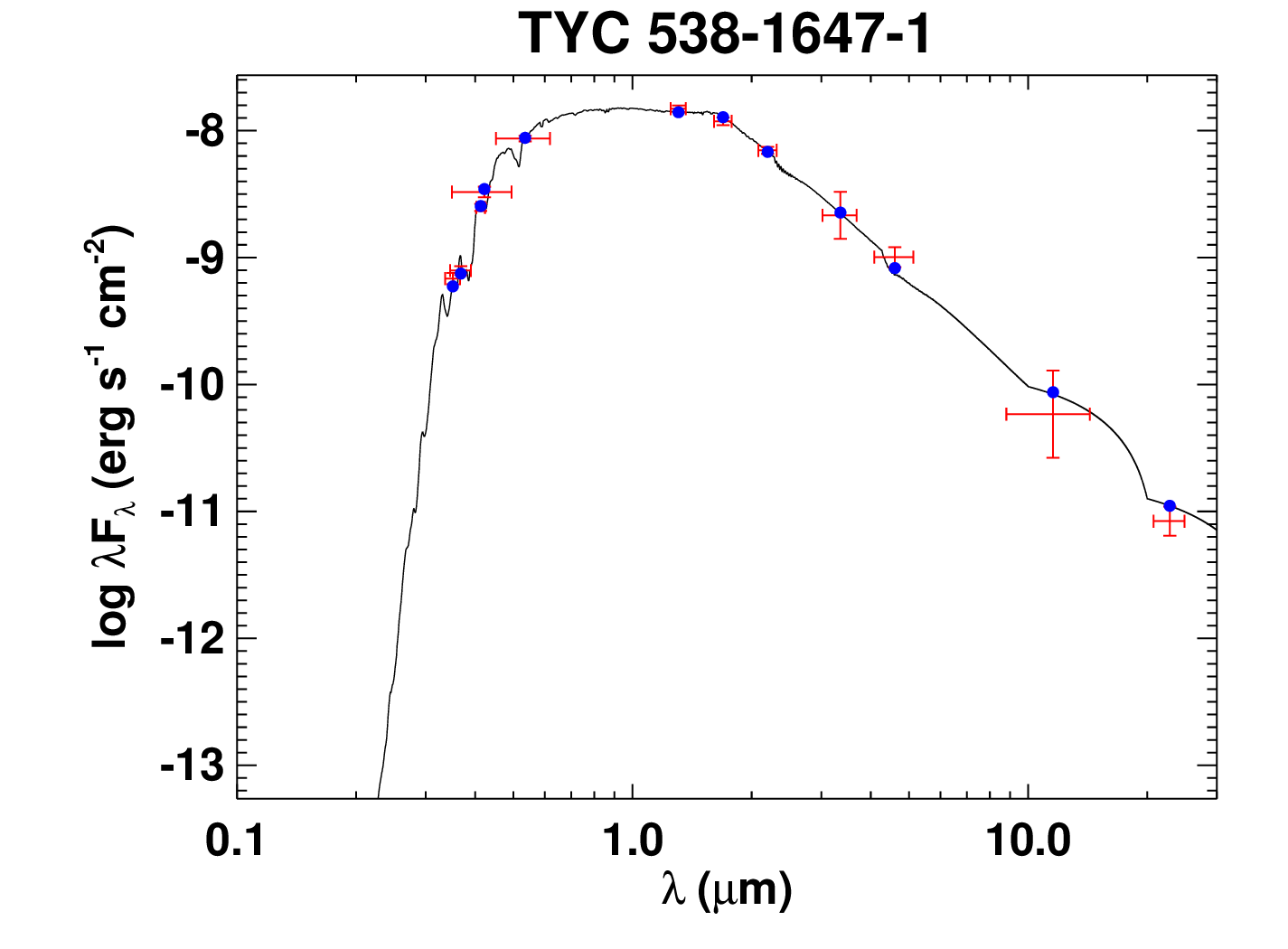}\includegraphics[width=0.333\linewidth]{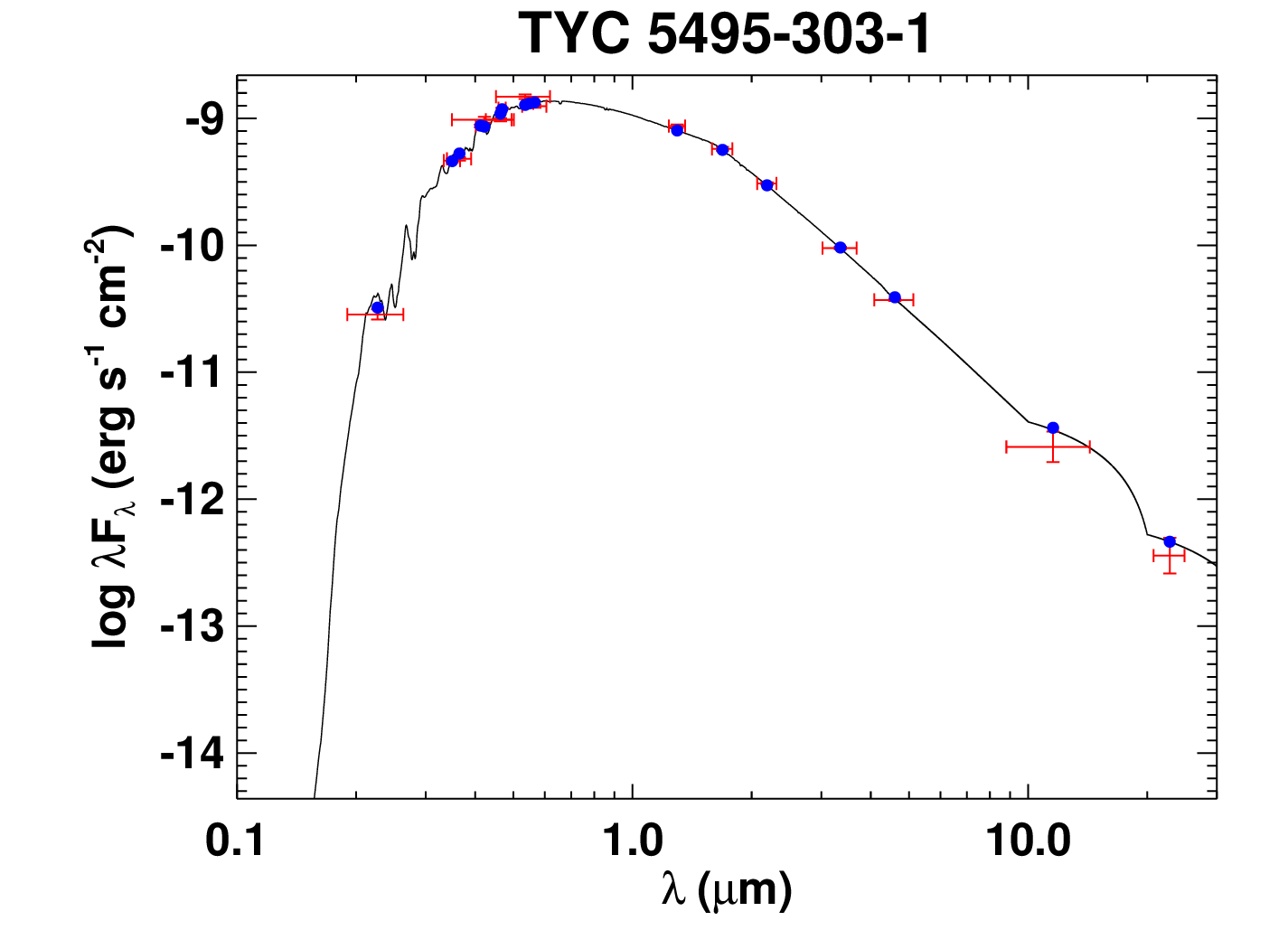}\includegraphics[width=0.333\linewidth]{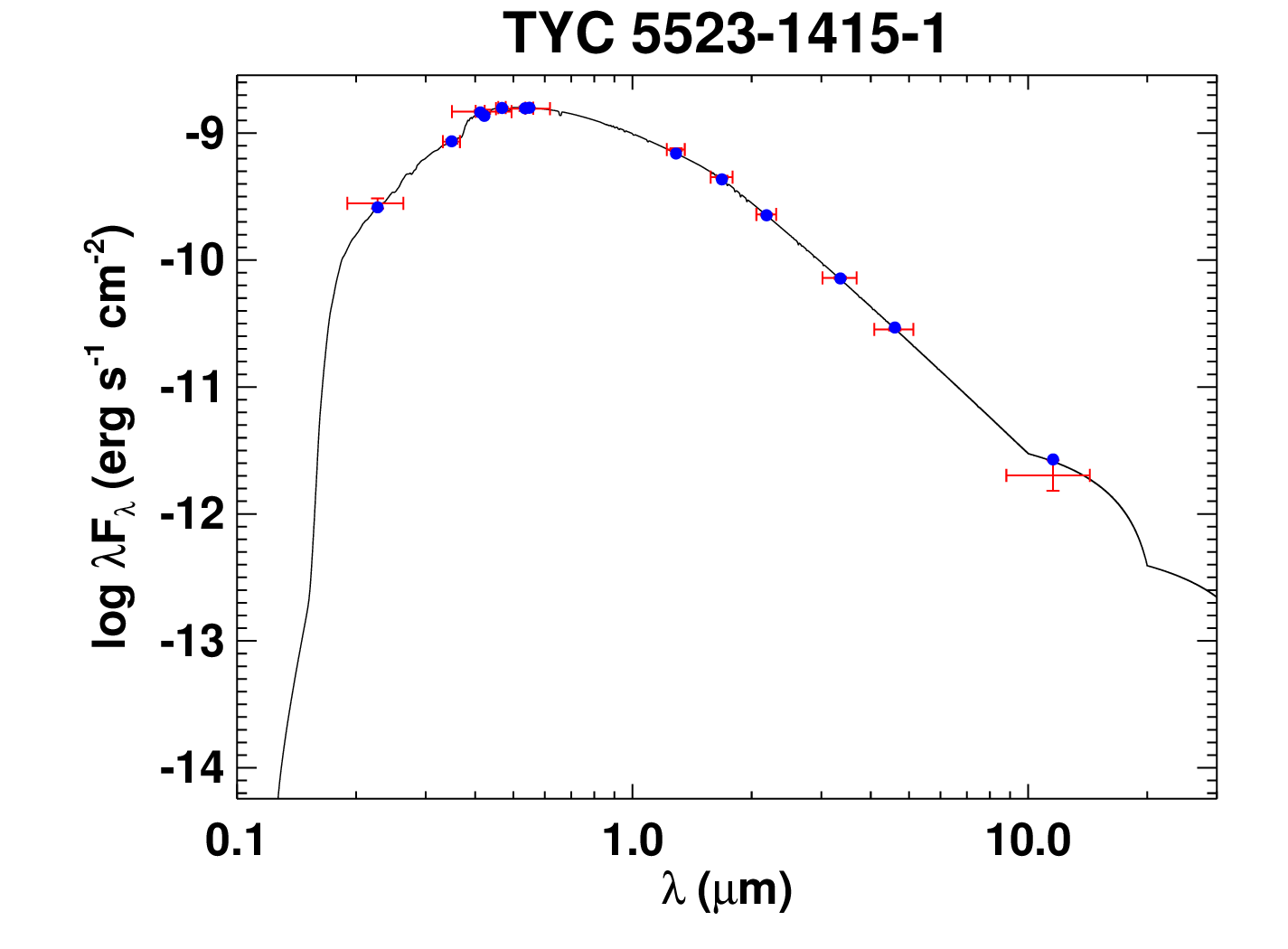}
\includegraphics[width=0.333\linewidth]{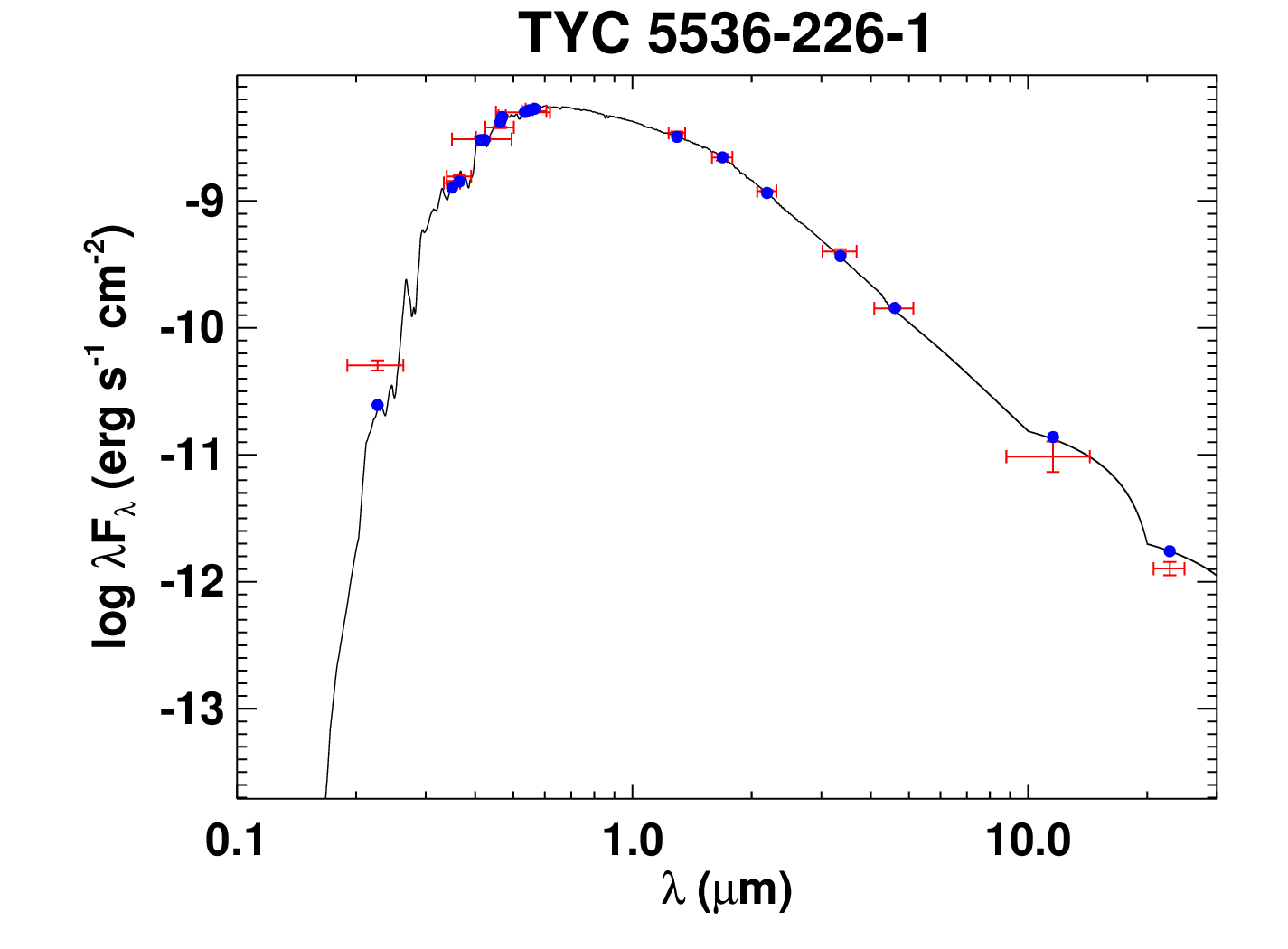}\includegraphics[width=0.333\linewidth]{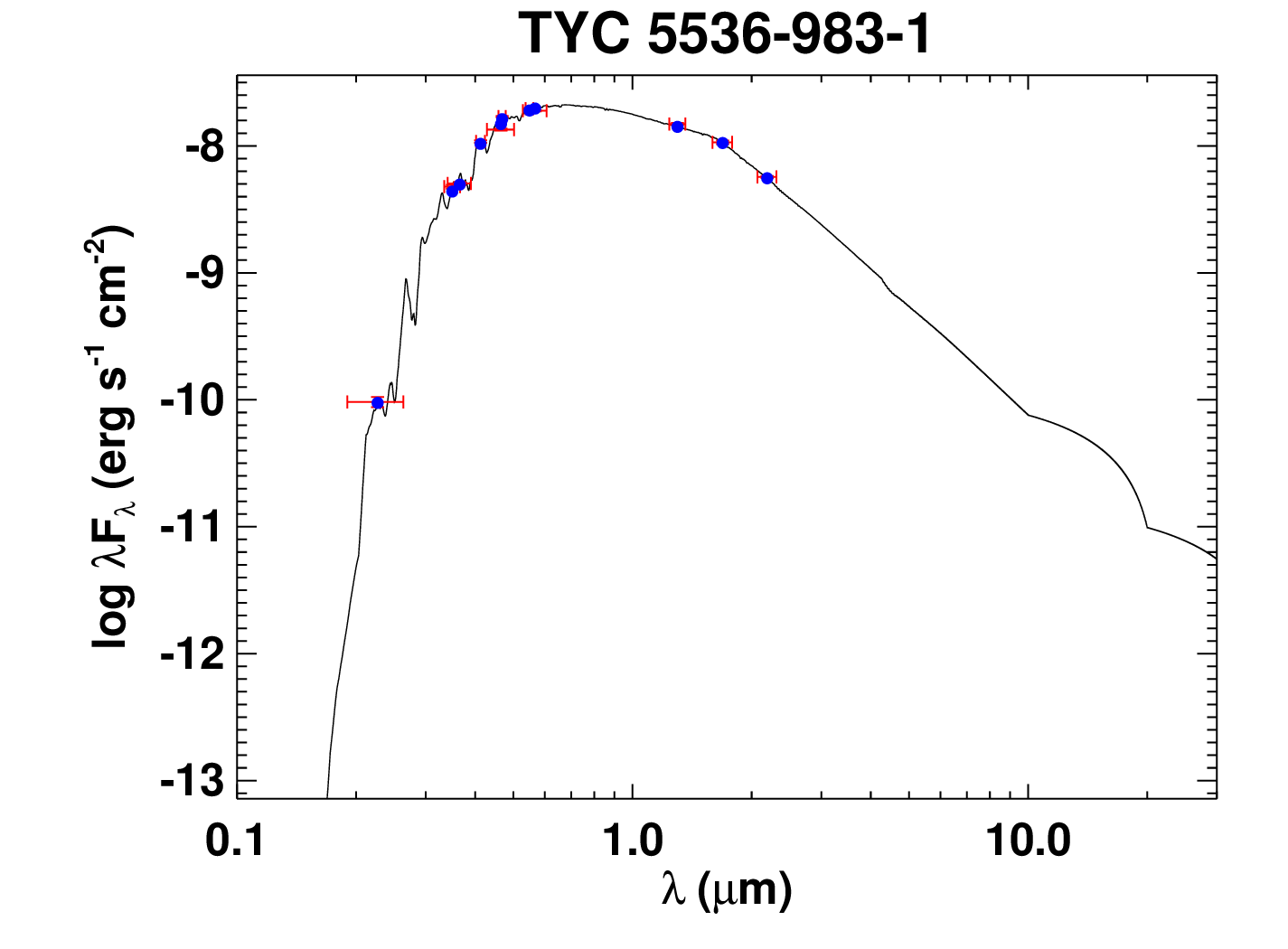}\includegraphics[width=0.333\linewidth]{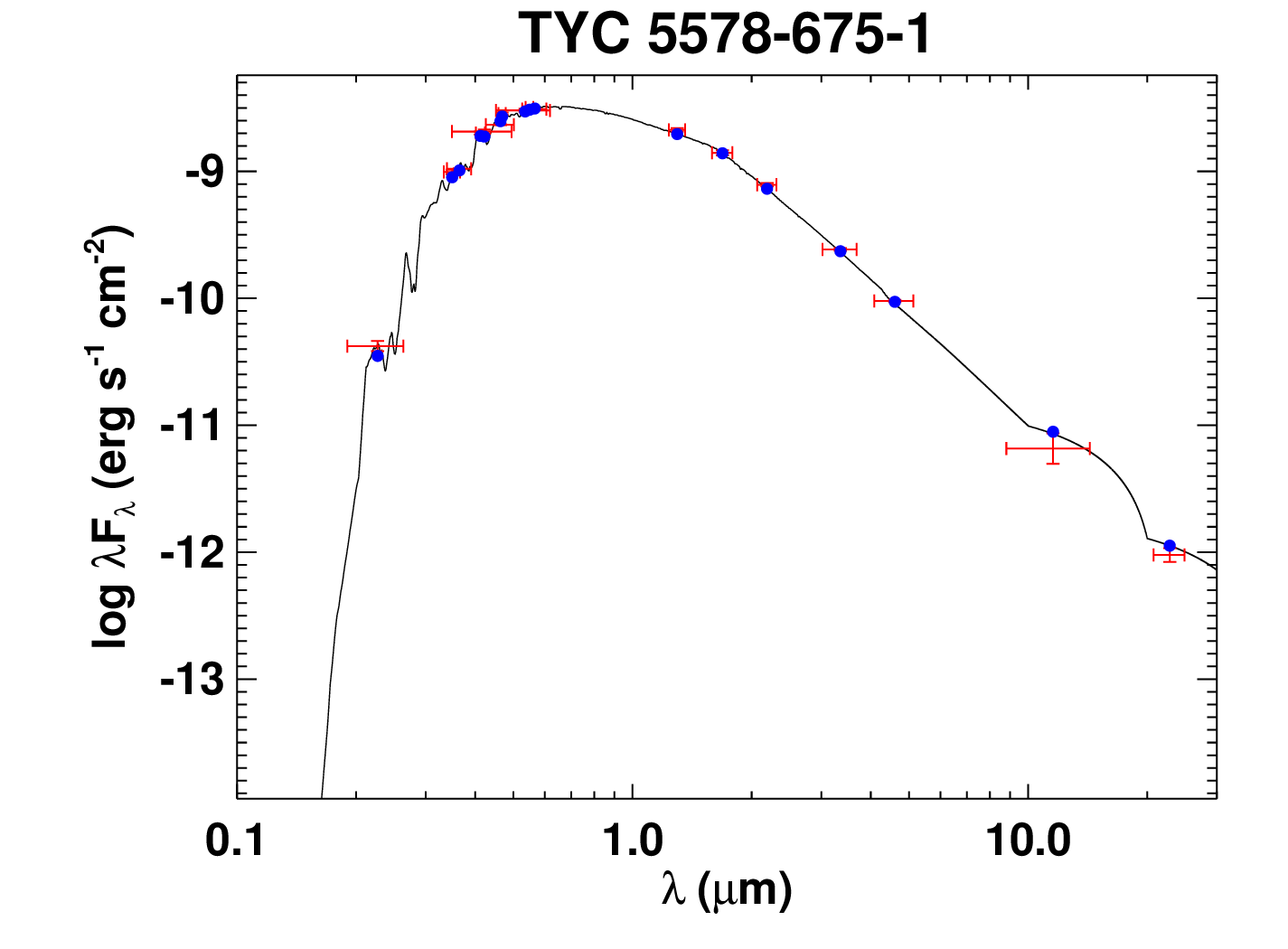}
\includegraphics[width=0.333\linewidth]{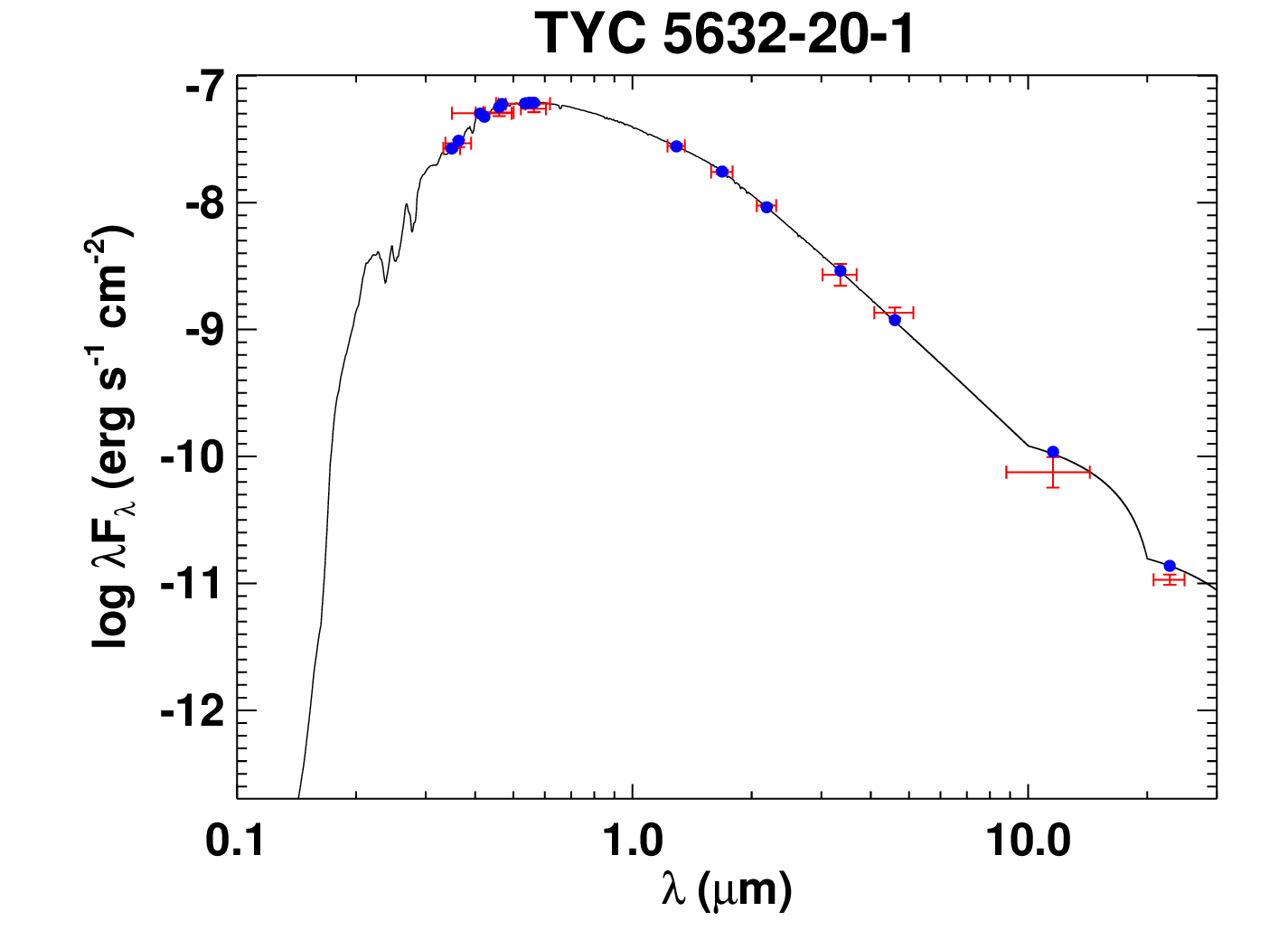}\includegraphics[width=0.333\linewidth]{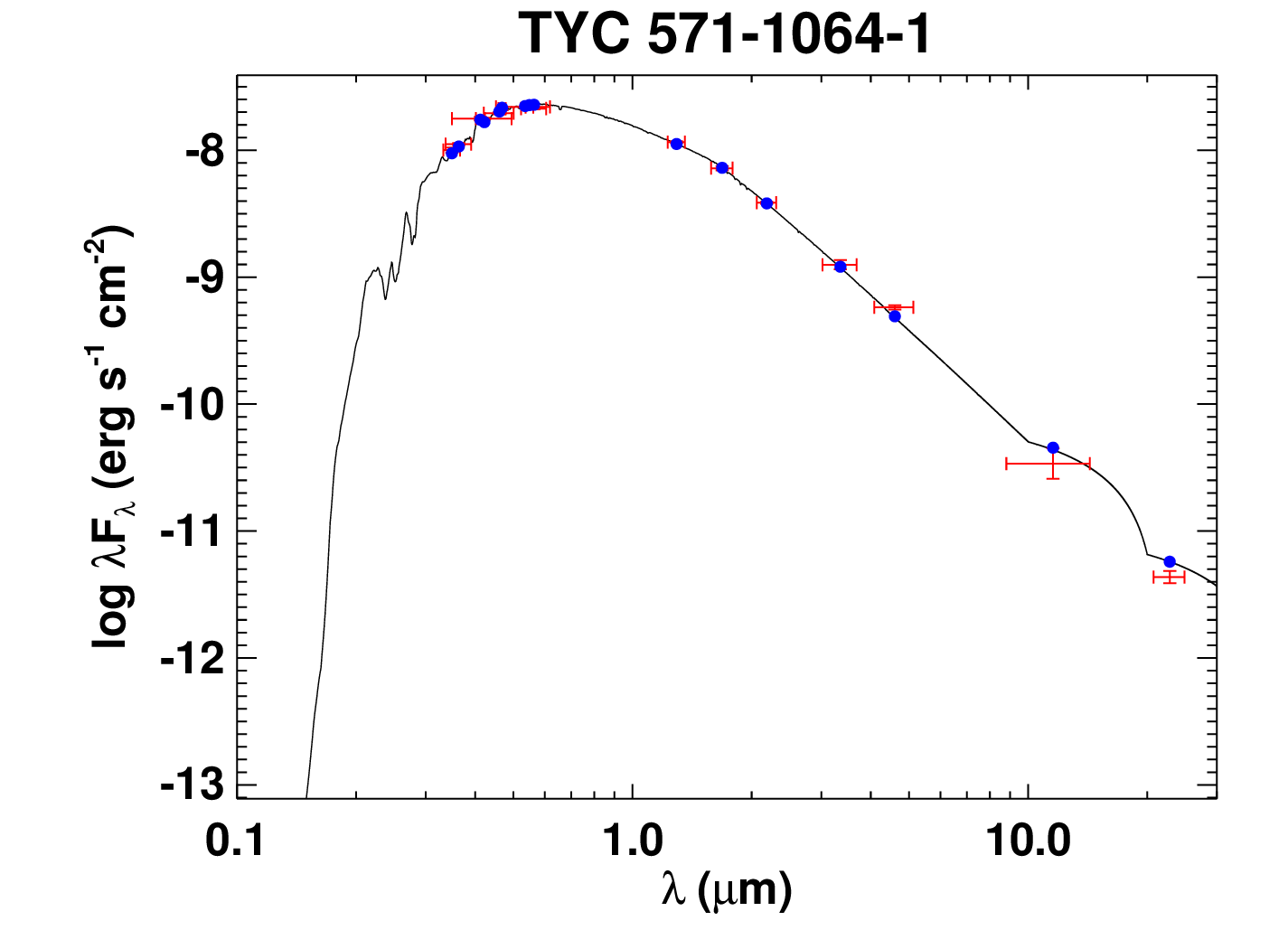}\includegraphics[width=0.333\linewidth]{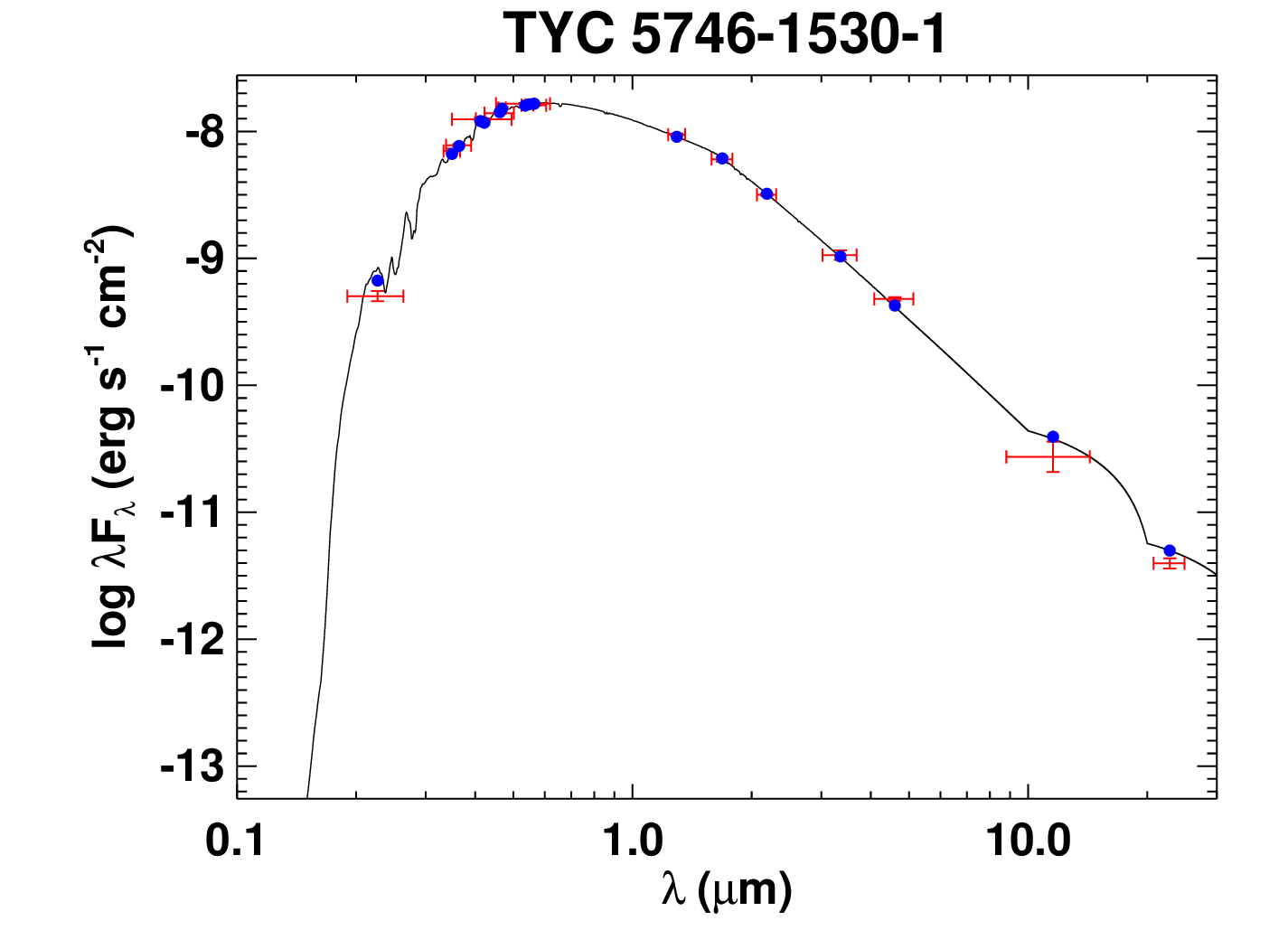}
\caption{\label{fig:seds13} All labels, lines, symbols, and colors as in Figure \ref{fig:seds}.}
\end{figure*}

\begin{figure*}
\includegraphics[width=0.333\linewidth]{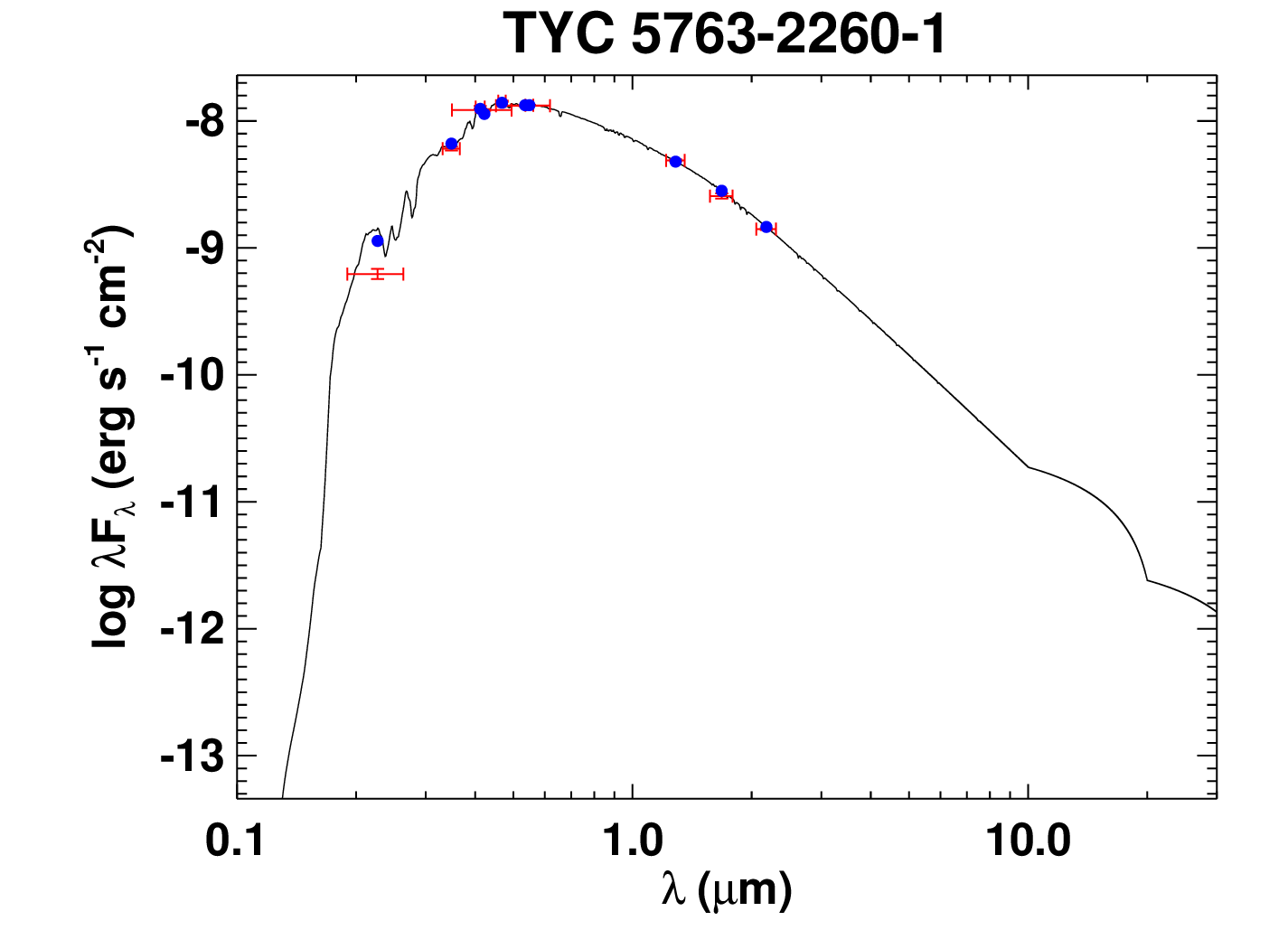}\includegraphics[width=0.333\linewidth]{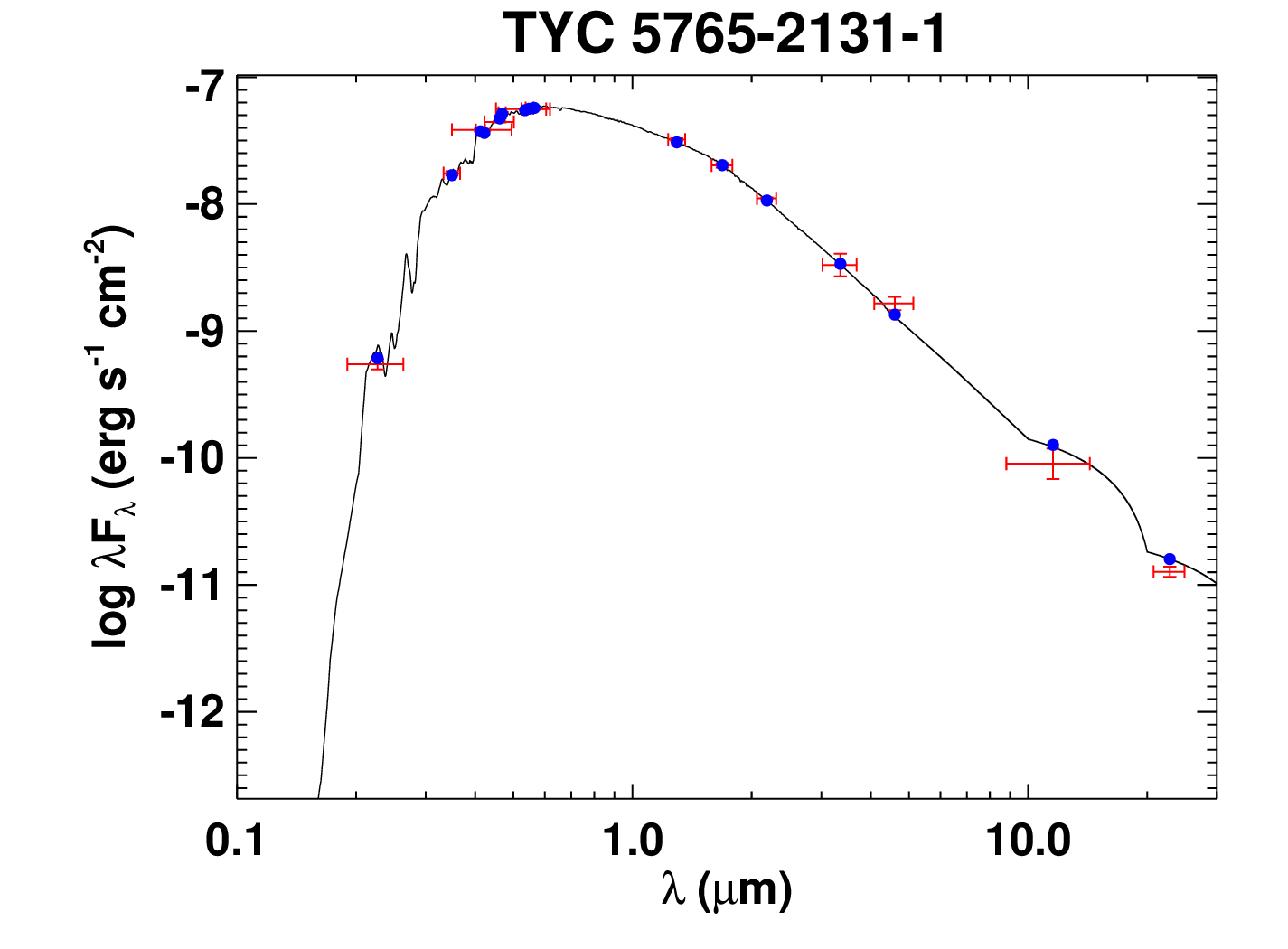}\includegraphics[width=0.333\linewidth]{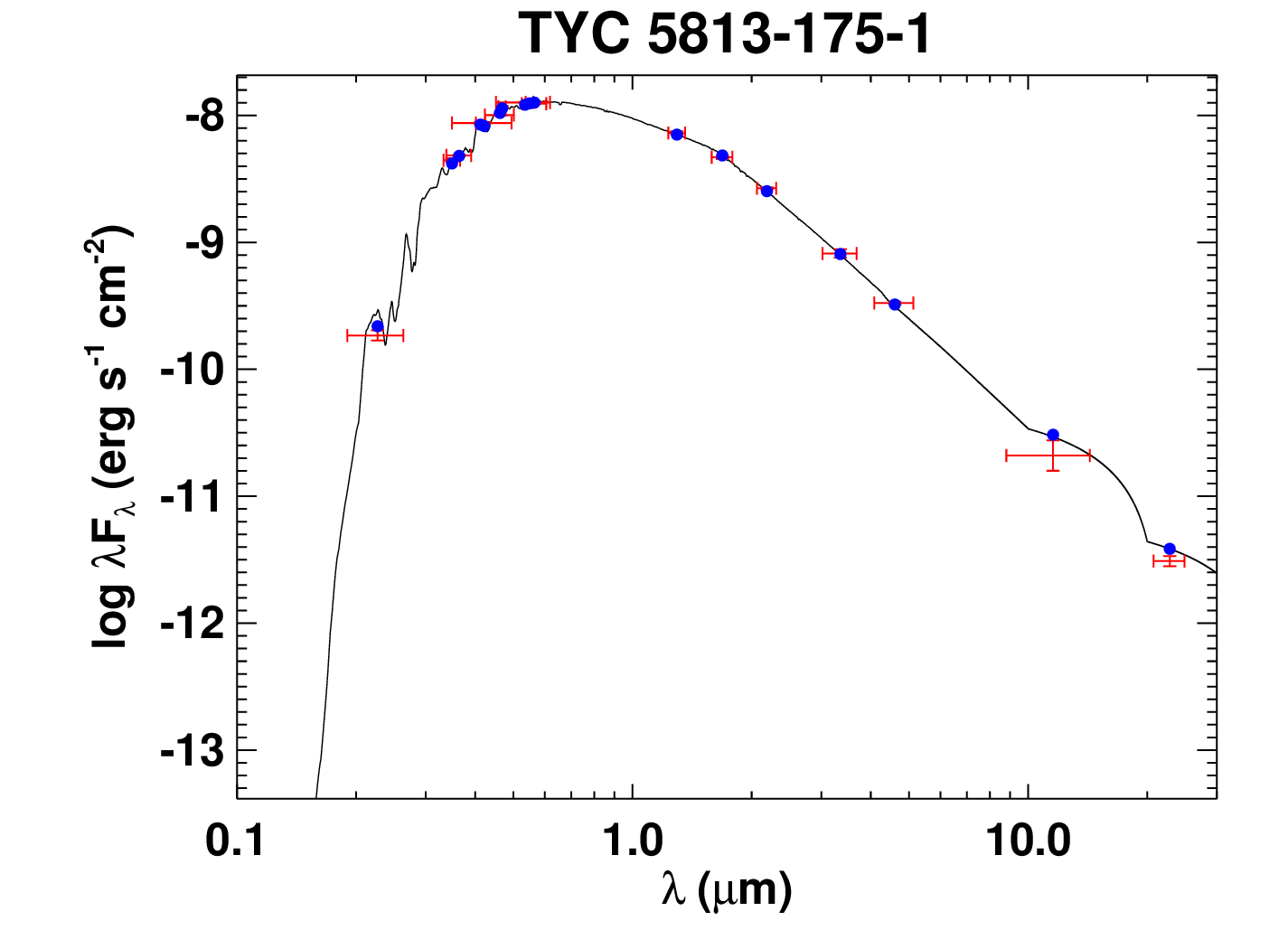}
\includegraphics[width=0.333\linewidth]{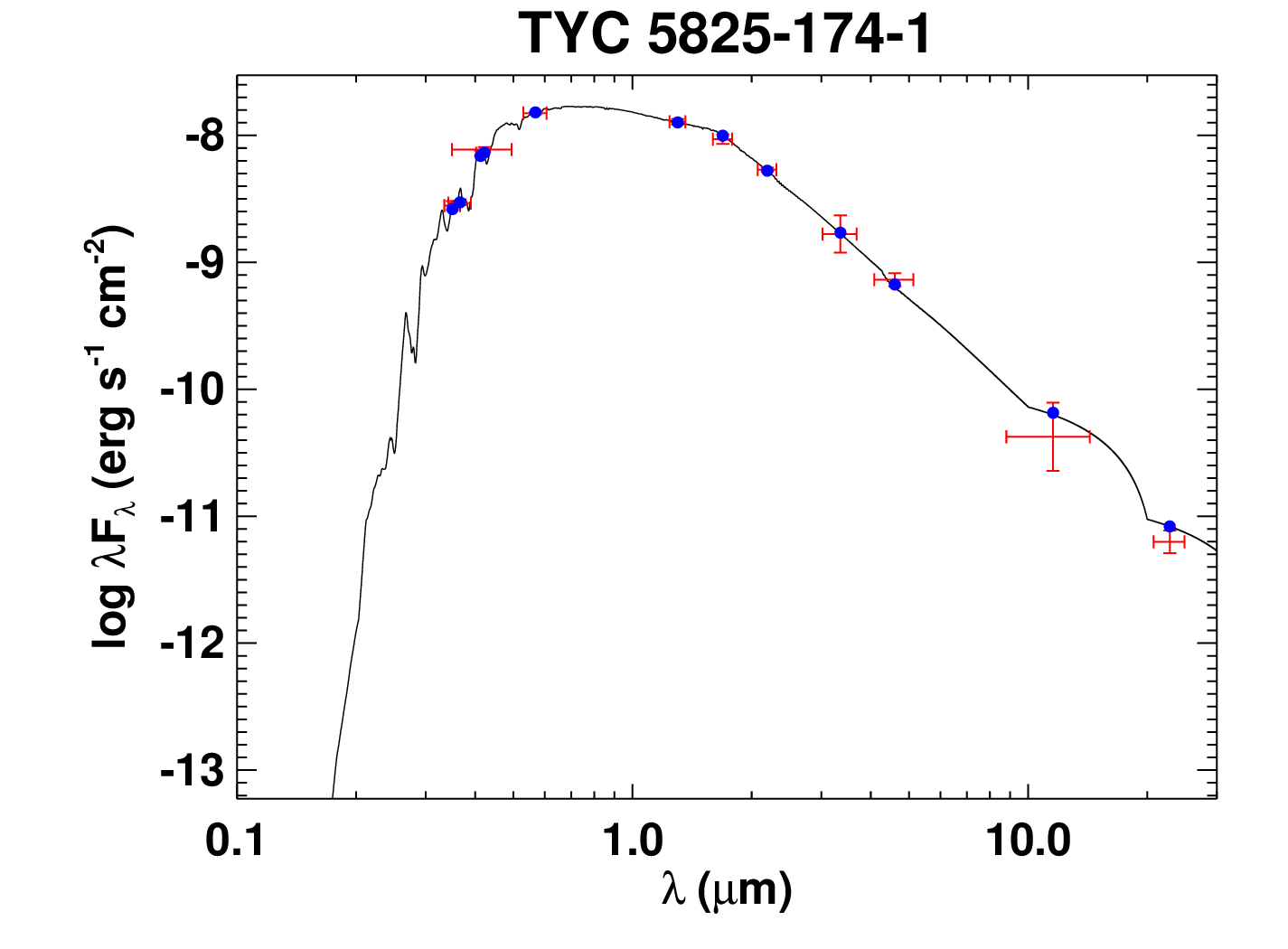}\includegraphics[width=0.333\linewidth]{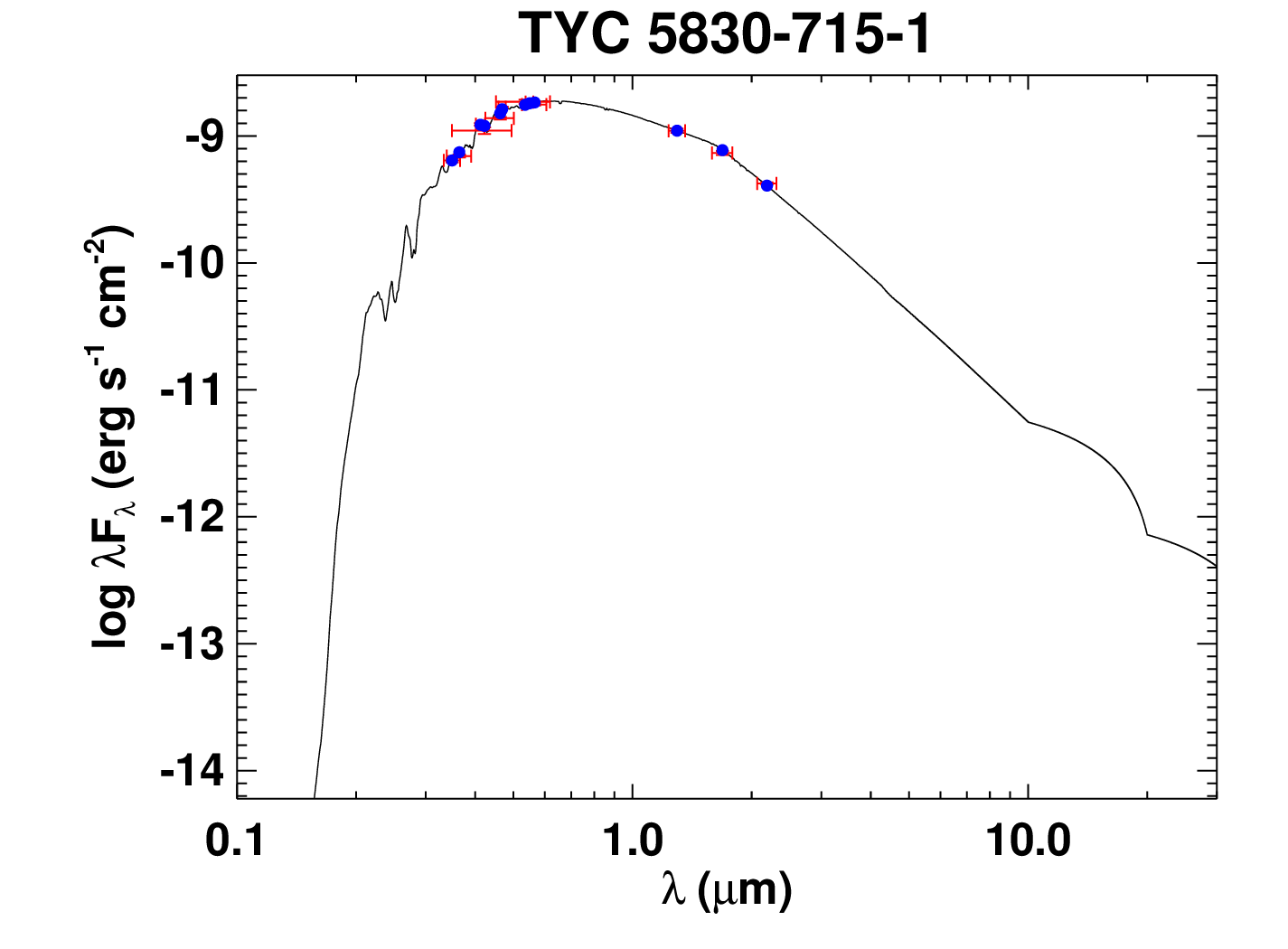}\includegraphics[width=0.333\linewidth]{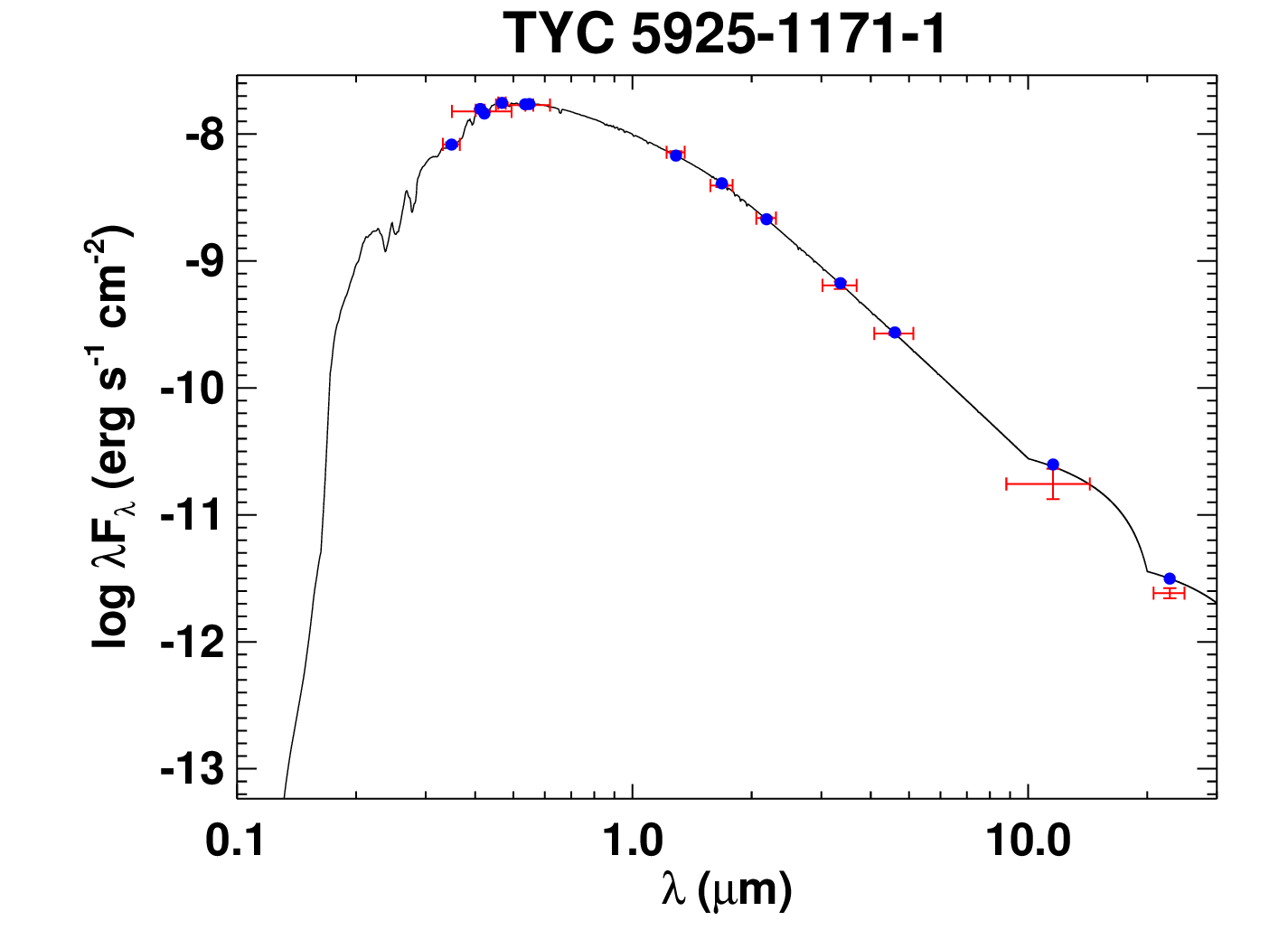}
\includegraphics[width=0.333\linewidth]{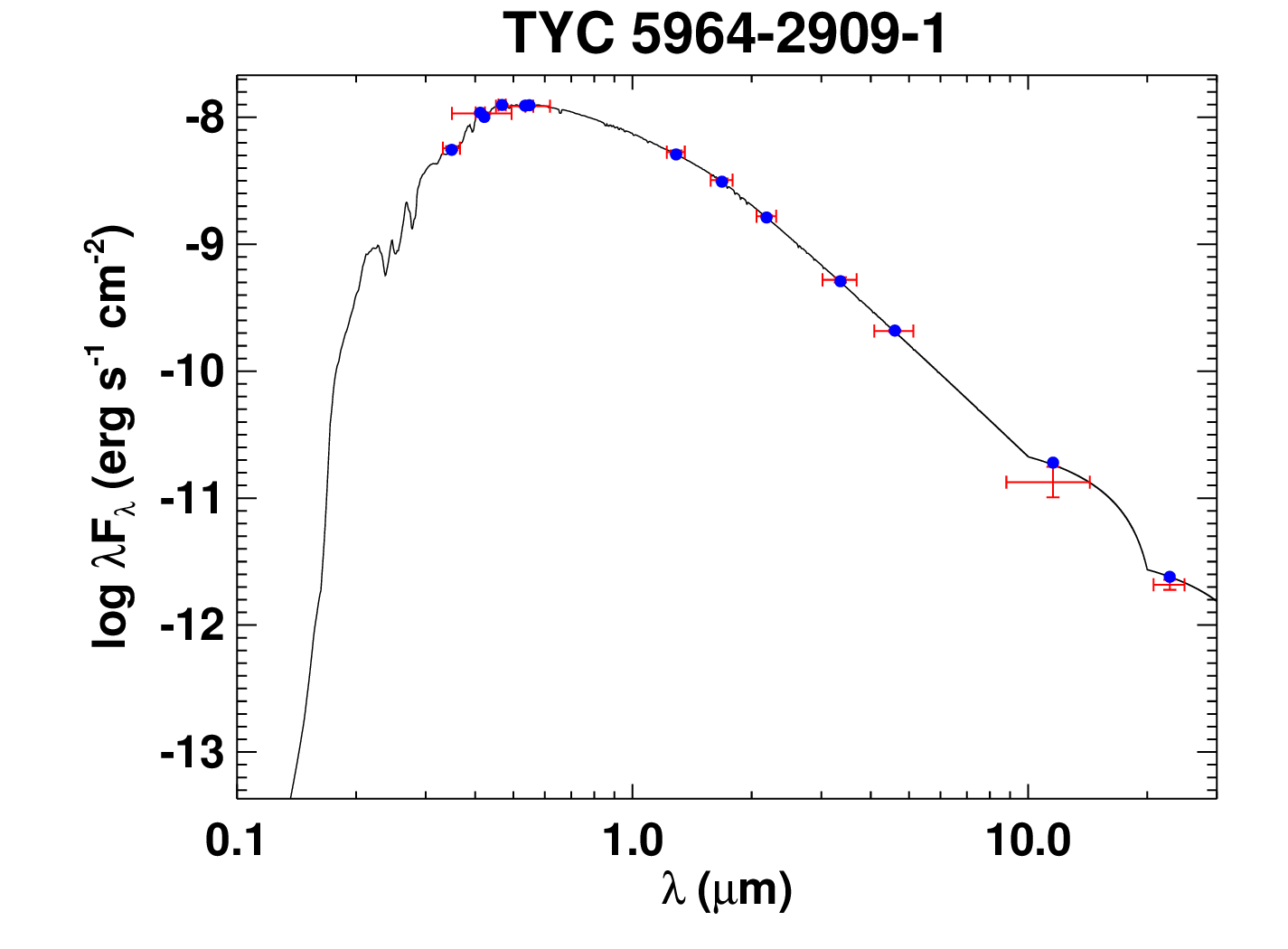}\includegraphics[width=0.333\linewidth]{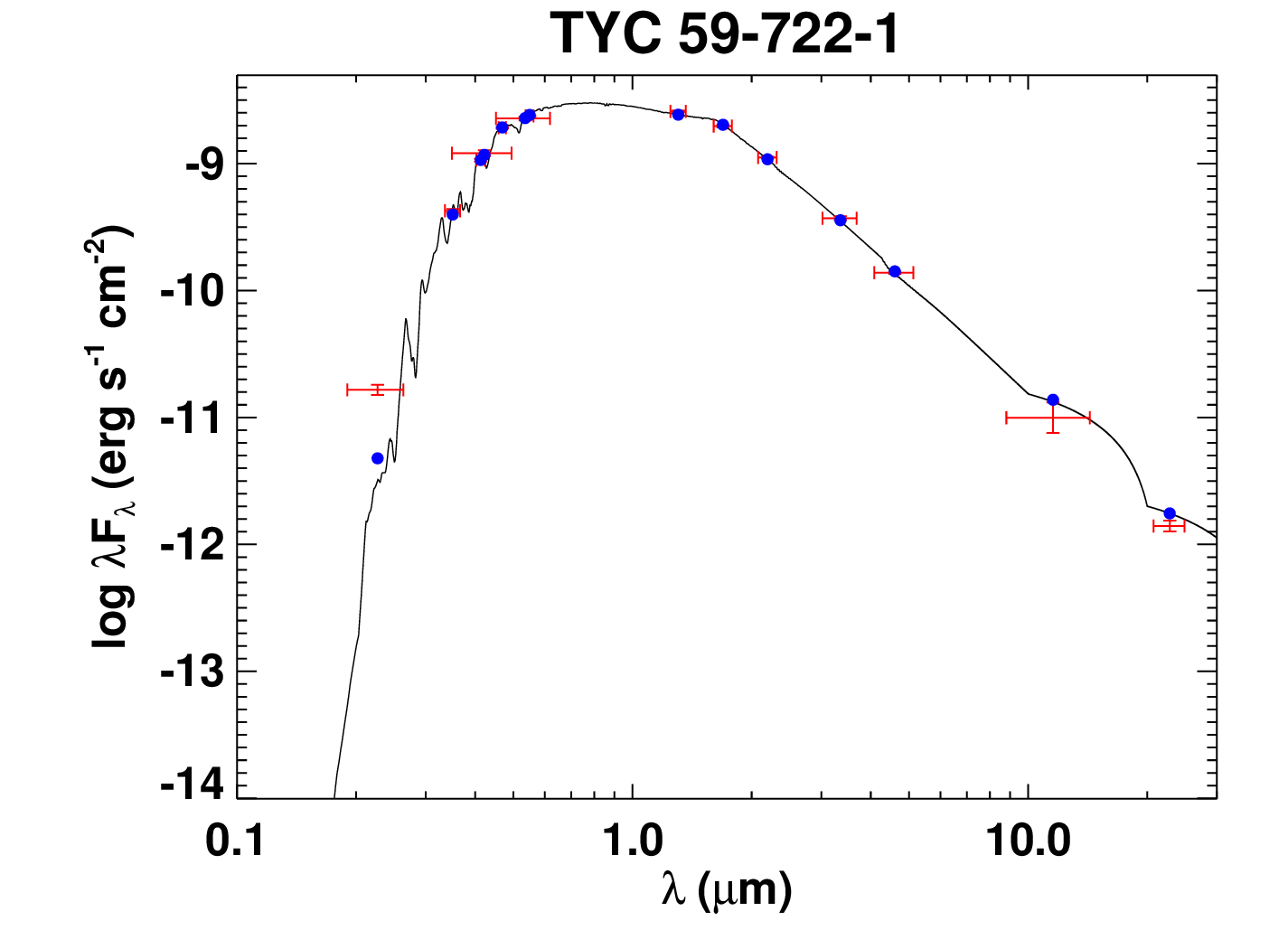}\includegraphics[width=0.333\linewidth]{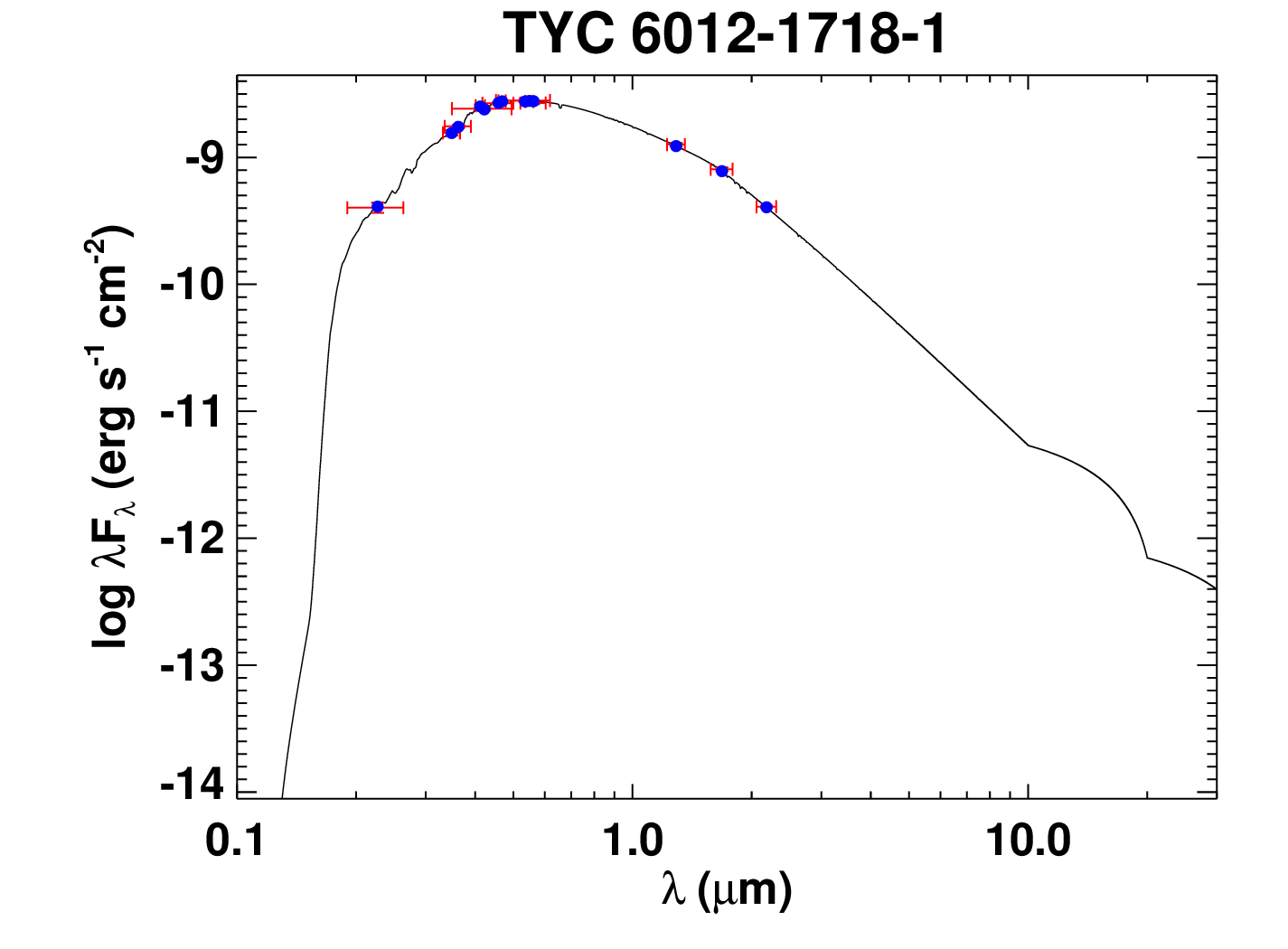}
\includegraphics[width=0.333\linewidth]{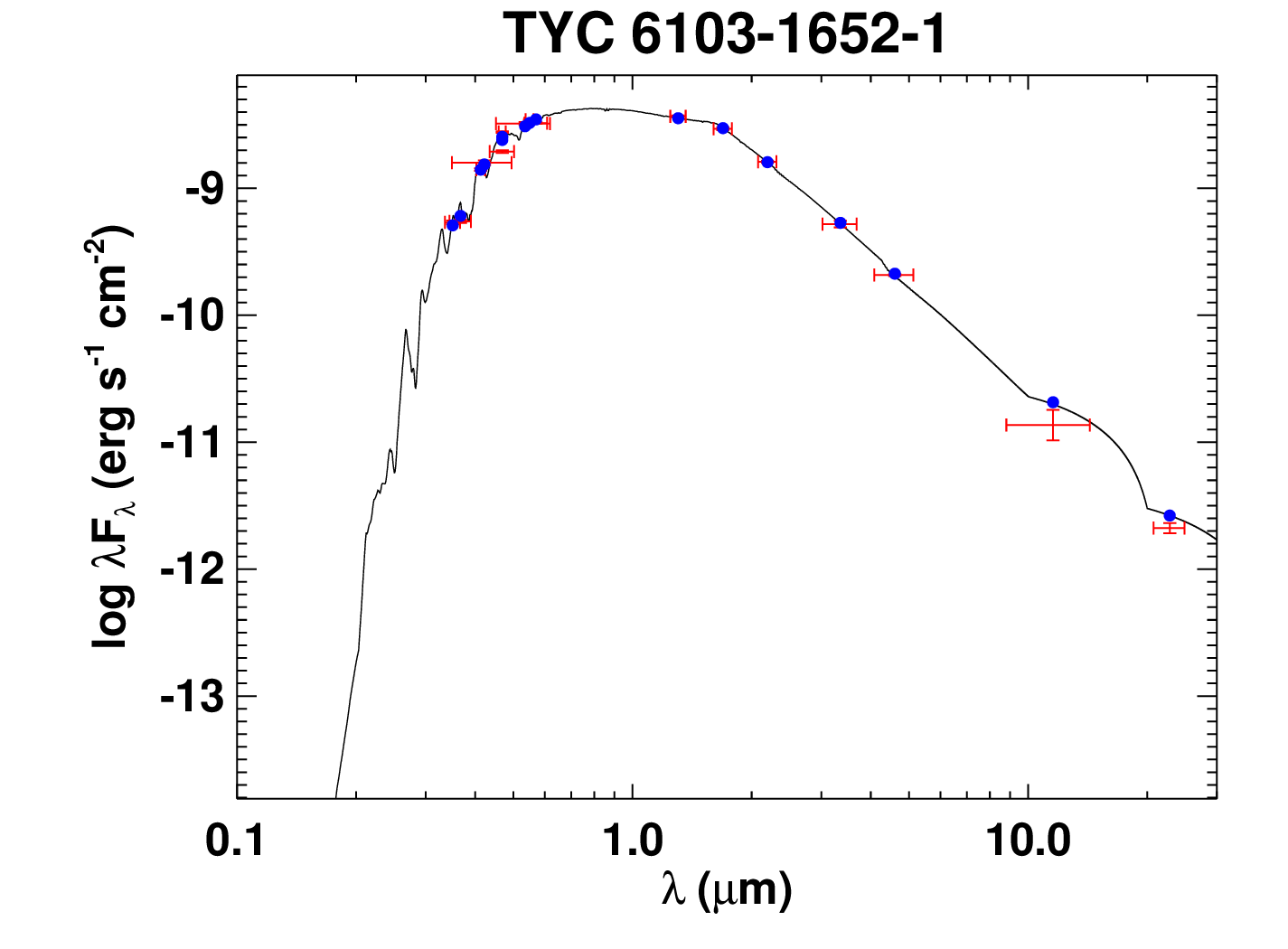}\includegraphics[width=0.333\linewidth]{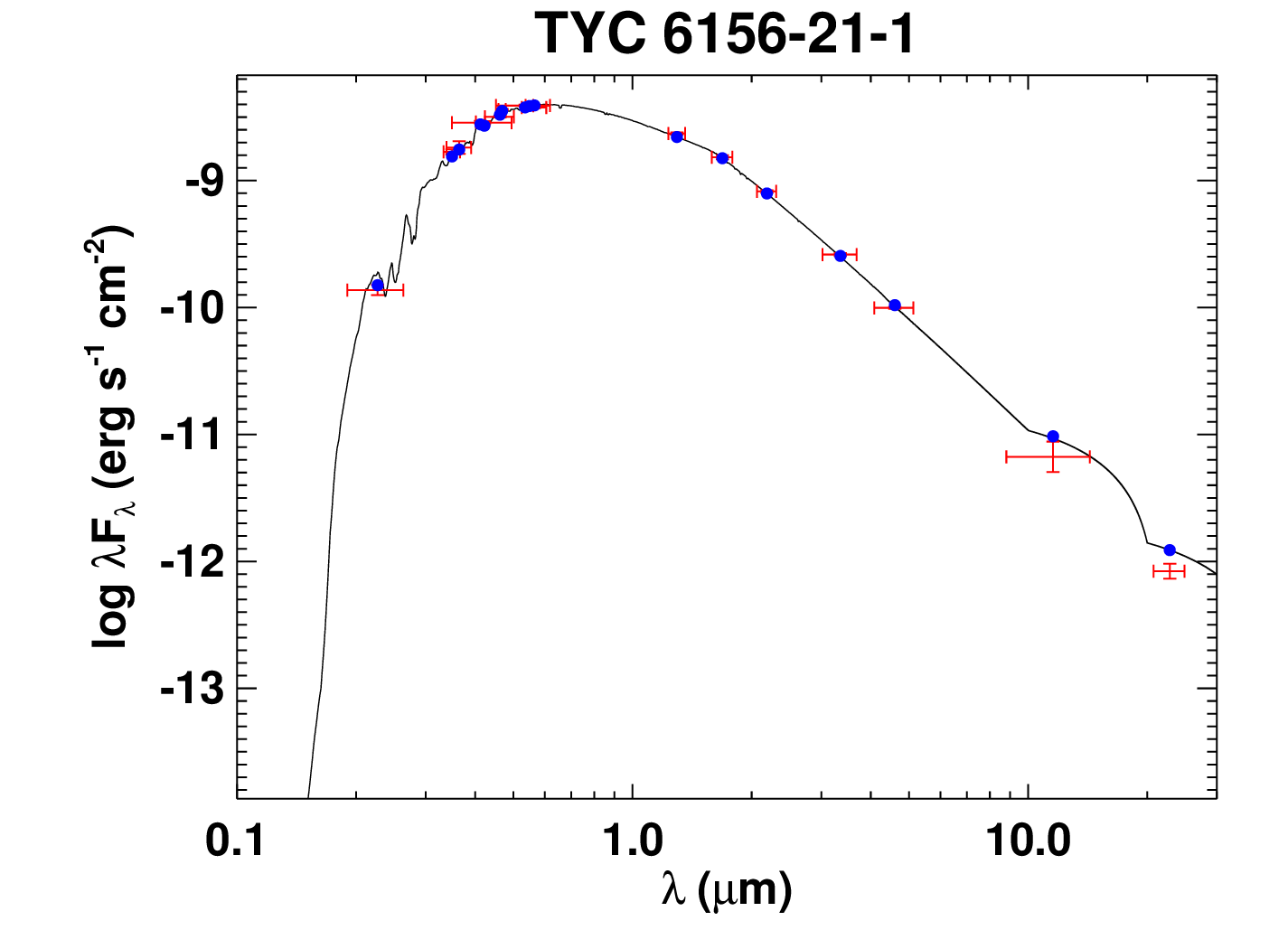}\includegraphics[width=0.333\linewidth]{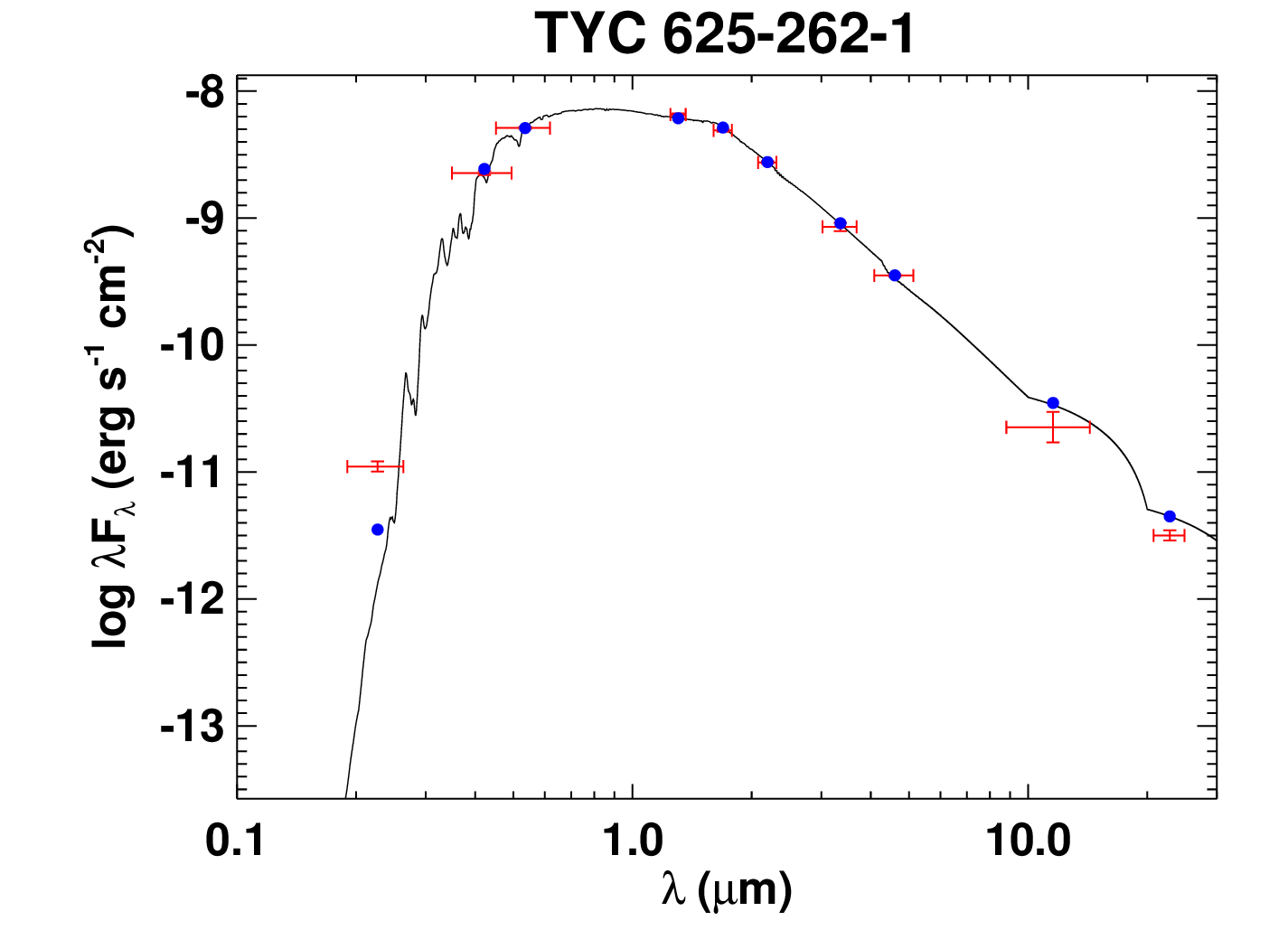}
\caption{\label{fig:seds14} All labels, lines, symbols, and colors as in Figure \ref{fig:seds}.}
\end{figure*}

\begin{figure*}
\includegraphics[width=0.333\linewidth]{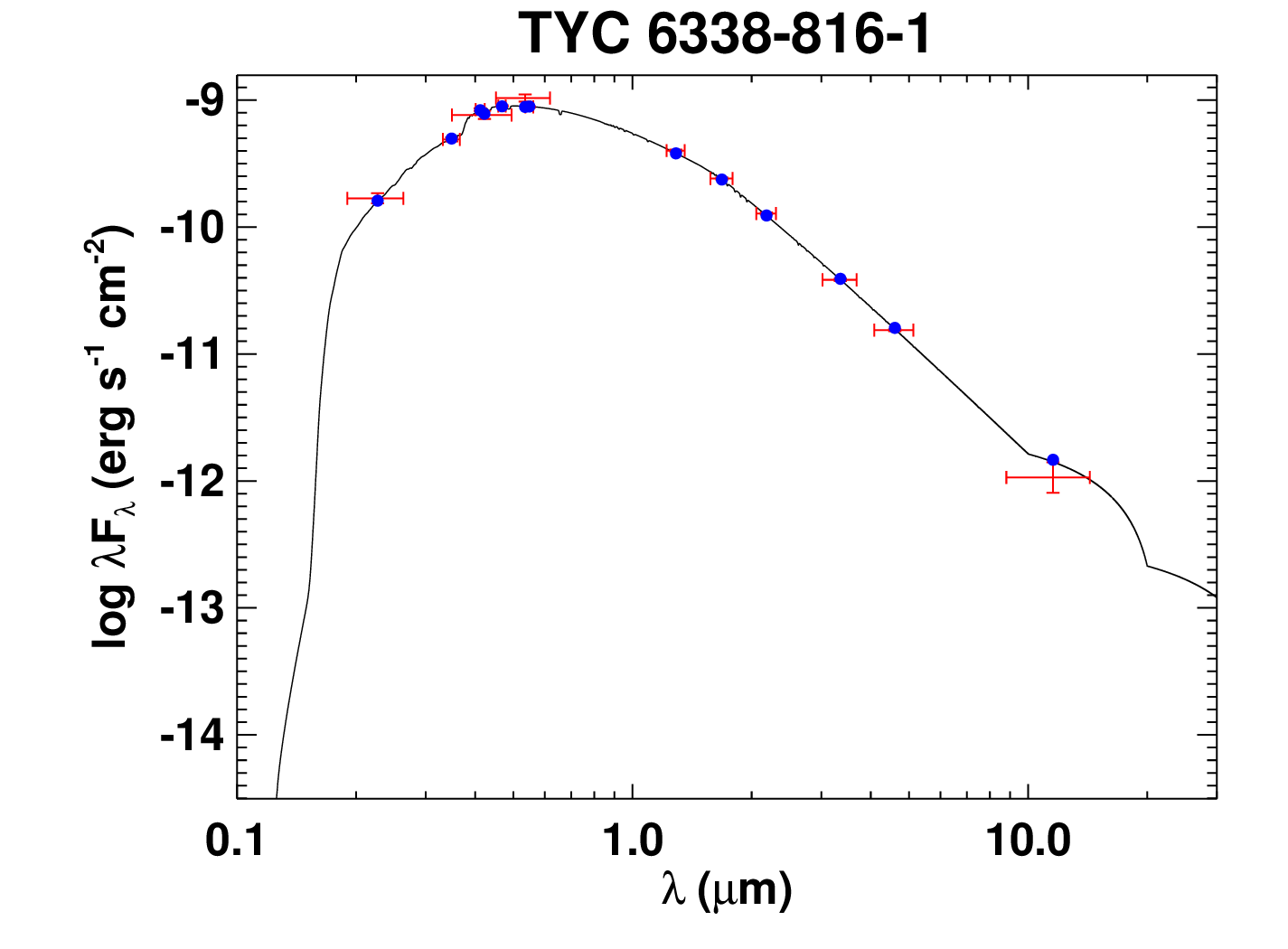}\includegraphics[width=0.333\linewidth]{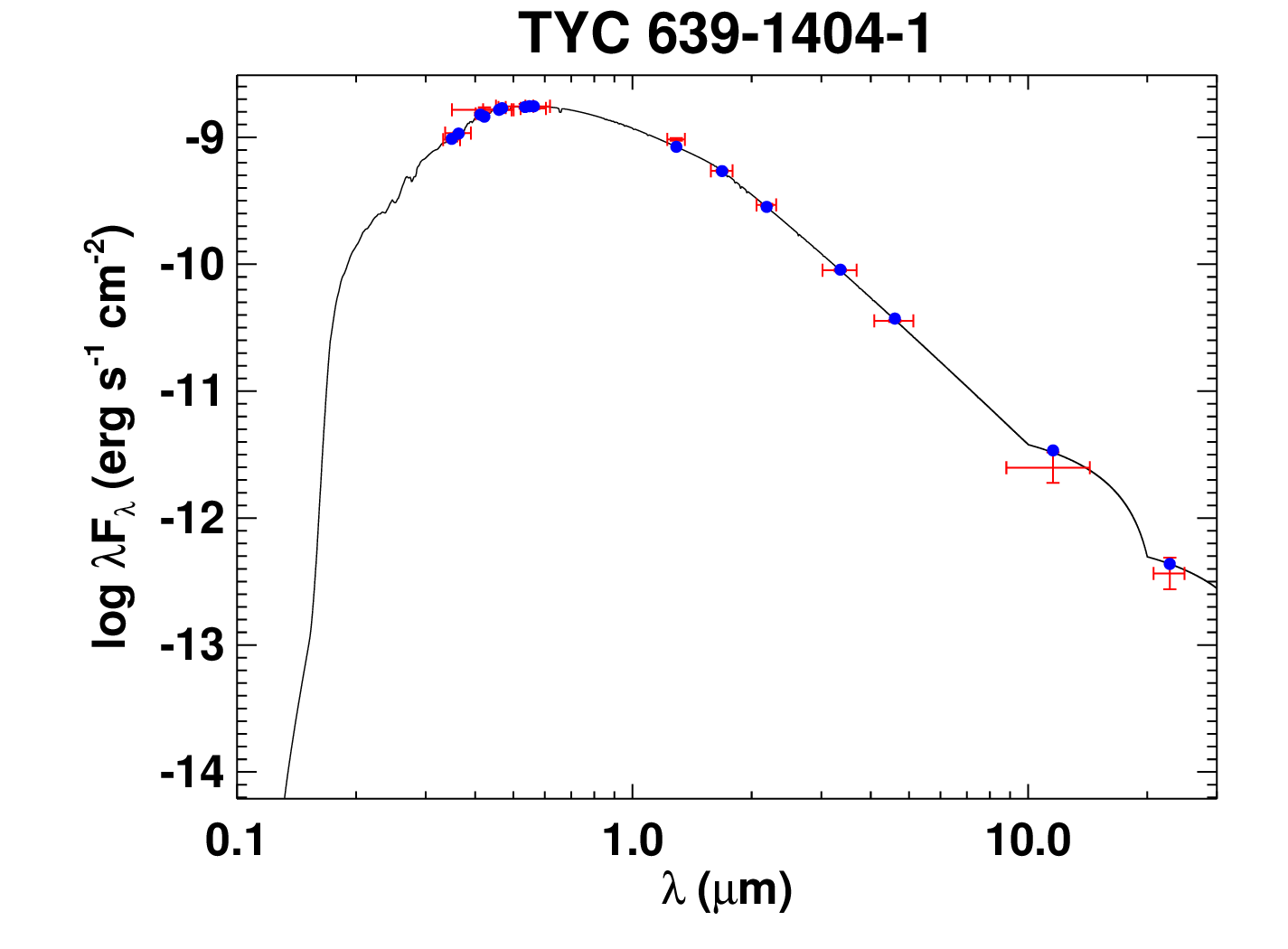}\includegraphics[width=0.333\linewidth]{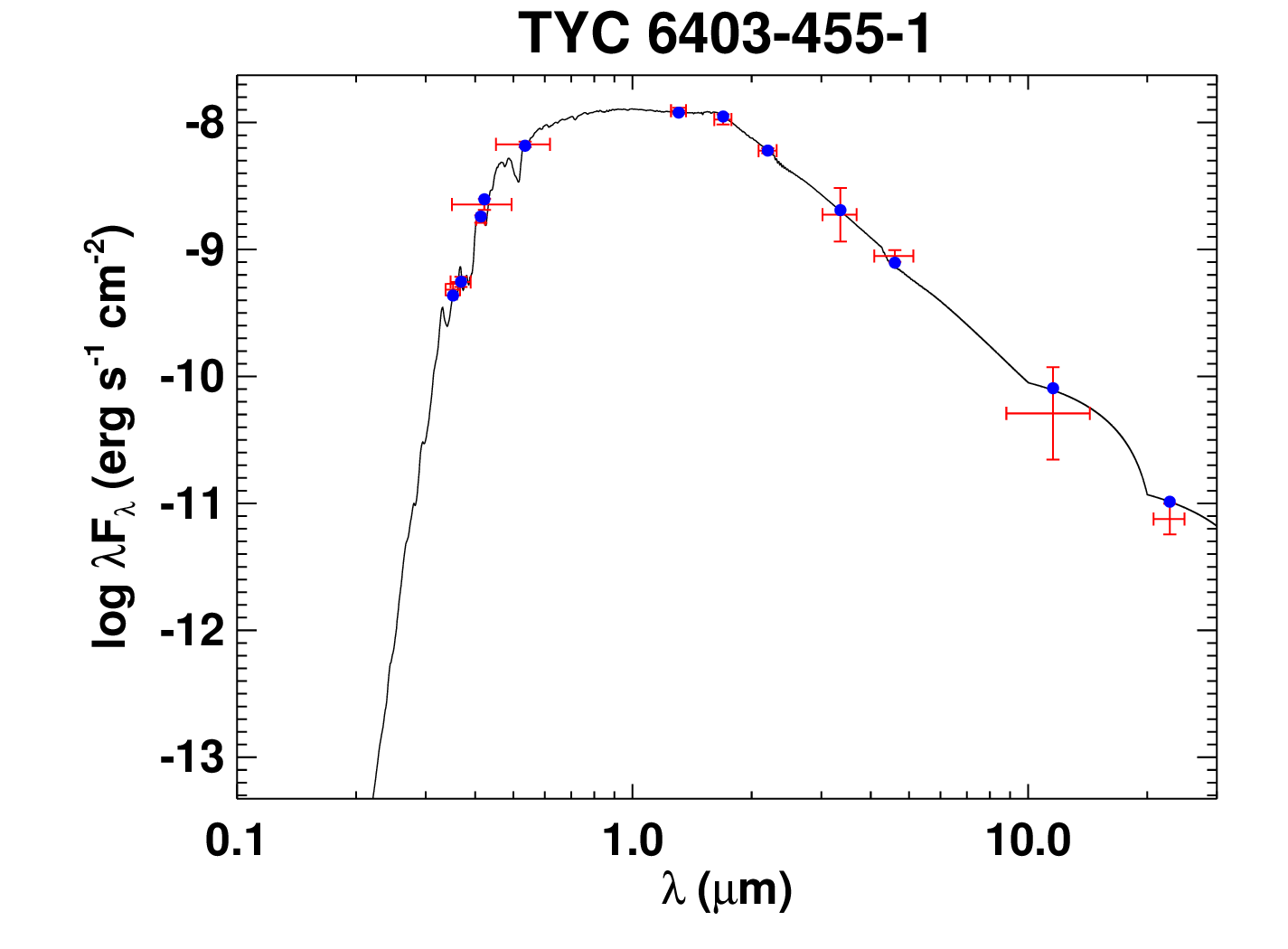}
\includegraphics[width=0.333\linewidth]{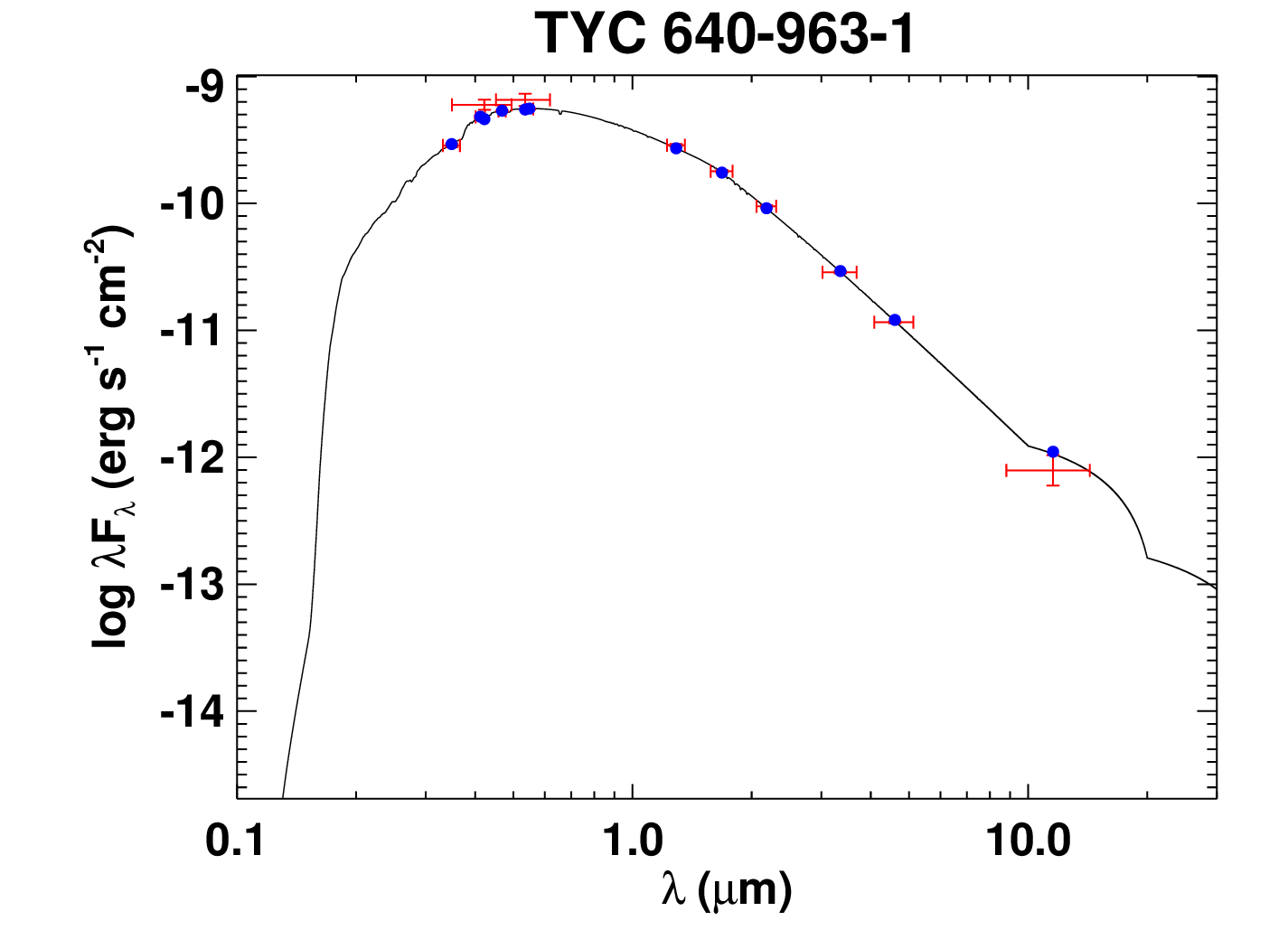}\includegraphics[width=0.333\linewidth]{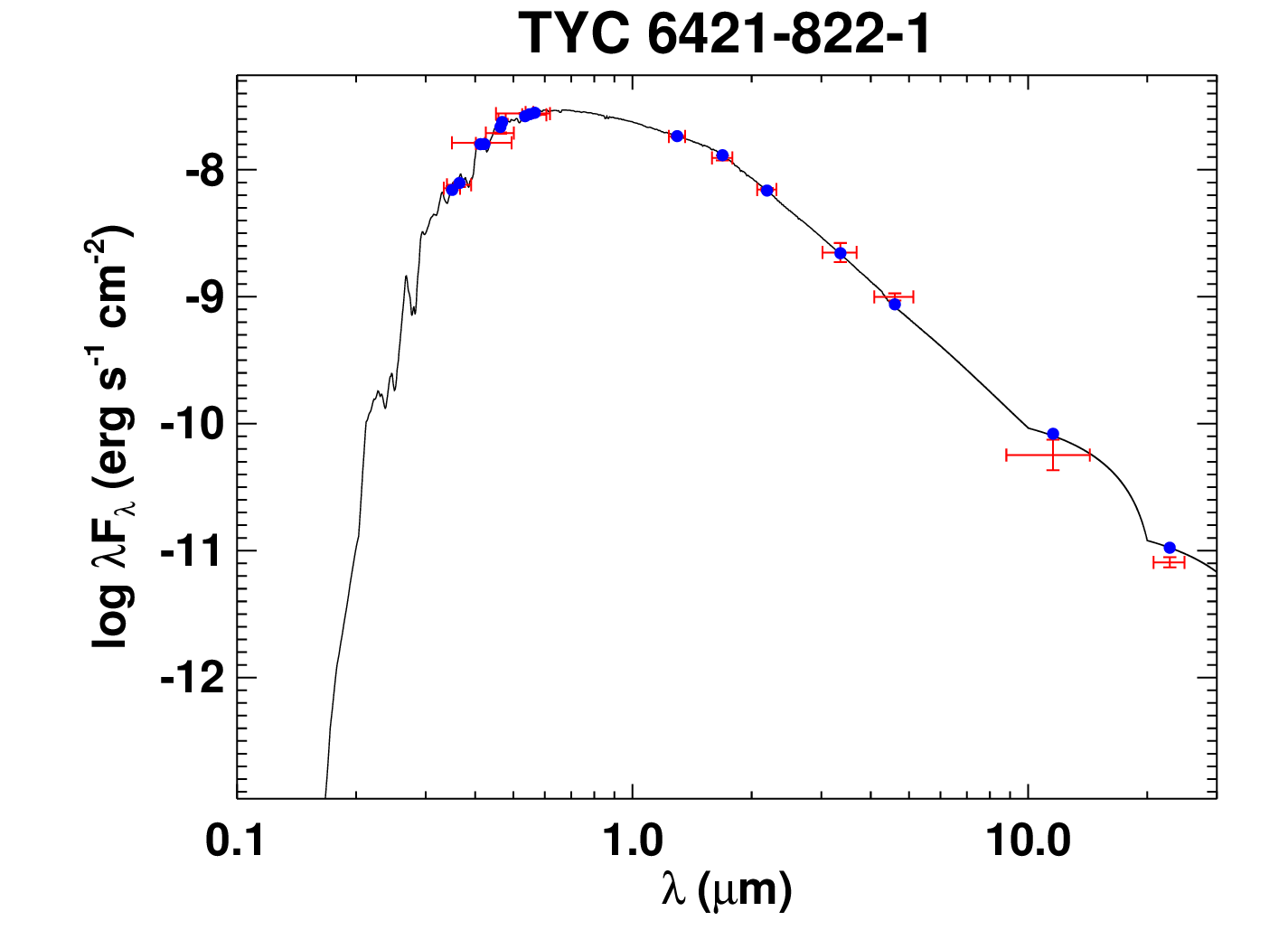}\includegraphics[width=0.333\linewidth]{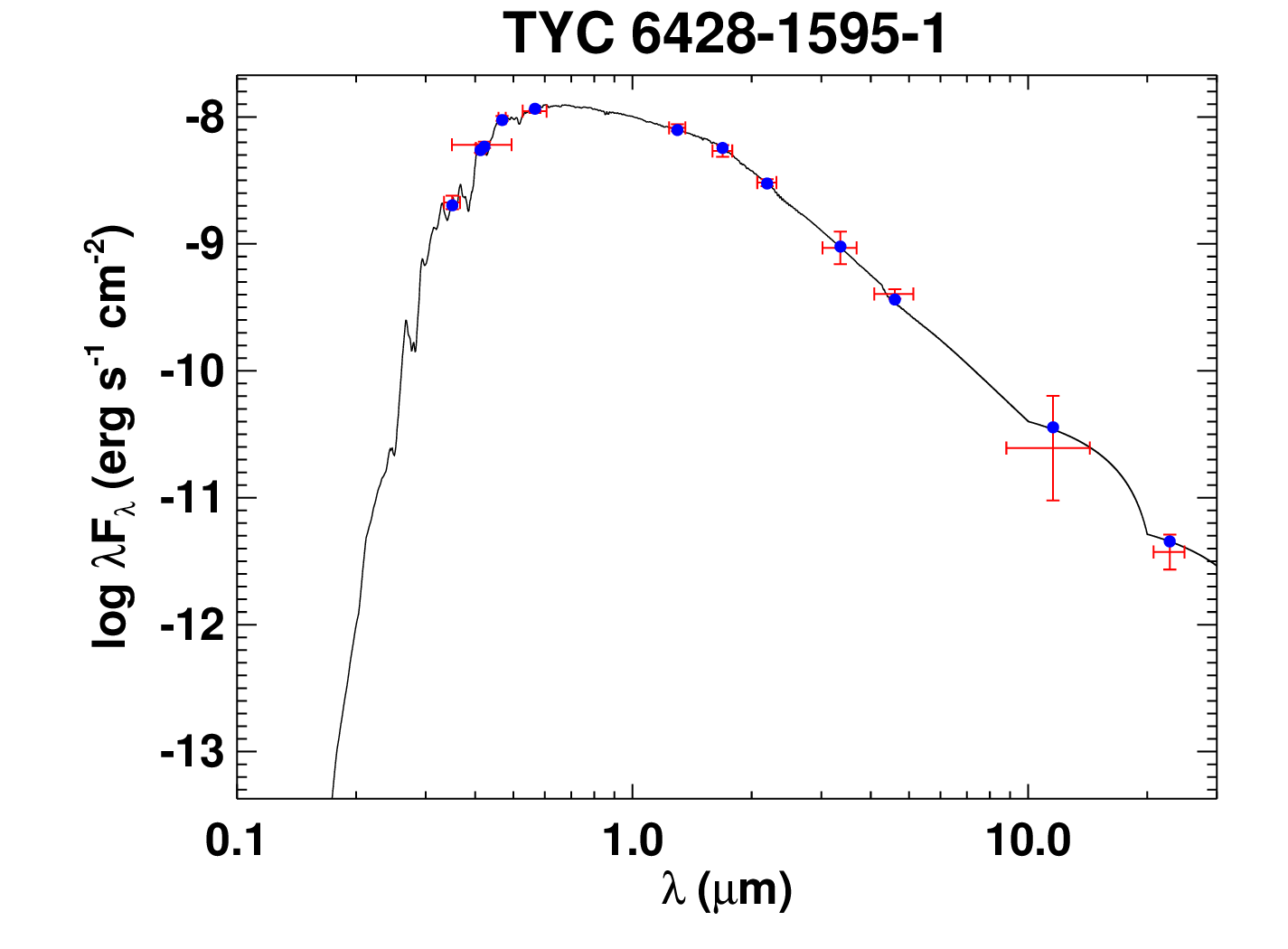}
\includegraphics[width=0.333\linewidth]{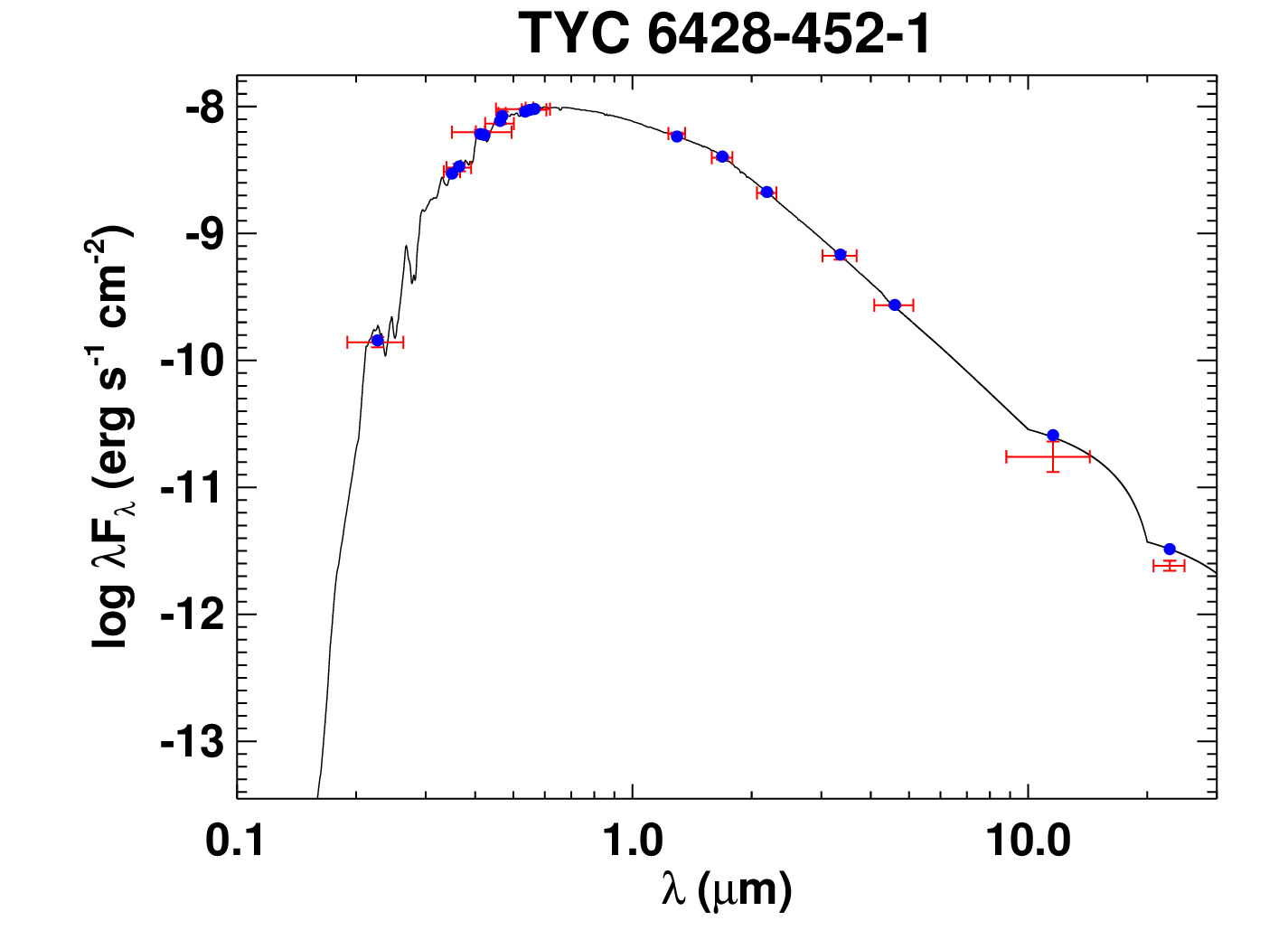}\includegraphics[width=0.333\linewidth]{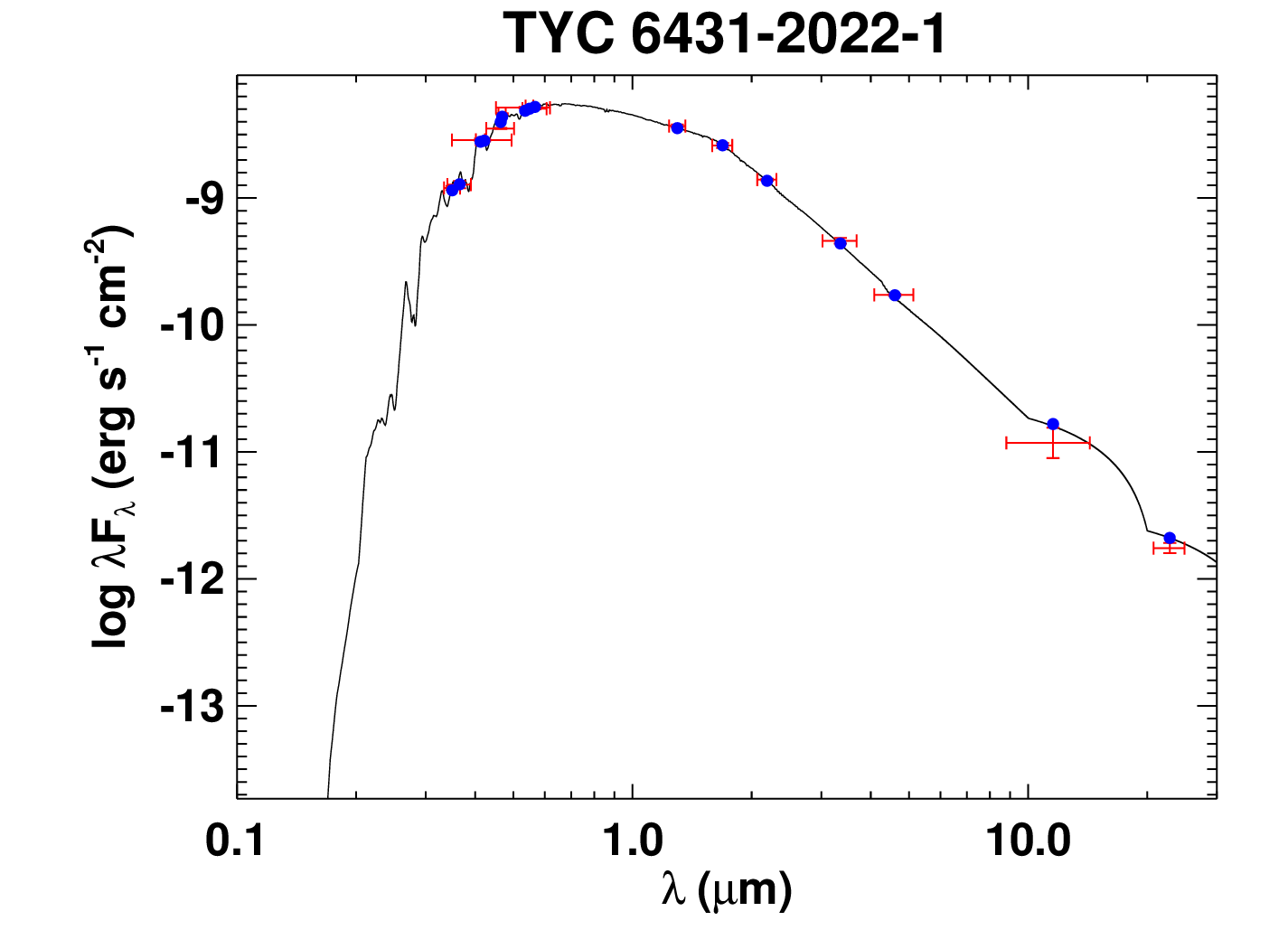}\includegraphics[width=0.333\linewidth]{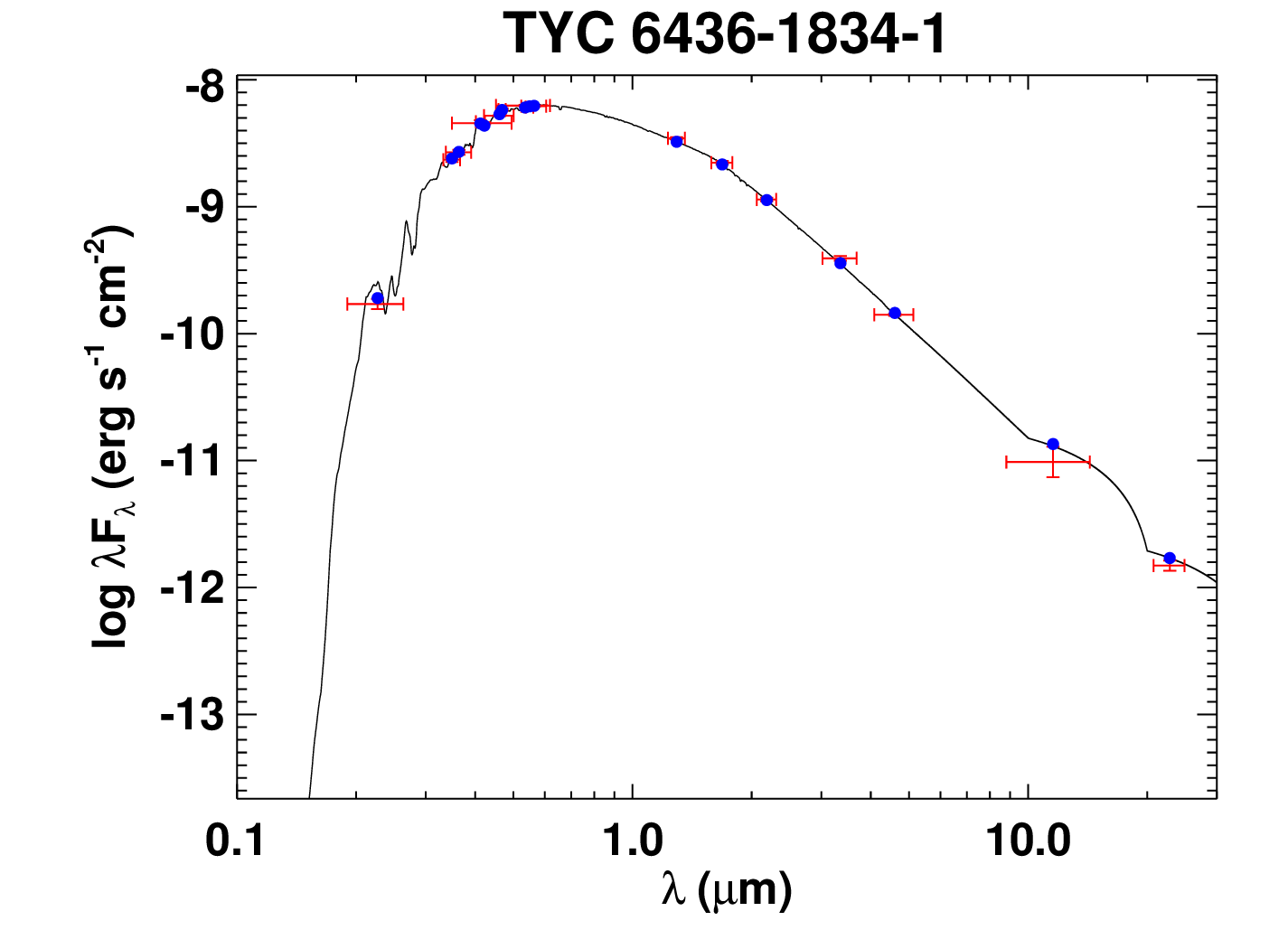}
\includegraphics[width=0.333\linewidth]{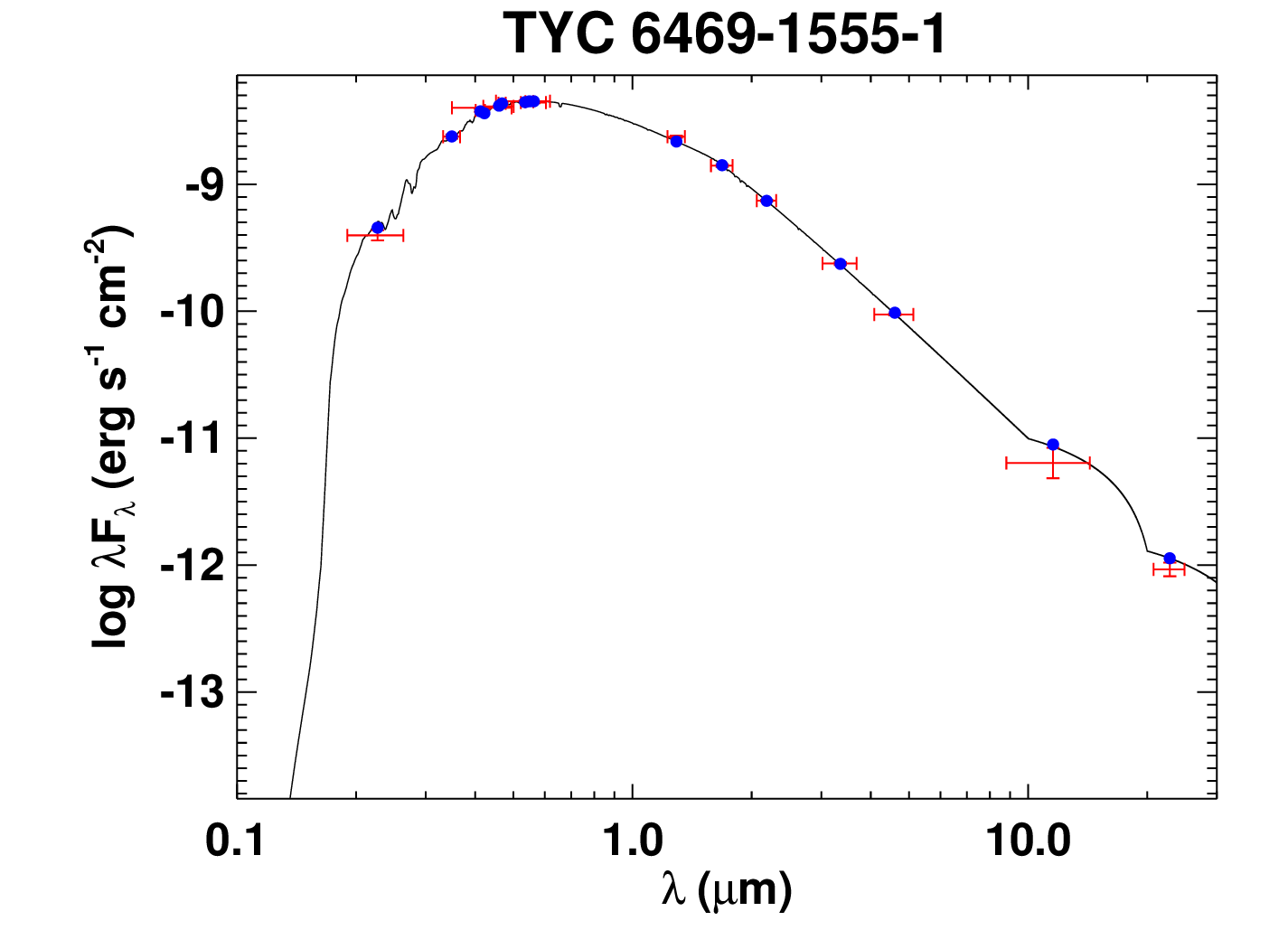}\includegraphics[width=0.333\linewidth]{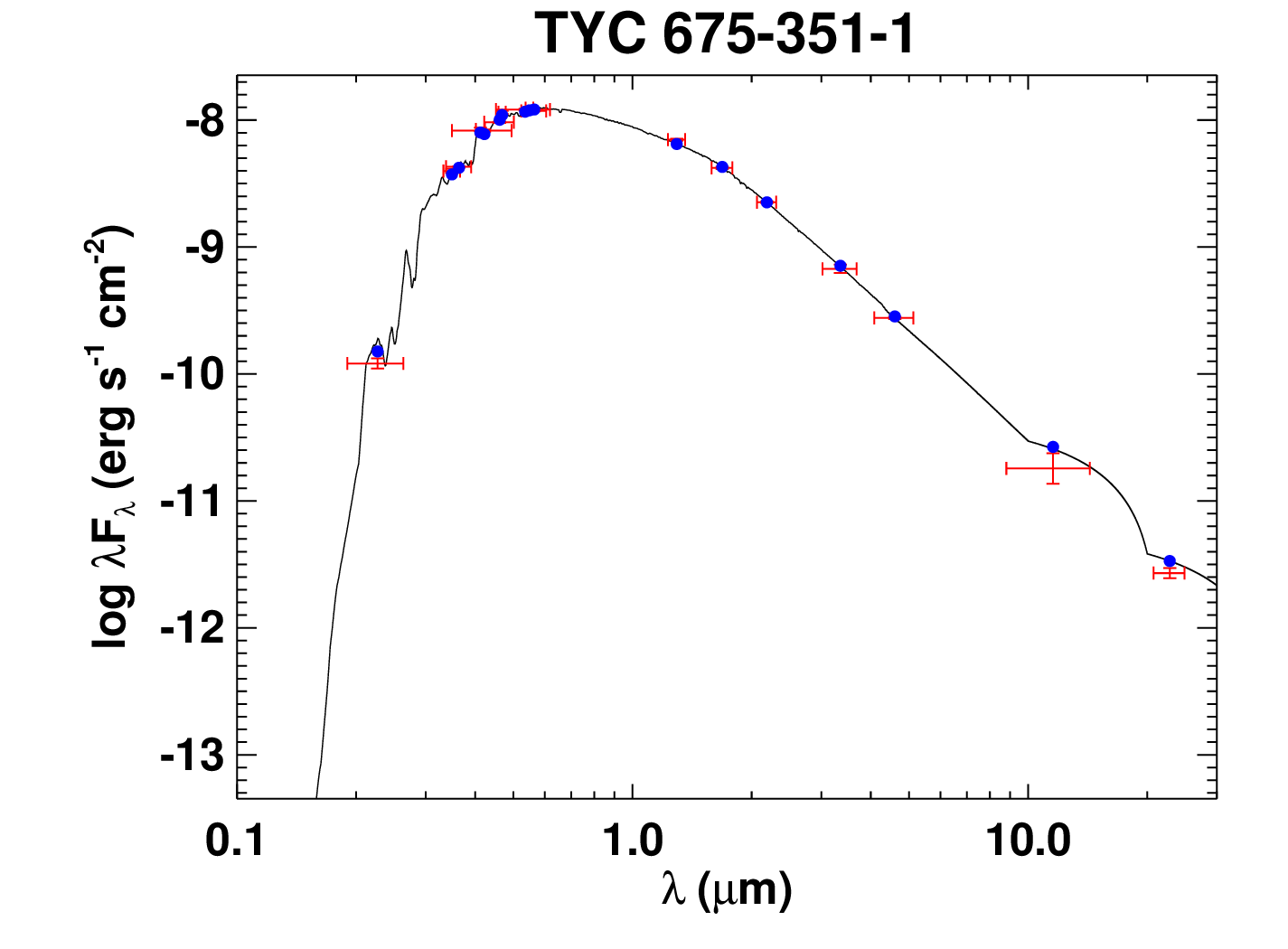}\includegraphics[width=0.333\linewidth]{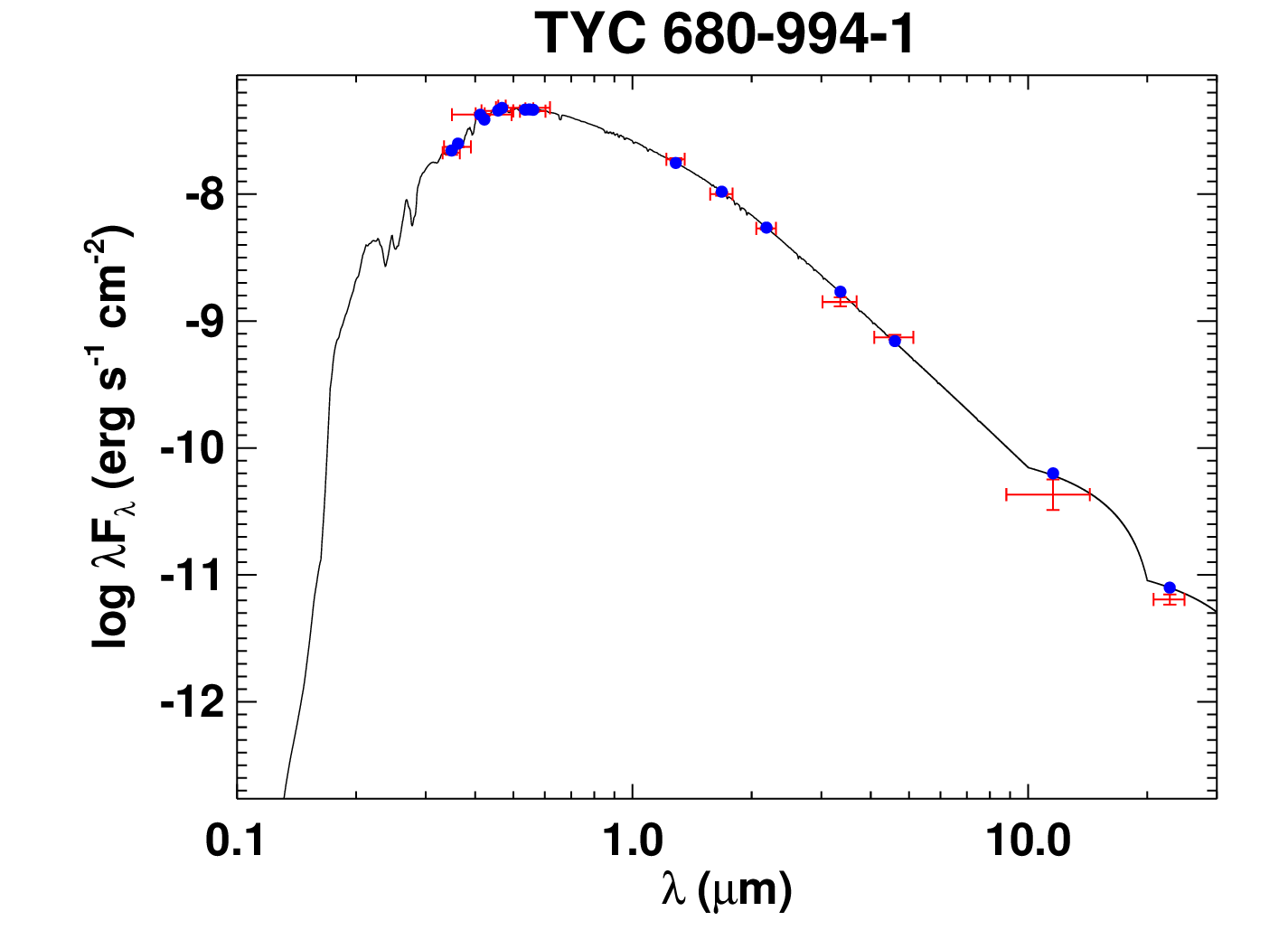}
\caption{\label{fig:seds15} All labels, lines, symbols, and colors as in Figure \ref{fig:seds}.}
\end{figure*}

\begin{figure*}
\includegraphics[width=0.333\linewidth]{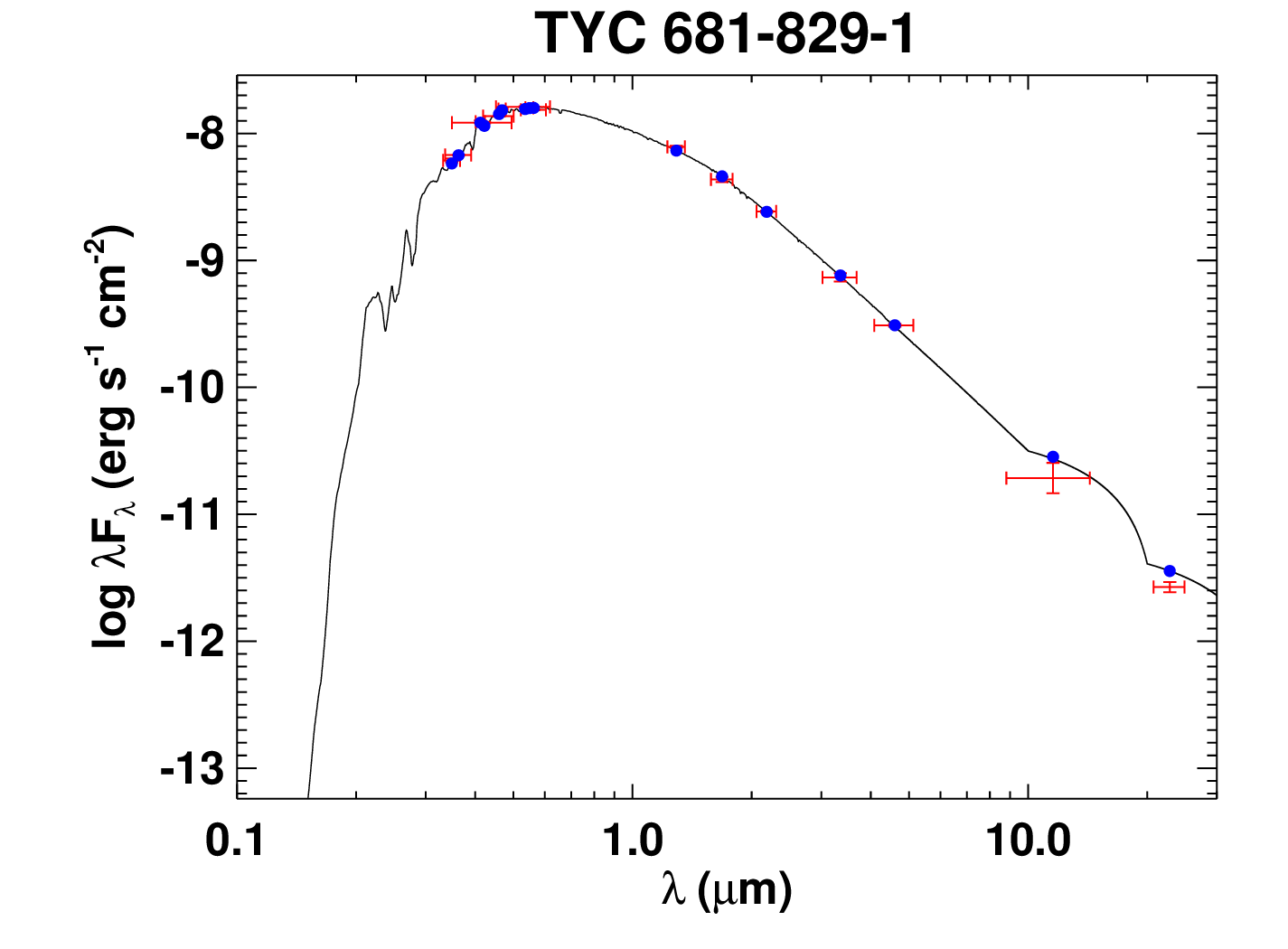}\includegraphics[width=0.333\linewidth]{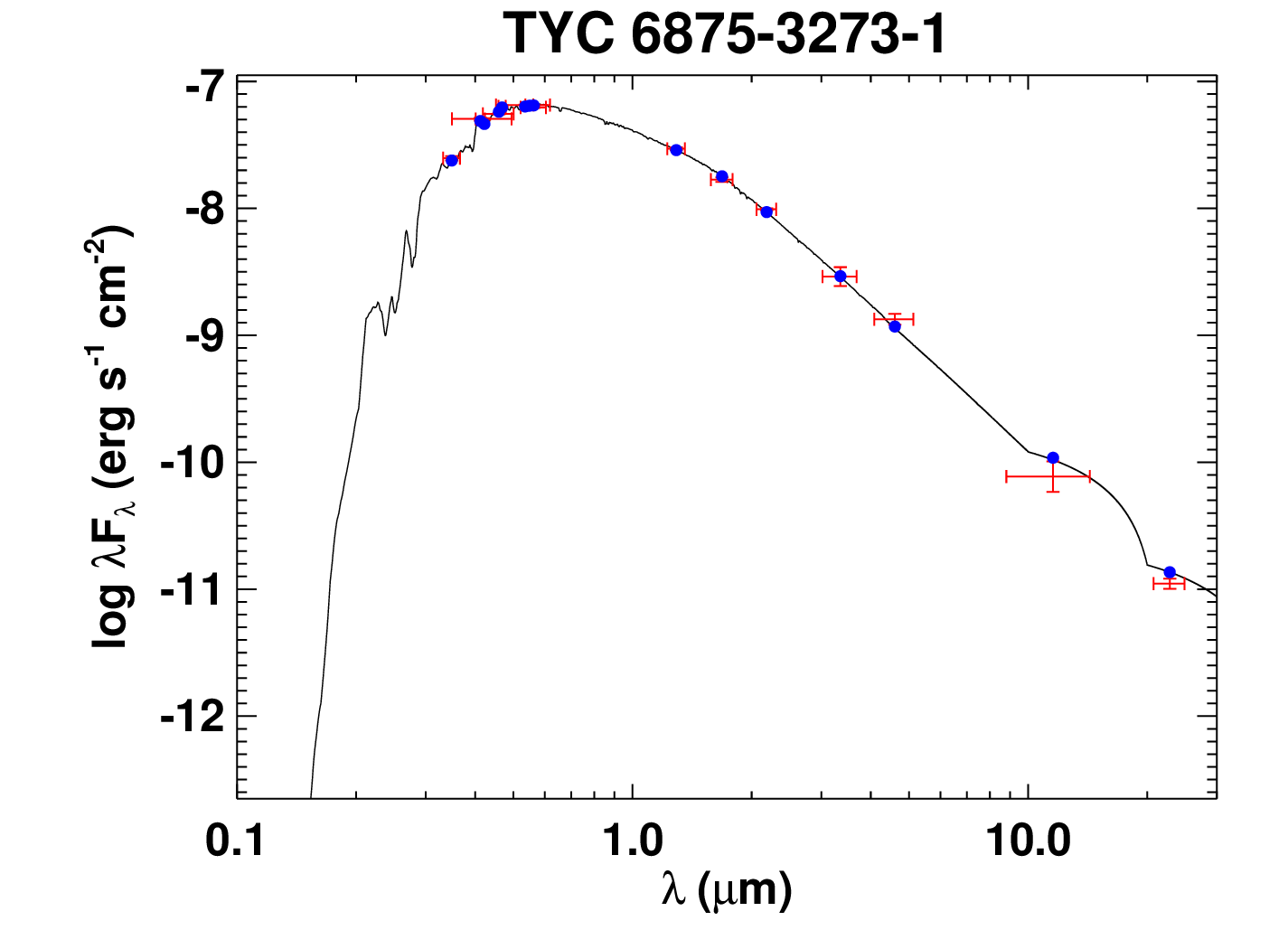}\includegraphics[width=0.333\linewidth]{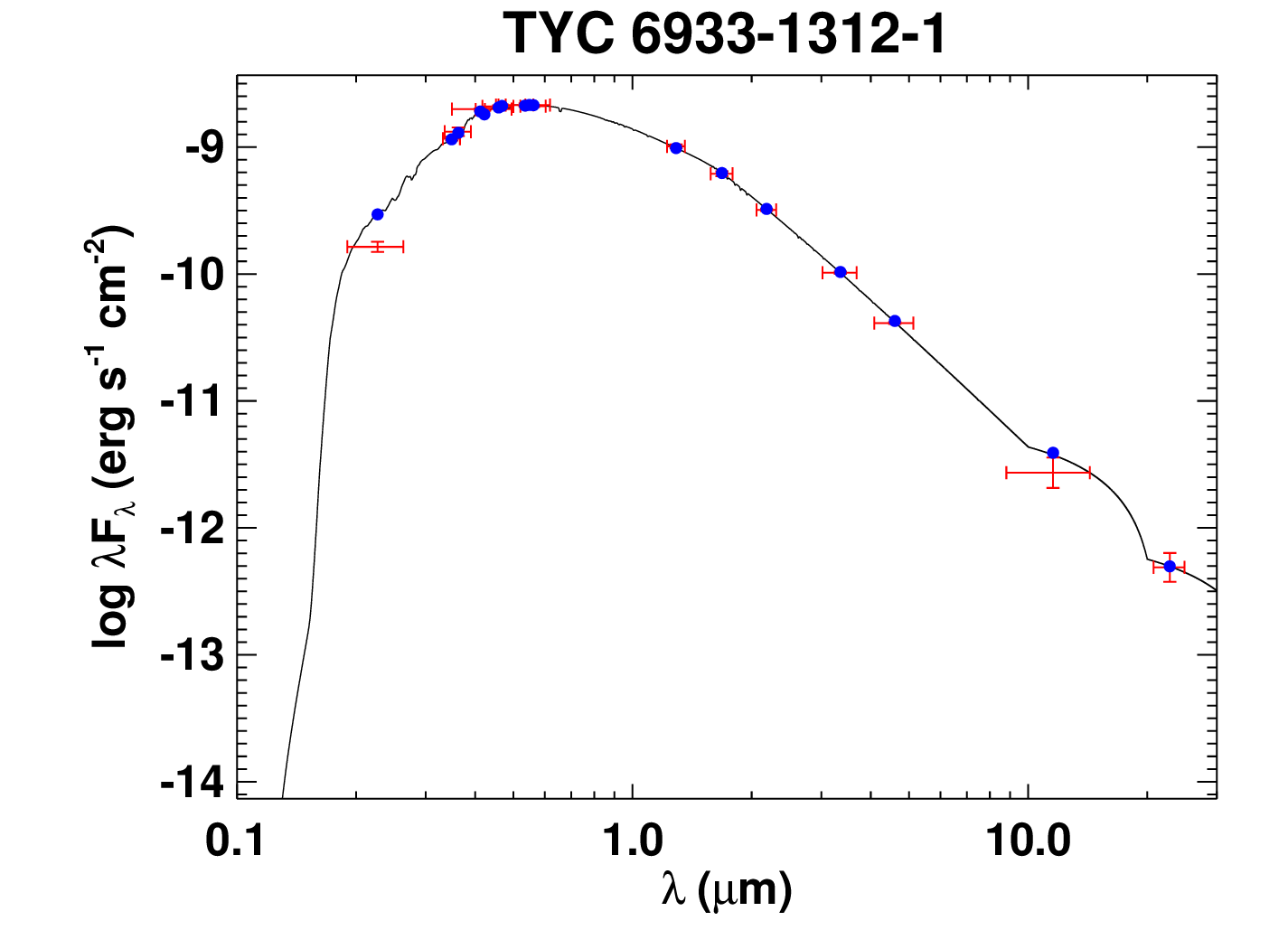}
\includegraphics[width=0.333\linewidth]{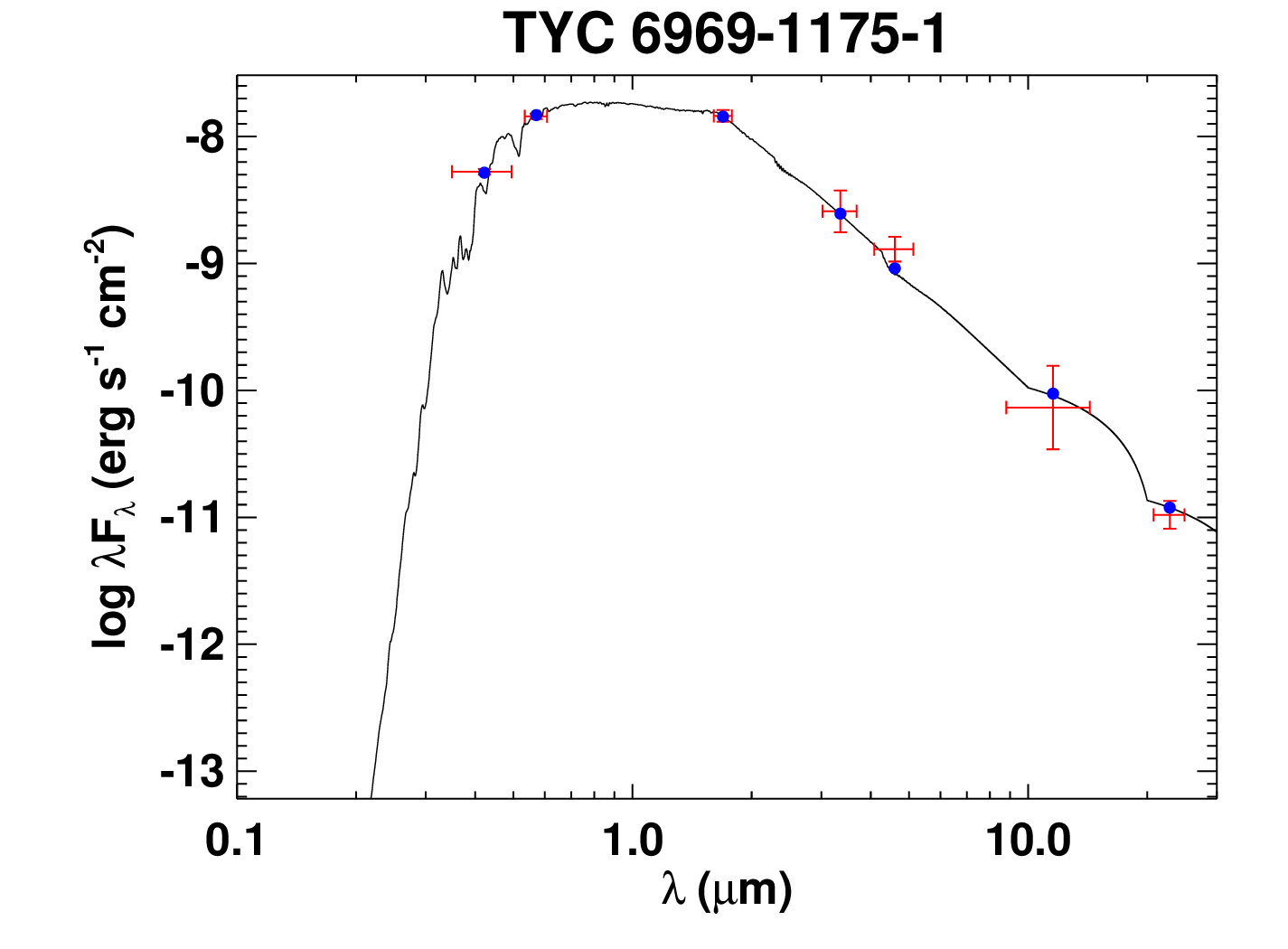}\includegraphics[width=0.333\linewidth]{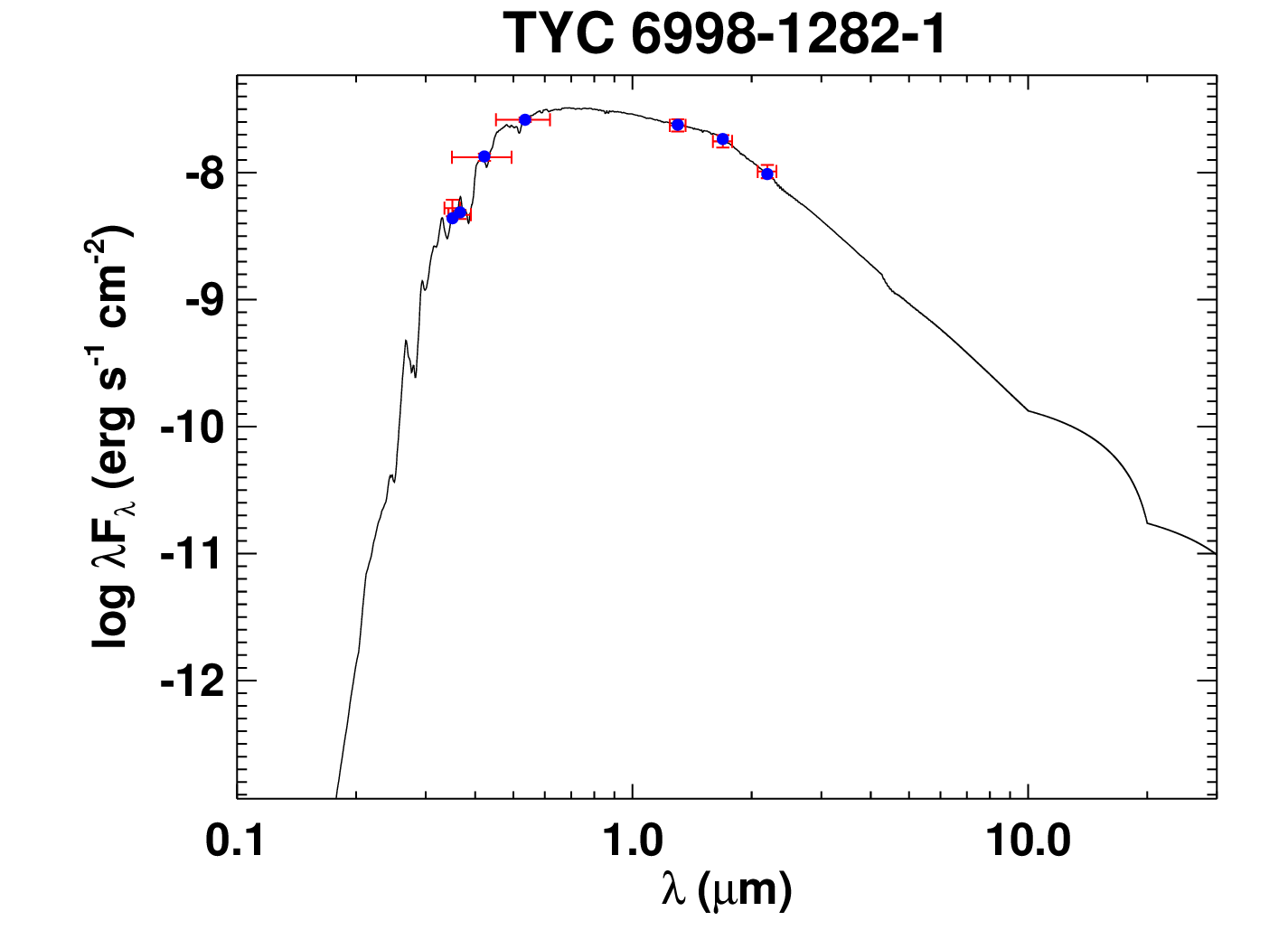}\includegraphics[width=0.333\linewidth]{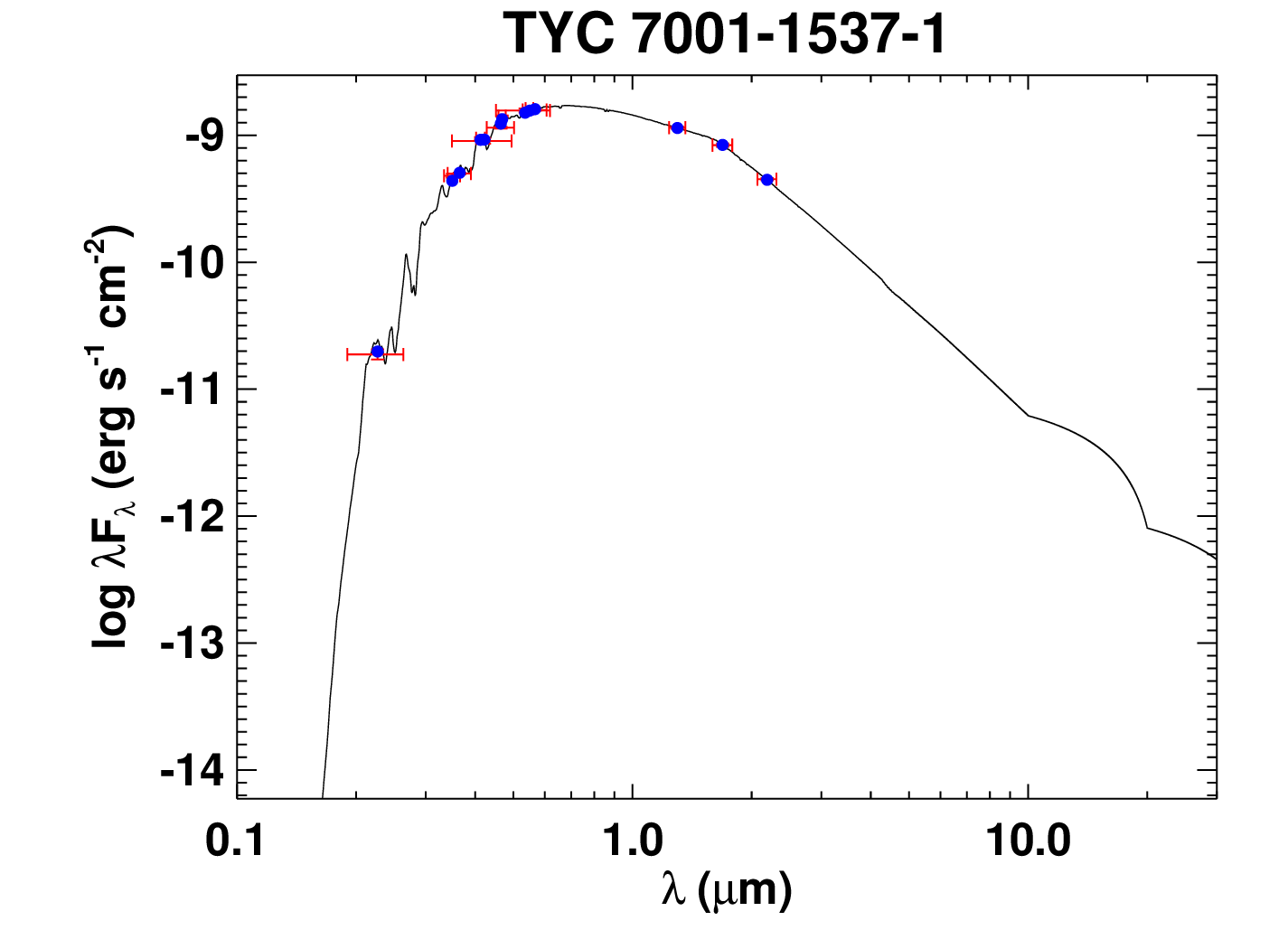}
\includegraphics[width=0.333\linewidth]{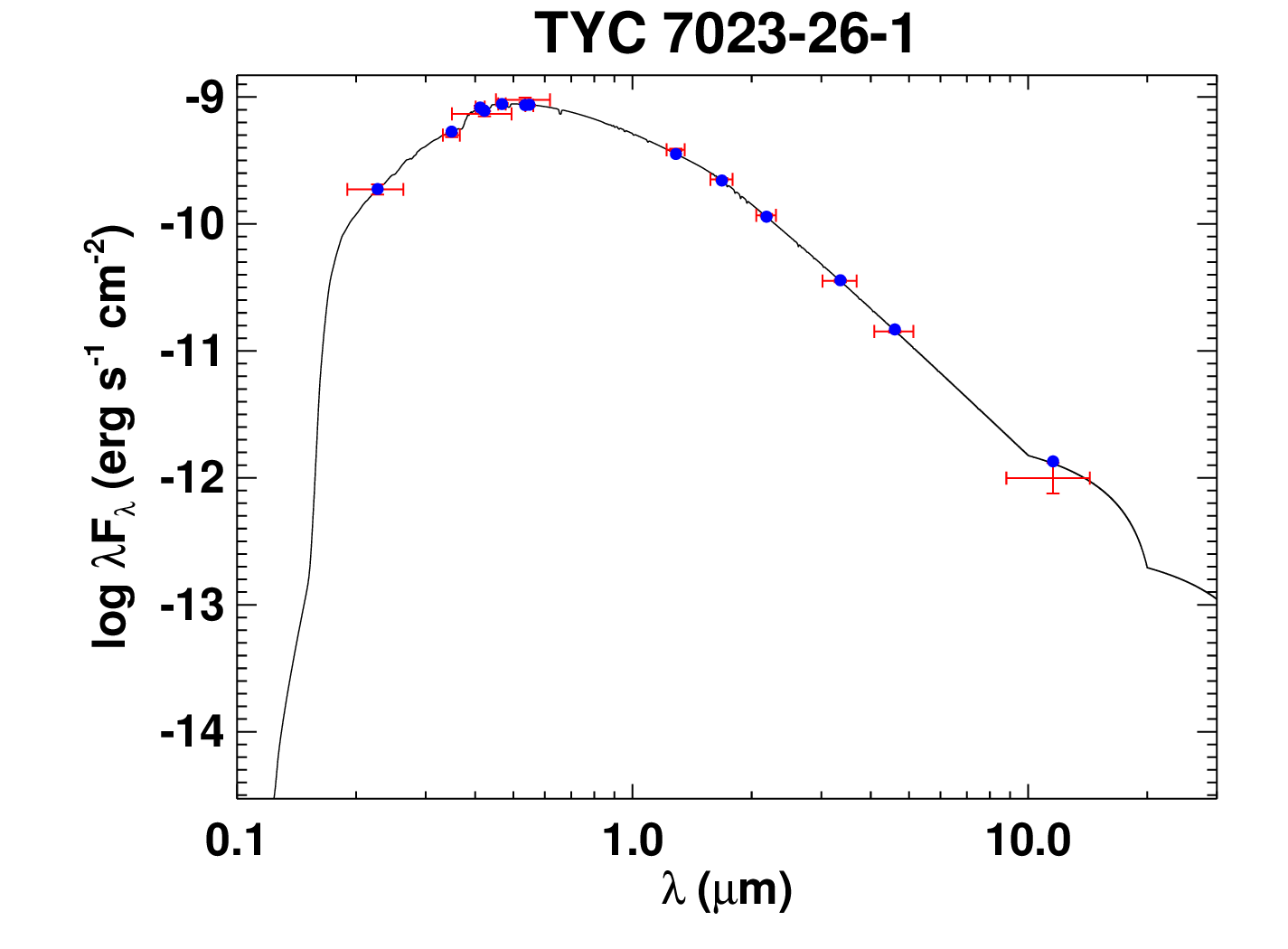}\includegraphics[width=0.333\linewidth]{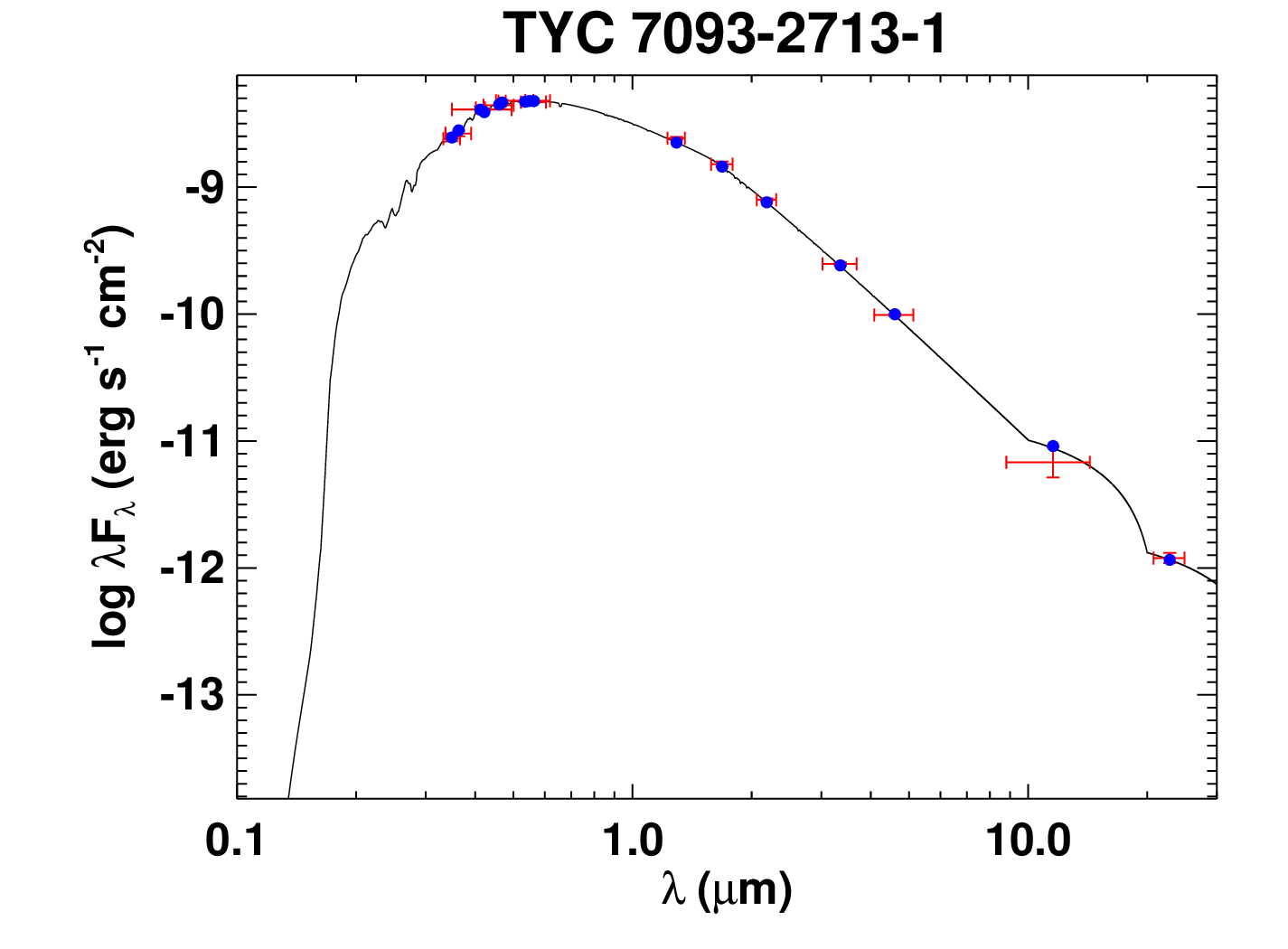}\includegraphics[width=0.333\linewidth]{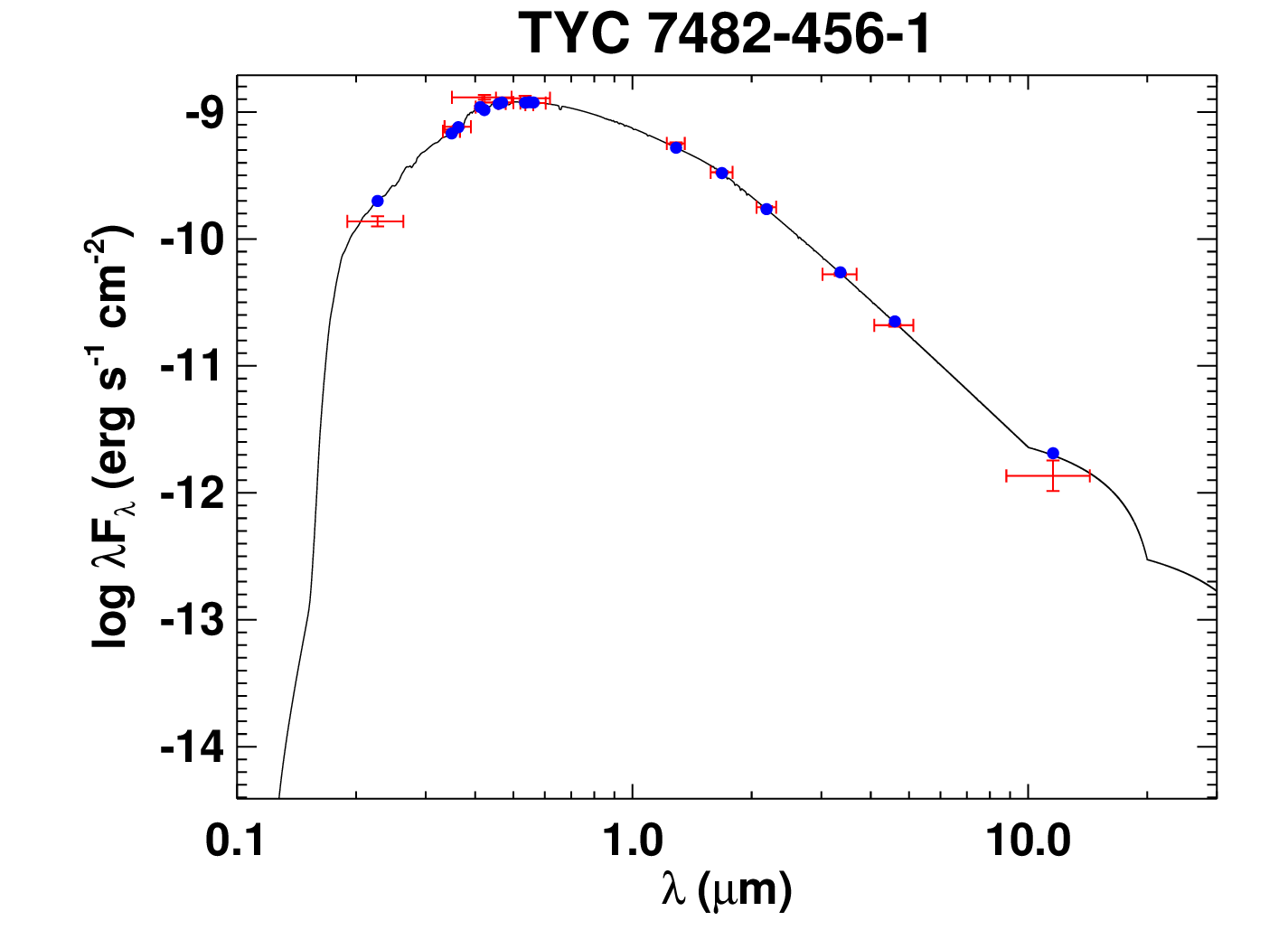}
\includegraphics[width=0.333\linewidth]{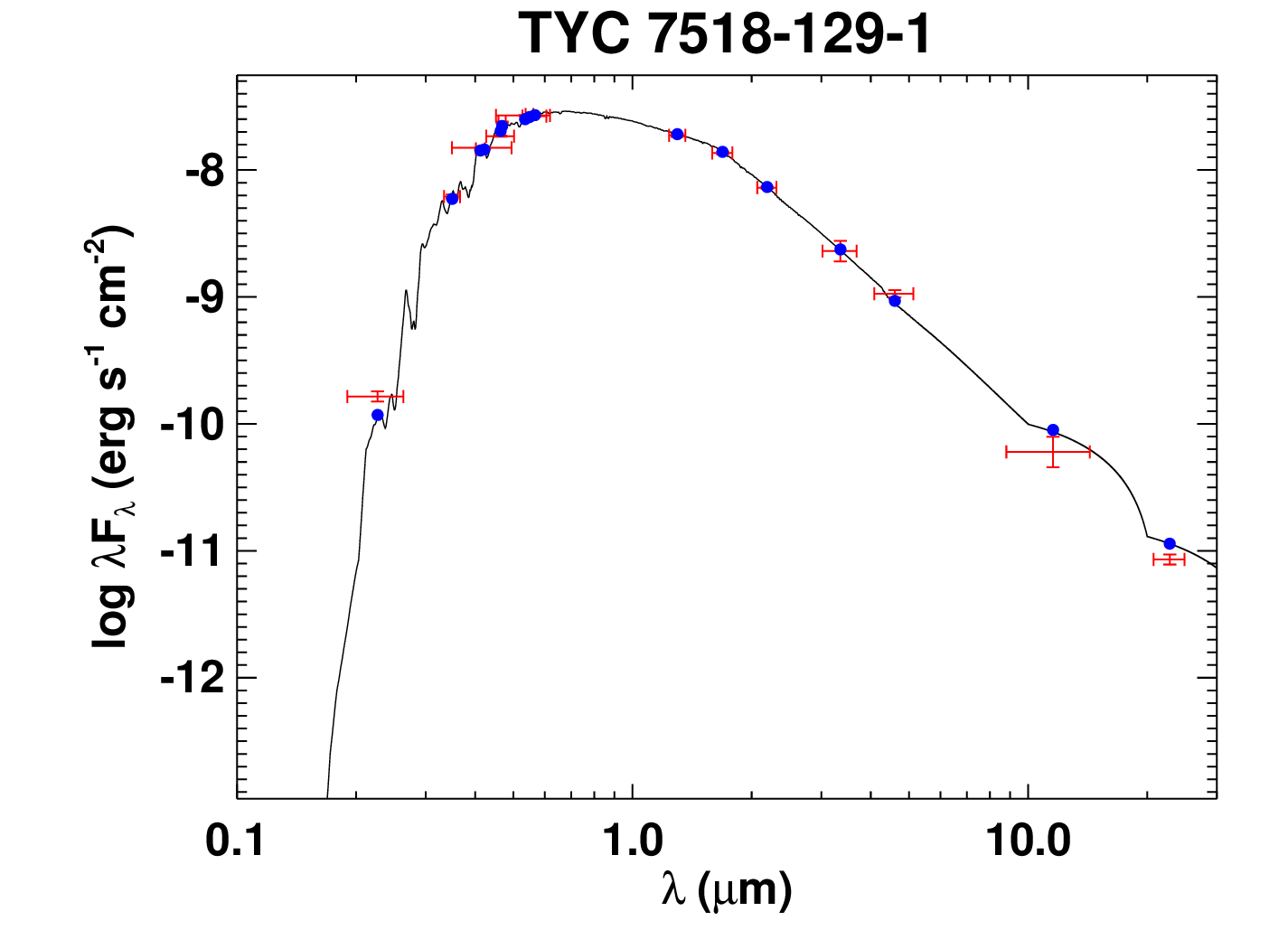}\includegraphics[width=0.333\linewidth]{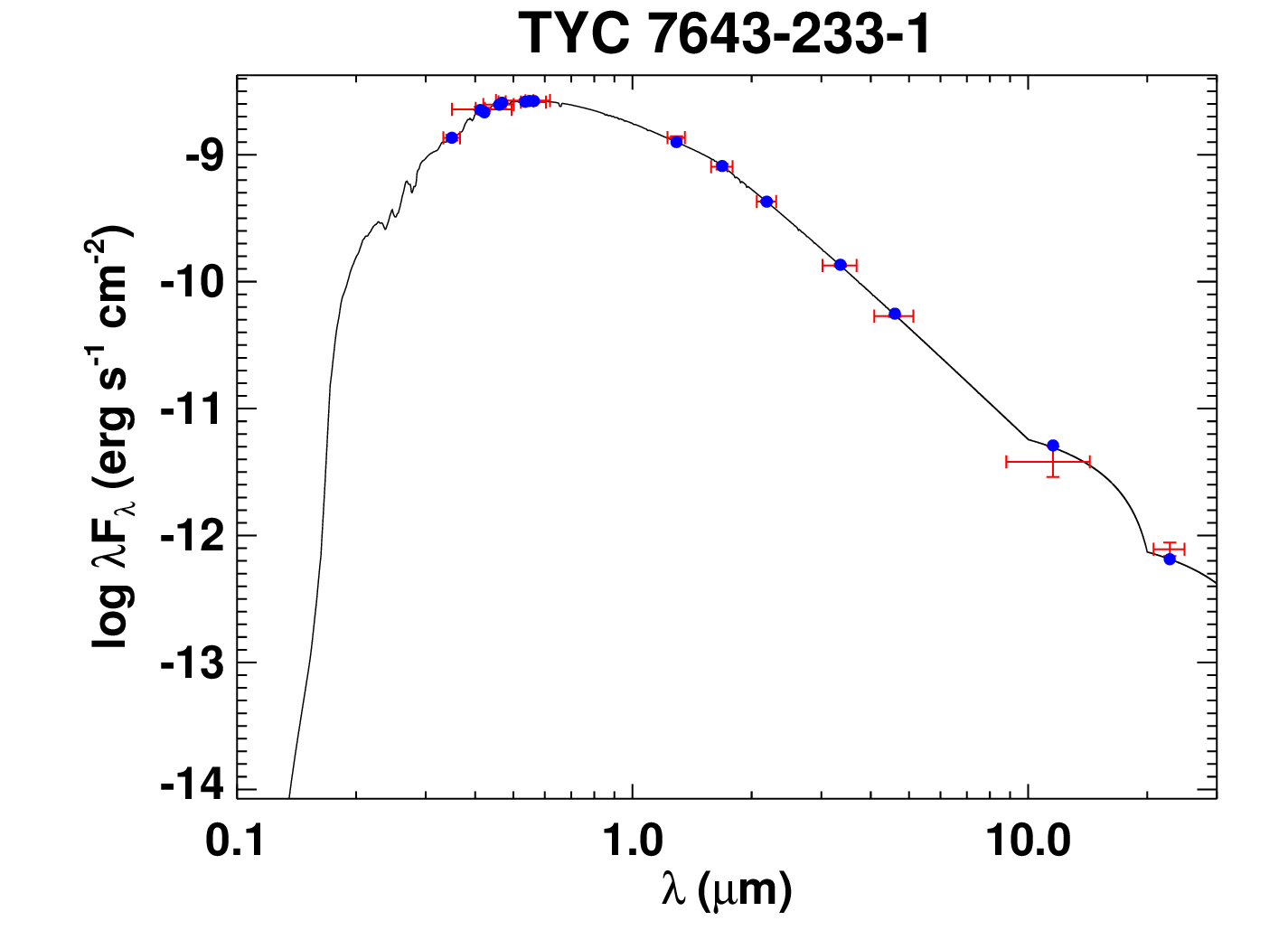}\includegraphics[width=0.333\linewidth]{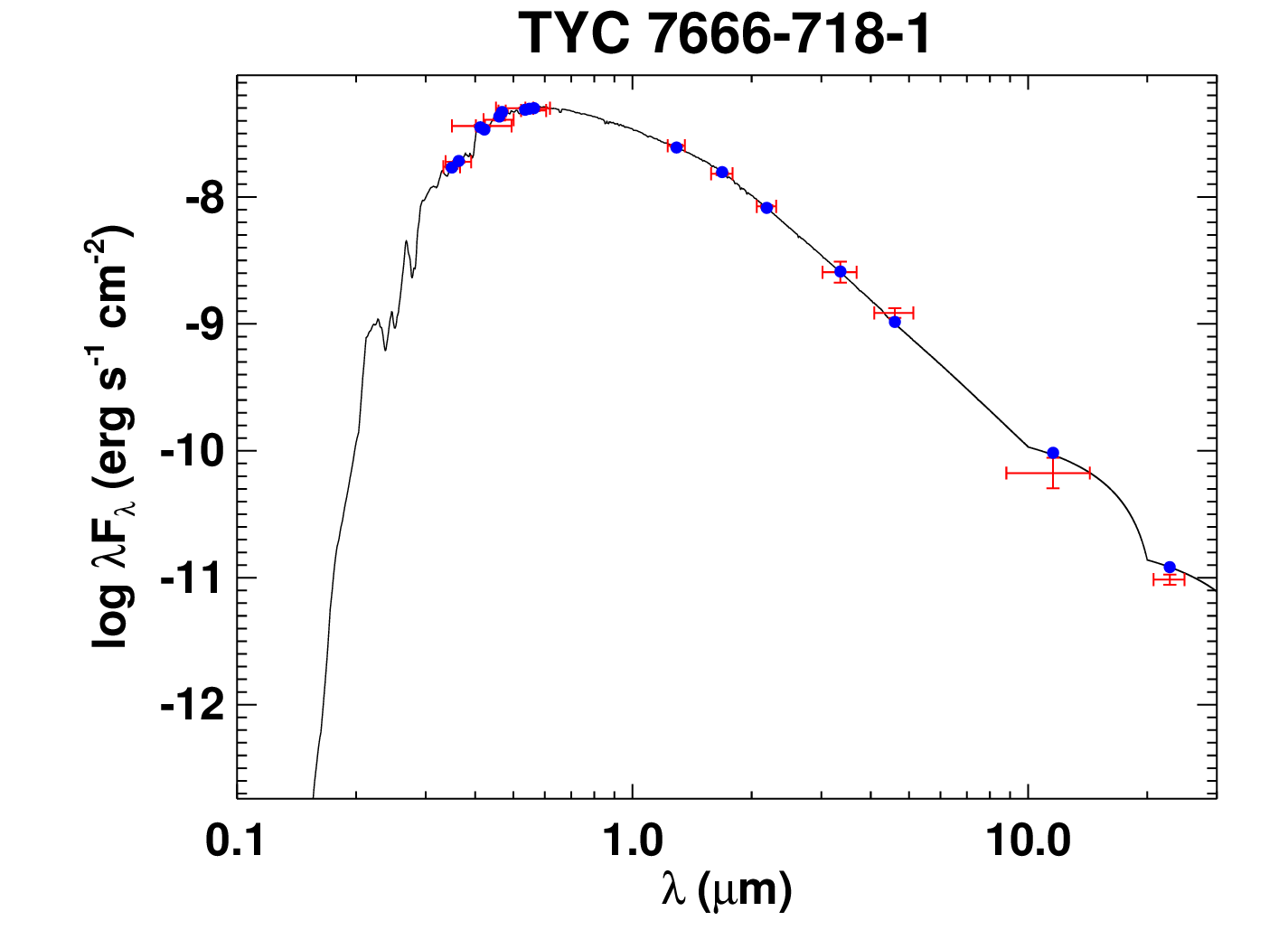}
\caption{\label{fig:seds16} All labels, lines, symbols, and colors as in Figure \ref{fig:seds}.}
\end{figure*}

\begin{figure*}
\includegraphics[width=0.333\linewidth]{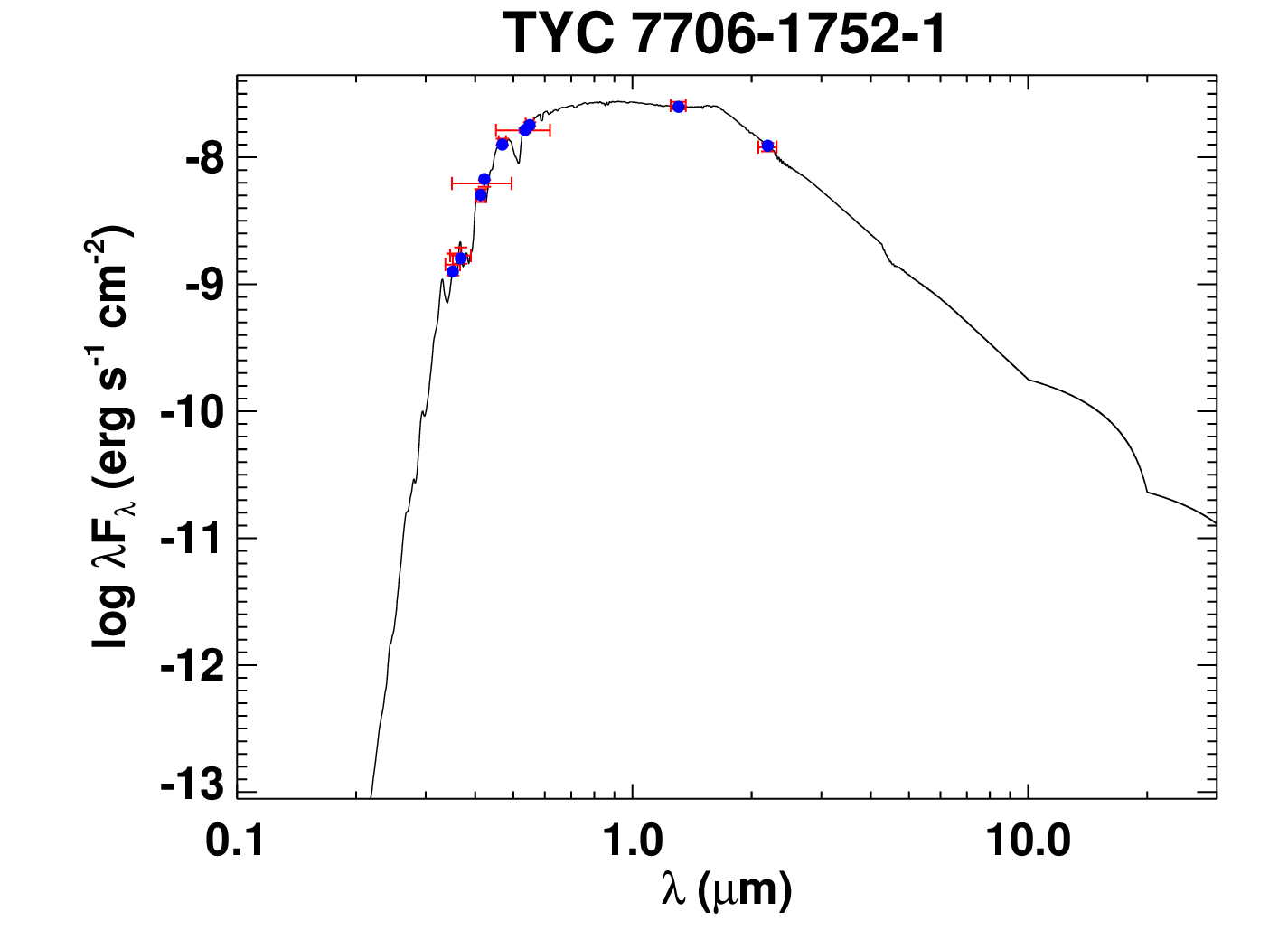}\includegraphics[width=0.333\linewidth]{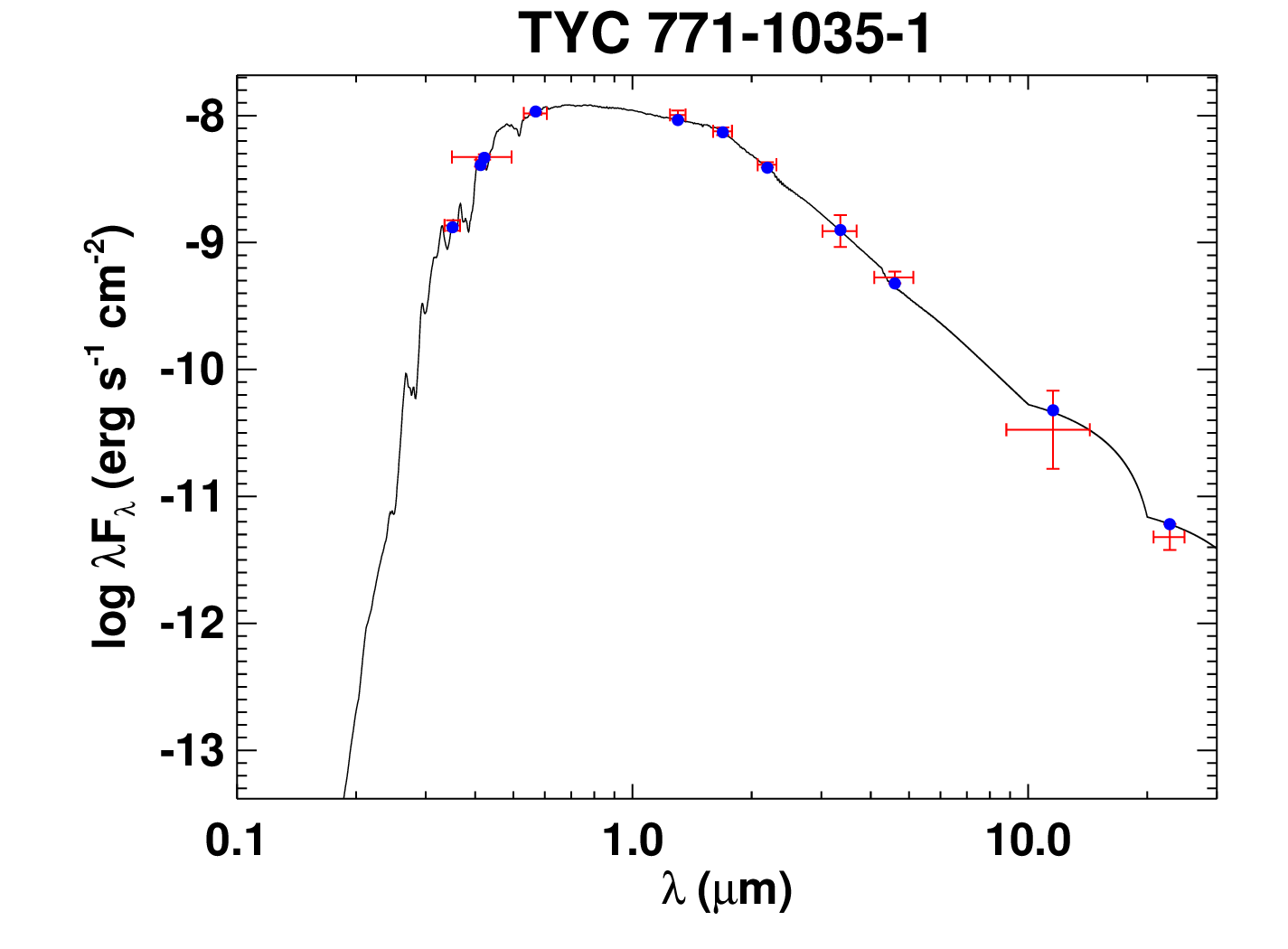}\includegraphics[width=0.333\linewidth]{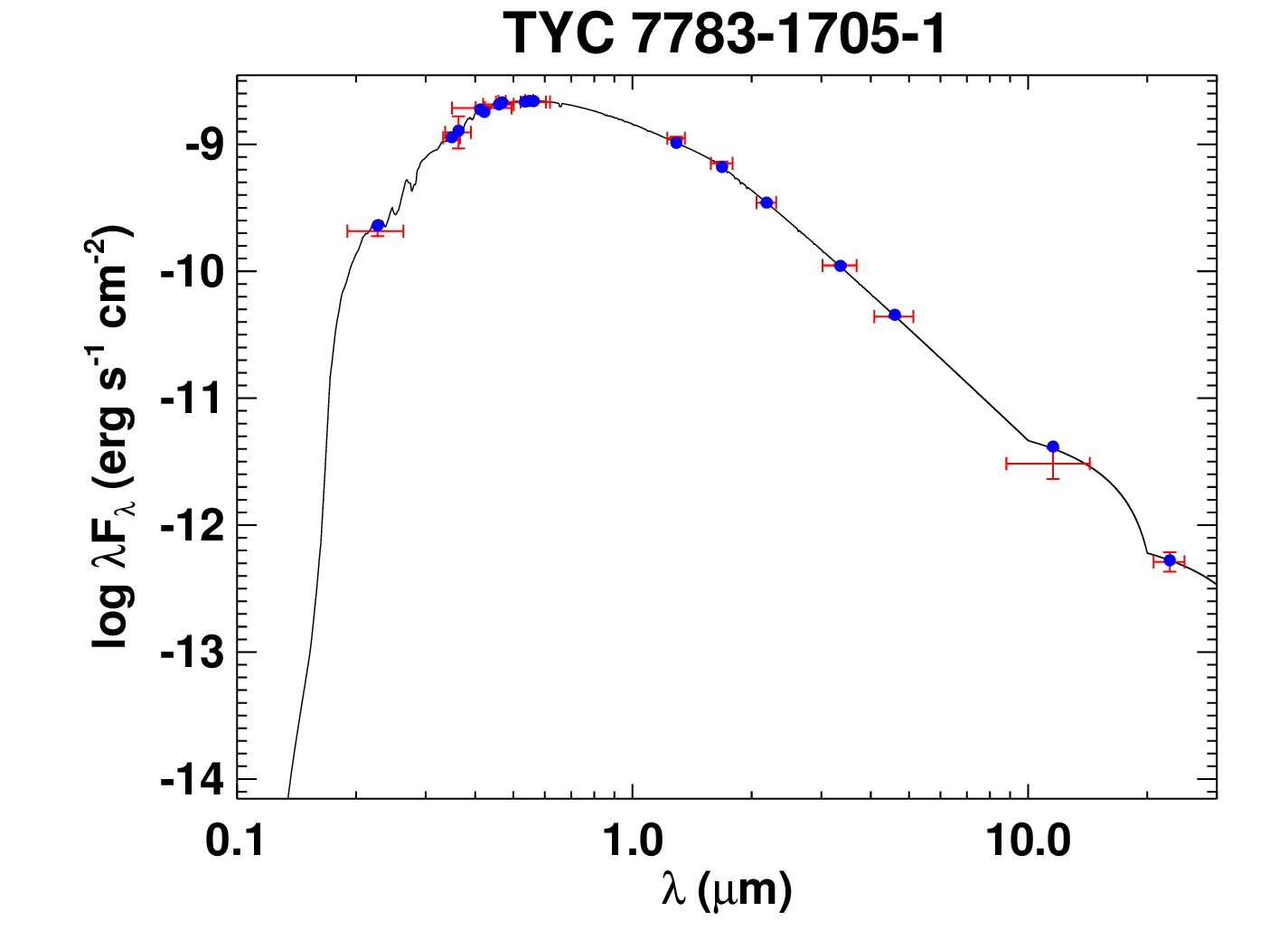}
\includegraphics[width=0.333\linewidth]{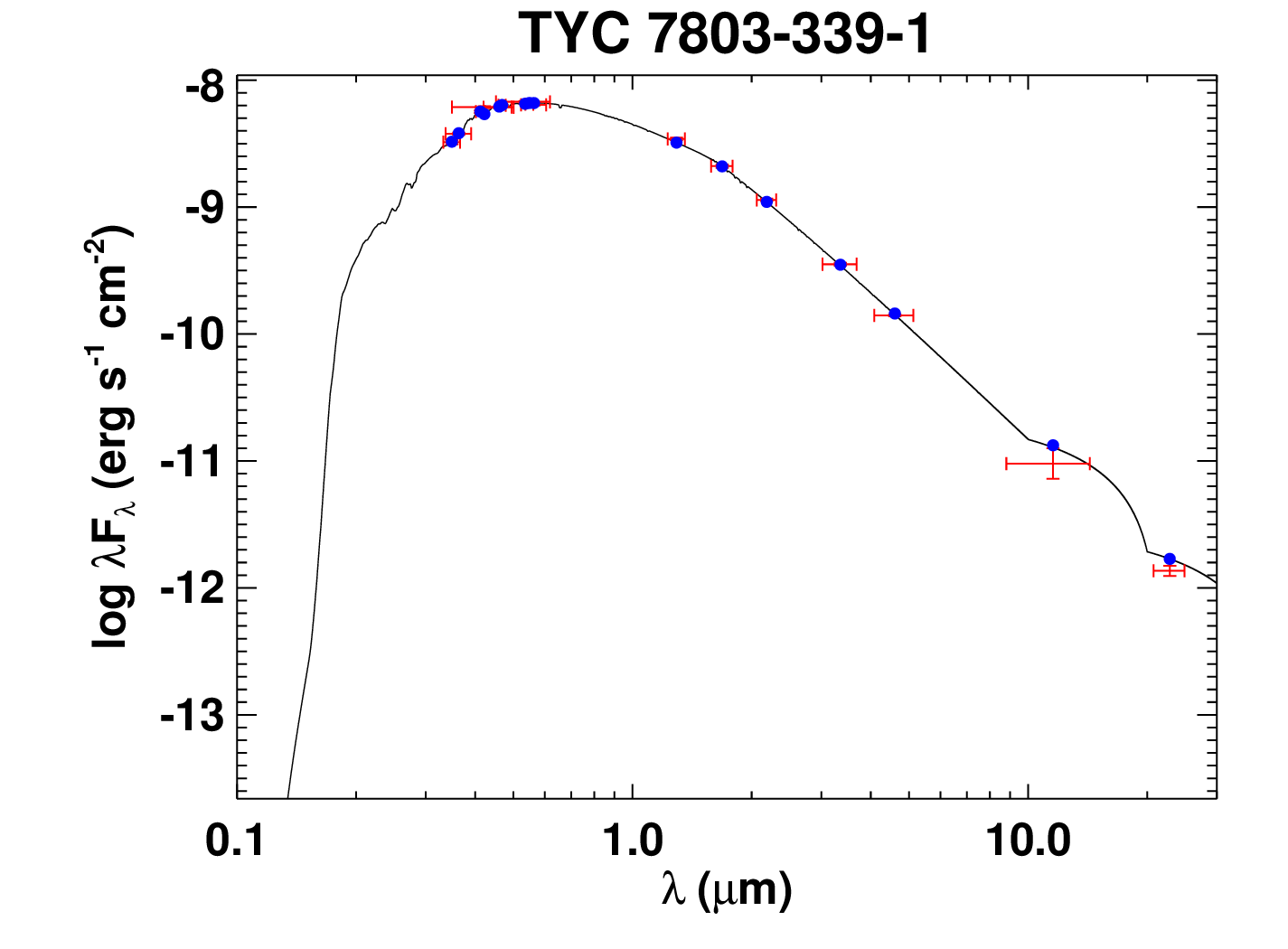}\includegraphics[width=0.333\linewidth]{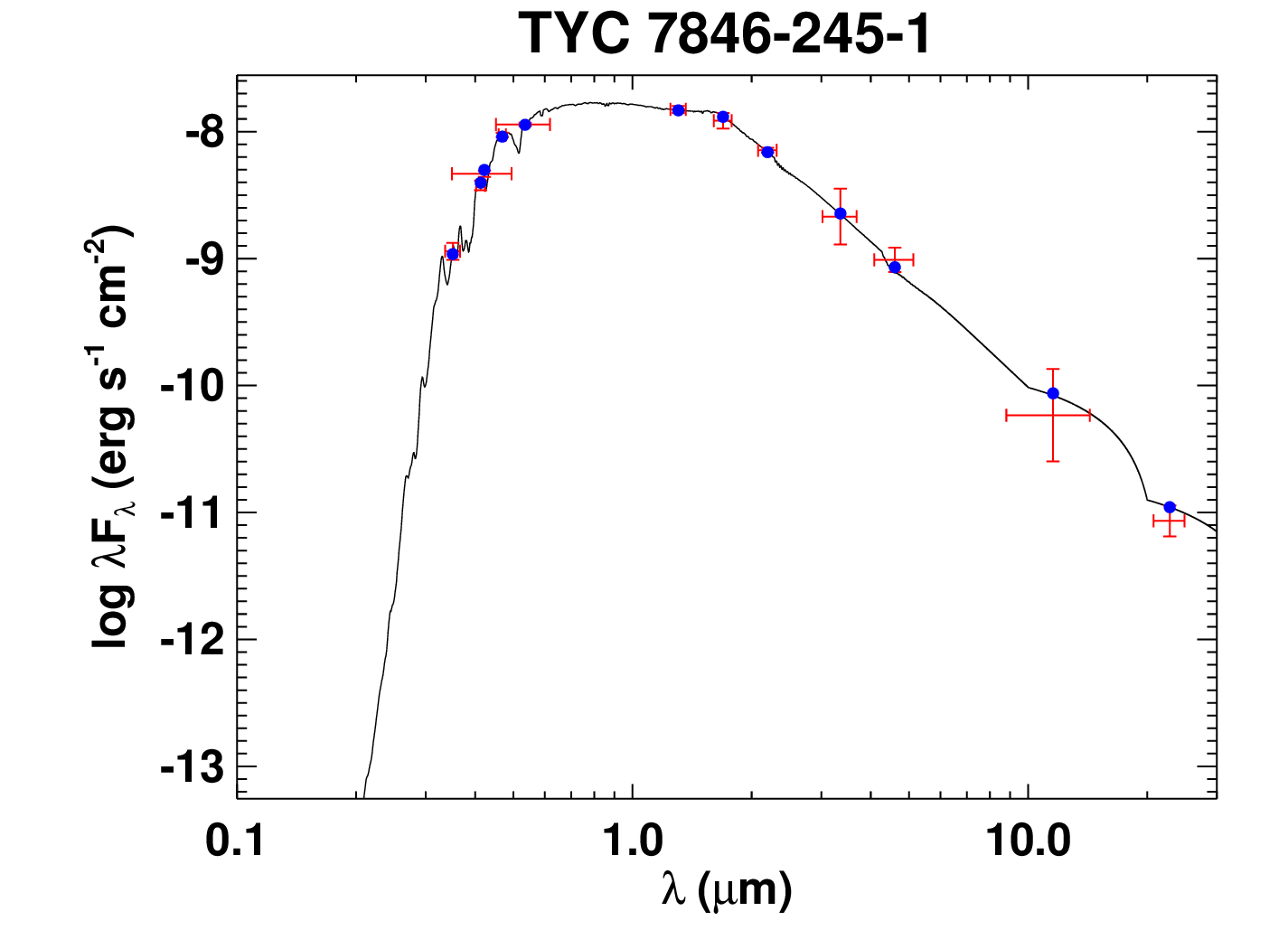}\includegraphics[width=0.333\linewidth]{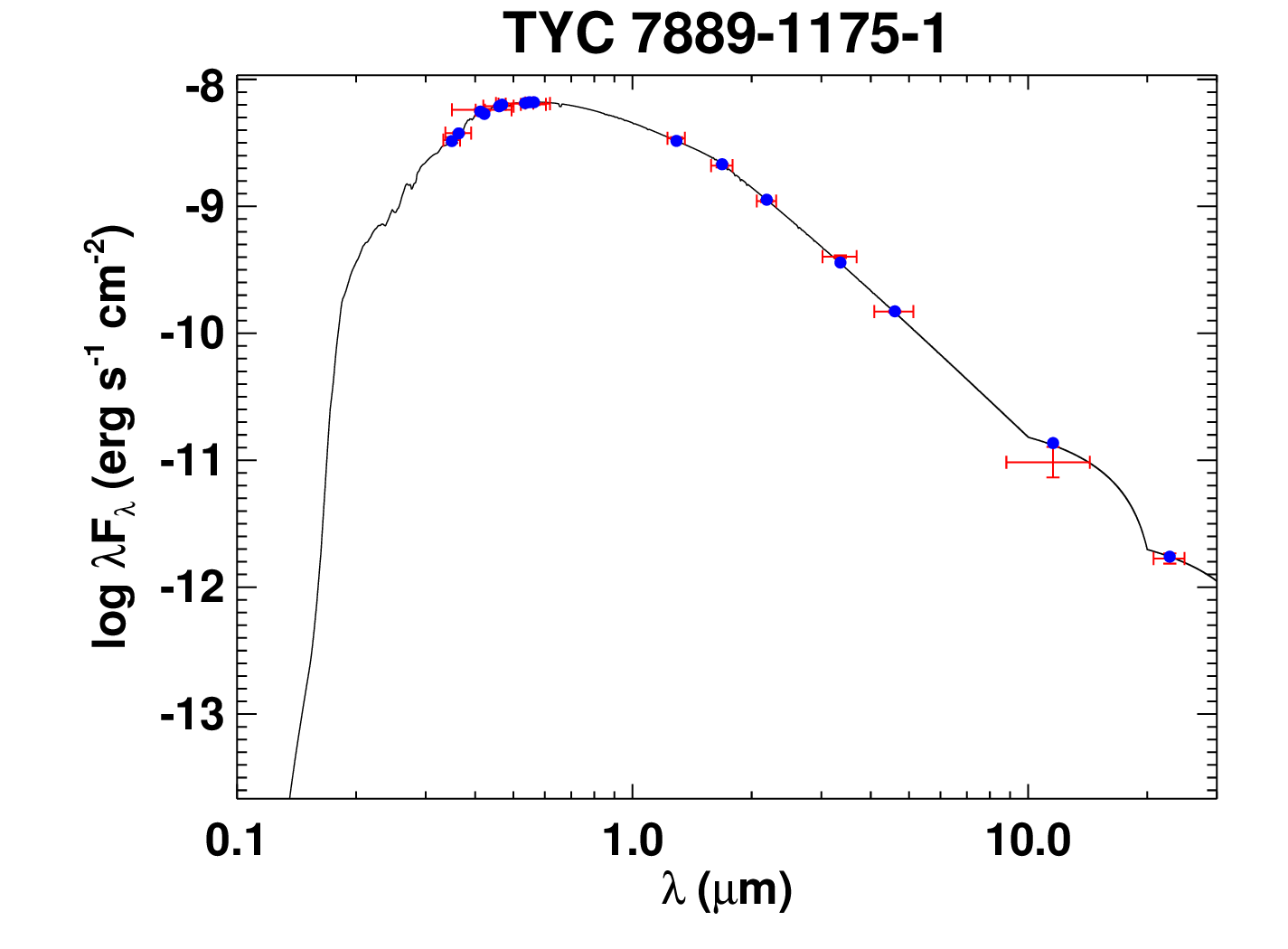}
\includegraphics[width=0.333\linewidth]{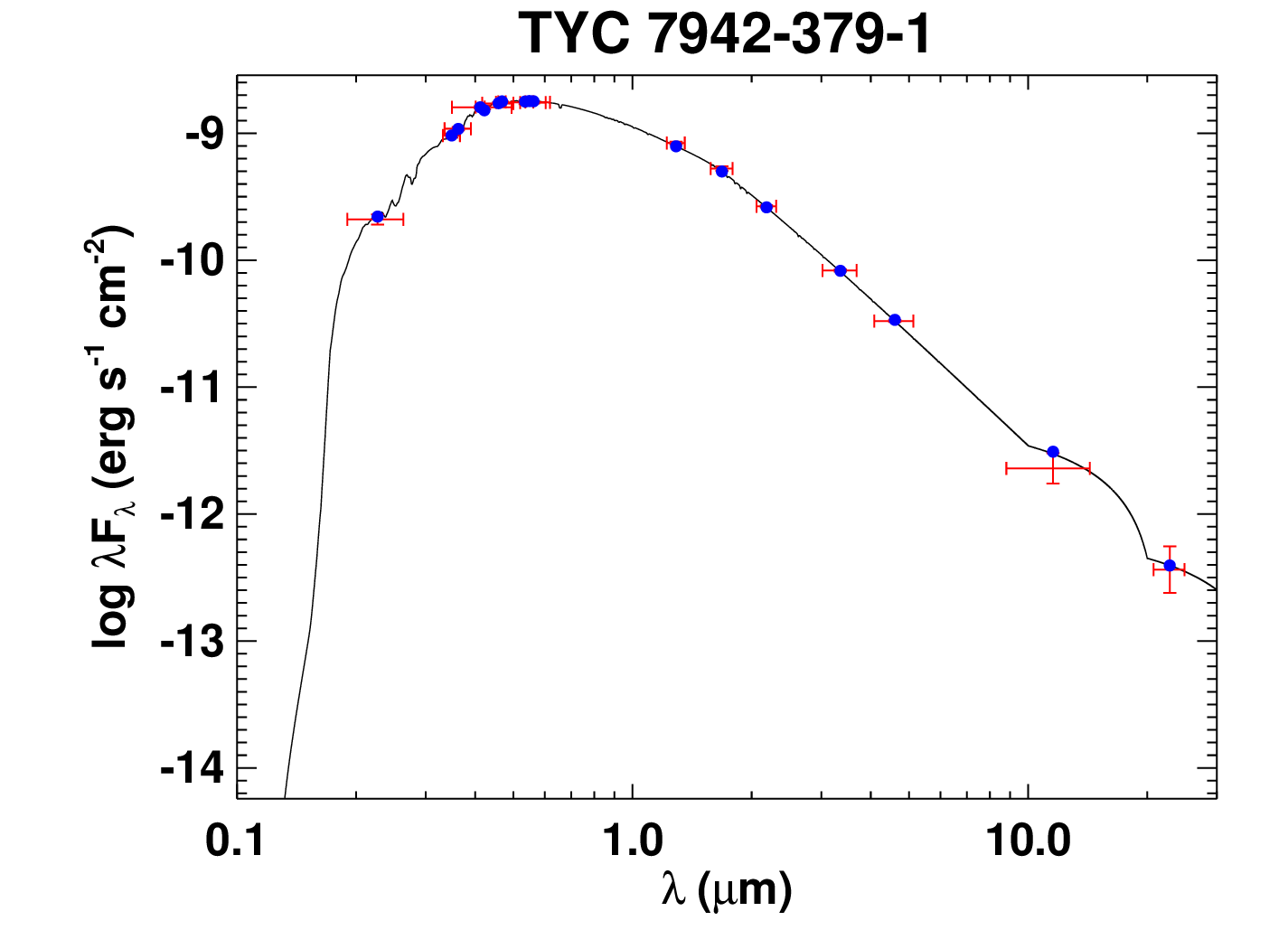}\includegraphics[width=0.333\linewidth]{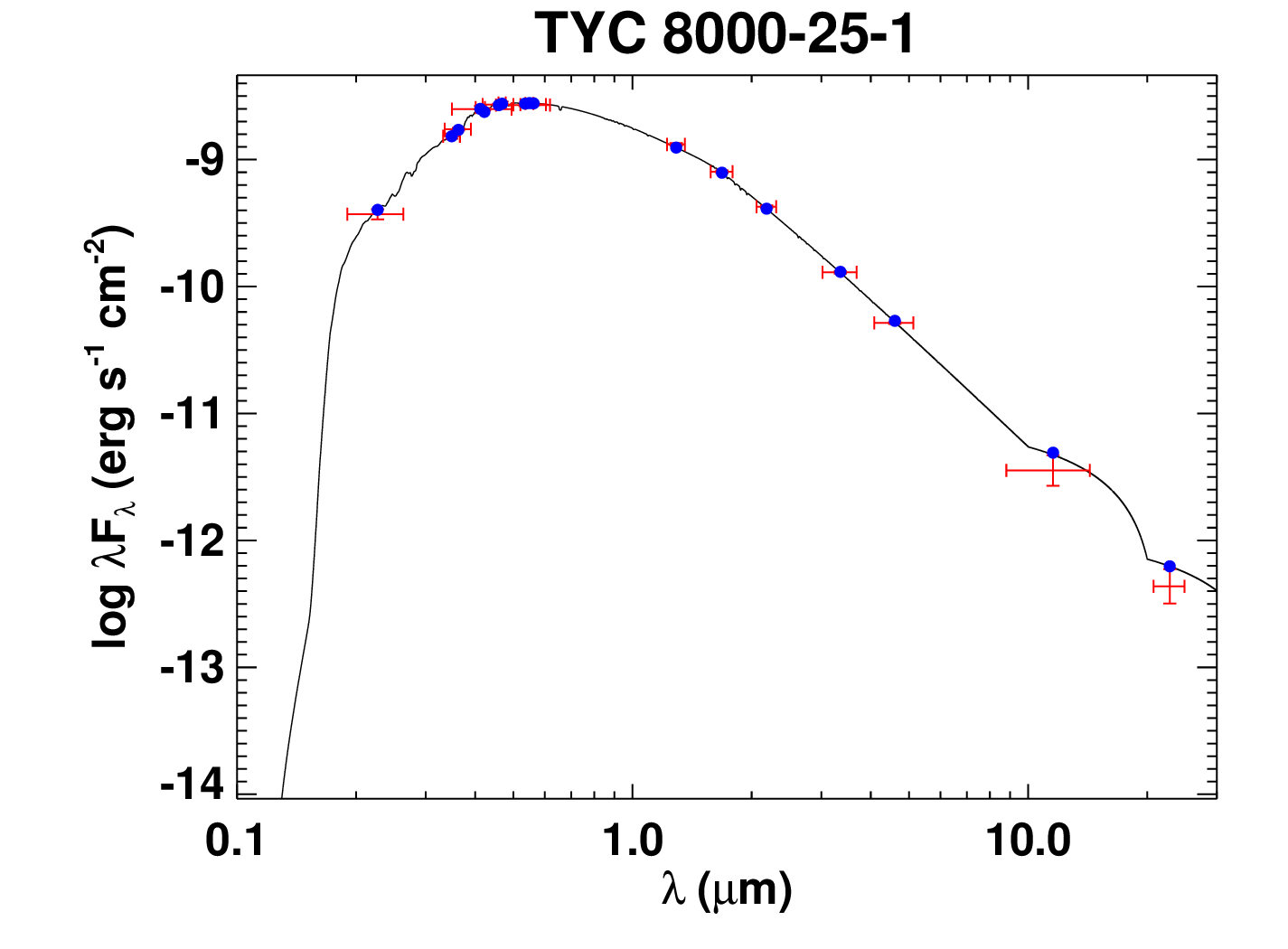}\includegraphics[width=0.333\linewidth]{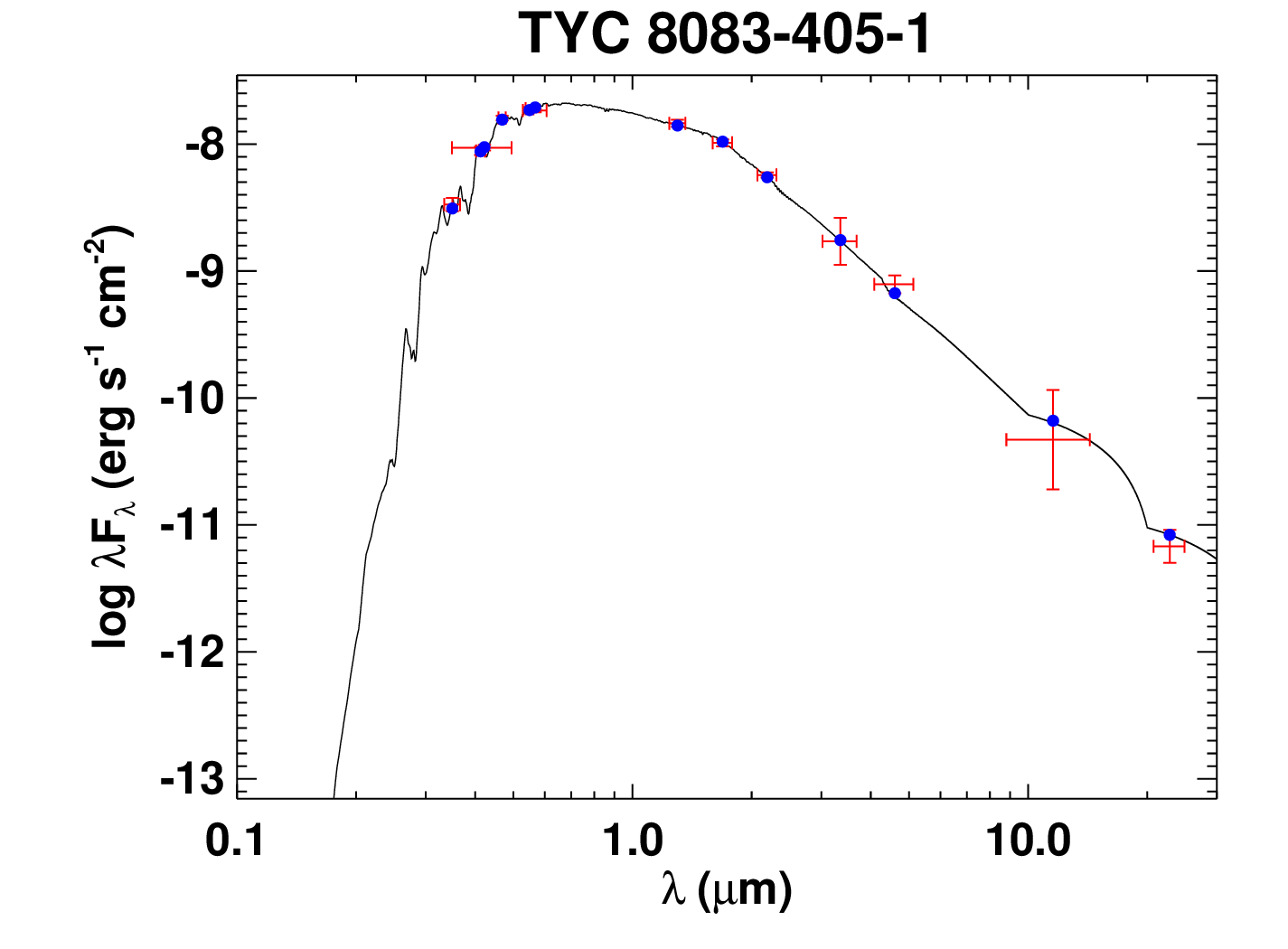}
\includegraphics[width=0.333\linewidth]{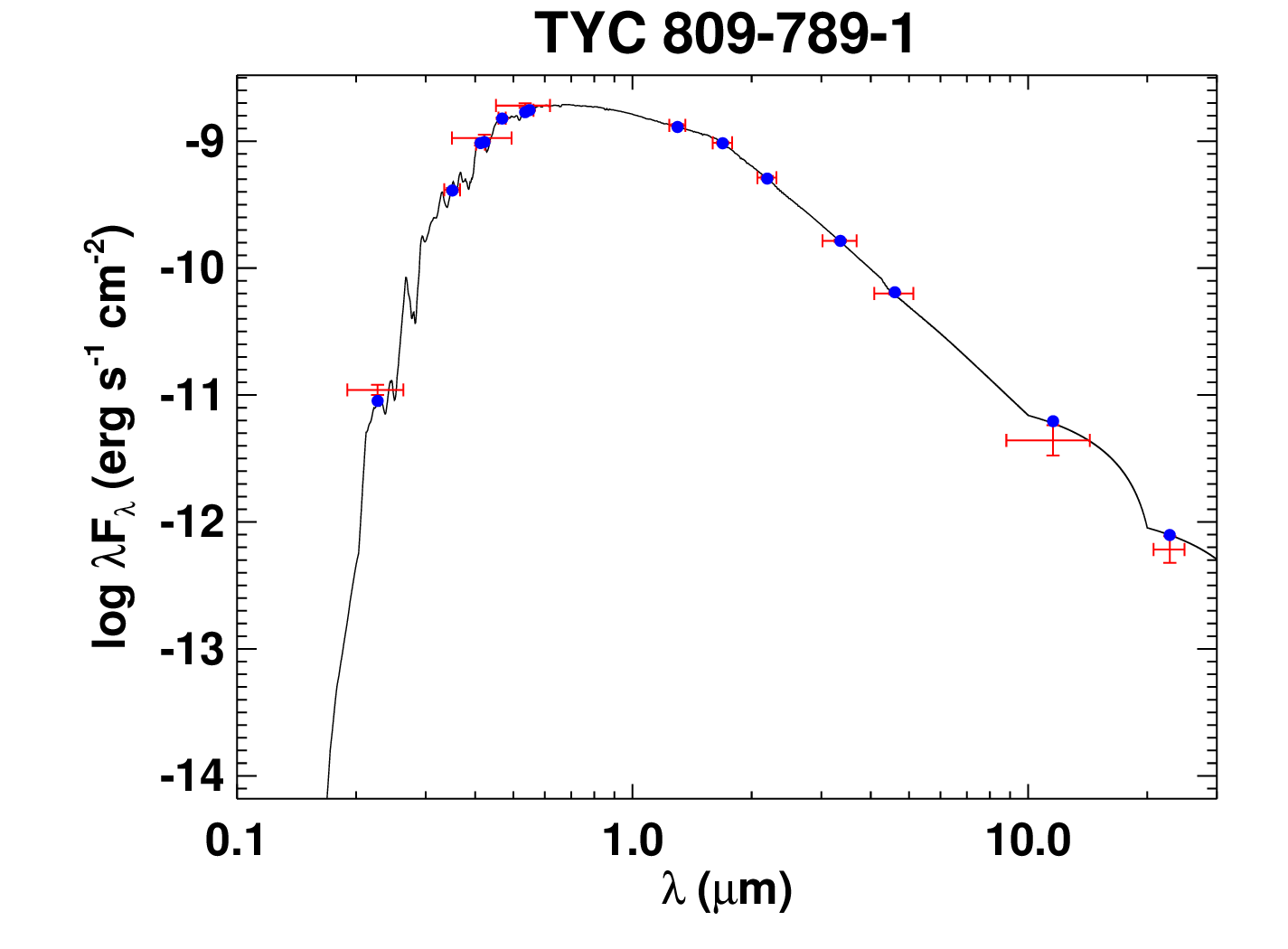}\includegraphics[width=0.333\linewidth]{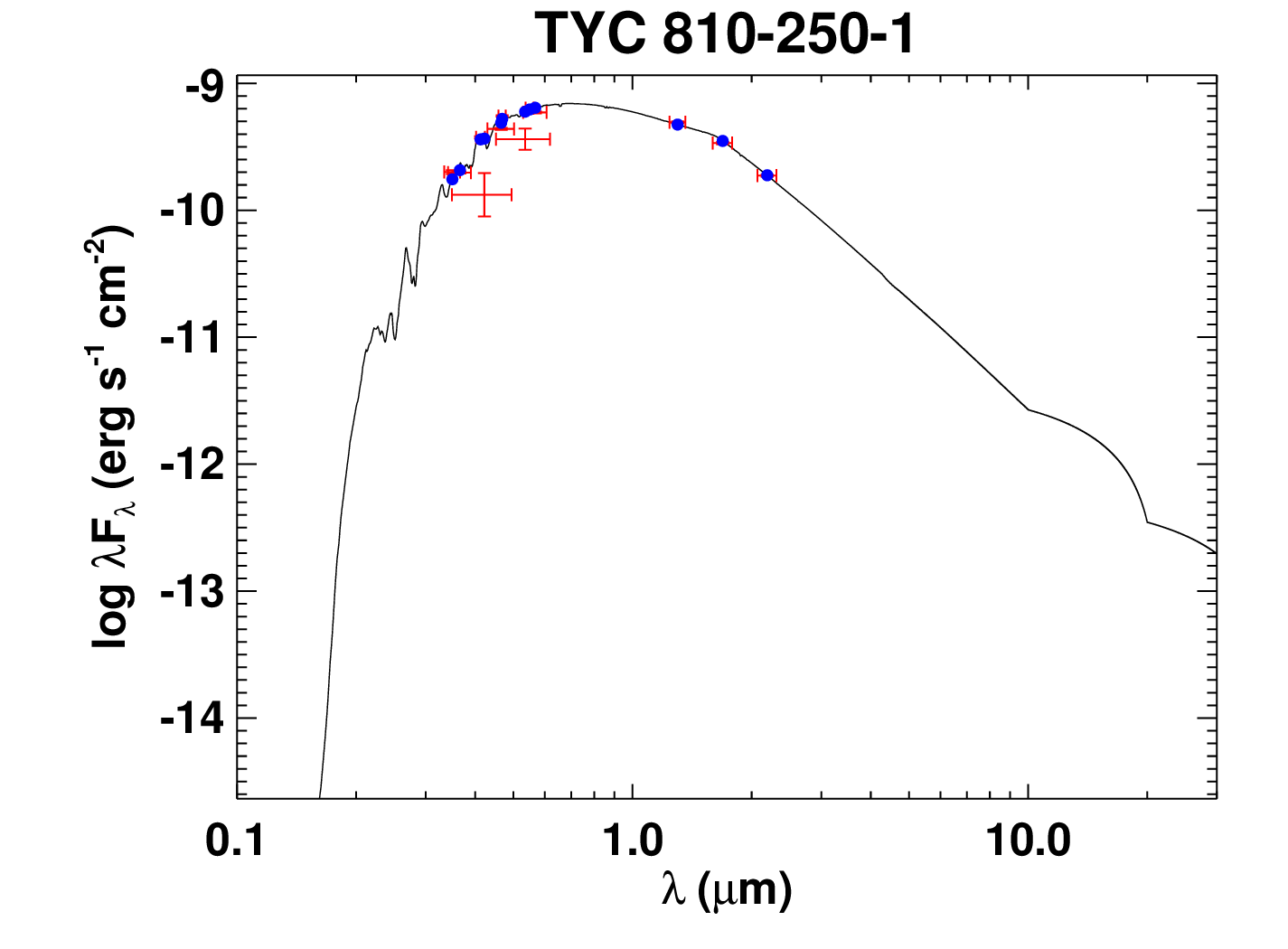}\includegraphics[width=0.333\linewidth]{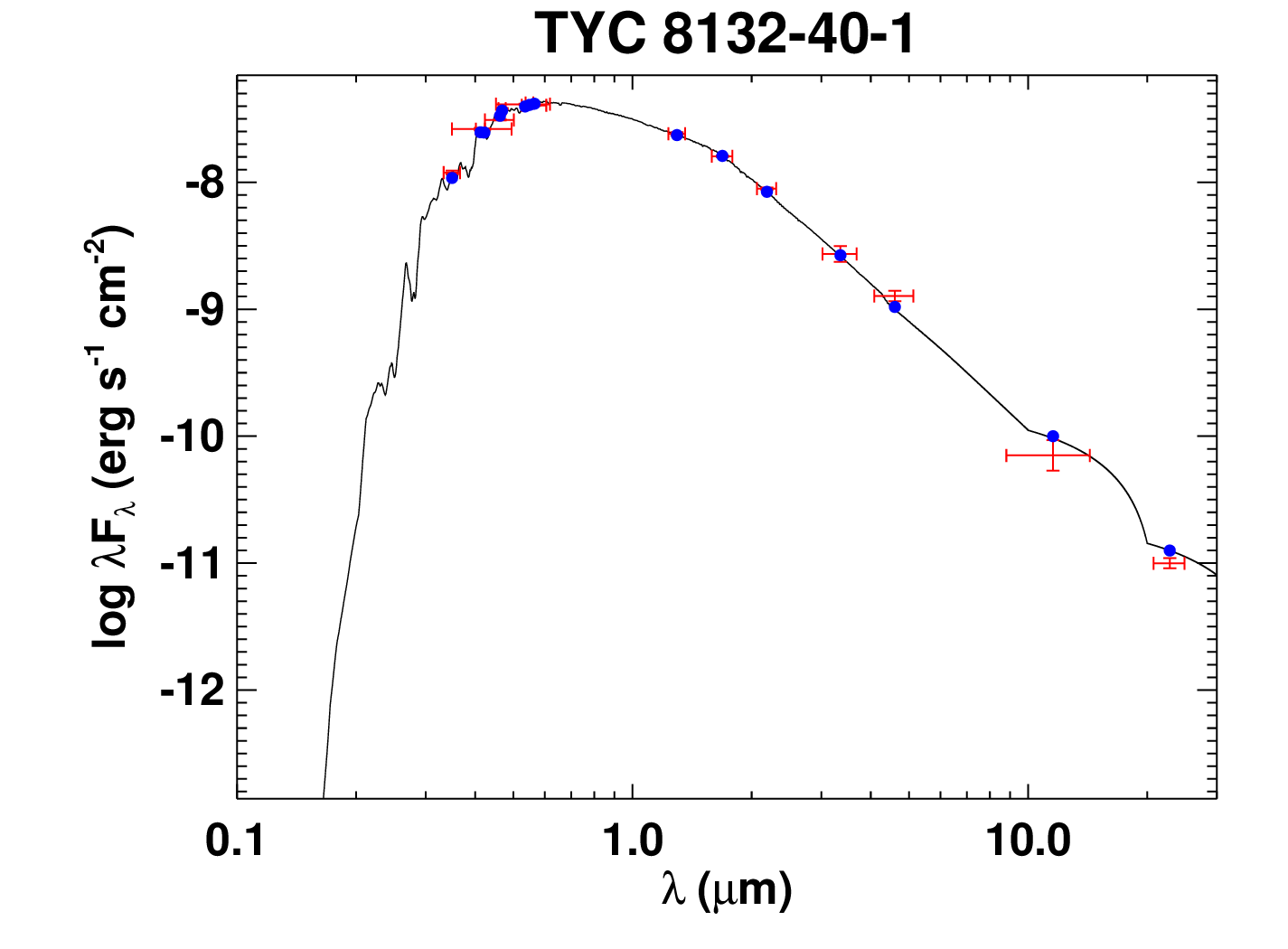}
\caption{\label{fig:seds17} All labels, lines, symbols, and colors as in Figure \ref{fig:seds}.}
\end{figure*}

\begin{figure*}
\includegraphics[width=0.333\linewidth]{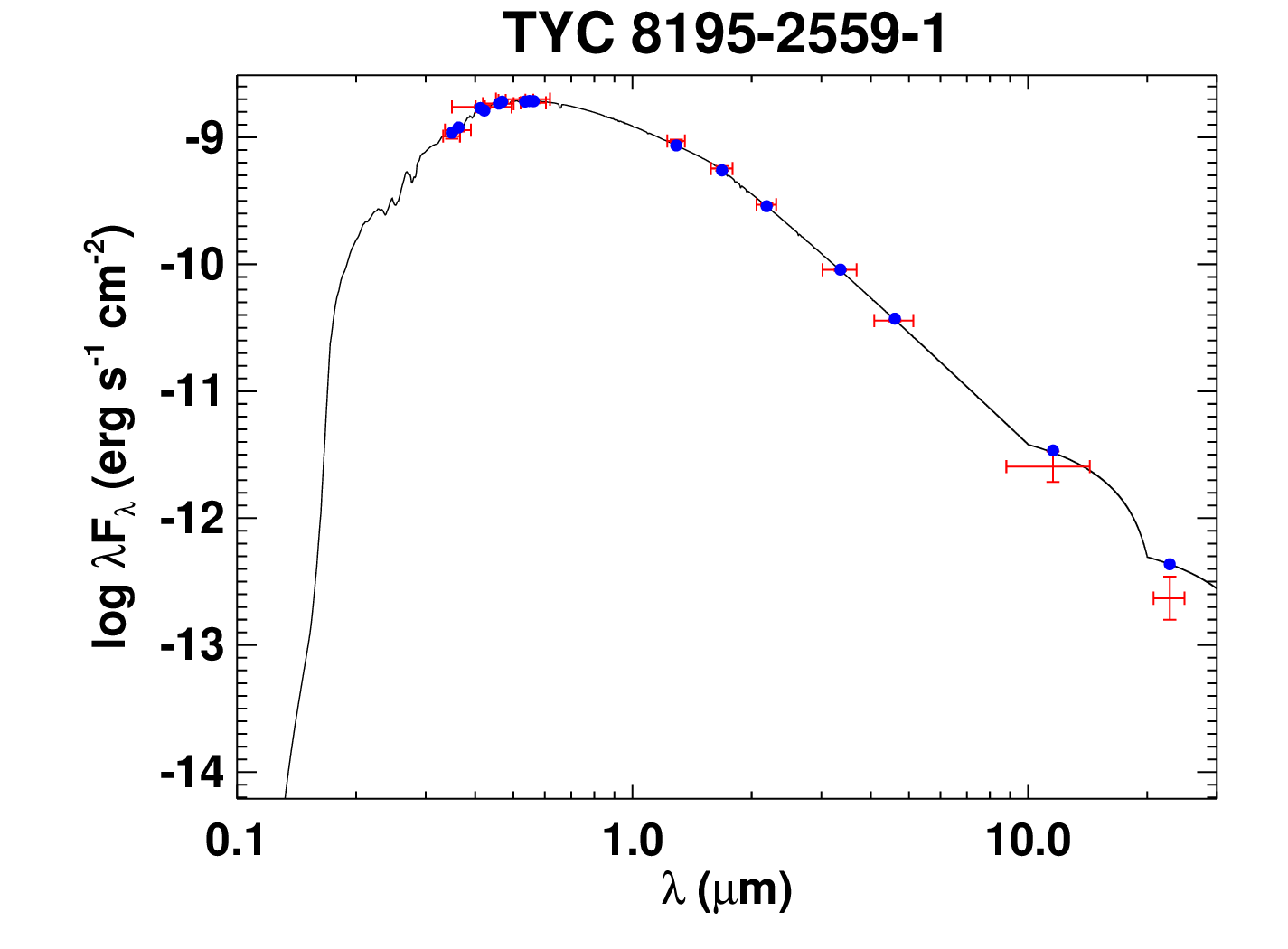}\includegraphics[width=0.333\linewidth]{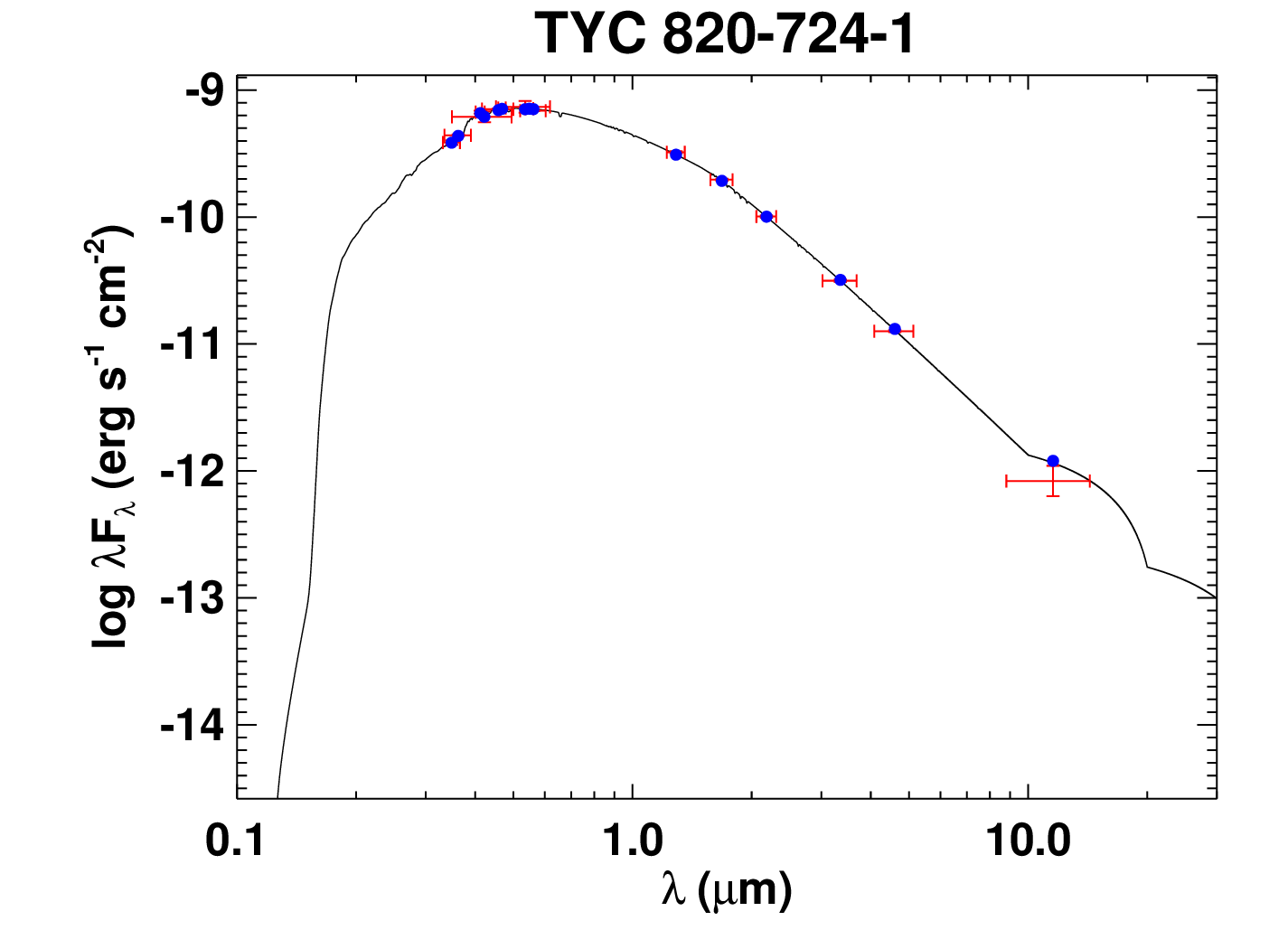}\includegraphics[width=0.333\linewidth]{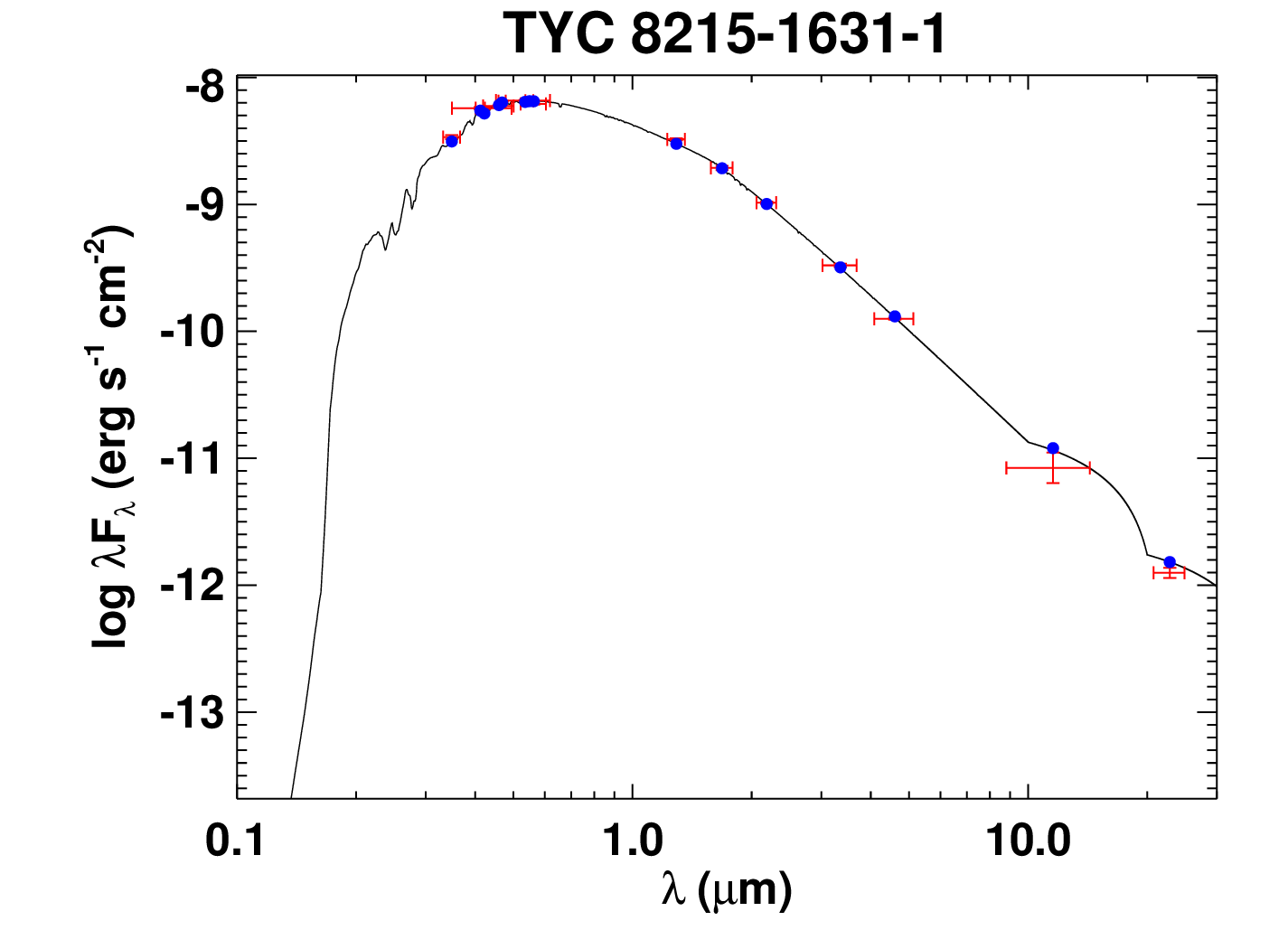}
\includegraphics[width=0.333\linewidth]{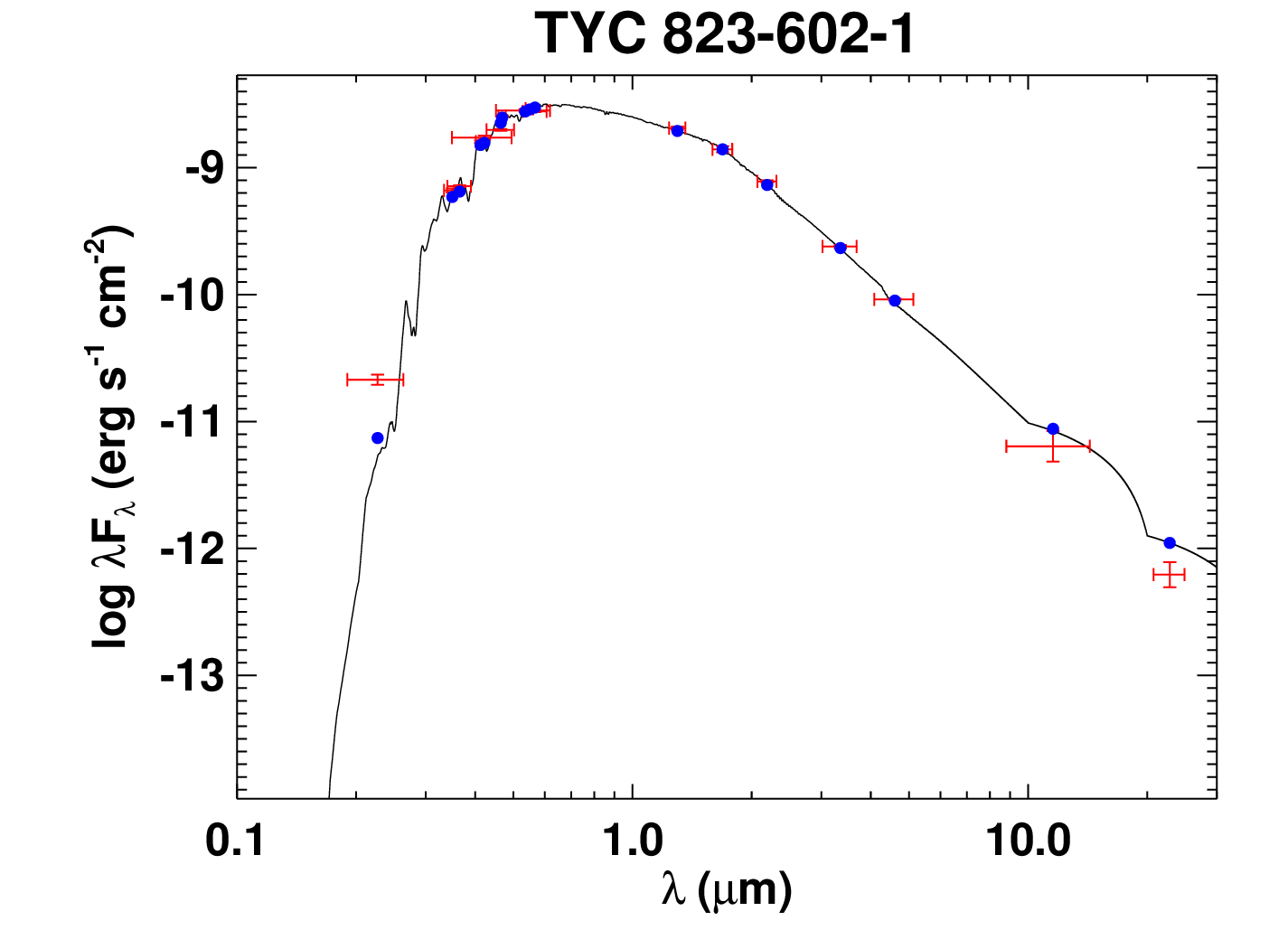}\includegraphics[width=0.333\linewidth]{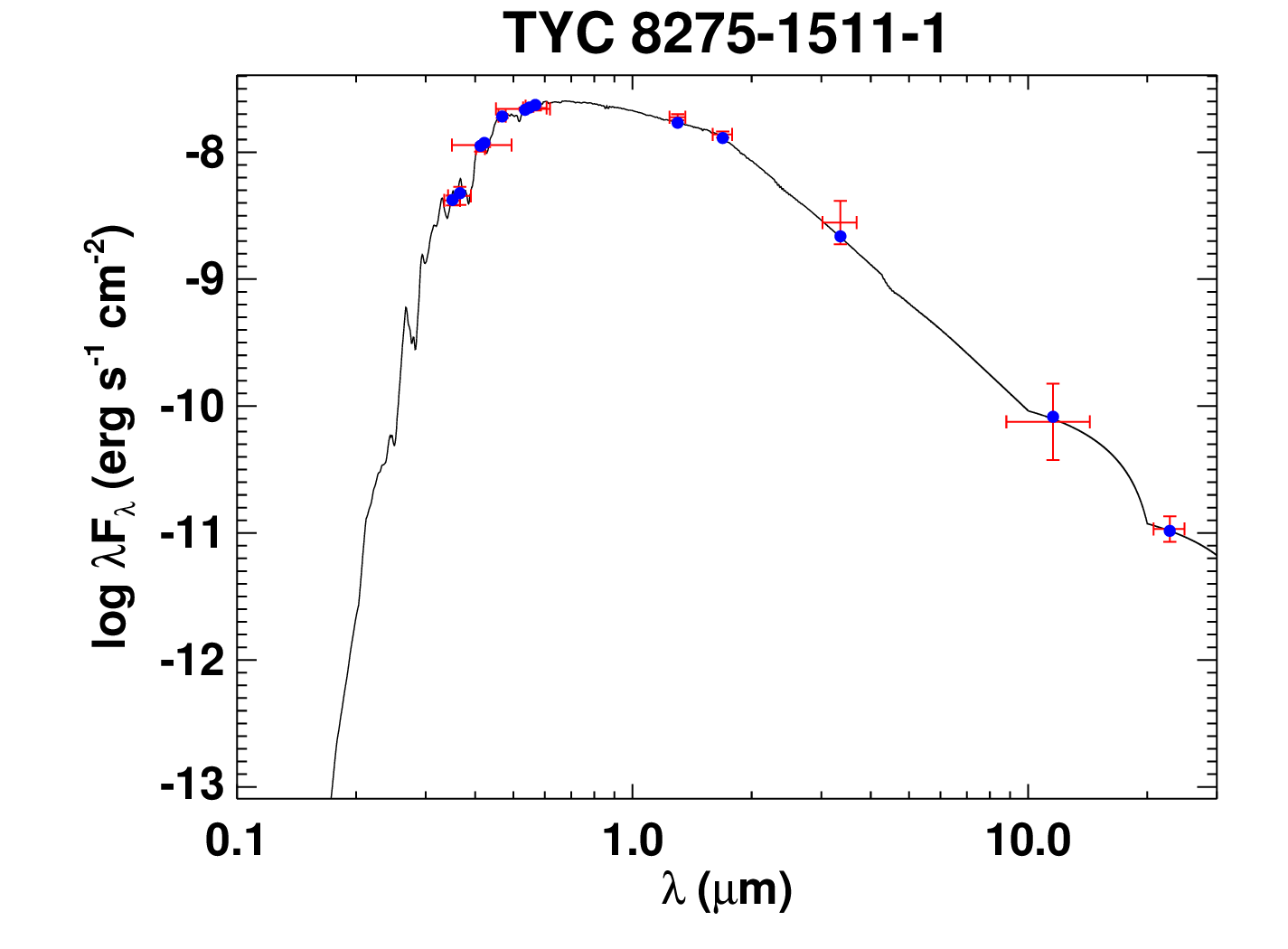}\includegraphics[width=0.333\linewidth]{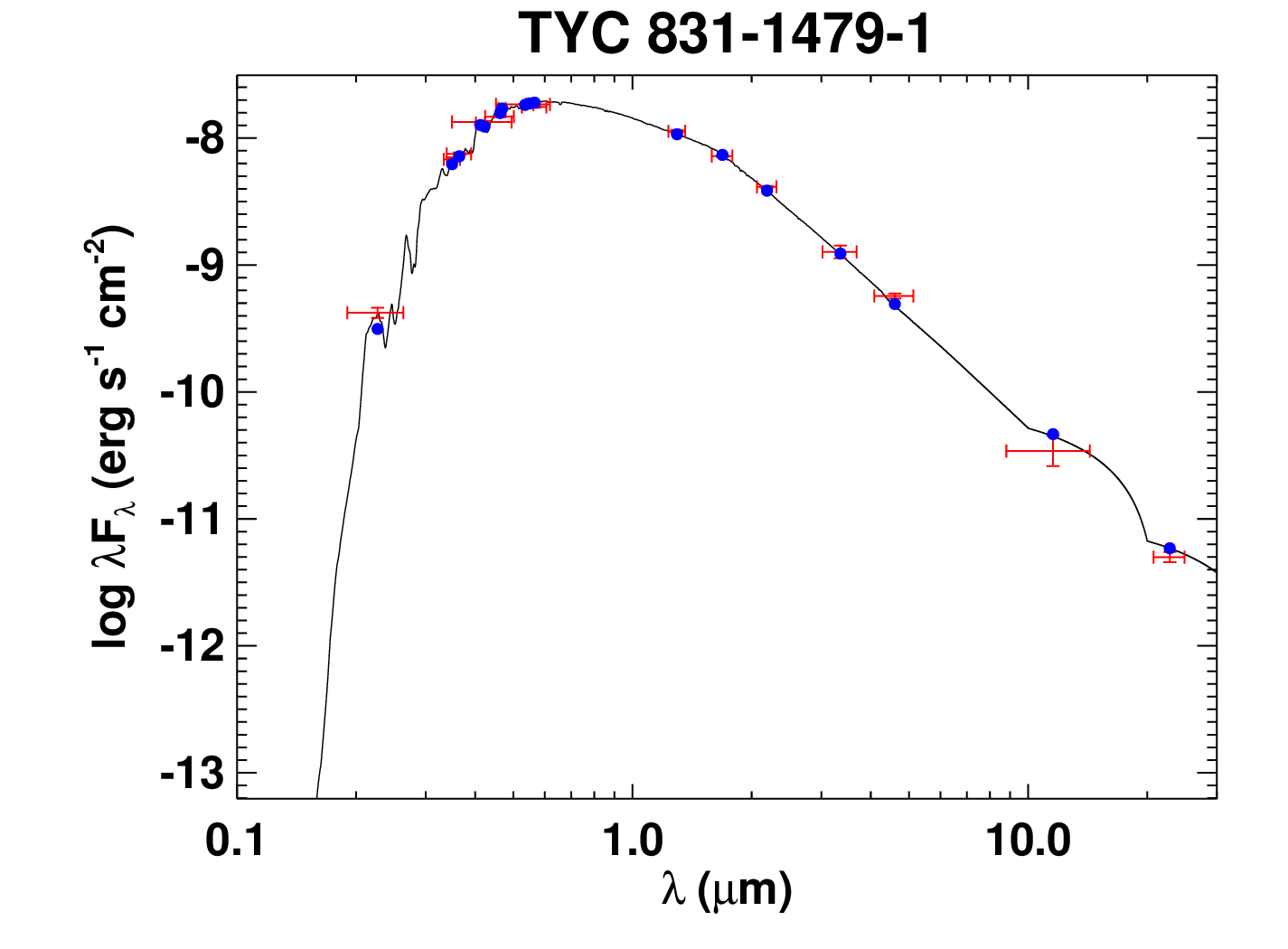}
\includegraphics[width=0.333\linewidth]{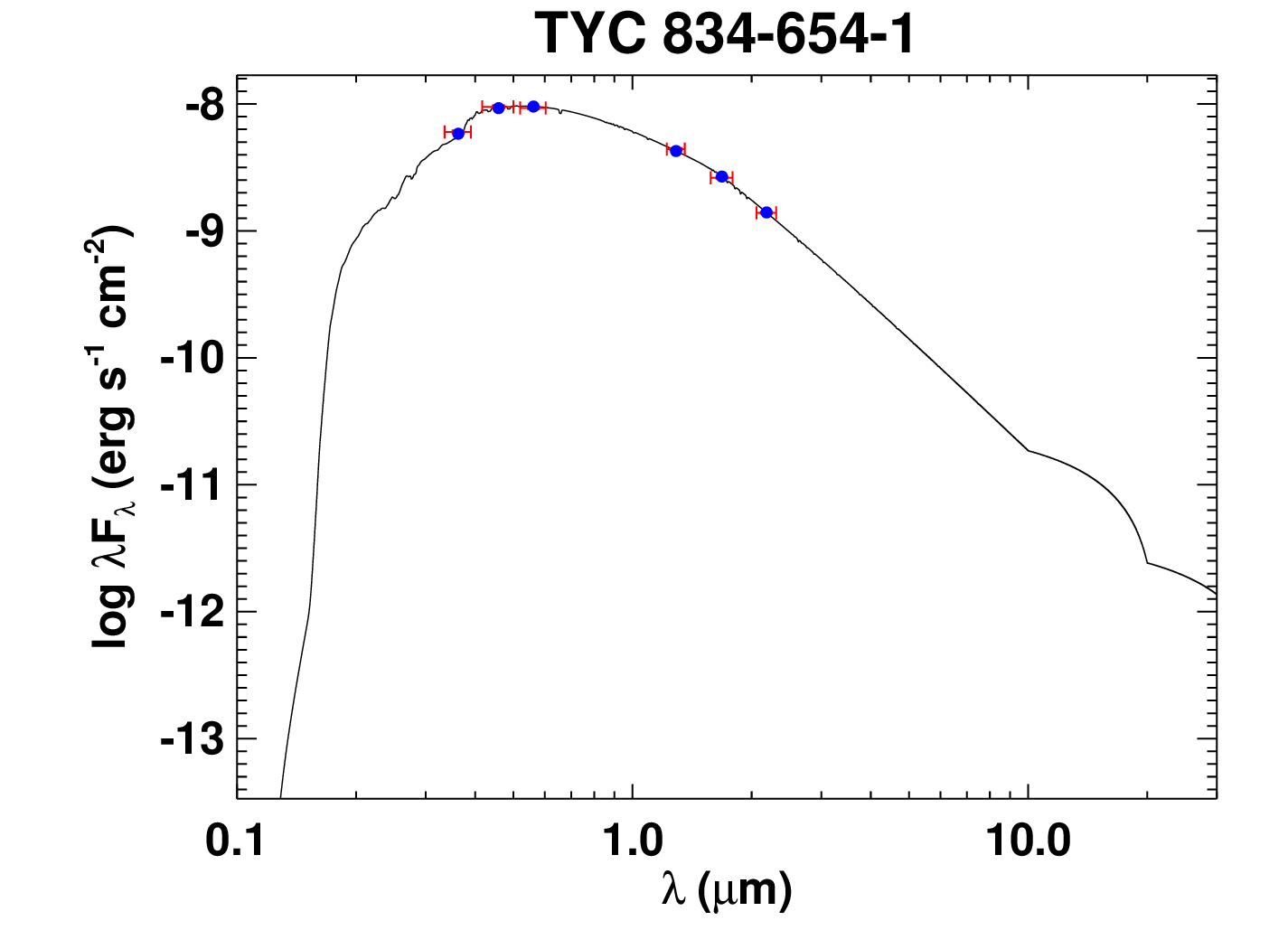}\includegraphics[width=0.333\linewidth]{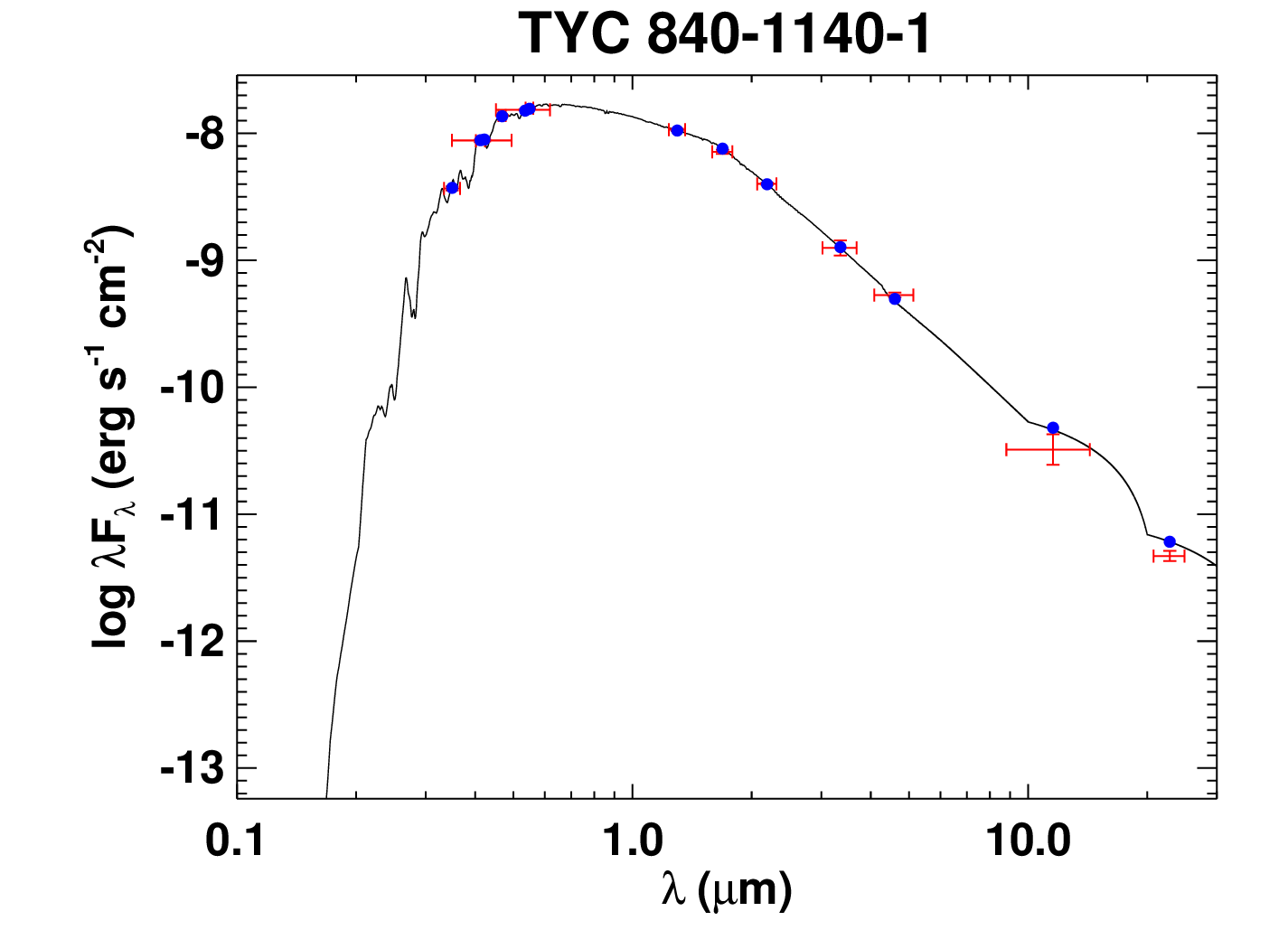}\includegraphics[width=0.333\linewidth]{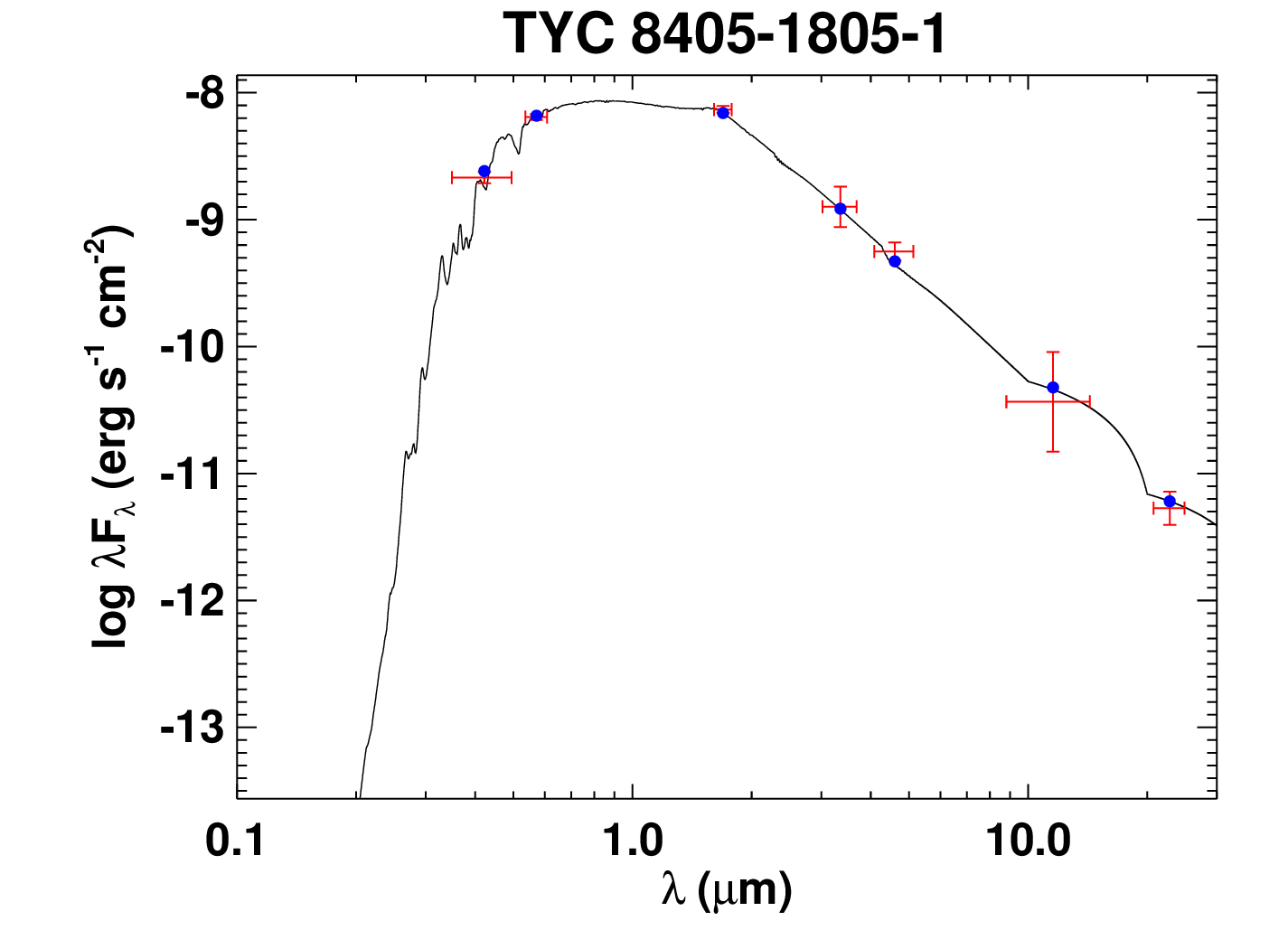}
\includegraphics[width=0.333\linewidth]{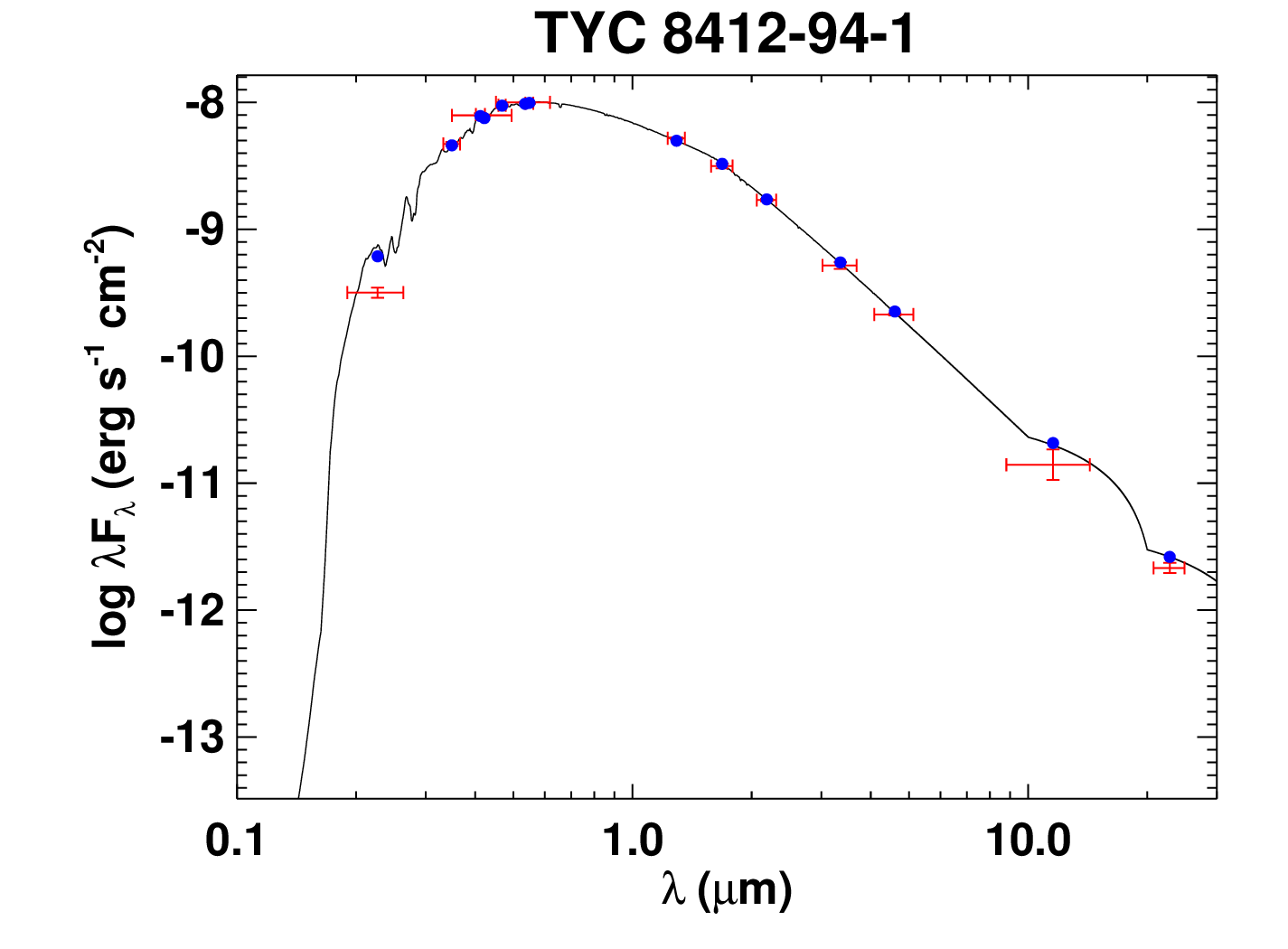}\includegraphics[width=0.333\linewidth]{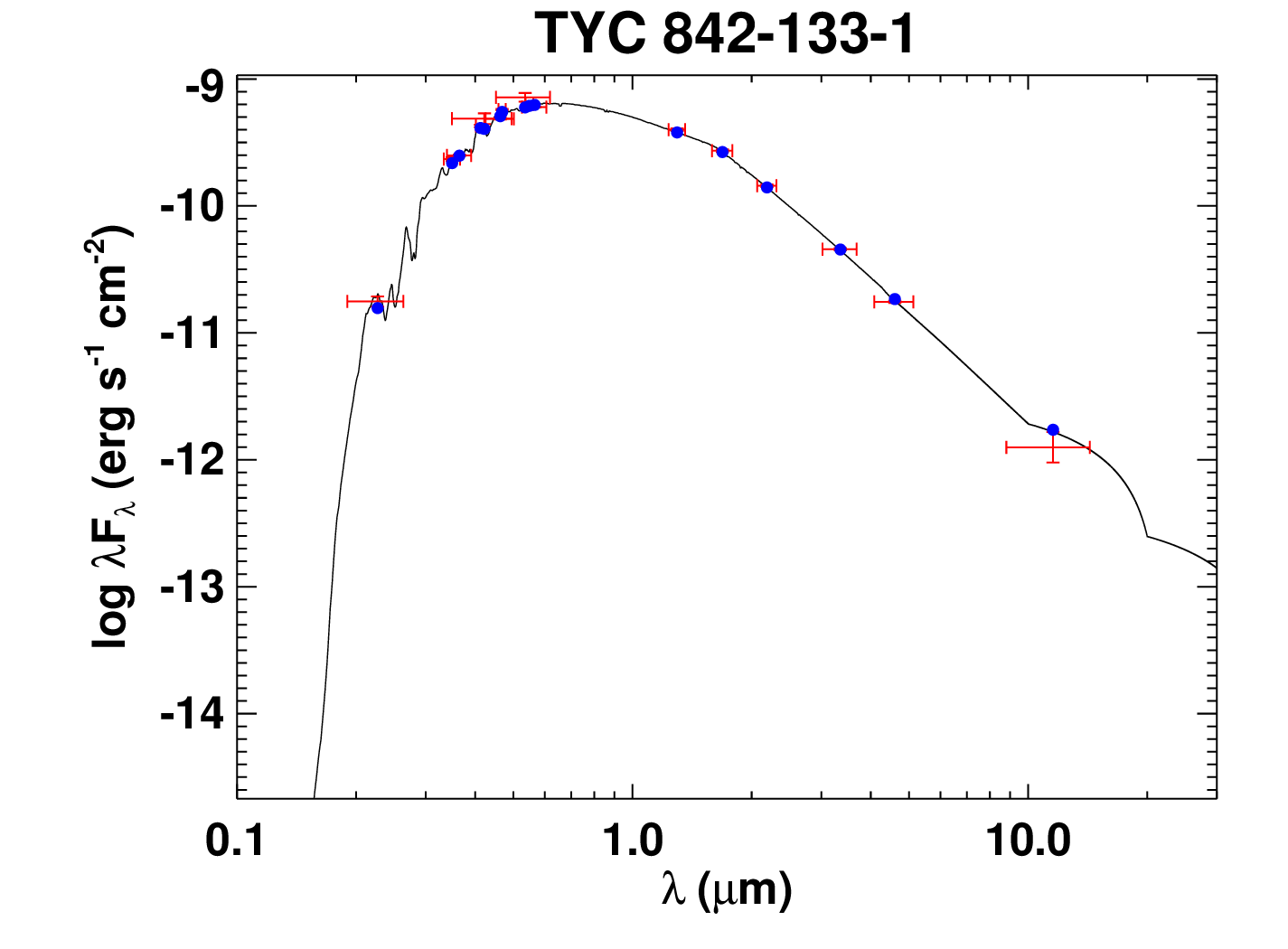}\includegraphics[width=0.333\linewidth]{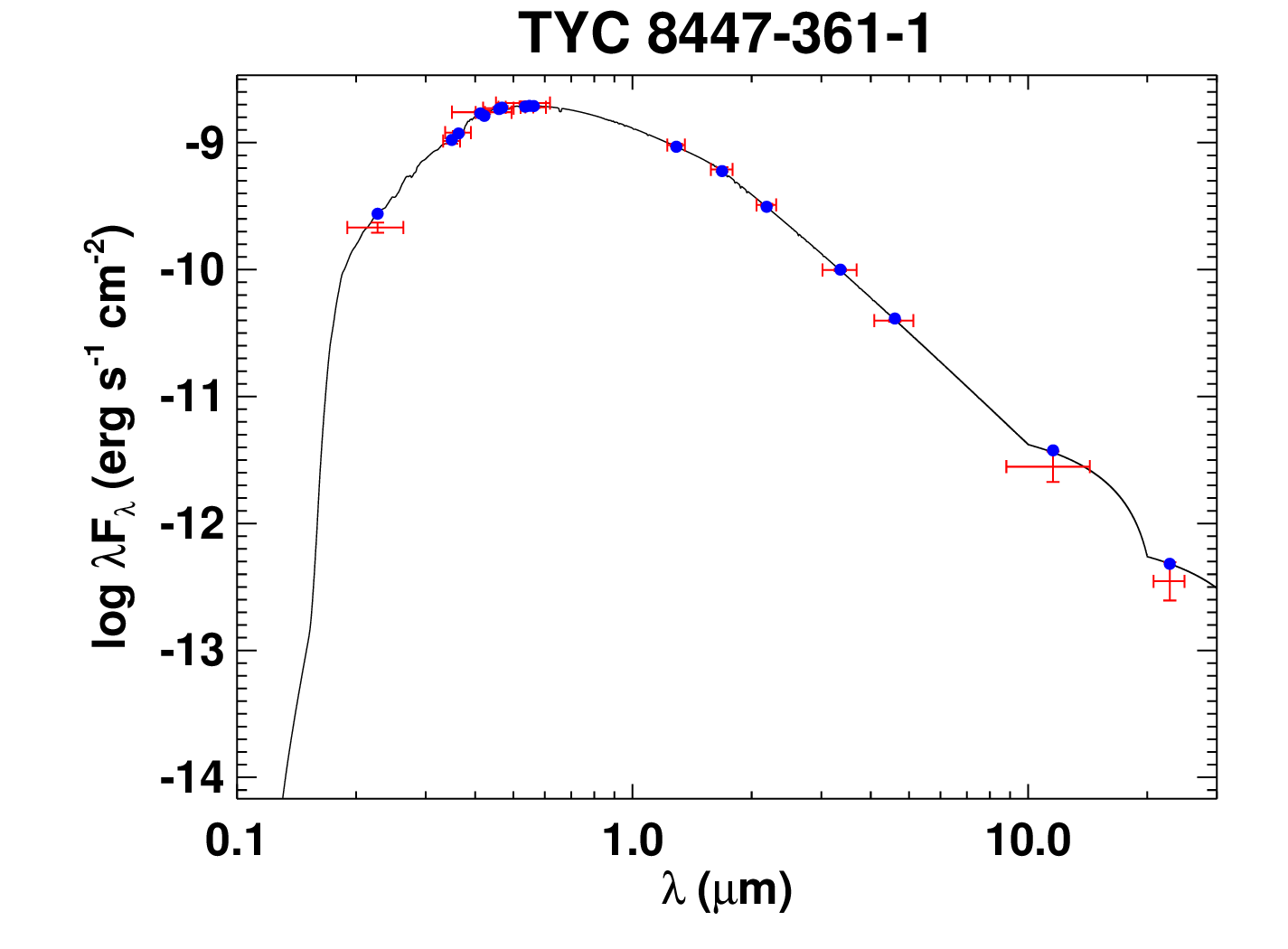}
\caption{\label{fig:seds18} All labels, lines, symbols, and colors as in Figure \ref{fig:seds}.}
\end{figure*}

\begin{figure*}
\includegraphics[width=0.333\linewidth]{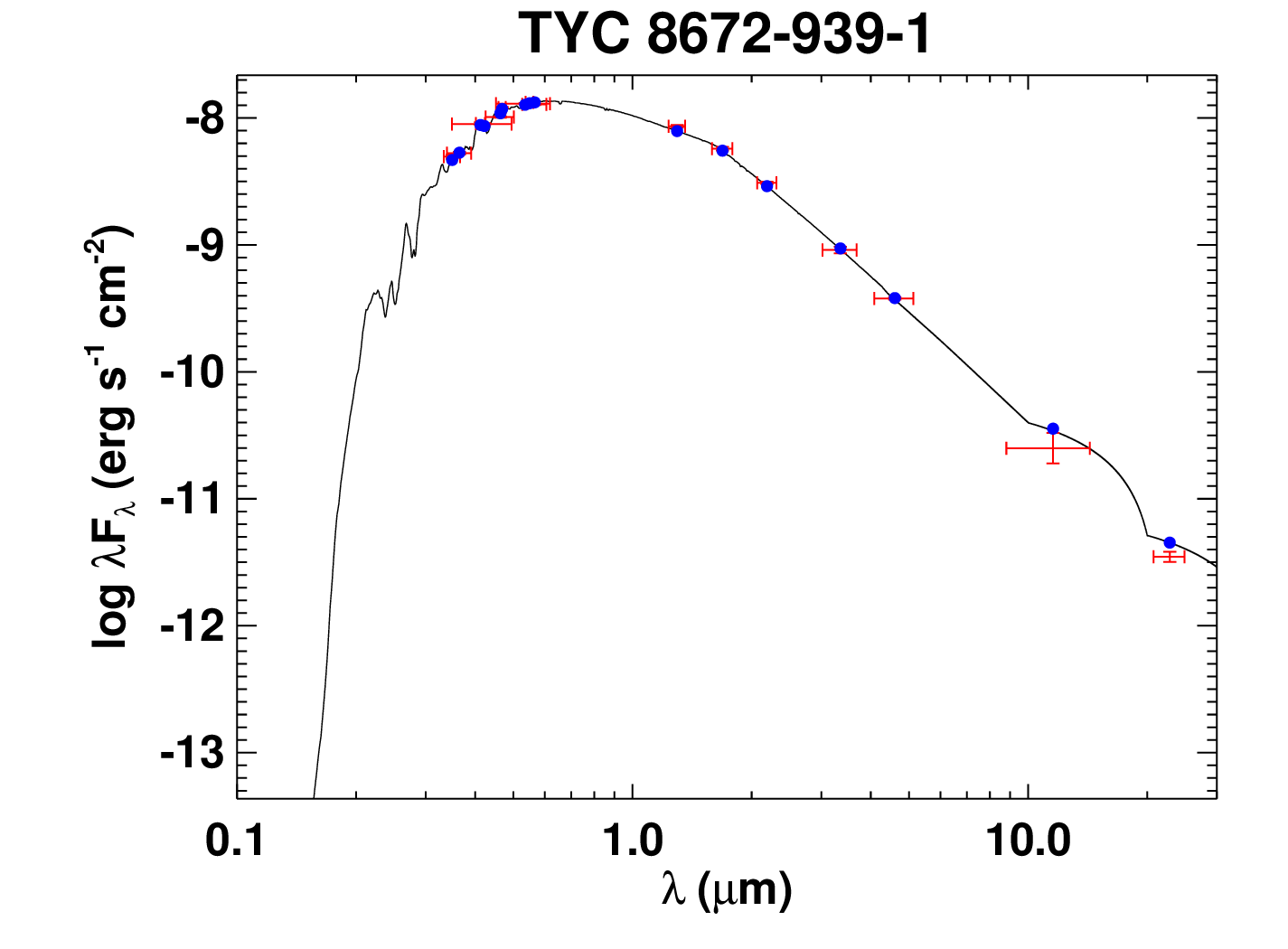}\includegraphics[width=0.333\linewidth]{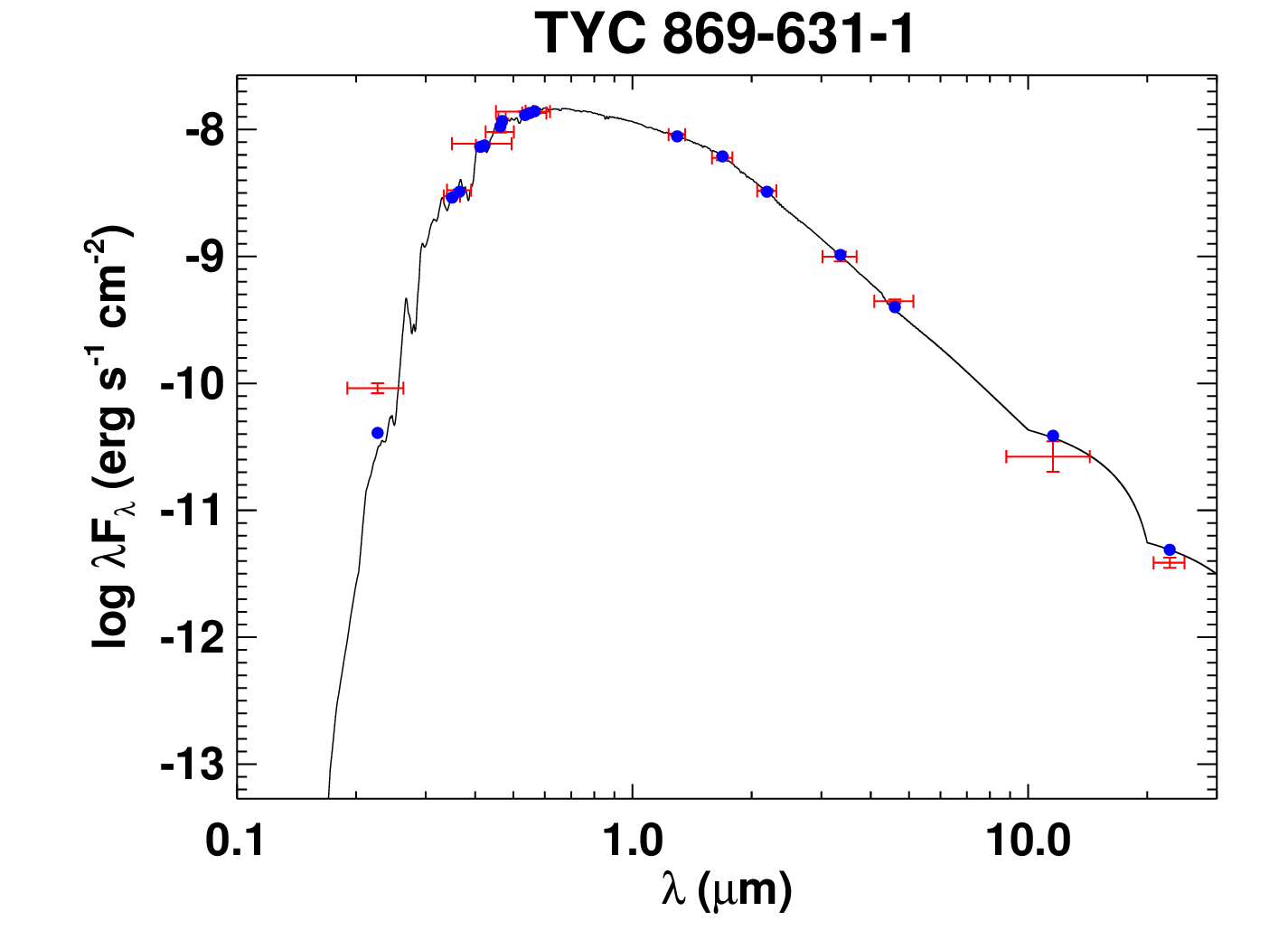}\includegraphics[width=0.333\linewidth]{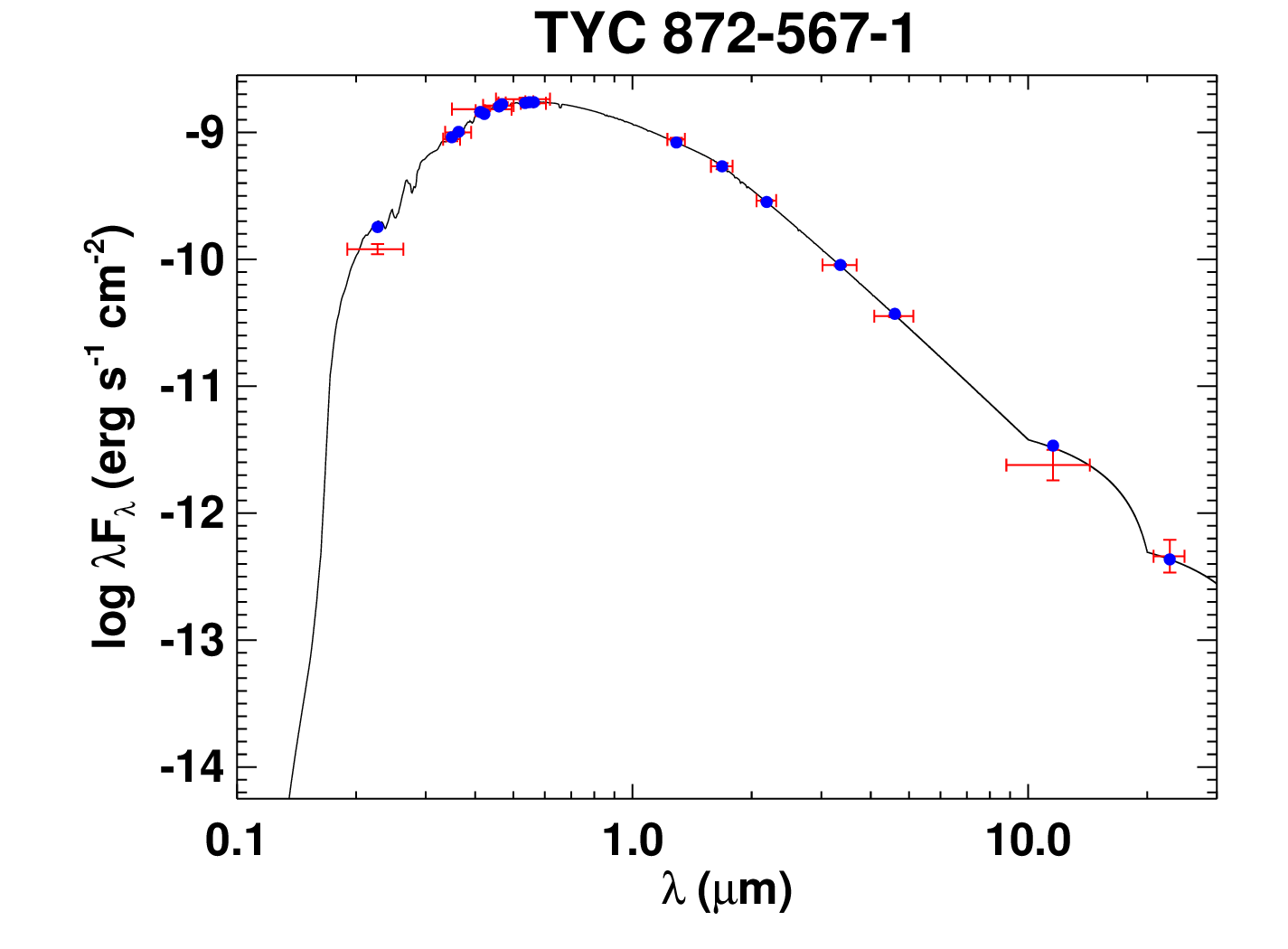}
\includegraphics[width=0.333\linewidth]{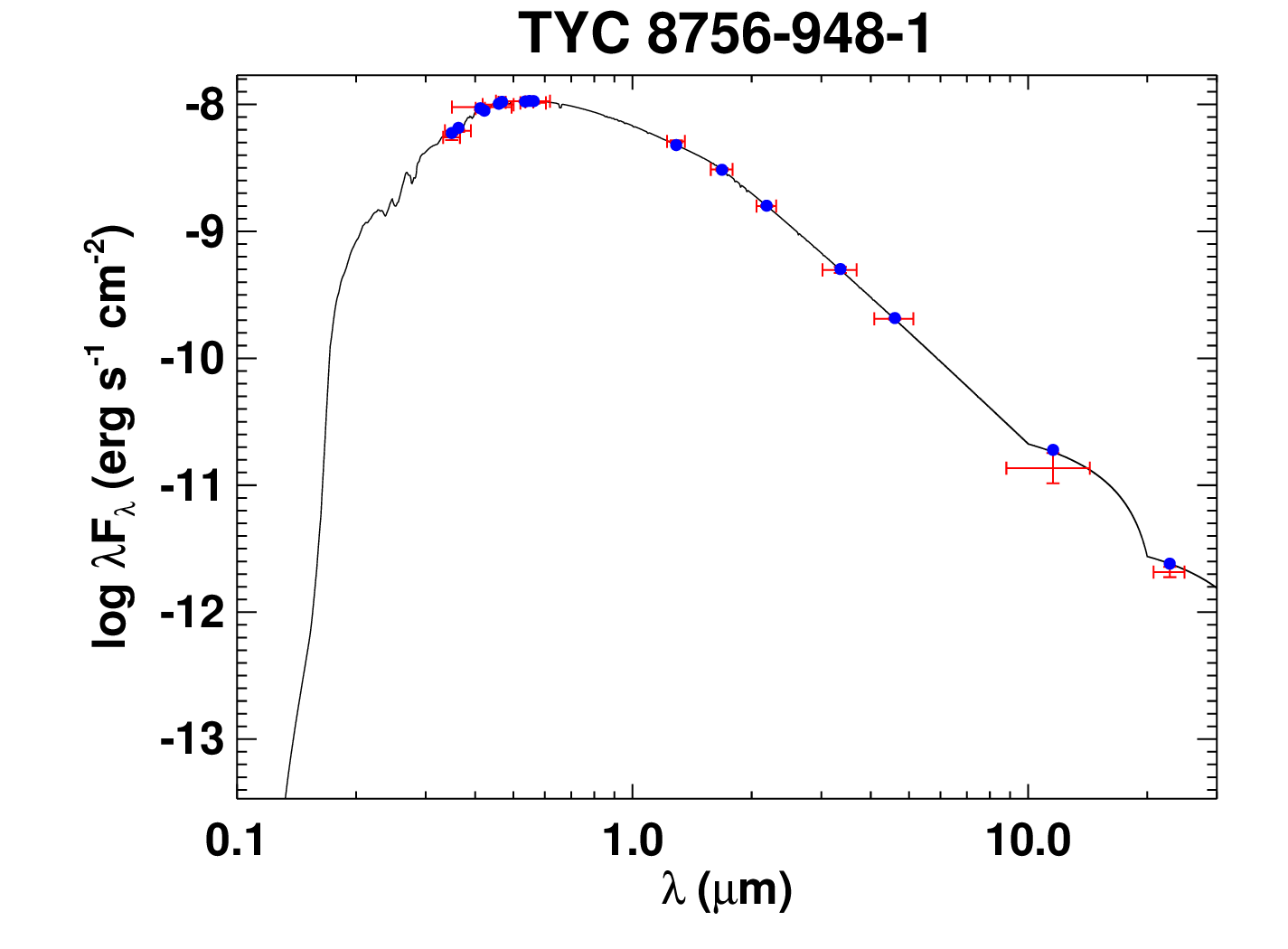}\includegraphics[width=0.333\linewidth]{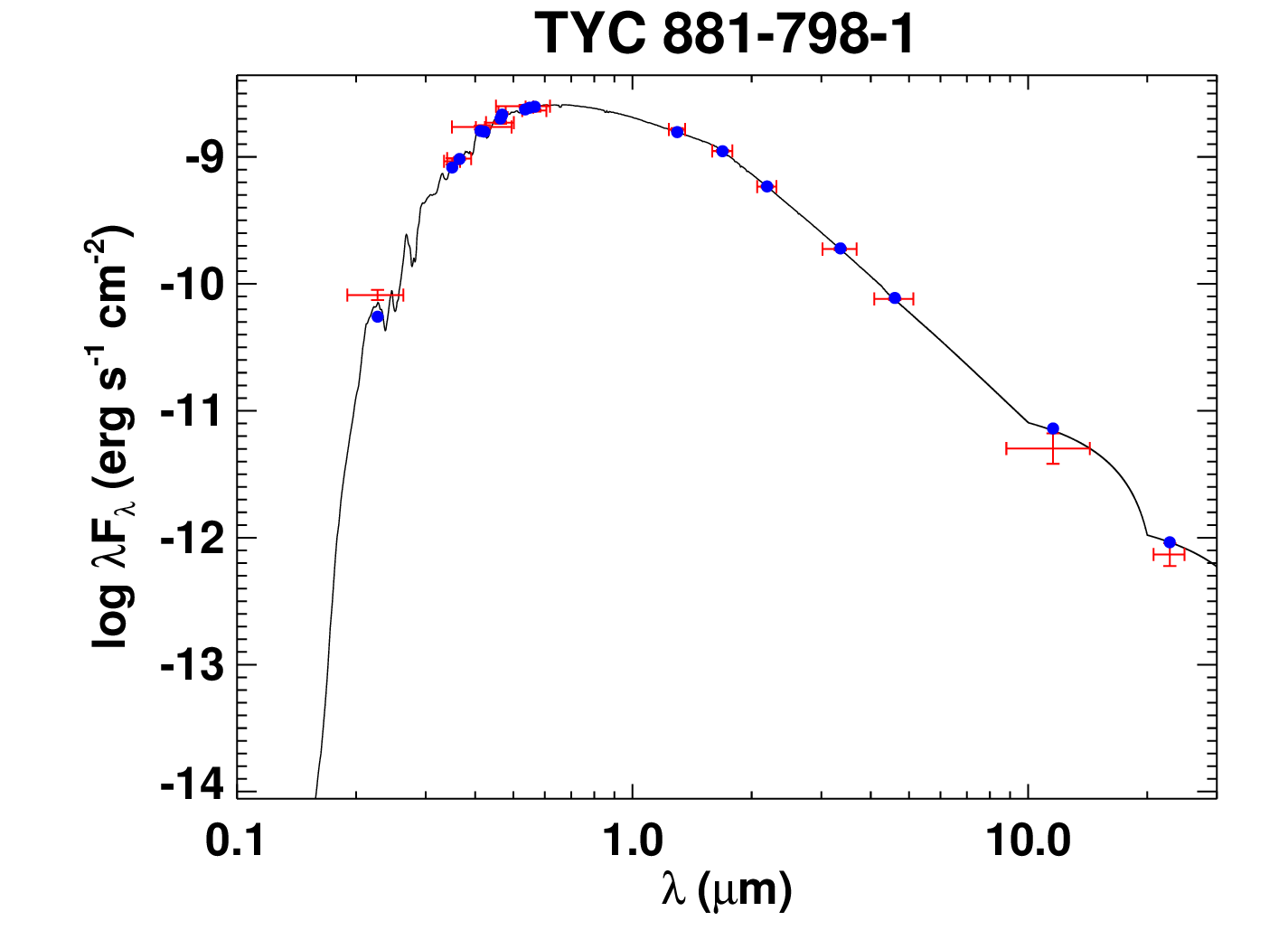}\includegraphics[width=0.333\linewidth]{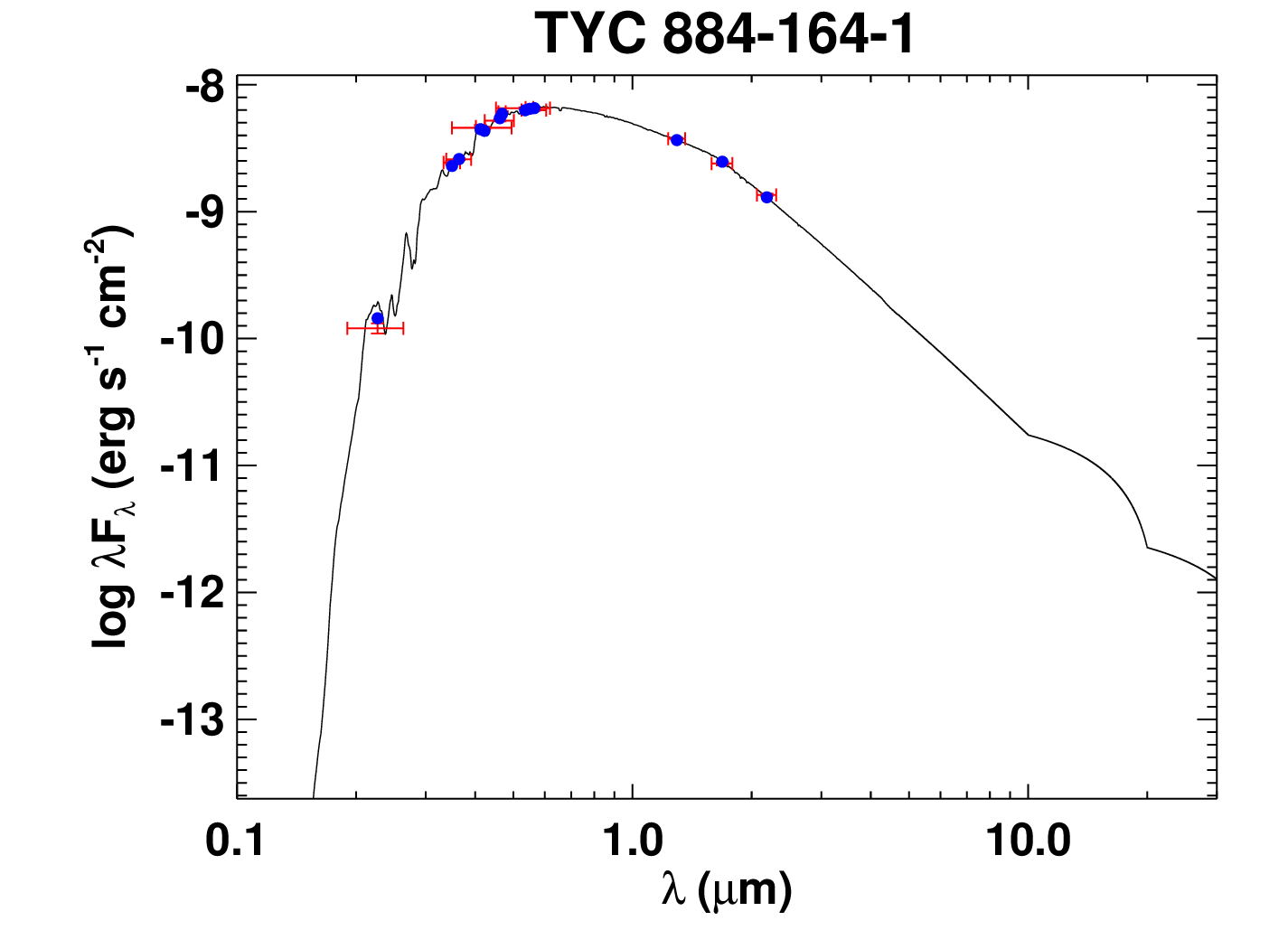}
\includegraphics[width=0.333\linewidth]{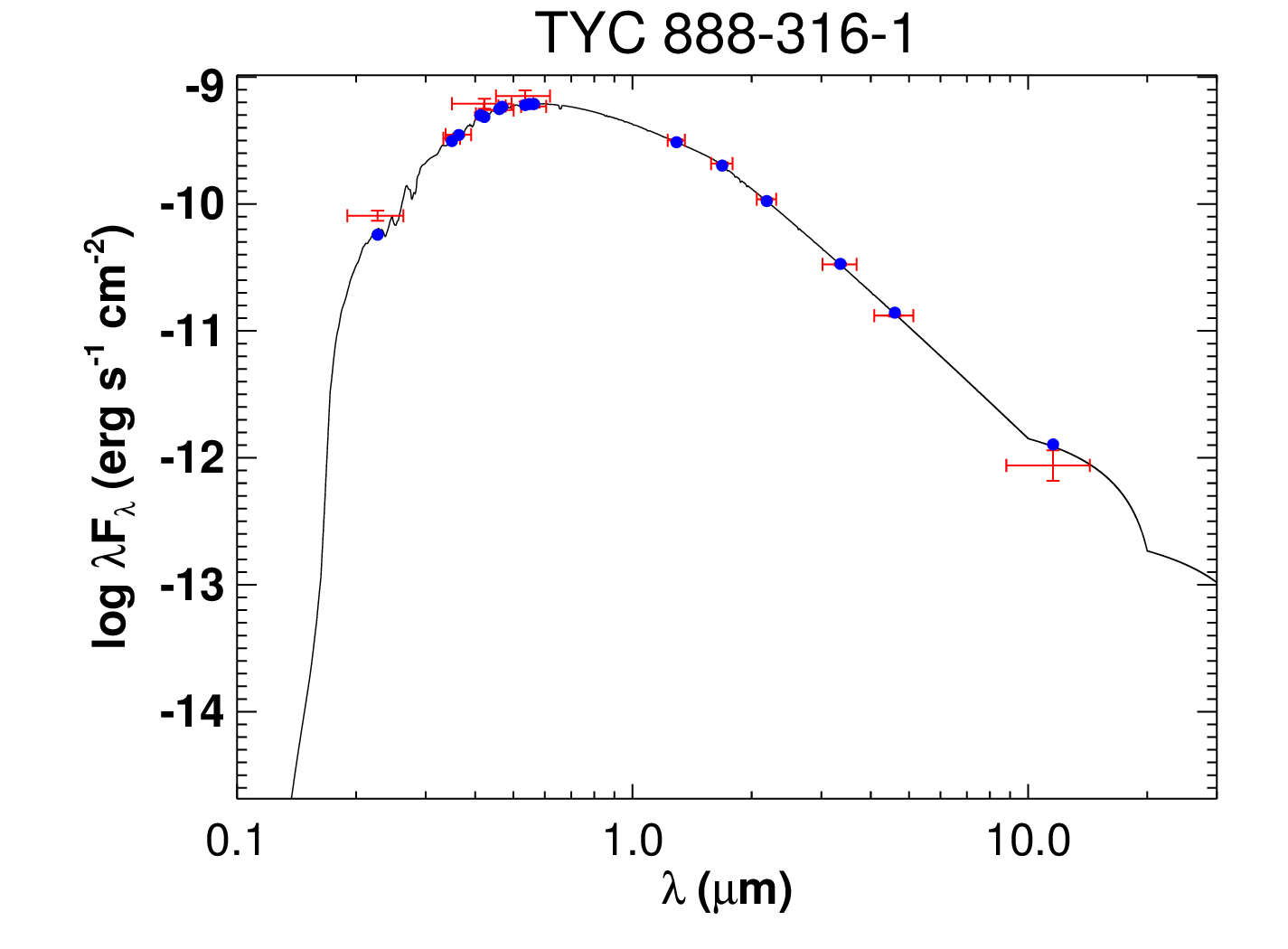}\includegraphics[width=0.333\linewidth]{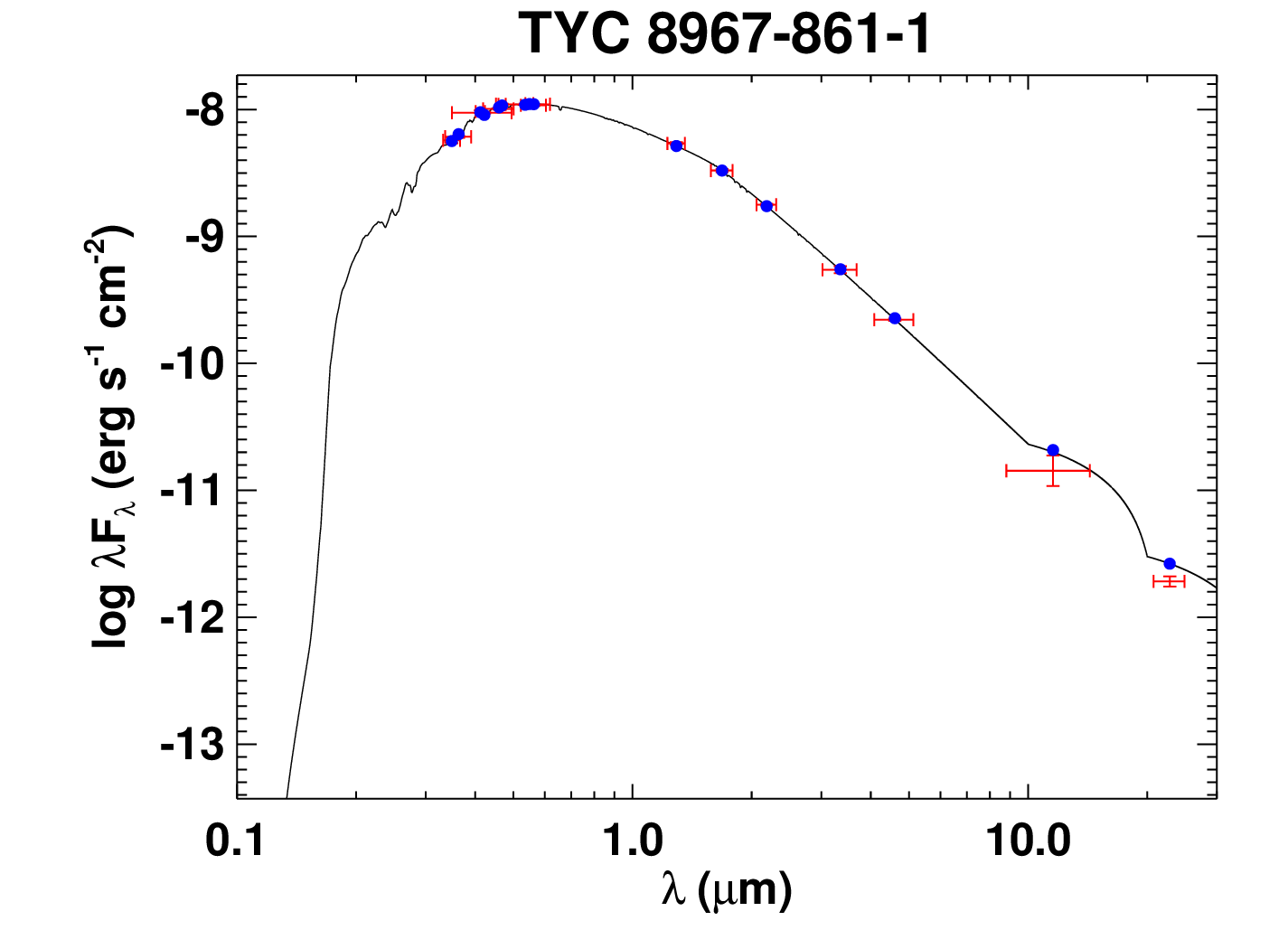}\includegraphics[width=0.333\linewidth]{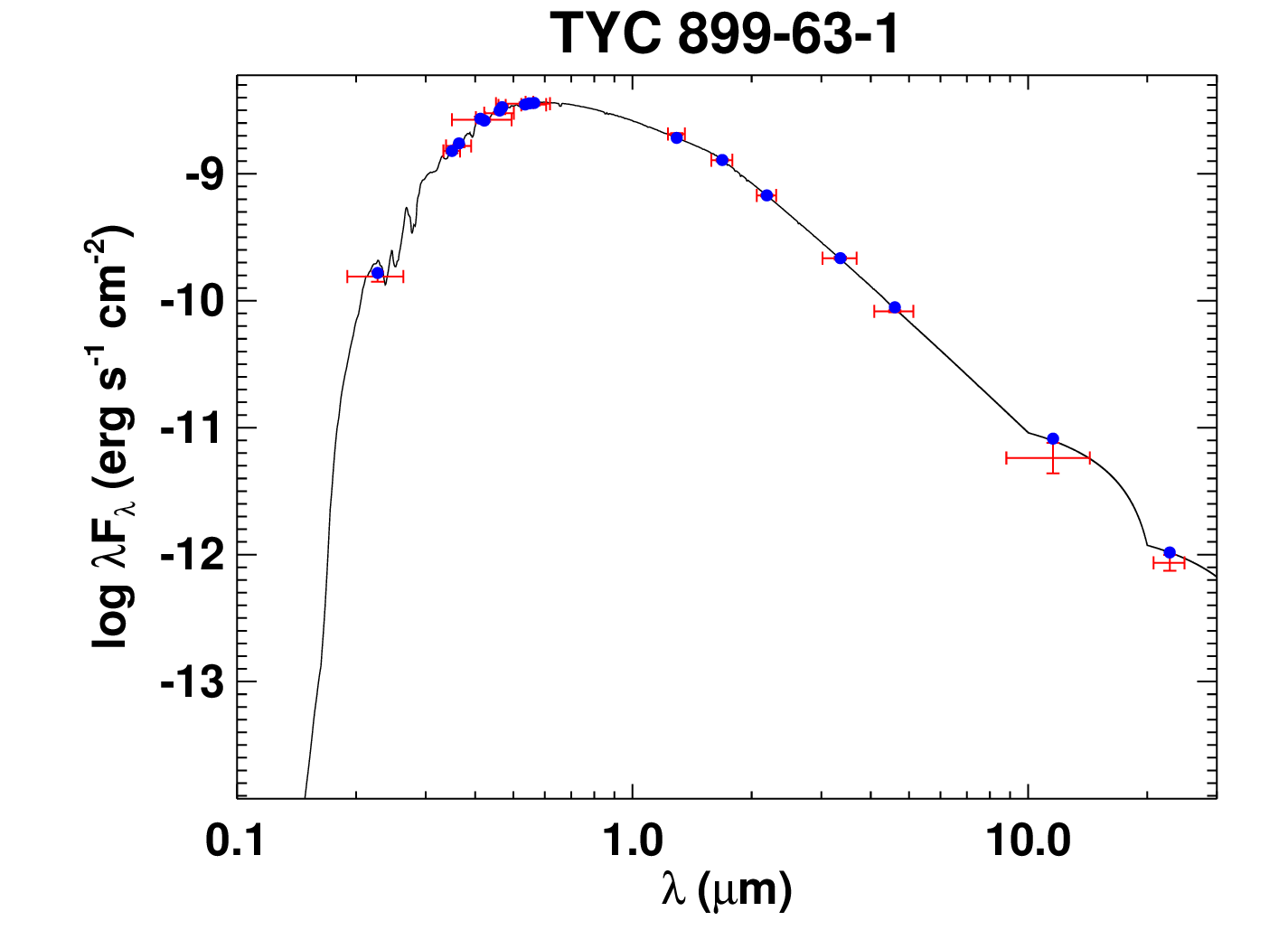}
\includegraphics[width=0.333\linewidth]{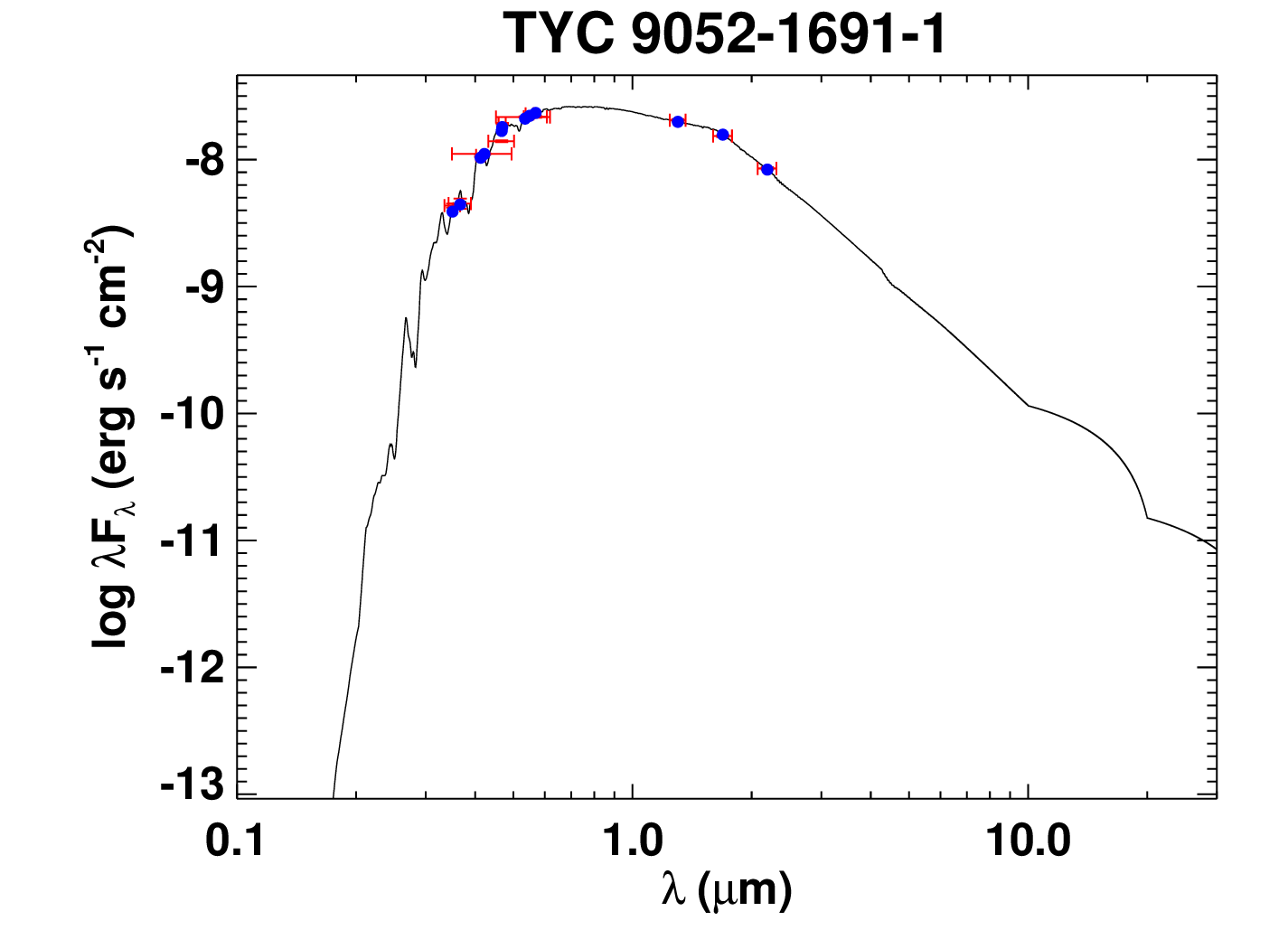}\includegraphics[width=0.333\linewidth]{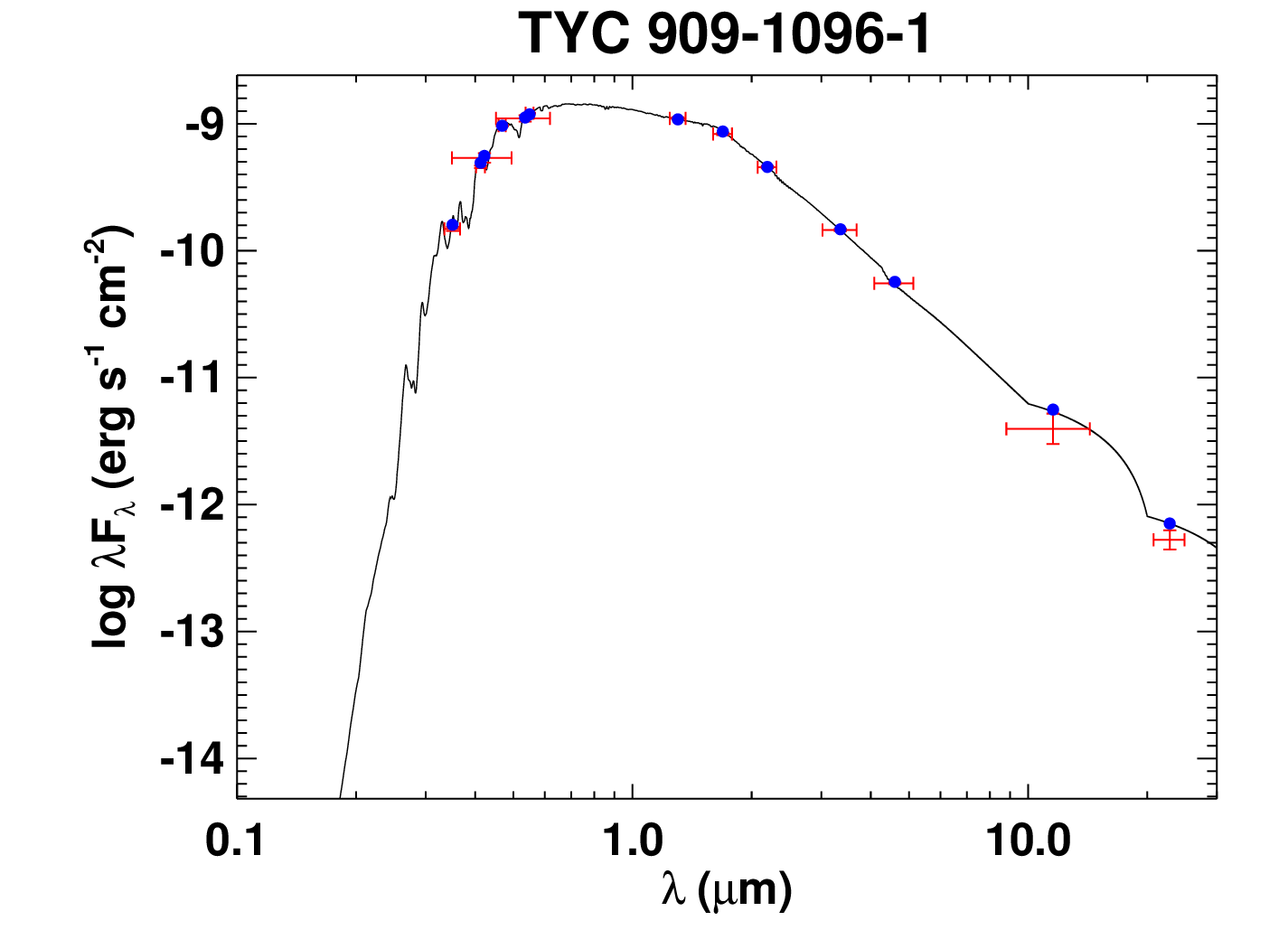}\includegraphics[width=0.333\linewidth]{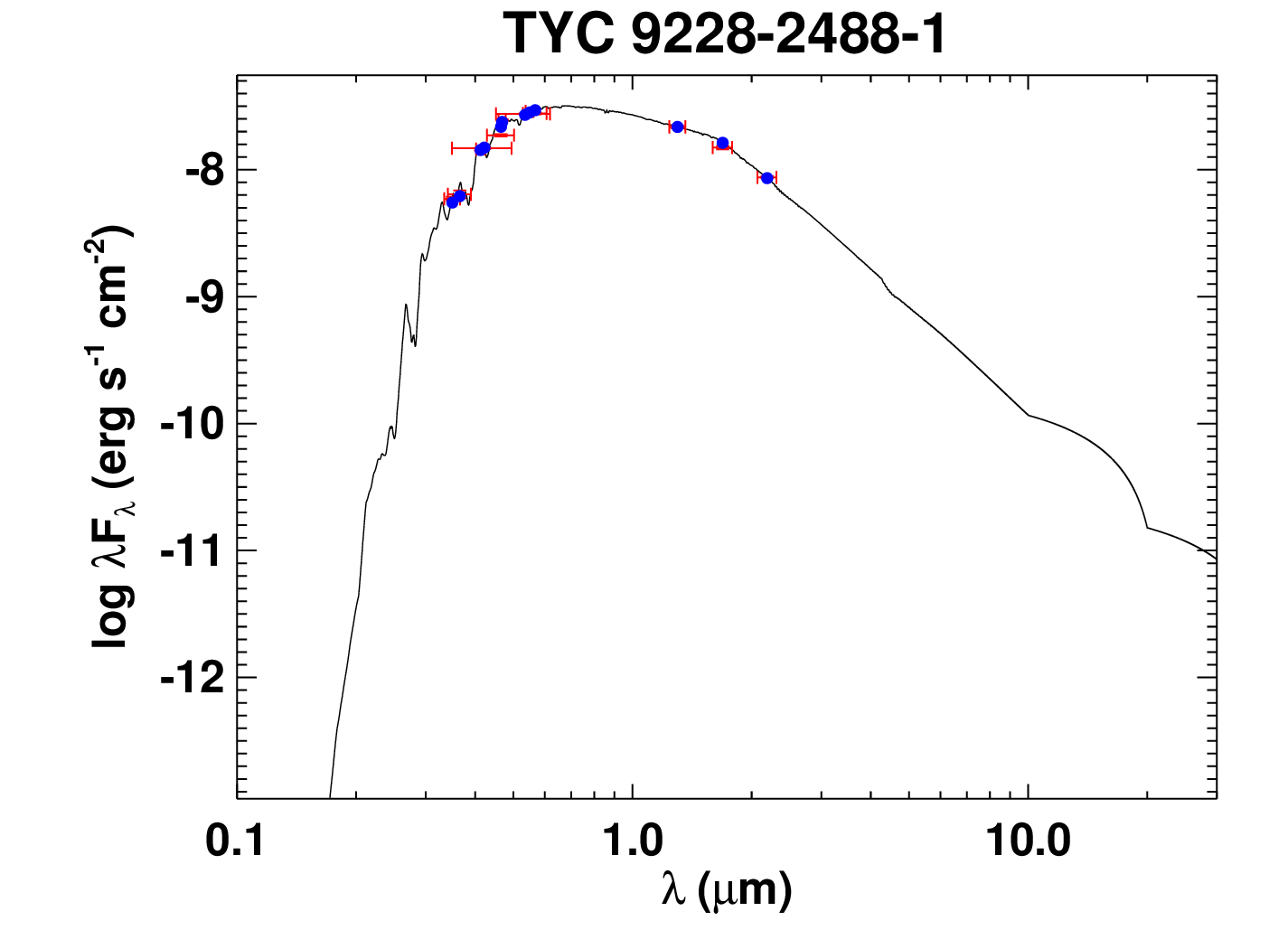}
\caption{\label{fig:seds19} All labels, lines, symbols, and colors as in Figure \ref{fig:seds}.}
\end{figure*}

\begin{figure*}
\includegraphics[width=0.333\linewidth]{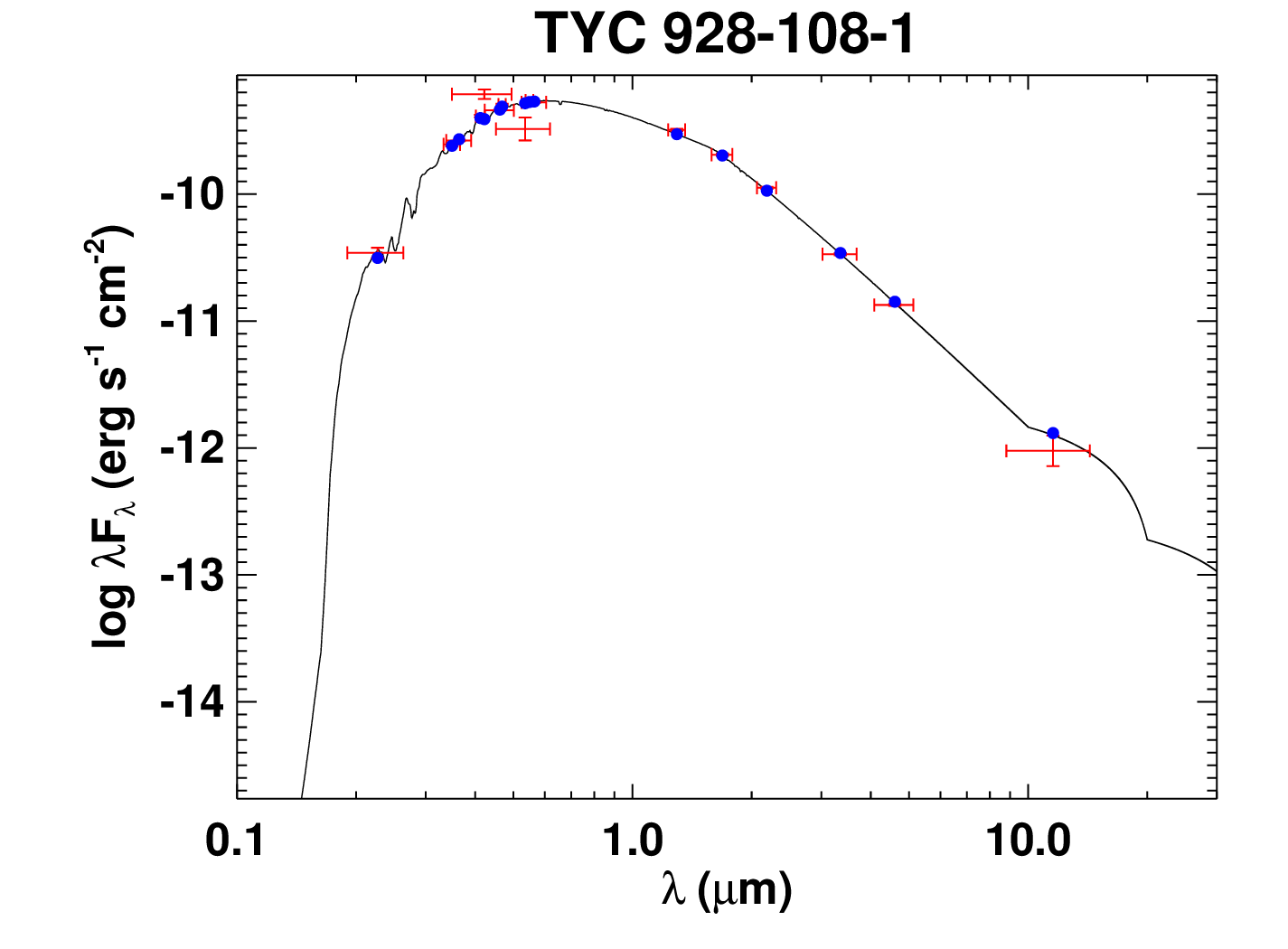}\includegraphics[width=0.333\linewidth]{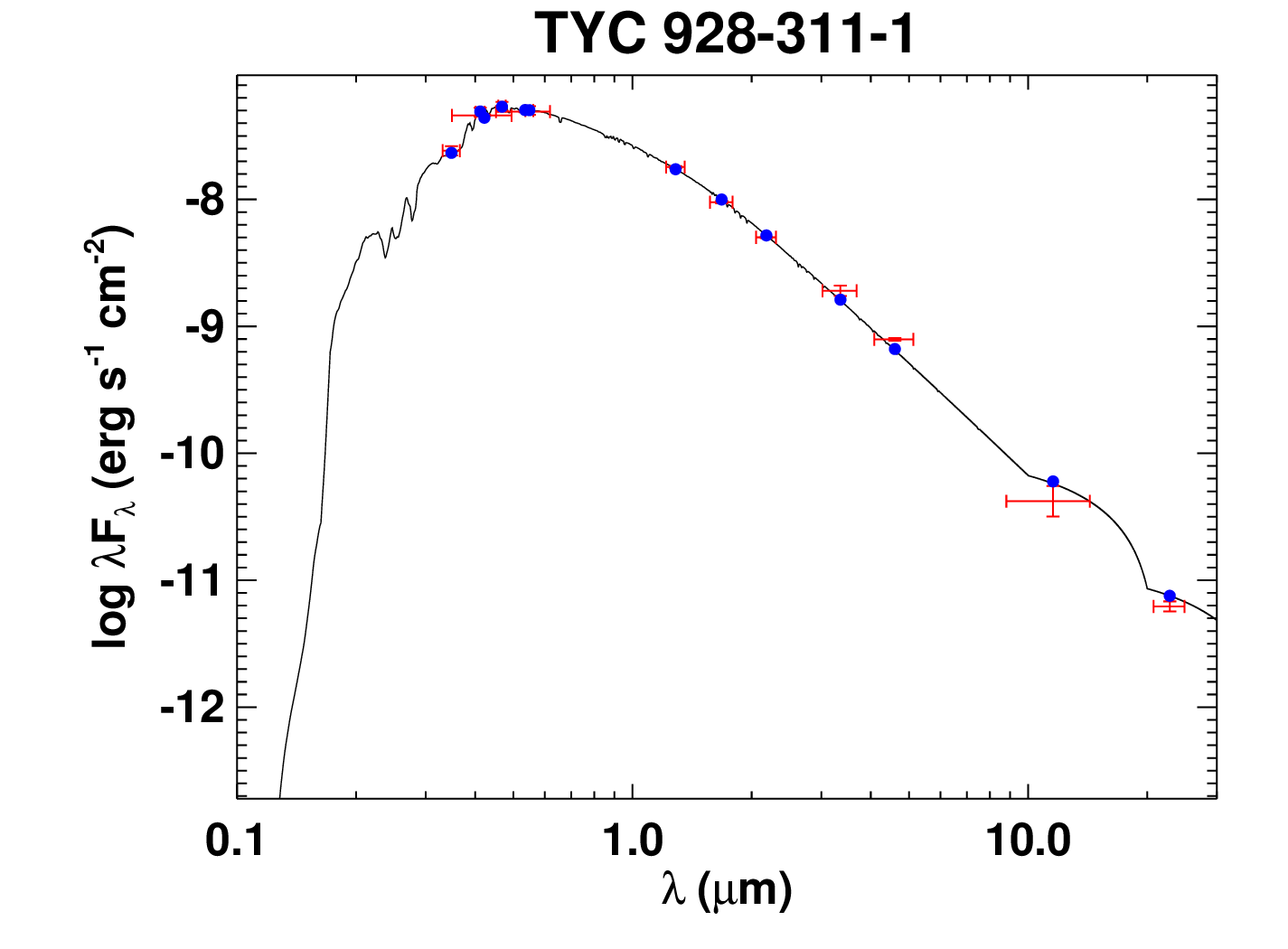}\includegraphics[width=0.333\linewidth]{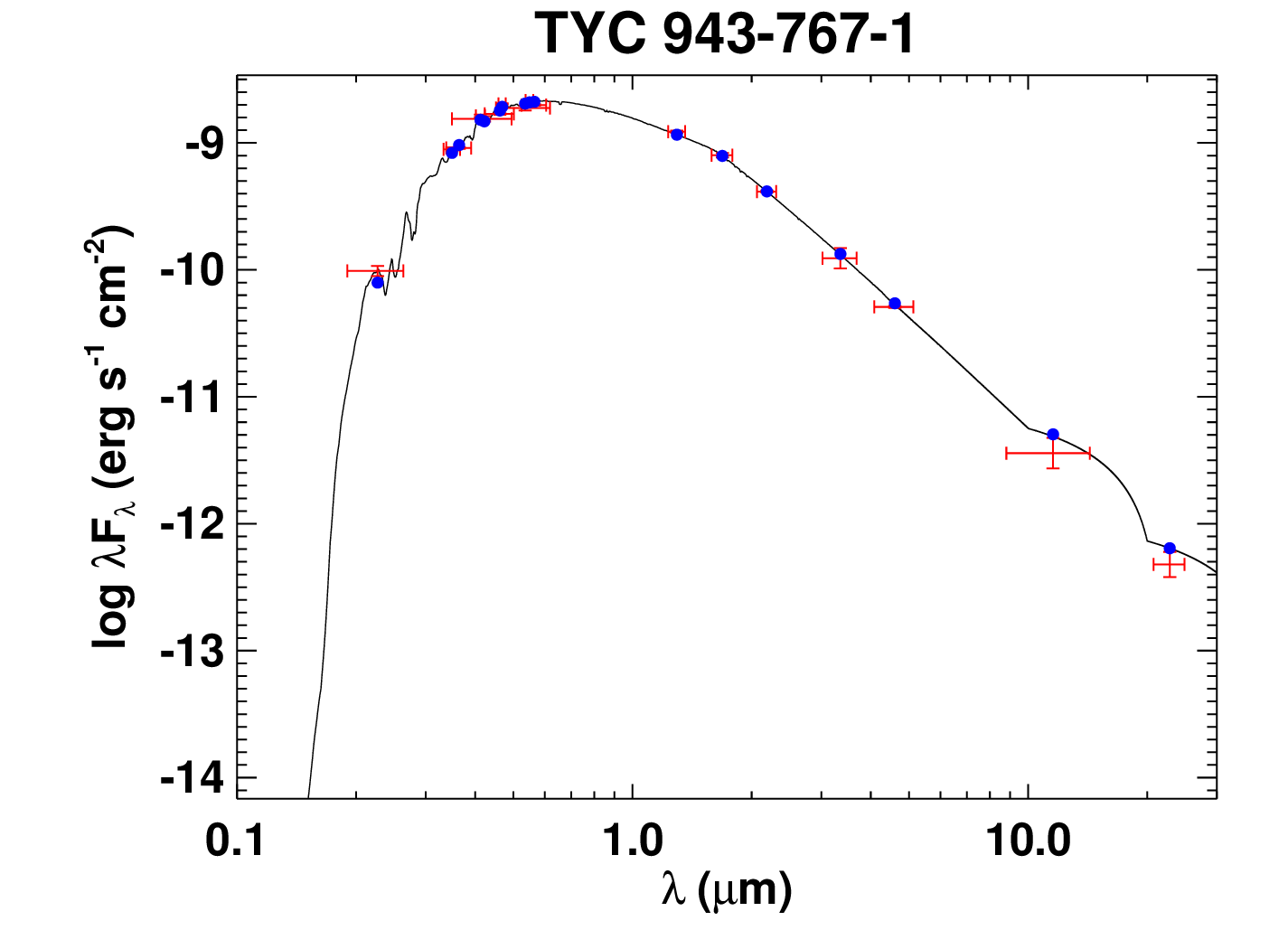}
\begin{center}
\includegraphics[width=0.333\linewidth]{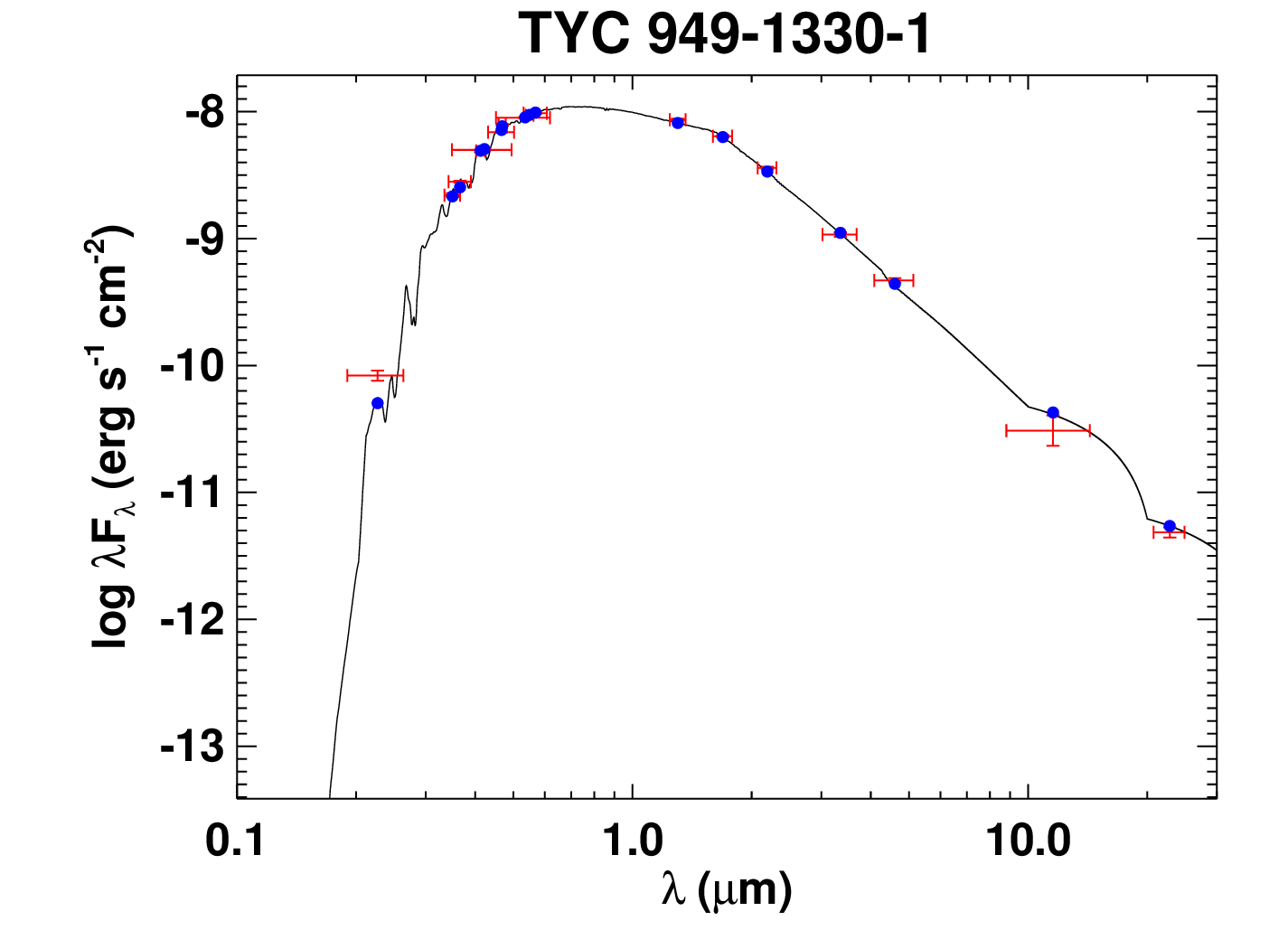}
\end{center}
\caption{\label{fig:seds20} All labels, lines, symbols, and colors as in Figure \ref{fig:seds}.}
\end{figure*}
\end{document}